\documentclass[twoside,12pt]{report} 
\usepackage{currvita}
\usepackage{picins}
\usepackage{a4} 
\usepackage{amsmath,amssymb} 
\usepackage{epsfig} 
\usepackage{setspace}          
\usepackage{fancyhdr}          
\usepackage{indentfirst}       
\usepackage{epsf,graphicx}     
\usepackage{ctable} 
\usepackage{wrapfig} 
\usepackage{rotate}
\usepackage[linktocpage,colorlinks]{hyperref} 
\usepackage[german,english]{babel}

\newcommand{\eq}[1]{\begin{align} #1 \end{align}} 
 
\begin{document} 
 
\pagestyle{fancy}\pagenumbering{arabic}\setcounter{page}{1} 
 
\thispagestyle{empty} 

\begin{center} 
\null\vspace*{1cm} 
{\Large\bf Statistical Fluctuations and Correlations\\
           in Hadronic Equilibrium Systems\\}%
  \vspace{1.5cm}  
  {\large  
  Dissertation\\ 
  zur Erlangung des Doktorgrades\\ 
  der Naturwissenschaften\\ 
  \vspace{1.5cm} 
  vorgelegt beim Fachbereich Physik\\ 
  der Johann Wolfgang Goethe--Universit{\"a}t\\ in 
  Frankfurt am Main\\ 
  \vspace{5cm} 
  von\\ 
  Michael Hauer\\ 
  aus M\"unchen, Deutschland\\ 
  \vspace{1.5cm} 
  Frankfurt am Main\\ 
  2010\\ 
  \vspace{0.5cm} 
  (D30)} 
\end{center} 
 
\newpage 
\thispagestyle{empty} 
\vspace*{1cm} 
{\large 
\noindent vom Fachbereich Physik der\\ Johann Wolfgang Goethe--Universit{\"a}t, Frankfurt am Main,\\ als Dissertation angenommen.\\\par 
 
\vspace*{0.6\textheight} 
\noindent Dekan: Prof. Dr. Dirk-Hermann Rischke\\ \par 
\noindent Gutachter: PD Dr. Elena Bratkovskaya\\ \par 
\noindent Datum der Disputation: 17.~6.~2010\\}

\selectlanguage{german}
\phantomsection
\addcontentsline{toc}{chapter}{Zusammenfassung}

\chapter*{Zusammenfassung}

\vspace{-0.5cm}
Diese Arbeit ist dem Studium von Fluktuationen und Korrelationen zwischen extensiven Observablen in hadronischen 
Gleichgewichtssystemen gewidmet. 
Als extensive Observablen bezeichnet man Me{\ss}gr\"o{\ss}en, die Aufschlu{\ss} \"uber die "`Gr\"o{\ss}e"' des zu 
beobachtenden Systems geben,  
wie etwa Teilchenanzahl oder Gesamtenergieinhalt, aber auch (elektrische) Nettoladung. 
Als Gleichgewichtssysteme bezeichnet man Systeme, die sich in einem Gleichgewichtszustand befinden. 
Also in einem Zustand, der mit (intensiven) Observablen zu beschreiben ist, die \"uber die Zeit betrachtet, 
ihren Wert nicht mehr \"andern. 
Hiermit k\"onnte die Temperatur oder die Ladungsdichte des Systems gemeint sein. 
Unter Hadronen versteht man Teilchen, denen eine gewisse Substruktur zugeschrieben werden kann, 
die also aus elementaren Bausteinen, den Quarks und Gluonen, aufgebaut sind. 
In der uns im Alltag vertrauten Welt, in der im Wesentlichen die Hadronenvertreter Neutronen und Protonen vorkommen, 
spielen diese elementaren Bausteinen jedoch keine eigene Rolle. 
Zusammen mit den Elektronen aus der Familie der Leptonen bilden diese beiden Hadronenarten die Atome und Molek\"ule 
aus den wir und unsere Umgebung bestehen.
 
Wenn zwei schwere Ionen (also von ihrer Elektronenh\"ulle befreite Atomkerne) kollidiert werden, entsteht f\"ur 
sehr kurze Zeit ein sehr hei{\ss}es und dichtes, m\"oglicherweise kollektives System. 
Die Beschreibung dieses Systems ausschlie{\ss}lich mit den uns vertrauten hadronischen (und leptonischen) 
Freiheitsgraden ist nun nicht mehr m\"oglich. 
Die Bausteine dieser Urmaterie sind die Quarks und die Feldteilchen der starken Wechselwirkung, die Gluonen. 
Dieser Zustand kann jedoch im Labor nicht direkt beobachtet werden. 
Nach einer kurzen, aber heftigen Expansionsphase wird die Teilchendichte zu gering, als da{ss} die Teilchen des Systems 
noch miteinander wechselwirken k\"onnten. 
Quarks und Gluonen "`hadronisieren"' zur\"uck zu den uns bekannten hadronischen Freiheitsgraden. 
Aus den Fluktuationen und Korrelationen zwischen den Teilchenanzahlen verschiedener Arten dieser Hadronen, so 
wird vermutet, k\"onnen Aussagen \"uber die dynamische Entwicklung, also etwa m\"ogliche Phasen\"uberg\"ange und 
die effektiven Freiheitsgrade des System gemacht werden. 
Aufgrund der offensichtlich kraftvollen Expansion des Systems kann hier wohl kaum von einem Gleichgewichtssystem 
ausgegangen werden.

Diese Arbeit besch\"aftigt sich mit trivialen Beitr\"agen zu Fluktuationen und Korrelationen, wie
sie durch globale Erhaltungss\"atze, eingeschr\"ankte Akzeptanz im Impulsraum, Resonanzzerfall, oder 
Quantenstatistik erzeugt werden.
Diese Effekte k\"onnen nicht vernachl\"assigt werden, da deren Auswirkungen von \"ahnlicher Gr\"o{\ss}enordnung sind, 
wie die, die aufgrund von Phasen\"uberg\"angen oder m\"oglicher kritischer Punkte im Phasendiagramm erwartet werden.   
Die statistischen Eigenschaften von idealen, relativistischen, hadronischen Gleichgewichtsensembles wurden 
untersucht. 
Neben den drei Standard-kanonischen Ensembles wurde auch eine Klasse von Ensembles mit endlichem thermodynamischen Bad 
eingef\"uhrt.  
Die Abh\"angkeit statistischer Eigenschaften auf (intensive) thermische Parameter wurde untersucht.
Es wurde argumentiert, dass das entstehende Bild zumindest qualitativ auf die Schwerionen-Physik anwendbar ist. 

In Kapitel \ref{chapter_clt} wurden gro{\ss}kanonische multivariate Verteilungen von extensiven Me{\ss}gr{\"o}{\ss}en, 
also Verteilungen von mehr als einer Zufallsvariable, durch Fourierintegration der gro{\ss}kanonische 
Zustandssumme erhalten. 
Eine analytische Entwicklungsmethode zur Berechnung der Verteilungsfunktionen von kanonischen sowie von mikrokanonischen 
Ensembles des idealen relativistischen Hadronen-Resonanz-Gases bei endlichem Volumen wurde vorgestellt. 
Die Einf\"uhrung der Temperatur in die mikrokanonischen Zustandssumme, und die von chemischen Potentialen in
die kanonische Zustandssumme, haben dazu gef\"uhrt, da{ss} die gro{\ss}kanonische Zustandssumme mit der
charakteristischen Funktion der damit verbundenen gro{\ss}kanonischen multivariaten Wahrscheinlichkeitsverteilung 
identifiziert werden konnte. 
Mikrokanonische und kanonische Teilchenanzahlverteilungen konnten somit durch bedingte gro{\ss}kanonische 
Wahrscheinlichkeitsverteilungen definiert werden. 
Unter bedingter Wahrscheinlichkeit versteht man die Wahrscheinlichkeit in einem gro{\ss}kanonischen Ensemble etwa 
eine bestimmte Teilchenanzahl zu beobachten, w\"ahrend andere extensive Me{\ss}gr\"o{\ss}en, wie etwa globale Ladung oder 
Gesamtenergie, als fest angenommen werden. 

In Kapitel \ref{chapter_montecarlo} wurden multivariate Verteilungen von extensiven Me{\ss}gr{\"o}{\ss}en 
f\"ur Systeme mit endlichem - an Stelle von einem unendlichem - thermodynamischen Bad eingef\"uhrt. 
Hierzu wurde ein mikrokanonisches System konzeptionell in zwei Subsysteme aufgeteilt. 
F\"ur diese Subsysteme wurde weiter angenommen, da{ss}  sie im thermodynamischen Gleichgewicht zueinander stehen. 
Des weiteren sollen sie gemeinsame Energie-, Impuls- und Ladungserhaltung respektieren. 
Teilchen k\"onnen nur in einem der beiden Teilsysteme gemessen werden, w\"ahrend das andere Teilsystem als 
thermodynamisches Bad fungiert. 
Wird die Gr\"o{\ss}e des ersten Teilsystems nun als fest angenommen, w\"ahrend die Gr\"o{\ss}e des zweiten variiert 
werden kann, so kann man die Abh\"angigkeit der statistischen Eigenschaften eines Ensembles von dem 
beobachtbaren Anteil des System ermitteln. 
Das hei{\ss}t, man untersucht deren Sensitivit\"at auf die Anwendung von globalen Erhaltungss\"atzen. 

Die erzeugten Ensembles sind thermodynamisch \"aquivalent im dem Sinne, da{ss} Mittelwerte extensiver Observablen 
im beobachteten Teilsystem unver\"andert bleiben, wenn die Gr\"o{\ss}e des thermodynamischen Bades variiert wird, 
sofern das kombinierte System hinreichend gro{\ss} ist. 
Die drei Standard-kanonischen Ensembles bleiben dabei spezielle Idealisierungen von physikalischen Systemen. 
Die allgemeineren Ensembles mit endlichem thermodynamischem Bad sollten daher ebenfalls von ph\"anomenologischem 
oder konzeptionellem Interesse sein.
Diese ersten beiden Kapitel bilden somit die mathematische Grundlage f\"ur die analytischen Berechnungen und 
Monte Carlo Simulationen, die f\"ur diese Arbeit durchgef\"uhrt wurden.

Die Analyse von Hadronen-Resonanz-Gas-Ereignissen beginnt mit dem Studium des gro{\ss}kanonischen Ensembles in 
Kapitel \ref{chapter_gce}. 
Das gro{\ss}kanonische Ensemble gilt als das am leichtesten zug\"angliche unter den Standard-kanonischen Ensembles. 
Aufgrund der Annahme eines unendlichen thermodynamischen Bades, sind die Besetzungszahlen in den einzelnen 
Impulszust\"anden der Teilchen miteinander unkorreliert. 
Somit erscheinen auch die Teilchenanzahlen von je zwei verschiedenen Gruppen von Teilchensorten miteinander unkorreliert. 
Aufgrund der Annahme von unkorrelierten Besetzungszahlen, ergibt sich, da{\ss} 
alle extensiven Me{\ss}gr{\"o}{\ss}en, mit Ausnahme des Volumens von Ereignis zu Ereignis (oder von Mikrozustand zu 
Mikrozustand) variieren. 
Der Energieinhalt und die Teilchenanzahl des Systems sind somit stark miteinander korreliert, 
w\"ahrend die durchschnittliche Energie pro Teilchen mit der Teilchenanzahl unkorreliert ist. 
Die elektrische Ladung, Baryonenzahl und Seltsamkeit des Systems sind korreliert mit der Hadronenanzahl, 
weil einige Teilchenarten mehrere dieser Ladungen tragen. 
Verschiedene Teilchenarten haben verschiedene Quantenzahlkonfigurationen und folgen aufgrund ihrer Masse 
unterschiedlichen Impulsspektren. 
Die Korrelation zwischen Baryonenzahl und Seltsamkeit im beobachteten Subsystem h\"angt dann, wie die 
Korrelation von Energie und Impuls oder von Energie und Teilchenanzahl, davon ab, 
welcher Teil des Impulsspektrums der Messung zug\"anglich ist.

In Kapitel \ref{chapter_extrapol} wurden multivariate Wahrscheinlichkeitsverteilungen von 
extensiven Me{\ss}gr{\"o}{\ss}en zu ihren mikrokanonischen Grenzwert extrapoliert. 
Zu diesem Zweck wurden iterativ Stichproben von Ereignissen erzeugt, und f\"ur abnehmende Gr\"o{\ss}e des thermodynamischen 
Bades analysiert. 
Die Verteilungsfunktion der extensiven Me{\ss}gr{\"o}{\ss}en, die gewichtet wurden, konvergiert zu einer $\delta$-Funktion, 
w\"ahrend die Positionen der Mittelwerte konstant geblieben sind. 
W\"ahrend der Transversalimpuls pro Teilchen und die Teilchenanzahl im gro{\ss}kanonischen Ensemble 
noch unkorreliert waren, so gilt diese Annahme nicht mehr f\"ur Systeme mit endlichem W\"armebad. 
Durch die sukzessive Konzentration auf Ereignisse in der unmittelbaren N\"ahe eines gew\"ahlten 
Gleichgewichtswertes zeigen sich die Auswirkungen globaler Erhaltungss\"atze. 
Dies verl\"auft in einer systematischen Weise, so da{ss} die Extrapolation von Observablen von ihren 
gro{\ss}kanonischen zu ihren mikrokanonischen Grenzwerten m\"oglich wurde. 
Ein Nachteil ist, dass die statistische Unsicherheit, verbunden mit endlichen Stichproben, w\"achst, wenn man sich dem 
(mikro-)kanonischen Grenzwerten n\"ahert.

In Kapitel \ref{chapter_multfluc_hrg} wurden Teilchenanzahlfluktuationen und Korrelationen f\"ur ein neutrales 
Hadronen-Resonanz-Gas mit begrenzter Teilchenakzeptanz im Impulsraum unter Ber\"ucksichtigung der Auswirkungen von 
Resonanzzerf\"allen untersucht. 
Die Extrapolationsmethode wurde angewandt, um kanonische und mikrokanonische Grenzwerte von Beobachtunggr\"o{\ss}en 
zu ermittlen. 
Ein Vergleich mit analytischen, asymptotischen L\"osungen f\"ur primordiale Verteilungen in begrenzter 
Teilchenakzeptanz zeigen eine sehr gute \"Ubereinstimmung der beiden Methoden. 
Je gr\"o{\ss}er die Anzahl der zu erhaltenen extensiven Me{\ss}gr{\"o}{\ss}en, desto gr\"o{\ss}er ist auch die 
statistische Unsicherheit, die mit endlichen Stichproben von Ereignissen verbunden ist. 
Mikrokanonische Effekte jedoch werden durch den Monte Carlo Ansatz korrekt wiedergegeben. 
Werden Erhaltungss\"atze eingeschaltet, so werden Teilchenanzahlfluktuationen und Korrelationen modifiziert. 
Impulsraumeffekte bei Teilchenanzahlfluktuationen und Korrelationen ergeben sich aufgrund von 
Erhaltungss\"atzen. 
F\"ur ein ideales, primordiales, gro{\ss}kanonisches Ensemble in der Boltzmann-N\"aherung (der Ausgangspunkt), 
sind die Teilchenanzahlverteilungen unkorrelierte Poissonverteilungen, unabh\"angig von der gew\"ahlten Akzeptanz 
im Impulsraum, da davon ausgegangen wurde, da{ss} Teilchen unabh\"angig voneinander produziert werden. 
Das Erfordernis der Energie-, Impuls- und Ladungserhaltungss\"atze f\"uhrt zu unterdr\"uckten Fluktuationen 
und verst\"arkten Korrelationen zwischen den Teilchenanzahlen von zwei verschiedenen Gruppen von Teilchen am 
"`oberen"' Ende des Impulspektrums, im Vergleich zum "`unteren"' Ende des
Impulsspektrum, vorausgesetzt, einen nicht vernachl\"assigbarer Teil eines isolierten Systems wird beobachtet. 
Resonanzzerf\"alle \"anderen diese Trends nicht.

Kapitel \ref{chapter_multfluc_pgas} ist dem mikrokanonischen Ensemble gewidmet. 
Ein vereinfachtes physikalisches System wurde gew\"ahlt, um eine einfachere Diskussion zu erm\"oglichen. 
Aufgrund der zur Verf\"ugung stehenden analytischen L\"osungen, konnten Fermi-Dirac und Bose-Einstein-Effekte 
in die Analyse miteinbezogen werden. 
Bose-Einstein-Verst\"arkung und Fermi-Dirac-Unterdr\"uckung der Teilchenanzahlfluktuationen sind besonders stark in 
Impulsraumsegmenten, in denen die Besetzungszahlen in den einzelnen Impulszust\"anden gro{\ss} sind. 
Dieser Effekt ist deutlich st\"arker als der, der in fr\"uheren Berechnungen von Fermi-Dirac und 
Bose-Einstein Effekten auf Teilchenanzahlfluktuationen im vollen Phasenraum ermittelt wurde. 
F\"ur Systeme in kollektiver Bewegung wurde festgestellt, da{ss} die Rolle, die kinematische Erhaltungss\"atze 
spielen, besonders wichtig ist. 
Fluktuation und Korrelation von Messgr\"o{\ss}en sind Lorentz-invariant, sofern Impulserhaltung entlang 
der kollektiven Bewegungsrichtung ber\"ucksichtigt wird. 
Schlie{\ss}lich wurde festgestellt, da{\ss} auch im thermodynamischen Grenzwert Langstreckenkorrelationen zwischen 
getrennten Regionen im Impulsraum verbleiben. 
Teilchenanzahlen in unterschiedlichen Intervallen in der Rapidit\"at, im transversalen Impulsraum, oder im Azimut, 
haben einen nicht verschwindenden Korrelationskoeffizienten.

In Kapitel \ref{chapter_phasediagram} wurde das Phasendiagramm f\"ur das Hadronen-Resonanz-Gas-Modell in seiner 
Abh\"angigkeit von Temperatur und baryonchemischem Potential untersucht. 
Gro{\ss}kanonische Ladungskorrelationen und Fluktuationen sind unterschiedlich in den vier verschiedenen Ecken des 
Phasendiagramms. 
Wie in der Akzeptanzanalyse ist die Korrelation zwischen zwei Ladungen stark, wenn Teilchen, die beide Ladungen tragen,
reichlich vorhanden sind. 
Bei niedriger Temperatur und niedrigem baryonchemischem Potential dominieren Mesonen \"uber Baryonen, bei hoher 
Temperatur und hohem baryonchemischem Potential hingegen tragen Baryonen einen Gro{\ss}teil der Gesamtentropie des Systems. 
Diesem Zusammenhang entsprechend verhalten sich Fluktuationen und Korrelationen der Ladungen systematisch.   
Auch Teilchenanzahlfluktuationen und Korrelationen im kanonischen und mikrokanonischen Ensembles tragen dem Rechnung. 
Der Einfluss eines bestimmten Erhaltungsatzes auf Teilchenanzahlfluktuationen ist stark, wenn die Teilchen der 
analysierten Arten reichlich vorhanden sind, und somit einen wesentlichen Teil der Gesamtladung und -energie tragen. 
Ein Vergleich zwischen Ensembles mit und ohne Energie-, Ladungs- oder Impulserhaltung zeigt subtile Unterschiede, 
die unter anderem am Beispiel von Resonanzzerf\"allen in kanonischen und mikrokanonischen Ensembles herausgearbeitet 
wurden. 
Resultierende primordialen Teilchenanzahlkorrelationen ergeben sich nicht aufgrund von lokalen Interaktionen 
zwischen den Bestandteilen, sondern aufgrund von global implementierten Erhaltungss\"atzen f\"ur Energie und Ladungen.

In Kapitel \ref{chapter_freezeoutline} wurden Teilchenanzahlfluktuationen und Korrelationen f\"ur 
thermische Parameter analysiert, die der chemischen Freeze-out Linie zentraler Schwerionenkollisionen folgen. 
Modellparameter wurden von fr\"uheren Hadronen-Resonanz-Gas-Modell-Vergleichen mit experimentellen Messungen 
von mittleren Hadronproduktionsraten \"ubernommen.
Diese Messungen stellen eine Verbindung zwischen den experimentellen Kontrollparametern Kollisionsenergie und 
Gr\"o{\ss}e der zu kollidierenden Ionen, und der Region im Phasendiagramm, die durch das Experiment sondiert wird, her. 
Einem ersten Vergleich mit experimentellen Daten zufolge ergibt sich eine gute \"Ubereinstimmung mit 
Hadronen-Resonanz-Gas-Berechnungen. 
Insbesondere die mikrokanonischen Formulierung des Modells scheint qualitative und quantitative Eigenschaften 
der Daten gut zu reproduzieren.

Die hier vorgestellten Berechnungen beschreiben qualitative Auswirkungen auf Teilchenanzahlfluktuationen und 
Korrelationen. 
Diese Auswirkungen ergeben sich allein aus den Prinzipien der statistischen Mechanik und aufgrund der Anwendung 
globaler Erhaltungss\"atze. 
Es wird interessant sein, zu sehen, ob diese qualitativen Auswirkungen in weiteren experimentellen Messungen 
von Fluktuationen und Korrelationen sichtbar werden. 
Wenn ja, k\"onnten diese Effekte wohl von \"ahnlicher Gr\"o{\ss}enordnung sein, wie die Signale "`neuer"' Physik. 
Das Trennen von dynamischer Evolution des Systems und statischen Erhaltungss\"atzen ist dann eine wichtige, 
wenn auch nicht-triviale Aufgabe. 
Gleichgewichtskorrelationen sind Restkorrelationen, die zur\"uck bleiben, nachdem das System seine 
(Entwicklungs) Geschichte "`vergessen"' hat. 
Wenn das System "`beobachtet"' oder "`gemessen"' wird, bevor es einen Gleichgewichtszustand erreicht hat, dann 
werden Korrelationen aufgrund der Anfangskonfiguration verbleiben. 
Aber selbst wenn sich das System w\"ahrend seiner Lebensdauer fern von jedem Gleichgewichtspunkt bewegt, so spielen
Korrelationen aufgrund der globalen Erhaltungss\"atze eine wichtige Rolle.


\selectlanguage{english}
\phantomsection
\addcontentsline{toc}{chapter}{Acknowledgments}

\chapter*{Acknowledgments}

First of all I would like to thank my supervisor Professor Elena Bratkovskaya for her patience and guidance.
Most of all for giving me the freedom to pursue every once in a while my own interest.
I am also grateful to Professor Mark Gorenstein and Professor Marek Ga\'zdzicki for constructive collaboration
and many stimulating discussions on, amongst other things, the philosophy of doing science.\\

I would like to acknowledge my scholarship of the Helmholtz Research School at the University of Frankfurt.
In particular I would like to thank our course administrator and coordinator Henner Busching for dedicating
so much of his energy to this first generation of Helmholtz-PhD students.\\

I would like to express my gratitude to the Professors Francesco Becattini, Marcus Bleicher, Christoph Blume, 
Wojciech Broniowski, Jean Cleymans, Krzysztof Redlich, and Peter Steinberg for always having an open ear or two.\\

I also would like to thank my collaborators and colleagues Viktor Begun, Volodymir Konchakovski, Benjamin Lungwitz, 
and Oleg Moroz, and for an enjoyable and fruitful time together.\\

Special thanks here goes to my friends Lorenzo Ferroni, Stephane H\"aussler, Torsten Kolleger, Jaakko Manninen, 
Michael Mitrovski, Sophie Nahrwold, Irina Sagert, Tim Schuster, Giorgio Torrieri, 
and Sascha Vogel for invaluable feedback, opinions, and (if applicable) beer drinking moments.\\

I am also grateful to Bruce Becker, Gareth de Vaux, Spencer Wheaton, and many many more of, at, and around the physics 
department of the University of Cape Town, my second home base. 
In particular I would like to gratefully acknowledge the financial support I have received for two trips 
to South Africa, which have been the most productive days of my PhD.\\
 
To family and friends, and all of you who have shared the one or the other cigarette with me. 

\newpage

\renewcommand{\chaptermark}[1]{\markboth{\chaptername \ \thechapter:\,\ #1}{}} 
\renewcommand{\sectionmark}[1]{\markright{\thesection\,\ #1}} 
 
\addtolength{\headheight}{3pt} 
\fancyhf{} 
\fancyhead[RE]{\sl\leftmark} 
\fancyhead[LO]{\sl\rightmark} 
\fancyhead[RO,LE]{\rm\thepage} 
 
\fancypagestyle{plain}{ 
	\addtolength{\headheight}{3pt} 
	\fancyhf{} 
	\fancyhead[RE]{\sl\leftmark} 
	\fancyhead[RO,LE]{\rm\thepage} 
} 
 
\onehalfspacing 
 
\phantomsection 
\addcontentsline{toc}{chapter}{Contents} 
\tableofcontents 
 
 
\chapter{Introduction}
\label{chapter_intro}
This thesis is dedictaed to the study of fluctuation and correlation observables of hadronic equilibrium systems. 
The statistical hadronization model of high energy physics, in its ideal, i.e. non-interacting, gas approximation 
will be investigated in different ensemble formulations. The hypothesis of thermal and chemical equilibrium in high 
energy interaction will be tested against qualitative and quantitative predictions. \\

The statistical hadronization model, first introduced by Fermi~\cite{Fermi:1950fix_1} and 
Hagedorn~\cite{Hagedorn:1970fix_2}, has been surprisingly successful during the last couple of decades in describing 
fundamental properties of systems created in heavy ion collisions, cosmic rays, and elementary particle reactions. 
In the context of heavy ion collisions it has been applied to an extensive set of data on hadron production, ranging 
form the center of mass energies of the experiments at the SIS~\cite{Cleymans:1997sw,Cleymans:1998yb,Averbeck:2000sn}, 
AGS~\cite{BraunMunzinger:1994xr}, SPS~\cite{BraunMunzinger:1995bp,BraunMunzinger:1999qy,Becattini:2003wp},and most 
recently, RHIC~\cite{Adams:2005dq} facilities. Model predictions for the upcoming LHC and future 
FAIR~\cite{Kraus:2007bg,Andronic:2007vh,Andronic:2007vg,Rafelski:2008an,Rafelski:2005jc,Becattini:2008yn,Andronic:2008gm} 
experiments largely follow these trends. A systematic evolution of thermodynamic parameters, as collision energy (and 
size of colliding ions) is changed~\cite{Cleymans:2005xv,Cleymans:1999st,Cleymans:1998fq,Becattini:2005xt,Andronic:2005yp}, 
has allowed to establish the `chemical freeze-out line', which is now a commonly accepted ingredient in the phase 
diagram of strongly interacting matter. 
More controversially this model has also been applied to a range of elementary collision 
systems~\cite{Becattini:1995if,Becattini:1997rv,Becattini:2001fg}, where only few particles are produced, and the picture 
of a gas of hadrons can hardly be suitable. The remarkable ability of the statistical model to explain these data has 
lead to the 
suggestion~\cite{Becattini:1995if,Becattini:1997rv,Becattini:2001fg,Stock:1999hm,Hagedorn:1994fix_1,Hagedorn:1970fix_1}, 
that thermal (or phase space dominated) particle production is a general property of the hadronization process itself, 
rather than the result of a long sequence of microscopic interactions. This thesis will not argue about possible physical 
interpretations~\cite{Becattini:2004td} of the partition function of statistical mechanics. It is, however, noted that 
in order to apply a semi-classical approximation a volume of $\mathcal{O}(10$~fm$^3)$ seems to be 
sufficient~\cite{Becattini:2007nx}.\\

Somewhere above this chemical freeze-out line in the phase diagram a phase transition from hadronic degrees of freedom 
to a phase of deconfined quarks and gluons, generally termed the quark-gluon plasma is conjectured; more specifically, 
a first order phase transition at low temperature and high baryon chemical potential, and a cross-over at high 
temperature and low baryon chemical potential. In between, a second order endpoint or a critical point might emerge. 
For recent reviews see~\cite{Karsch:2001nf,Fodor:2001pe,Fukushima:2008wg,Fodor:2004nz,Hatta:2002sj,deForcrand:2003hx,Schaefer:2006ds,Bowman:2008kc}.
One of the answers still outstanding in high energy physics is then the one of a possible formation of a deconfined state 
of matter, and the nature of the transition between phases. The growing interest in the study of event-by-event 
fluctuations in strong interactions is, thus, motivated by expectations of anomalies in the vicinity of the onset of 
deconfinement~\cite{Jeon:2000wg,Asakawa:2000wh,Gazdzicki:2003bb,Gorenstein:2003hk,Jeon:2003gk} and in the case when the 
expanding system goes through the transition line between quark-gluon plasma and hadron 
gas~\cite{Mishustin:1998eq,Mishustin:2001fix_1,Heiselberg:2000ti}. In particular, a critical point of strongly 
interacting matter may be accompanied by a characteristic power-law pattern in 
fluctuations~\cite{Stephanov:1998dy,Stephanov:1999zu,Stephanov:2004xs}. Recently, it has been suggested that 
correlations across a large interval of rapidity could also arise from color glass condensate initial 
conditions~\cite{Armesto:2006bv,Lappi:2006fp}.\\

In recent years a wide range of experimental measurements of fluctuations of particle 
multiplicities~\cite{Afanasev:2000fu,Aggarwal:2001aa,Adams:2003st,Roland:2004pu,Chai:2005fj,Rybczynski:2004yw,Lungwitz:2006cx,Alt:2007jq}, 
transverse momenta~\cite{Appelshauser:1999ft,Adamova:2003pz,Anticic:2003fd,Adler:2003xq,Adams:2003uw} 
and multiplicity correlations in rapidity~\cite{Tarnowsky:2008am,Adler:2007fj,Back:2006id}   
have been reported, leading to a lively discussion regarding their physical interpretation. 
The most promising region in the phase diagram for observation of critical phenomena seems to be accessible to
the SPS accelerator~\cite{Afanasiev:2002mx,Gazdzicki:2004ef,Alt:2007fe}. 
A new SPS scan program~\cite{Gazdzicki:2006fy} for different ion sizes as well as center-of-mass energies 
has been proposed to study strongly interacting systems at different energy and net-baryon densities, and life times.
Also a second scan program~\cite{Stephans:2006tg}, lowering the RHIC colliding beam energy to probe the same domain
of the phase diagram, is under discussion.
This should be as well the main motivation for further investigation of properties of statistical ensembles. \\

Fluctuations of, and correlations between, various experimental observables are believed to have the potential to 
reveal new physics. They are amongst the most promising candidates suggested to be suitable for signaling the 
formation of new states of matter, and transitions between them. For recent reviews here 
see~\cite{Gazdzicki:2003bb,Gorenstein:2003hk,Jeon:2003gk,Mishustin:1998eq,Heiselberg:2000ti,Stephanov:1998dy,Stephanov:1999zu,Stephanov:2004xs,Stephanov:2004wx,Koch:2008ia}. 
In particular, multiplicity and charge fluctuations have been proposed to be a good discriminating tool between 
quark-gluon plasma and hadron gas~\cite{Jeon:2000wg,Asakawa:2000wh}. Provided the signal survives the phase 
transition~\cite{Haussler:2007un}, and subsequent evolution of the system. Hence, in order to properly  assess the 
discriminating power of such observables, one might firstly want to asses the magnitude of `trivial' physical effects, 
such as the ones induced by global conservation laws, quantum statistics, resonance decays, kinematical cuts, 
finite spatial extension, etc. The statistical properties of a sample of events are certainly not solely determined 
by critical phenomena. To get a reliable indication of new physics, it is therefore important to note that most   
fluctuation and correlation observables are also sensitive to some `baseline' contributions   
that, nevertheless, can have non-trivial behavior.  \\

A rather general observation regarding statistical samples can be made:
The statistical properties of a sample of events depend on the rules chosen to
select this sample from an even larger sample, and on the degree of completeness of
the information available about the sample.
In the context of heavy ion collision physics these two aspects roughly translate into
centrality class construction and particle acceptance in momentum space. 

Centrality selection is discussed first. Two heavy ions collide with relativistic momenta.
Being extended objects, they can do so in many different ways. Roughly the following rule 
should apply (in the average sense): The larger the interaction region, the more particles 
of each particle species are produced. The problem is now that the initial state of a collision 
cannot be observed directly. All that can be observed is its final state. From this,  
one can then infer the likelihood of a certain initial state. Yet, each single possible final 
state observable will generally suggest a slightly different initial state. Hence, the need to
average over centrality classes. The problem is then, within any such centrality class will be
events with rather different initial states, altering the true correlation between two observables.
Results will depend, in general, on which trigger was chosen to construct the centrality class or sample of 
events~\cite{Konchakovski:2008cf,Bzdak:2009xq}.
Centrality selection, although being a crucial experimental issue, 
is foreign to the statistical hadronization model, and will not be considered in this thesis.

The second problem is particle acceptance. This term should denote particle identification and momentum measurement.
Centrality selection is now ignored and a perfectly prepared initial state is assumed. Two limiting cases can be explored. 
The first one being the ideal detector. All final state particles are observed.  Any
correlation is measured to any degree. The opposite limit would be a very bad detector. Capable of 
only detecting a particle every once in a while. This detector could surely measure the ratios 
of the occurrence of particles of different species. But it would be completely unable to inform one on 
how particle species within one event are correlated. Any realistic detector is in between these limits.
And, hence, will disagree with either limit, in basically any observable. 
The statistical hadronization model provides a natural framework for such a discussion.\\

Strictly speaking the model, at least in the from presented here, is not directly comparable to data of 
heavy ion collisions. It does certainly no justice to the complexity and the dynamical evolution of the system it seeks to 
describe. The model exhibits no collective flow, or expansion. The size (or volume) of the system is assumed
to be the same for all events. Furthermore, no distinction is made between chemical and kinetic freeze-out.
On the other hand, the medium created during the collisions of two heavy ions 
is rapidly expanding, while the initial state can only be accessed indirectly. 
Yet, essentially the model seeks to describe properties of many-particle systems, and, here in this thesis, 
their statistical properties. The statistical hadronization model, might not be a bad place to start such a task. \\

The purpose of this thesis is the calculation of `baseline' contributions, on top of which one hopes to find 
unambiguous signals of a phase transition~\cite{Mishustin:1998eq,Mishustin:2001fix_1,Heiselberg:2000ti}, a critical 
point~\cite{Stephanov:1998dy,Stephanov:1999zu,Stephanov:2004xs}, or thermal/chemical (local or global) 
non-equilibrium~\cite{Ko:2000vp,Fochler:2006et,Torrieri:2001ue}.
I.e. to study these baseline correlations in a limiting case: that of a thermalized relativistic ideal (no inter-particle 
interactions) quantum gas, for which we want to assess the importance of globally applied conservation laws, quantum 
statistics, resonance decays, and kinematical cuts for fluctuation and correlation observables. In this case, all 
observables are calculable simply using statistical mechanics techniques. Such an approach has a long and distinguished 
history of calculating particle multiplicities in hadronic 
collisions~\cite{Fermi:1950fix_1,Becattini:2003wp,Pomeranchuk:1951fix_1,Landau:1953fix_1,Hagedorn:1965fix_1,BraunMunzinger:2001ip,Cleymans:2006qe,Letessier:2002fix_1,Cleymans:2004pp,Torrieri:2004zz,Torrieri:2006xi,Wheaton:2004qb}.
Given its success in describing experimentally measured average hadron yields, and its ability to reproduce low 
temperature lattice susceptibilities~\cite{Cheng:2008zh}, the question arises as to whether fluctuation and correlation 
observables also follow its main line. Critical phenomena (and many more), however, remain beyond the present study.\\

Conventionally in statistical mechanics three standard ensembles are discussed; the micro canonical ensemble (MCE), the 
canonical ensemble (CE), and the grand canonical ensemble (GCE). In the MCE\footnote{The term MCE is also often applied 
to ensembles with energy but not momentum conservation.} one considers an ensemble of micro states with exactly fixed 
values of extensive conserved quantities (energy, momentum, electric charge, etc.), with `a priori equal probabilities` 
of all micro states (see e.g.~\cite{Patriha:1996fix_1}). The CE introduces the concept of temperature by introduction of 
an infinite thermal bath, which can exchange energy (and momentum) with the system. The GCE introduces further chemical 
potentials by attaching the system under consideration to an infinite charge bath\footnote{Note that a system with many 
charges can have some charges described via the CE and others via the GCE.}. Only if the experimentally accessible system 
is just a small fraction of the total, and all parts have had the opportunity to mutually equilibrate, can the 
appropriate ensemble be the grand canonical ensemble.  \\

The main focus of the past study of the statistical hadronization model has been on the mean multiplicities of 
produced hadrons. However, there is a qualitative difference in the properties of mean values and event-by-event 
fluctuations about these mean values in statistical mechanics. In the case of the ensemble averages, results obtained in 
the GCE, CE, and MCE approach each other in the large volume limit. One refers here to as the thermodynamical equivalence 
of statistical ensembles. However, even in this limit, these ensembles have different properties with respect to 
fluctuations and correlations~\cite{Begun:2004zb,Begun:2004gs,Begun:2004pk}. In the MCE, energy and charge are exactly fixed. 
In the CE, charge remains fixed, while energy is allowed to fluctuate about some average value. Finally, in the GCE the 
requirement of exact charge conservation is dropped, too. One may also consider isobaric 
ensembles~\cite{Gorenstein:2008nr}, or even more general `extended Gaussian 
ensembles`~\cite{Challa:1988zz,Gorenstein:2008th}. In previous 
articles~\cite{Begun:2004zb,Begun:2004gs,Begun:2004pk,Gorenstein:2008nr,Gorenstein:2008th,Begun:2005ah,Begun:2005iv,Begun:2006jf,Begun:2006gj,Begun:2006uu,Gorenstein:2007ep,Hauer:2007ju,Hauer:2007im}  
it was shown that these differences mean that, in particular, multiplicity fluctuations and correlations are 
ultimately ensemble specific. I.e. depend on how the system under investigation is prepared.\\

The observation that fluctuations of certain quantities are constraint by conservation laws is not new, nor restricted 
to heavy ion physics. The Italian physicist Ugo Fano wrote back in 1947 in the abstract of his article~\cite{Fano:1947zz} 
`Ionization Yield of Radiations. II. The Fluctuations of the Number of Ions`:\\

\parbox{13cm}{ 
\it The ionization produced by individual fast charged particles is frequently used as a measure of their initial energy; 
fluctuation effects set a theoretical limit to the accuracy of this method. Formulas are derived here to estimate the 
statistical fluctuations of the number of ions produced by constant amounts of radiation energy. The variance of the 
number of ionizations is found to be two or three times smaller than if this number were governed by a Poisson 
distribution. An improved understanding is gained of the statistical treatment of fluctuation phenomena.}
\vspace{0.7cm}

So, it has been well understood, already some 60 years ago, that global conservation laws affect the statistical 
properties of a system. Similar observation was made in the fluctuation of the number of atoms forming a Bose-Einstein 
condensate. The `grand canonical fluctuation catastrophe' at the Bose-Einstein condensation point~\cite{Ziff:1977} 
is avoided by a micro canonical formulation~\cite{PhysRevE.54.3495}. \\

 

This thesis is organized as follows: 
In Chapter~\ref{chapter_clt} grand canonical joint distributions of extensive quantities are obtained by Fourier 
integration of the grand canonical partition function. An analytical expansion method for calculation of distributions 
at finite volume for the canonical as well as the micro canonical ensembles of the ideal relativistic hadron resonance gas
will be presented.
In Chapter~\ref{chapter_montecarlo} joint distributions of extensive quantities are then considered for statistical 
systems with finite, rather than infinite, thermodynamic bath. 
These two chapters form the mathematical basis for analytical calculations and Monte Carlo simulations performed in 
this thesis.
The analysis of hadron resonance gas events will start with the study of the grand canonical ensemble in 
Chapter~\ref{chapter_gce}. The grand canonical ensemble is considered to be the most accessible amongst the 
standard canonical ensembles. 
In Chapter~\ref{chapter_extrapol} joint distributions of extensive quantities will be extrapolated to the micro canonical 
limit.
In Chapter~\ref{chapter_multfluc_hrg} multiplicity fluctuations and correlations are studied for a neutral and static 
hadron resonance gas with limited acceptance in momentum space. The effects of resonance decay will be considered,
and the extrapolation scheme will be applied to obtain canonical and micro canonical ensemble limits. 
Chapter~\ref{chapter_multfluc_pgas} is dedicated to the micro canonical ensemble itself. A simplified physical system is 
chosen to allow for smoother discussion. Owing to available analytical solutions, Fermi-Dirac and Bose-Einstein effects, 
as well as collective motion, are included into the analysis.
In Chapter~\ref{chapter_phasediagram} the temperature and baryon chemical potential phase diagram of the hadron 
resonance gas model is explored. 
Lastly, in Chapter~\ref{chapter_freezeoutline}, multiplicity fluctuations and correlations will be analyzed for thermal 
parameter sets following the chemical freeze-out line. A first comparison to available experimental data suggests good 
agreement with hadrons resonance gas calculations. 
A summary, Chapter~\ref{chapter_summary}, will close the thesis. Further technical details of the calculations and 
Monte Carlo simulations, are presented in the Appendix.

\chapter{Grand Canonical Partition Function}
\label{chapter_clt}
Aim of this chapter is to introduce a technique for calculation of grand canonical probability 
distributions\footnote{The term `distribution' is used for simplicity, rather than the more 
correct `probability density function`.} of various extensive quantities at finite volume. 
The method is based on Fourier analysis of the grand canonical partition function. 
Taylor expansion of the generating function is used to separate contributions to the partition function in
their power in volume. Laplace's asymptotic expansion is employed to show that any equilibrium distribution 
of multiplicity, charge, energy, etc. tends to a multivariate normal distribution  in the thermodynamic limit. 
Gram-Charlier expansion allows additionally for calculation of finite volume corrections. 
Analytical formulas facilitate inclusion of resonance decay (in full acceptance), or inclusion of 
finite acceptance effects (without resonance decay) directly into the system partition function. 
Multiplicity distributions in (micro) canonical ensembles can then be defined through conditional 
grand canonical distributions. 

The discussion of equilibrium systems with finite, rather than infinite, thermodynamic bath, 
in Chapter~\ref{chapter_montecarlo}, will provide a description for a Monte Carlo approach,
capable of assessing the effects resonance decay in finite acceptance.
These two chapters will prepare the mathematical framework for the following discussions of statistical 
fluctuations and correlations in hadronic equilibrium systems.

\section{Probability Distributions}
\label{sec_clt_probdist}

In textbooks on statistical mechanics (see e.g., Ref.\cite{Greiner:1997fix_1,Reichl:1998fix_1,Landau:2001fix_1}) 
often  first the~MCE is introduced, where exact conservation laws for energy-momentum and particle number are
imposed on a collection of micro states. Relaxing the constraints for energy and momentum constitutes the~CE, 
while allowing additionally particle number to fluctuate about some mean value introduces the~GCE. 
In a relativistic gas of hadrons  quantum numbers (charges), rather than particle numbers, will be the 
conserved quantities. 

Here it will prove to be of considerable advantage to start off with the~GCE formulation and 
imposing exact conservation laws thereafter.  Generally, the (micro) canonical partition  function is
obtained from the grand canonical one by  multiplication with Kronecker (or Dirac) $\delta$-functions which 
pick out a set of micro states consistent with a particular  conservation law. 
It is often  more economical to use Fourier representations of $\delta$-functions, rather than 
the $\delta$-function themselves.

The basic idea is to define the probability of finding the system in a state with a given number of 
particles~$N_A$ of some species~$A$ at some fixed value of conserved (electric) charge,~$Q$, i.e. the~CE 
distribution~$P_{ce}(N_A)$, in terms of the~GCE distributions,~$P_{gce}(N_A,Q)$ and~$P_{gce}(Q)$. 
In general one may write for the multiplicity distribution~$P_{ce} (N_A)$ of a~CE with conserved electric charge~$Q$:  
\begin{equation}   
P_{ce} (N_A)~=~ 
\frac{\textrm{number of all micro states with~$Q$ and~$N_A$}}{
      \textrm{number of all micro states with~$Q$ }}~.  
\end{equation}     
Likewise, one can write for the~CE joint multiplicity distribution $P_{ce} (N_A,N_B)$ of particle species~$A$ and~$B$:   
\begin{equation}   
P_{ce} (N_A,N_B)~=~ 
\frac{\textrm{number of all micro states with~$Q$,~$N_A$ and~$N_B$}}{
      \textrm{number of all micro states with~$Q$ }}~.    
\end{equation}   
The number of all micro states with electric charge~$Q$, and multiplicities~$N_A$ and~$N_B$ of a system 
with temperature~$T=\beta^{-1}$ and volume~$V$ is given by the~CE partition function~$Z(V,\beta,Q,N_A,N_B)$. 
Similarly,~$Z(V,\beta,Q)$ denotes the number of micro states with fixed electric charge~$Q$, but arbitrary 
multiplicities~$N_A$ and~$N_B$, for the same physical system.  

Throughout this thesis the following conventions are used:
Ensembles are to be identified by the arguments of their partition functions.
Extensive quantities, denoted by capital letters in the argument, are to be considered as exactly conserved.
Intensive quantities, denoted by small letters in the argument, indicate that the corresponding extensive 
quantity is conserved only in the average sense.
The partition function~$Z(V,E,Q)$ denotes a~MCE with conserved energy~$E$ and electric charge~$Q$.
The partition function~$Z(V,\beta,Q)$ belongs the a~CE with an infinite heat bath at 
temperature~$\beta^{-1}$. The energy content of the volume~$V$ fluctuates then about some mean value~$\langle E \rangle$. 
Lastly,~$Z(V,\beta,\mu)$, is the number of micro states available to a system with an infinite heat and charge
bath at temperature~$\beta^{-1}$ and chemical potential~$\mu$.
All constraints on the micro states are dropped, and a weighted average over all charge and energy configurations is taken.

Particle number is here included in the argument of partition functions, despite the fact that no (non-vanishing) 
particle number specific chemical potential is introduced. Fourier integrals associated with particle multiplicity 
are going to be solved along with those associated with conserved quantities.
In the next chapter, particle number is deligated to an index of the partition function, as its integrals do not 
need to be solved for the Monte Carlo approach.

The strategy to calculate joint multiplicity distributions could thus be the following 
(in principle also valid at finite volume):  
\begin{eqnarray}   
\label{P_one}  
P_{ce} (N_A,N_B) &=& \frac{Z(V,\beta,Q,N_A,N_B)}{Z(V,\beta,Q)}~, \\   
\label{P_two}  
&=& \frac{e^{Q\mu \beta}~Z(V,\beta,Q,N_A,N_B)}{Z(V,\beta,\mu)} ~   
\frac{Z(V,\beta,\mu)}{e^{Q \mu \beta}~Z(V,\beta,Q)}~,  \\    
\label{P_three}  
&=& P_{gce}(Q,N_A,N_B) ~P_{gce}^{-1}(Q)  
~=~ P_{gce}(N_A,N_B|Q) ~.  
\end{eqnarray}   
In order to get from Eq.(\ref{P_one}) to Eq.(\ref{P_three}) both canonical partition 
functions~$Z(V,\beta,Q,N_A,N_B)$ and~$Z(V,\beta,Q)$ are divided by their~GCE 
counterpart~$Z(V,\beta,\mu)$ and multiplied by the Boltzmann weight~$e^{Q \mu \beta}$.  
The first term on the right hand side of Eq.(\ref{P_two}) then equals the~GCE joint 
distribution~$P_{gce}(Q,N_A,N_B)$, while the second term is just the inverse of the~GCE charge 
distribution~$P_{gce}(Q)$. Their ratio is the (normalized)~GCE conditional distribution of particle 
multiplicities $N_A$ and~$N_B$ at fixed electric charge~$Q$,~$P_{gce}(N_A,N_B|Q)$, and equals the~CE 
distribution~$P_{ce}(N_A,N_B)$ at the same value of~$Q$. 
This result is independent of the choice of chemical potential~$\mu$.  
  
The problem of finding a solution, or a (large volume) approximation, to the~CE 
distribution~$P_{ce}(N_A,N_B)$ is now turned into the problem of finding a solution or approximation to 
the~GCE distribution of multiplicities~$N_A$ and~$N_B$, and charge~$Q$,~$P_{gce}(Q,N_A,N_B)$.   

It is worth noting that Eqs.(\ref{P_one}-\ref{P_three}) are as well the basis for any Monte Carlo 
approach~\cite{Becattini:2003ft,Becattini:2004rq}. A sampling distribution, usually taken from a 
Boltzmann~GCE system, is used to generate a~$\lbrace N_i\rbrace$-tuple of particle multiplicities of 
all species~$i$ considered. All `events' consistent with certain constraints, like a set of conserved charges, 
are accepted, while the rest is rejected. On the basis of this set of all accepted `events` one constructs 
an ensemble by using a suitable re-weighting scheme to account for quantum statistics and proper normalization. 
For the Monte Carlo approach presented in the next chapter an unconstrained, i.e. grand canonical, sample of 
events,~$P_{gce}(Q,\lbrace N_i\rbrace)$, is generated. Constrained distributions could then be defined through:
\begin{equation}
P \big(Q,\lbrace N_i \rbrace \big) ~=~ W \big(Q \big)~ P_{gce} \big( Q,\lbrace N_i \rbrace \big)~, 
\end{equation}
where a calculated weight factor~$W(Q)$ is employed to project out a set of events in the vicinity of an 
equilibrium value~$Q_{eq}$.

An immediate consequence of Eqs.(\ref{P_one}-\ref{P_three}) is that temperature and
chemical potentials appear in this formulation of (micro) canonical distributions (as well as in the Monte 
Carlo~\cite{Becattini:2003ft,Becattini:2004rq}, microscopic correlator~\cite{Begun:2005ah} and saddle point 
expansion~\cite{Becattini:2005cc} approaches). At first sight this seems to be a serious problem and an unnecessary 
complication of the initial task of finding a reasonable approximation to~CE and~MCE partition functions. 
However, the main technical challenge when numerically integrating the original version of the (micro) canonical 
partition function arises from a heavily oscillating integrand. Auxiliary parameters~$\beta$ and~$\mu$ will
produce a very smooth function, for which approximation schemes can be used. In taking the ratio, Eq.(\ref{P_two}), 
artificially introduced temperature and chemical potential drop out. The quality of the approximation on the other 
hand will crucially depend on their choice. 

In Section~\ref{sec_clt_GFCD} the generating function of the~GCE distribution of extensive quantities is introduced.
In Section~\ref{sec_clt_finitevolcorr} it is shown that the requirement of maximizing the generating function 
of the charge distribution at some given equilibrium point leads to a unique determination of
thermal parameters, and moreover constitutes the optimal choice for the approximation scheme. 
Emerging thermodynamic relations are discussed in Section~\ref{sec_clt_tempandchem}.
  
\section{Generating Function of the Charge Distribution}
\label{sec_clt_GFCD}
The~GCE partition function of an ideal  relativistic gas with volume~$V$, local temperature~$T=\beta^{-1}$, 
chemical potentials~$\mu_j$ and collective four velocity~$u_{\mu}$ reads (the system 
four-temperature~\cite{Becattini:2003ft} is~$\beta_{\mu}~=~\beta  u_{\mu}$):    
\begin{equation}\label{ZGCE}  
Z(V,\beta, u_{\mu}, \mu_j)~=~ \exp \Bigg[V \Psi   
\left(\beta,u_{\mu},\mu_j \right) \Bigg]~,   
\end{equation}   
where~$\Psi \left( \beta,u_{\mu},\mu_j\right)$ is a sum over the single particle partition 
functions~$\psi_i\left(\beta,u_{\mu},\mu_j\right)$ of all particle species~$i$ considered in the model:  
\begin{equation} \label{Psi}  
\Psi \left( \beta, u_{\mu},\mu_j \right) ~=~ \sum_{i} ~\psi_i   
\left( \beta,u_{\mu},\mu_j \right)~.  
\end{equation}   
The single particle partition function~$\psi_i \left(\beta,u_{\mu},\mu_j \right)$   
of particle species~$i$ is given by a J\"uttner distribution:  
\begin{equation} \label{psi}  
\psi_i \left(\beta,u_{\mu},\mu_j \right) ~=~   
\frac{g_i}{\left(2 \pi \right)^3} ~\int d^3p ~  
\ln \left(1~\pm~e^{-\beta ~p_i^{\mu}~u_{\mu} ~+~ \beta ~q_i^j~ \mu_j}   
\right)^{\pm1}~,  
\end{equation}   
where~$p_i^{\mu}$ are the components of the four momentum,~$q_i^j$ are the components of the charge vector, 
and~$g_i$ is the degeneracy factor. The upper sign refers to Fermi-Dirac statistics, while the lower sign refers 
to Bose-Einstein statistics. The case of Maxwell-Boltzmann statistics is analogous.  
  
For a hadron resonance gas, including hadrons and their resonances up to a mass of about~$2.5$~GeV, i.e. 
excluding the charm quark sector, the vector of chemical potentials~$\mu_j~$ and the `charge` vector~$q^j_i$ 
of particle species~$i$ are introduced\footnote{Finite acceptance effects are discussed here. Modifications 
to Eqs.(\ref{psi},\ref{chempot_partchr_vec}) which allow for inclusion of resonance decay in full acceptance 
are presented in Appendix \ref{appendix_cumulant}.}:  
\begin{equation} \label{chempot_partchr_vec}
\mu_j ~=~ (\mu_B,~\mu_S,~\mu_Q,~ \mu_{N_A},~ \mu_{N_B}) \qquad   
q^j_i = (b_i,~s_i,~q_i,~ n_A(\Omega),~ n_B(\Omega)) ~,  
\end{equation}  
where~$\mu_B$,~$\mu_S$, and~$\mu_Q$ are the baryon, strangeness, and electric charge chemical potentials, 
respectively. 
The parameters~$\mu_{N_A}$ and~$\mu_{N_B}$ are particle-specific chemical potentials, and could denote out of chemical   
equilibrium multiplicities of species `A` and `B`, similar to phase space occupancy 
factors~$\gamma_s$~\cite{Koch:1986ud} and~$\gamma_q$ \cite{Letessier:1999rm,Rafelski:2000by}. 
Throughout this thesis out of equilibrium effects are neglected, and thus~$\mu_{N_A}=\mu_{N_B}=0$.  
  
In addition,~$b_i$,~$s_i$, and~$q_i$ are the baryonic charge, the strangeness, and the electric charge of a particle 
of species~$i$.~$\Omega$ is the momentum space bin in which one is set to measure particle 
multiplicity.~$n_A(\Omega)=1$ if the momentum vector of the particle is within the acceptance,~$n_A(\Omega)=0$ if not. 
The charge vector~$q_i^j$ also contains, to maintain a common notation for all particle species considered 
in Eq.(\ref{Psi}), the `quantum` number~$n_B(\Omega)$. If one were set to determine the joint distribution of 
positively versus negatively charged hadrons, for instance, the~$\pi^+$ particle would have~${q_{\pi^+}=(0,0,1,1,0)}$, 
while the~$K^-$ particle would have~${q_{K^-}=(0,-1,-1,0,1)}$, 
see Chapters~\ref{chapter_multfluc_hrg} and~\ref{chapter_multfluc_pgas}.
  
One may also be interested in correlations of, for instance, the systems  net-baryon number~$B$ and net-strangeness~$S$, 
as e.g. in Refs.\cite{Koch:2008ia,Cheng:2008zh}. In this case, the~$\Lambda$ particle, 
with~${q_{\Lambda}=(1,-1,0,1,-1)}$, would be counted in groups~$A$ and~$B$, provided the momentum vector is within 
the acceptance~$\Omega$. The~$\Xi^{-}$ particle on the other hand carries two strange quarks and would 
have~${q_{\Xi^{-}}=(1,-2,-1,1,-2)}$, see Chapters~\ref{chapter_gce} and~\ref{chapter_extrapol}.
    
The generating function of the charge distribution in the~GCE is introduced by the substitutions in Eq.(\ref{psi}):  
\begin{eqnarray} \label{subst_1}  
\beta ~\mu_j &\rightarrow& \beta ~\mu_j ~+~ i \phi_{j}~, \\ \label{subst_2}  
\beta ~u_{\mu} &\rightarrow& \beta ~u_{\mu} ~-~ i \alpha_{\mu} ~.  
\end{eqnarray}   
The yet un-normalized joint probability distribution of extensive quantities~$Q^j, P^{\mu}$ in the~GCE is then 
given by the Fourier transform of Eq.(\ref{ZGCE}) after substitutions Eqs.(\ref{subst_1},\ref{subst_2}):  
\begin{eqnarray} \label{Curly}  
\mathcal{Z}^{P^{\mu},Q^j}(V,\beta,u_{\mu},\mu_j) &=&   
\int \limits_{-\pi}^{\pi} \frac{d^J\phi}{\left( 2\pi \right)^J} ~e^{-iQ^j \phi_{j}} ~   
\int \limits_{-\infty}^{\infty} \frac{d^4\alpha }{\left( 2\pi \right)^4}    
~ e^{-iP^{\mu} \alpha_{\mu}}  \nonumber \\   \label{Curly}
&& \times ~
~\exp \Bigg[V \Psi \left(\beta, u_{\mu},\mu_j;\alpha_{\mu},\phi_j \right) \Bigg]~.   
\end{eqnarray}   
Depending on the system under consideration, the vector of extensive quantities~$Q^j$ and corresponding Wick 
rotated fugacities~$\phi_j$ could read:   
\begin{equation}   
 Q^j ~=~ (B,~S,~Q, N_A, N_B) \qquad   
\phi_j = (\phi_B,~\phi_S,~\phi_Q,~ \phi_{N_A},~ \phi_{N_B}) ~.  
\end{equation}   
Here~$B$ is the net-baryon number,~$S$ is the net-strangeness, and~$Q$ is the electric net-charge of the system. 
Together with particle numbers~$N_A$ and~$N_B$ this would be a 5-dimensional distribution in the case of a~CE hadron 
resonance gas. Additionally for four-momentum conservation, yielding a 9-dimensional Fourier transform 
Eq.(\ref{Curly}) for a MCE hadron resonance gas, the vectors~$P^{\mu}$ and~$\alpha_{\mu}$ are introduced:  
\begin{equation}  
P^{\mu} ~=~ (E,~P_x,~P_y,~P_z) \qquad  
\alpha_{\mu} ~=~ (\alpha_{E},~\alpha_{P_x},~\alpha_{P_y},~\alpha_{P_z})~,  
\end{equation}   
where~$E$ is the energy and~$P_{x}$,~$P_{y}$, and~$P_{z}$ are the components of the   
collective momentum of the system, while~$\alpha_{\mu}$ are the corresponding fugacities.  

The distinction between discrete (Kronecker~$\delta$) and continuous quantities (Dirac~$\delta$) is not relevant 
for the large volume approximation, where particle number is a continuous variable to be integrated over. 
Proceeding by Taylor expansion of Eq.(\ref{Psi}), it is convenient to include discrete and continuous quantities 
into a common vector notation:  
\begin{equation}\label{AllQVector}
\mathcal{Q}^l ~=~ \left( Q^j,~  P^{\mu} \right) \qquad \textrm{and} \qquad  
\theta_l ~=~ \left( \phi_j,~ \alpha_{\mu} \right)~.  
\end{equation}  
The dimensionality of the vector~$\mathcal{Q}^l$ is denoted as~$L=2+3+4=9$ for a~MCE hadron resonance gas. Now, expanding 
the cumulant generating function,~$\Psi \left(\beta, u_{\mu},\mu_j;\theta_l \right)$, in a Taylor series yields:  
\begin{equation}\label{Psi_expand}  
\Psi \left(\beta,u_{\mu},\mu_j;\theta_l \right)   
~\simeq~ \sum \limits_{n=0}^{\infty} ~ \frac{i^n}{n!}~  
\kappa_n^{l_1,l_2, \dots, l_n}~   
\theta_{l_1} ~ \theta_{l_2} \dots  \theta_{l_n}~,  
\end{equation}  
where the elements of the cumulant tensor,~$\kappa_n^{l_1,l_2,\dots,l_n}$, are defined by:  
\begin{equation}\label{kappa}  
\kappa_n^{l_1,l_2,\dots, l_n}~ = \left( -i \right)^n~  
\frac{\partial^n \Psi}{\partial \theta_{l_1} \partial \theta_{l_2} \dots \partial \theta_{l_n}}   
\Bigg|_{\theta_l =  0_l}~.  
\end{equation}  
Generally cumulants are tensors of dimension~$L$ and order~$n$. The first cumulant is then a vector, while the 
second cumulant is a symmetric~$L \times L$ matrix. A good approximation to Eq.(\ref{Curly}) around the  
point~$\mathcal{Q}^l_{eq}=(Q^j_{eq}$,~$ P^{\mu}_{eq})$, can be found in terms of a Taylor expansion of Eq.(\ref{Psi}) 
in~$\theta_l=(\phi_j,\alpha_{\mu})$, if (see Section~\ref{sec_clt_finitevolcorr}):  
\begin{equation} \label{LM_mu}   
\frac{\partial \mathcal{Z}^{\mathcal{Q}^l}(V,\beta,u_{\mu}, \mu_j)}{\partial   
  \mathcal{Q}^l} \Bigg|_{\mathcal{Q}^l=\mathcal{Q}^l_{eq}}~=~ 0_l ~.  
\end{equation}   
Implicitly, Eq.(\ref{LM_mu}) does not define chemical potentials~$\mu_j$ and four-temperature $\beta_{\mu}=\beta u_{\mu}$, 
but corresponding Lagrange multipliers, which maximize the amplitude of the Fourier 
spectrum~$\mathcal{Z}^{\mathcal{Q}^l}(V,\beta,u_{\mu},\mu_j)$ of the generating function for a desired value 
of~$\mathcal{Q}^l_{eq}=(Q^j_{eq},P^{\mu}_{eq})$. Their values generally differ from the~GCE 
set~$(\beta,u_{\mu},\mu_j)$, however they  coincide in the thermodynamic limit. 
Lagrange multipliers are relevant for finite volume corrections in  Section~\ref{sec_clt_finitevolcorr}. 
The temperature~$\beta$ and the four-velocity~$u_{\mu}$ are not five independent Lagrange multipliers, 
as the four-velocity vector has unit length~$u_{\mu}u^{\mu}=1$. The integrand of Eq.(\ref{Curly}) is sharply peaked at the 
origin~$\phi_j=\alpha_{\mu}=0$ in the thermodynamic limit. The main contribution therefore comes from a very small 
region~\cite{Becattini:2005cc}. To see this, a second derivative test can be done on the integrand of 
Eq.(\ref{Curly}) taking into account the first two terms of Eq.(\ref{kappa}), see Appendix~\ref{appendix_sdt}.
   
The cumulant of~$0^{th}$ order,~$\kappa_0$, is just the logarithm of the~GCE partition function divided by the 
volume,~$Z\left(V,\beta, u_{\mu},\mu_j\right) \equiv \exp (V \kappa_0)$. Hence, after extending the limits of
integration to~$\pm \infty$, which will introduce a negligible error, one finds:
\begin{eqnarray}\label{CE_PF_1} 
\mathcal{Z}^{\mathcal{Q}^l}(V,\beta, u_{\mu},\mu_j) &\simeq& 
Z(V,\beta, u_{\mu},\mu_j) ~ 
\int \limits_{-\infty}^{\infty} \frac{d^L\theta}{\left( 2\pi \right)^L} 
~\exp \Bigg[-i \mathcal{Q}^l \theta_l 
\nonumber \\
&+&V \sum_{n=1}^{\infty} \frac{i^n}{n!} \; \kappa_n^{l_1,l_2,\dots,l_n } \;
\theta_{l_1} \theta_{l_2} \dots \theta_{l_n} \Bigg] ~.  
\end{eqnarray}
It is worth noting that the parts of the integrand of Eq.(\ref{Curly}) related to discrete quantities were~$2\pi$-periodic, 
while the integrand of Eq.(\ref{CE_PF_1}) is a superposition of oscillating and decaying modes.
Spelling out the first two terms of the summation yields: 
\begin{eqnarray} 
\mathcal{Z}^{\mathcal{Q}^l}(V,\beta, u_{\mu},\mu_j) &\simeq& 
Z(V,\beta, u_{\mu},\mu_j) ~ 
\int \limits_{-\infty}^{\infty} \frac{d^L\theta}{\left( 2\pi \right)^L}   
~\exp  \Bigg[ -i\mathcal{Q}^l \theta_l  \\
&+& iV \kappa_1^l \theta_l~-~V \frac{\kappa_2^{l_1,l_2}}{2!} \theta_{l_1} \theta_{l_2} 
~+~ V \sum_{n=3}^{\infty} \frac{i^n}{n!} \; \kappa_n^{l_1,l_2,\dots,l_n } \;
\theta_{l_1} \theta_{l_2} \dots \theta_{l_n} \Bigg]. \nonumber 
\end{eqnarray} 
Performing now a change of variables will simplify this integral:
\begin{equation}\label{theta} 
\vartheta_l ~=~ \sqrt{V} ~\sigma_{l}^{~k} ~\theta_k~, 
\end{equation} 
where~$\sigma_{l}^{~k}$ is the square root of the second rank tensor~$\kappa_2$:  
\begin{equation}\label{sigmatensor}
\sigma_{l}^{~k} ~\equiv~ \left( \kappa_2^{~~1/2}\right)_l^{~~k}.  
\end{equation} 
The new integral measure~$d^L \vartheta$ then equals to: 
\begin{equation}\label{dtheta} 
d^L \vartheta ~=~ \det | \sqrt{V} ~\sigma | ~d^L \theta ~=~ 
V^{L/2}~ \det | \sigma | ~d^L \theta~.
\end{equation} 
Lastly in terms of this transformation normalized cumulant tensors $\lambda_n$ with components:
\begin{equation}\label{normcum} 
\lambda_n^{l_1,l_2,\dots,l_n} ~ \equiv ~ \kappa_n^{k_1,k_2,\dots,k_n} 
\left(
  \sigma^{-1}\right)_{k_1}^{\;\;\;l_1} \left( 
  \sigma^{-1}\right)_{k_2}^{\;\;\;l_2} \dots \left(
  \sigma^{-1}\right)_{k_n}^{\;\;\;l_n}  ~ 
\end{equation} 
are introduced. The new variable~$\xi^l$ will be a measure for the distance of the actual charge 
vector~$\mathcal{Q}^k$ to the peak $V \kappa_1^k$ of the distribution:
\begin{equation}\label{xi-var}
\xi^l ~=~ \left( \mathcal{Q}^k - V \kappa_1^k \right) \left( 
  \sigma^{-1}\right)_{k}^{\;\;l}  V^{-1/2}~. 
\end{equation}
Including above steps at once yields:  
\begin{eqnarray} \label{CLTfinalZ} 
\mathcal{Z}^{\mathcal{Q}^l}(V\beta,u_{\mu},\mu_j) &\simeq& 
\frac{Z(V,\beta,u_{\mu},\mu_j)}{V^{L/2} \det|\sigma|}  
\int \limits_{-\infty}^{\infty} \frac{d^L \vartheta}{\left( 2\pi \right)^L}  
~ \exp \Bigg[ - i \xi^l \vartheta_l - \frac{\vartheta^l \vartheta_l}{2!} \nonumber \\
&+& \sum \limits_{n=3}^{\infty} i^n  
 V^{-\frac{n}{2}+1}~ \frac{\lambda_n^{l_1,l_2,\dots,l_n}}{n!}
 \vartheta_{l_1}\vartheta_{l_2} \dots \vartheta_{l_n}   \Bigg]~.  
\end{eqnarray}
Eq.(\ref{CLTfinalZ}) is the starting point for obtaining an asymptotic solution in this section, as well as 
for finite volume corrections in Section~\ref{sec_clt_finitevolcorr}. Through coordinate transformation 
Eq.(\ref{theta}) terms were separated in their power in volume. Thus, as system size is increased, influence of 
higher order normalized cumulants~$\lambda_n$ decreases, allowing for truncation of the summation for 
sufficiently large volume.

A few words on physical units are in order. Discussion of the canonical ensemble shall suffice.
The single particle partition function Eq.(\ref{psi})~$\psi_i[$fm$^{-3}]$, and therefore all cumulant elements
Eq.(\ref{kappa})~$\kappa_n^{j_1,j_2,\dots,j_n }$~$[$fm$^{-3}]$ in the~CE. 
Consequently entries in Eq.(\ref{sigmatensor}) are~$\sigma_{j}^{~k}$~$[$fm$^{-3/2}]$. 
The normalization in Eq.(\ref{CLTfinalZ}), $V^{J/2} \det|\sigma|$, for J-dimensional~$\sigma$, 
as well as the new variable of integration Eq.(\ref{theta})~$\theta_l$, are hence dimensionless. 
The elements of the inverse sigma tensor are~$\left( \sigma^{-1}\right)_{k}^{~j}$~$[$fm$^{3/2}]$
and, thus, the elements~$\xi^j$ of the vector Eq.(\ref{xi-var}) will be dimensionless. 
Finally, the elements of the normalized cumulants, Eq.(\ref{normcum}), are~$\lambda_n^{j_1,j_2,\dots,j_n}[$fm$^{-3+3n/2}]$, 
which is canceled by the factor~$V^{-n/2+1}$ in the summation in Eq.(\ref{CLTfinalZ}). 
Thus all terms involved in Eq.(\ref{CLTfinalZ}) are dimensionless. 

As discussed, for~$V\rightarrow \infty$ one can discard terms of~$V^{-1/2}$ and higher in Eq.(\ref{CLTfinalZ}), 
and consider only the first two cumulants for the asymptotic solution: 
\begin{equation}\label{CE_Square}
\mathcal{Z}^{\mathcal{Q}^l}(V\beta,u_{\mu},\mu_j) ~\simeq~ 
\frac{Z(V,\beta,u_{\mu},\mu_j)}{V^{L/2} \det|\sigma|}  
\int \limits_{-\infty}^{\infty} \frac{d^L \vartheta}{\left( 2\pi \right)^L}  
~ \exp \Bigg[ - i \xi^l \vartheta_l - \frac{\vartheta^l \vartheta_l}{2!}
 \Bigg]~.
\end{equation}
This is the characteristic function of a multivariate normal distribution~\cite{Abramowitz:1965fix_1}.
Completing the square the integral (\ref{CE_Square}) can be solved:
\begin{equation}\label{MND}   
 \mathcal{Z}^{\mathcal{Q}^l}(V,\beta,u_{\mu},\mu_j) 
~\simeq~ 
~ \frac{Z(V,\beta,u_{\mu},\mu_j)  }{\left(2\pi V\right)^{L/2}\det|\sigma|} 
~\exp \left[-\frac{1}{2}~ \xi^l \xi_l \right]~.
\end{equation}   
The asymptotic solution of the~GCE joint distribution of extensive quantities~$\mathcal{Q}^l$ is then given by:  
\begin{equation}\label{GCE_dist}  
P_{gce}(\mathcal{Q}^l) ~=~ 
\frac{ \mathcal{Z}^{\mathcal{Q}^l}(V,\beta,u_{\mu},\mu_j)}{
Z(V,\beta,u_{\mu},\mu_j)} 
~\simeq~ ~ \frac{1}{\left(2\pi V\right)^{L/2} \det|\sigma|} 
~\exp \left[-\frac{1}{2}~ \xi^l \xi_l \right]~.  
\end{equation}  
Mean values in the thermodynamic limit are given by the first Taylor expansion 
terms,~$\langle N_A\rangle=V\kappa_1^{N_A}$,~$\langle Q\rangle= V \kappa_1^{Q}$,~$\langle E \rangle= V \kappa_1^{E}$,~etc. 
and converge to~GCE values. To obtain a joint~(two-dimensional) particle multiplicity distribution one has to take a 
two-dimensional slice of the ($L$ dimensional)~GCE distribution, Eq.(\ref{GCE_dist}), around the peak of the 
extensive quantities which one is considering as exactly fixed. Please note that Eq.(\ref{MND}) (albeit in  different 
notation) was used as an assumption in the microscopic correlator 
approach~\cite{Begun:2005ah,Begun:2005iv,Begun:2006jf,Begun:2006gj}. 
More details of the calculation, in particular on the connection between the partition 
functions~$\mathcal{Z}^{P^{\mu},Q^j}(V,\beta,u_{\mu},\mu_j)$ and the conventional   
version~$Z(V,P^{\mu},Q^j)$~\cite{Becattini:2003ft,Becattini:2004rq}, can be found in Appendix~\ref{appendix_partfunc}.  
Joint particle multiplicity distributions are constructed from Eq.(\ref{GCE_dist}) in Appendix~\ref{appendix_multmist}. 
Details of the cumulant tensor are discussed in Appendix~\ref{appendix_cumulant}. A closed formula for the 
scaled variance of CE or MCE particle multiplicity fluctuations can be found in~\cite{Hauer:2007ju}.

\section{Finite Volume Corrections}
\label{sec_clt_finitevolcorr}
In considering finite system size effects on distributions, the region where the thermodynamic limit approximation 
is valid is left. Chemical potentials~$\mu_j$ and four-temperature~$\beta_{\mu}$ do not correspond anymore to the physical 
ones, which would be found in the~GCE, but have to be thought of as Lagrange multipliers, used to maximize the 
partition function for a given (micro) canonical state. First  some volume dependent correction terms are derived. 
A condition is obtained that defines the values of~$\mu_j$ and~$\beta_{\mu}$. 
The correct choice allows to write down the thermodynamical potentials, the Helmholtz free energy~$F$ for the~CE, and the 
entropy~$S$ for the MCE, in terms of the generalized partition function. Some general criterion for the validity of the 
expansion is given. Approximations will be compared to analytical~CE and~MCE solutions of multiplicity distributions.

\subsection{Gram-Charlier Expansion}
In Section~\ref{sec_clt_GFCD} it was shown that in the thermodynamic limit any equilibrium distribution can be 
approximated by a multivariate normal distribution, Eq.(\ref{MND}). Further parameters, describing the shape of the
distribution, like skewness ($\kappa_3$), or excess/kurtosis ($\kappa_4$), tend to zero as volume is increased. 
Returning to Eq.(\ref{CLTfinalZ}) with a number~$L$ of extensive quantities:
\begin{eqnarray}
\mathcal{Z}^{\mathcal{Q}^l}(V,\beta,u_{\mu},\mu_j) ~&\simeq&~ 
\frac{Z(V,\beta,u_{\mu},\mu_j)}{ V^{L/2} \det|\sigma |} 
\int \limits_{-\infty}^{\infty} \frac{d^L \vartheta}{\left( 2\pi \right)^L } 
~ \exp \Bigg[ - i  \xi^l \vartheta_l - \frac{\vartheta^l  \vartheta_l}{2!} \nonumber \\
   && \quad +~\sum \limits_{n=3}^{\infty} ~i^n~ V^{-\frac{n}{2}+1}
  ~\frac{\lambda_n^{l_1,l_2,\dots,l_n}}{n!} 
  ~\vartheta_{l_1}\vartheta_{l_2} \dots \vartheta_{l_n}   \Bigg]~.
\end{eqnarray}
Finite volume corrections will be obtained by Gram-Charlier 
expansion~\cite{Feller:1968fix_1,Feller:1970fix_1,Hughes:1995fix_1,Hughes:1996fix_1}.
The charge vector~$\mathcal{Q}^l$ denotes a vector of~$L$ Abelian charges, and could read for a hadron resonance 
gas~$\mathcal{Q}^l = (B,S,Q,E,P_x,P_y,P_z)$, but could, in principle, also include particles multiplicities.
Expanding the exponential in terms of powers in volume, one finds:
\begin{eqnarray}\label{Vcorrection}
&&\mathcal{Z}^{\mathcal{Q}^l}(V,\beta,u_{\mu},\mu_j) \simeq 
\frac{ Z(V,\beta,u_{\mu},\mu_j)}{ V^{L/2} \det|\sigma |} 
\int \limits_{-\infty}^{\infty} \frac{d^L \vartheta}{\left( 2\pi \right)^L} 
~ \exp \left[ - i \xi^l \vartheta_l - \frac{\vartheta^l \vartheta_l}{2!} \right]  
\nonumber \\
 &&\qquad \qquad \times \Bigg[ 1 
 + \frac{\lambda_3^{l_1,l_2,l_3}}{3!} 
 ~\frac{ i^3 \vartheta_{l_1} \vartheta_{l_2} \vartheta_{l_3} }{ V^{1/2}} 
 + \frac{\lambda_4^{l_1,l_2,l_3,l_4}}{4!}
 ~\frac{ i^4 \vartheta_{l_1} \vartheta_{l_2} \vartheta_{l_3} \vartheta_{l_4} }{ V} 
\nonumber \\
 &&\qquad\qquad\qquad
+ ~\frac{1}{2!}\frac{\lambda_3^{l_1,l_2,l_3}}{3!} ~\frac{\lambda_3^{l_4,l_5,l_6}}{3!}~
  \frac{i^6 \vartheta_{l_1} \dots \vartheta_{l_6} }{V} 
 + \mathcal{O} \left( V^{-3/2}\right) \Bigg]~.
\end{eqnarray}
Correction terms in Eq.~(\ref{Vcorrection}) can be obtained by differentiation 
of~$\exp \left[ - i \xi^l \vartheta_l \right]$ with respect to the elements~$\xi_l$. One can thus reverse the 
order by first integrating and then again differentiating. Using generalized Hermite polynomials:
\begin{equation} \label{Hermite}
\Big[ H_n \left( \xi \right)\Big]_{l_1,l_2,\dots,l_n} = 
\left(-1 \right)^n ~ \exp \left[ ~\frac{\xi^l \xi_l}{2}~ \right]~
\frac{d^n}{d\xi_{l_1} \; d\xi_{l_2} \dots d\xi_{l_n} }~
\exp \left[ -\frac{\xi^l\xi_l}{2} \right]~,
\end{equation}
with the adjusted shorthand notation for the contractions:
\begin{eqnarray}
h_3 \left( \xi \right) &=&
\frac{\lambda_3^{l_1,l_2,l_3}}{3!} ~ 
\Big[  H_3 \left( \xi \right) \Big]_{l_1,l_2,l_3}~,  \\
h_4 \left( \xi \right) &=&
\frac{\lambda_4^{l_1,l_2,l_3,l_4}}{4!} 
~ \Big[ H_4 \left(\xi \right) \Big]_{l_1,l_2,l_3,l_4} \nonumber \\
&&~~~~+~ \frac{1}{2!}~ \frac{\lambda_3^{l_1,l_2,l_3}}{3!} ~
\frac{\lambda_3^{l_4,l_5,l_6}}{3!} 
~\Big[ H_6 \left( \xi \right) \Big]_{l_1,\dots l_6}~,  \\
h_5 \left( \xi \right) &=&
\frac{\lambda_5^{l_1,\dots,l_5}}{5!} 
~\Big[ H_5 \left( \xi \right) \Big]_{l_1,\dots, l_5} ~+~
\frac{\lambda_3^{l_1,l_2,l_3}}{3!} ~
\frac{\lambda_4^{l_1,l_2,l_3,l_4}}{4!}~
\Big[ H_7 \left( \xi \right) \Big]_{l_1,\dots,l_7} \nonumber \\
&&~~~~ +~ \frac{1}{3!}~ \frac{\lambda_3^{l_1,l_2,l_3}}{3!}
~\frac{\lambda_3^{l_4,l_5,l_6}}{3!}
~\frac{\lambda_3^{l_7,l_8,l_9}}{3!}~ 
\Big[ H_9 \left( \xi \right) \Big]_{l_1,\dots,l_9}~,
\end{eqnarray}
the partition function for finite volume can be approximated by:
\begin{eqnarray} \label{GramCharlier}
&&\mathcal{Z}^{\mathcal{Q}^l}(V,\beta,u_{\mu},\mu_j)
 ~\simeq~ \frac{Z(V,\beta,u_{\mu},\mu_j)}{\left(2 \pi V  \right)^{L/2}\det| \sigma| } 
 ~~\exp \left[ - \frac{\xi^l \xi_l}{2} \right] \nonumber \\  
 && \qquad \times ~\Bigg[ 1
 ~+~\frac{h_3 \left( \xi \right)}{V^{1/2}}
 ~+~\frac{h_4 \left( \xi \right)}{V} 
 ~+~\frac{h_5 \left( \xi \right)}{V^{3/2}} 
 ~+~\mathcal{O} \left( V^{-2}\right) \Bigg]~.
\end{eqnarray}
Considering the simplest case of only one conserved charge, it is evident from Eq.(\ref{Hermite}), that the first 
order correction term in Eq.(\ref{GramCharlier}) is a polynomial of order 3 in~$\xi$, while the second 
order correction term is a polynomial of order 4, etc. Hence for large values of~$\xi^l$, e.g. a charge, energy, 
momentum, and multiplicity state far from the peak of the distribution will lead to a bad approximation, and even 
to negative values for~$P(\mathcal{Q}^l)$. 
The validity of this approximation is thus restricted to the central region of the distribution.~CE and~MCE results 
will be compared to scenarios which are accessible to analytical methods in Section~\ref{QofA}. In order to 
distinguish approximations which include corrections up to different orders in volume in Eq.(\ref{GramCharlier}), 
the asymptotic solution is denoted as CLT (central limit theorem), including terms up 
to~$\mathcal{O} \left( V^{-1/2}\right)$ as GC3 (Gram-Charlier 3), including terms
up to~$\mathcal{O} \left( V^{-1}\right)$ as GC4, and including terms up 
to~$\mathcal{O} \left( V^{-3/2}\right)$ as~GC5.

\subsection{Chemical and Thermal Equilibrium}
Here  the question is addressed of how to choose the optimal values for~$\beta_{\mu}$ and~$\mu_j$. 
The postulate is that a (micro) canonical equilibrium state should be as well the most likely state in the~GCE. 
On the other hand, it is apparent that the expansion works best around the peak of the distribution. 
Hence~$\beta_{\mu}$ and~$\mu_j$ are chosen such that the partition function is maximized at some 
equilibrium point~$\mathcal{Q}^l_{eq}$. Taking  terms up to~$\mathcal O (V^{-1/2})$ into account, the first
derivative of the partition function Eq.(\ref{CE_PF_1}) reads:
\begin{eqnarray}\label{DiffPF}
&&\frac{\partial \mathcal{Z}^{\mathcal{Q}^l}(V,\beta,u_{\mu},\mu_j)}{\partial \mathcal{Q}^l} 
~=~\frac{1}{\left(2\pi \right)^{L/2} V^{\left(L+1\right)/2} \det| \sigma|} 
~\exp \left [-\frac{1}{2}~\xi^l \xi_l \right]~~~~~~~~~~~~~~~~~~\\
&& ~~~ \times ~\Bigg[\xi_k \left(\sigma^{-1} \right)^{ \; k}_{\;\; l} 
 ~ +~ \frac{\lambda_3^{k_1,k_2,k_3}}{3!\sqrt{V}} \left(\sigma^{-1}
\right)^{ \; k_4}_{\;\; l} \Big[ H_4\left( \xi\right)\Big]_{k_1,k_2,k_3,k_4}
+ \mathcal{O} \left( V^{-1}\right) \Bigg]~.\nonumber 
\end{eqnarray}
Lagrange multipliers (or chemical potentials) should be chosen such that the first derivative Eq.(\ref{DiffPF}) 
of~$\mathcal{Z}^{\mathcal{Q}^l}$ with respect to the conserved quantities~$\mathcal{Q}^l$ vanishes, hence 
Eq.(\ref{GramCharlier}) is maximized at the point~$\mathcal{Q}^l_{eq.}$:
\begin{equation}\label{gChemPot} 
\frac{\partial \mathcal{Z}^{\mathcal{Q}^l}(V,\beta,u_{\mu},\mu_j)}{\partial \mathcal{Q}^l}
~\Bigg|_{\mathcal{Q}^l_{eq}}= ~ 0_l~. 
\end{equation}
Using only the asymptotic solution, valid in the thermodynamic limit, this condition leads to:
\begin{equation}\label{CLTmu}
\xi_l = \left( \mathcal{Q}_k - V \kappa_{1,k} \right) \left(
  \sigma^{-1}\right)^{k}_{\;\;l} ~ =~ 0_l~.
\end{equation}
Hence the partition function is maximal at the point~$Q_{k,eq}= V\kappa_{1,k}$. Charge and energy density 
correspond thus to the~GCE values, and each component of~$\mu^j \rightarrow \mu^j_{gce}$, 
and~$\beta^{\mu} \rightarrow \beta^{\mu}_{gce}$. While, when taking the first finite volume correction term in 
Eq.(\ref{DiffPF}) into account, one obtains:
\begin{equation}\label{GCmu}
\xi_k \left(\sigma^{-1} \right)^{ \; k}_{\;\; l} +
 \frac{\lambda_3^{k_1,k_2,k_3}}{3!\sqrt{V}}  \left(\sigma^{-1}
\right)^{  k_4}_{\;\; l} \Big[ H_4\left( \xi \right)\Big]_{k_1,k_2,k_3,k_4} = 0_l~,
\end{equation}
rather than Eq.(\ref{CLTmu}), and~$\mu^j \not= \mu^j_{gce}$, and~$\beta^{\mu} \not= \beta^{\mu}_{gce}$. 
For calculation of distributions for systems of finite volume one should therefore find chemical potentials that satisfy
condition (\ref{gChemPot}). A technical comment is in order. From Eq.(\ref{DiffPF}) it is evident that the first order 
correction term to the derivative of the partition function is a polynomial of order 4 in $\xi$, while the second 
one is of order 5, etc. It is therefore crucial to find in numerical calculations the correct maximum.

\subsection{Quality of Approximation}
\label{QofA}
To test the quality of the approximation for multiplicity distributions at finite volume for (very) small systems, 
the analytical solution for a classical~CE particle-anti-particle gas~\cite{Begun:2004gs}, and a classical~MCE 
(without momentum conservation) ultra-relativistic gas~\cite{Fermi:1950fix_1,Begun:2004pk} are compared to 
Eq.(\ref{GramCharlier}). 

\begin{figure}[ht!]
\epsfig{file=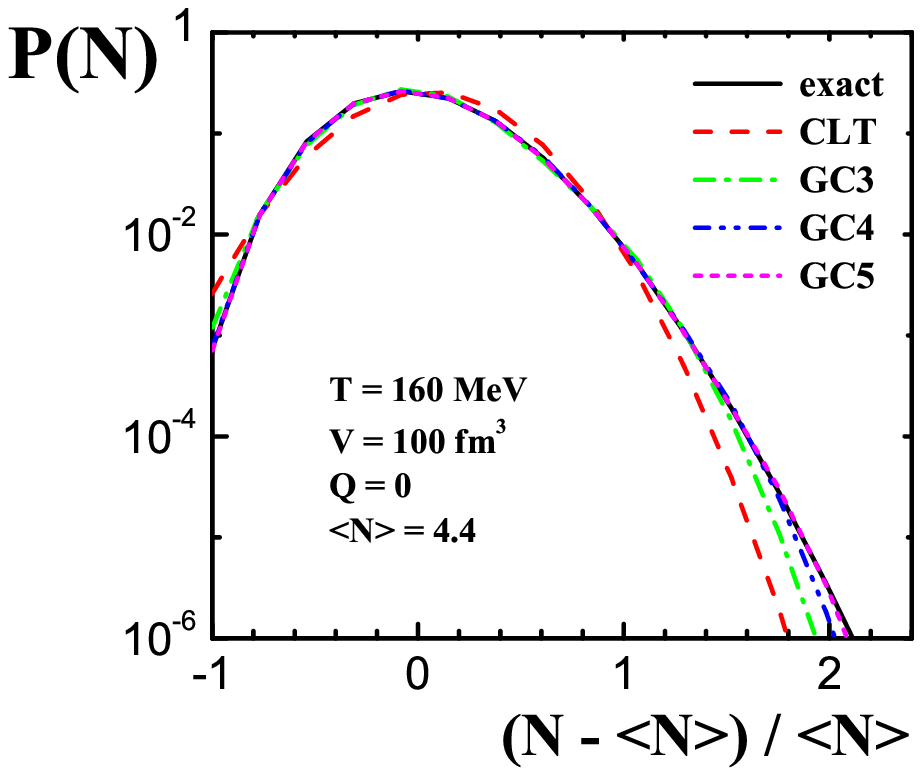,width=7.5cm}
\epsfig{file=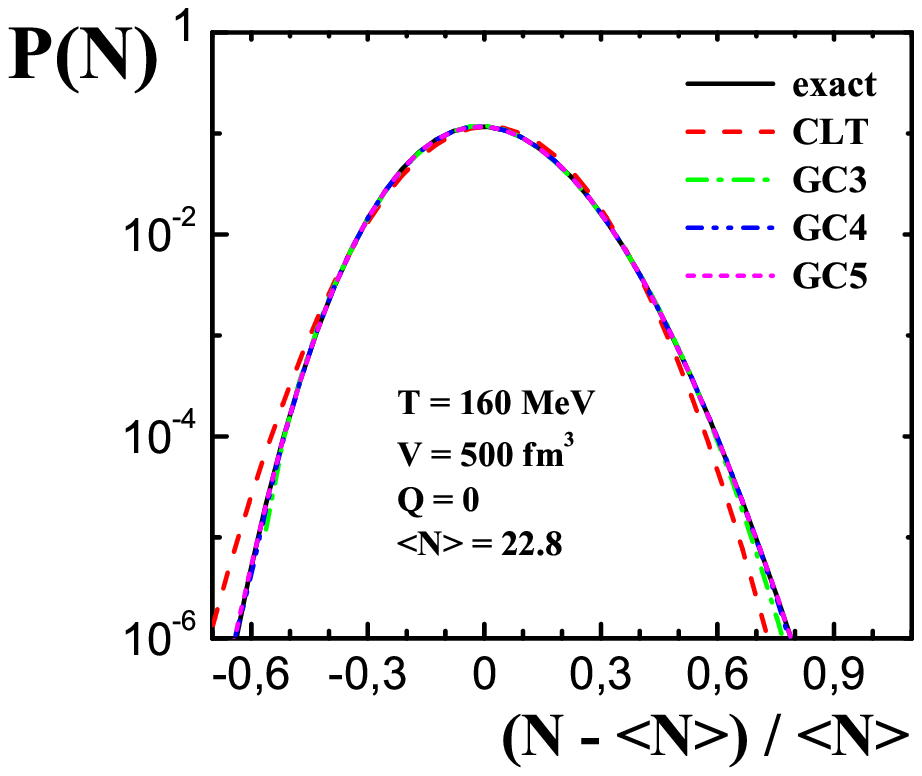,width=7.5cm}
\epsfig{file=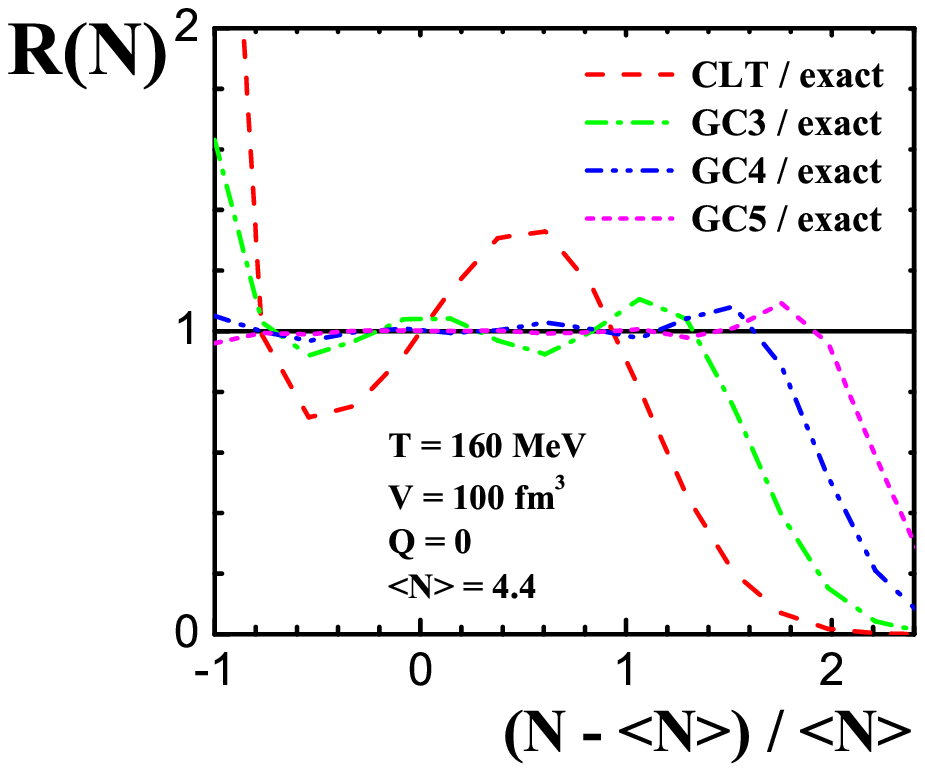,width=7.5cm}
\epsfig{file=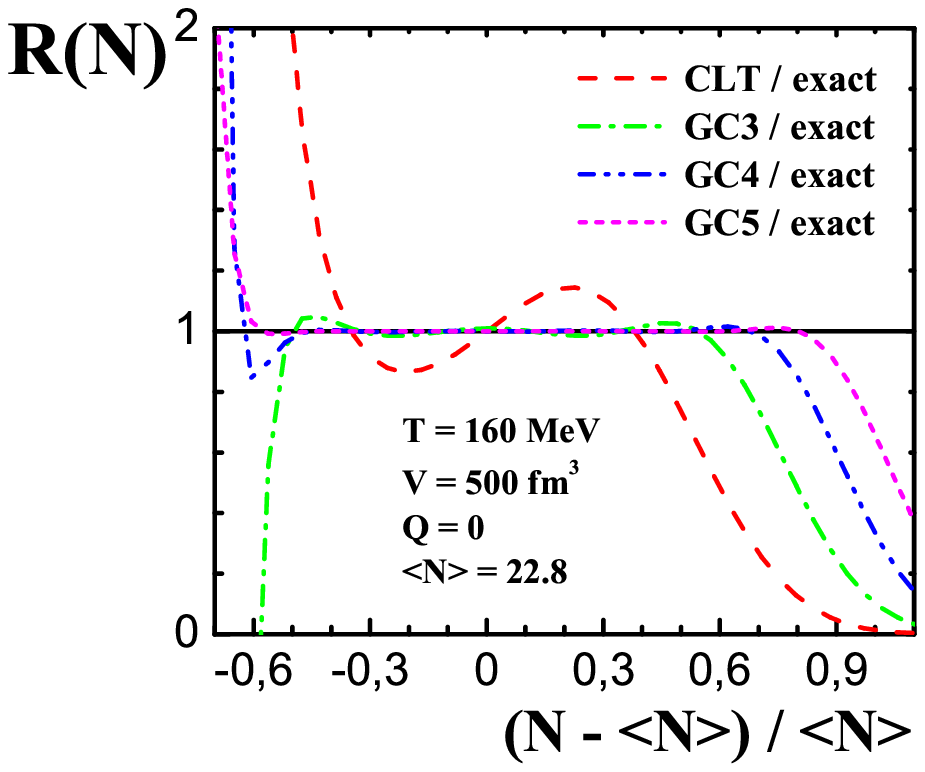,width=7.5cm}
\caption{({\it Top}): The~CE~$\pi^{+}$ multiplicity distribution in Boltzmann approximation for temperature~{$T=160$~MeV},
        electric net-charge~$Q=0$, and volume~$V=100$~fm$^3$~({\it left}), or volume~$V=500$~fm$^3$~({\it right}). 
        Exact solutions (solid), and various approximations, CLT (dash), GC3 (dash-dot), GC4 (dash-dot-dot), and GC5 (dot) 
        are shown. 
        ({\it  Bottom}): same as~({\it top}), but the ratios of exact solution to approximations are shown.}
\label{PN1}
\end{figure}

\begin{figure}[ht!]
\epsfig{file=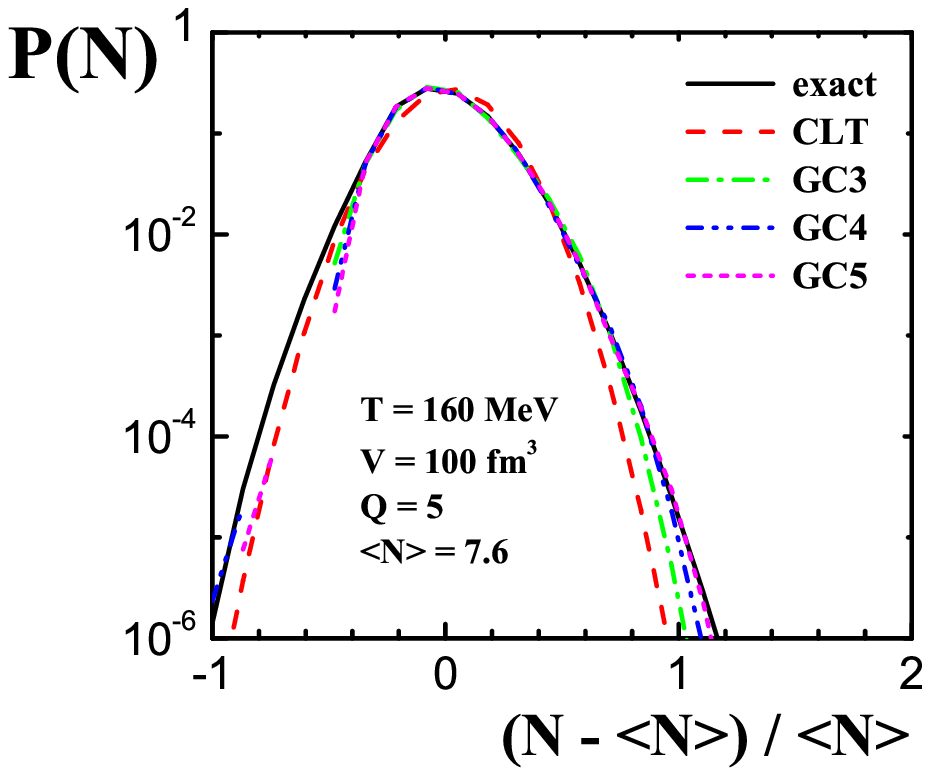,width=7.5cm}
\epsfig{file=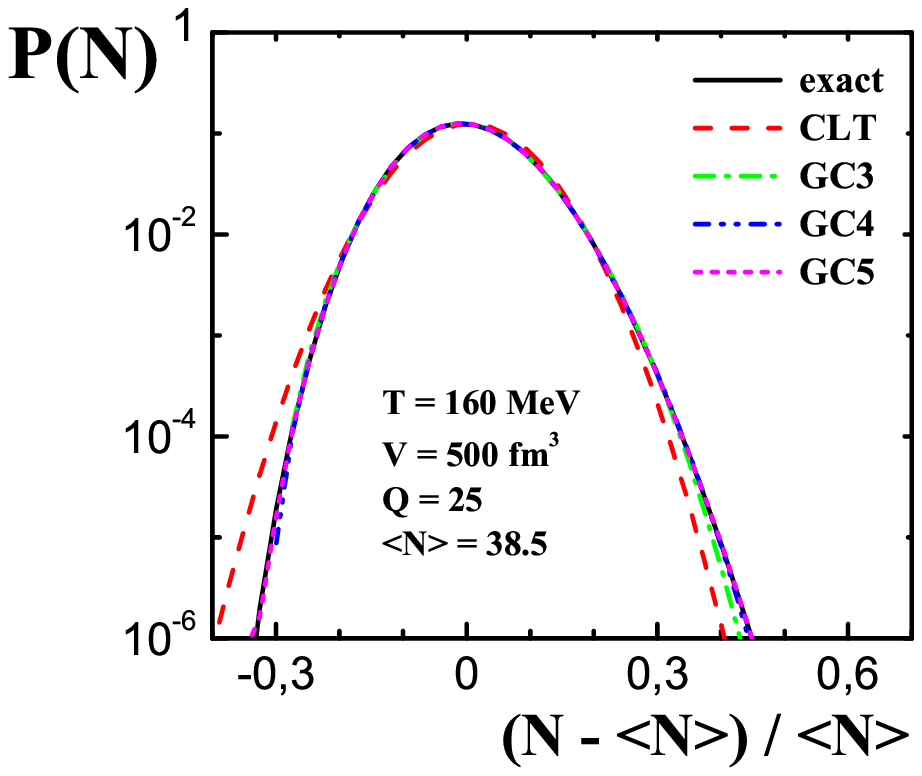,width=7.5cm}
\epsfig{file=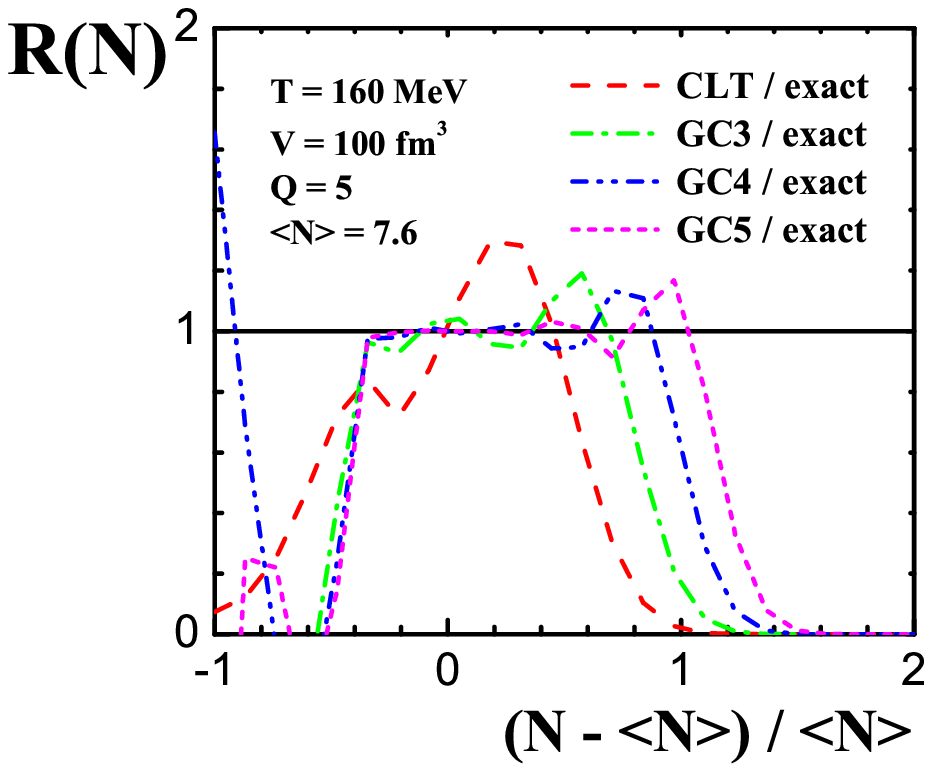,width=7.5cm}
\epsfig{file=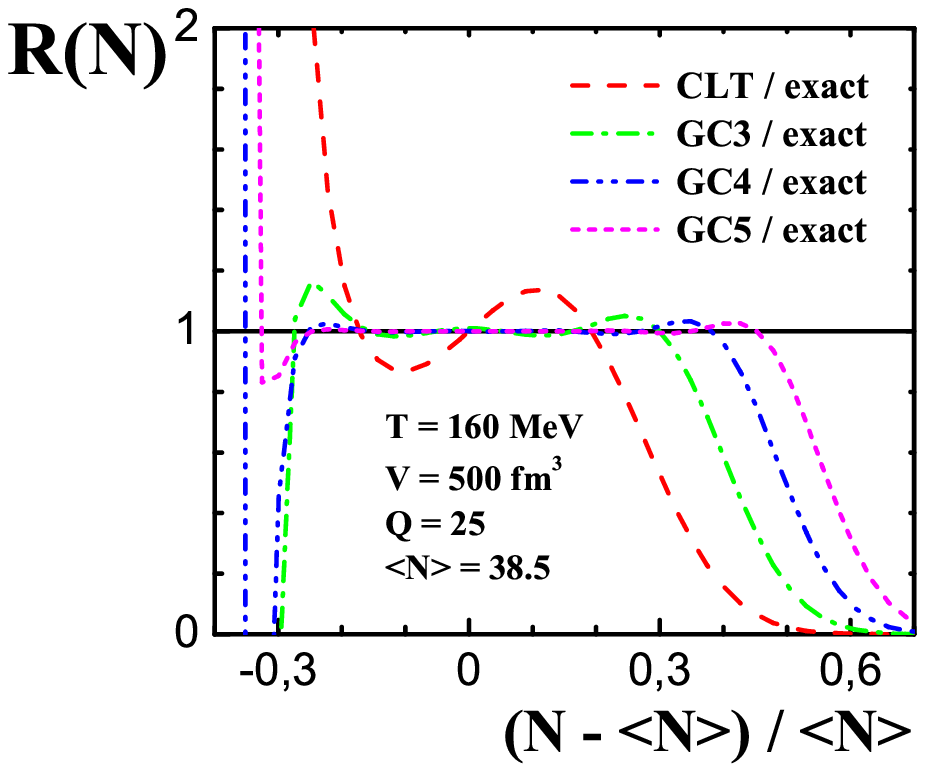,width=7.5cm}
\caption{Same as~Fig.(\ref{PN1}), but for an electric net-charge of~$Q=5$ ({\it left}), 
         or an electric net-charge of~$Q=25$~({\it right}).}
\label{PN2}
\end{figure}

\begin{figure}[ht!]
\epsfig{file=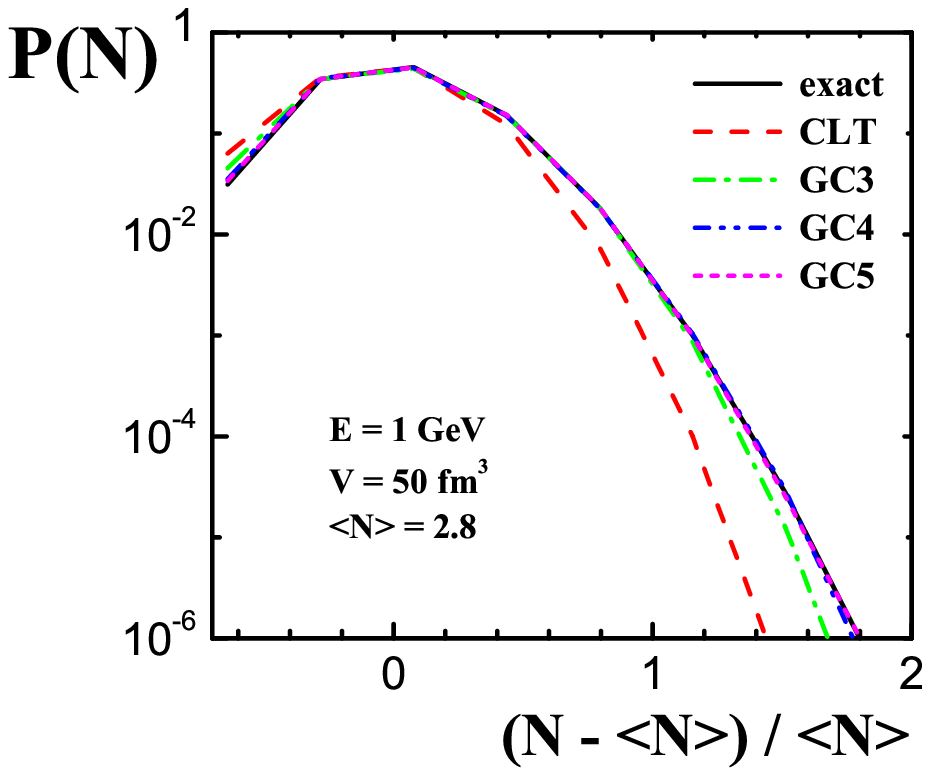,width=7.5cm}
\epsfig{file=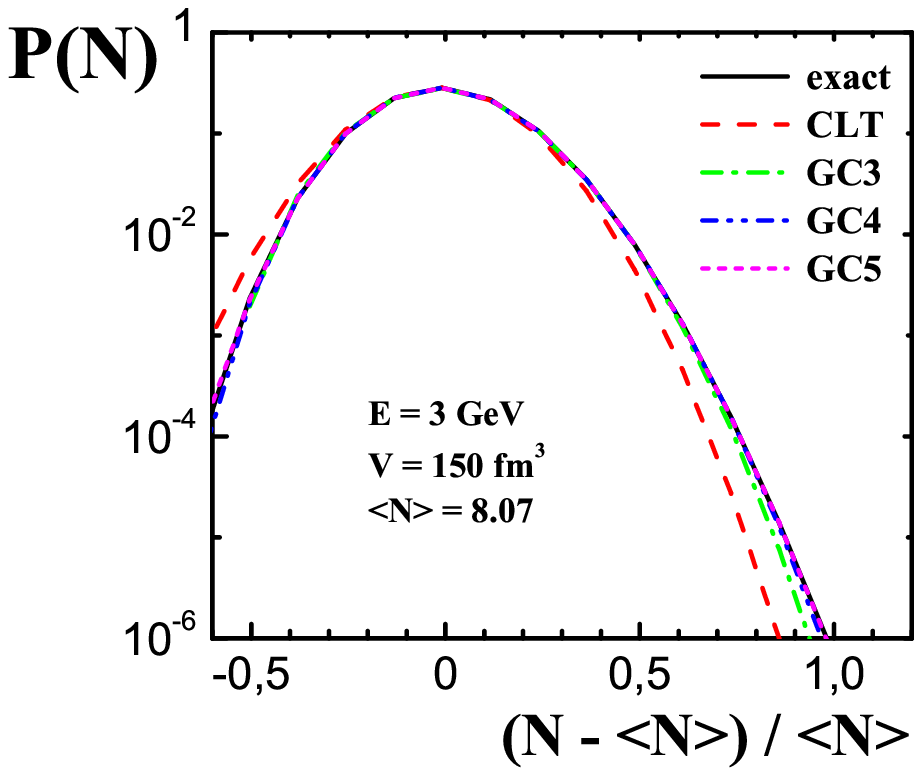,width=7.5cm}
\epsfig{file=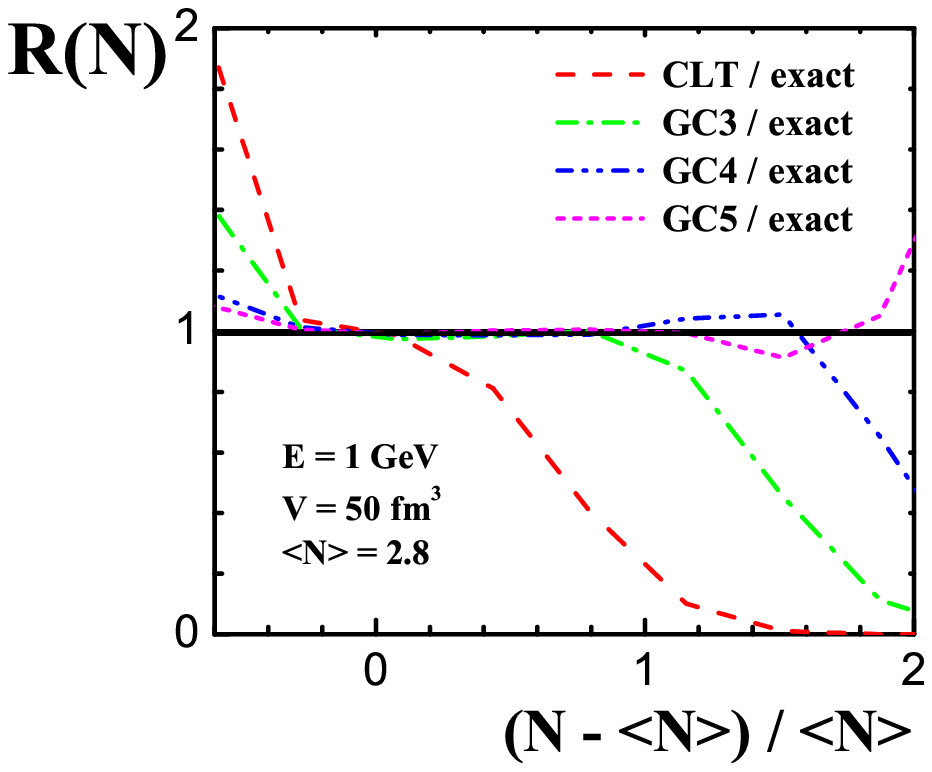,width=7.5cm}
\epsfig{file=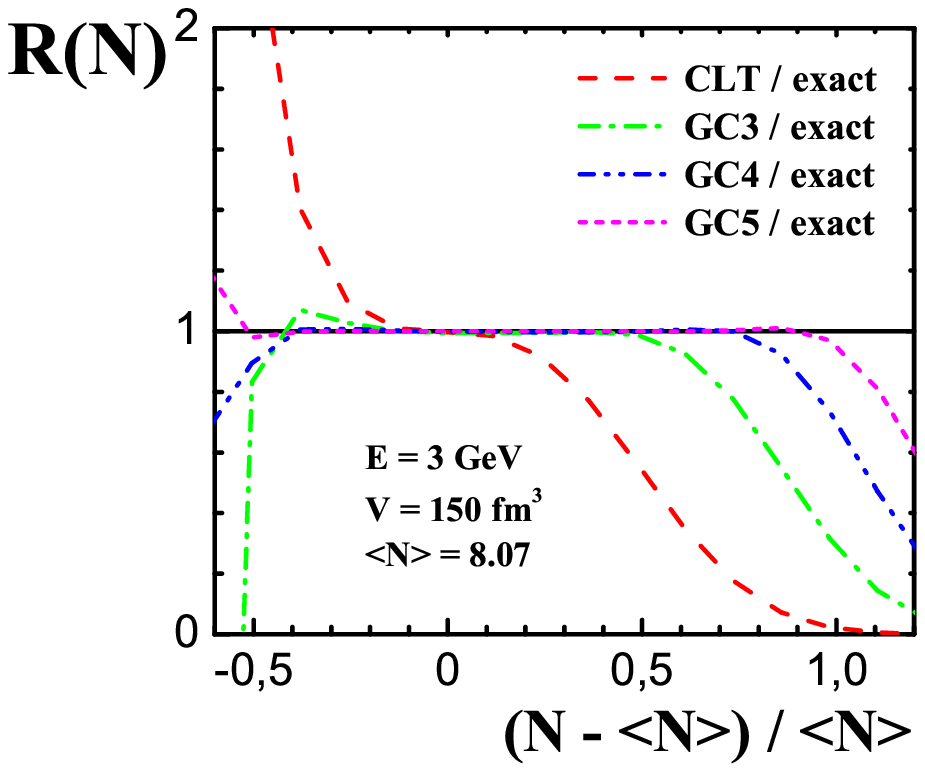,width=7.5cm}
\caption{({\it Top}): MCE multiplicity distribution for an ultra-relativistic gas in Boltzmann approximation. 
        For energy~$E=1$~GeV, and volume~$V=50$~fm$^3$~({\it left}), or for~$E=3$~GeV, and~$V=150$~fm$^3$~({\it right}).
        Exact solutions (solid), and various approximations, CLT (dash), GC3 (dash-dot), GC4 (dash-dot-dot), 
        and GC5 (dot) are shown. 
        ({\it Bottom}): same as~({\it top}),  but the ratios of exact solution to approximations are shown.}
\label{PN3}
\end{figure}

Fig.(\ref{PN1}) shows the multiplicity distribution of positively charged particles for two system sizes in the 
exact form and in different orders of approximation Eq.(\ref{GramCharlier})~({\it top}), and the ratio of 
approximation to exact solution~({\it bottom}). In~Fig.(\ref{PN2}) the same physical system is shown for a 
(relatively large) positive electric net-charge. Due to a one-to-one correspondence between the distributions 
of negatively (suppressed) and positively (enhanced) particles one finds the multiplicity distribution~$P(N_+)$ 
generally more narrow than in the case of a neutral system. In particular towards the edge of the body of the 
distribution the approximation becomes worse. For the~MCE  massless gas approximations to multiplicity 
distributions~({\it top}) and ratios to the exact solution~({\it bottom}) are again compared in~Fig.(\ref{PN3}) for two 
system sizes.

A few general comments attempt to summarize. The first observation is that indeed as system size is increased 
a better description of the central region is found in terms of the asymptotic solution CLT. 
The second observation is that even for systems with a very small number (in the order of~$5$) of produced particles 
one finds a good approximation in terms of Gram-Charlier expansion. In particular GC5 provides a very accurate 
description of the central region with deviations in the order of a few percent. 
This is quite remarkable given the fact that multiplicity distributions for such small systems are not smooth 
and continuous functions of multiplicity, while the approximation Eq.(\ref{GramCharlier}) is. 
Furthermore, implicitly the concepts of chemical potential and temperature are introduced for systems with small 
particle number, which may be in contradiction to the common believe that these parameters can only be meaningful 
when the number of involved particles becomes very large, i.e. in the thermodynamic limit. The last observation is
that, indeed, see bottom rows of Figs.(\ref{PN1}-\ref{PN3}), finite volume corrections (given in terms of polynomials) 
lead only to good results for the central region of the distribution.

To give an estimate for a region in which the approximation is reliable, one notes that the finite volume 
approximation scheme begins to break down when the first expansion term in Eq. (\ref{GramCharlier}) becomes unity. 
In the one-dimensional case this would be:
\begin{equation}
\frac{h_3 \left( \xi_{max} \right)}{\sqrt{V}} \sim \mathcal{O}
\left(1 \right)~. \label{unityCond}
\end{equation}
Approximating the Hermite polynomial~$H_3 \sim \xi^3$, one can get an estimate for~$\xi_{max}$:
\begin{equation}
\xi_{max} \simeq \left(\frac{3!}{\lambda_3} \right)^{1/3} V^{1/6}~.
\end{equation}
While, when switching back to the definition of~$\xi= \frac{Q - V \kappa_1 }{\sigma \sqrt{V}}~$, Eq.(\ref{xi-var}), 
the width of the central region can be estimated by:
\begin{equation}
\frac{|Q - V \kappa_1|_{max} }{\sigma} \simeq
\left(\frac{3!}{\lambda_3} \right)^{1/3} V^{2/3}~.
\end{equation}
Hence the width of the central region scales as~$V^{2/3}$, while the width of the distribution should 
scale as~$V^{1/2}$ and the approximation should be quite good. Even though larger volumes work better, they will 
still be sufficiently small  enough to allow for calculation of distributions relevant for heavy ion collisions. 
However, it is stressed that there is no simple criterion for what is a `small` or a `large` volume for a 
particular physical system. Formally the existence and finiteness of (at least) the first three cumulants~$\kappa$ 
is sufficient for application of the asymptotic 
solution~\cite{Feller:1968fix_1,Feller:1970fix_1,Hughes:1995fix_1,Hughes:1996fix_1}. 
Considering the simple case of a multiplicity distribution of Bose-Einstein particles in the~GCE one finds from 
Eq.(\ref{psi})~$\kappa_1^N < \kappa_2^{N,N} < \kappa_3^{N,N,N}<\cdots$. 
Hence, in particular when finite chemical potentials are involved, cumulants are growing with order, implying 
that apart from mean value and variance further parameters like skewness and excess (or 
kurtosis)~\cite{Abramowitz:1965fix_1,Gradshteyn:2000fix_1} of the distribution will remain important quantities. 
For chemical potentials approaching the Bose-Einstein condensation point, one finds for instance the 
cumulant~$\kappa_2^{N,N}$ diverging. The expansion discussed here is hence not valid in this regime.

\section{Temperature and Chemical Potential}
\label{sec_clt_tempandchem}
The introduction of chemical potentials in the~CE and temperature in the~MCE was first and foremost a mathematical 
trick which allowed to conveniently integrate partition functions for which otherwise no analytical solution could 
be obtained. However the generalized partition function is self-consistent and not in contradiction to the common 
definition of temperature and chemical potential. It is  shown in the following that the definition of~$\beta$ 
and~$\mu$ through Eq.(\ref{gChemPot}) coincides with expressions well known from 
textbooks~\cite{Greiner:1997fix_1,Reichl:1998fix_1,Landau:2001fix_1}. 

\subsubsection{Canonical Ensemble}
The canonical partition function known from textbooks and the generalized version discussed here are connected as follows:
\begin{equation}
{Z}\left(V,Q^j,\beta\right)~\equiv~ 
\mathcal{Z}^{Q^j}\left(V,\beta,\mu_j\right) ~ 
e^{-Q^j\mu_j \beta}~.
\end{equation}
The Helmholtz free energy~$F$ is the thermodynamic potential relevant for~CE,
\begin{equation}
F ~\equiv~ - T \ln {Z}^{CE}~.
\end{equation}
Employing the first law of thermodynamics~$dE = TdS - PdV + \mu_j dQ^j$, and using~${F=E-TS}$ for the free energy, 
where~$P$ is the pressure, and~$E$ and~$S$ are total energy and entropy, respectively, one can write for the 
differential of the free energy~$dF = -S dT - P dV + \mu_j dQ^j$. 
The Lagrange multiplier~$\mu_j$ associated with conserved charge~$Q^j$ is thus defined by:
\begin{equation} \label{CEchemPot}
\left( \frac{\partial F}{\partial Q_j} \right)_{V,\beta}~=~ 
-\beta^{-1}~\frac{ \frac{\partial \mathcal{Z}^{Q^j}}{\partial Q_j} ~ e^{-Q^j\mu_j\beta}~ 
-~ \mu_j \beta ~\mathcal{Z}^{Q} ~ e^{-Q^j\mu_j\beta}}{\mathcal{Z}^{Q^j} ~
  e^{-Q^j \mu_j\beta }} ~=~ \mu_j~,
\end{equation}
where advantage of condition (\ref{gChemPot}) was taken. Thus, the correct choice for the effective chemical potential 
is given by the extremum,~${\frac{\partial \mathcal{Z}^{Q^j}}{\partial Q_j}= 0}$, of the charge distribution, which 
coincides with~$\mu_j = \left( \frac{\partial F}{\partial Q_j} \right)_{V,\beta}$, Eq.( \ref{CEchemPot}). 
In the thermodynamic limit this is equivalent to, Eq.(\ref{CLTmu}),~$\mu_j \rightarrow \mu_{j,gce}$. 
The subscripts indicate that the derivative with respect to the conserved charge has to be taken at fixed values 
of~$V$ and~$\beta$. In the non-relativistic case, where particle number~$N$, rather than quantum numbers,
is conserved, the corresponding relation to Eq.(\ref{CEchemPot}) would 
be~$\left( \frac{\partial F}{\partial N} \right)_{V,\beta} =\mu_N$~\cite{Greiner:1997fix_1,Reichl:1998fix_1,Landau:2001fix_1}. 
\subsubsection{Micro Canonical Ensemble}
The standard~MCE partition function can be obtain by Fourier integration of the generating function of the~GCE 
four-momentum distribution and multiplication with the inverse Boltzmann factor,
\begin{equation}
Z\left(V,P^{\mu}\right) ~\equiv~ 
\mathcal{Z}^{P^{\mu}}\left(V,\beta,u_{\mu} \right) ~
e^{P_{\mu}u^{\mu} \beta}~.
\end{equation}
The relevant thermodynamic potential in the~MCE is the entropy~$S$,
\begin{equation}
S ~\equiv~ \ln Z\left(V,P^{\mu} \right)~.
\end{equation}
The Lagrange multiplier for the temperature 
is~\cite{Greiner:1997fix_1,Reichl:1998fix_1,Landau:2001fix_1,Becattini:2003ft,Pathria:1957fix_1,Touschek:1968fix_1}:
\begin{equation} \label{MCETemp}
\left( \frac{\partial S}{\partial P_{\mu}} \right)_{V}~=~ 
\frac{ 
\frac{\partial \mathcal{Z}^{P^{\mu}}}{\partial P_{\mu}}  ~e^{P_{\mu}u^{\mu} \beta}
~ +~ u^{\mu} \beta  ~\mathcal{Z}^{P^{\mu}} ~ e^{P_{\mu}u^{\mu}\beta}}{
  \mathcal{Z}^{P^{\mu}} ~ e^{P_{\mu}u^{\mu} \beta}}
  ~=~ u^{\mu} \beta~,
\end{equation}
where again condition (\ref{gChemPot}) was employed, $ \frac{\partial \mathcal{Z}^{P^{\mu}}}{\partial P_{\mu}} = 0$. 
The subscript in Eq.(\ref{MCETemp}) is used to indicate that the derivative with respect to the components 
of~$P_{\mu}$ has to be taken at fixed volume~$V$. Thus Eq.(\ref{MCETemp}) resembles the optimal choice of an
effective temperature for the approximation scheme. In the thermodynamic limit,~$V\rightarrow \infty $, one finds 
${\beta^{\mu} \rightarrow \beta^{\mu}_{gce}}$, due to Eq.(\ref{CLTmu}).
\subsubsection{Grand Canonical Ensemble}
Conventionally, e.g. in textbooks, first the~MCE is introduced. Summation over energy - with temperature being a 
Lagrange multiplier - which is used to maximize the entropy, introduces the~CE. Additionally dropping the 
constraint of exact charge conservation leads to the~GCE. Here the chemical potentials~$\mu_j$ are the
Lagrange multipliers. The~GCE partition function can be written as:
\begin{eqnarray}
Z(V,\beta,u_{\mu},\mu_j) 
&=& \sum \limits_{\{Q^j\}} ~e^{Q^j \mu_j \beta}~ Z(V,\beta,u_{\mu},Q^j) \\
&=& \sum \limits_{\{Q^j\}} \sum \limits_{\{P^{\mu}\}}~~e^{Q^j \mu_j \beta} ~e^{-P_{\mu}u^{\mu} \beta}
~ Z(V,P^{\mu},Q^j)~.
\end{eqnarray}
While, in the notation used here, this line would read:
\begin{eqnarray}
Z(V,\beta,u_{\mu},\mu_j)  
&=& \sum \limits_{\{Q^j\}}  \mathcal{Z}^{Q^j}(V,\beta,u_{\mu},\mu_j) \\
&=& \sum \limits_{\{Q^j\}} \sum \limits_{\{P^{\mu}\}}~ \mathcal{Z}^{Q^j,P^{\mu}}(V,\beta,u_{\mu},\mu_j)~.
\end{eqnarray}
The thermodynamic potential for the~GCE is the grand potential~$\Omega$:
\begin{equation}
\Omega ~\equiv~ -T \ln Z(V,\beta,u_{\mu},\mu_j) 
~=~ -T \ln \mathcal{Z}(V,\beta,u_{\mu},\mu_j,\theta_l) \Big|_{\theta_l=0_l}~.
\end{equation}
The relations (\ref{CEchemPot}) and (\ref{MCETemp}) are essentially Legendre transformations from~$F$ and~$S$ 
to~${Z}\left(V,Q^j,\beta\right)$ and~$Z\left(V,P^{\mu}\right)$, respectively. The inverse Legendre transformations 
were obtained in the saddle-point expansion method (see Refs.~\cite{Becattini:2003ft,Becattini:2004rq,Becattini:2005cc}). 
Lastly it is emphasized that both methods are complementary from a thermodynamic point of view. 

Thus, if exact solutions of the canonical or micro canonical partition functions were available this reversal would 
not have been necessary. However, this redefinition of the~GCE partition function is entirely consistent and
simplifies calculations considerably. Whenever an exact solution to the generalized partition function is possible, 
all relations above would hold exactly. This interpretation of the~GCE partition function as the generating 
(or characteristic) function of a statistical system should be quite useful, even in more general cases than the one
presented here.

\section{Discussion}
\label{sec_clt_discussion}

An analytical expansion method for calculation of distributions at finite volume for the canonical as well as the
micro canonical ensembles of the ideal relativistic hadron resonance gas has been presented. The
introduction of temperature into the micro canonical partition function and chemical potentials into the canonical 
partition function have lead to the identification of the grand canonical partition function with the characteristic 
function of associated joint probability distributions. Micro canonical and canonical multiplicity distributions 
could, thus, be defined through conditional probability distributions. Hence the probability of finding a certain 
multiplicity, while other parameters (global charge or energy) were taken to be fixed. 

In considering finite volume corrections to the system partition function, thus relaxing the assumption of
thermodynamic equivalence of different statistical ensembles, one is lead to demand that the partition function 
should be maximized for a particular set of conserved charges. It turned out that this requirement is entirely 
equivalent to the well known textbook definitions of chemical potential in the canonical ensemble as the derivative 
of Helmholtz free energy with respect to conserved charge and temperature in the micro canonical ensemble through 
differentiation of entropy with respect to conserved energy.

This method is based on Fourier analysis of the grand canonical partition function. Conventionally one would not 
introduce chemical potentials and temperature into these calculations. However, one then faces the problem of a
heavily oscillating (or even irregular) integrand, making numerical integration unpractical. Artificially 
introduced temperature and chemical potentials, correctly chosen, produce a very smooth integrand allowing for expansion 
of the integrand in powers of volume. Analytical solutions to asymptotic distributions could thus be found in terms of 
Laplace's expansion, while finite volume corrections could be obtained from Gram-Charlier expansion. A comparison with 
available analytical solutions to simple statistical systems suggests that good results can be expected even for rather 
small system size. One drawback is that the results can only be applied to the central region of the distribution, 
owning to the fact that finite volume correction terms appear in the form of Hermite polynomial of low order.

Another, rather practical, drawback is that the effects of resonance decay~\cite{Begun:2006jf,Begun:2006uu,Hauer:2007ju} 
and limited acceptance~\cite{Hauer:2007im,Hauer:2009nv} are hard to consider simultaneously with the approach presented 
here. This will be solved by the Monte Carlo approach, considered in the next chapter.

\chapter{Monte Carlo Approach}
\label{chapter_montecarlo}
A statistical hadronization model Monte Carlo event generator provides the means for studying fluctuation and 
correlation observables in equilibrium systems. Data analysis can be done in close relation to experimental analysis 
techniques. Imposing global constraints on a sample is always technically more challenging.
Direct sampling of~MCE events (or micro states) has only been done in 
the non-relativistic limit~\cite{Randrup:1990fix_1,Randrup:1991fix_1}. 
Sample and reject procedures, suitable for relativistic systems, become rapidly 
inefficient with increasing system size. However, they have the advantage of being very 
successful for small system sizes~\cite{Becattini:2003ft,Becattini:2004rq,Becattini:2005cc}. 

In this chapter a different approach is taken: the~GCE is sampled, events are then re-weighted 
according to their values of extensive quantities, and the sample-reject limit 
(MCE) is approached in a controlled manner. In this way one can study the statistical properties 
of a global equilibrium system in their dependence on the size of its thermodynamic bath.
As any of the three standard ensembles remain idealizations of physical systems,
one might find these intermediate ensembles to be of phenomenological or conceptual interest too.

In Section~\ref{sec_montecarlo_finbath} statistical ensembles with finite thermodynamic bath are discussed. 
The~GCE Monte Carlo sampling procedure is described in Section~\ref{sec_montecarlo_gcesampling}.

\section{Statistical Ensembles with Finite Bath}
\label{sec_montecarlo_finbath}
The starting point taken is similar to the one chosen by Patriha~\cite{Patriha:1996fix_1}, and Challa and 
Hetherington~\cite{Challa:1988zz}. However quickly a different route is taken. Two systems, described by their micro 
canonical partition functions, i.e. the number of micro states for two separate systems, are assumed. The first system 
is further assumed to be enclosed in a volume~$V_1$ and to have fixed values of extensive 
quantities~${P_1^{\mu}=(E_1,P_{x,1},P_{y,1},P_{z,1})}$, and~$Q_1^j=(B_1,S_1,Q_1)$, while the second system is assumed to 
be enclosed in a volume~$V_2$ and to have fixed values of extensive quantities~${P_2^{\mu}=(E_2,P_{x,2},P_{y,2},P_{z,2})}$, 
and~$Q_2^j=(B_2,S_2,Q_2)$, where~$E$ is the energy of the system,~$P_{x,y,z}$ are the components of its three-momentum, 
and~$B$,~$S$, and~$Q$, are baryon number, strangeness and electric charge, respectively. Thus:
\begin{equation}\label{eq_one}
Z(V_1,P_1^{\mu},Q_1^j) ~=~ \sum_{\{N_1^i \}} ~ Z_{N_1^i}(V_1,P_1^{\mu},Q_1^j)~, ~~
~~\textrm{and}~~~~Z(V_2,P_2^{\mu},Q_2^j)~, 
\end{equation}
where~$Z_{N_1^i}(V_1,P_1^{\mu},Q_1^j)$ denotes the number of micro states of system 1 with additionally fixed 
multiplicities~$N_1^i$ of particles of all species~$i$ considered in the model. Supposing that system~1 and system~2 
are subject to the following constraints:
\begin{eqnarray}
V_g &=& V_1 ~+~ V_2 ~,\label{constraint_V}\\
P_g^{\mu} &=& P_1^{\mu} ~+~ P_2^{\mu}~,\label{constraint_P} \\
Q_g^{j} &=& Q_1^{j} ~+~ Q_2^{j}~. \label{constraint_Q}
\end{eqnarray}
Then, the partition function~$Z(V_g,P_g^{\mu},Q_g^j)$ of the joint system is constructed as the sums over all possible 
charge and energy-momentum split-ups:
\begin{eqnarray} \label{PF_combined}
Z(V_g,P_g^{\mu},Q_g^j)\!\!\! &=& \!\!\!\!\!
\sum \limits_{\{Q_1^j\}} \sum \limits_{\{P_1^{\mu}\}}
Z(V_g-V_1,P_g^{\mu}-P_1^{\mu},Q_g^j-Q_1^j)~ Z(V_1,P_1^{\mu},Q_1^j).
\end{eqnarray}
The distribution of extensive quantities in the subsystem~$V_1$ is given by the ratio of the number of all micro 
states consistent with a given charge and energy-momentum split-up and a given set of particle multiplicities 
to the number of all possible configurations:
\begin{equation}
P(P_1^{\mu},Q_1^j,N_1^i) ~=~ 
\frac{Z(V_g-V_1,P_g^{\mu}-P_1^{\mu},Q_g^j-Q_1^j)}{Z(V_g,P_g^{\mu},Q_g^j)} ~
~Z_{N_1^i}(V_1,P_1^{\mu},Q_1^j)~.
\end{equation}
Lastly, a weight factor~$W(V_1,P_1^{\mu},Q_1^j;V_g,P_g^{\mu},Q_g^j)$ is defined such that:
\begin{equation}\label{basic}
P(P_1^{\mu},Q_1^j,N_1^i) ~=~ W(V_1,P_1^{\mu},Q_1^j;V_g,P_g^{\mu},Q_g^j) ~
~Z_{N_1^i}(V_1,P_1^{\mu},Q_1^j)~.
\end{equation}
By construction, the first moment of the weight factor is equal to unity:
\begin{eqnarray}
\langle W \rangle &=&
\sum \limits_{\{Q_1^j\}} \sum \limits_{\{P_1^{\mu}\}} 
W(V_1,P_1^{\mu},Q_1^j;V_g,P_g^{\mu},Q_g^j) ~
~Z_{N_1^i}(V_1,P_1^{\mu},Q_1^j) = 1~,
\end{eqnarray}
as the distribution is properly normalized.

The weight factor~$W(V_1,P_1^{\mu},Q_1^j;V_g,P_g^{\mu},Q_g^j)~$ generates an ensemble with statistical properties 
different from the limiting cases of vanishing,~$V_g \rightarrow V_1$~(MCE), and of an infinite thermodynamic 
bath,~$V_g \rightarrow \infty$ (GCE).
This effectively allows for extrapolation of~GCE results to the~MCE limit. In principle any other (arbitrary) choice 
of~$W(V_1,P_1^{\mu},Q_1^j;V_g,P_g^{\mu},Q_g^j)~$ could be taken. This thesis confines itself to the situation discussed 
above. It is worth noting that all micro states consistent with the same set of extensive quantities~$(P_1^{\mu},Q_1^j)$ 
still have `a priori equal probabilities`. 

In the large volume limit, ensembles are equivalent in the sense that densities are the same. The ensembles defined 
by Eq.(\ref{basic}) and later on by Eq.(\ref{thetrick}) are no exception. If both~$V_1$ and~$V_g$ are sufficiently 
large, then the average densities in both systems will be the same,~$Q^j_g / V_g$ and~$P^{\mu}_g / V_g$ respectively. 
The system in~$V_1$ will hence carry on average a certain fraction: 
\begin{equation}\label{lambda_def}
\lambda ~\equiv~ V_1/V_g~,
\end{equation}
 of the total charge~$Q^j_g$ and four-momentum~$P^{\mu}_g$, i.e.: 
\begin{equation}
\langle Q^j_1 \rangle ~=~ \lambda~ Q^j_g ~,\qquad \textrm{and} 
\qquad \langle P_1^{\mu} \rangle ~=~ \lambda~ P^{\mu}_g~.
\end{equation} 
By varying the ratio~$\lambda = V_1 /V_g$, while keeping~$\langle Q^j_1 \rangle$ and~$\langle P_1^{\mu} \rangle$ constant, 
one can thus study a class of systems with the same average charge content and four-momentum, but different statistical 
properties. In the thermodynamic limit (it is enough to demand that $V_1$ is sufficiently large) a family of 
thermodynamically equivalent (same densities) ensembles is generated. 

\subsection{Introducing the Monte Carlo Weight~$\mathcal{W}$}
Since Eq.(\ref{basic}) poses a formidable challenge, both mathematically and numerically, one may write instead:
\begin{equation}\label{thetrick}
P(P_1^{\mu},Q_1^j,N_1^i) ~=~ \mathcal{W}^{P_1^{\mu},Q_1^j;P_g^{\mu},Q_g^j}(V_1;V_g|\beta,u_{\mu},\mu_j) ~
~P_{gce}(P_1^{\mu},Q_1^j,N_1^i|\beta,u_{\mu},\mu_j)~,
\end{equation}
where the distribution of extensive quantities~$P_1^{\mu}$,~$Q_1^j$ and particle multiplicities~$N_1^i$ of a~GCE system 
with temperature~$T=\beta^{-1}$, volume~$V_1$, chemical potentials~$\mu_j$ and collective four-velocity~$u_{\mu}$ is 
given by:
\begin{equation}\label{Pgce}
P_{gce}(P_1^{\mu},Q_1^j,N_1^i|\beta,u_{\mu},\mu_j) ~\equiv~ \frac{e^{-P_1^{\mu} u_{\mu} \beta }~ 
e^{Q_1^j \mu_j \beta}}{Z(V_1,\beta,u_{\mu},\mu_j)} ~ Z_{N_1^i}(V_1,P_1^{\mu},Q_1^j)~,
\end{equation}
where~$\mu_j = (\mu_B,\mu_S,\mu_Q)$ summarizes the chemical potentials associated with baryon number, strangeness and 
electric charge in a vector. The normalization in Eq.(\ref{Pgce}) is given by the~GCE partition 
function~$Z(V_1,\beta,u_{\mu},\mu_j)$, i.e. the number of all micro states averaged over the Boltzmann 
weights~$e^{-P_1^{\mu} u_{\mu} \beta }$ and~$e^{Q_1^j \mu_j \beta}$:
\begin{equation}
Z(V_1,\beta,u_{\mu},\mu_j)~=~ \sum_{\{P_1^{\mu}\}} \sum_{\{Q_1^{j} \}} \sum_{\{N_1^i \}}
~ e^{-P_1^{\mu} u_{\mu} \beta }~e^{Q_1^j \mu_j \beta}~  Z_{N_1^i}(V_1,P_1^{\mu},Q_1^j)~.
\end{equation}
The new  weight factor~$ \mathcal{W}^{P_1^{\mu},Q_1^j;P_g^{\mu},Q_g^j}(V_1;V_g|\beta,u_{\mu},\mu_j)$ now reads:
\begin{eqnarray}\label{newW}
 \mathcal{W}^{P_1^{\mu},Q_1^j;P_g^{\mu},Q_g^j}(V_1;V_g|\beta,u_{\mu},\mu_j) &=& 
Z(V_1,\beta,u_{\mu},\mu_j)~
\frac{e^{-(P_g^{\mu}-P_1^{\mu}) u_{\mu} \beta }~ 
e^{(Q_g^j-Q_1^j) \mu_j \beta} }{e^{-P_g^{\mu} u_{\mu} \beta}~ e^{Q_g^j \mu_j \beta} }  \nonumber \\
&\times&\frac{Z(V_g-V_1,P_g^{\mu}-P_1^{\mu},Q_g^j-Q_1^j)}{Z(V_g,P_g^{\mu},Q_g^j)}~.
\end{eqnarray}
In the case of an ideal (non-interacting) gas, Eq.(\ref{newW}) can be written, see Chapter~\ref{chapter_clt}, and 
Appendix~\ref{appendix_partfunc}, as:
\begin{eqnarray}\label{simpleW}
&& \mathcal{W}^{P_1^{\mu},Q_1^j;P_g^{\mu},Q_g^j}(V_1;V_g|\beta,u_{\mu},\mu_j) ~=~
Z(V_1,\beta,u_{\mu},\mu_j)~ \nonumber \\
&&\qquad \qquad \qquad \qquad\qquad \times~
\frac{\mathcal{Z}^{P_g^{\mu}-P_1^{\mu},Q_g^j-Q_1^j}(V_g-V_1,\beta,u_{\mu},\mu_j)}{
\mathcal{Z}^{P_g^{\mu},Q_g^j} (V_g,\beta,u_{\mu},\mu_j)} ~.
\end{eqnarray}
The advantage of Eq.(\ref{thetrick}), compared to Eq.(\ref{basic}), is that the 
distribution $P_{gce}(P_1^{\mu},Q_1^j,N_1^i|\beta,u_{\mu},\mu_j)$ can easily be sampled for Boltzmann particles, while a 
suitable approximation for the weight~$\mathcal{W}^{P_1^{\mu},Q_1^j;P_g^{\mu},Q_g^j}(V_1;V_g|\beta,u_{\mu},\mu_j)$ is available.
Again, by construction, the first moment of the new weight factor is equal to unity:
\begin{equation}
\langle \mathcal{W} \rangle ~= \sum_{\{P_1^{\mu}\}} \sum_{\{Q_1^j \}} \sum_{\{N_1^i \}} ~  
\mathcal{W}^{P_1^{\mu},Q_1^j;P_g^{\mu},Q_g^j}(V_1;V_g|\beta,u_{\mu},\mu_j) 
~P_{gce}(P_1^{\mu},Q_1^j,N_1^i|\beta,u_{\mu},\mu_j)  = 1~.
\end{equation}
In principle, Eq.(\ref{basic}) and Eq.(\ref{thetrick}) are equivalent. In fact, Eq.(\ref{basic}) can be obtained by 
taking the limit $(\mu_B,\mu_S,\mu_Q) = (0,0,0)$,~$u_{\mu}=(1,0,0,0)$, and~$\beta \rightarrow 0$ of Eq.(\ref{thetrick}). 
However, as one can easily see,~$\langle \mathcal{W}^n \rangle \not= \langle W^n \rangle $. Higher, and in particular 
the second, moments of the weight factors~$W$ and~$\mathcal{W}$ are a measure of the statistical error to be expected 
for a finite sample of events. The larger the higher moments of the weight factor, the larger the statistical error, and 
the slower the convergence with sample size. Please see also Appendices~\ref{appendix_convstudy} and~\ref{appendix_cbg}.

As~GCE and~MCE densities are the same in the system~$V_g$, these values are effectively regulated by intensive 
parameters~$\beta$,~$\mu_j$ and~$u_{\mu}$. In order to study a system with average~$\langle Q^j_1 \rangle$, 
it is most economical to sample the~GCE with~$\langle Q^j_1 \rangle$ and calculate weights according to 
Eq.(\ref{simpleW}). This will result in a low statistical error for finite samples (as shown in later sections), and 
allow for extrapolation to the~MCE limit.

Firstly the weight factor Eq.(\ref{simpleW}) will be calculated, and then the relevant limits are taken. With the 
appropriate choice of~$\beta$,~$\mu_j$ and~$u_{\mu}$ the calculation of Eq.(\ref{simpleW}) can be done with 
the method presented in Chapter~\ref{chapter_clt}.

\subsection{Calculating the Monte Carlo Weight~$\mathcal{W}$}
For the Monte Carlo approach in this chapter, the total number of (potentially) conserved extensive quantities in
a hadron resonance gas is~$L=J+4 = 3+4=7$, where~$J=3$ is the number of charges~$(B,S,Q)$ and there are four components 
of the four-momentum. Particle number integrals are not going to be solved here. Including all extensive quantities 
into a single vector:
\begin{equation}
\mathcal{Q}^l = (Q^j,P^{\mu}) = (B,S,Q,E,P_x,P_y,P_z)~,
\end{equation}
the weight Eq.(\ref{simpleW}) can be expressed as:
\begin{eqnarray}\label{curly_W}
\mathcal{W}^{\mathcal{Q}_1^l;\mathcal{Q}_g^l}(V_1;V_g|\beta,u_{\mu},\mu_j) =
Z(V_1,\beta,u_{\mu},\mu_j) 
\times
\frac{\mathcal{Z}^{\mathcal{Q}_g^l-\mathcal{Q}_1^l}(V_g-V_1,\beta,u_{\mu},\mu_j)}{
\mathcal{Z}^{\mathcal{Q}_g^l} (V_g,\beta,u_{\mu},\mu_j)} ~.
\end{eqnarray}
The general expression for the partition function~$\mathcal{Z}^{\mathcal{Q}^l}(V,\beta,u_{\mu},\mu_j)$ 
in the large volume limit is given by Eq.(\ref{MND}):
\begin{equation}\label{clt_approx}
\mathcal{Z}^{\mathcal{Q}^l}(V,\beta,u_{\mu},\mu_j) ~\simeq~ 
\frac{Z(V,\beta,u_{\mu},\mu_j)}{(2 \pi V)^{L/2} \det|\sigma|}~ 
\exp \left[ - \frac{1}{2}~ \frac{1}{V}~ \xi^l \xi_l \right]~,
\end{equation}
where, unlike in Eq.(\ref{xi-var}), the volume is split off:
\begin{equation}\label{xi_noV}
\xi^l ~=~ \left(\mathcal{Q}^k - V \kappa_1^k \right) ~ \left( \sigma^{-1} \right)_k^l~,
\end{equation}
while the definition of the sigma tensor, Eq.(\ref{sigmatensor}), remains:
\begin{equation}
\sigma_k^l ~=~ \left( \kappa_2^{1/2} \right)_k^l ~.
\end{equation}
Here~$\kappa_1$ and~$\kappa_2$ are the~GCE vector of mean values and the GCE covariance matrix respectively. 
The values of~$\beta$,~$\mu_j$ and~$u_{\mu}$ are chosen such (Eq.(\ref{gChemPot})) that:
\begin{equation}
\frac{\partial \mathcal{Z}^{\mathcal{Q}^l}}{\partial \mathcal{Q}^l}
\Big|_{\mathcal{Q}^l=\mathcal{Q}^l_{eq}} ~=~ 0_l.
\end{equation}
The approximation (\ref{clt_approx}) gives then a reliable description of~$\mathcal{Z}^{\mathcal{Q}_g^l}$ around the 
equilibrium value~$\mathcal{Q}_g^l = V_g \kappa_1^l$, provided~$V_g$ is sufficiently large. The charge vector, 
Eq.(\ref{xi-var}), is then equal to the null-vector~$\xi_l = 0_l$ ($\mathcal{Q}_g^l = V_g \kappa_1^l$).
For the denominator in Eq.(\ref{curly_W}) one then finds:
\begin{equation} \label{W_norm}
\mathcal{Z}^{\mathcal{Q}_g^l}(V_g,\beta,u_{\mu},\mu_j)\Big|_{\mathcal{Q}_g^l=\mathcal{Q}^l_{g,eq}} 
~\simeq~ \frac{Z(V_g,\beta,u_{\mu},\mu_j)}{(2 \pi V_g)^{L/2}\det|\sigma|}~ 
\exp \left[ ~0 ~\right]~,
\end{equation}
while for the numerator one obtains:
\begin{eqnarray}  \label{W_numm}
&&\mathcal{Z}^{\mathcal{Q}_g^l-\mathcal{Q}_1^l}(V_g-V_1,\beta,u_{\mu},\mu_j)
\Big|_{\mathcal{Q}_g^l=\mathcal{Q}^l_{g,eq}} \simeq
\frac{Z(V_g-V_1,\beta,u_{\mu},\mu_j) }{(2 \pi \left(V_g -V_1 \right))^{L/2} \det|\sigma|}~
\qquad\qquad\qquad\nonumber \\
&&\qquad \qquad\qquad \qquad \qquad \qquad\qquad \qquad \times 
\exp \Bigg[ -\frac{1}{2}~ \frac{1}{(V_g-V_1) }~ \xi^l \xi_l\Bigg]~,
\end{eqnarray}
where the charge vector~$\xi^l$, Eq.(\ref{xi_noV}), in Eq.(\ref{W_numm}) is:
\begin{equation}
\xi^l = \left( \Delta \mathcal{Q}_2 \right)^k \left( \sigma^{-1} \right)_k^l~.
\end{equation}
Then, using the condition~$\mathcal{Q}^k_g = \mathcal{Q}^k_{g,eq} = V_g \kappa_1^k$ yields:
\begin{equation}
\left( \Delta \mathcal{Q}_2 \right)^k = \left(\mathcal{Q}_g - \mathcal{Q}_1 \right)^k - 
\left( V_g-V_1\right) \kappa_1^k ~=~  -\left( \mathcal{Q}_1 -V_1\kappa_1 \right)^k~.
\end{equation}
Substituting Eq.(\ref{W_norm}) and Eq.(\ref{W_numm}) into Eq.(\ref{curly_W}) results in:
\begin{eqnarray}\label{almost_solved_curly_W}
 \mathcal{W}^{\mathcal{Q}_1^l;\mathcal{Q}_g^l}(V_1;V_g|\beta,u_{\mu},\mu_j)
\Big|_{\mathcal{Q}_g^l=\mathcal{Q}^l_{g,eq}}
 &\simeq& \frac{Z(V_1,\beta,u_{\mu},\mu_j)~Z(V_g-V_1,\beta,u_{\mu},\mu_j)}{
Z(V_g,\beta,u_{\mu},\mu_j)} \nonumber \\
 &&\times ~\frac{(2 \pi V_g )^{L/2} \det|\sigma|}{(2 \pi \left(V_g -V_1 \right))^{L/2} \det|\sigma|} 
\nonumber \\
&&\times ~\exp \Bigg[ -\frac{1}{2}~ \frac{1}{(V_g-V_1) }~ \xi^l \xi_l\Bigg]~.
\end{eqnarray}
The~GCE partition functions are multiplicative in the sense that their product is 
$Z(V_1,\beta,u_{\mu},\mu_j)~Z(V_g-V_1,\beta,u_{\mu},\mu_j) = Z(V_g,\beta,u_{\mu},\mu_j)$, and thus the first term in 
Eq.(\ref{almost_solved_curly_W}) is equal to unity. Now using Eq.(\ref{lambda_def}),~$\lambda = V_1 / V_g$, one can 
re-write Eq.(\ref{almost_solved_curly_W}) as:
\begin{eqnarray}\label{solved_curly_W}
 \mathcal{W}^{\mathcal{Q}_1^l;\mathcal{Q}_g^l}(V_1;V_g|\beta,u_{\mu},\mu_j)
\Big|_{\mathcal{Q}_g^l=\mathcal{Q}^l_{g,eq}}
 &\simeq& \frac{1}{(1-\lambda)^{L/2} }   \\
&&\times~ \exp \Bigg[ -\frac{1}{2} \left( \frac{\lambda}{1-\lambda}\right) 
\frac{1}{V_1}~ \xi^l \xi_l\Bigg].\nonumber
\end{eqnarray}
Model parameters are hence the intensive variables inverse temperature~$\beta$, four-velocity $u^{\mu}$ and chemical 
potentials~$\mu^j$, which regulate energy and charge densities, and collective motion. Provided~$V_1$ is sufficiently 
large, a family of thermodynamically equivalent ensembles is defined, which can now be studied in their dependence of 
fluctuation and correlation observables on the size of the bath~$V_2 = V_g - V_1$. Hence, one can test the sensitivity of
such observables, for example, to globally applied conservation laws. The expectation values are then identical 
to~GCE expectation values, while higher moments will depend  crucially on the choice of~$\lambda$.

\subsection{The Limits of~$\mathcal{W}$}
The largest weight is given to states for which~$\xi^l \xi_l = 0$, i.e. with extensive 
quantities~$\mathcal{Q}_1^l = \mathcal{Q}_{1,eq.}^l$. Hence, the maximal weight a micro state (or event) at a given 
value of~$\lambda = V_1/V_g$ can assume 
is $\mathcal{W}_{max}^{\mathcal{Q}_1^l;\mathcal{Q}_g^l}(V_1;V_g|\beta,u_{\mu},\mu_j) = (1-\lambda)^{-L/2}$. 
Taking the limits of Eq.(\ref{solved_curly_W}), it is easy to see that:
\begin{equation}
\lim_{\lambda \rightarrow 0} ~ \mathcal{W}^{\mathcal{Q}_1^l;\mathcal{Q}_g^l}
(V_1;V_g|\beta,u_{\mu},\mu_j) ~=~ 1~.
\end{equation}
I.e. for~$\lambda = 0$ the~GCE is sampled, and all events have a weight equal to unity. Hence, one also 
finds~$\langle \mathcal{W}^2 \rangle = 1$ and therefore $\langle (\Delta \mathcal{W})^2 \rangle = 0$, implying a low 
statistical error. For~$\lambda \rightarrow 1$, effectively  the sample-reject 
limit~\cite{Becattini:2003ft,Becattini:2004rq,Becattini:2005cc} is approached. Accordingly:
\begin{equation}
\lim_{\lambda \rightarrow 1} ~ \mathcal{W}^{\mathcal{Q}_1^l;\mathcal{Q}_g^l}
(V_1;V_g|\beta,u_{\mu},\mu_j) ~\propto~ \delta(\mathcal{Q}_1^l - V_1 \kappa_1^l)~.
\end{equation}
However, as now not all events have equal weight,~$\langle (\Delta \mathcal{W})^2 \rangle$ grows 
and so too the statistical error of finite samples. Also, the larger the number $L$ of extensive quantities considered 
for re-weighting, the larger will be the statistical uncertainty.

\section{The~GCE Sampling Procedure}
\label{sec_montecarlo_gcesampling}
The Monte Carlo sampling procedure for a~GCE system in the Boltzmann approximation is now explained.
The system to be sampled is assumed to be in an equilibrium state enclosed in a volume $V_1$ with 
temperature~$T = \beta^{-1}$ and chemical potentials~$\mu_j = (\mu_B,\mu_S,\mu_Q)$. Additionally, the system is assumed 
to be at rest. The four-velocity is then~$u^{\mu} = (1,0,0,0)$ and the four-temperature is~$\beta^{\mu}= (\beta,0,0,0)$. 
In this case, multiplicity distributions are Poissonian, while momentum spectra are of Boltzmann type. 
The~GCE sampling process is composed of four steps, each discussed below.

\subsubsection{Multiplicity Generation}
In the first step, multiplicities~$N_1^i$ are randomly sampled of all particle species~$i$ considered in the model. 
The expectation value of the multiplicity of thermal Boltzmann particles in the~GCE is given by:
\begin{equation}\label{psi_meanN}
\langle N_1^i \rangle ~=~ \frac{g_i V_1}{2 \pi^2} ~m_i^2 ~\beta^{-1}~ 
K_2\left( m_i \beta \right)~ e^{\mu_i\beta}~.
\end{equation}
Multiplicities~$\lbrace N_1^i \rbrace_n$ are randomly generated for each event~$n$ according to Poissonians with 
mean values~$\langle N_1^i \rangle$:
\begin{equation}
P(N_1^i) ~=~ \frac{\langle N_1^i \rangle^{N_1^i}}{N_1^i!} ~ e^{-\langle N_1^i \rangle}~.
\end{equation}
In the above,~$m_i$ and~$g_i$ are the mass and spin-degeneracy factor of a particle of species~$i$ respectively.
The chemical potential~$\mu_i = \mu_j q_i^j = \mu_B b_i + \mu_S s_i + \mu_Q q_i$, where $q_i^j = (b_i,s_i,q_i)$ 
represents the quantum number content of a particle of species~$i$.

\subsubsection{Momentum Spectra}
In the second step, momenta are generated for each particle according to a Boltzmann spectrum. For a static thermal 
source spherical coordinates are convenient:
\begin{equation}
\frac{dN_i}{d|p|} ~=~ \frac{g_i V_1}{2 \pi^2}~ \beta^{-3} ~|p|^2~ e^{-\varepsilon \beta}~.
\end{equation}
These momenta are then isotropically distributed in momentum space. Hence:
\begin{eqnarray}
p_x &=& |p| ~ \sin \theta ~\cos \phi ~,\\
p_y &=& |p| ~ \sin \theta ~\sin \phi ~,\\
p_z &=& |p| ~ \cos \theta  ~,\\
\varepsilon &=& \sqrt{|p|^2 + m_i^2}~,
\end{eqnarray}
where~$p_x$,~$p_y$, and~$p_z$ are the components of the three-momentum,~$\varepsilon$ is the energy, 
and~$|p| = \sqrt{p_x^2+p_y^2+p_z^2}$ is the total momentum. The polar and azimuthal angles are sampled according to:
\begin{eqnarray}
\theta &=& \cos^{-1} \left[ 2 \left(x-0.5 \right) \right] ~,\\
\phi &=& 2~\pi \left(x-0.5 \right)~,
\end{eqnarray}
where~$x$ is uniformly distributed between~$0$ and~$1$. Additionally, the transverse momentum~$p_T$ and rapidity~$y$ 
are calculated for each particle:
\begin{eqnarray}
p_T &=& \sqrt{p_x^2 + p_y^2}~,\\
y &=& \frac{1}{2} \ln \left(\frac{\varepsilon+p_z}{\varepsilon-p_z} \right)~.
\end{eqnarray}
Finally, particles are distributed homogeneously in a sphere of radius~$r_1$ and decay times are calculated based 
on the Breit-Wigner width of the resonances.

\subsubsection{Resonance Decay}
The third step (if applicable) is resonance decay. Following the prescription used by the authors of the 
THERMINATOR package~\cite{Kisiel:2005hn}, only 2 and 3 body decays are performed, while unstable daughter particles 
are allowed to decay in a successive manner. Only strong and electromagnetic decays are considered, while weak decays 
are omitted. Particle decay is first calculated in the parent's rest frame, with daughter momenta then boosted into 
the lab frame. Finally, decay positions are generated based on the parent's production point, momentum and life time.

Throughout this thesis, always only the lightest states of the following baryons:
\begin{equation}
\textrm{p} \qquad \textrm{n} \qquad \Lambda \qquad \Sigma^+ \qquad \Sigma^- \qquad \Xi^- \qquad 
\Xi^0 \qquad \Omega^-
 \end{equation}
and their respective anti-baryons, as well as following mesons:
\begin{equation}
\pi^+ \qquad \pi^- \qquad \pi^0 \qquad K^+ \qquad K^- \qquad K^0 \qquad \overline{K}^0 
 \end{equation}
are considered as stable. The system could now be given collective velocity~$u^{\mu}$.

\subsubsection{Re-weighting}
In the fourth step, the values of extensive quantities are calculated for the events generated by iteration over the 
particle list of each event. For the values of extensive 
quantities~$\mathcal{Q}^l_{1,n}~=~(B_{1,n},S_{1,n},Q_{1,n},E_{1,n},P_{x,1,n},P_{y,1,n},P_{z,1,n})$ in 
subsystem~$V_1$ of event~$n$ one may write: 
\begin{equation}\label{code_charge}
\mathcal{Q}^l_{1,n} ~=~ \sum_{\textrm{particles } i_n } \mathfrak{q}^l_{i_n}~,
\end{equation}
where~$\mathfrak{q}^l_{i_n} = \left( b_{i_n}, s_{i_n}, q_{i_n}, \varepsilon_{i_n}, p_{x,i_n}, p_{y,i_n}, p_{z,i_n}\right)$ is 
the `charge vector' of particle~$i$  in event~$n$. Based on~$\mathcal{Q}^l_{1,n}$ the weight~$w_n$ is calculated for 
the event:
\begin{equation}\label{code_weight}
w_n = \mathcal{W}^{\mathcal{Q}_{1,n}^l;\mathcal{Q}_g^l}(V_1;V_g|\beta,u_{\mu},\mu_j)~,
\end{equation}
according to Eq.(\ref{solved_curly_W}). All micro states (or events) with the same set of extensive 
quantities~$\mathcal{Q}^l_{1,n}$ are still counted equally.

\subsubsection{Monte Carlo Distributions}

The Monte Carlo output is essentially a distribution~$P_{MC}(X_1,X_2,X_3,...)$ of a set of 
observables~$X_1$,~$X_2$,~$X_3$, etc. For all practical purposes this distribution is obtained by histograming all 
events~$n$ according to their values of $X_{1,n}$,~$X_{2,n}$,~$X_{3,n}$, etc. and their weight~$w_n$. 
One can then define moments of two observables~$X_i$ and~$X_j$ through:
\begin{equation}\label{MCmoment}
  \langle X_i^n X_j^m \rangle ~\equiv~ \sum_{X_i,X_j} X_i^n~ X_j^m ~P_{MC}(X_i,X_j)~.
\end{equation}
Additionally, the variance~$\langle \left( \Delta X_i \right)^2 \rangle$ and the 
covariance~$\langle \Delta X_i \Delta X_j \rangle$ respectively are, defined as:
\begin{eqnarray}\label{variance}
\langle \left( \Delta X_i \right)^2 \rangle &\equiv&  \langle X_i^2\rangle ~-~ 
\langle X_i  \rangle^2~, \qquad \textrm{and} \\
\label{covariance}
\langle \Delta X_i \Delta X_j \rangle &\equiv&  \langle X_i  X_j \rangle 
~-~ \langle X_i \rangle  \langle X_j \rangle~.
\end{eqnarray}
In the following, the correlation coefficient~$\rho_{ij}$, defined as:
\begin{equation}
\label{rho}
\rho_{ij} ~\equiv~ \frac{\langle \Delta X_i \Delta X_j \rangle}{
\sqrt{\langle \left( \Delta X_i \right)^2 \rangle 
\langle \left( \Delta X_j \right)^2 \rangle }}~,
\end{equation}
and later on the scaled variance~$\omega_i$, given by:
\begin{equation}
\label{omega}
\omega_i ~\equiv~ \frac{\langle \left( \Delta X_i \right)^2 \rangle}{\langle X_i \rangle}~,
\end{equation}
are used for quantification of fluctuations and correlations. The scaled variance~$\omega_i$, is often also called
the `Fano Factor'~\cite{Fano:1947zz}, and is a useful measure for the width of a distribution, provided $X_i>0$, 
such as in the case of particle multiplicity. 

\section{Discussion}
\label{sec_montecarlo_discussion}
In this chapter distributions of extensive quantities in equilibrium systems with finite thermodynamic bath have 
been discussed. A recipe for a thermal model Monte Carlo event generator has been presented. The event generator is
capable of extrapolating fluctuation and correlation observables for Boltzmann systems of large volume from 
their~GCE values to the~MCE limit. This approach has a strong advantage compared to analytical approaches or
standard micro canonical sample-and-reject Monte Carlo techniques, in that resonance decays as well as (very) large 
system sizes can handled at the same time.

To introduce the scheme,  a micro canonical system has been conceptually divided into two subsystems.
These subsystems are assumed to be in equilibrium with each other, and subject to the constraints of joint 
energy-momentum and charge conservation. Particles are only measured in one subsystem, while the second subsystem
provides a thermodynamic bath. By keeping the size of the first subsystem fixed, while varying the size of the second, 
one can thus study the dependence of statistical properties of an ensemble on the fraction of the system 
observed (i.e. assess their sensitivity to globally applied conservation laws). The ensembles generated are 
thermodynamically equivalent in the sense that mean values in the observed subsystem remain unchanged
when the size of the bath is varied, provided the combined system is sufficiently large. 

The Monte Carlo process can be divided into four steps. In the first two steps primordial particle multiplicities 
for each species, and momenta for each particle, are generated for each event by sampling the grand canonical 
partition function. In the third step resonance decay of unstable particles is performed. Lastly, the values of 
extensive quantities are calculated for each event and a corresponding weight factor is assigned. All events with 
the same set of extensive quantities hence still have `a priori equal probabilities'. In the limit of an infinite bath, 
all events have a weight equal to unity. In the opposite limit of a vanishing bath, only events with an exactly 
specified set of extensive quantities have non-vanishing weight. In between, extrapolation is done in a controlled manner. 
The method is particularly efficient for systems of large volume, inaccessible to sample-and-reject procedures, and 
agrees well, where available, with analytic asymptotic micro canonical solutions. 

Given the success of  the hadron resonance gas model in describing experimentally measured average hadron yields, 
and its ability to reproduce low temperature lattice susceptibilities, the question arises as to whether fluctuation 
and correlation observables also follow its main line. The first and, in particular, second moments of joint 
distributions of extensive quantities will be studied in the following. The focus will be mainly on particle multiplicity 
distributions and distributions of `conserved' charges. In particular, the effects of resonance decay, conservation laws, 
and limited acceptance in momentum space are discussed. Due to the Monte Carlo nature, data can be analyzed 
in close relation to experimental analysis techniques. The hadron resonance gas is an ideal testbed 
for this type of study, in that it is simple and intuitive.

\chapter{Grand Canonical Ensemble}
\label{chapter_gce}
The grand canonical ensemble of the hadron resonance gas model will be the starting point of the discussion of 
statistical properties of equilibrium systems. The GCE is considered to be the most accessible amongst the standard 
ensembles. Its discussion will provide the basis for the study of canonical and micro canonical ensembles in the 
following chapters. The GCE is the correct ensemble to choose, if the observed subsystem is embedded into a much 
larger heat and charge bath. Primordial momenta of particles within a small subvolume are then uncorrelated with 
each other in the Boltzmann approximation. 

Extensive quantities are fluctuating, while corresponding intensive variables are constant. Hence, in a subvolume 
one finds the energy content (extensive) to fluctuate, while temperature (intensive) is constant. 
Chemical potentials are constants, but the corresponding net-charge content is fluctuating. 
And naturally in a relativistic gas, the systems total particle number and integrated transverse momentum 
(no Lagrange multiplier associated with them) also fluctuate. 
However, extensive quantities do not merely fluctuate, but are also correlated with each other.

On the other hand, correlations observables do not depend on where they are measured in coordinate space, due to 
the infinite bath assumption. Observable correlation and fluctuation signals are, however, sensitive on the 
location and size of the available acceptance window in momentum space within which particles can be measured. 

In Section~\ref{sec_gce_jdeq} joint distributions of fully phase space integrated extensive quantities are considered.
In Section~\ref{sec_gce_MomSpect} momentum spectra of primordial and final state hadrons are analyzed. 
Correlations between extensive quantities are studied in their dependence on the acceptance window in momentum space
in Sections~\ref{sec_gce_LCfluc} and~\ref{sec_gce_CorrFunc}.

\section{Joint Distributions of Extensive Quantities}
\label{sec_gce_jdeq}

\begin{figure}[ht!]
  \epsfig{file=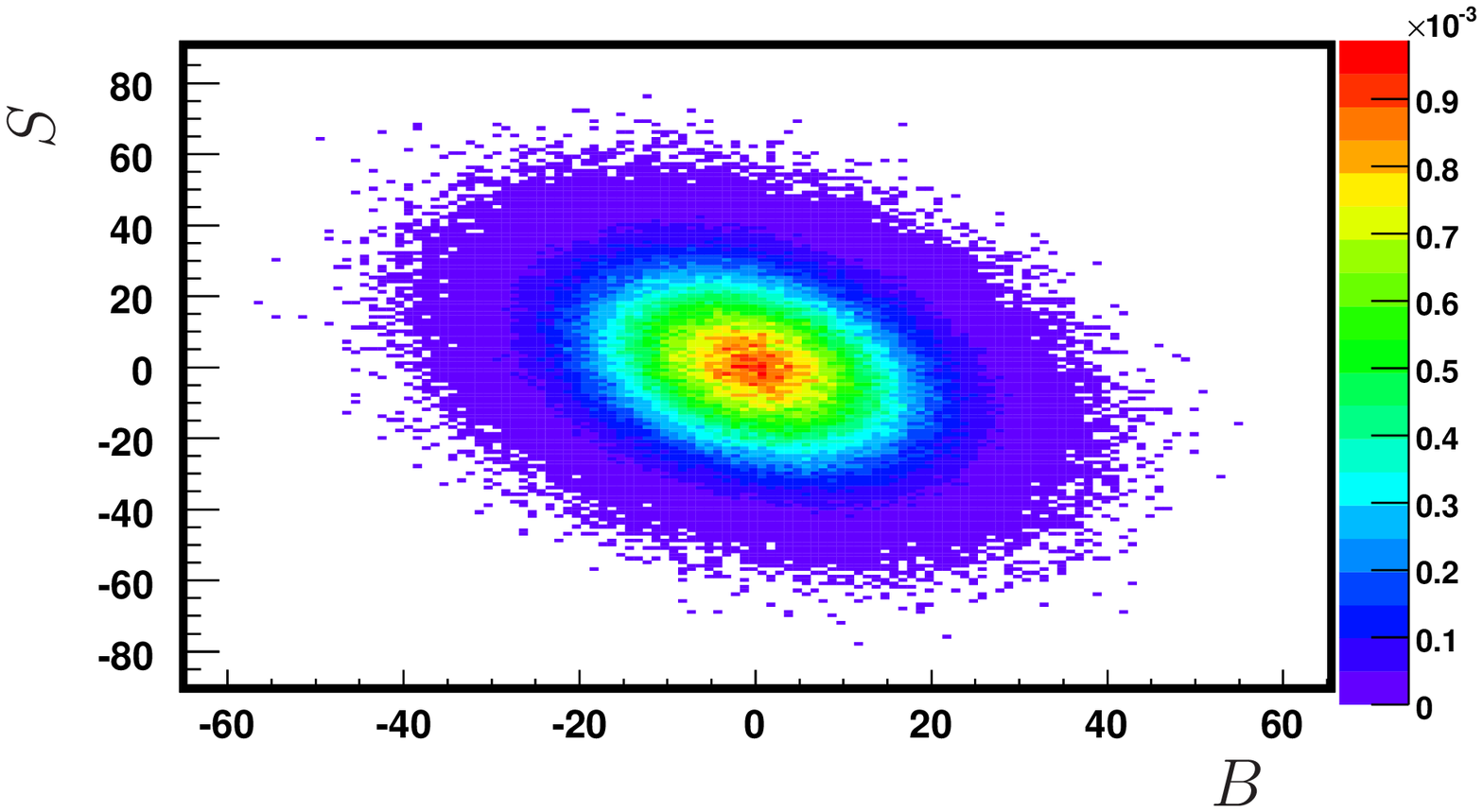,width=7.2cm,height=6.5cm}
  \epsfig{file=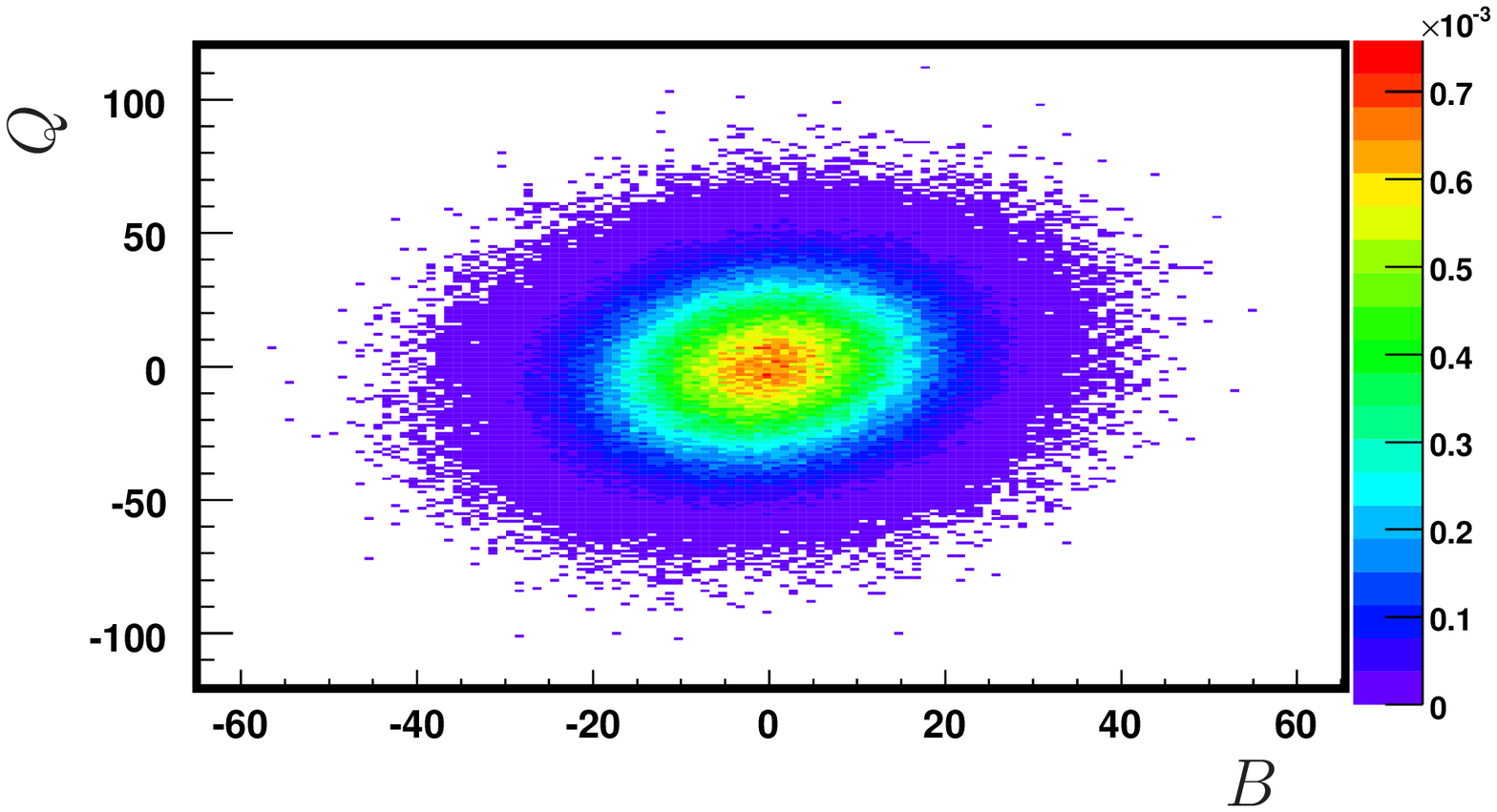,width=7.2cm,height=6.5cm}
  \caption{Grand canonical joint distribution of baryon number~$B$ and strangeness~$S$~({\it left}), and 
  grand canonical joint distribution of baryon number~$B$ and electric charge~$Q$~({\it right}), 
  for a neutral hadron resonance gas at~$T=0.160$~GeV and~$V=2000$~fm$^3$. 
  Here~$5 \cdot 10^5$~events have been sampled.}  
  \label{P_0000_neu_1}
\end{figure}

\begin{figure}[ht!]
  \epsfig{file=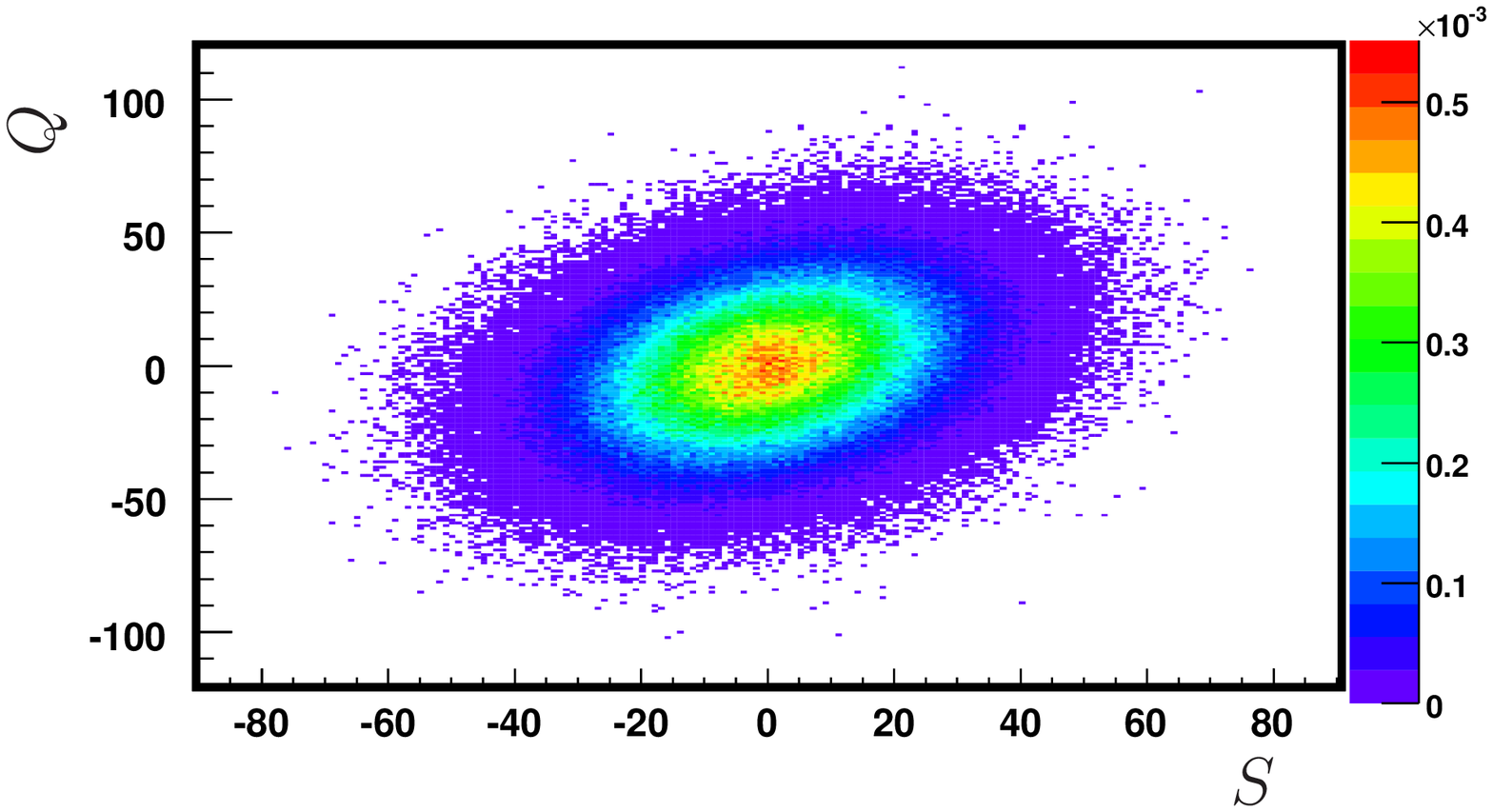,width=7.2cm,height=6.5cm}
  \epsfig{file=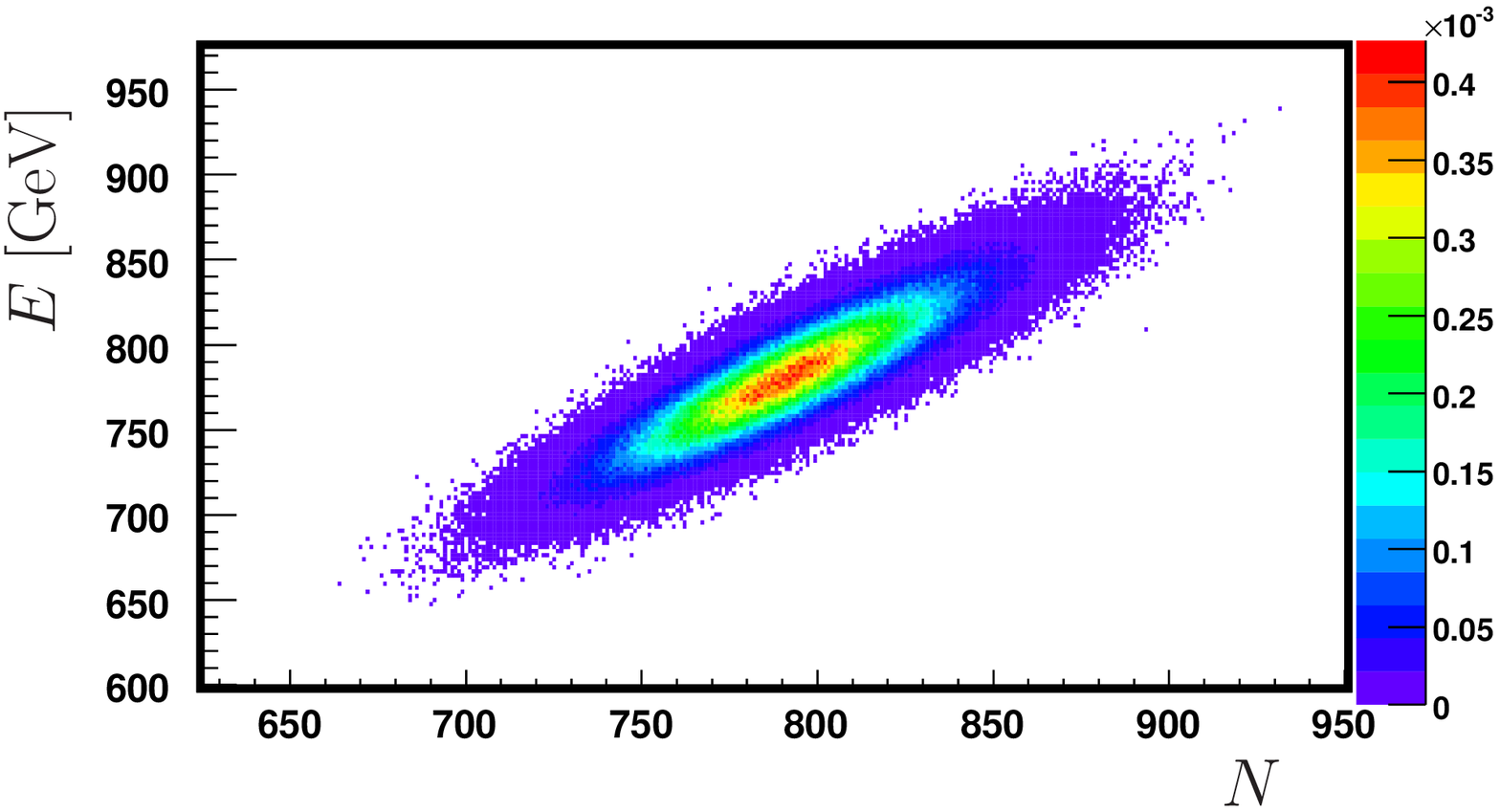,width=7.2cm,height=6.5cm}
  \caption{Grand canonical joint distribution of strangeness~$S$ and 
  electric charge~$Q$~({\it left}), and grand canonical joint distribution 
  of energy~$E$ and total particle number~$N$~({\it right}).
  The rest as in Fig.(\ref{P_0000_neu_1}).
  }  
  \label{P_0000_neu_2}
\end{figure}

\begin{figure}[ht!]
  \epsfig{file=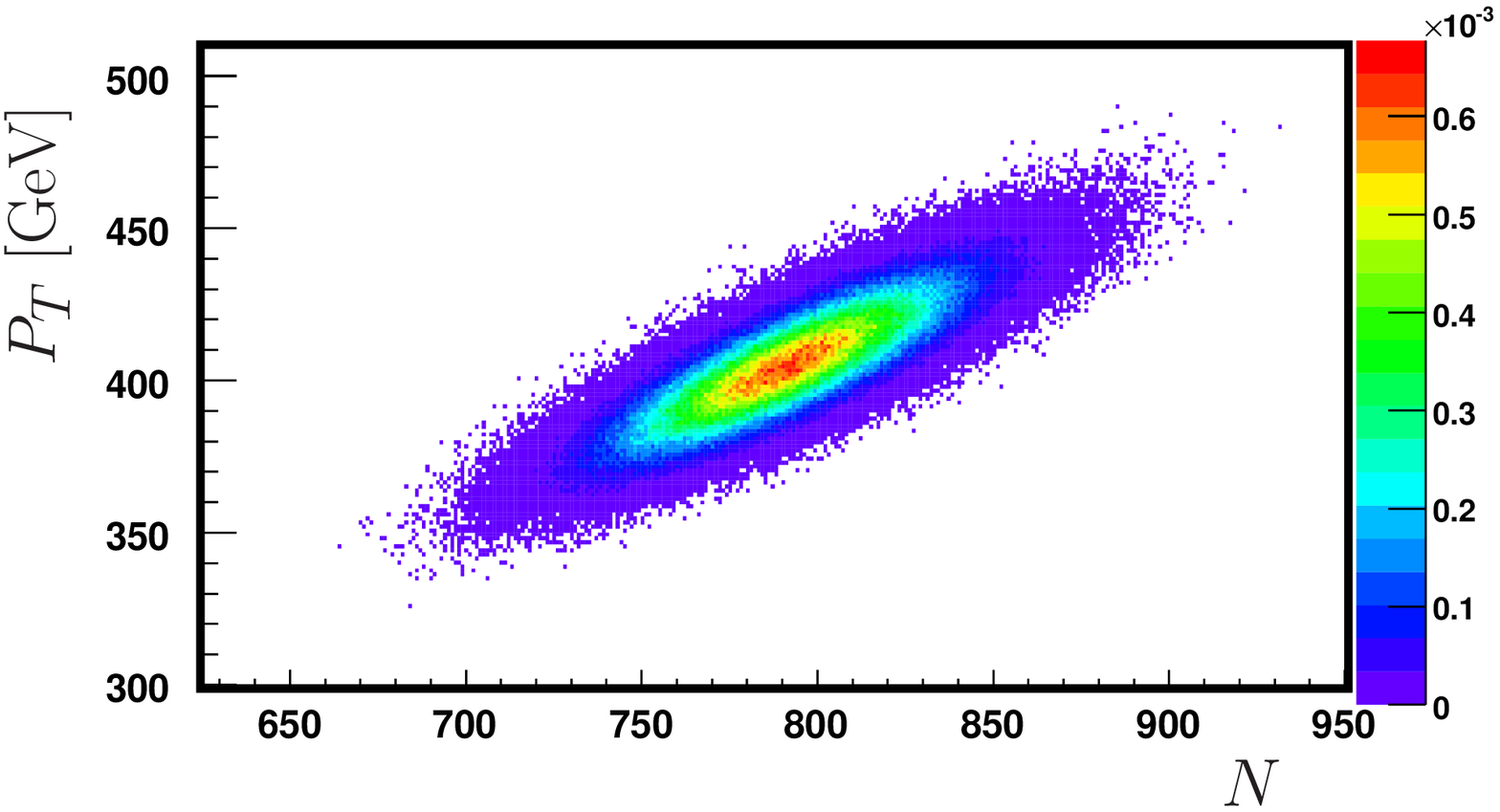,width=7.2cm,height=6.5cm}
  \epsfig{file=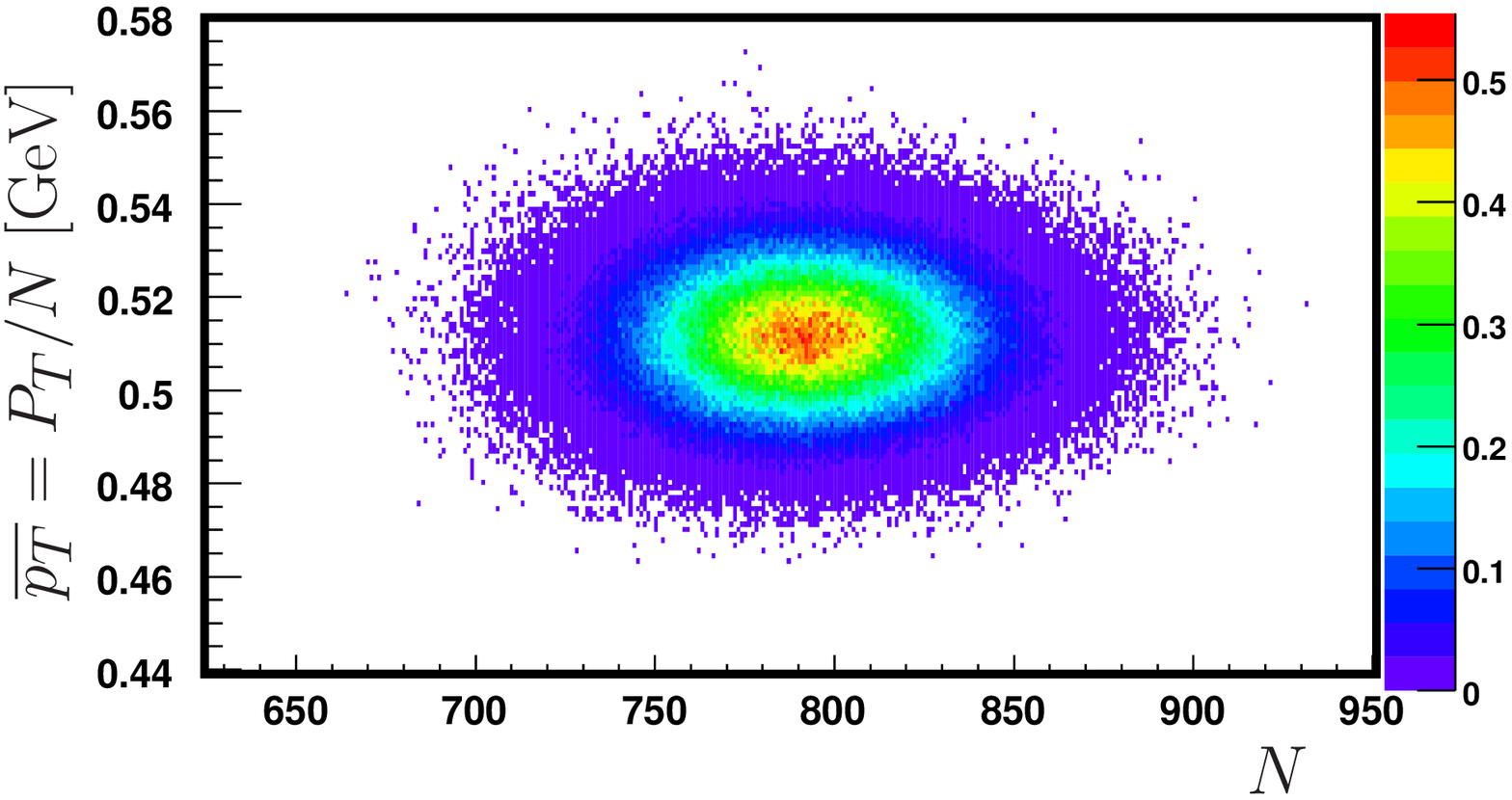,width=7.2cm,height=6.5cm}
  \caption{Grand canonical joint distribution~$P(N,P_T)$ of total particle number~$N$ and total transverse 
  momentum~$P_T$~({\it left}), and grand canonical joint distribution $P(N,\overline{p_T})$  of total particle 
  number~$N$ and mean transverse momentum $\overline{p_T} = P_T/N$~({\it right}).
  The rest as in Fig.(\ref{P_0000_neu_1}).
  }  
  \label{P_0000_neu_3}
\end{figure}

A neutral and static GCE hadron resonance gas with local temperature $T=0.160$~GeV and volume~$V=2000$~fm$^3$ has 
been sampled according to the procedure described in Section~\ref{sec_montecarlo_gcesampling}.
The values of net-charges and energy and momentum~$\mathcal{Q}^l_{1,n}$ are calculated by iteration for the 
particle list of each event, according to~Eq.(\ref{code_charge}). 
In the GCE all events have weight~$w_n=1$, Eq.(\ref{code_weight}).
In Figs.(\ref{P_0000_neu_1}-\ref{P_0000_neu_3}) the event output is summarized.
GCE joint distributions of extensive quantities are shown for samples of~$5 \cdot 10^5$~events. 

Charge fluctuations directly probe the degrees of freedom of a system, i.e. they are sensitive to its particle 
mass spectrum (and its quantum number configurations). The correlation coefficient and variances carry information 
about orientation and elongation of the probability distributions in Figs.(\ref{P_0000_neu_1}-\ref{P_0000_neu_3}). 
Firstly, charge correlations and the contributions of different particle species to the 
covariance~$\langle \Delta X_i \Delta X_j \rangle$, Eq.(\ref{covariance}), and therefore to the correlation 
coefficient~$\rho_{ij}$, Eq.(\ref{rho}), are considered.

All baryons have baryon number~$b=+1$. Baryons can only carry strange (valence) quarks, i.e. their strangeness is 
always~$s\le0$. Anti-baryons have~$b=-1$, and~$s\ge0$. I.e., both groups contribute negatively to the 
baryon-strangeness covariance, and so~$\langle \Delta B \Delta S \rangle < 0$, and therefore~${\rho_{BS}<0}$, 
as seen in Fig.(\ref{P_0000_neu_1})~({\it left}) and indicated by the solid lines in Fig.(\ref{lc_bs_0000}).

Positively charged baryons and their anti-particles contribute positively to the baryon-electric charge 
covariance~$\langle \Delta B \Delta Q \rangle$, while negatively charged baryons (and their anti-particles) 
contribute negatively. Two observations can be made on the hadron resonance gas mass spectrum: there are more 
positively charged baryons than negatively charged ones, and their average mass is lower. I.e., in a neutral 
gas ($\mu_B=\mu_Q=\mu_S = 0$) the contribution of positively charged baryons (and negatively charged anti-baryons) 
dominates and therefore~$\langle \Delta B \Delta Q \rangle >0$ and~${\rho_{BQ}>0}$, as seen in 
Fig.(\ref{P_0000_neu_1})~({\it right}) and indicated by the solid lines in Fig.(\ref{lc_bq_0000}).

Mesons and their anti-particles always contribute positively to the strangeness-electric charge correlation 
coefficient~$\rho_{SQ}$. Electrically charged strange mesons are either composed of an~$u$-quark and a {$\bar{s}$-quark}, 
or of an~$\bar{u}$-quark and a~$s$-quark (and superpositions thereof). 
Their contribution to~$\langle \Delta S \Delta Q \rangle$ is in either case positive. 
On the baryonic side, only the~$\Sigma^+$ (as well as its degenerate states and their respective anti-particles) 
has a negative contribution to~$\langle \Delta S \Delta Q \rangle~$, while all other strangeness carrying baryons 
have either electric charge~$q=-1$, or~$q=0$. Therefore, one finds~${\rho_{SQ} >0}$, as seen in 
Fig.(\ref{P_0000_neu_2})~({\it left}) and indicated by the solid lines in Fig.(\ref{lc_sq_0000}).

Since there are more particles (at~$\mu_B=0$, and~$T=0.160$~GeV) which carry electric charge than particles which 
carry strangeness, one finds the electric charge variance to be larger than the strangeness 
variance $\langle \left( \Delta Q \right)^2\rangle > \langle \left( \Delta S \right)^2\rangle$.
Similarly, as there are more strangeness or anti-strangeness carrying particles than thermal baryon and anti-baryons, 
strangeness fluctuations are found to be stronger than baryon number 
fluctuation, $\langle \left( \Delta S \right)^2\rangle > \langle \left( \Delta B \right)^2\rangle$.

The correlation between particle number~$N$ and energy~$E$ in Fig.(\ref{P_0000_neu_2})~({\it right}) is more obvious. 
The more particles are in a subvolume, the more energy it will contain at a certain temperature~$\beta^{-1}$. 
The total transverse momentum~$P_T$ can be seen as a measure for the kinetic energy in the system, 
Fig.(\ref{P_0000_neu_3})~({\it left}). And naturally, the more particles are inside the subvolume, the 
more kinetic energy is stored in them. These extensive quantities are strongly correlated in the GCE. 
Due to the independent sampling of primordial particle momenta in the GCE, however, the mean energy per 
particle~$\overline{\varepsilon} = E/N$ and the mean transverse momentum~$\overline{p_T}=P_T/N$, are 
event-by-event un-correlated with the particle number~$N$ of the events. Here, the mean transverse momentum versus 
particle number distribution is shown in Fig.(\ref{P_0000_neu_3})~({\it right}).

\section{Momentum Spectra}
\label{sec_gce_MomSpect}
\begin{figure}[ht!]
  \epsfig{file=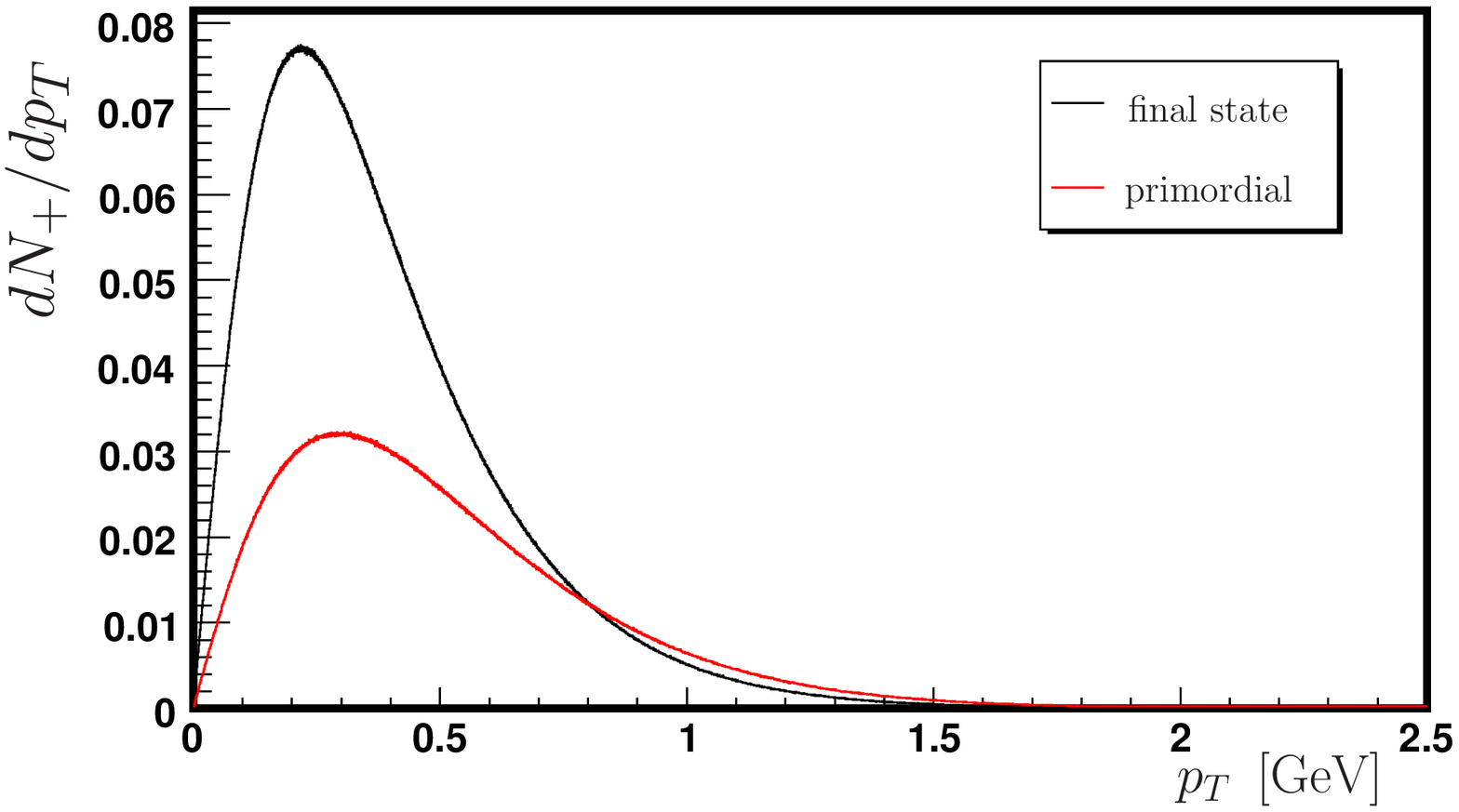,width=7.2cm,height=6.5cm}
  \epsfig{file=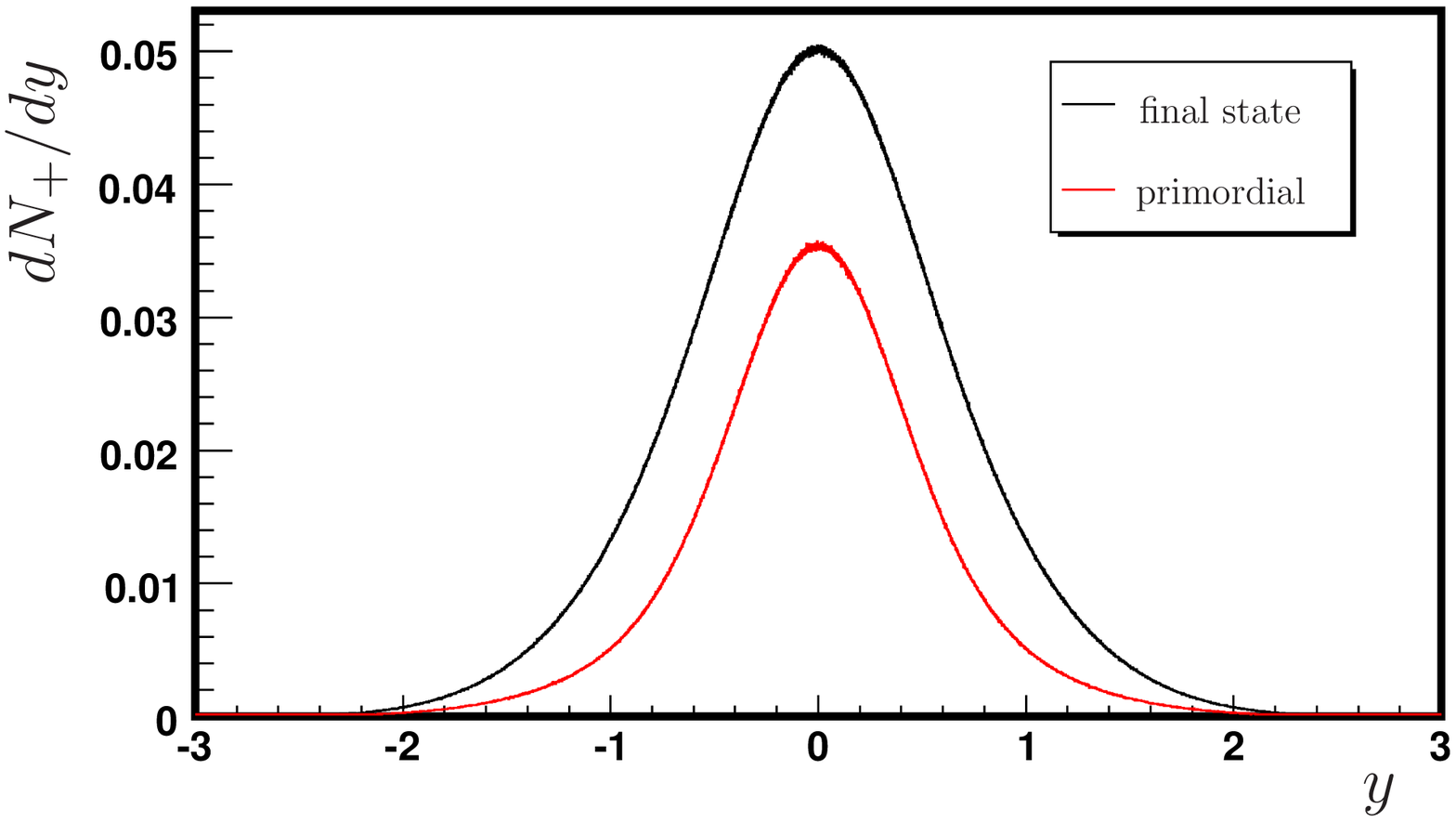,width=7.2cm,height=6.5cm}
  \caption{({\it Left:}) Transverse momentum spectrum of positively charged hadrons, both primordial and final state. 
  ({\it Right:})~Rapidity spectrum of positively charged hadrons, both primordial and final state.
  Here~$2 \cdot 10^6$ events have been sampled for a static neutral hadron resonance gas in Boltzmann approximation 
  with temperature $T=0.160$~GeV.}  
  \label{mom_spect}
\end{figure}

To prepare a more detailed analysis of charge fluctuations and correlations, inclusive momentum spectra are analyzed. 
In Fig.(\ref{mom_spect}) transverse momentum and rapidity spectra of positively charged hadrons, both primordial 
and final state, are shown for a static thermal system. 

Based on these momentum spectra acceptance bins~$\Delta p_{T,i}$ and~$\Delta y_{i}$ are constructed 
following~\cite{Gorenstein:2008th,Hauer:2007im,Hauer:2009nv} and~\cite{Alt:2007jq,Lungwitz:2007uc}. 
Momentum bins are chosen such that each of the five bins contains on average one fifth of the total 
yield of positively charged particles. The values defining the bounds of the momentum 
space bins~$\Delta p_{T,i}$ and~$\Delta y_i$ are summarized in Table~\ref{accbins}.

Resonance decay shifts the transverse momentum distribution to lower average transverse 
momentum~$\langle p_T \rangle$ and widens the rapidity distribution of thermal 
`fireballs`~\cite{Sollfrank:1990qz,Sollfrank:1991xm}. Transverse momentum bins of final state spectra are 
hence slightly `contracted`, while rapidity bins get slightly `wider`, when compared to their respective 
primordial counterparts.

\begin{table}[h!]\footnotesize
  \begin{center}
    \begin{tabular}{ccccccc}
    \toprule
      & $p_{T,1}$ [GeV] & $p_{T,2}$ [GeV]  & $p_{T,3}$ [GeV] & $p_{T,4}$ [GeV]  
      & $p_{T,5}$ [GeV]  & $p_{T,6}$ [GeV]  \\
    \midrule
      Primordial~ & 0.0 & 0.22795 & 0.36475 & 0.51825 & 0.73995 & 5.0 \\ 
      Final state & 0.0 & 0.17105 & 0.27215 & 0.38785 & 0.56245 & 5.0 \\ 
    \midrule[0.09em]
      & $y_1$  & $y_2$  & $y_3$  & $y_4$  & $y_5$  & $y_6$  \\
    \midrule 
      Primordial~ & -5.0 & -0.4275 & -0.1241 & 0.1241 & 0.4273 & 5.0 \\ 
      Final state & -5.0 & -0.5289 & -0.1553 & 0.1551 & 0.5289 & 5.0 \\
    \bottomrule
    \end{tabular}
    \caption{Transverse momentum and rapidity bins $\Delta p_{T,i} = \left[p_{T,i},p_{T,i+1} \right]$
    and $\Delta y_{i} = \left[y_{i},y_{i+1} \right]$, both primordial and final state,
    for a static neutral hadron resonance gas in Boltzmann approximation with temperature $T=0.160$~GeV. } 
    \label{accbins}
  \end{center}
\end{table}

Resonance decay combined with transverse as well as longitudinal flow provides a rather good description of 
experimentally observed momentum spectra in relativistic heavy ion collisions at SPS and RHIC 
energies~\cite{Kisiel:2005hn,Schnedermann:1993ws,Becattini:2007qr}. The momentum spectra discussed here, on the 
other hand, contain no flow, and results, thus, cannot be directly compared to experimental data or transport simulations.
However, qualitatively one might observe the effects discussed in the following sections.

\section{Correlations between Charges}
\label{sec_gce_LCfluc}

An interesting example of quantities - for which the measured value depends on the observed part of the 
momentum spectrum - are the correlation coefficients between the charges baryon number~$B$, strangeness~$S$ and 
electric charge~$Q$. Also variances and covariances of the baryon number, strangeness, and electric charge 
distribution are strongly sensitive to the acceptance cuts applied. Their values are additionally rather sensitive to 
the effects of globally enforced conservation laws. If the size of the `bath` is reduced, a change in one interval of 
phase space will have to be balanced (preferably) by a change in another interval, and not by the (finite) `bath`.

The correlation coefficients~$\rho_{BS}$, $\rho_{BQ}$, and~$\rho_{SQ}$ in limited acceptance 
bins~$\Delta p_{T,i}$ and~$\Delta y_i$, as defined in Table~\ref{accbins}, are considered in the grand canonical ensemble. 
Primordial particles in one momentum bin are then sampled independently from particles in any other momentum space 
segment, due to the `infinite bath` assumption.  Nevertheless, the way in which quantum numbers are correlated is 
different in different momentum bins\footnote{For the extensive quantities energy~$E$ and longitudinal momentum~$P_z$
the correlation in limited acceptance bins  $\Delta p_{T,i}$ and~$\Delta y_i$ is more apparent than for quantum numbers.}, 
as different particle species have, due to their different masses, different momentum spectra. 
Resulting correlation functions are hence spectra, rather than simply values.

\begin{figure}[ht!]
  \epsfig{file=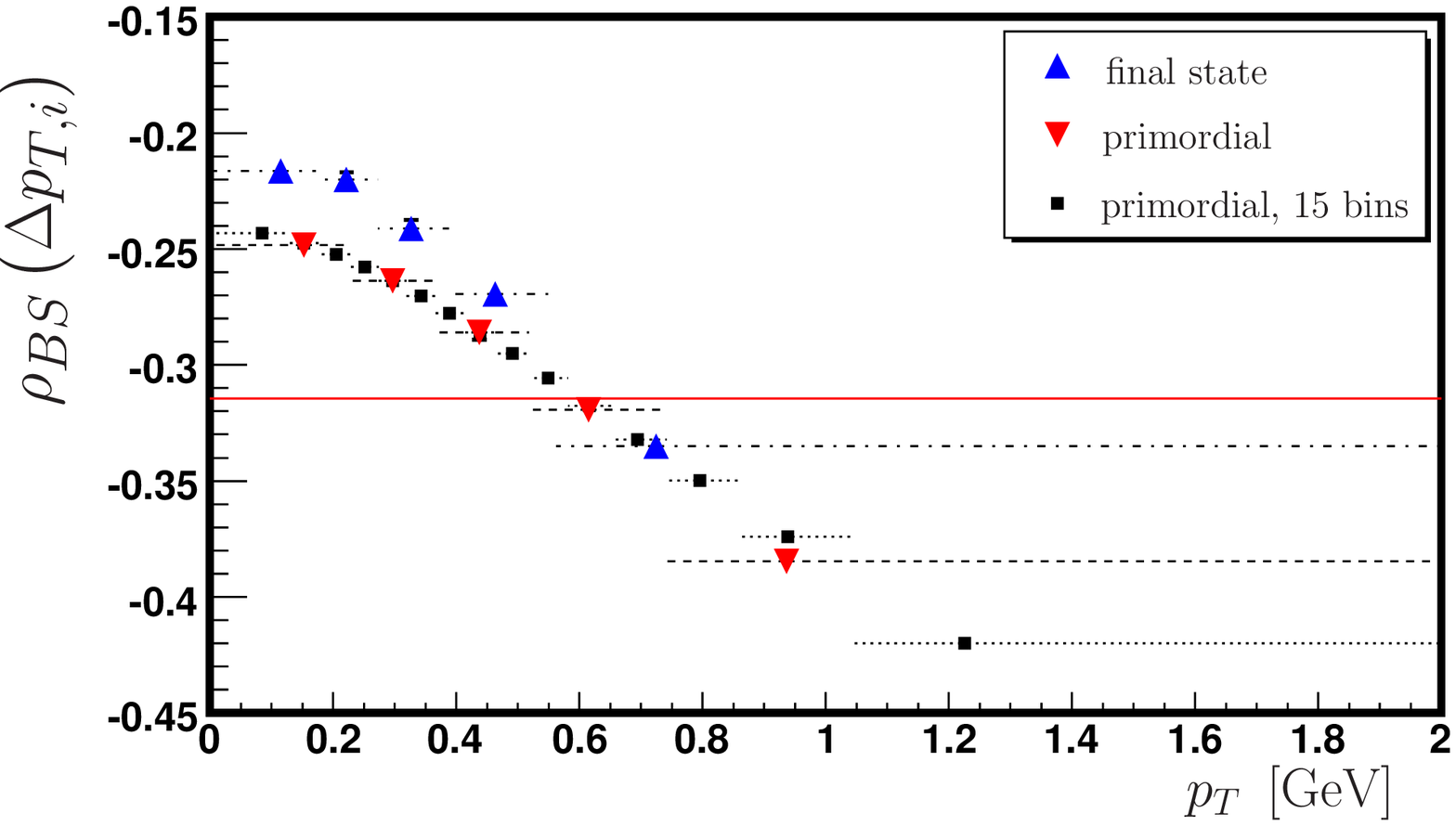,width=7.4cm,height=6.5cm}
  \epsfig{file=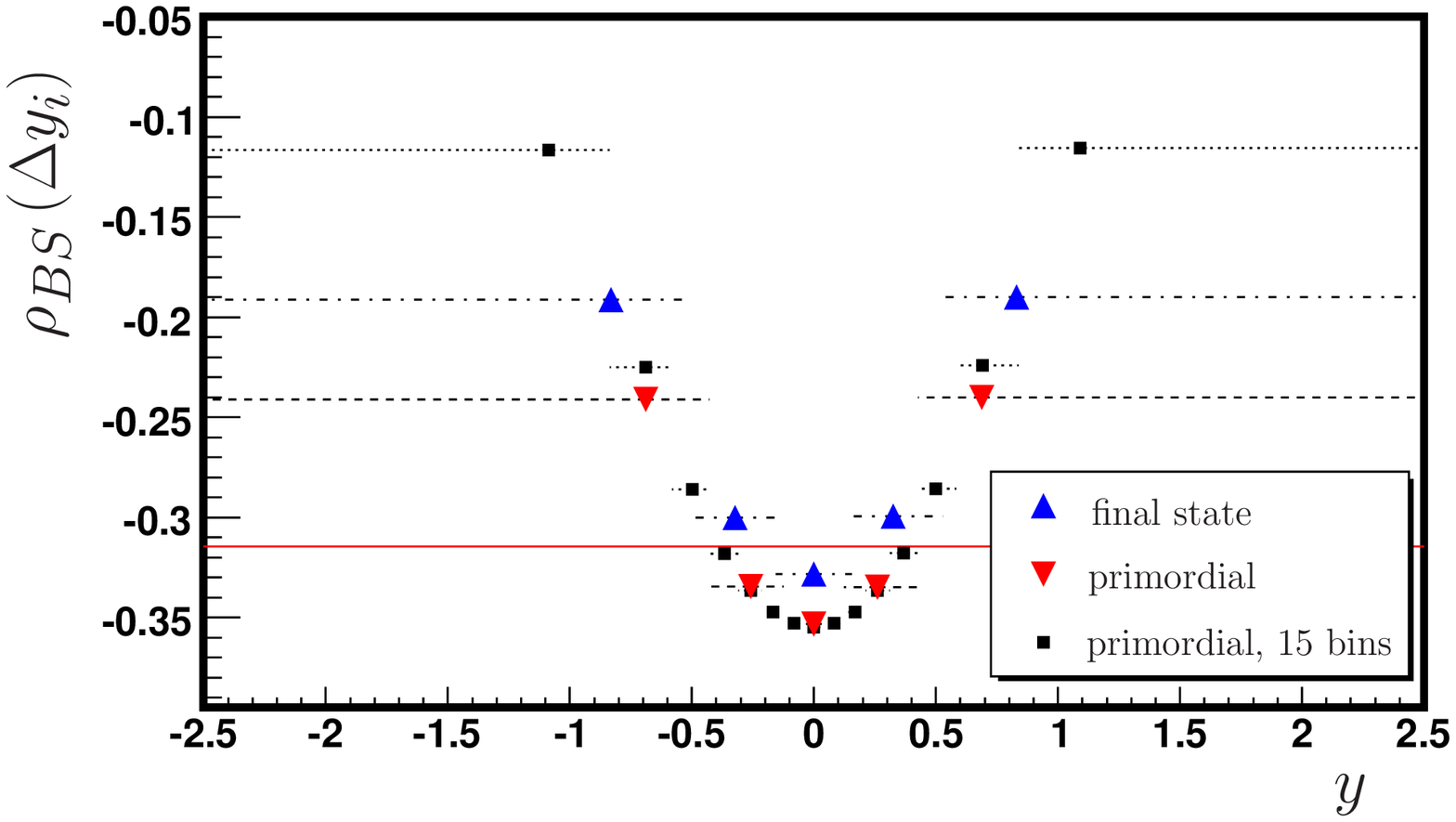,width=7.4cm,height=6.5cm}
  \caption{Baryon-strangeness correlation coefficient~$\rho_{BS}$ in the GCE in limited acceptance windows,
  both primordial and final state in transverse momentum bins  $\Delta p_{T,i}$ ({\it left}), and in 
  rapidity bins~$\Delta y_i$ ({\it right}).  
  Horizontal error bars indicate the width and position of the momentum bins (And not an uncertainty!).
  Vertical error bars indicate the statistical uncertainty of~$20$ Monte Carlo runs of~$10^5$ events each.
  The marker indicates the center of gravity of the corresponding bin.
  The solid lines show the fully phase space integrated GCE result.}  
  \label{lc_bs_0000}
\end{figure}

\begin{figure}[ht!]
  \epsfig{file=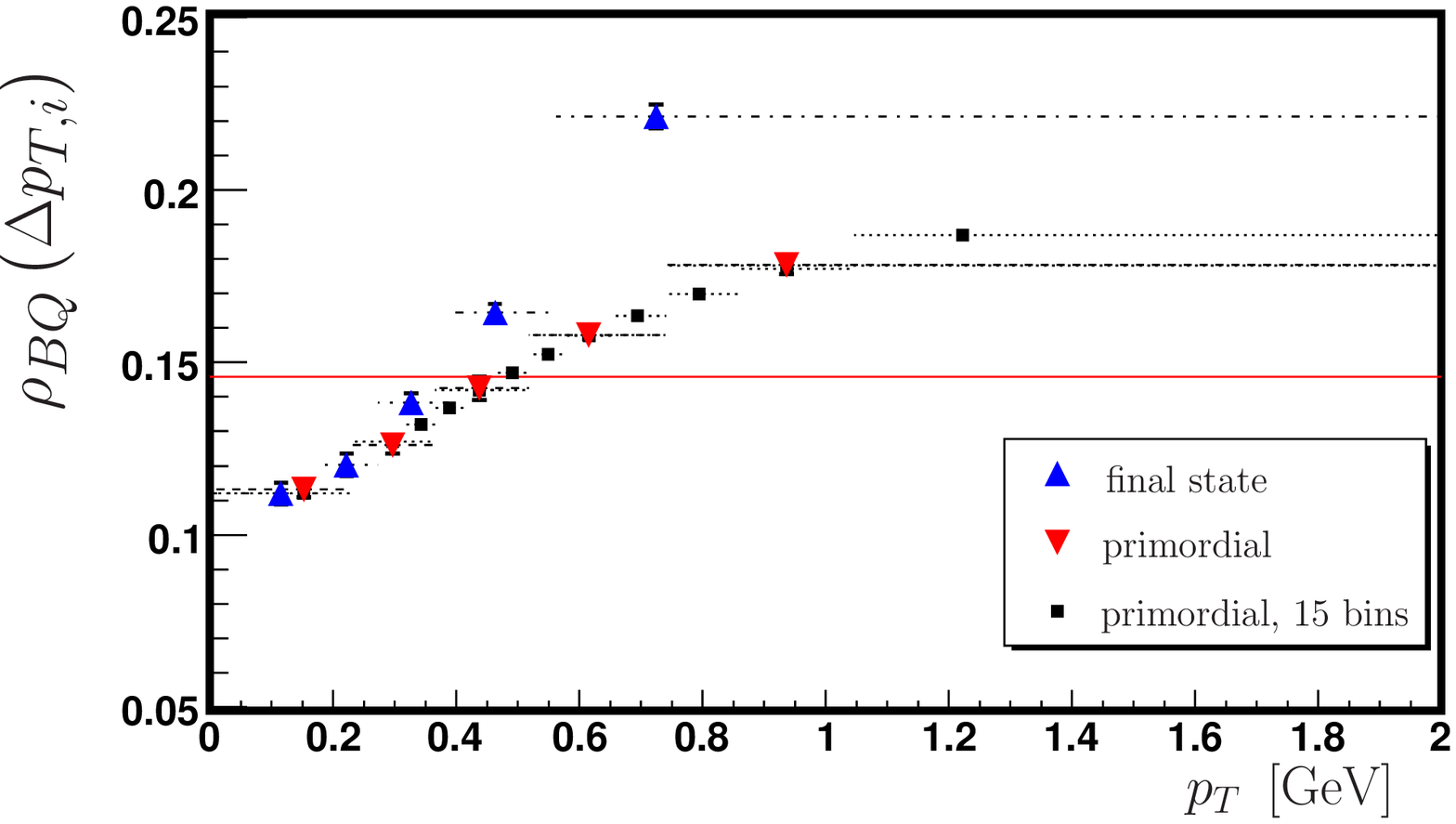,width=7.4cm,height=6.5cm}
  \epsfig{file=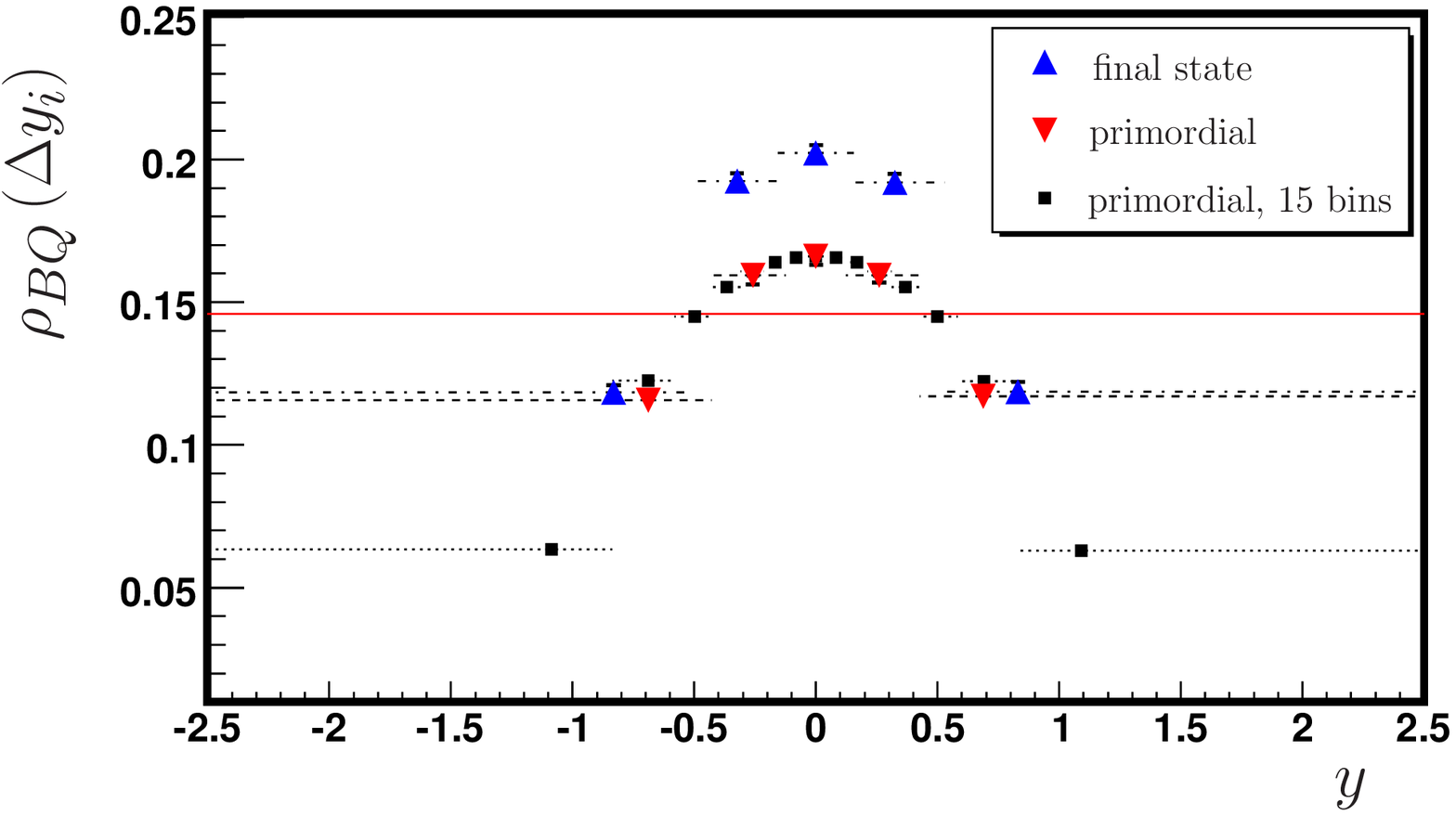,width=7.4cm,height=6.5cm}
  \caption{Baryon-electric charge correlation coefficient~$\rho_{BQ}$ in the GCE in limited acceptance windows, 
  both primordial and final state in transverse momentum bins~$\Delta p_{T,i}$ ({\it left}), and in 
  rapidity bins~$\Delta y_i$ ({\it right}). The rest as in Fig.(\ref{lc_bs_0000}).  }  
  \label{lc_bq_0000}
\end{figure}

\begin{figure}[ht!]
  \epsfig{file=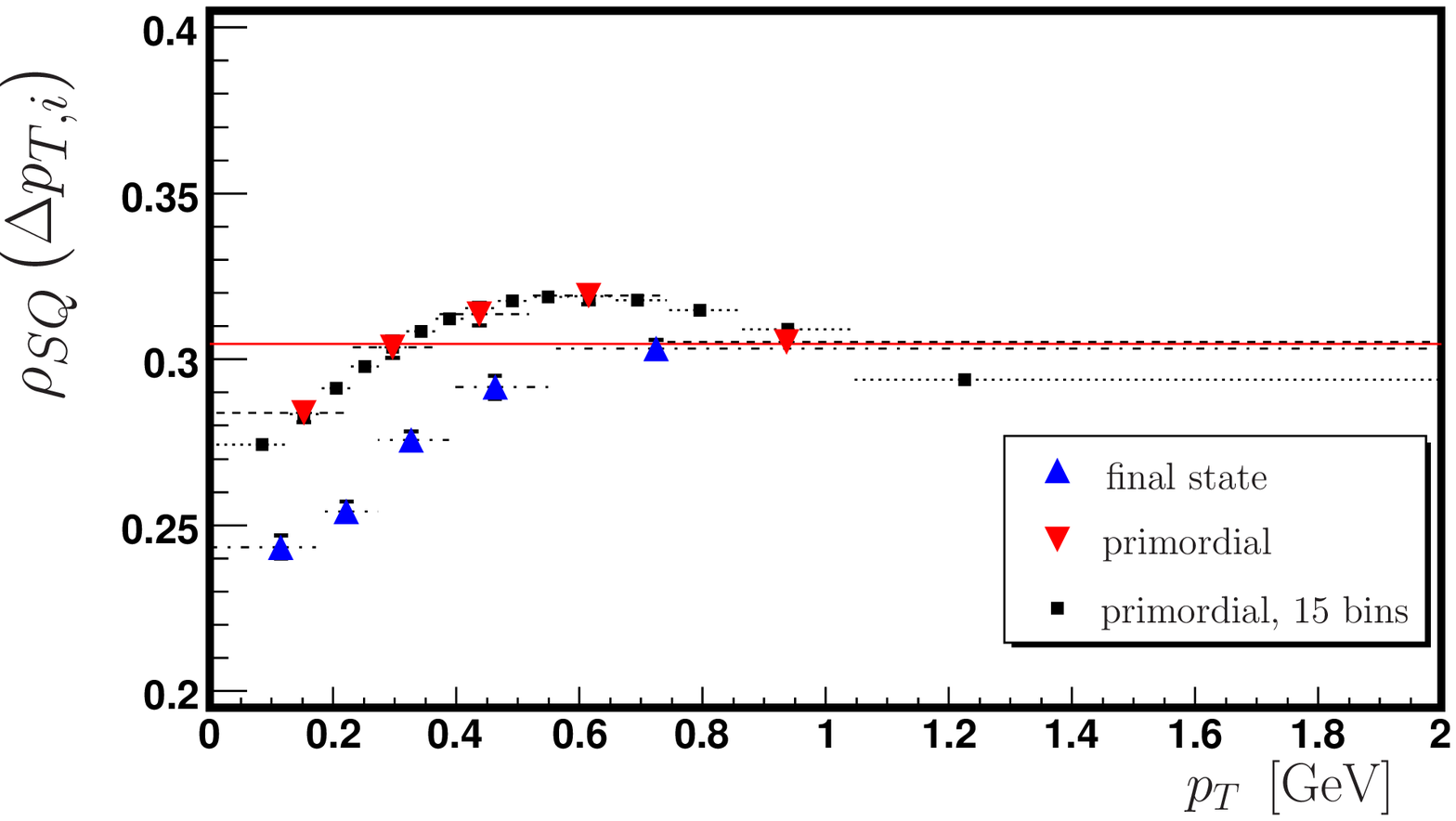,width=7.4cm,height=6.5cm}
  \epsfig{file=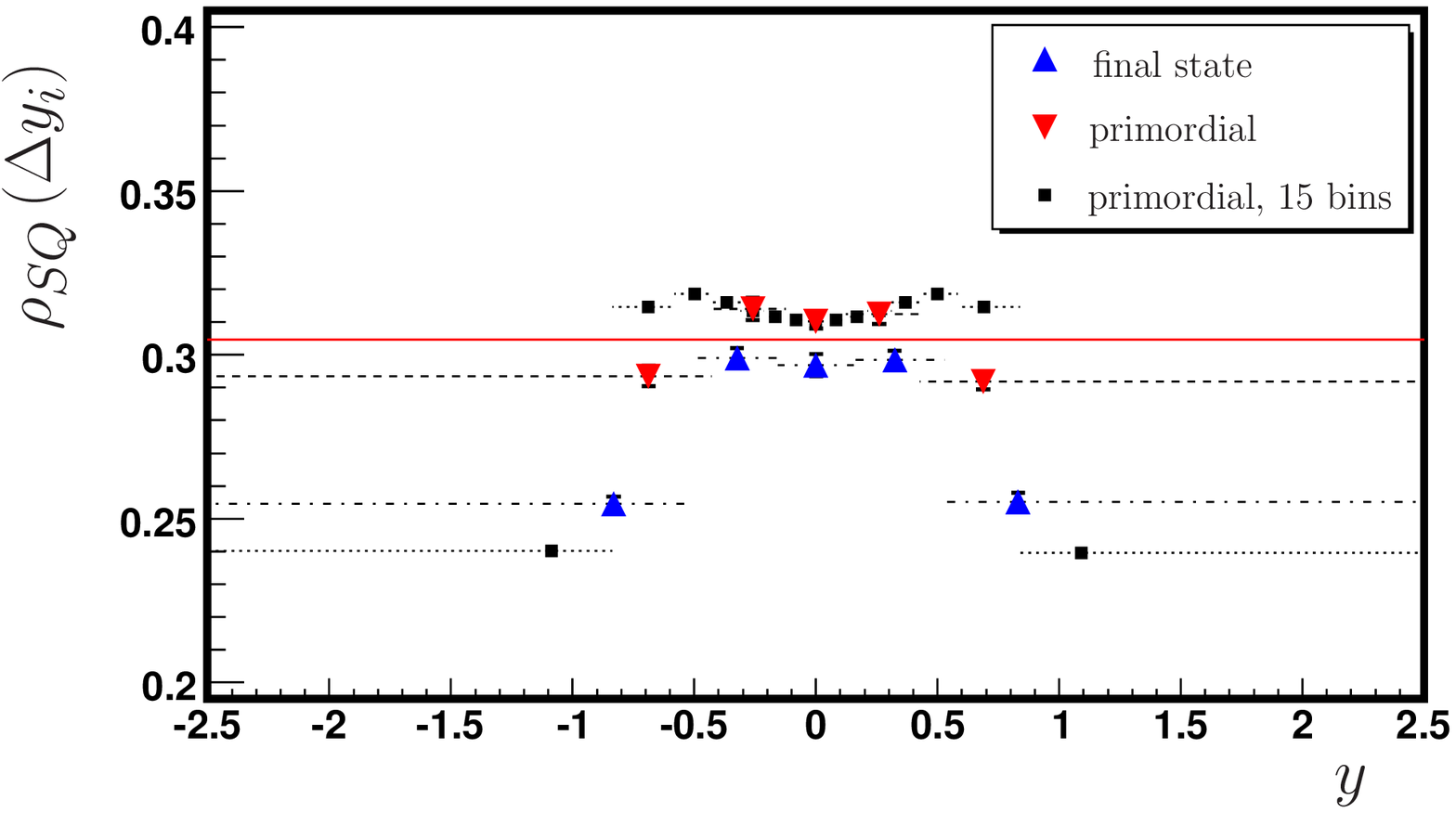,width=7.4cm,height=6.5cm}
  \caption{Strangeness-electric charge correlation coefficient~$\rho_{SQ}$ in the GCE in limited acceptance windows, 
  both primordial and final state in transverse momentum bins~$\Delta p_{T,i}$ ({\it left}), and in 
  rapidity bins  $\Delta y_i$ ({\it right}). The rest as in Fig.(\ref{lc_bs_0000}).}
  \label{lc_sq_0000}
\end{figure}

Figs.(\ref{lc_bs_0000}-\ref{lc_sq_0000}) show the correlation coefficients between baryon number and 
strangeness, $\rho_{BS}$, between baryon number and electric charge, $\rho_{BQ}$, and between strangeness and electric 
charge, $\rho_{SQ}$, as measured in the acceptance bins~$\Delta p_{T,i}$ and~$\Delta y_i$ defined in 
Table~\ref{accbins}, both primordial and final state. The average baryon number, strangeness, and electric charge in 
each bin is equal to zero, since the system was assumed to be neutral. The primordial values (15 bins) shown in 
Figs.(\ref{lc_bs_0000}-\ref{lc_sq_0000}) are calculated using analytical solutions. 
The error bars on many data points are smaller than the symbol used.

\begin{table}[h!]\footnotesize
  \begin{center}
    \begin{tabular}{cccccc}
    \toprule
      $\rho_{BS}$\!\!\! 
      & $\Delta p_{T,1}$ & $\Delta p_{T,2}$ & $\Delta p_{T,3}$ & $\Delta p_{T,4}$ 
      & $\Delta p_{T,5}$   \\
    \midrule
      $\rho_{prim}^{calc}$\!\!\! 
      & $-0.2479$ &$-0.2641$ &$-0.2864$ &$-0.3188$ &$-0.3839$  \\
      $\rho_{prim}$\!\!\! 
      & $-0.248 \pm 0.003$\! &$-0.264 \pm 0.003$\! &$-0.286 \pm 0.003$\! 
      &$-0.319 \pm 0.002$\! &$-0.385 \pm 0.002$\!  \\
      $\rho_{final}$\!\!\! 
      & $-0.216 \pm 0.002$\! &$-0.220 \pm 0.003$\! 
      &$-0.241 \pm 0.004$\! &$-0.269 \pm 0.003$\! &$-0.335 \pm 0.003$\!  \\
    \midrule[0.09em]
      $\rho_{BS}$\!\!\!\! 
      & $\Delta y_1$  & $\Delta y_2$  & $\Delta y_3$  & $\Delta y_4$  
      & $\Delta y_5$ \\
    \midrule 
      $\rho_{prim}^{calc}$\!\!\! 
      & $-0.2407$ &$-0.3345$ &$-0.3536$ &$-0.3345$ &$-0.2408$  \\
      $\rho_{prim}$\!\!\! 
      & $-0.241 \pm 0.003$\! &$-0.334 \pm 0.003$\! 
      &$-0.353 \pm 0.003$\! &$-0.335 \pm 0.003$\! &$-0.240 \pm 0.003$\!  \\
      $\rho_{final}$\!\!\! 
      & $-0.191 \pm 0.002$\! &$-0.300 \pm 0.002$\!
      &$-0.328 \pm 0.002$\! &$-0.299 \pm 0.002$\! &$-0.190 \pm 0.002$\!  \\
    \bottomrule
    \end{tabular}
    \caption{Baryon-strangeness correlation coefficient $\rho_{BS}$ in the GCE in transverse momentum 
    bins $\Delta p_{T,i}$ and rapidity bins $\Delta y_i$, both primordial and final state. For comparison, analytical 
    values $\rho_{prim}^{calc}$ for primordial correlations are included.
    The statistical uncertainty corresponds to $20$ Monte Carlo runs of $10^5$ events each. }
    \label{accbins_LC_BS}
  \end{center}
\end{table}

\begin{table}[h!]\footnotesize
  \begin{center}
    \begin{tabular}{cccccc}
    \toprule
      $\rho_{BQ}$ & $\Delta p_{T,1}$ & $\Delta p_{T,2}$ & $\Delta p_{T,3}$ & $\Delta p_{T,4}$ 
      & $\Delta p_{T,5}$   \\
    \midrule
      ~$\rho_{prim}^{calc}$ & $0.1120$ &$0.1271$ &$0.1420$ &$0.1579$ &$0.1781$ ~ \\
      ~$\rho_{prim}$ & $0.113 \pm 0.002$ &$0.126 \pm 0.002$ &$0.143 \pm 0.003$ ~
      &~$0.158 \pm 0.002$ &$0.178 \pm 0.003$ ~ \\
      ~$\rho_{final}$ & $0.112 \pm 0.003$ &$0.120 \pm 0.003$ &$0.138 \pm 0.003$ ~
      &~$0.164 \pm 0.003$ &$0.221 \pm 0.003$ ~ \\
    \midrule[0.09em]
      $\rho_{BQ}$ & $\Delta y_1$  & $\Delta y_2$  & $\Delta y_3$  & $\Delta y_4$  
      & $\Delta y_5$ \\
    \midrule
      ~$\rho_{prim}^{calc}$ & $0.1160$ &$0.1601$ &$0.1658$ &$0.1601$ &$0.1160$ ~ \\
      ~$\rho_{prim}$ & $0.116 \pm 0.002$ &$0.160 \pm 0.003$ &$0.166 \pm 0.003$ ~
      &~$0.159 \pm 0.003$ &$0.117 \pm 0.002$ ~ \\
      ~$\rho_{final}$ & $0.118 \pm 0.003$ &$0.192 \pm 0.003$ &$0.202 \pm 0.003$ ~
      &~$0.192 \pm 0.003$ &$0.119 \pm 0.003$ ~ \\
    \bottomrule
    \end{tabular}
    \caption{Baryon-electric charge correlation coefficient~$\rho_{BQ}$ in the GCE in transverse momentum 
    bins~$\Delta p_{T,i}$ and rapidity bins~$\Delta y_i$, both primordial and final state. }
    \label{accbins_LC_BQ}
  \end{center}
\end{table}

\begin{table}[h!]\footnotesize
  \begin{center}
    \begin{tabular}{cccccc}
    \toprule
      $\rho_{SQ}$ & $\Delta p_{T,1}$ & $\Delta p_{T,2}$ & $\Delta p_{T,3}$ & $\Delta p_{T,4}$ 
      & $\Delta p_{T,5}$   \\
    \midrule
      ~$\rho_{prim}^{calc}$ & $0.2831$ &$0.3033$ &$0.3150$ &$0.3185$ &$0.3055$ ~ \\
      ~$\rho_{prim}$ & $0.284 \pm 0.003$ &$0.304 \pm 0.003$ &$0.314 \pm 0.003$ ~
      &~$0.319 \pm 0.002$ &$0.305 \pm 0.002$ ~ \\
      ~$\rho_{final}$ & $0.243 \pm 0.003$ &$0.254 \pm 0.003$ &$0.276 \pm 0.003$ ~
      &~$0.292 \pm 0.003$ &$0.303 \pm 0.002$ ~ \\
    \midrule[0.09em]
      $\rho_{SQ}$ & $\Delta y_1$  & $\Delta y_2$  & $\Delta y_3$  & $\Delta y_4$  
      & $\Delta y_5$ \\
    \midrule 
      ~$\rho_{prim}^{calc}$ & $0.2934$ &$0.3137$ &$0.3104$ &$0.3137$ &$0.2934$ ~ \\
      ~$\rho_{prim}$ & $0.294 \pm 0.003$ &$0.314 \pm 0.003$ &$0.310 \pm 0.002$ ~
      &~$0.312 \pm 0.003$ &$0.292 \pm 0.002$ ~ \\
      ~$\rho_{final}$ & $0.255 \pm 0.002$ &$0.299 \pm 0.003$ &$0.297 \pm 0.003$ ~
      &~$0.298 \pm 0.003$ &$0.255 \pm 0.003$ ~ \\
    \bottomrule
    \end{tabular}
    \caption{Strangeness-electric charge correlation coefficient $\rho_{SQ}$ in the GCE in transverse momentum 
    bins~$\Delta p_{T,i}$ and rapidity bins~$\Delta y_i$, both primordial and final state. }
    \label{accbins_LC_SQ}
  \end{center}
\end{table}

Tables~\ref{accbins_LC_BS}~to~\ref{accbins_LC_SQ} summarize the transverse momentum and rapidity dependence of the 
correlation coefficients~$\rho_{BS}$, $\rho_{BQ}$, and~$\rho_{SQ}$. The statistical error quoted corresponds to $20$ 
Monte Carlo runs of~$10^5$ events each. The analytical values (5 bins) listed in the tables are calculated using the 
momentum bins defined in Table~\ref{accbins}. Mild differences between Monte Carlo and analytical results are 
unavoidable. The calculated values are also not exactly symmetric in~$\Delta y_i$, as the exact size of the acceptance 
bins constructed is sensitive to the number of bins used for the calculation of the momentum spectra. Correlation 
coefficients~$\rho$ are also rather sensitive to exact bin size, and the fourth digit becomes somewhat unreliable.

The rapidity dependence of~$\rho_{BS}$,~$\rho_{BQ}$, and~$\rho_{SQ}$ is now considered. Strange baryons are, on average, 
heavier than non-strange baryons, so their thermal rapidity distributions are narrower. The kaon rapidity distribution 
is then, compared to baryons, again wider. A change in baryon number (strangeness) at high~$|y|$ is less likely to be 
accompanied by a change in strangeness (baryon number) than at low~$|y|$. The value of~$\rho_{BS}$, therefore, drops 
toward higher rapidity, as shown in Fig.(\ref{lc_bs_0000})~({\it right}). 

By the same argument, one finds a weakening of the baryon-electric charge correlation~$\rho_{BQ}$ at higher 
rapidity, Fig.(\ref{lc_bq_0000})~({\it right})), as the rapidity distribution of electrically charged particles is wider 
than the one of baryons. The strangeness-electric charge correlation coefficient~$\rho_{SQ}$ first rises mildly 
and then drops somewhat stronger towards higher rapidity. As one shifts ones acceptance window to higher
values of~$|y|$, first the contribution of baryons (in particular~$\Sigma^+$) decreases and, as the meson contribution 
grows, $\rho_{SQ}$ rises slightly. Towards the highest~$|y|$, pions again dominate and de-correlate the quantum numbers.

The transverse momentum  dependence can be understood as follows: heavier particles have higher average transverse 
momentum~$\langle p_T \rangle$ and, hence, their influence increases towards higher~$p_T$. Heavy particles, in 
particular baryonic resonances, often carry several quantum numbers, causing the correlation coefficients to grow.

The contribution of strange baryons compared to non-strange baryons grows for higher transverse momentum, since strange 
baryons have on average larger mass than non-strange baryons. The correlation coefficient~$\rho_{BS}$, thus, becomes 
strongly negative at high~$p_T$. As the contribution of baryons compared to mesons grows stronger towards larger~$p_T$, 
a change in baryon number (electric charge) is now more likely to be accompanied by a change in electric charge 
(baryon number) than at low~$p_T$, and~$\rho_{BQ}$ increases with~$p_T$ (the~$\Delta$ resonances\footnote{Included in 
the THERMUS particle table up to the~$\Delta(2420)$ .} ensure it keeps rising). For the~$\Delta p_{T,i}$ dependence 
of~$\rho_{SQ}$ one notes that one of the strongest contributors at higher~$p_T$ is the~$\Omega^-$, with a relatively low 
mass of {$m_{\Omega^-}=1.672$~GeV.} So after a rise,~$\rho_{SQ}$ drops again towards highest~$p_T$, due to an 
increasing~$\Sigma^+$ contribution\footnote{Included in the THERMUS particle table up to the~$\Sigma(2030)$.}. 

Since resonance decay has the habit of dropping the lighter particles (mesons) at low~$p_T$ and higher~$|y|$, while 
keeping heavier particles (baryons) at higher~$p_T$ and at mid-rapidity, none of the above arguments about the 
transverse momentum and rapidity dependence are essentially changed by resonance decay. The correlation 
coefficient~$\rho_{BS}$ becomes more negative towards higher~$p_T$, while becoming weaker towards higher~$|y|$. 
Similarly,~$\rho_{BQ}$ grows larger at high~$p_T$ and drops with increasing~$y$. The larger contributions of baryons 
to the high~$p_T$ tail of the transverse momentum spectrum, and their decreased contribution to the tails of the 
rapidity distribution, compared to mesons, are to blame. The bump in the~$p_T$ dependence of~$\rho_{SQ}$, caused by 
the~$\Sigma^+$, has vanished, as the~$\Sigma^+$ is only considered as stable in its lightest version with 
mass~$m_{\Sigma^+}=1.189$~GeV. The small bump in the~$y$ dependence of~$\rho_{SQ}$, however, stays. The correlation is 
first increased by a growing kaon contribution and then again decreased by a growing pion contribution at larger rapidities.

\section{Correlation Functions}
\label{sec_gce_CorrFunc}
Above results are now applied to measures of correlations used in heavy ion phenomenology. The measurement of charge 
correlations appears to be a good discriminating tool between different physics scenarios, such as the ones provided by 
lattice QCD~\cite{Gavai:2005yk}, effective field theories~\cite{Ratti:2007jf}, or non-equilibrium transport 
models~\cite{Haussler:2008gx}. Acceptance effects are studied in this section. A discussion of the dependence of 
fully phase 
space integrated quantities on temperature and chemical potential can be found in Chapters~\ref{chapter_phasediagram} 
and~\ref{chapter_freezeoutline}. In this context variances of marginal distributions are studied.

Following the definitions of Ref.~\cite{Koch:2005vg} one can relate the correlation coefficients $\rho_{BS}$ 
and $\rho_{SQ}$ to the baryon number strangeness correlation measure~$C_{BS}$:
\begin{equation}
\label{CBS_BS}
C_{BS} ~\equiv~ -3 ~\frac{\langle \Delta B \Delta S \rangle}{\langle 
       \left( \Delta S \right)^2\rangle}~=~ -3~ \rho_{BS}~\sqrt{\frac{\langle 
       \left( \Delta B \right)^2\rangle}{\langle 
       \left( \Delta S \right)^2\rangle}}~,
\end{equation}
and the electric charge strangeness correlation measure~$C_{QS}$: 
\begin{equation}
\label{CBS_QS}
C_{QS} ~\equiv~ 3 ~\frac{\langle \Delta Q \Delta S \rangle}{\langle 
       \left( \Delta S \right)^2\rangle}~=~ 3~ \rho_{SQ}~\sqrt{\frac{\langle 
       \left( \Delta Q \right)^2\rangle}{\langle 
       \left( \Delta S \right)^2\rangle}}~.
\end{equation}
In the ideally weakly coupled QGP the fully phase space integrated values are for the baryon number strangeness 
correlation measure~$C_{BS}^{QGP}=1$, and for the electric charge strangeness correlation 
measure~$C_{QS}^{QGP}=1$~\cite{Koch:2005vg,Gavai:2002jt}. On the other hand, if correlations were determined by 
hadronic degrees of freedom, the fully phase space integrated correlation measures would be ${C_{BS}^{HRG}\simeq0.66}$, 
and~${C_{QS}^{HRG}\simeq 1.2}$ respectively, for a hadron resonance gas at~$T=0.170$~GeV 
and~$\mu_B=0.0$~GeV~\cite{Koch:2005vg}. The `quark Molecular Dynamics' (qMD) model~\cite{Haussler:2008gx} is in agreement 
with hadron resonance gas results below the critical temperature~$T_c$, while following lattice QCD calculations 
above~$T_c$. Additionally, the baryon number electric charge correlation measure~$C_{BQ}$ is defined to be:
\begin{equation}
\label{CBS_BQ}
C_{BQ} ~=~ 3 ~\frac{\langle \Delta B \Delta Q \rangle}{\langle 
       \left( \Delta Q \right)^2\rangle}~=~ 3~ \rho_{BQ}~\sqrt{\frac{\langle 
       \left( \Delta B \right)^2\rangle}{\langle 
       \left( \Delta Q \right)^2\rangle}}
\end{equation}

Since the motivation in~\cite{Koch:2005vg} for having the baryon number 
variance $\langle \left( \Delta B \right)^2\rangle$ not in the denominator, was that the electrically neutral neutron 
is hard to measure, the covariance in Eq.(\ref{CBS_BQ}) is divided by the electric charge 
variance~$\langle \left( \Delta Q \right)^2\rangle$, which is experimentally more accessible. The fully phase space 
integrated hadron resonance gas estimate, Fig.(\ref{cbs_bq_0000}), is~$C_{BQ}^{HRG} \simeq 0.23$. 
A QGP estimate is not reported.

\begin{figure}[ht!]
  \epsfig{file=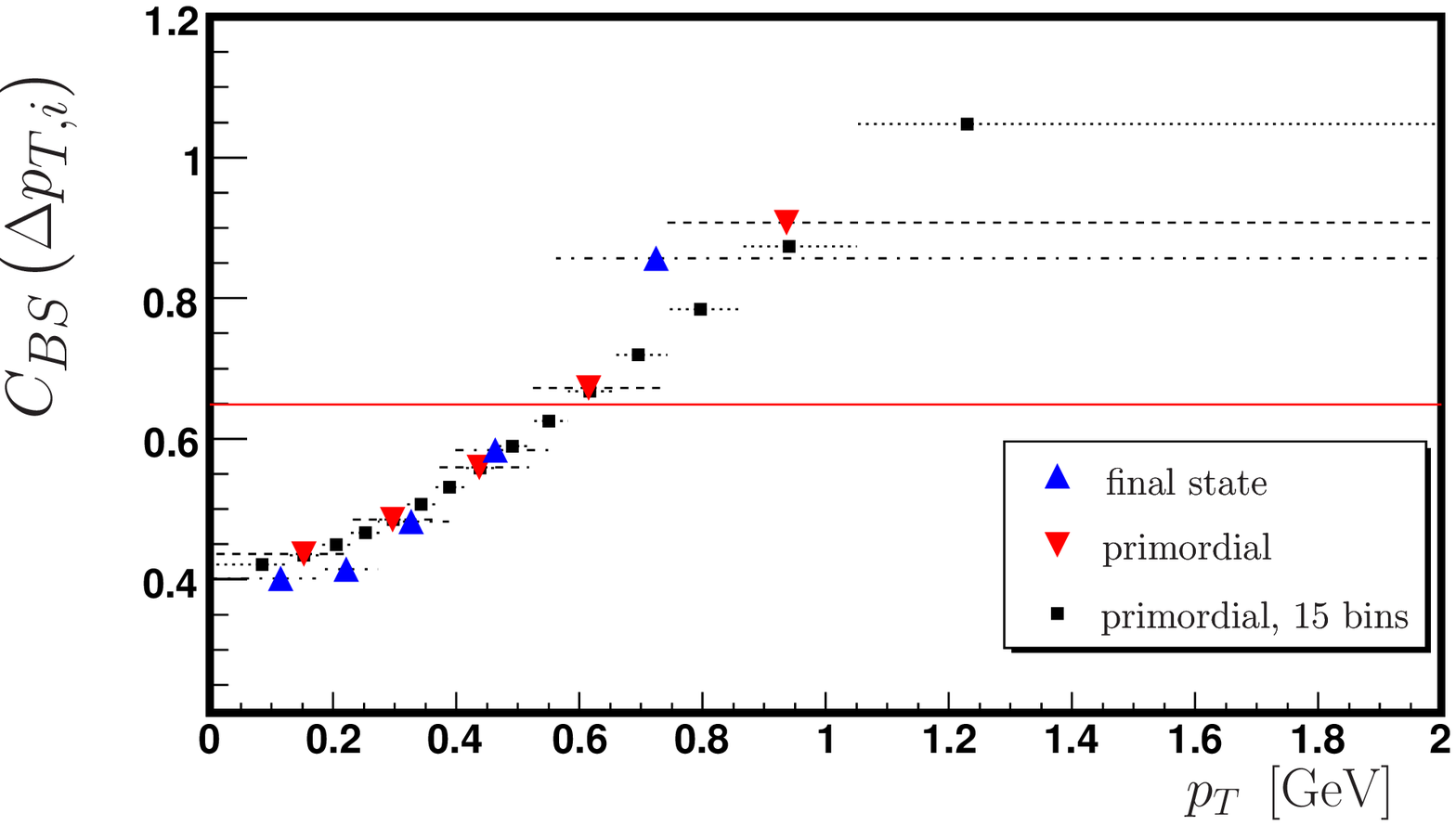,width=7.4cm,height=6.5cm}
  \epsfig{file=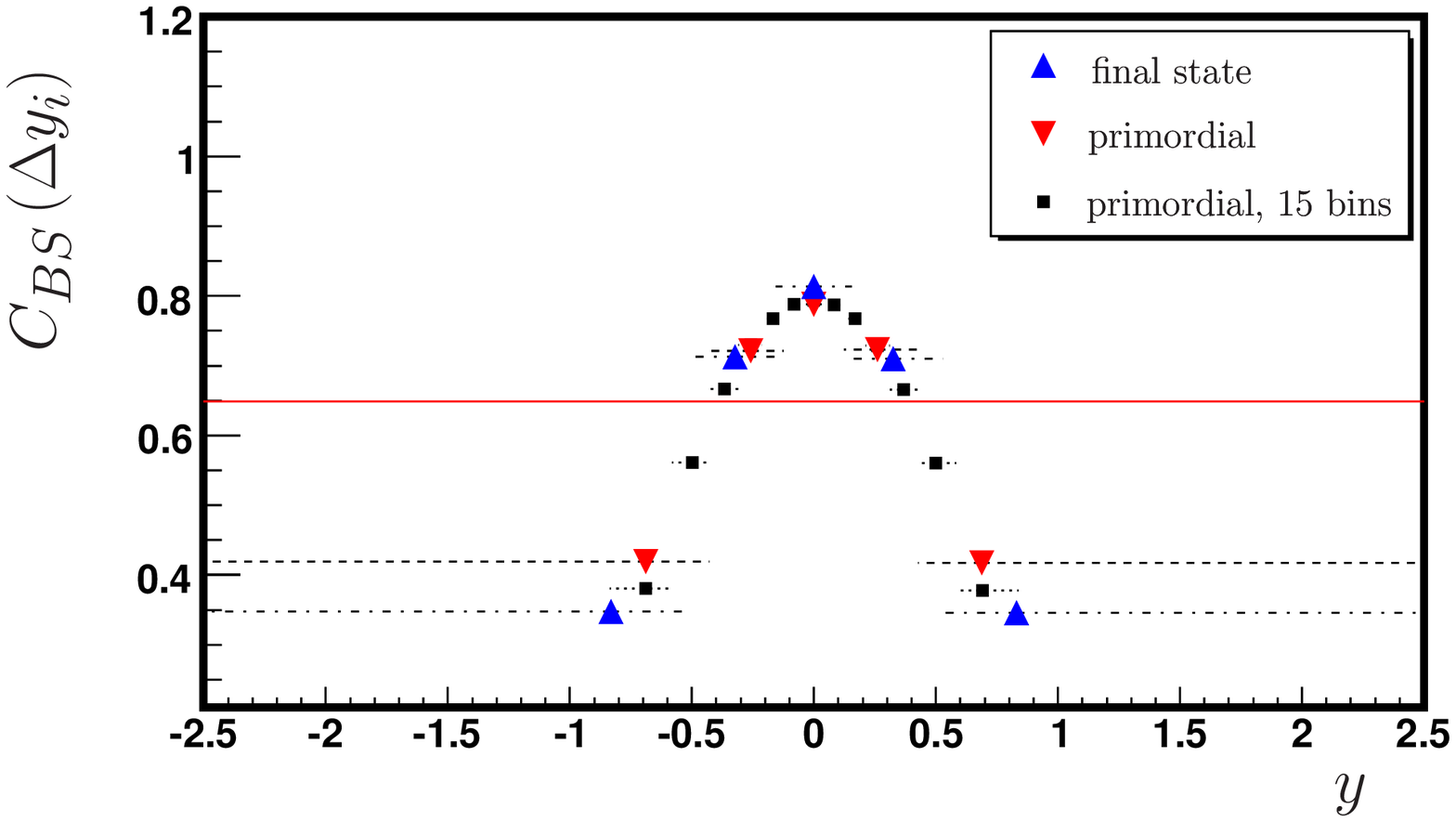,width=7.4cm,height=6.5cm}
  \caption{Baryon-strangeness correlation measure~$C_{BS}$ in the GCE in limited acceptance windows, both primordial 
  and final state, in transverse momentum bins~$\Delta p_{T,i}$ ({\it left}), and in 
  rapidity bins~$\Delta y_i$ ({\it right}).  
  Horizontal error bars indicate the width and position of the momentum bins (And not an uncertainty!).
  Vertical error bars indicate the statistical uncertainty of~$20$ Monte Carlo runs of~$10^5$ events each.
  The marker indicates the center of gravity of the corresponding bin.
  The solid lines show the fully phase space integrated GCE result.}  
  \label{cbs_bs_0000}
\end{figure}

\begin{figure}[ht!]
  \epsfig{file=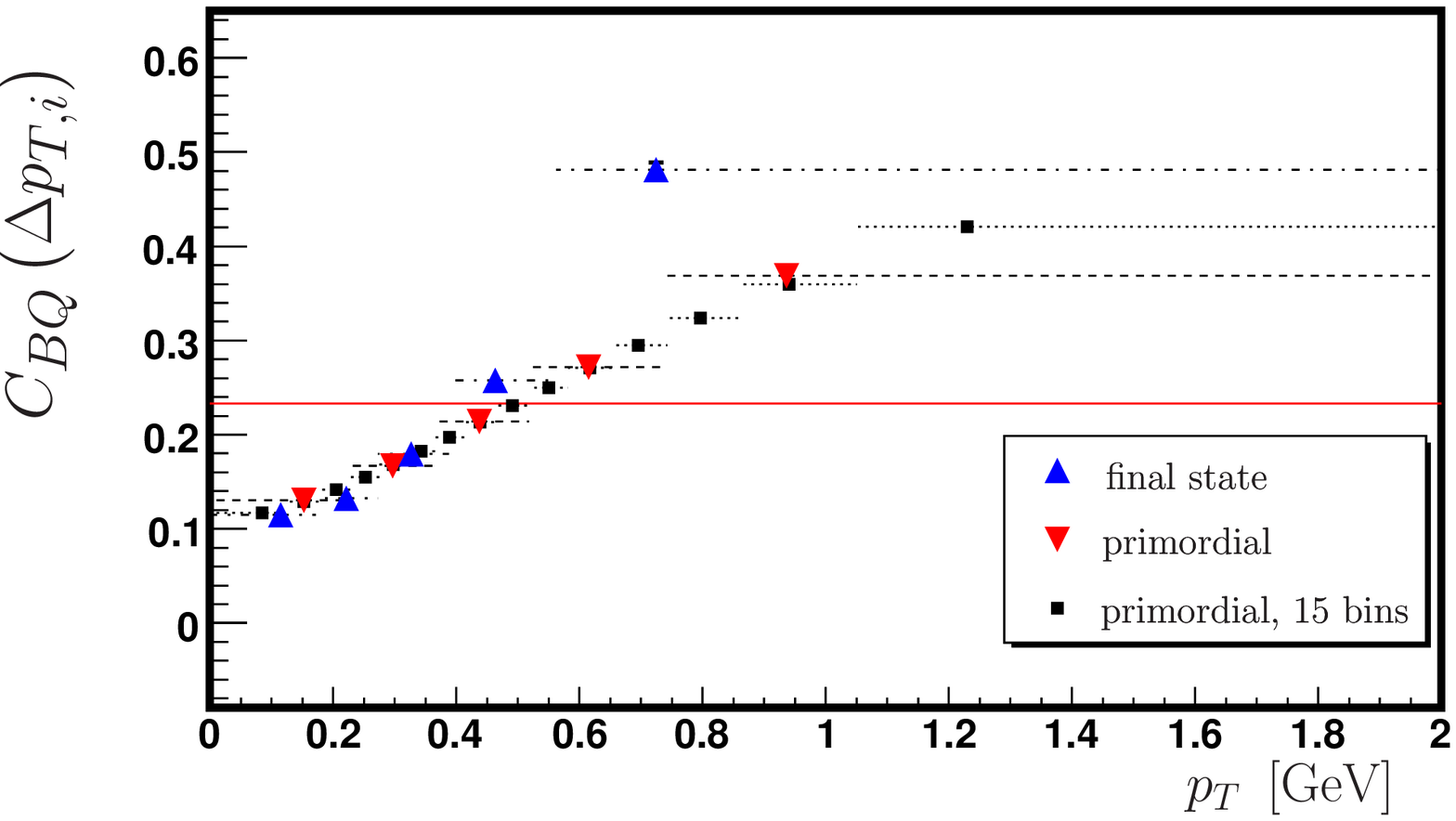,width=7.4cm,height=6.5cm}
  \epsfig{file=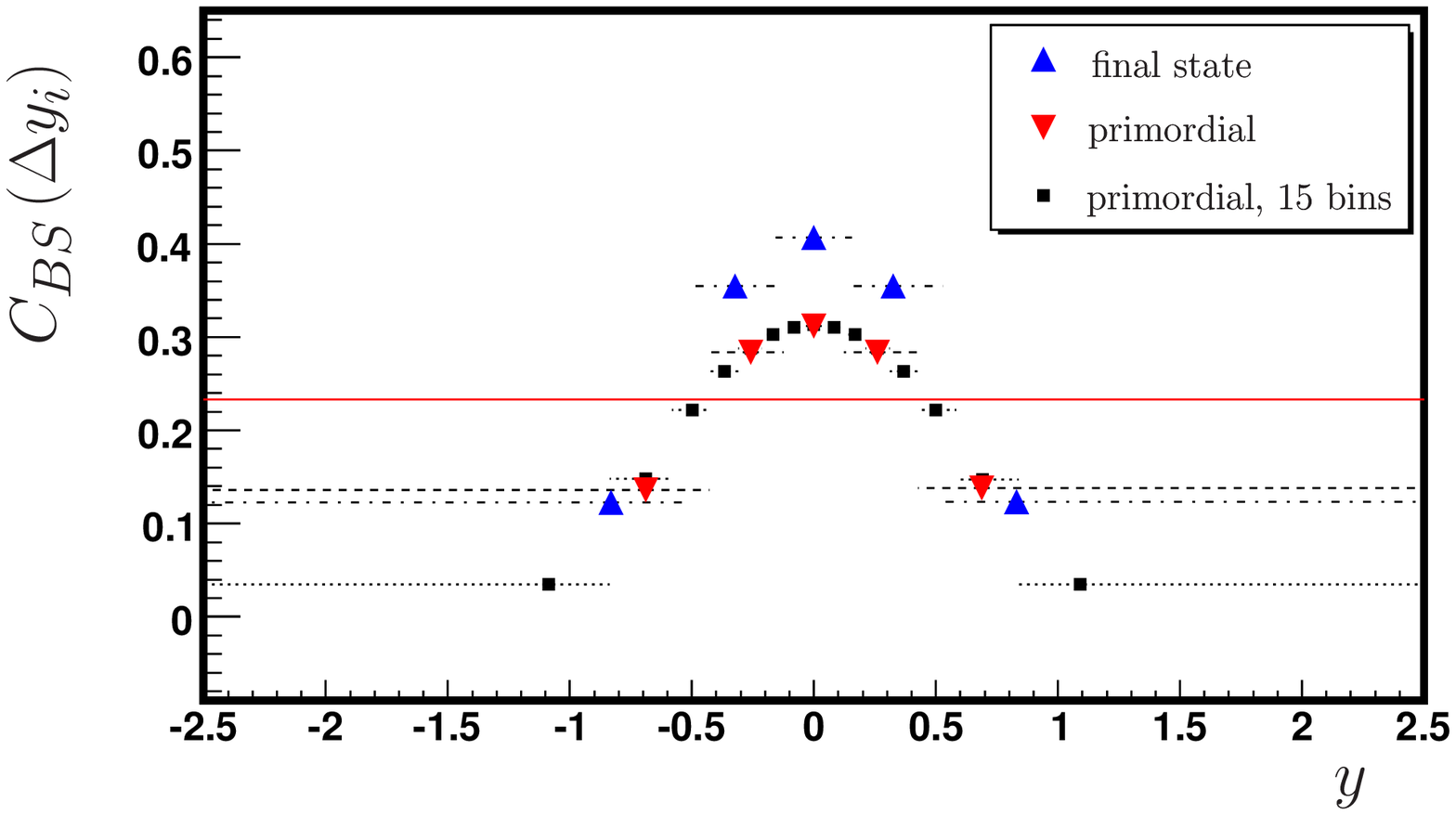,width=7.4cm,height=6.5cm}
  \caption{Baryon-electric charge correlation measure~$C_{BQ}$ in the GCE in limited acceptance windows, both primordial 
  and final state, in transverse momentum bins~$\Delta p_{T,i}$ ({\it left}), and in 
  rapidity bins~$\Delta y_i$ ({\it right}).
  The rest as in Fig.(\ref{cbs_bs_0000}).  }  
  \label{cbs_bq_0000}
\end{figure}

\begin{figure}[ht!]
  \epsfig{file=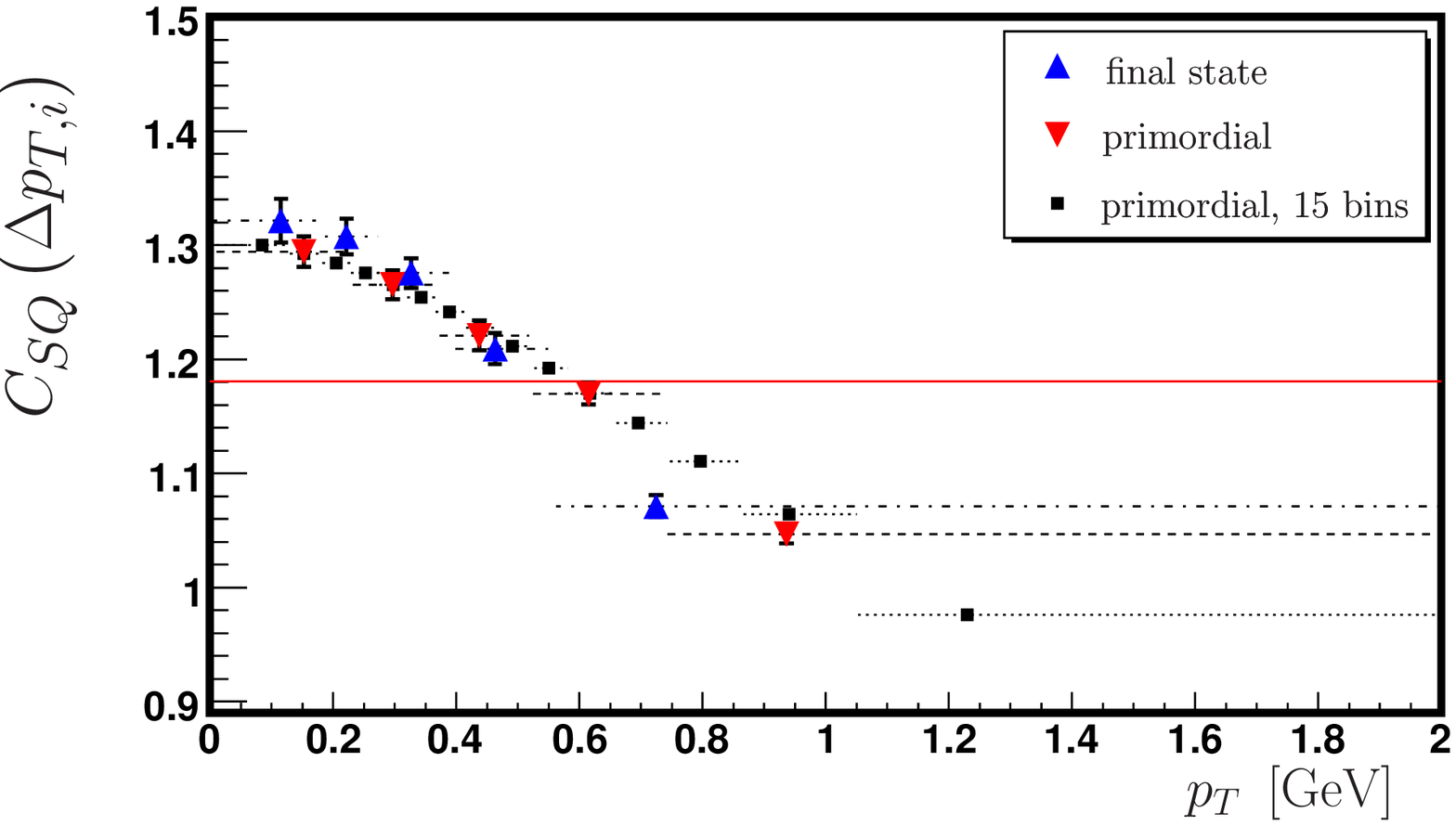,width=7.4cm,height=6.5cm}
  \epsfig{file=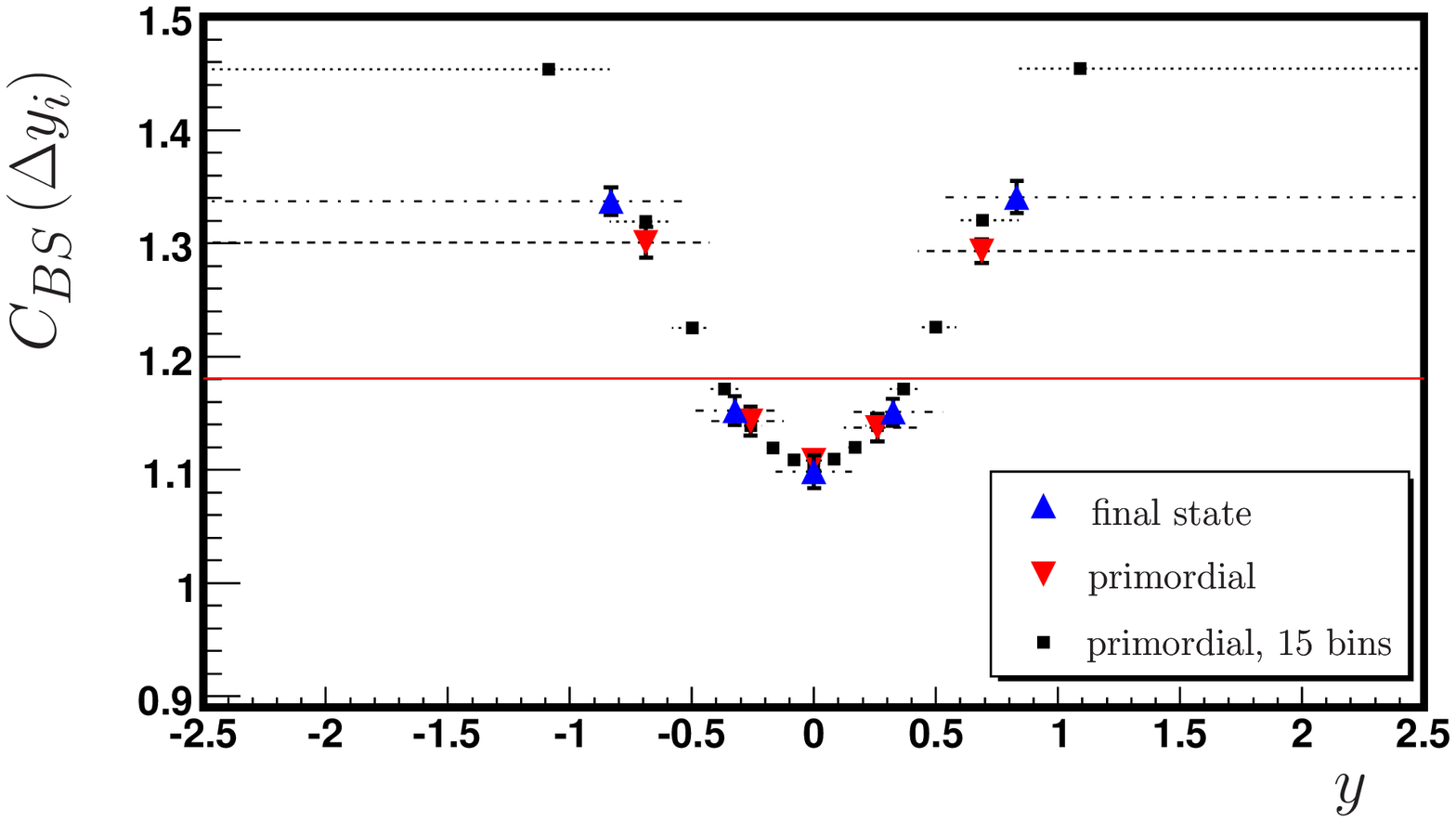,width=7.4cm,height=6.5cm}
  \caption{Strangeness-electric charge correlation measure~$C_{SQ}$ in the GCE in limited acceptance windows, both 
  primordial and final state, in transverse momentum bins~$\Delta p_{T,i}$ ({\it left}), and in 
  rapidity bins~$\Delta y_i$ ({\it right}).
  The rest as in Fig.(\ref{cbs_bs_0000}).}
  \label{cbs_sq_0000}
\end{figure}

In Figs.(\ref{cbs_bs_0000}-\ref{cbs_sq_0000}) the correlation measures~$C_{BS}$ (baryon number - 
strangeness), $C_{BQ}$ (baryon number - electric charge), and~$C_{SQ}$ (strangeness - electric charge), are shown as 
measured in the acceptance bins~$\Delta p_{T,i}$ and~$\Delta y_i$ defined in Table~\ref{accbins}, both primordial 
and final state. The average baryon number, strangeness, and electric charge in each bin is equal to zero, since the 
system the assumed to be neutral. The analytical primordial values  (15 bins) shown in 
Figs.(\ref{cbs_bs_0000}-\ref{cbs_sq_0000}) are calculated using analytical methods. It is mentioned that, again, the 
error bars on many data points are smaller than the symbol used. Fully phase space integrated values, indicated by solid 
lines, are compatible of hadron resonance gas results of Ref.~\cite{Koch:2005vg}. Nevertheless the results here show 
a strong dependence on the acceptance cuts applied. The fully phase space integrated values give a good estimate of 
the average. The data in limited acceptance are consistently above the weakly coupled QGP 
estimates~\cite{Koch:2005vg,Gavai:2002jt} in the case of~$C_{BS}$, or below as for~$C_{QS}$. 
The momentum space dependence of charge correlations in the QGP has not yet been calculated. In the hadron resonance 
gas the momentum space dependence is caused by different momentum spectra due to different hadron masses.

\begin{sidewaysfigure}
{
  \epsfig{file=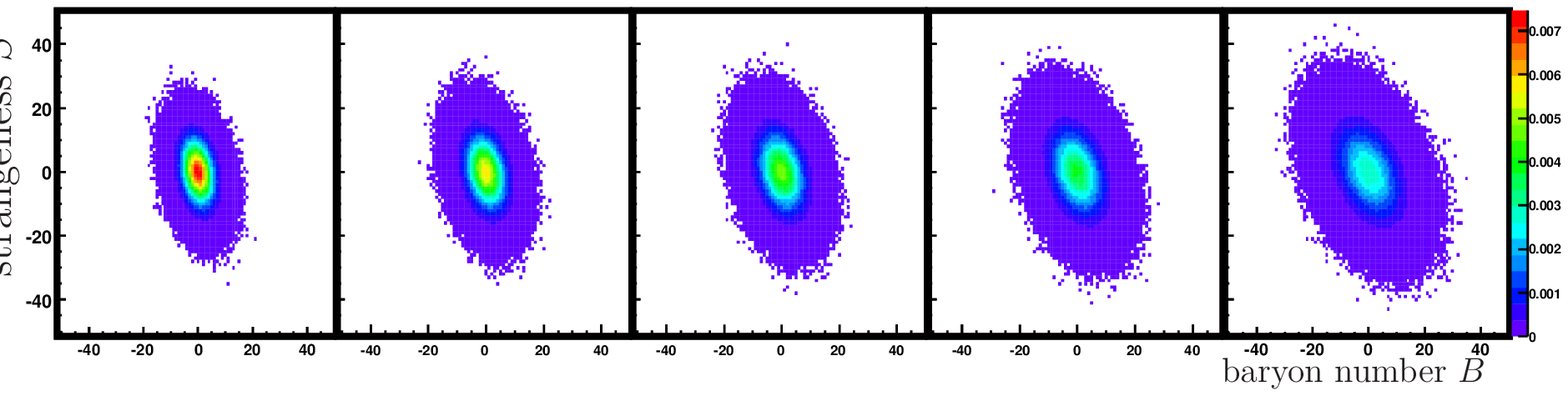,width=22.5cm,height=6.8cm}
  \epsfig{file=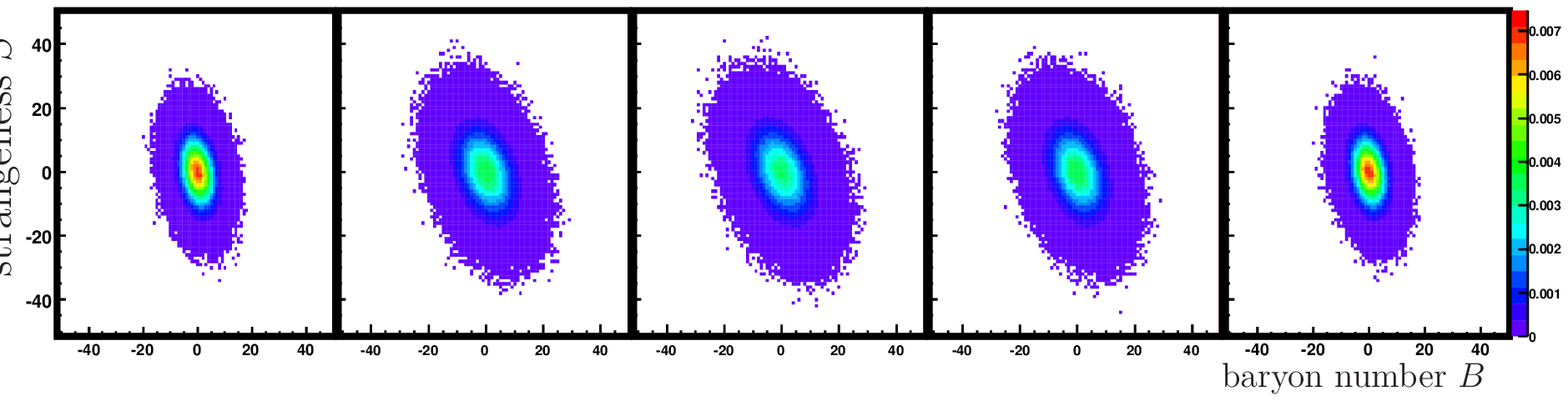,width=22.5cm,height=6.8cm}      
  \caption{({\it Top:})~Joint baryon number strangeness distribution in transverse momentum bins, 
  from left to right~$\Delta p_{T,1}$ to~$\Delta p_{T,5}$. 
  ({\it Bottom:})~Joint baryon number strangeness distribution in rapidity bins, 
  from left to right~$\Delta y_{1}$ to~$\Delta y_{5}$.}
  \label{dist_bs_0000}
} 
\end{sidewaysfigure}

One notes a qualitatively similar behavior of the correlation coefficients~$\rho$ shown in 
Figs.(\ref{lc_bs_0000}-\ref{lc_sq_0000}) to the correlation measure~$C$. To illustrate their differences, joint 
distributions are considered in limited acceptance. The correlation measures~$\rho$ and~$C$ are similar in 
their~$\Delta p_{T,i}$ and~$\Delta y_i$ dependence, and are essentially explained by arguments above. Additionally, 
the momentum space dependence of variances in Eqs.(\ref{CBS_BS}-\ref{CBS_QS}) needs to be taken into account.

\begin{figure}[ht!]
  \epsfig{file=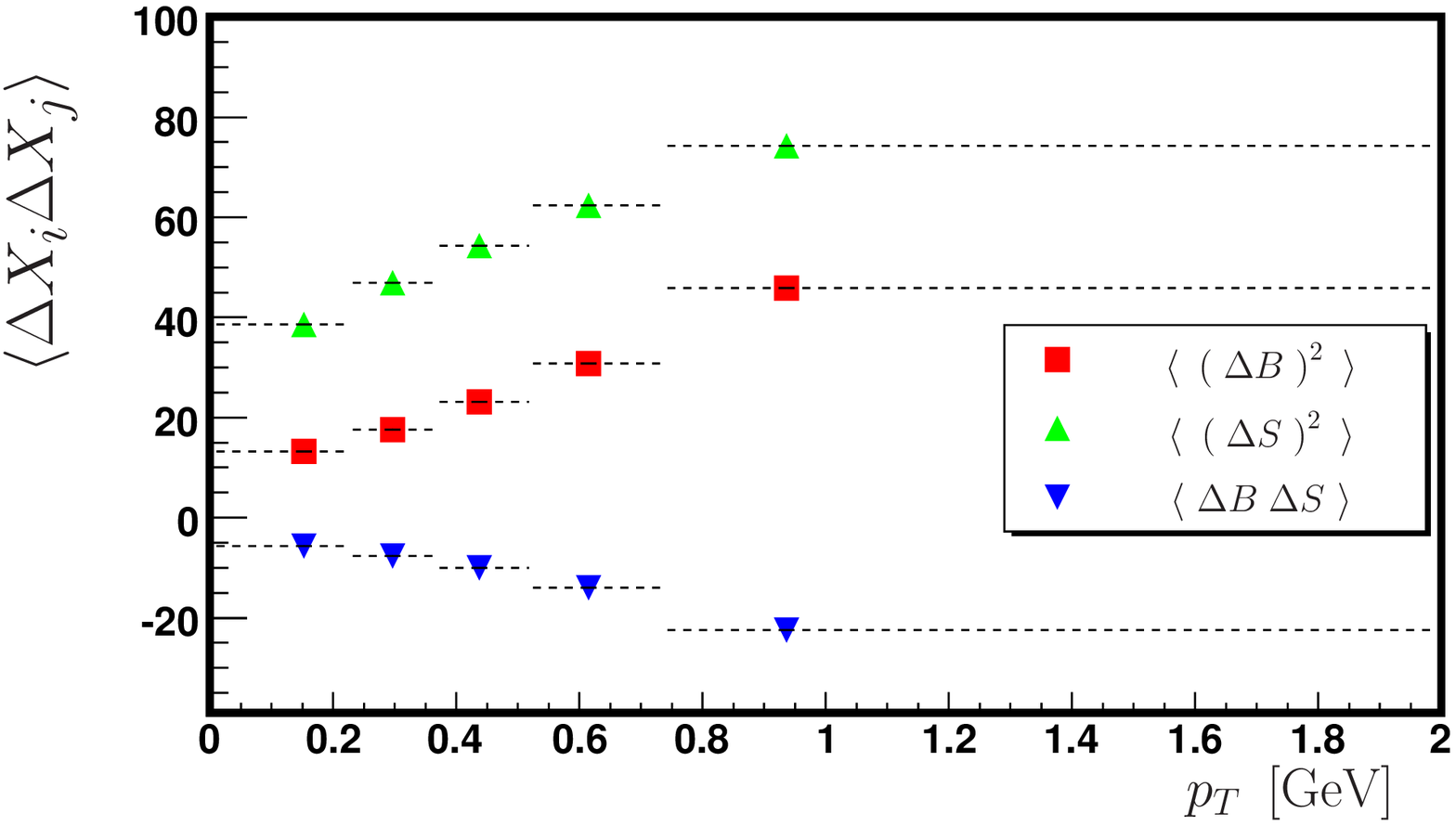,width=7.4cm,height=6.5cm}
  \epsfig{file=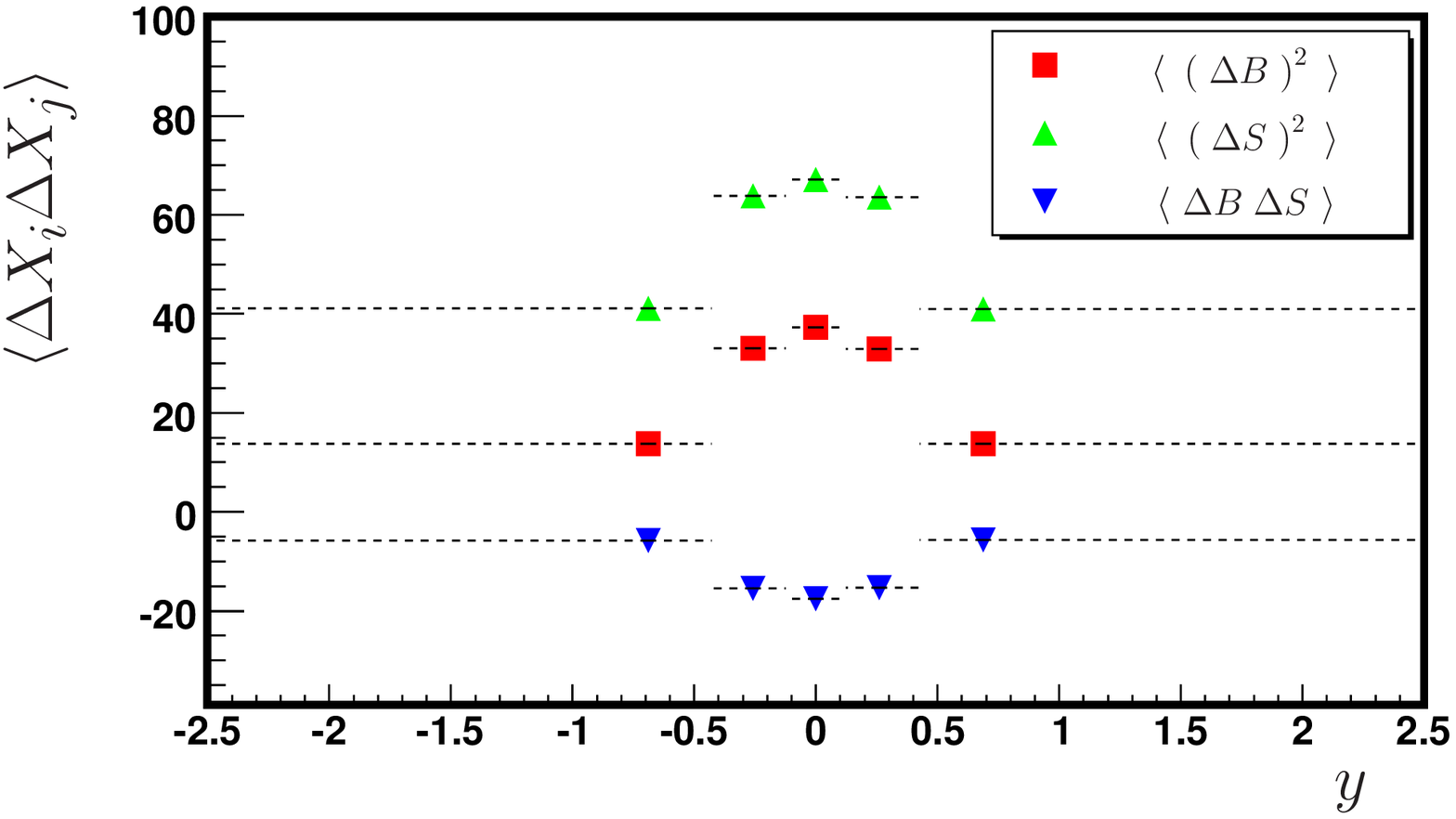,width=7.4cm,height=6.5cm}
  \caption{Primordial baryon number variance $\langle \left( \Delta B \right)^2\rangle$, strangeness 
  variance $\langle \left( \Delta B \right)^2\rangle$, and the baryon number strangeness 
  covariance~$\langle \Delta B \Delta S \rangle$ in the GCE in limited acceptance windows: 
  ({\it Left:}) transverse momentum bins~$\Delta p_{T,i}$. 
  ({\it Right:}) rapidity bins~$\Delta y_i$.  
  Horizontal error bars indicate the width and position of the momentum bins (And not an uncertainty!).
  Vertical error bars are smaller than the statistical uncertainty of~$20$ Monte Carlo runs of~$10^5$ events each.
  The marker indicates the center of gravity of the corresponding bin. }  
  \label{covar_bs_0000}
\end{figure}

Fig.(\ref{dist_bs_0000}) shows the GCE joint baryon number strangeness distribution as calculated in the acceptance 
windows $\Delta p_{T,i}$~({\it top}) and~$\Delta y_i$~({\it bottom}), while Fig.(\ref{covar_bs_0000}) summarizes 
the primordial baryon number variance $\langle \left( \Delta B \right)^2\rangle$, strangeness 
variance $\langle \left( \Delta S \right)^2\rangle$, and the baryon number strangeness 
covariance~$\langle \Delta B \Delta S \rangle$ in the GCE in transverse momentum bins~$\Delta p_{T,i}$~({\it left}) 
and rapidity bins~$\Delta y_i$~({\it right}).

The marginal variances~$\langle \left( \Delta B \right)^2\rangle$ and~$\langle \left( \Delta S \right)^2\rangle$ get wider 
in the transverse momentum direction~$\Delta p_{T,i}$, Fig.(\ref{dist_bs_0000})~({\it top}), while the 
correlation~$\rho_{BS}$ gets stronger. The contribution of baryons (compared to all hadrons) increases towards higher 
transverse momentum, which causes the baryon distribution to widen. Strangeness carrying particles are heavier than 
non-strange particles, so their variance increases, too. Together they drive up their 
covariance~$\langle \Delta B \Delta S \rangle$. 
In rapidity~$\Delta y_i$ direction, Fig.(\ref{dist_bs_0000})~({\it bottom}), the arguments are somewhat reversed. 
Particles carrying baryon number and strangeness are predominantly produced (thermally) around mid-rapidity. 
At larger rapidity, the pions take over, and the baryon number and strangeness variances decrease. And as the kaon 
rapidity distribution is wider than the one of the~$\Lambda$, one finds the correlation~$\rho_{BS}$ to decrease 
towards higher~$|y|$. 

Similarly, in full acceptance one finds for the variances of the marginal charge distributions in limited 
acceptance~$\langle \left( \Delta B \right)^2\rangle < \langle \left( \Delta S \right)^2\rangle < 
\langle \left( \Delta Q \right)^2\rangle$. Yet, the dependence of their ratios on transverse momentum and rapidity cuts 
is stronger for some and weaker for others. Baryons dominate at high~$p_T$; strangeness carrying particles are heavier;
accordingly both variances increase at large~$p_T$ compared to the electric charge variance (not shown). Heavier 
particles live around mid-rapidity. The ratios~$\langle \left( \Delta B \right)^2\rangle / 
\langle \left( \Delta Q \right)^2\rangle$ and~$\langle \left( \Delta B \right)^2\rangle / 
\langle \left( \Delta S \right)^2\rangle$ get larger at larger~$p_T$ and~$|y|$, while the 
ratio~$\langle \left( \Delta Q \right)^2\rangle / \langle \left( \Delta S \right)^2\rangle$ gets smaller at larger 
momentum. Therefore, the momentum dependence is stronger for~$C_{BS}$ and~$C_{BQ}$ than for $\rho_{BS}$ 
and~$\rho_{BQ}$, while being weaker for~$C_{QS}$ than for~$\rho_{SQ}$.

\section{Discussion}
\label{sec_gce_Discussion}

In this chapter samples of grand canonical Monte Carlo events have been analyzed. 
Joint distributions of extensive quantities have been considered with full acceptance, or with limited acceptance,
in momentum space assumed. Attention was given to the effects of resonance decay.

The correlations and fluctuations of extensive quantities of a statistical system are determined by the degrees of 
freedom, i.e. particles and their masses and quantum numbers, available to the system. So is, for instance, the 
correlation between energy and particle multiplicity dependent on the type of particle species measured and the 
experimental acceptance window assumed.

Due to the assumption of an infinite thermodynamic bath, occupation numbers of individual momentum levels are 
un-correlated with each other. Likewise, particle multiplicities of two distinct groups of particles appear 
un-correlated. Together with occupation number fluctuations unconstrained by global constraints, all extensive 
quantities, bar the volume, fluctuate on an event-by-event or micro state-by-micro state basis. Therefore, the energy 
content and particle multiplicity are strongly correlated, while the average energy per particle is un-correlated with 
multiplicity. The net-charge content of baryon number and strangeness in the system are correlated because some hadron 
species carry both charges. Different hadron species have different quantum numbers and different spectra.
The correlation of baryon number and strangeness depends, like the one of energy and momentum, or energy and particle 
multiplicity, on which part of the spectrum is accessible for measurement.

The correlation coefficients between quantum numbers increase towards higher~$p_T$, since heavier particles have higher 
average transverse momentum~$\langle p_T \rangle$. Heavy particles, in particular baryonic resonances, often carry 
several quantum numbers, causing the correlation coefficients to grow. For a thermal fireball this effect is strongest
in a mid-rapidity window.

The values of~$\rho$ or~$C$ after resonance decay are directly sensitive to how the data is analyzed. 
In the above study only final state particles (stable against strong decays) are analyzed. One could, however, also 
reconstruct decay positions and momenta of parent resonances and could then count them as belonging to the acceptance 
bin the parent momentum would fall into. In the situation above, however, this would again yield the primordial scenario. 
If reconstruction of resonances is not done, one is sensitive to charge correlations carried by final state particles. 
As in the primordial case, a larger acceptance bin effectively averages over smaller bins. However, the smaller the 
acceptance bin, the more information is lost due to resonance decay. In full acceptance, final state and primordial 
correlation coefficients ought to be the same, since quantum numbers (and energy-momentum) are conserved
in the decays of resonances (weak decays omitted).

\chapter{Extrapolating to the MCE}
\label{chapter_extrapol}
In the previous chapter the grand canonical ensemble and limited acceptance effects have been studied. 
In this chapter, fully phase space integrated extensive quantities are considered and studied in their dependence 
on the size of the thermodynamic bath for a neutral and static system. Firstly, mean values, co-variances, and 
correlations coefficients of joint distributions of extensive quantities are analyzed in Section~\ref{sec_extrapol_FPSIQ}. 
Then, in Section~\ref{sec_extrapol_dist}, the joint distributions themselves, and, in Section~\ref{sec_extrapol_weight}, 
the weight factor are considered more closely. This discussion will prepare the analysis of multiplicity fluctuations 
and correlations, Chapter~\ref{chapter_multfluc_hrg}, and the effect of conservation laws on them.
 
\section{Fully Phase Space Integrated Quantities}
\label{sec_extrapol_FPSIQ}

Fully phase space integrated grand canonical results are extrapolated to the micro canonical limit. 
For this purpose hadron resonance gas events for various values of~$\lambda = V_1/V_g$ are iteratively generated, 
re-weighted, and analyzed.
By construction of the weight factor~$\mathcal{W}$, Eq.(\ref{solved_curly_W}), the extrapolation proceeds in a systematic 
fashion such that, for instance, particle momentum spectra as well as mean values of extensive quantities remain 
unchanged. On the other hand, all variances and covariances of extensive quantities subject to re-weighting converge 
linearly to their micro canonical values.

This can be seen from the form of the analytical large volume approximation to the grand canonical distribution 
of (fully phase space integrated) extensive quantities~$P_{gce}(\mathcal{Q}^l_1)$ (from Eq.(\ref{GCE_dist})):
\begin{equation}
P_{gce}(\mathcal{Q}^l_1)~\simeq~\frac{1}{(2 \pi V_1)^{L/2} \det| \sigma|}~ 
\exp \left[ - \frac{1}{2}~ \frac{1}{V_1}~ \xi^l \xi_l \right]~,
\end{equation}      
where the variable~$\xi^l$ is given by Eq.(\ref{xi_noV}). Now taking the weight factor~$\mathcal{W}_{\lambda}$, 
Eq.(\ref{solved_curly_W}), ($\sigma$ and~$\xi_l$ are the same in both equations) one obtains for the 
distribution~$P_{\lambda}(\mathcal{Q}^l_1)$ of extensive quantities~$\mathcal{Q}^l_1$ in subsystem~$1$:
\begin{eqnarray}
P_{\lambda}(\mathcal{Q}^l_1) &\simeq&  \mathcal{W}_{\lambda}^{\mathcal{Q}_1^l;\mathcal{Q}_g^l}
~P_{gce}(\mathcal{Q}^l_1) \nonumber\\
\label{Pbath}
&\simeq& \frac{1}{(2 \pi(1-\lambda) V_1)^{L/2} \det| \sigma|}~ 
\exp \left[ - \frac{1}{2}~ \frac{1}{(1-\lambda)~V_1}~ \xi^l \xi_l \right]~.
\end{eqnarray}
This is essentially the same multivariate normal distribution as the grand canonical version~$P_{gce}(\mathcal{Q}^l_1)$, 
however linearly contracted. Monte Carlo results will then be compared to Eq.(\ref{Pbath}).

Again a static and neutral hadron resonance gas with collective four-velocity~${u_{\mu}=(1,0,0,0)}$, chemical potential 
vector~${\mu_j=(0,0,0)}$, local temperature~$T=\beta^{-1}=0.160$~GeV, and volume~$V_1=2000$~fm$^3$ is considered.
This is a system large enough\footnote{Generally it is not easy to define when a system is `large enough` for the 
large volume approximation to be valid. Here, good agreement is found with asymptotic analytic solutions. 
Charged systems, or Bose-Einstein/Fermi-Dirac systems, usually converge more slowly to their asymptotic solution.}
for using the large volume approximation worked out in Section~\ref{sec_clt_GFCD}.

\begin{figure}[ht!]
  \epsfig{file=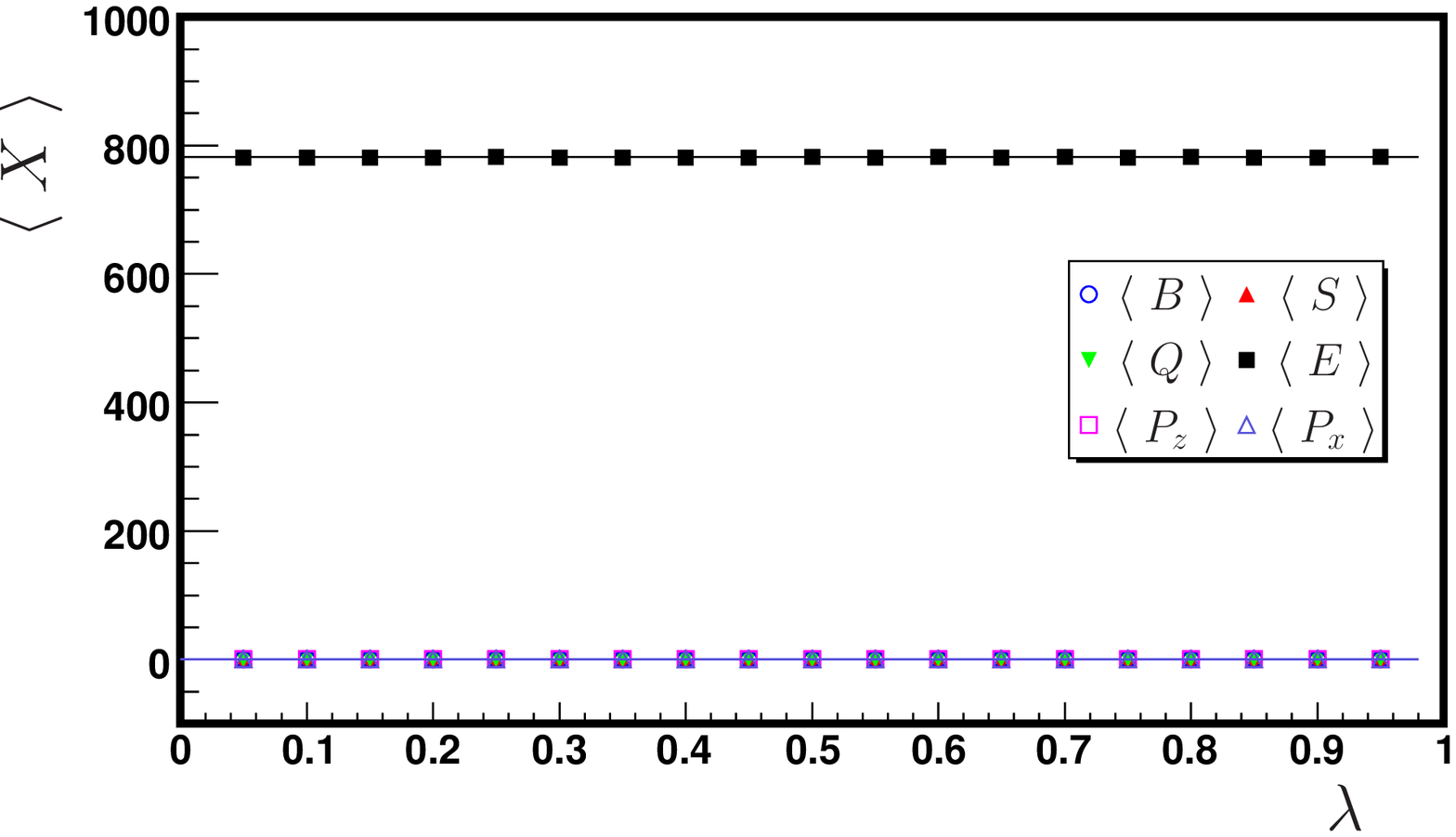,width=7.2cm,height=6.5cm}
  \epsfig{file=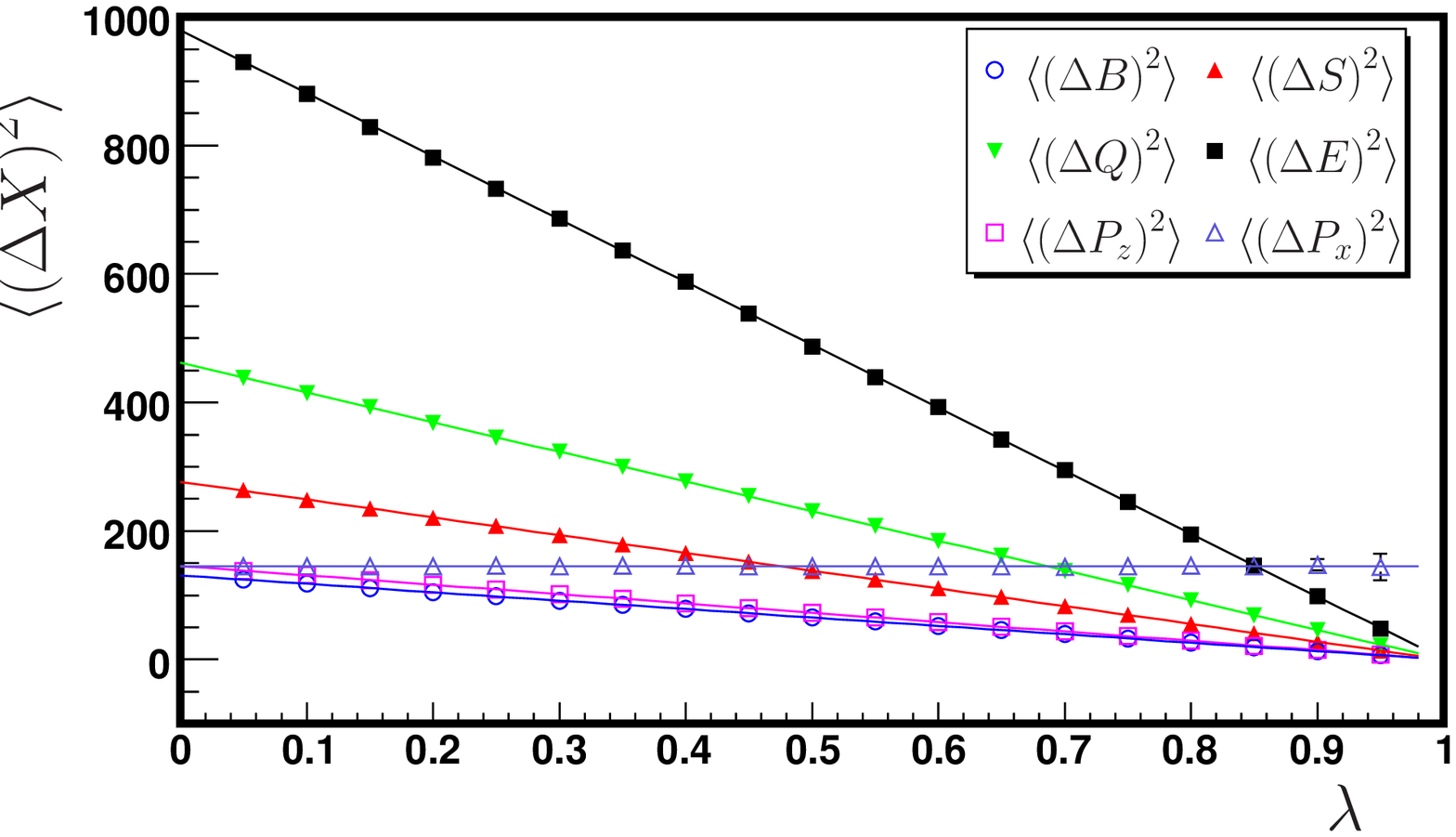,width=7.2cm,height=6.5cm}
  \caption{Mean values~({\it left}) and variances~({\it right}) of various extensive quantities, as listed in the 
  legends, as a function of~$\lambda$. Each marker and its error bar represents the result of 20 Monte Carlo runs 
  of~$2.5 \cdot 10^4$ events each. 19 different equally spaced values of~$\lambda$ have been investigated.
  Solid lines indicate GCE values~({\it left}), or linear extrapolations from the GCE value to the MCE 
  limit~({\it right}).}  
  \label{CwL_plot_fluc}
\end{figure}

\begin{figure}[ht!]
  \epsfig{file=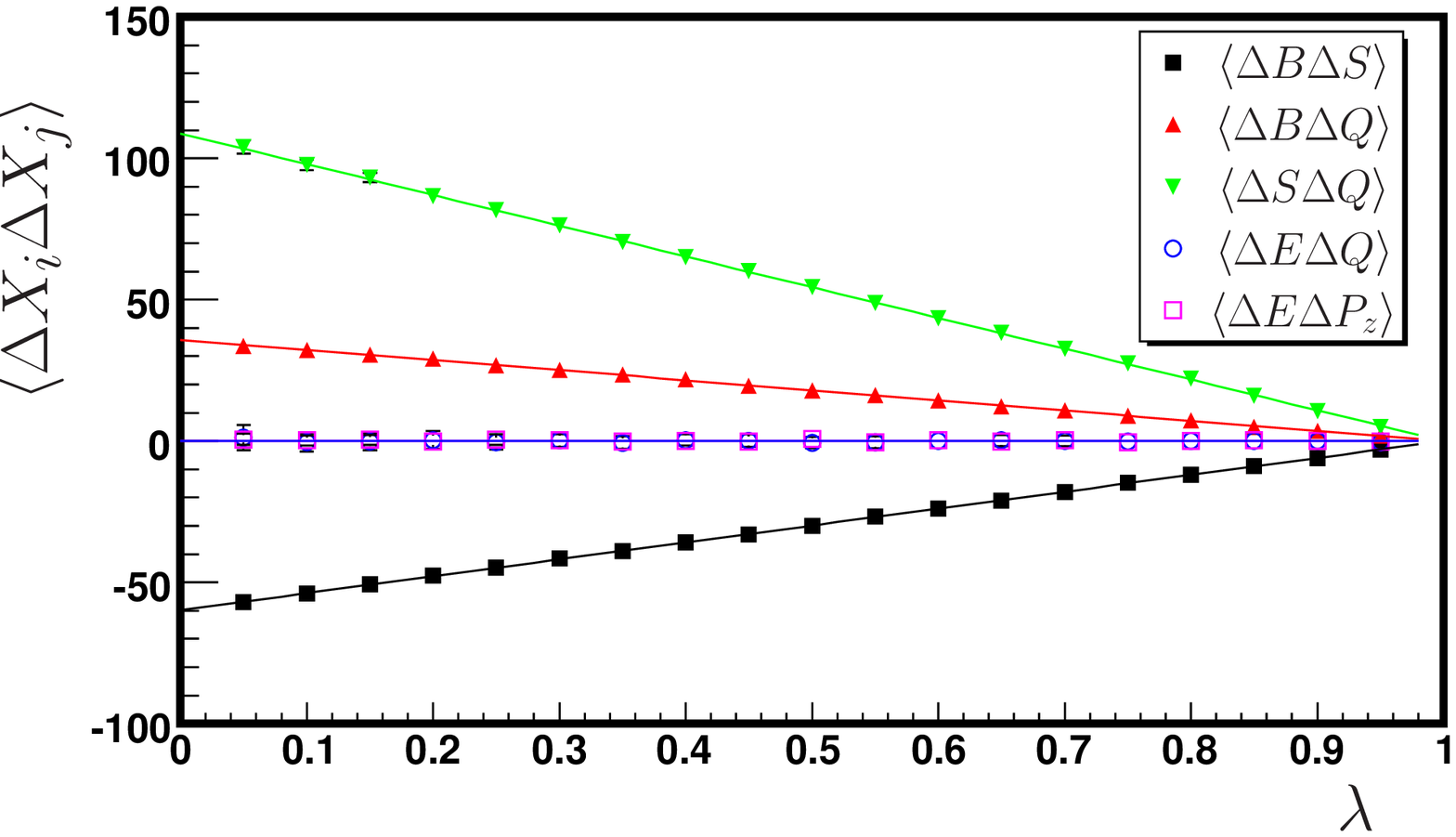,width=7.2cm,height=6.5cm}
  \epsfig{file=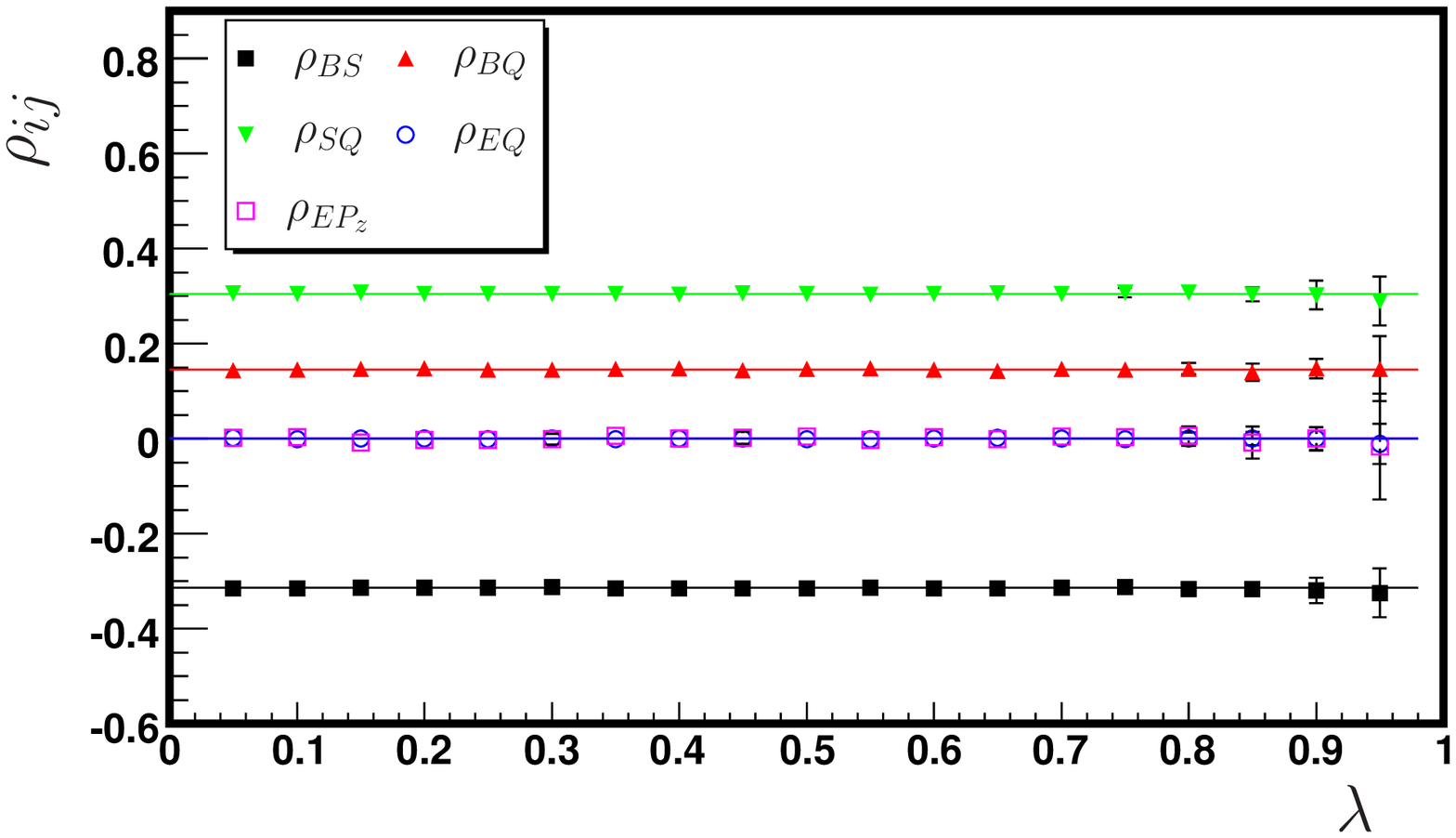,width=7.2cm,height=6.5cm}
  \caption{Covariances~({\it left}) and correlation coefficients~({\it right}) between various extensive quantities, 
  as listed in the legends, as a function of~$\lambda$. 
  Solid lines indicate  linear extrapolations from the GCE value to the MCE limit~({\it left}), or GCE values~({\it right}).
  The rest as in Fig.(\ref{CwL_plot_fluc}).}  
  \label{CwL_plot_corr}
\end{figure}

In Figs.(\ref{CwL_plot_fluc}) and (\ref{CwL_plot_corr}) the results of Monte Carlo runs of~$2.5 \cdot 10^4$ events each 
are shown. Each value of~$\lambda$ has been sampled~$20$ times to allow for calculation of a statistical uncertainty 
estimate. 19 different values of~$\lambda$ have been studied. In this case study, the extensive quantities baryon 
number~$B$, strangeness~$S$, electric charge~$Q$, energy~$E$, and longitudinal momentum~$P_z$ are considered for 
re-weighting. Conservation of transverse momenta~$P_x$ and~$P_y$ can be shown not to affect the~$\Delta p_{T,i}$ 
and~$\Delta y_i$ dependence of multiplicity fluctuations and correlations studied in Chapter~\ref{chapter_multfluc_hrg}.
Their~$\Delta y_i$ dependence is, however, rather sensitive to~$P_z$ conservation. Angular correlations, considered in 
Chapter~\ref{chapter_multfluc_pgas}, on the other hand, are strongly sensitive to joint~$P_x$ and~$P_y$ 
conservation~\cite{Hauer:2007im,Hauer:2009nv}.

Fig.(\ref{CwL_plot_fluc})~({\it left}) summarizes the results for mean values of baryon number~$\langle B \rangle$, 
strangeness~$\langle S \rangle$, electric charge~$\langle Q\rangle$, energy~$\langle E \rangle$, and the
momenta~$\langle P_x \rangle$ and~$\langle P_z \rangle$. The solid lines represent GCE values. Only the expectation 
value of energy is not equal to~$0$, since the system sampled is assumed to be static and neutral with~$T \not= 0$. 
The evolution of the respective variances is shown in Fig.(\ref{CwL_plot_fluc})~({\it right}). 
Variances of extensive quantities subject to re-weighting converge linearly to~$0$ as~$\lambda$ goes to~$1$. One notes 
that~$\langle \left( \Delta P_x \right)^2 \rangle$ remains constant (within error bars), as this quantity is not 
re-weighted in this case study. Error bars on many data points are smaller than the symbol used.

In Fig.(\ref{CwL_plot_corr})~({\it left}) the evolution of 
covariances~$\langle \Delta B \Delta S \rangle$,~$\langle \Delta B \Delta Q \rangle$, 
$\langle \Delta S \Delta Q \rangle$, and~$\langle \Delta E \Delta Q \rangle$ is shown as a function of the `size of 
the bath'. As seen, the covariances between quantities considered for re-weighting converge linearly to~$0$. In a neutral 
system, covariances between energy and charges are equal to 0. As an example, $\langle \Delta E \Delta Q \rangle$ is shown. 
In a static system, also the covariances between momenta and any other extensive quantity are equal to~$0$. As an example 
here~$\langle \Delta E \Delta P_z \rangle$ is shown. The correlation coefficients, Eq.(\ref{rho}), on the other hand, 
remain constant as a function of~$\lambda$, as shown in Fig.(\ref{CwL_plot_corr})~({\it right}). The values of fully 
phase space integrated correlation coefficients~$\rho_{BS}$,~$\rho_{BQ}$, and~$\rho_{SQ}$ can be compared to the GCE 
results denoted by the solid lines shown in {Figs.(\ref{lc_bs_0000}-\ref{lc_sq_0000})} in Section~\ref{sec_gce_LCfluc}.

\section{Probability Distributions}
\label{sec_extrapol_dist}

\begin{figure}[ht!]  
  \epsfig{file=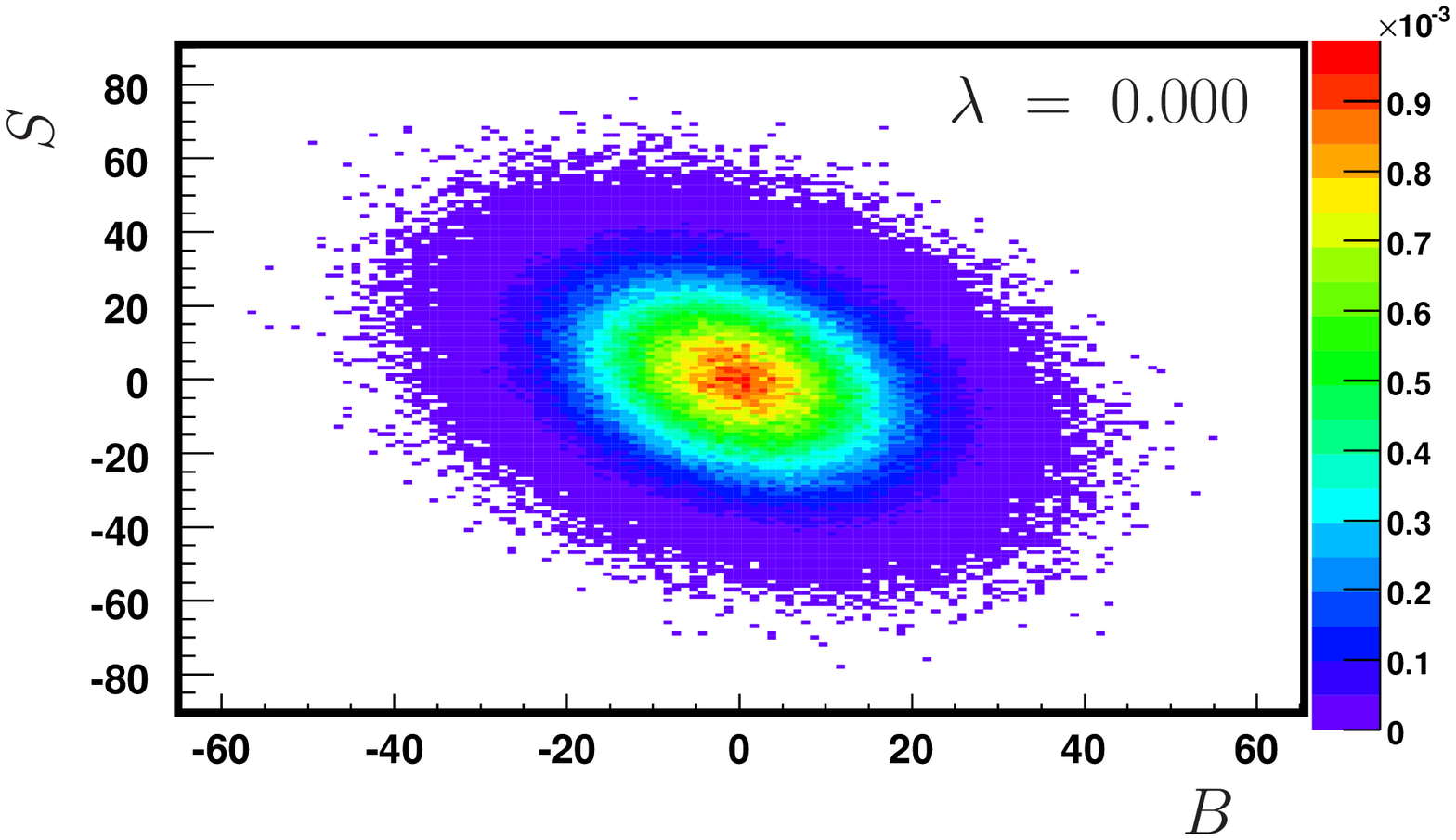,width=4.75cm,height=3.0cm}
  \epsfig{file=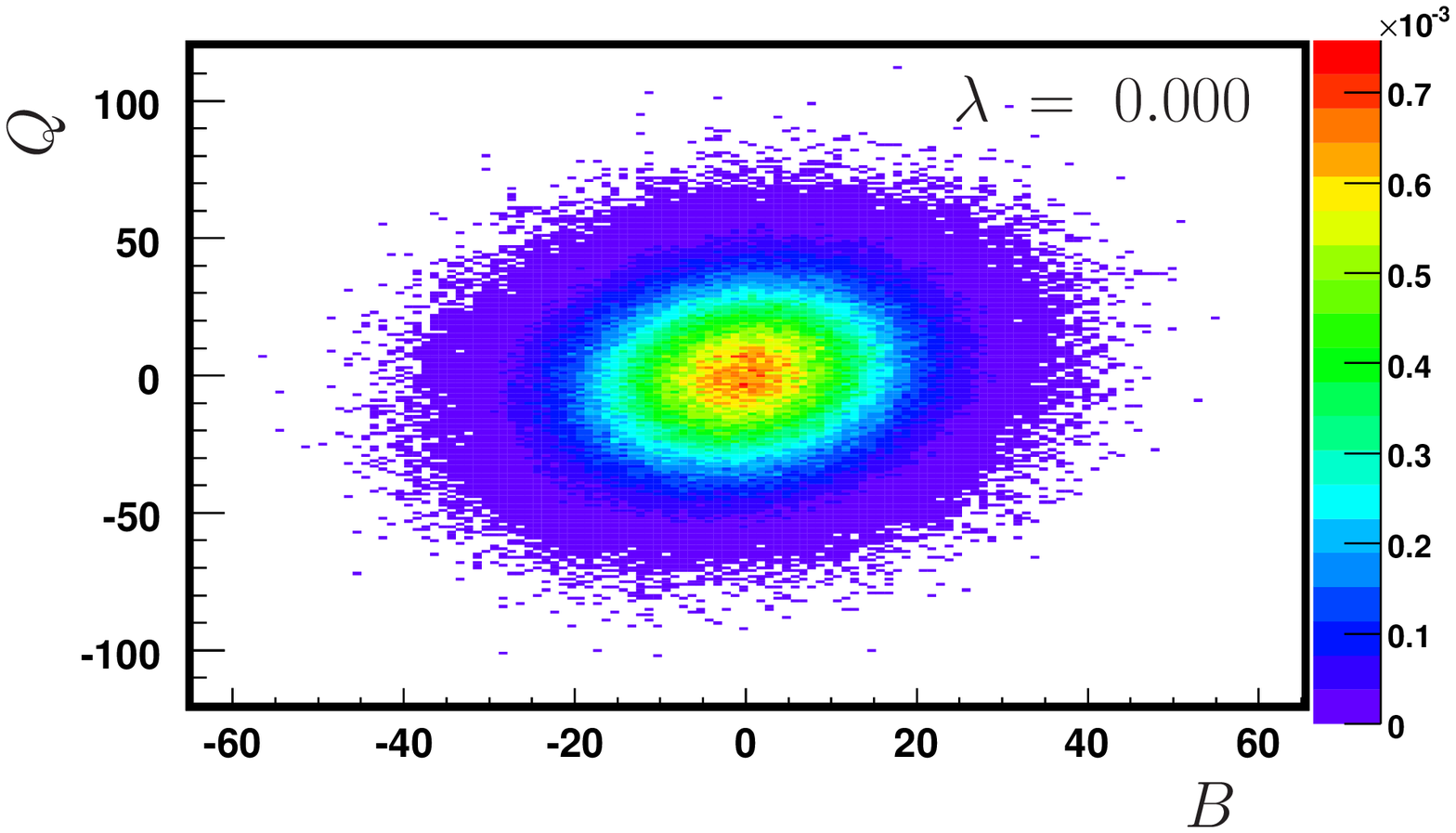,width=4.75cm,height=3.0cm}
  \epsfig{file=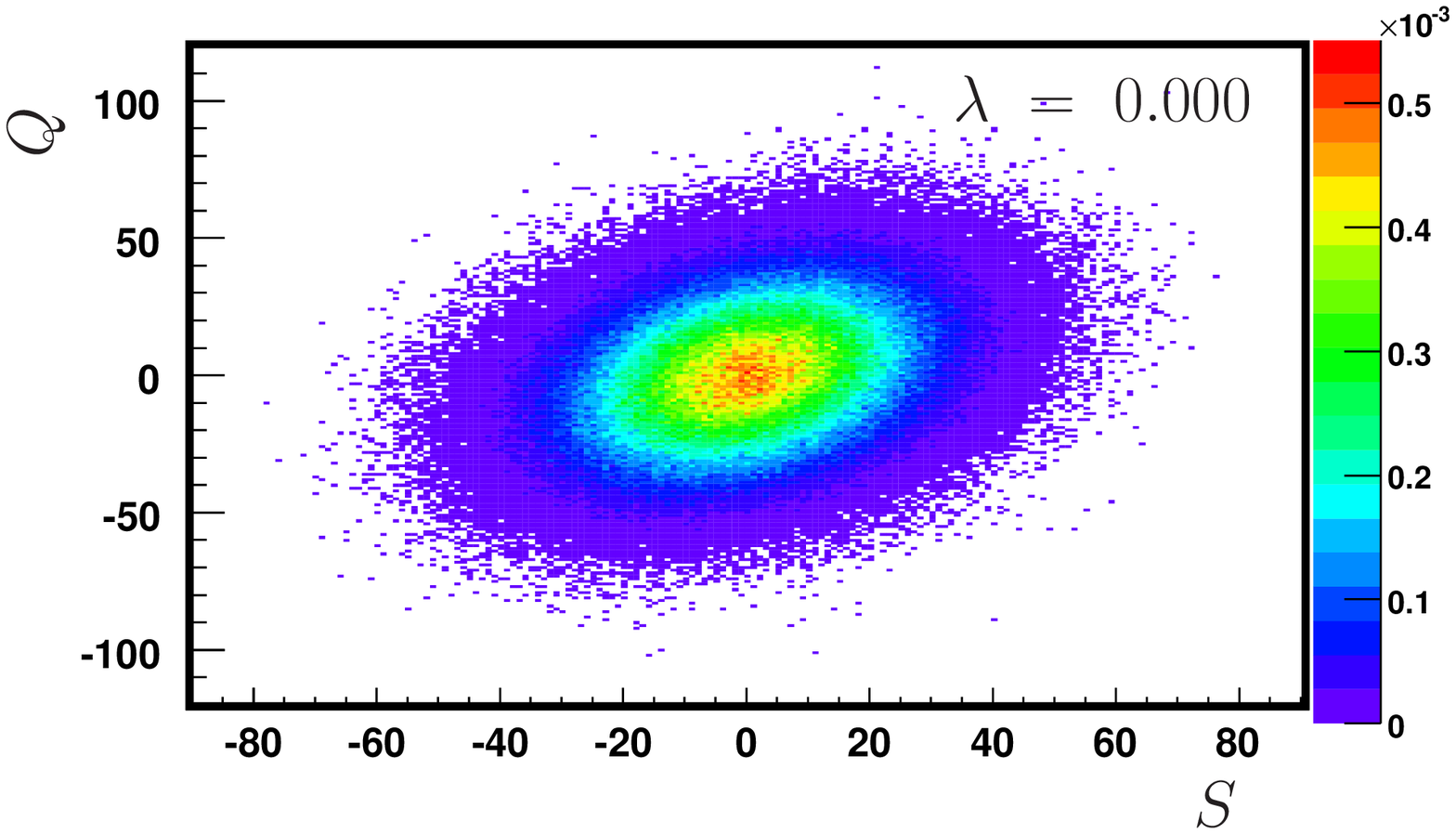,width=4.75cm,height=3.0cm}

  \epsfig{file=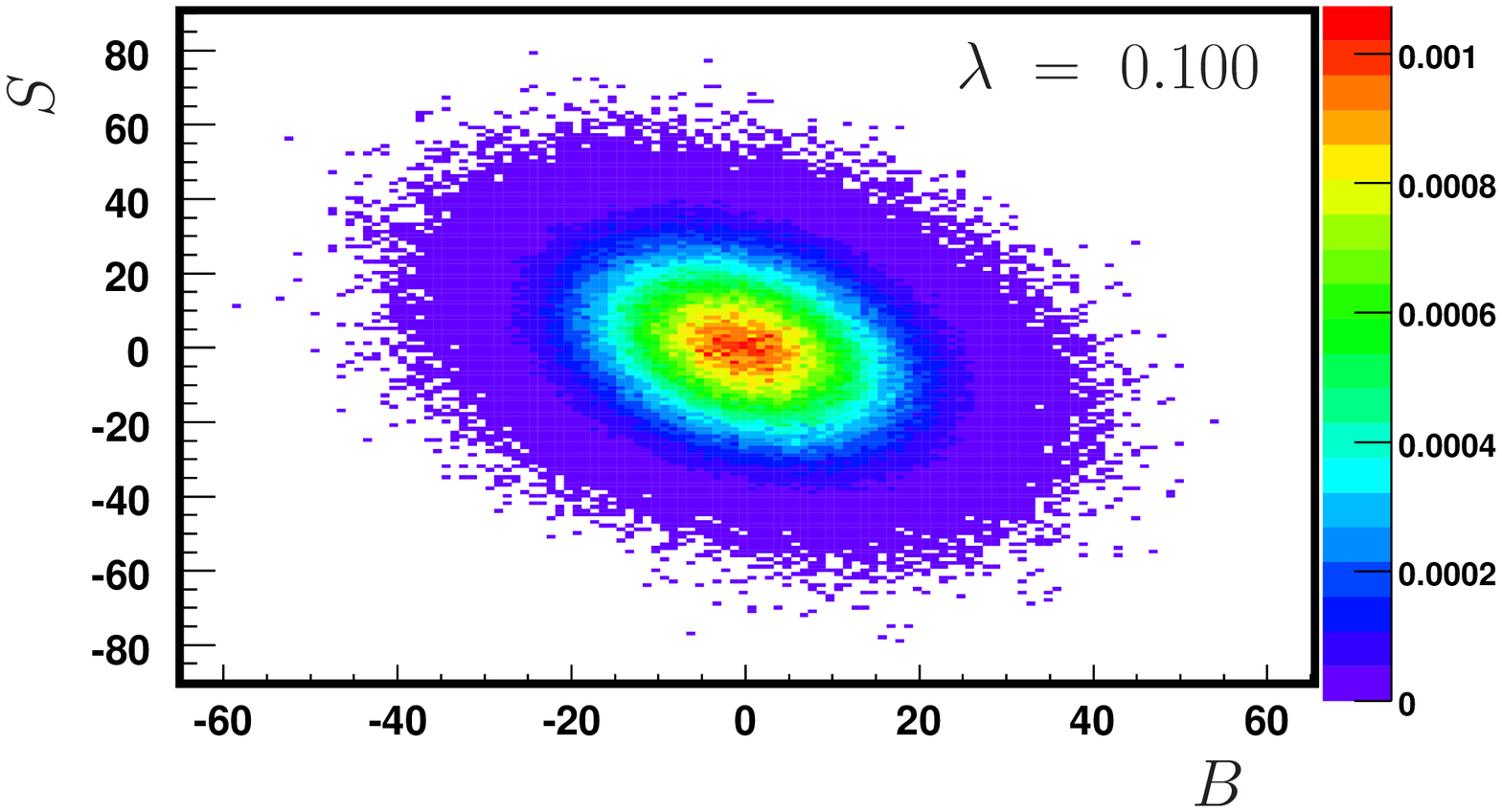,width=4.75cm,height=3.0cm}
  \epsfig{file=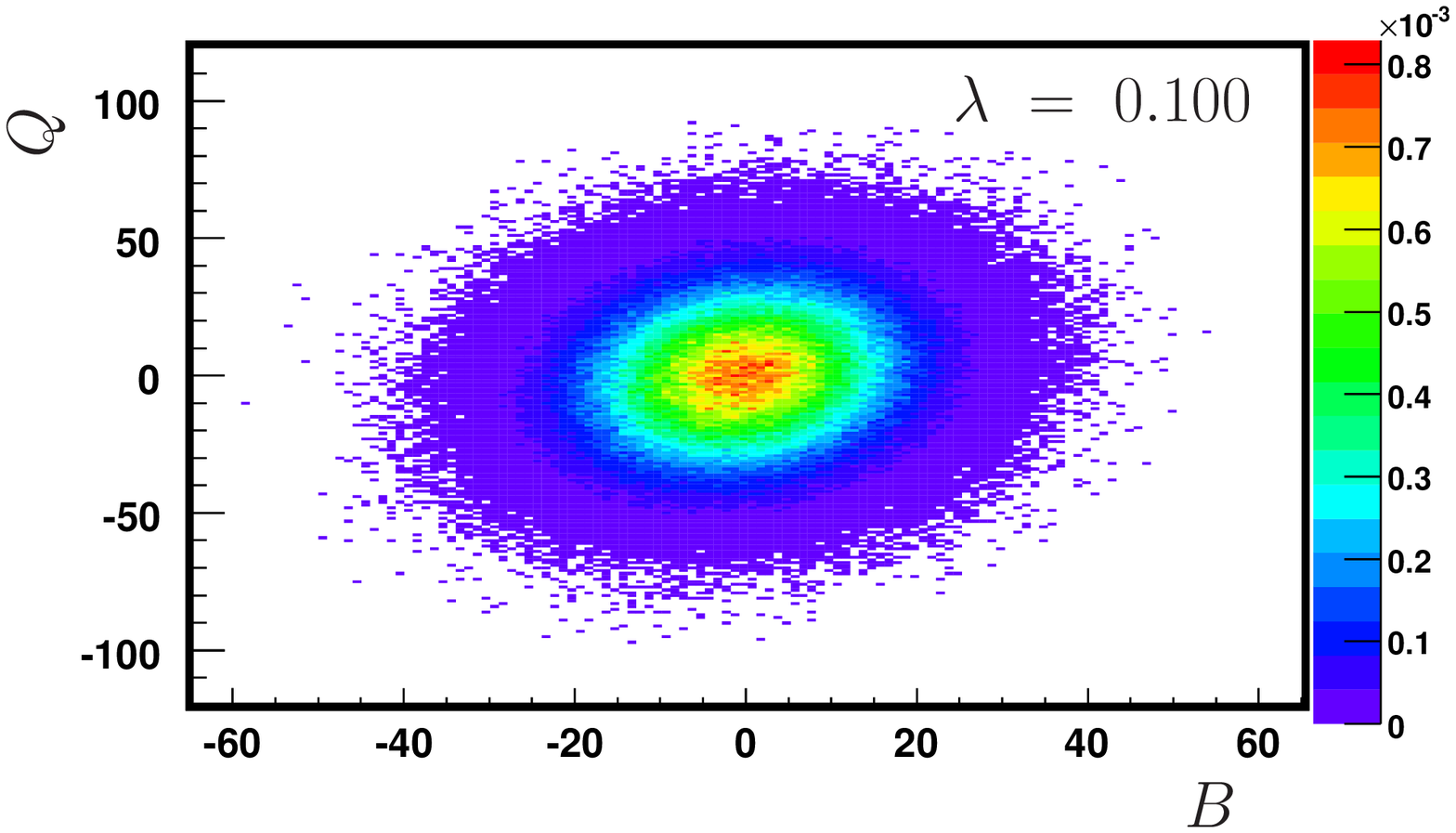,width=4.75cm,height=3.0cm}
  \epsfig{file=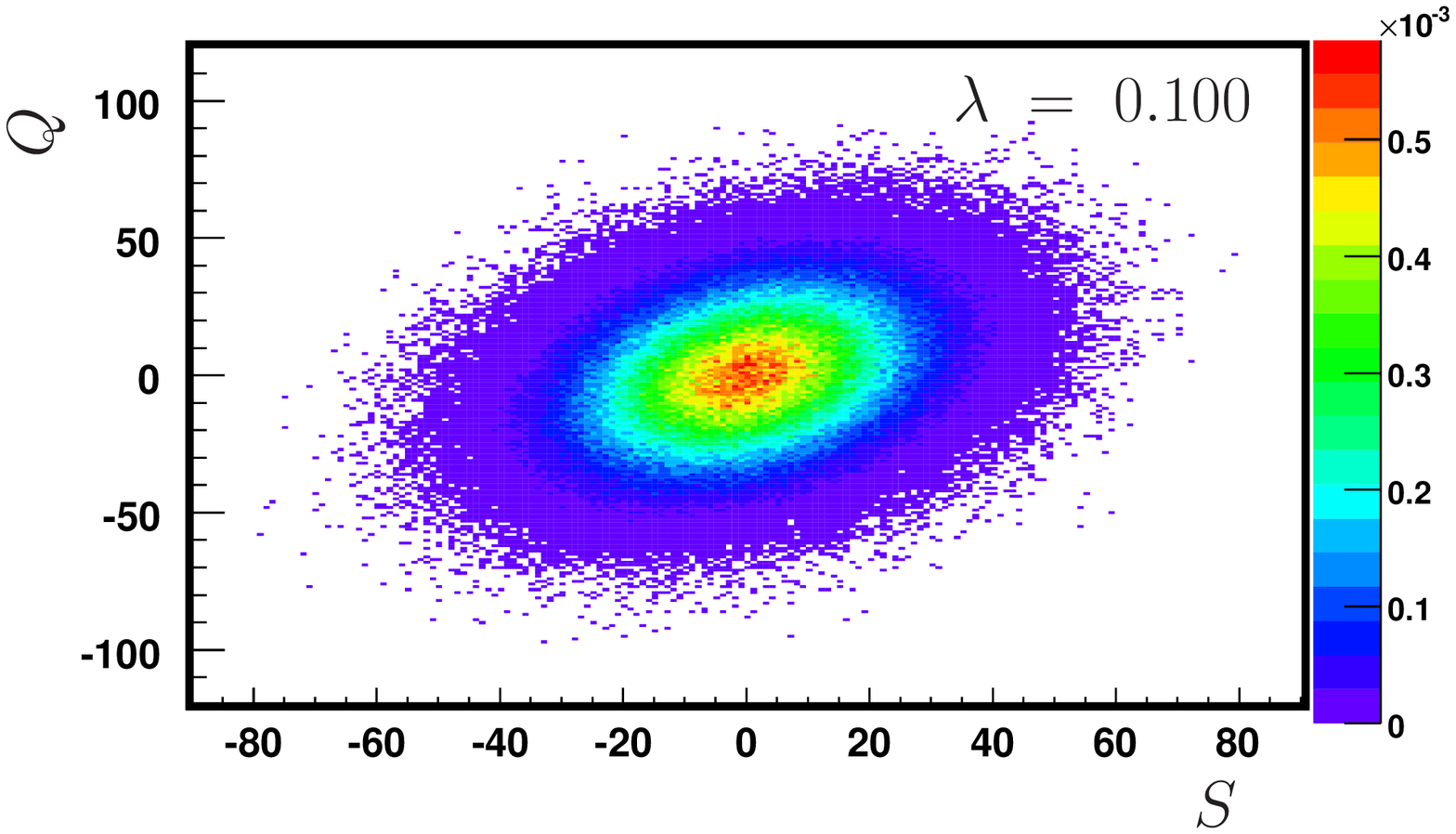,width=4.75cm,height=3.0cm}

  \epsfig{file=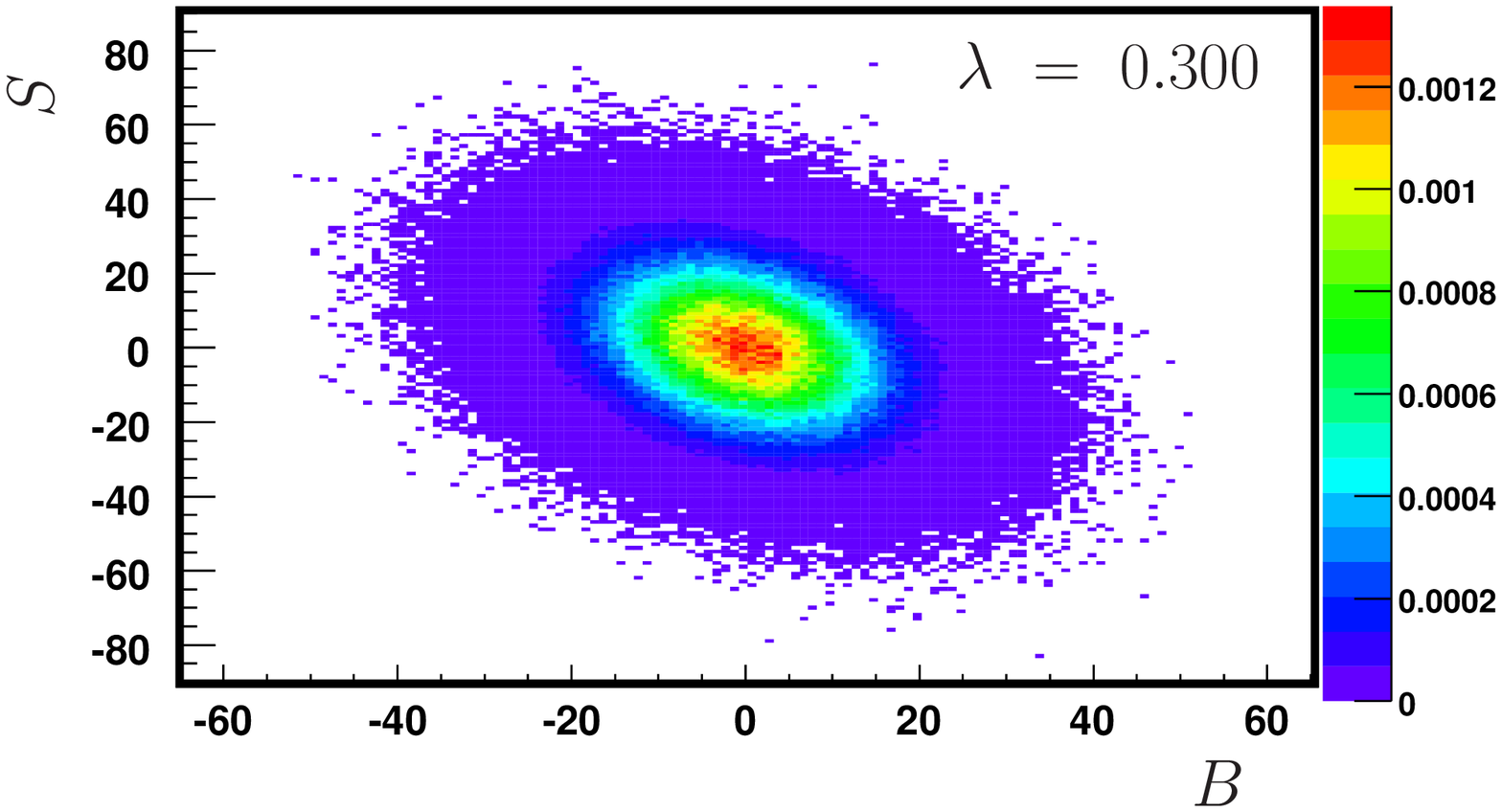,width=4.75cm,height=3.0cm}
  \epsfig{file=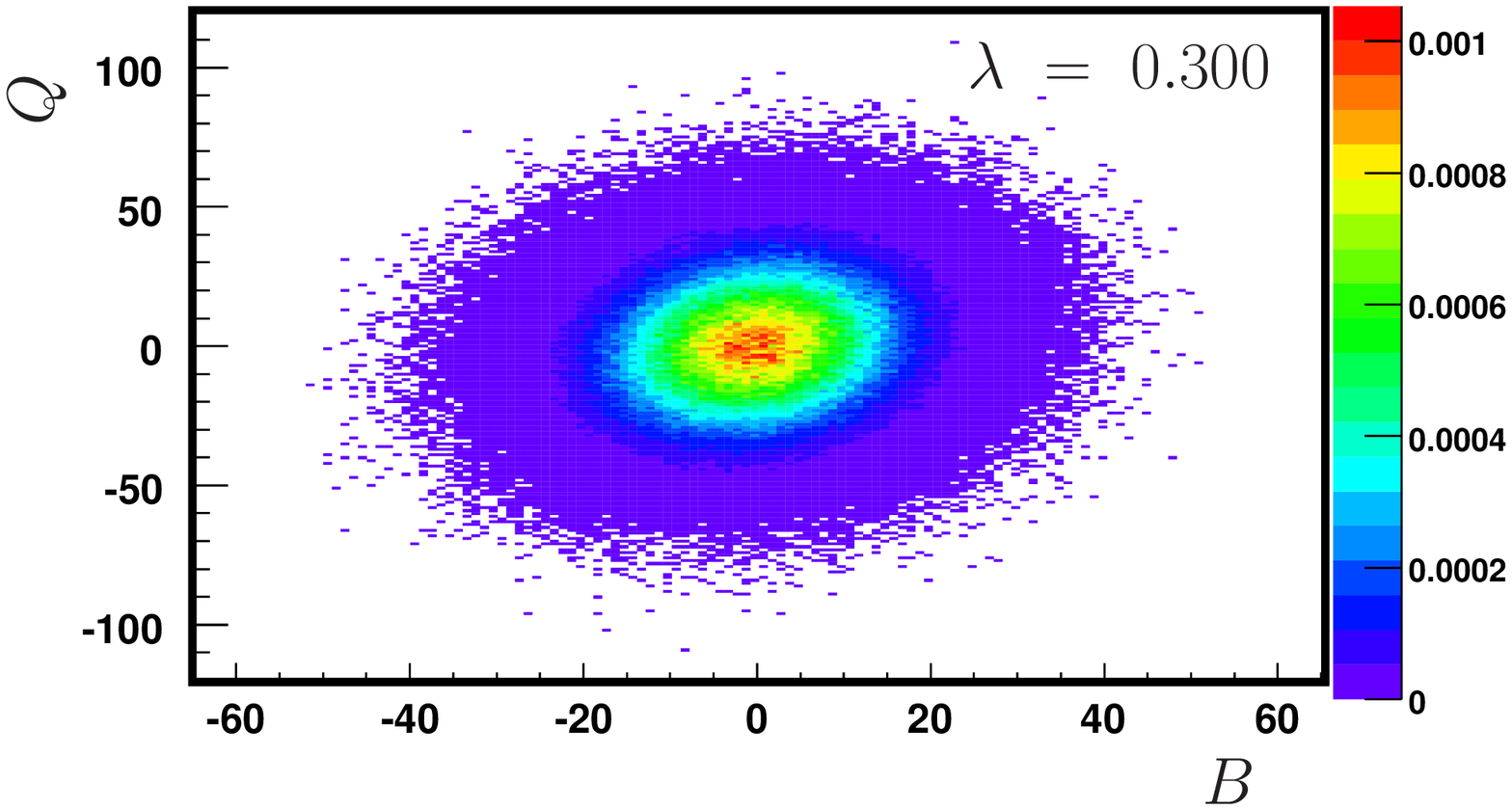,width=4.75cm,height=3.0cm}
  \epsfig{file=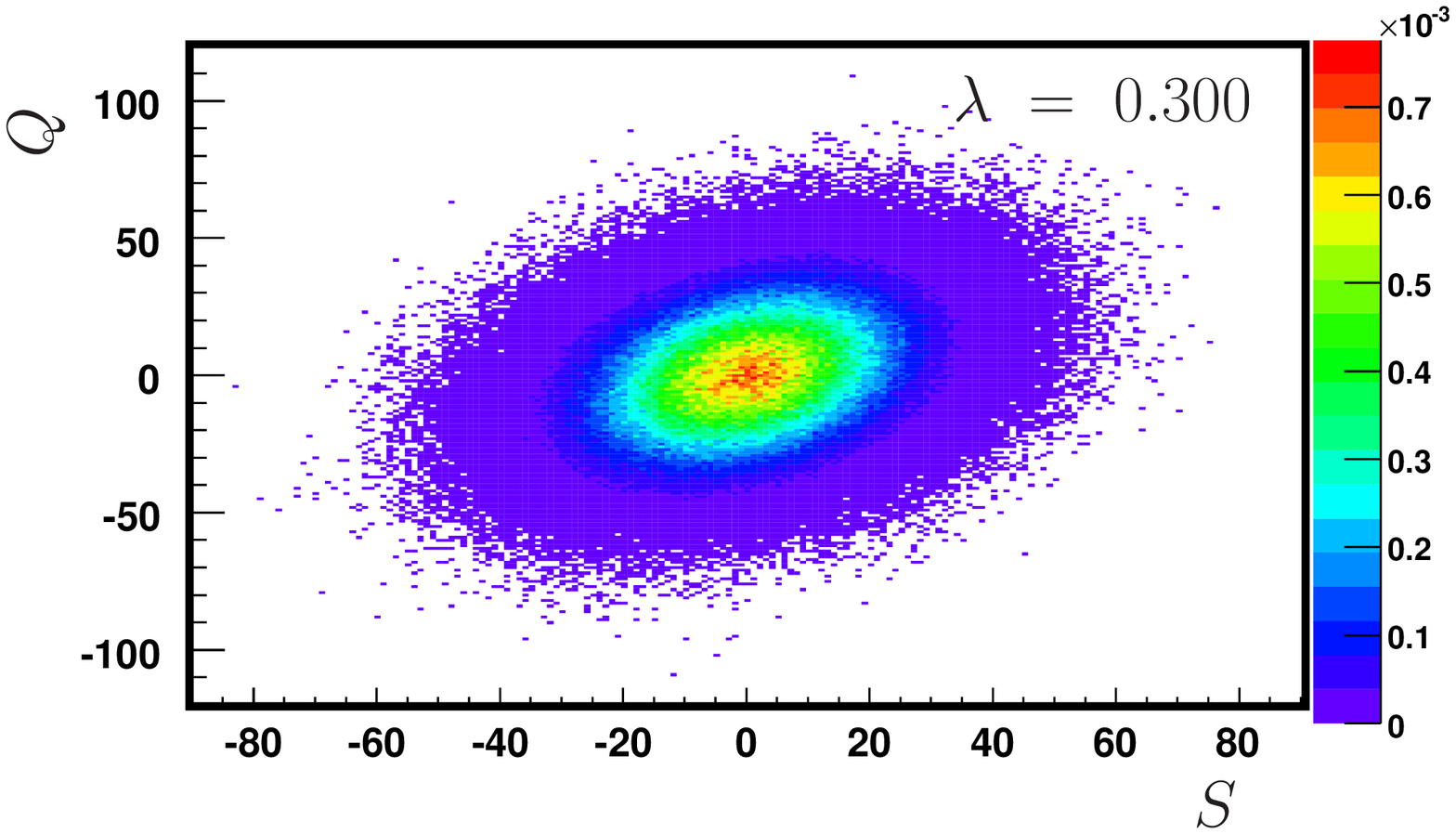,width=4.75cm,height=3.0cm}

  \epsfig{file=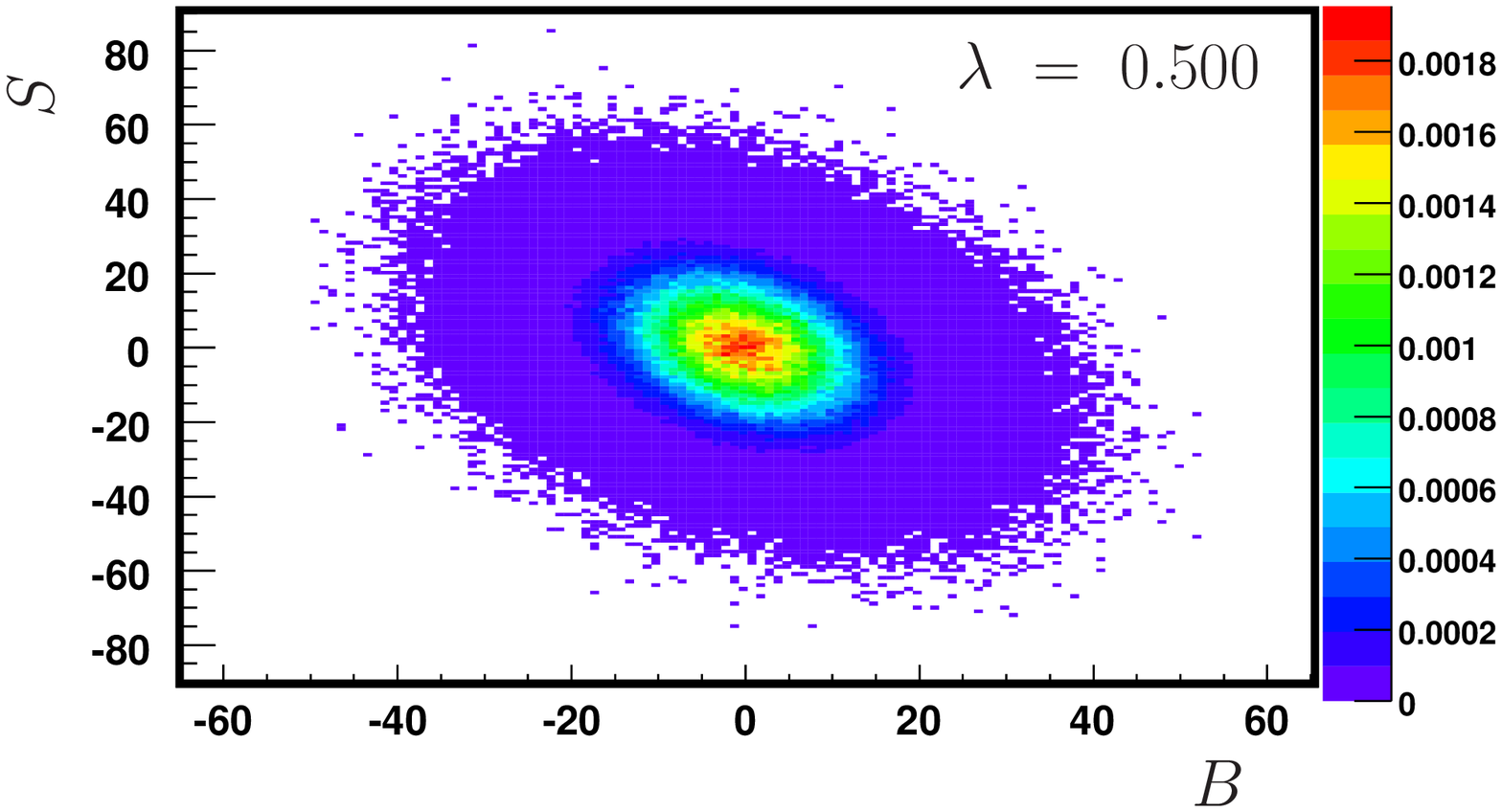,width=4.75cm,height=3.0cm}
  \epsfig{file=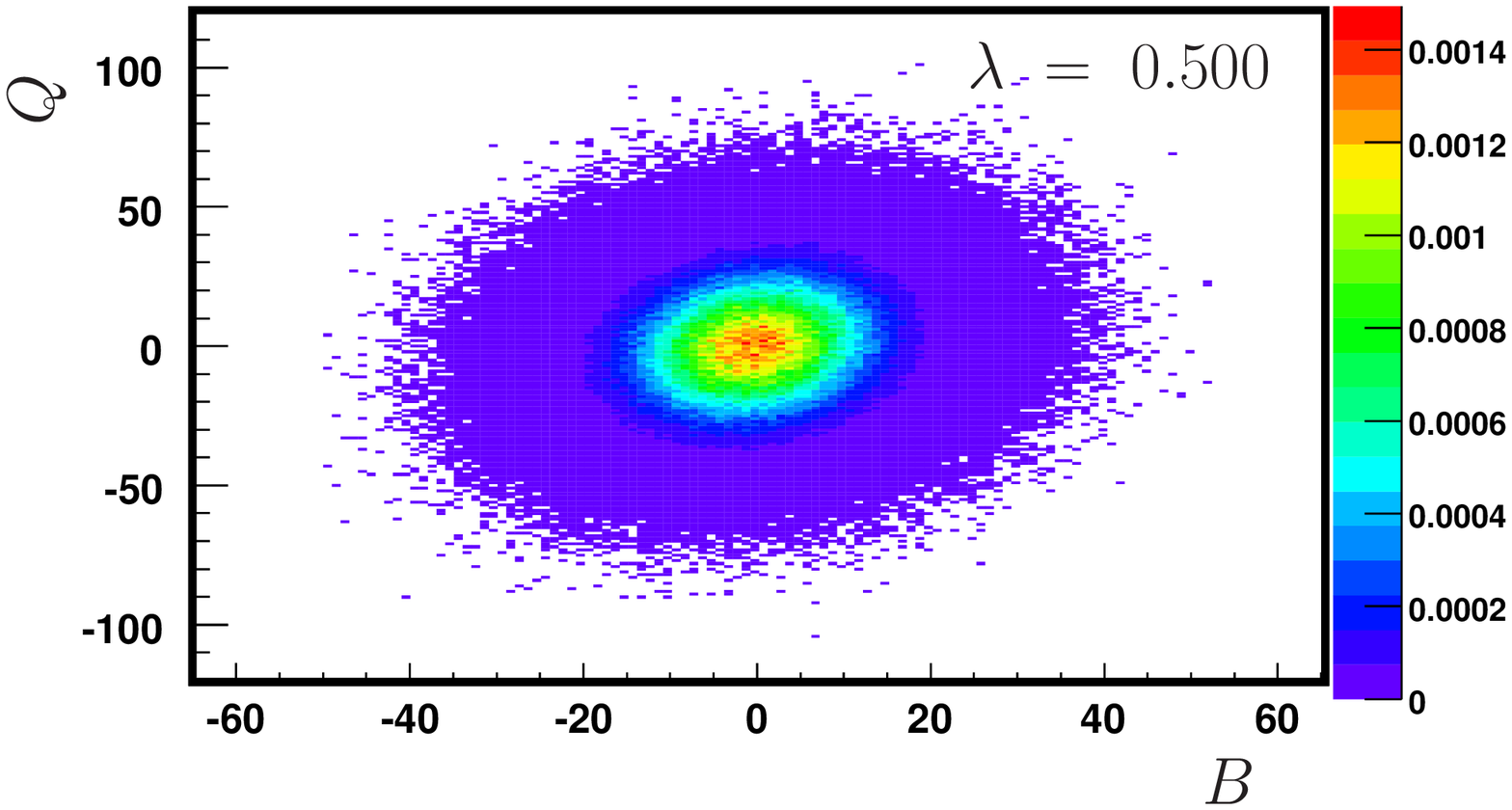,width=4.75cm,height=3.0cm}
  \epsfig{file=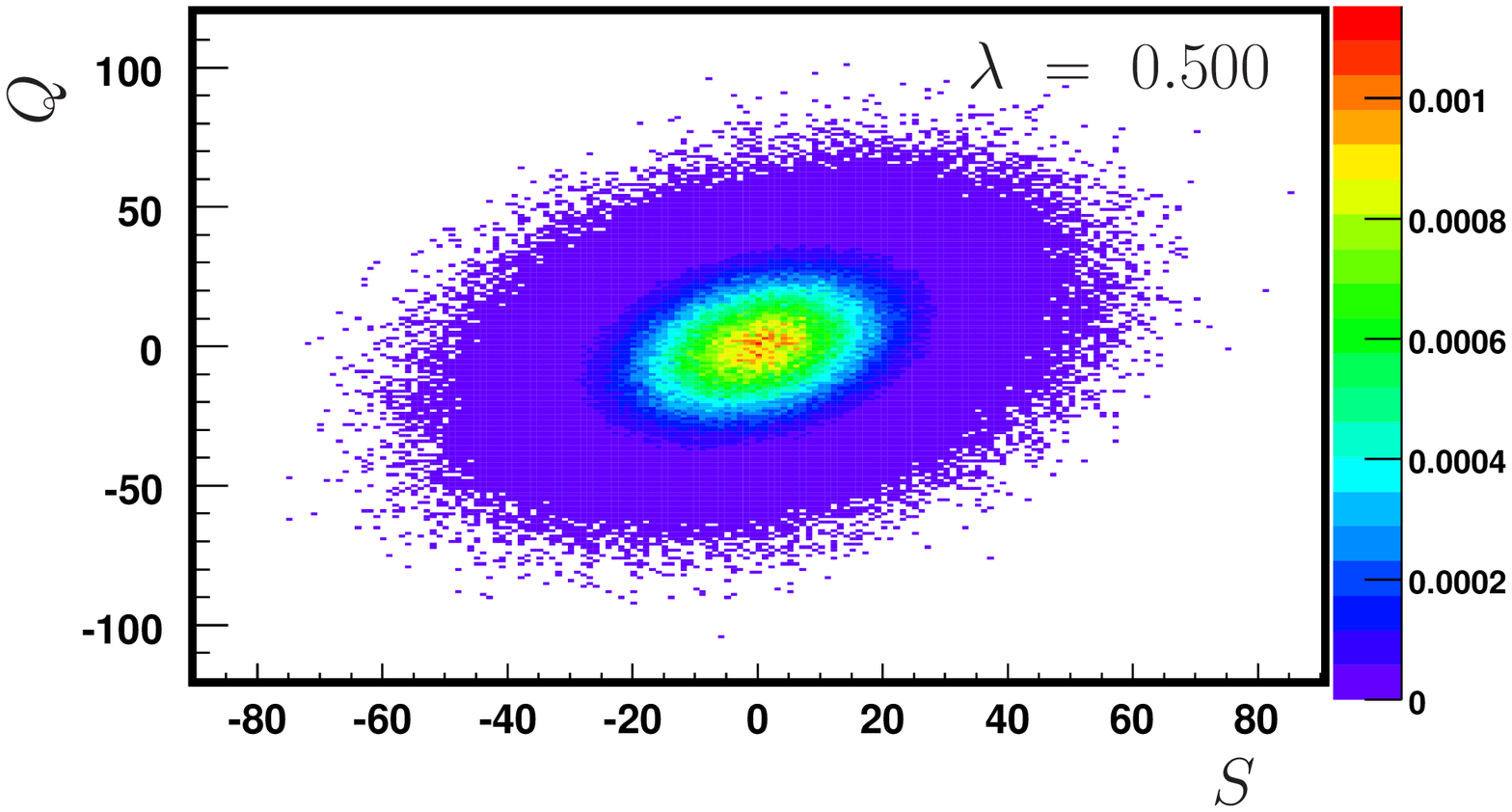,width=4.75cm,height=3.0cm}

  \epsfig{file=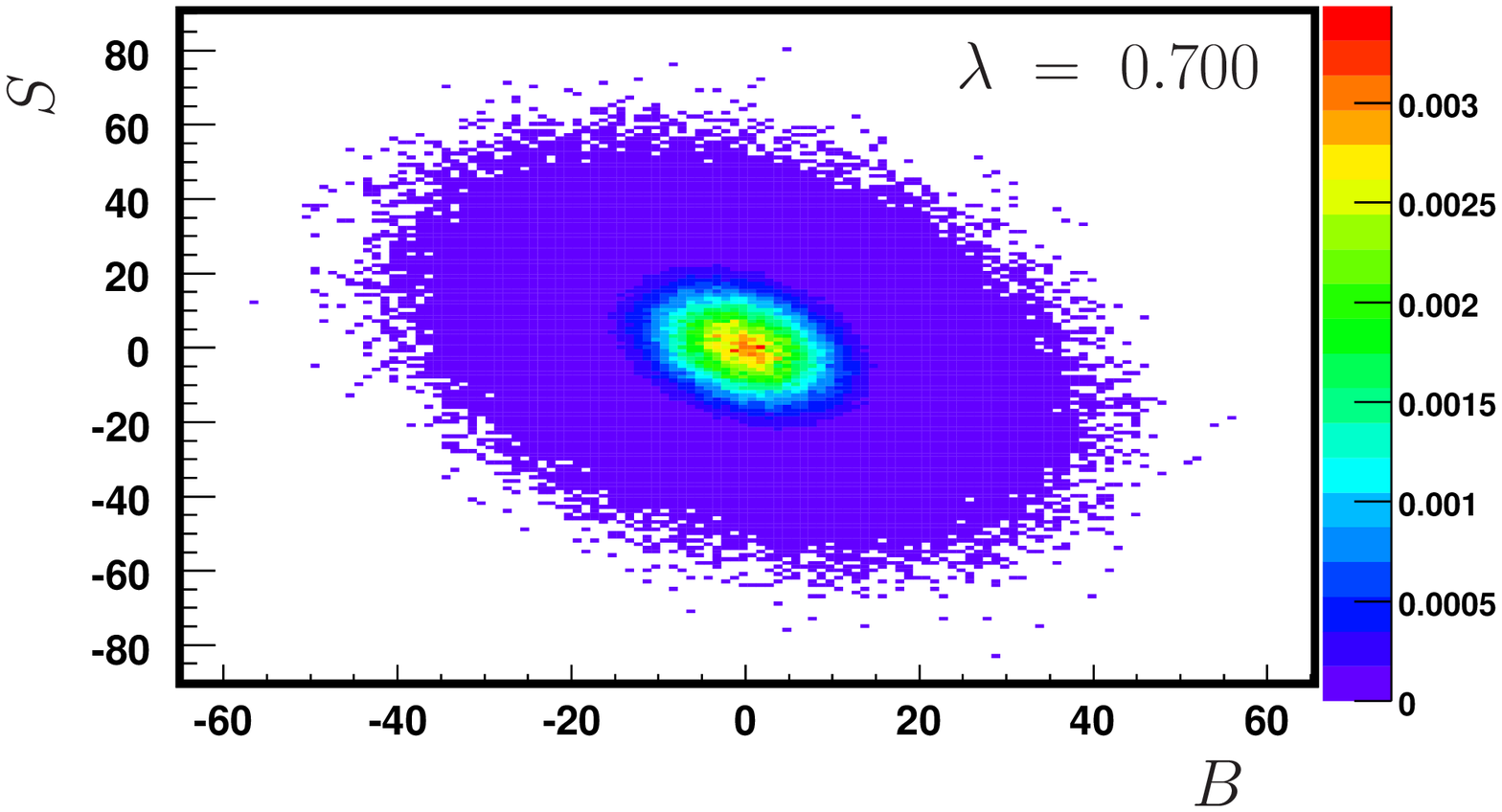,width=4.75cm,height=3.0cm}
  \epsfig{file=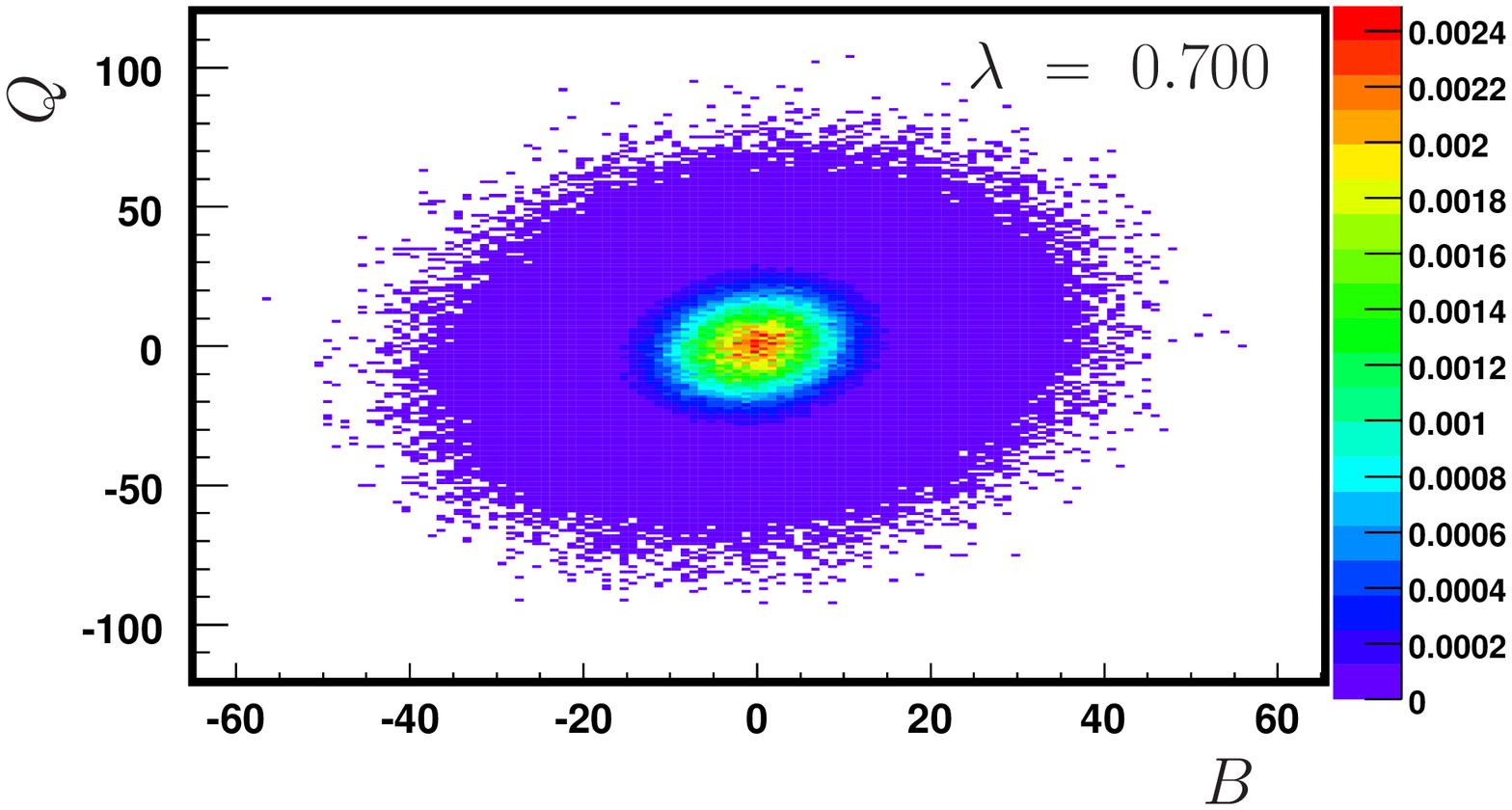,width=4.75cm,height=3.0cm}
  \epsfig{file=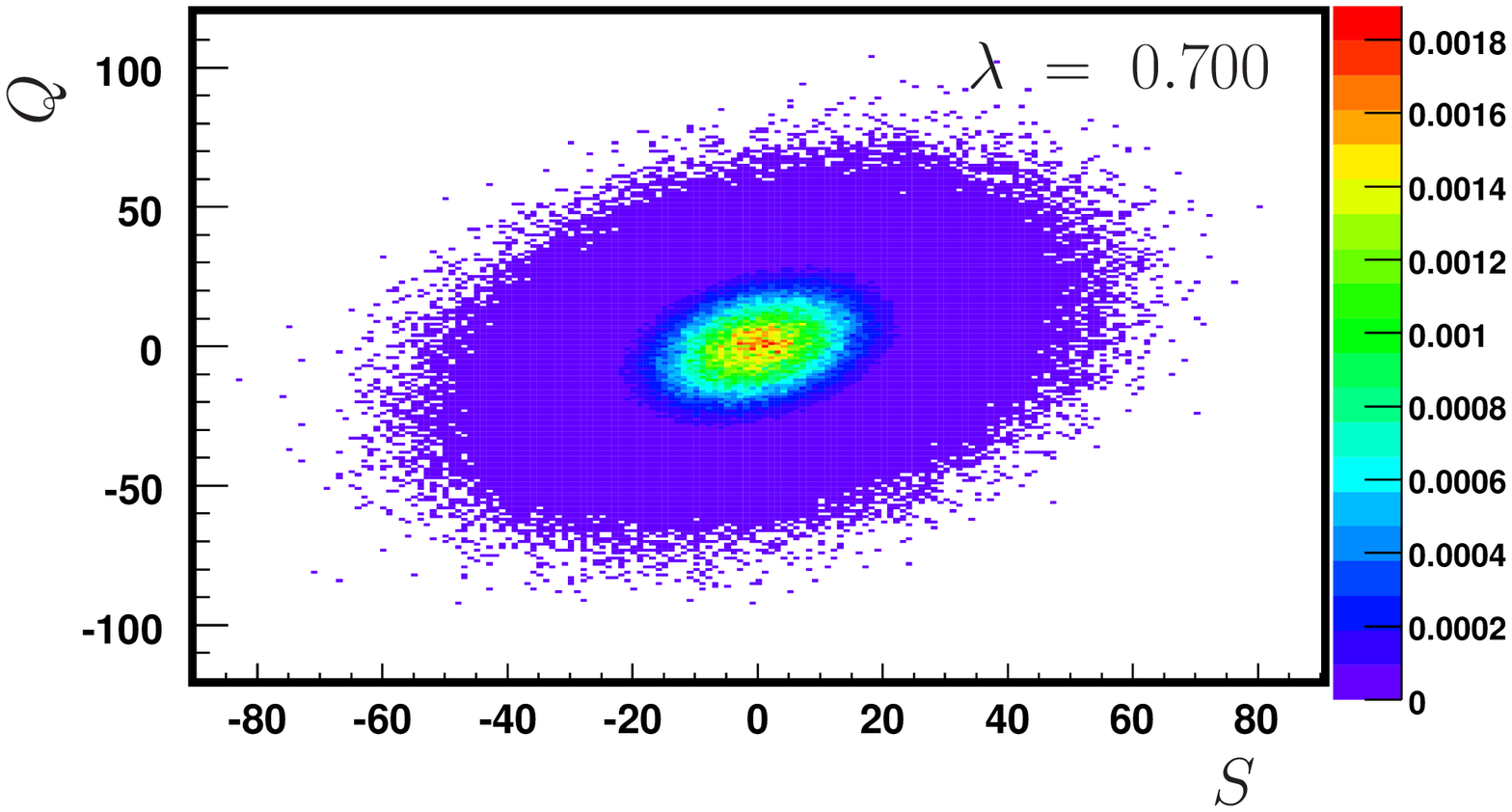,width=4.75cm,height=3.0cm}

  \epsfig{file=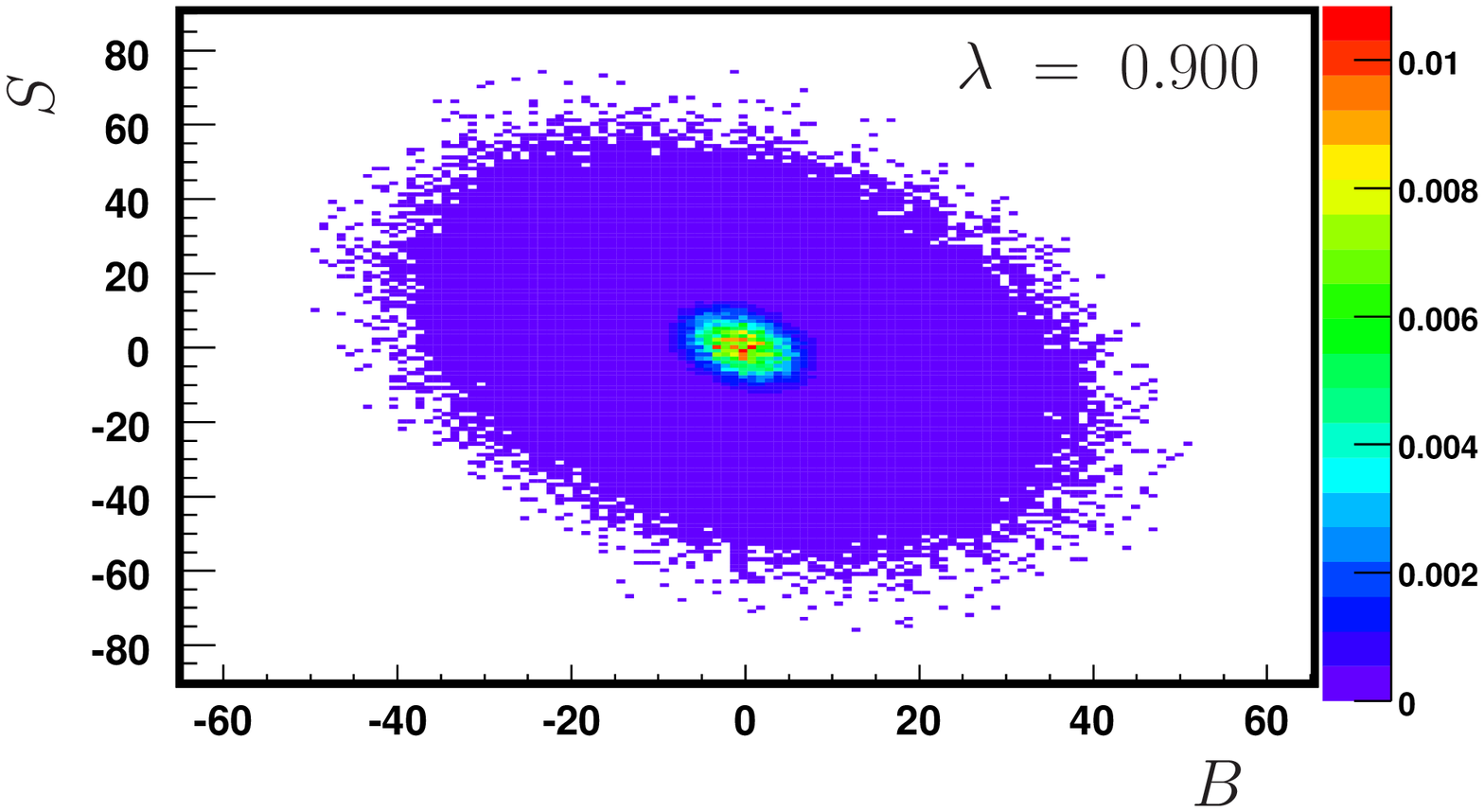,width=4.75cm,height=3.0cm}
  \epsfig{file=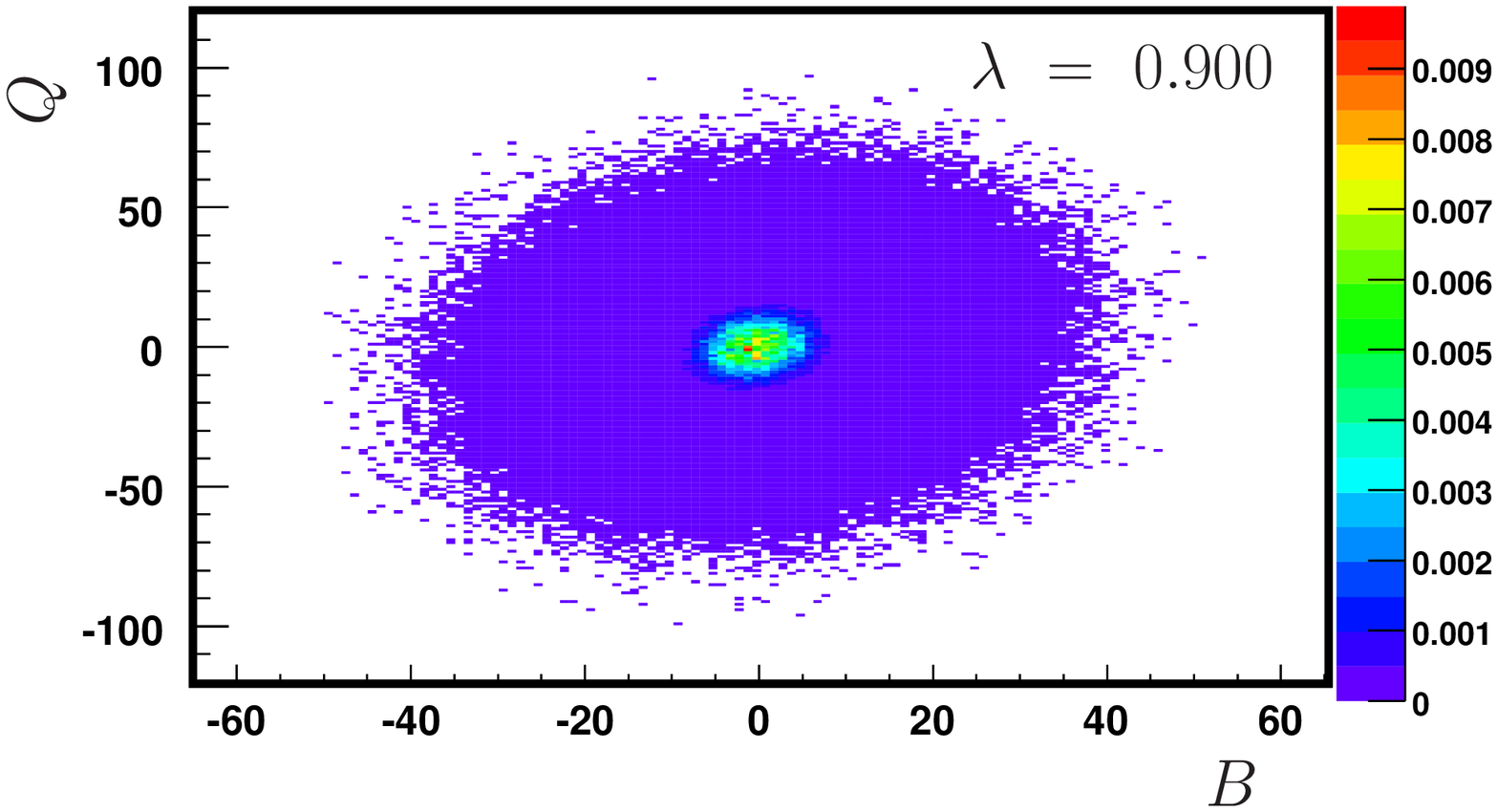,width=4.75cm,height=3.0cm}
  \epsfig{file=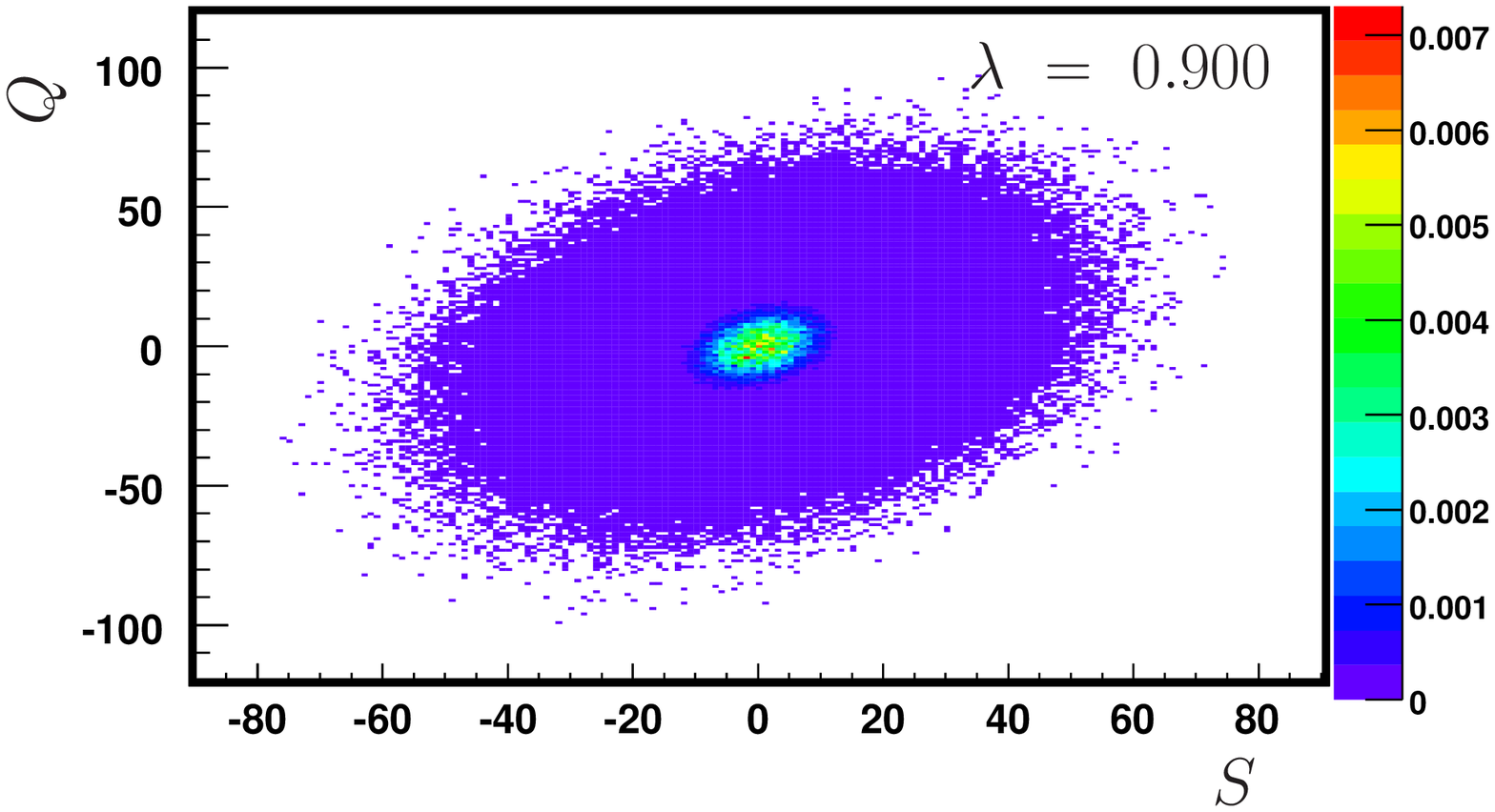,width=4.75cm,height=3.0cm}
 
  \caption{Evolution of the joint distribution of baryon number and strangeness,~$P_{\lambda}(B,S)$~({\it left}), 
  baryon number and electric charge,~$P_{\lambda}(B,Q)$~({\it center}), and strangeness and electric 
  charge,~$P_{\lambda}(S,Q)$~({\it right}), with the size of the bath~$\lambda = V_1/V_g$. 
  Here~$5 \cdot 10^5$ events have been sampled for each value of~$\lambda$.}           
  \label{P_lambda_charge}
\end{figure}

It is also interesting to study probability distributions as a function of the size of the bath~$\lambda$. Firstly 
joint distributions of charges,~$P_{\lambda}(B,S)$,~$P_{\lambda}(B,Q)$, and~$P_{\lambda}(S,Q)$ from~({\it left}) 
to~({\it right}), are attended to in Fig.(\ref{P_lambda_charge}). Mean values of net-charges stay constant throughout the 
extrapolation, see also Fig.(\ref{CwL_plot_fluc})~({\it left}). Their variances and covariances, compare 
Fig.(\ref{CwL_plot_fluc})~({\it right}) and Fig.(\ref{CwL_plot_corr})~({\it left}), converge to~$0$ as the 
size of the bath is reduced. All three quantities, $B$, $S$, and $Q$ are re-weighted, and thus correlation 
coefficients amongst them, Fig.(\ref{CwL_plot_corr})~({\it right}), stay constant. Hence events in a small region 
around~$\langle B_{1} \rangle$, $\langle S_{1} \rangle$, and~$\langle Q_{1} \rangle$, are highlighted, 
while correlations between them are left unchanged\footnote{Sometimes it may not be clear which ensemble one should 
choose to apply to a given statistical system. The three standard ensembles remain particular idealizations.
Also the scenario discussed here is a particular limit. The system under investigation could have a bath which might 
impose its own correlations. Or in practical terms, have a more general form of the weight~$\mathcal{W}^{Q_1^l}$ then 
the one discussed here.}.

\begin{figure}[ht!]
  \epsfig{file=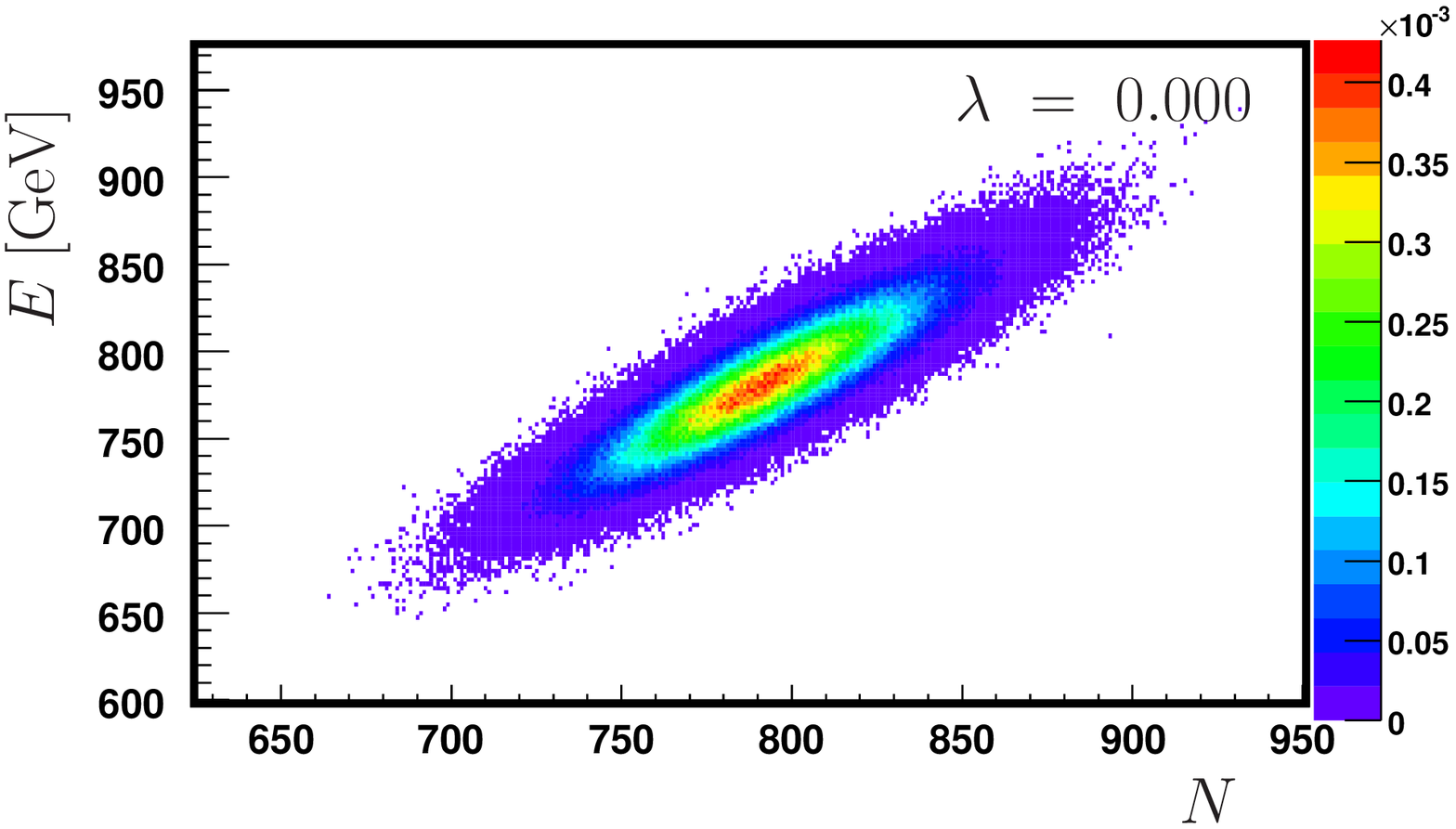,width=4.75cm,height=3.0cm}
  \epsfig{file=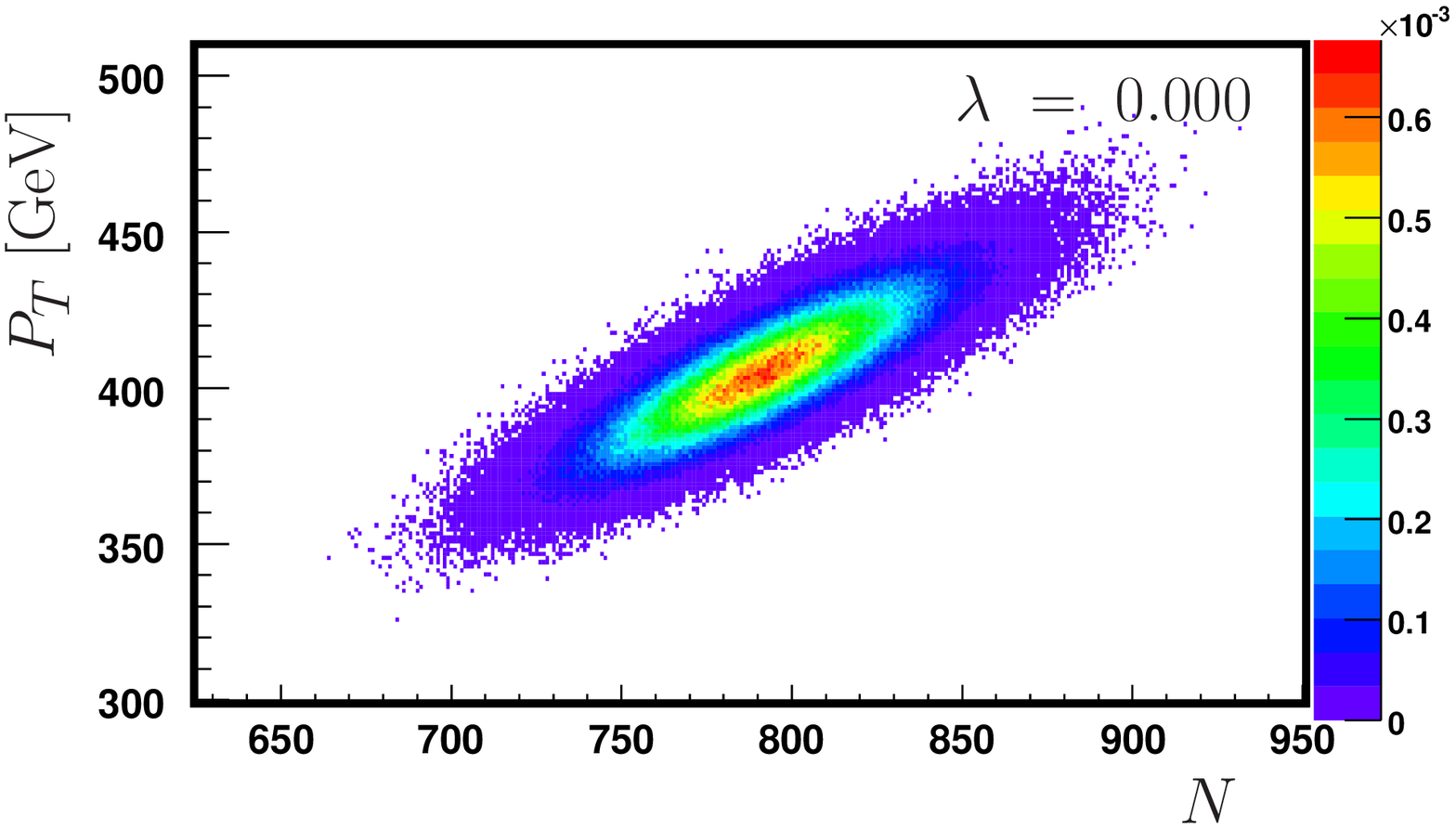,width=4.75cm,height=3.0cm}
  \epsfig{file=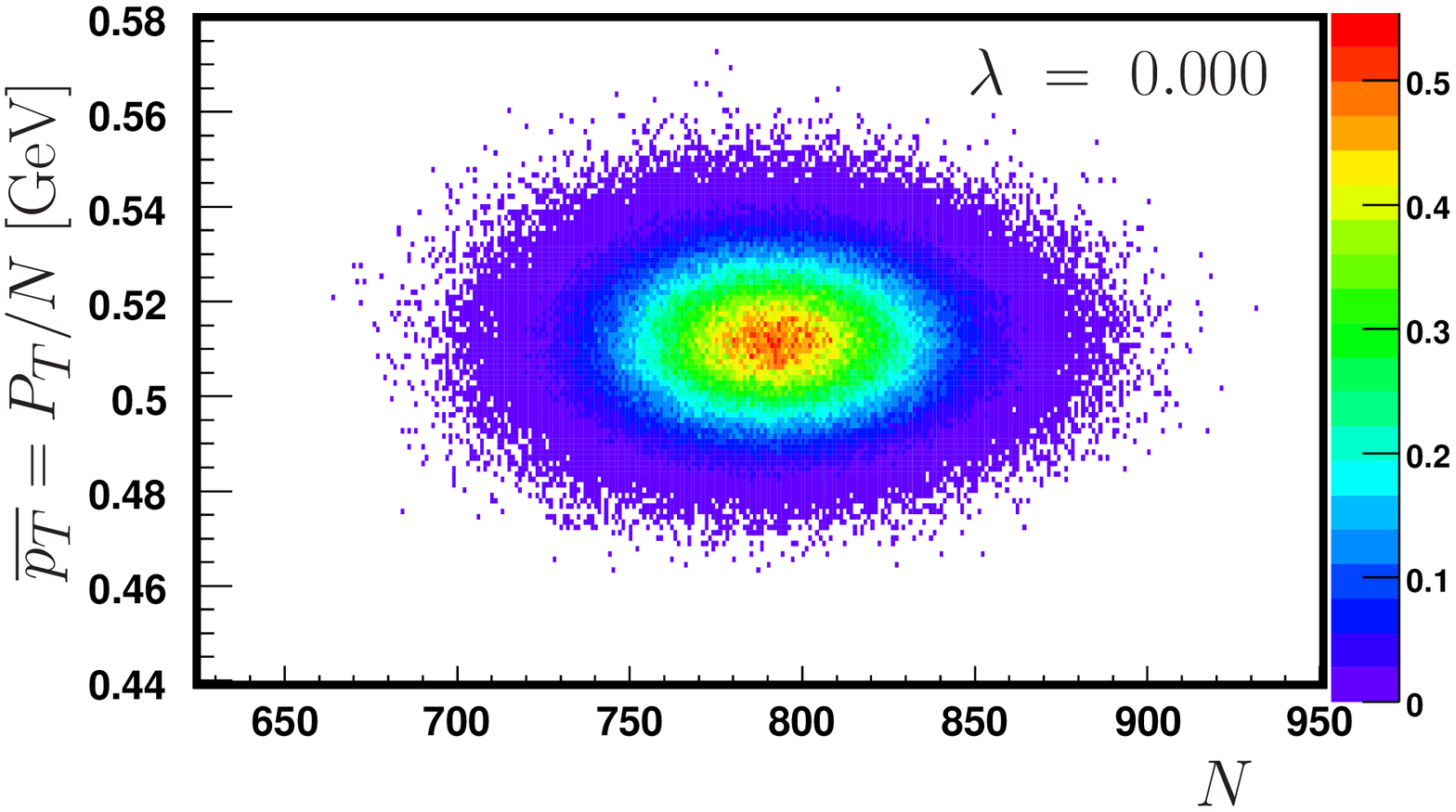,width=4.75cm,height=3.0cm}

  \epsfig{file=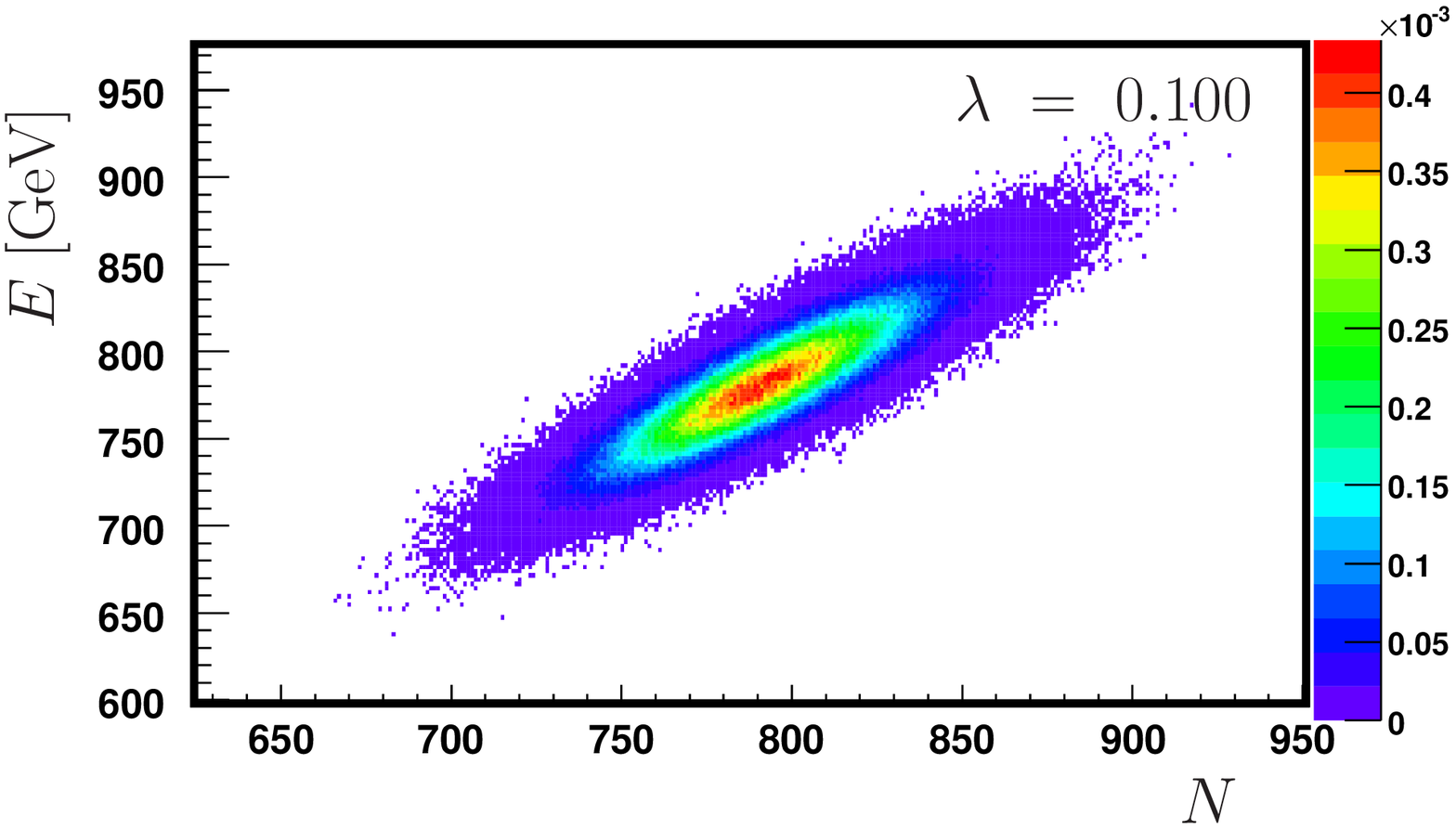,width=4.75cm,height=3.0cm}
  \epsfig{file=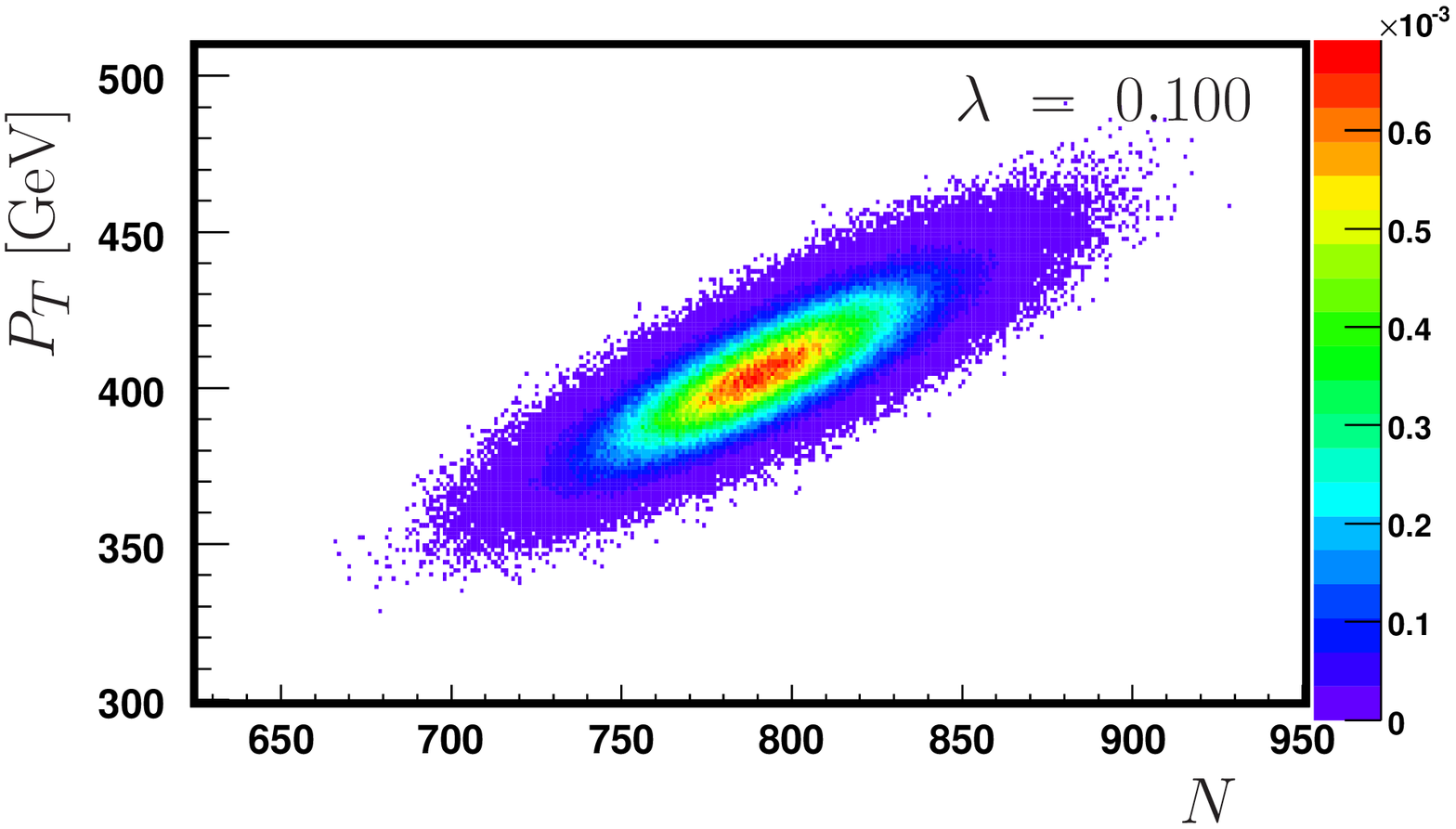,width=4.75cm,height=3.0cm}
  \epsfig{file=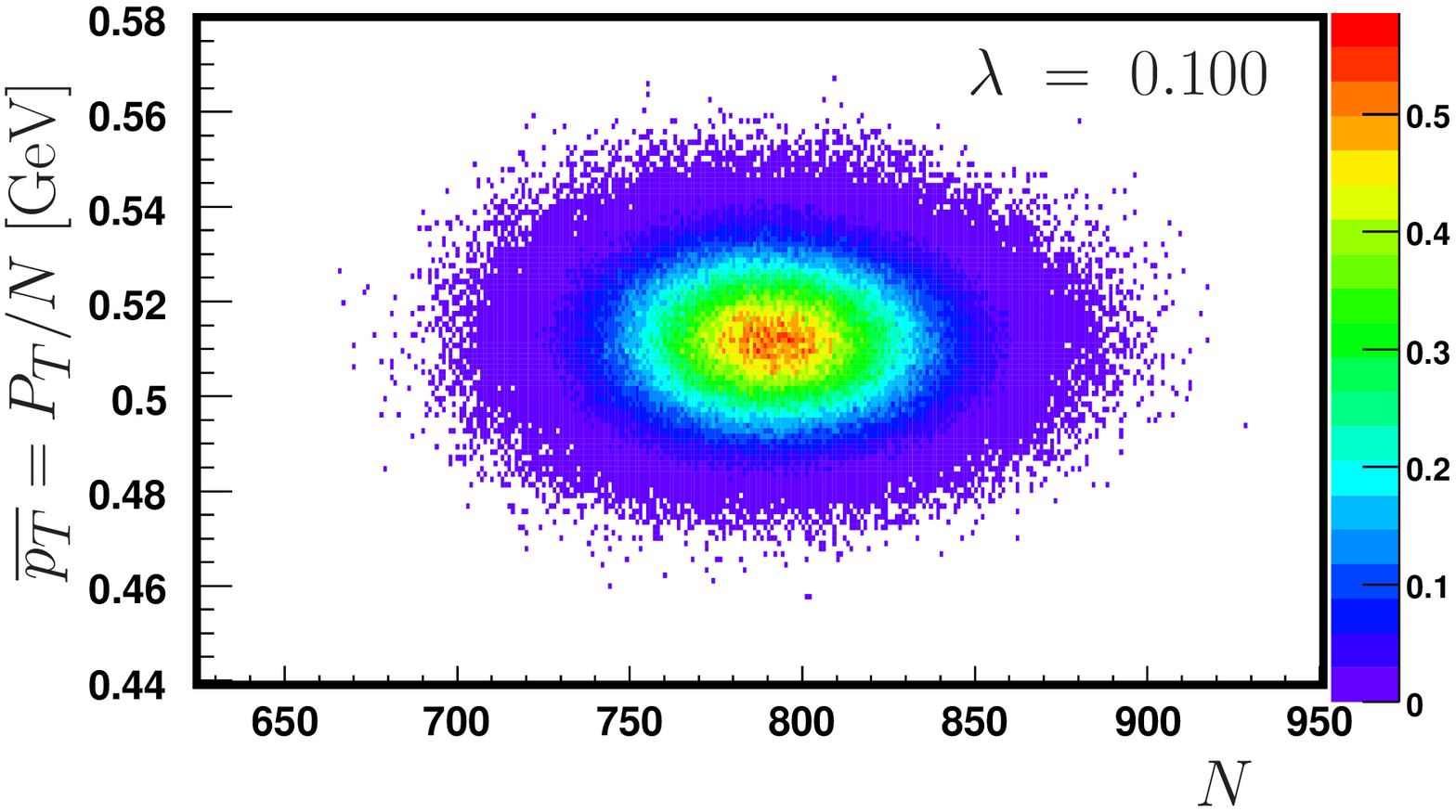,width=4.75cm,height=3.0cm}

  \epsfig{file=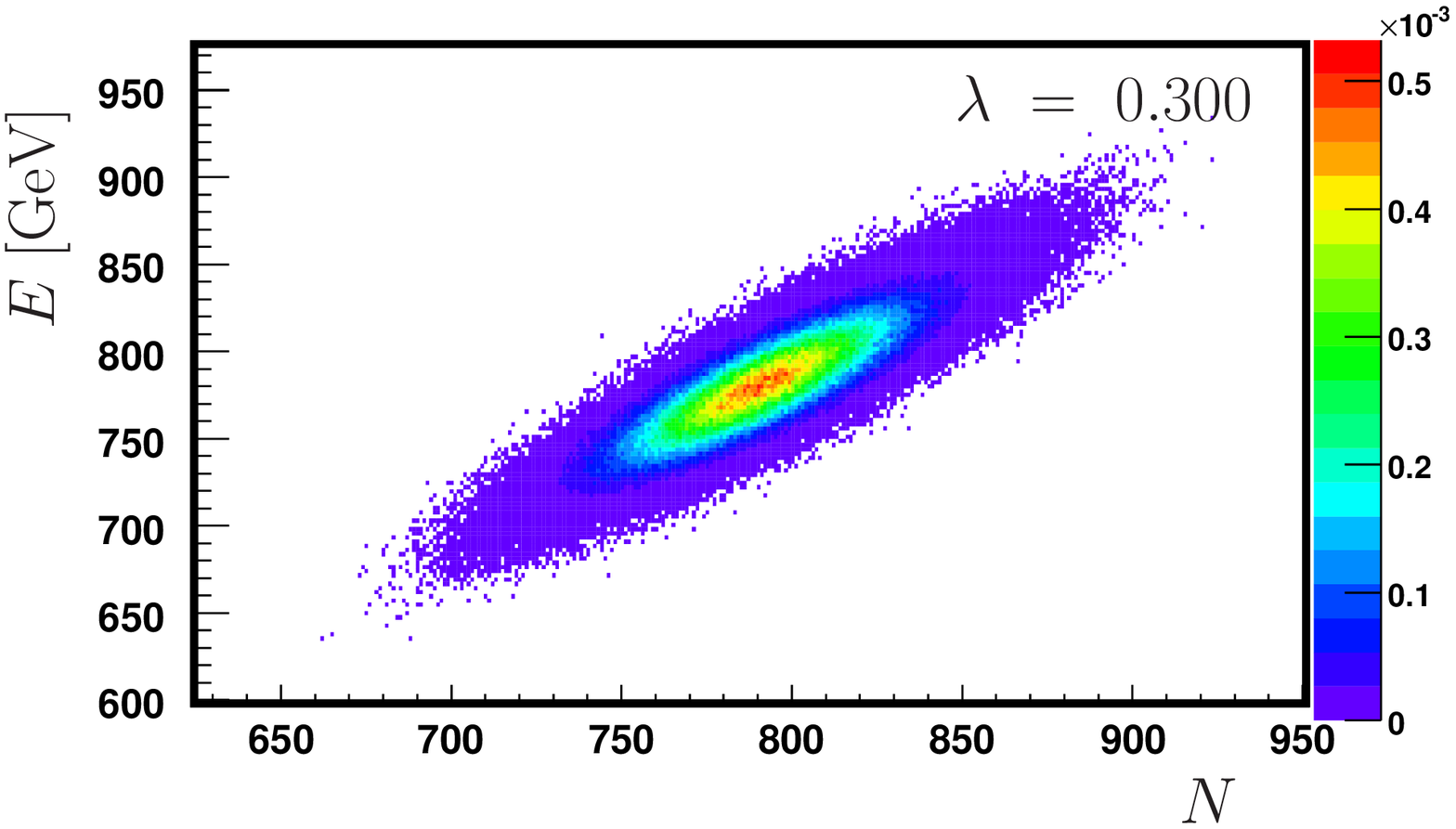,width=4.75cm,height=3.0cm}
  \epsfig{file=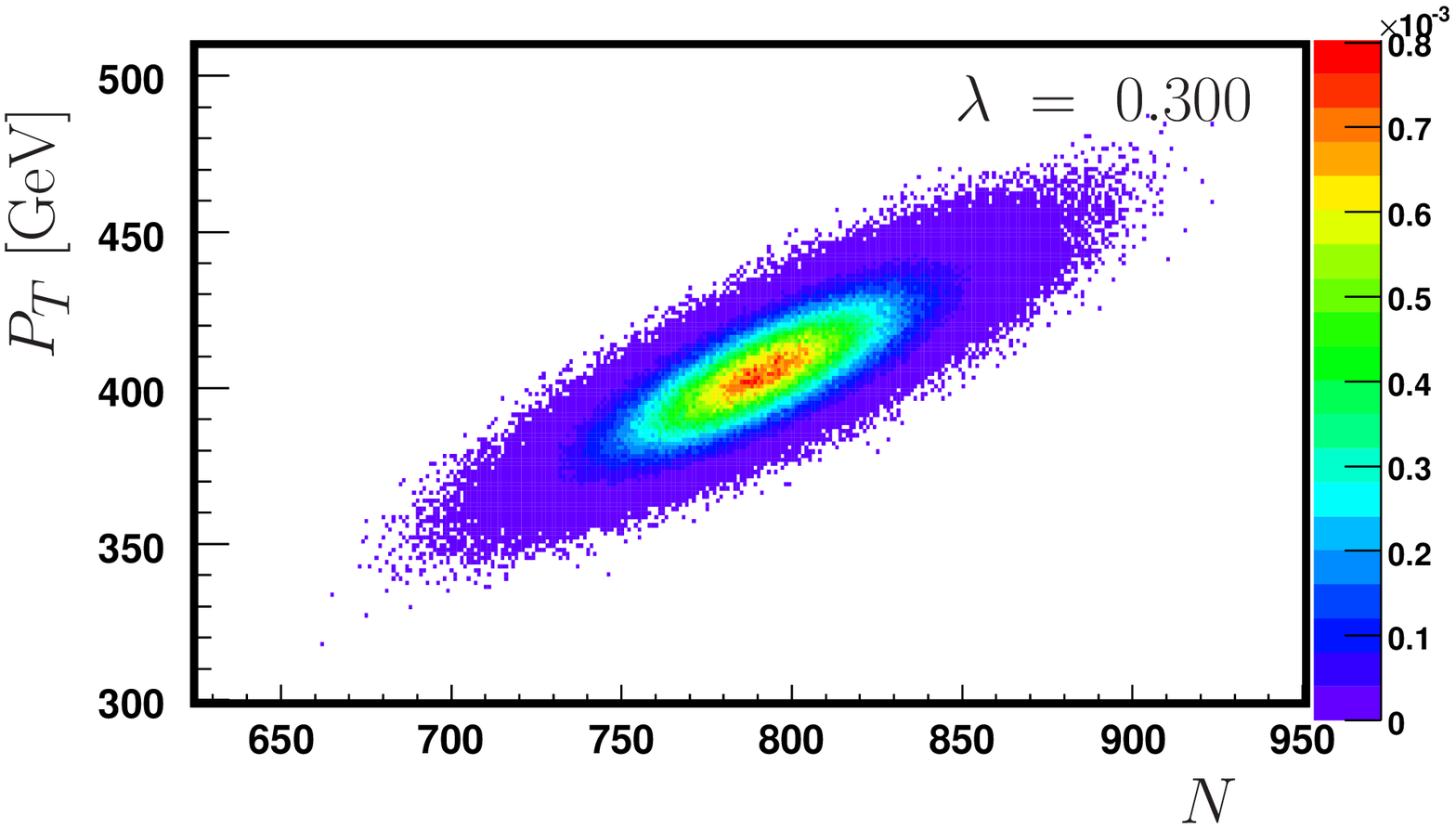,width=4.75cm,height=3.0cm}
  \epsfig{file=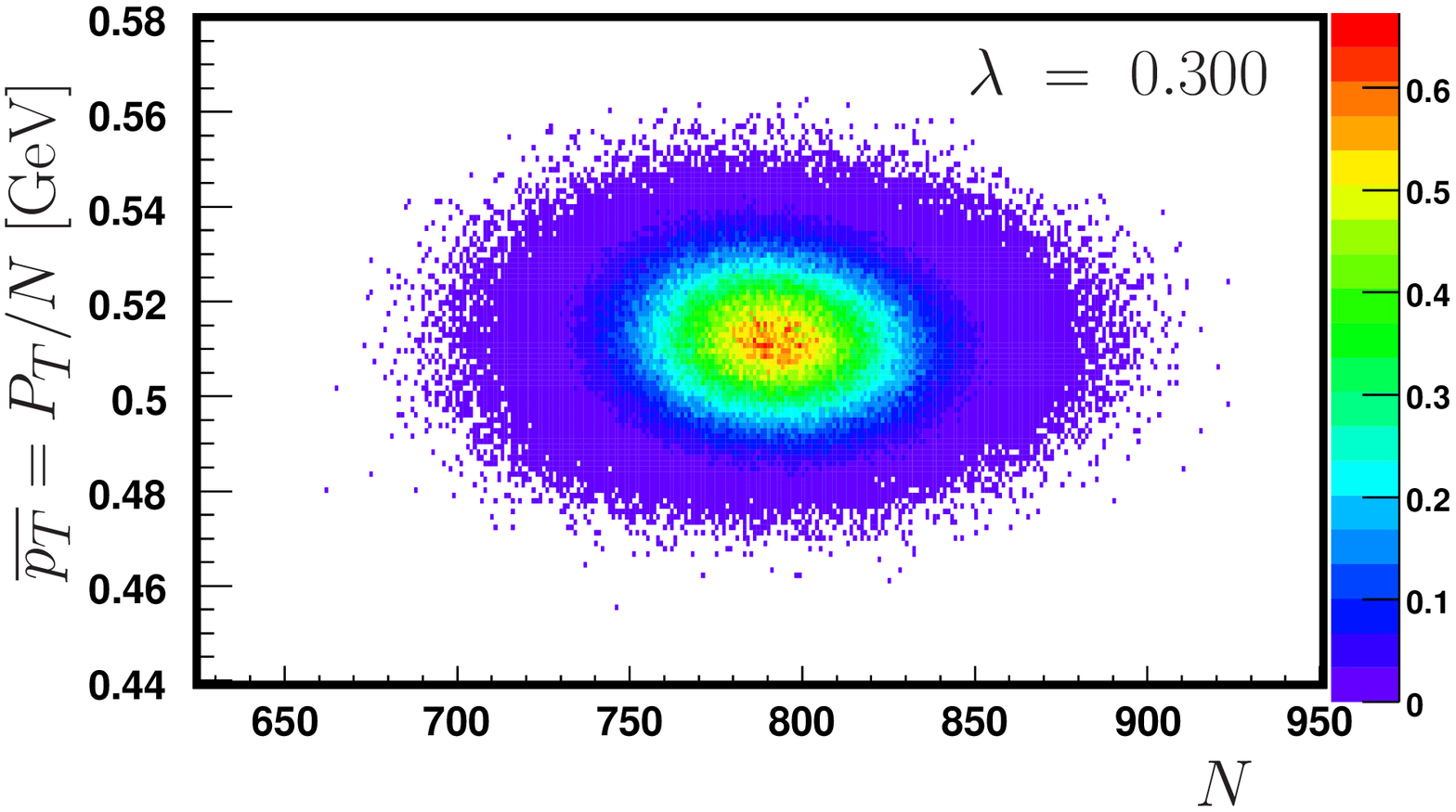,width=4.75cm,height=3.0cm}

  \epsfig{file=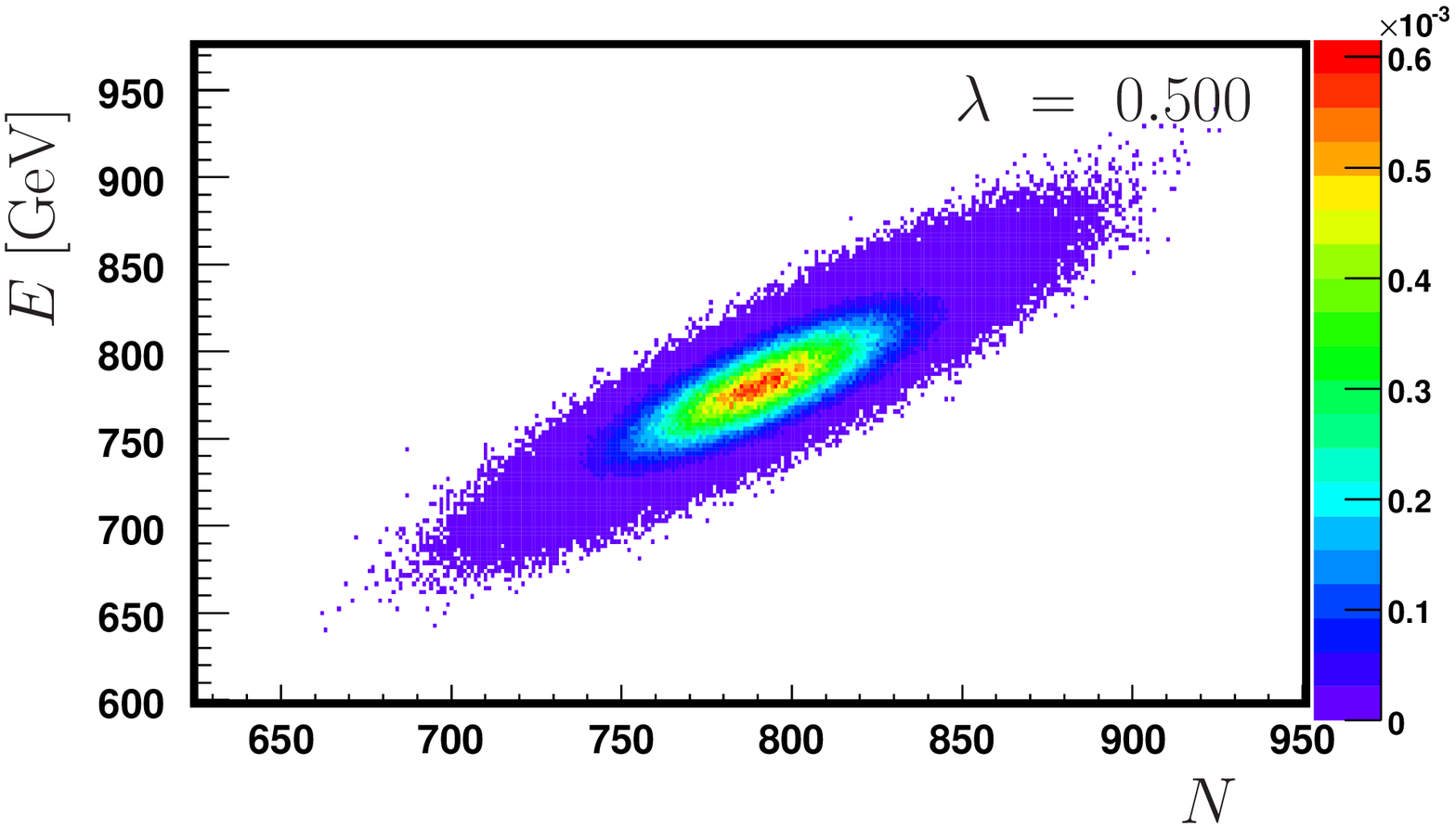,width=4.75cm,height=3.0cm}
  \epsfig{file=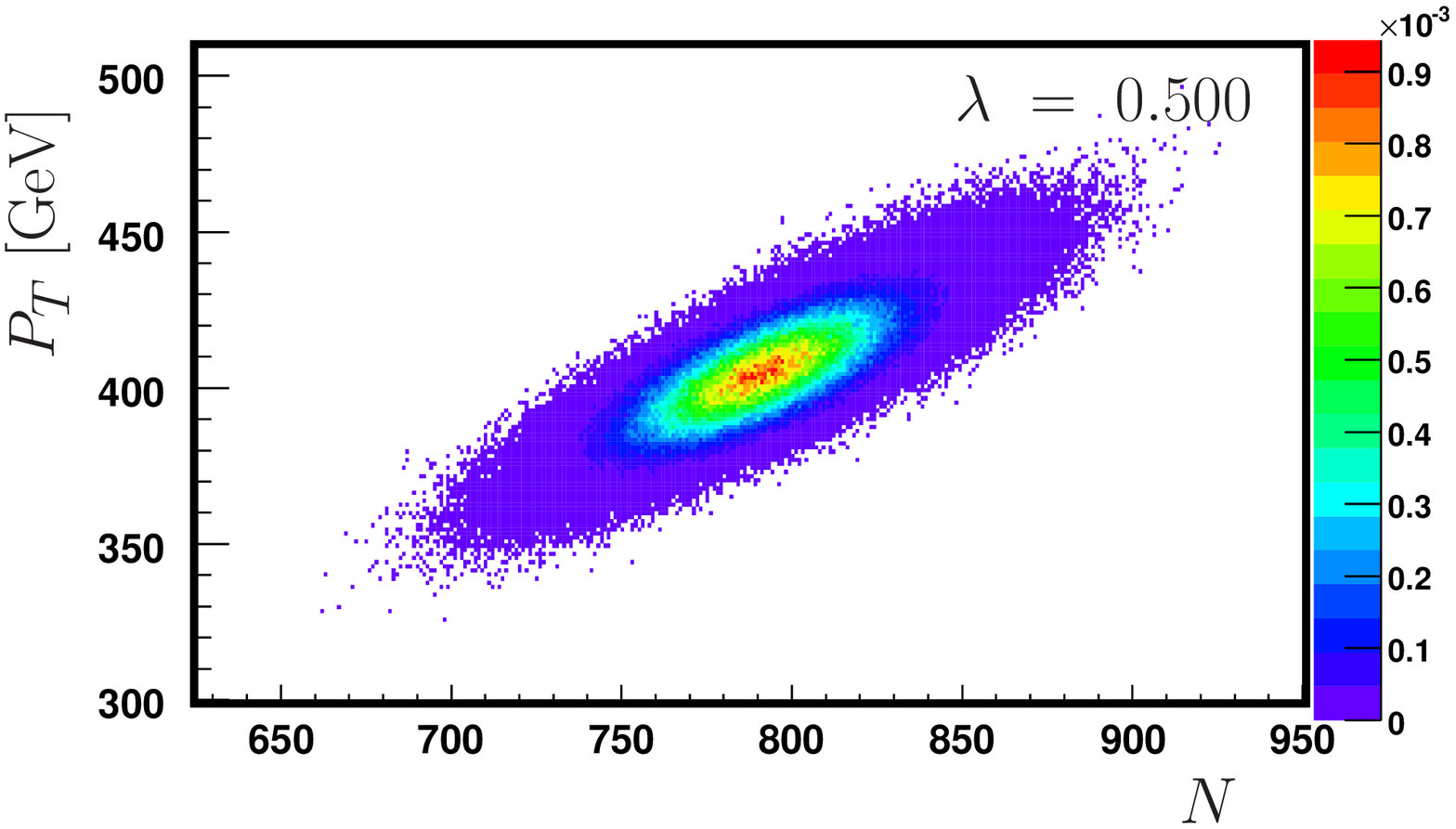,width=4.75cm,height=3.0cm}
  \epsfig{file=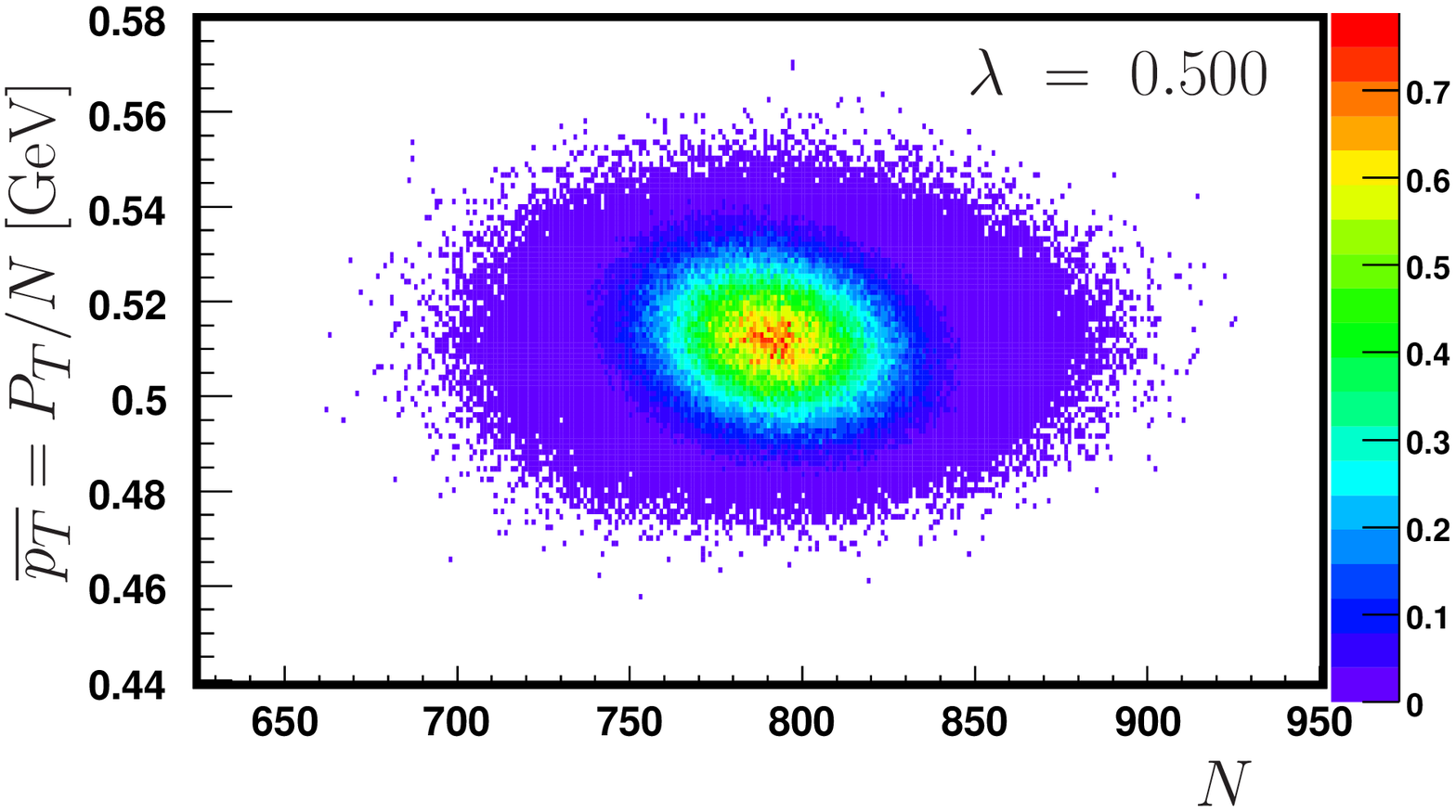,width=4.75cm,height=3.0cm}

  \epsfig{file=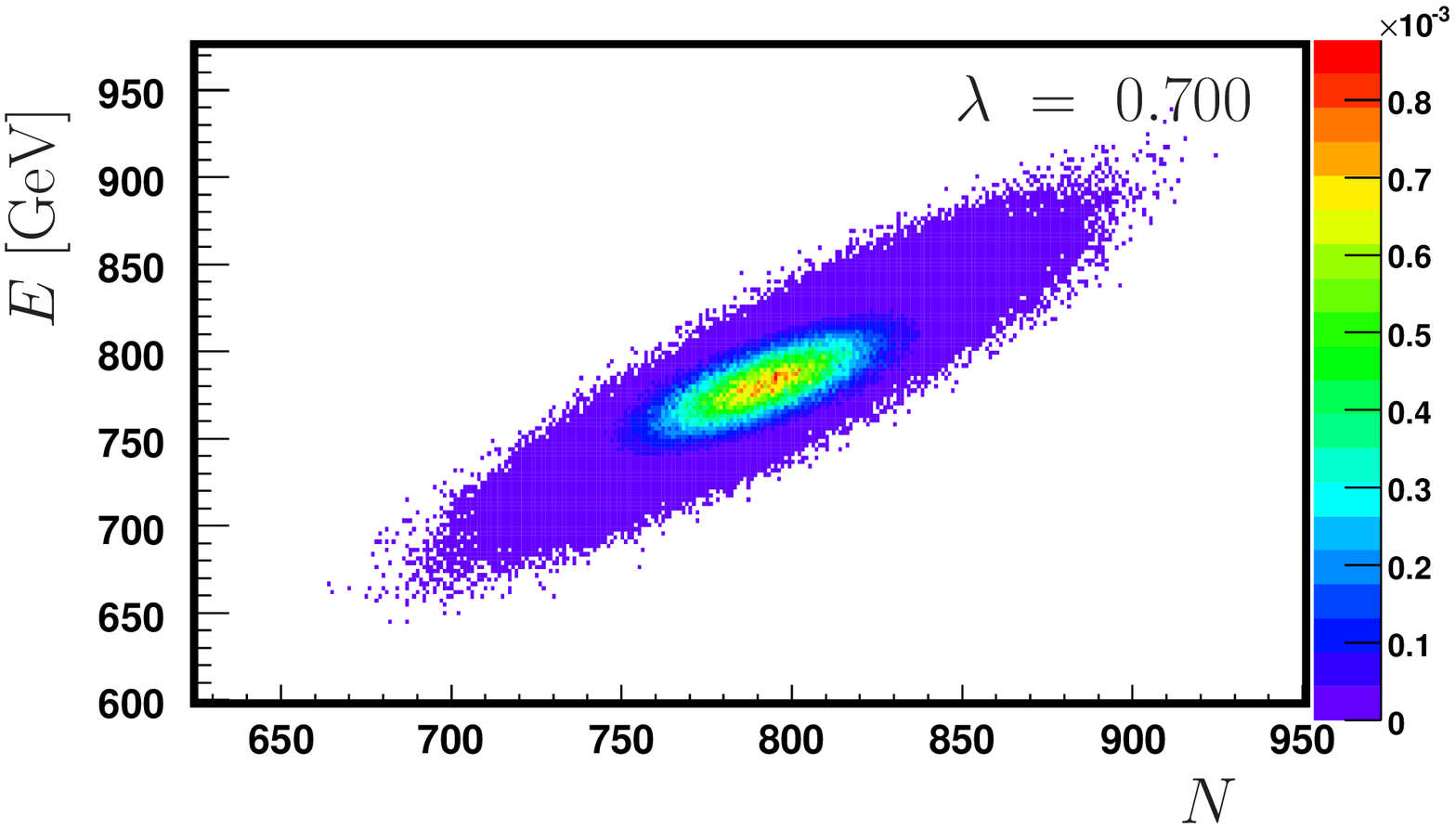,width=4.75cm,height=3.0cm}
  \epsfig{file=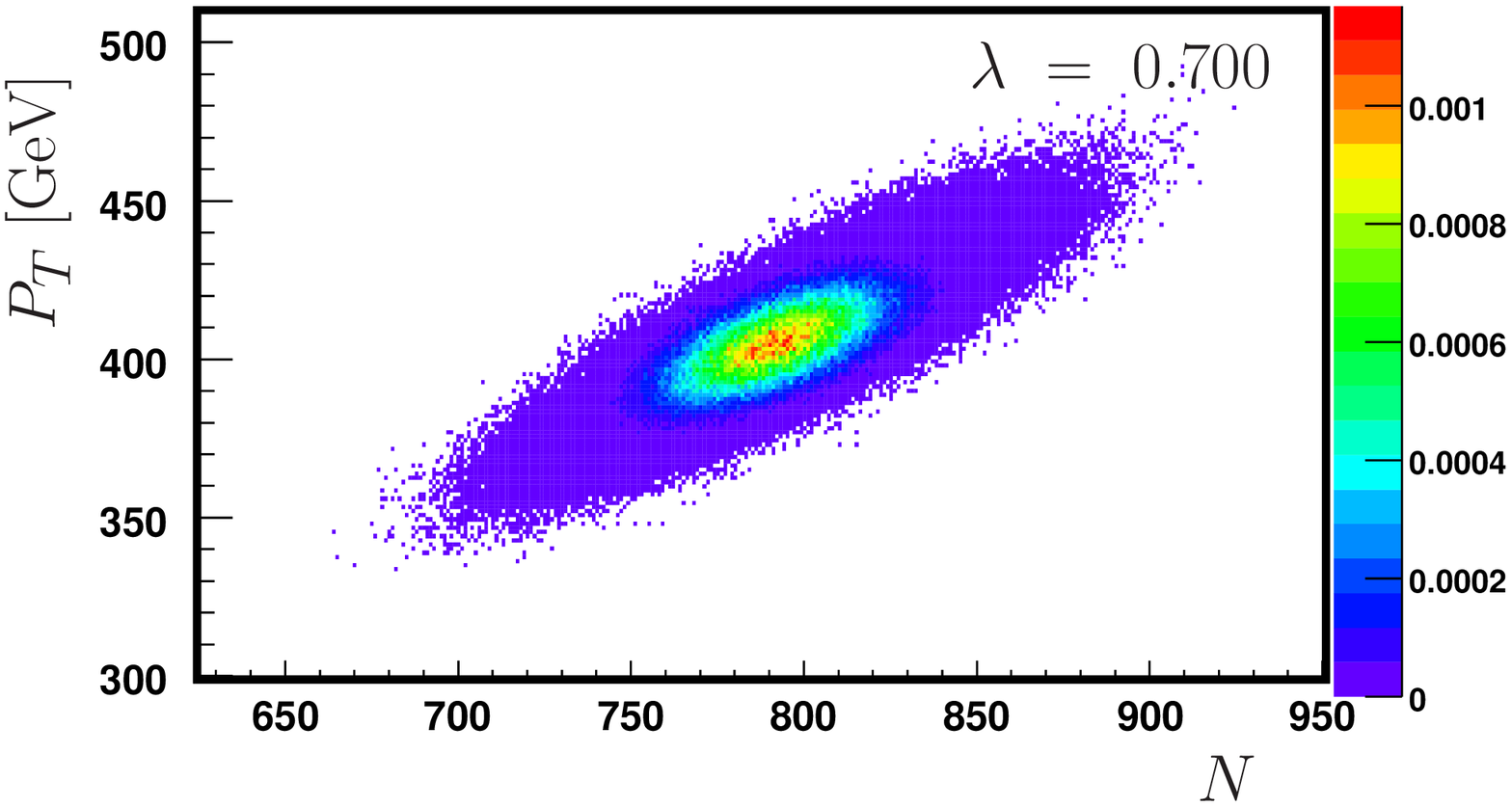,width=4.75cm,height=3.0cm}
  \epsfig{file=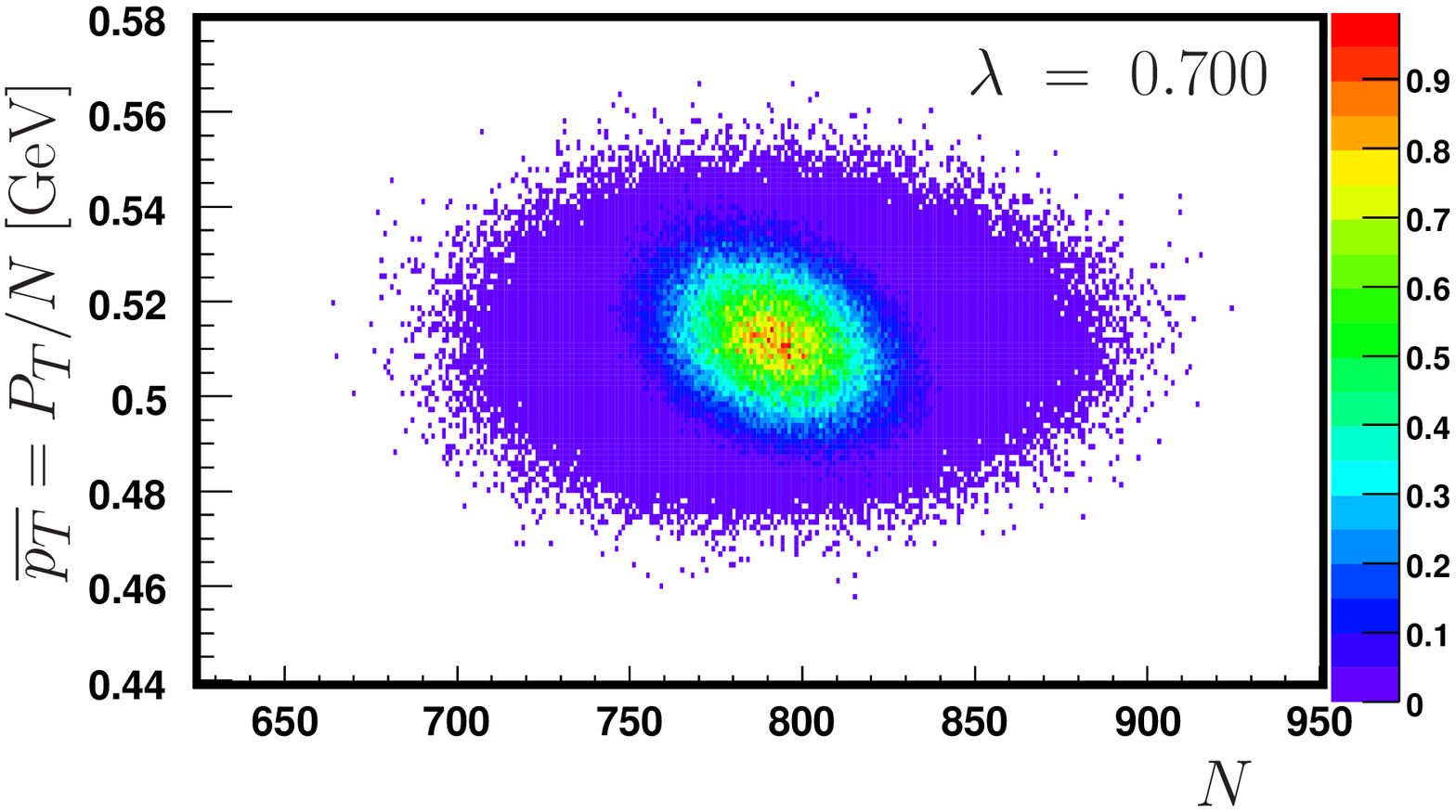,width=4.75cm,height=3.0cm}

  \epsfig{file=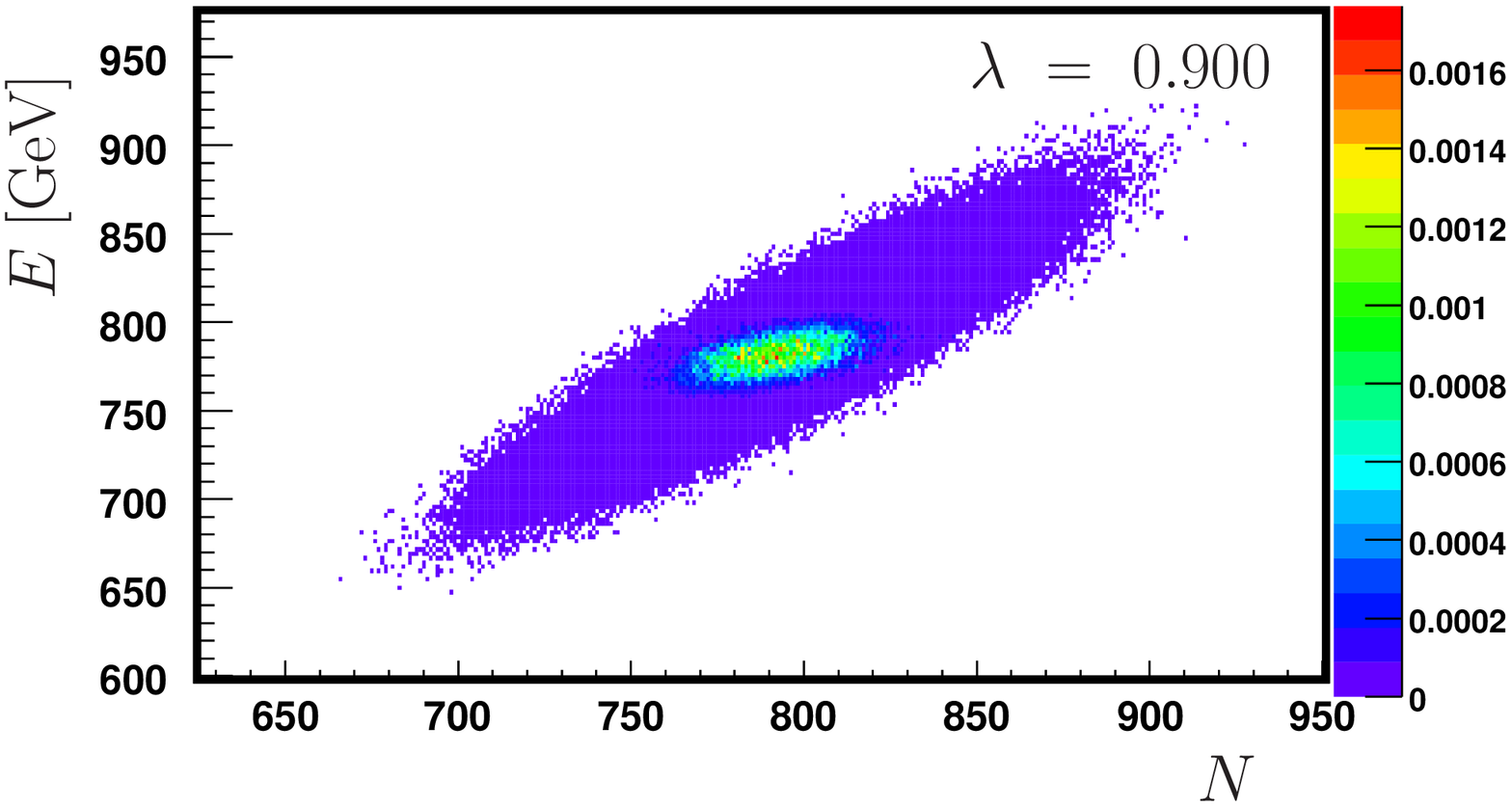,width=4.75cm,height=3.0cm}
  \epsfig{file=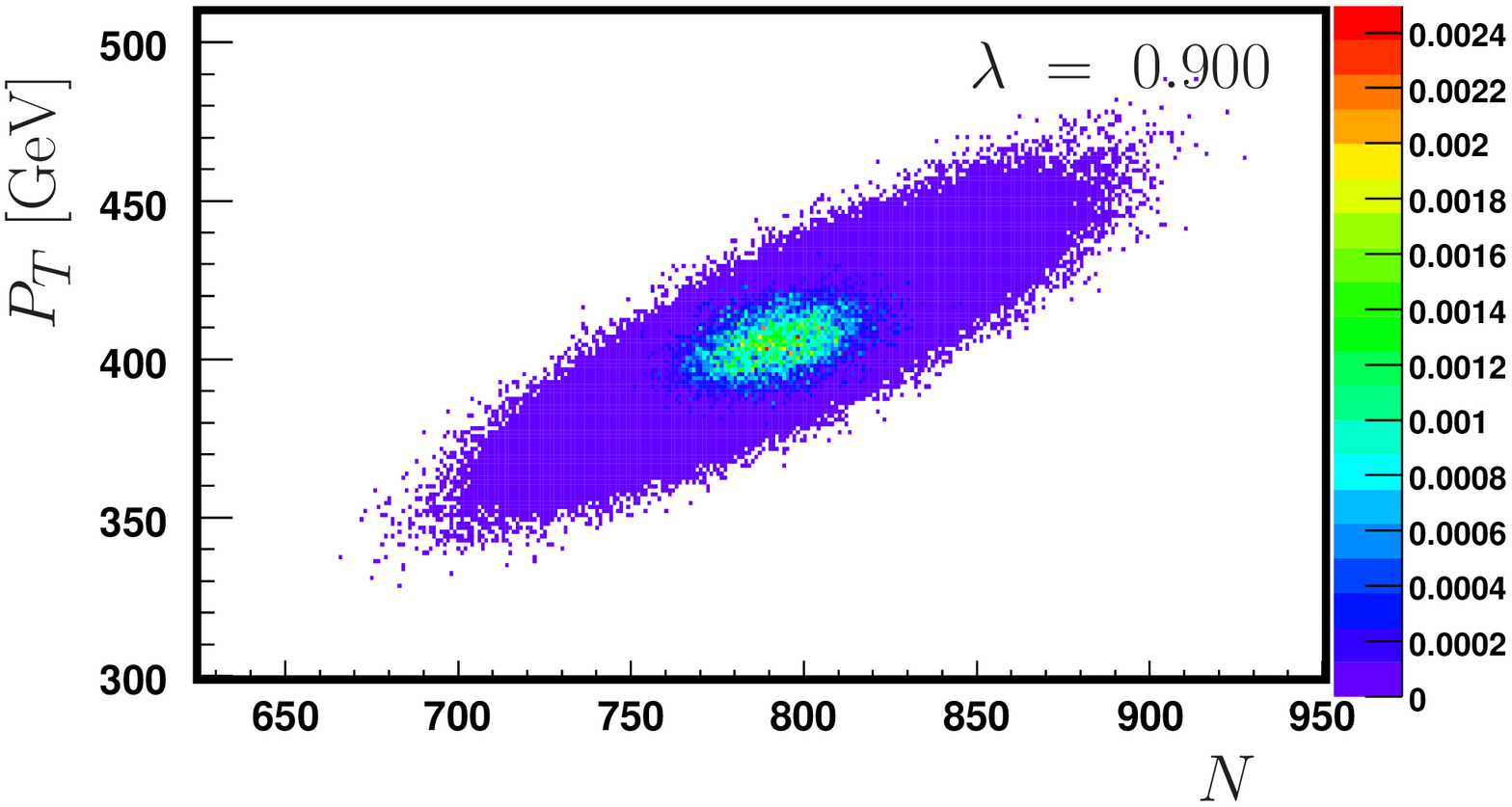,width=4.75cm,height=3.0cm}
  \epsfig{file=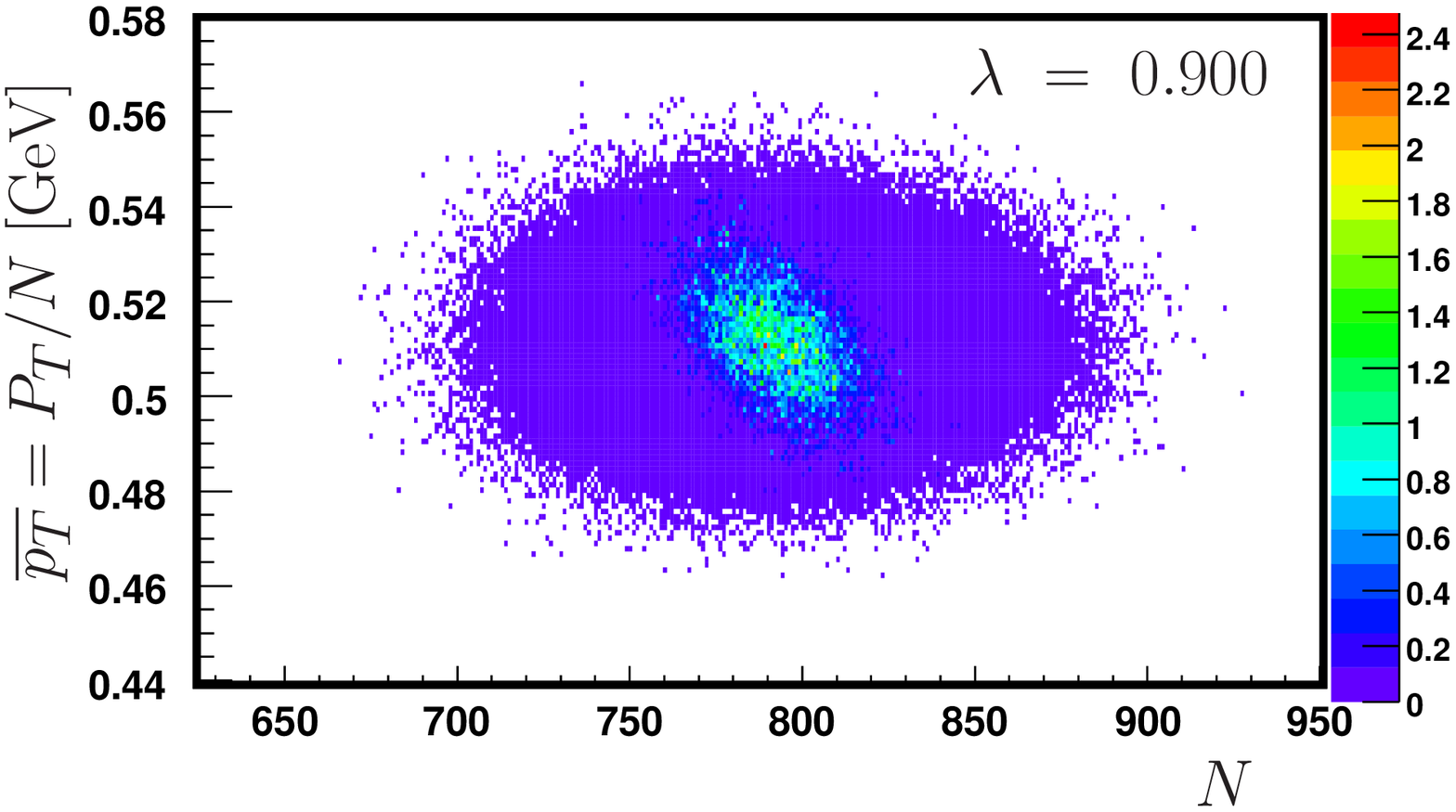,width=4.75cm,height=3.0cm}
 
  \caption{Evolution of the joint distributions of energy and particle number, $P_{\lambda}(E,N)$~({\it left}), particle 
  number and transverse momentum,~$P_{\lambda}(N,P_T)$~({\it center}), and particle number and mean transverse 
  momentum,~$P_{\lambda}(N,\overline{p_T})$~({\it right}), with the size of the bath~$\lambda = V_1/V_g$. 
  Here~$5 \cdot 10^5$ events have been sampled for each value of $\lambda$.}
  \label{P_lambda_kinetic}
\end{figure}

In Fig.(\ref{P_lambda_kinetic})~({\it left}) the joint energy and particle number distribution~$P_{\lambda}(E,N)$ is shown. 
Here only one of the two quantities was re-weighted. In the GCE one finds particle number to be strongly correlated with 
energy. The more particles are in a box of volume~$V_1$, the more energy is contained inside this box. This distribution 
does obviously not converge to a $\delta$-function, as the size of the bath is reduced. In direction of energy the 
distribution gets successively narrower, while retaining some width along the particle number direction. Along the way, 
the correlation coefficient for finite size of the bath changes. 
Therefore, the correlation coefficient~$\rho_{E N} \rightarrow 0$ as~$\lambda \rightarrow 1$.

Finally, distributions with neither quantity being re-weighted are considered. The distributions of particle number 
and total transverse momentum~$P_{\lambda}(N,P_T)$, Fig.(\ref{P_lambda_kinetic})~({\it center}), and of particle number 
versus mean transverse momentum $P_{\lambda}(N,\overline{p_T}=P_T/N)$, Fig.(\ref{P_lambda_kinetic})~({\it right}), 
will serve as examples. Although the total transverse momentum~$P_T$ is strongly correlated with particle number~$N$ in 
the GCE, the mean transverse momentum~$\overline{p_T} = P_T/N$ is not. The more particles are inside a box, the more 
kinetic energy is contained within its boundaries. Due to the infinite heat bath assumption, individual particle momenta 
are however un-correlated with each other. Measuring a particle with a certain momentum vector, does not constrain the
remaining system in any way or shape. As the size of the bath is now reduced, this does not hold true anymore. 
The total transverse momentum is still positively correlated with particle number, yet weaker as~$\lambda~\rightarrow~1$. 
The mean transverse momentum distribution makes this rather plain. At fixed energy, the larger the particle multiplicity, 
the more the energy goes into mass, and the less energy into thermal motion, or vice versa. 
The correlation coefficient is hence negative~$\rho_{N \overline{p_T}} <0$, and the more particles are measured in an 
event, the less kinetic energy the individual particle will carry on average. Please note, that 
although the distribution~$P_{\lambda}(N,\overline{p_T})$ changes, neither of the mean values does; 
$\langle N \rangle$ and~$\langle \overline{p_T} \rangle$ stay constant. 
The expectation value of mean transverse momentum at fixed particle number~$\langle \overline{p_T} \rangle_N$, 
obtained from the conditional distribution~$P(\overline{p_T}|N)$, however, decreases as~$N$ increases.

\section{Monte Carlo Weight Factor}
\label{sec_extrapol_weight}

\begin{figure}[ht!]
  \epsfig{file=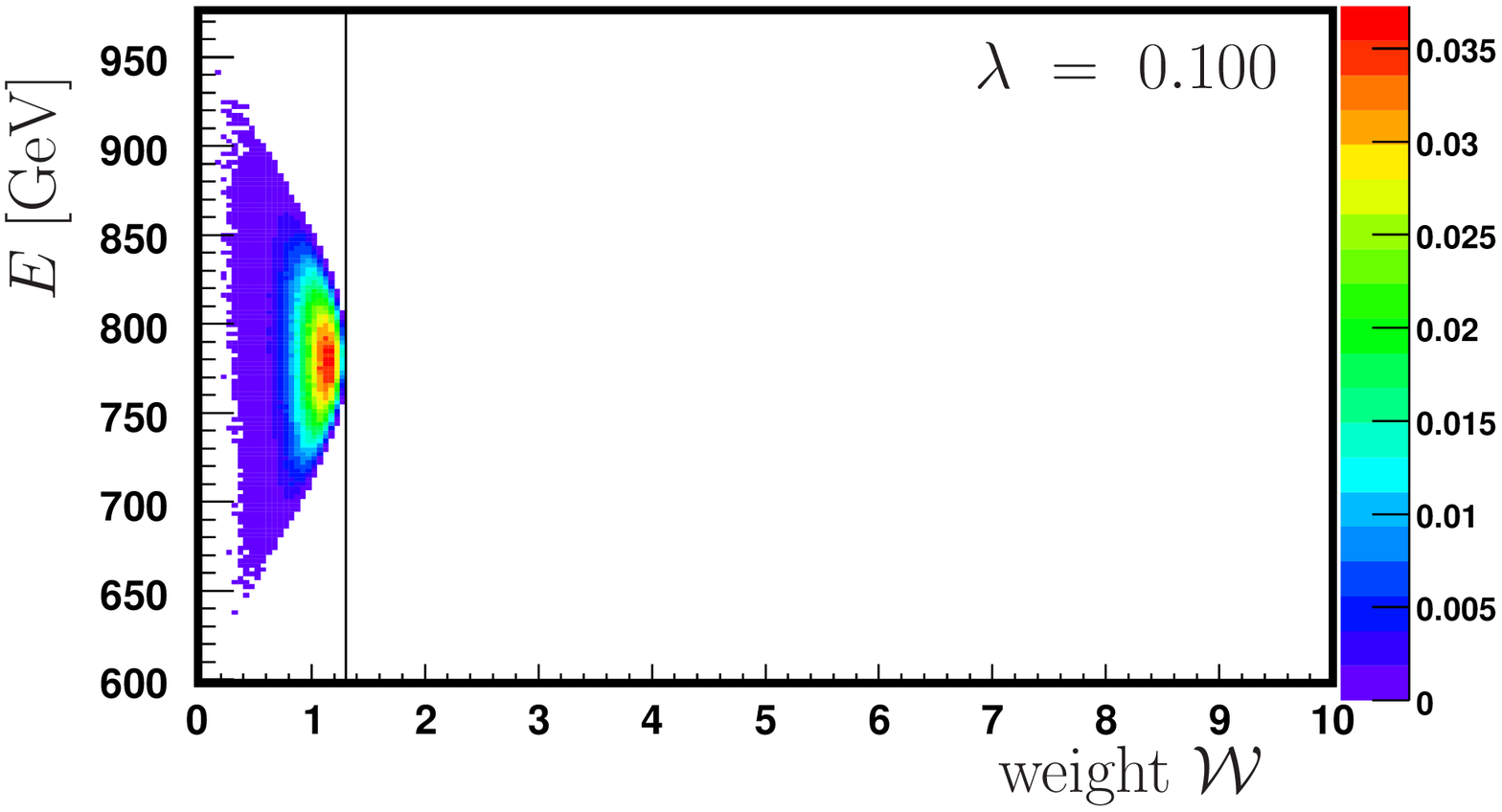,width=4.75cm,height=2.85cm}
  \epsfig{file=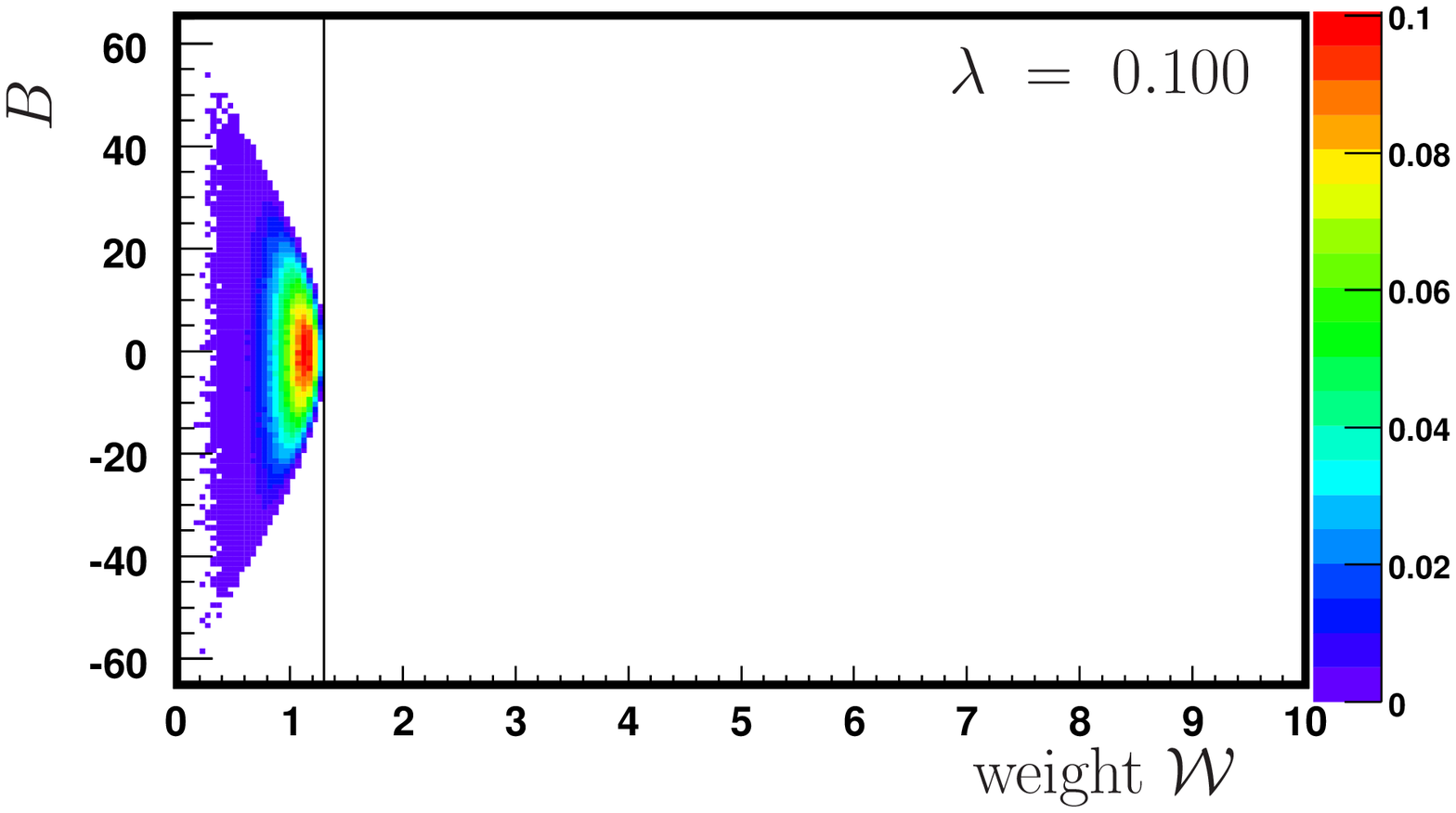,width=4.75cm,height=2.85cm}
  \epsfig{file=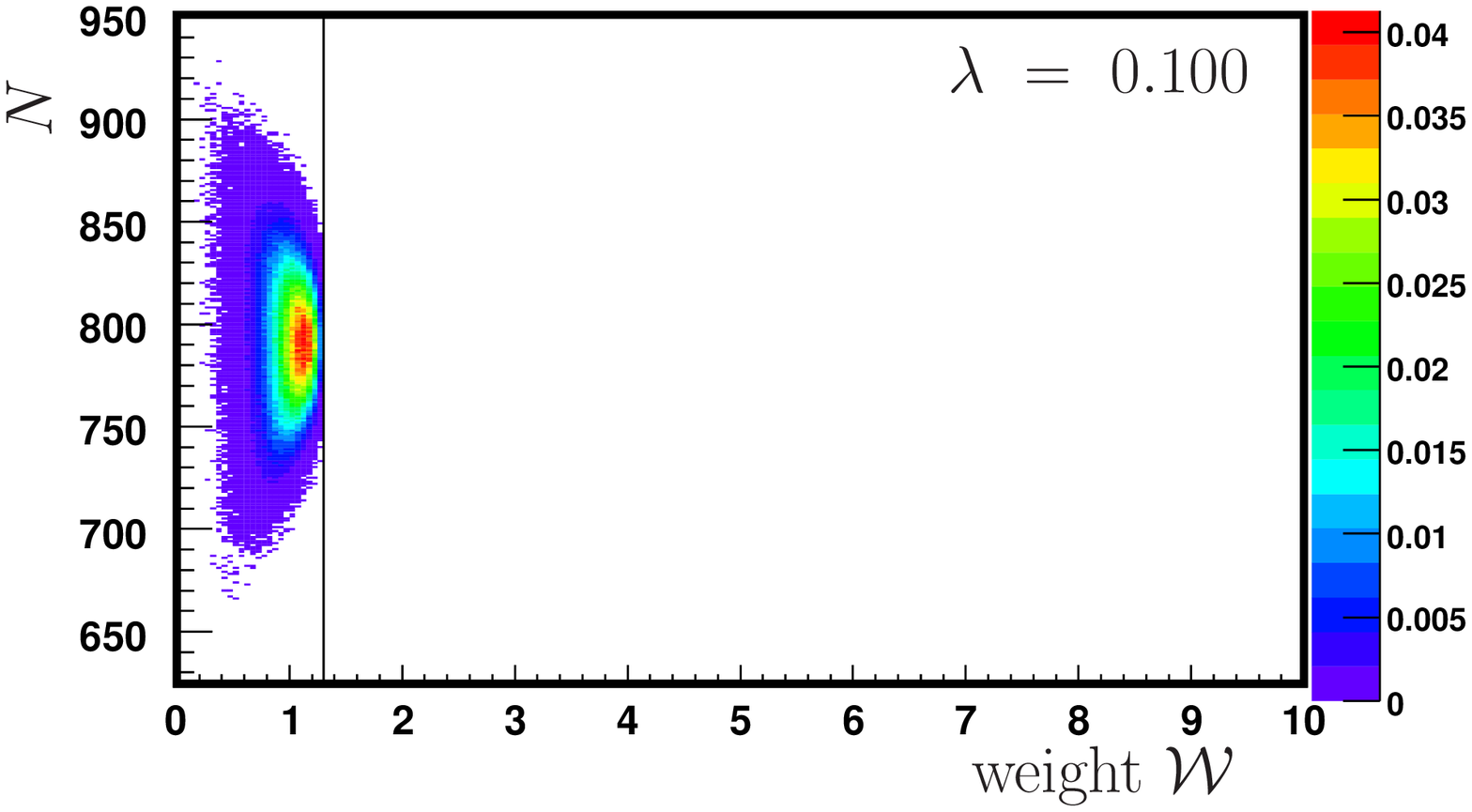,width=4.75cm,height=2.85cm}
  
  \epsfig{file=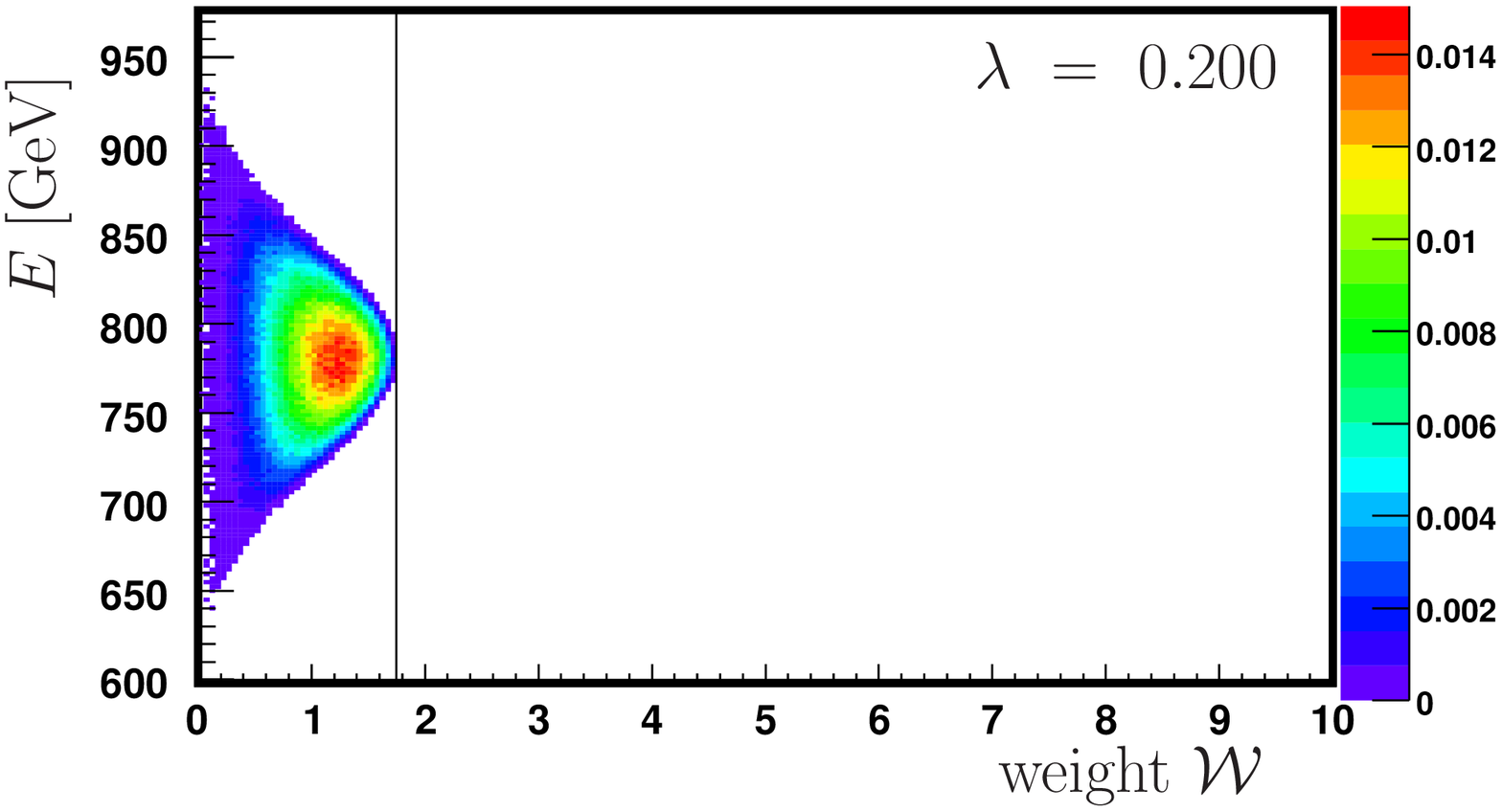,width=4.75cm,height=2.85cm}
  \epsfig{file=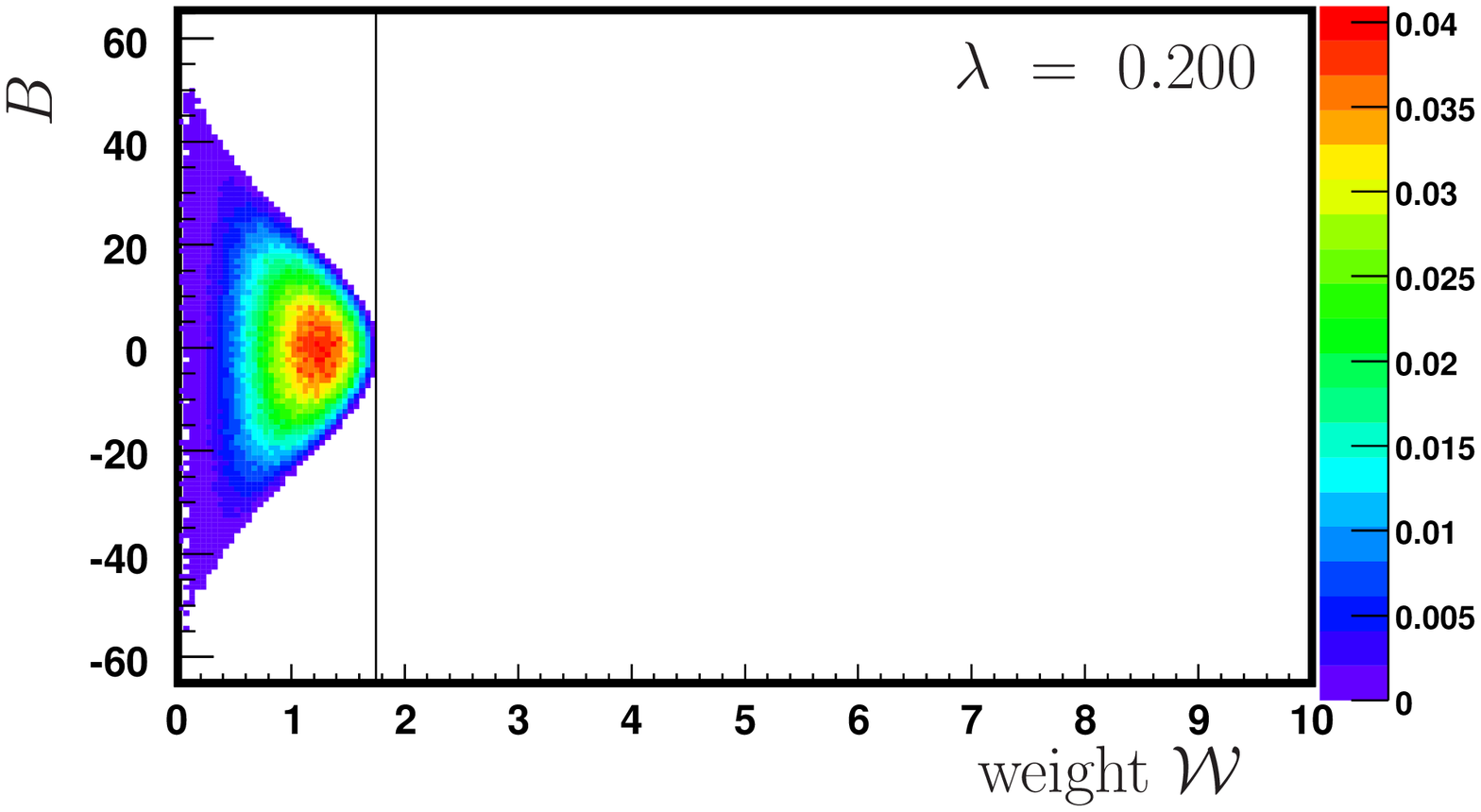,width=4.75cm,height=2.85cm}
  \epsfig{file=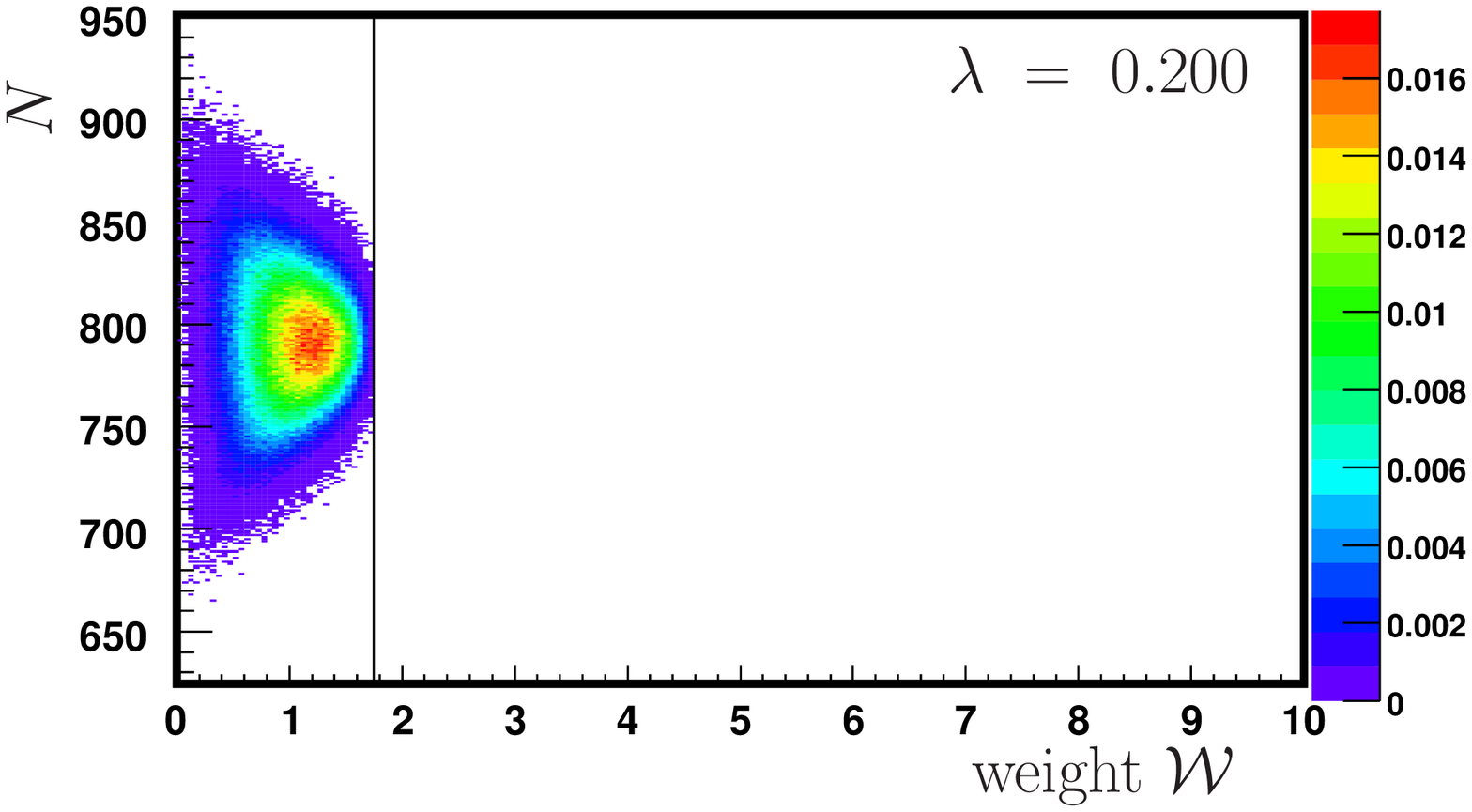,width=4.75cm,height=2.85cm}

  \epsfig{file=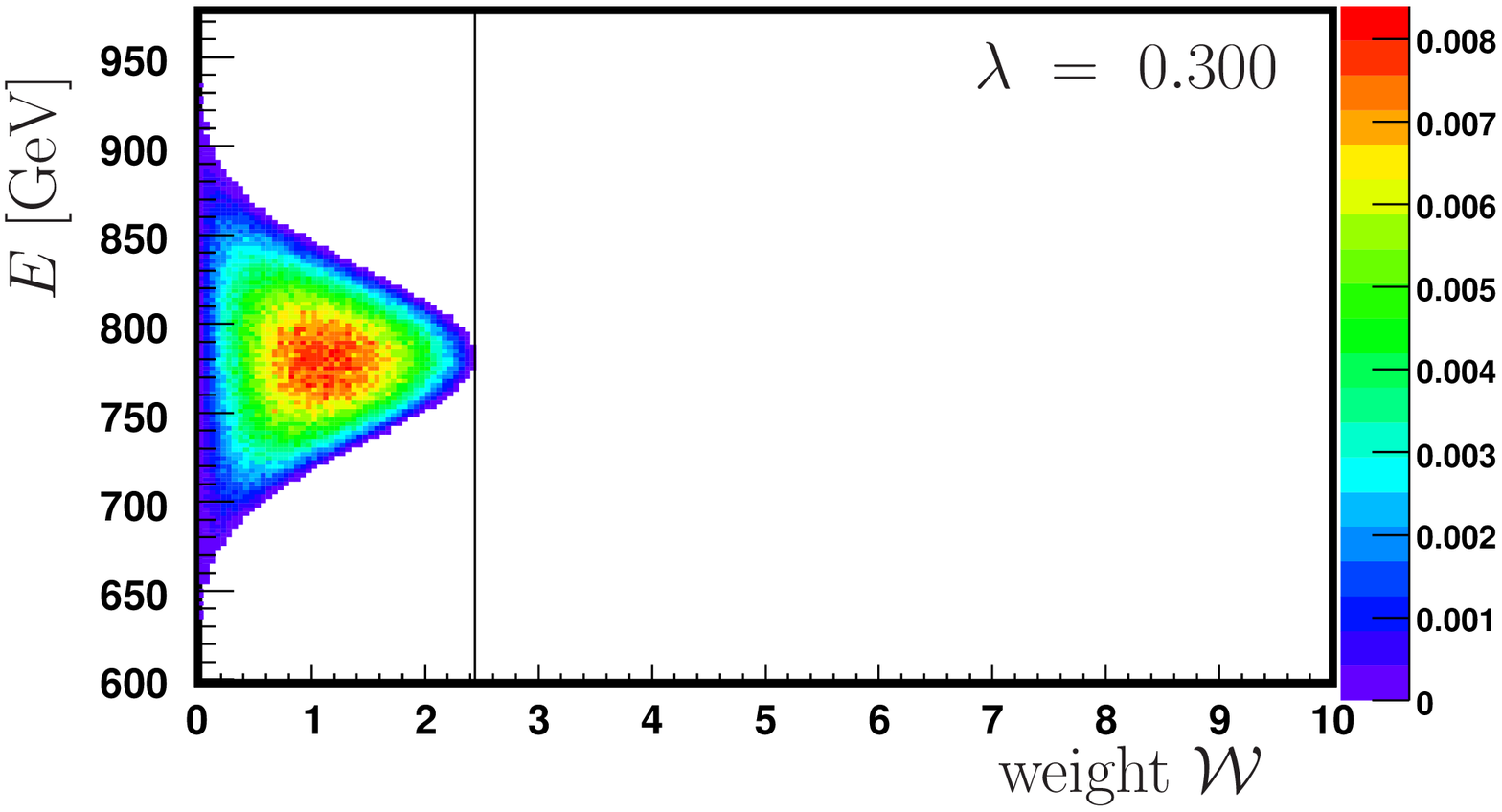,width=4.75cm,height=2.85cm}
  \epsfig{file=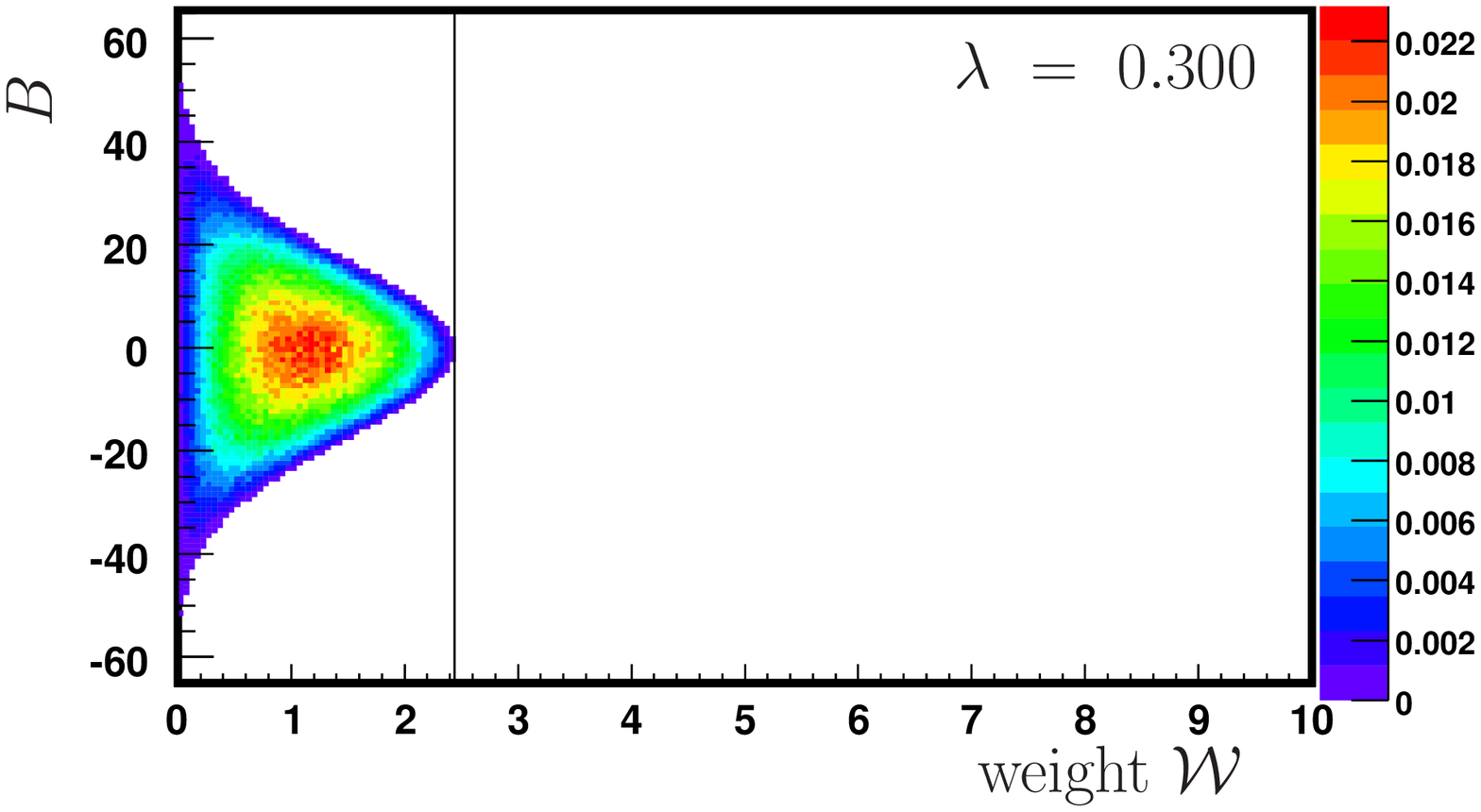,width=4.75cm,height=2.85cm}
  \epsfig{file=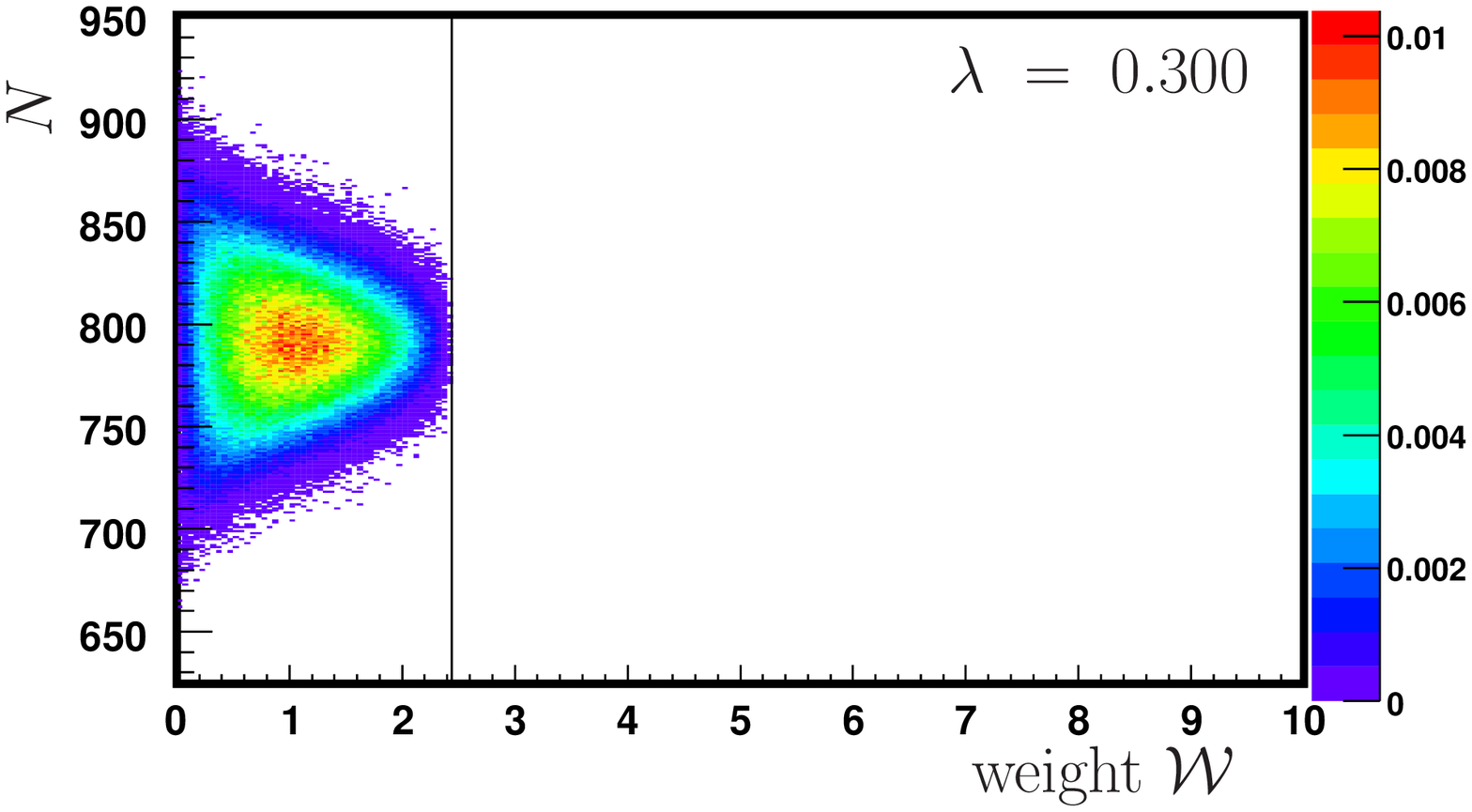,width=4.75cm,height=2.85cm}

  \epsfig{file=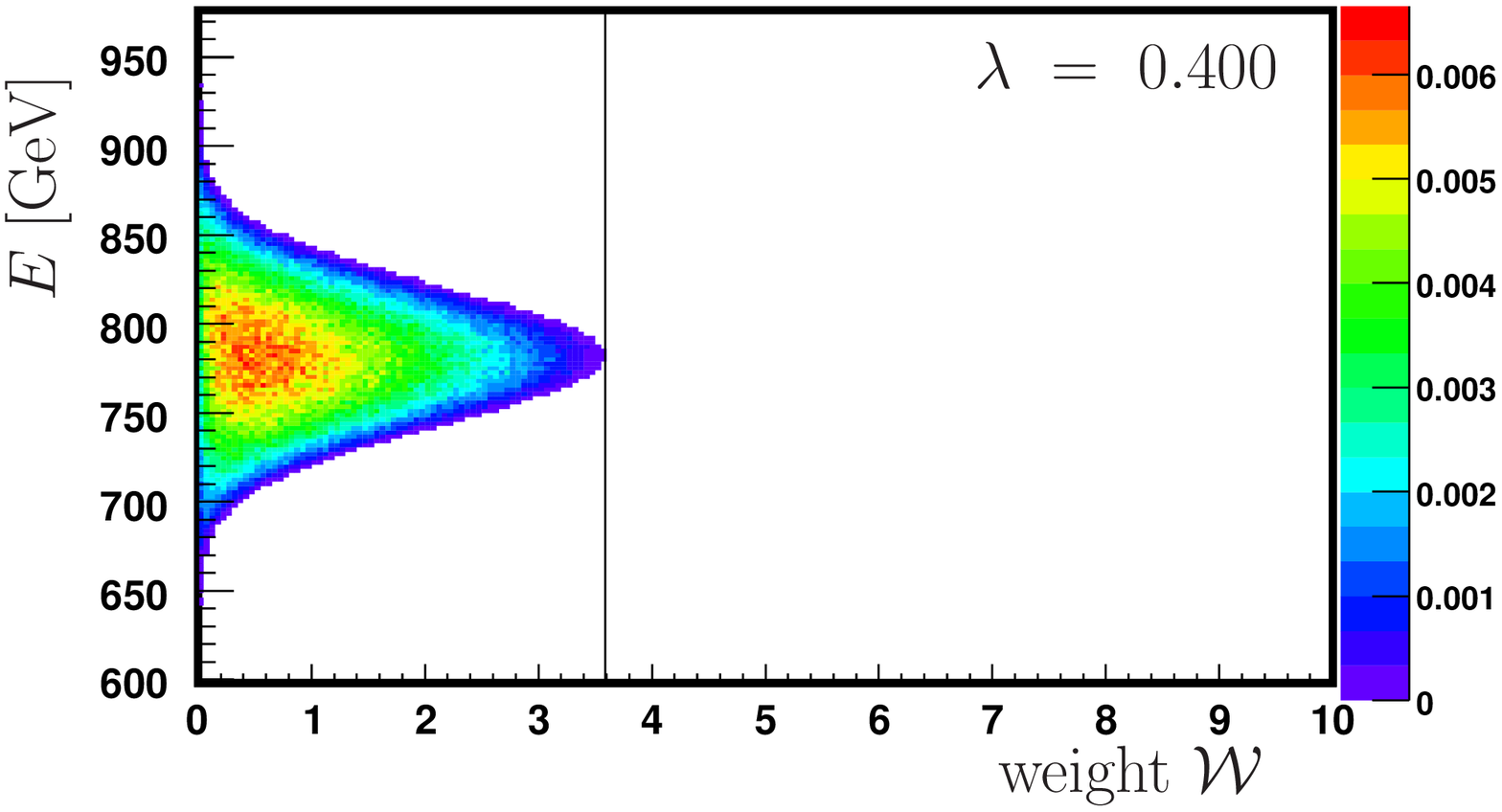,width=4.75cm,height=2.85cm}
  \epsfig{file=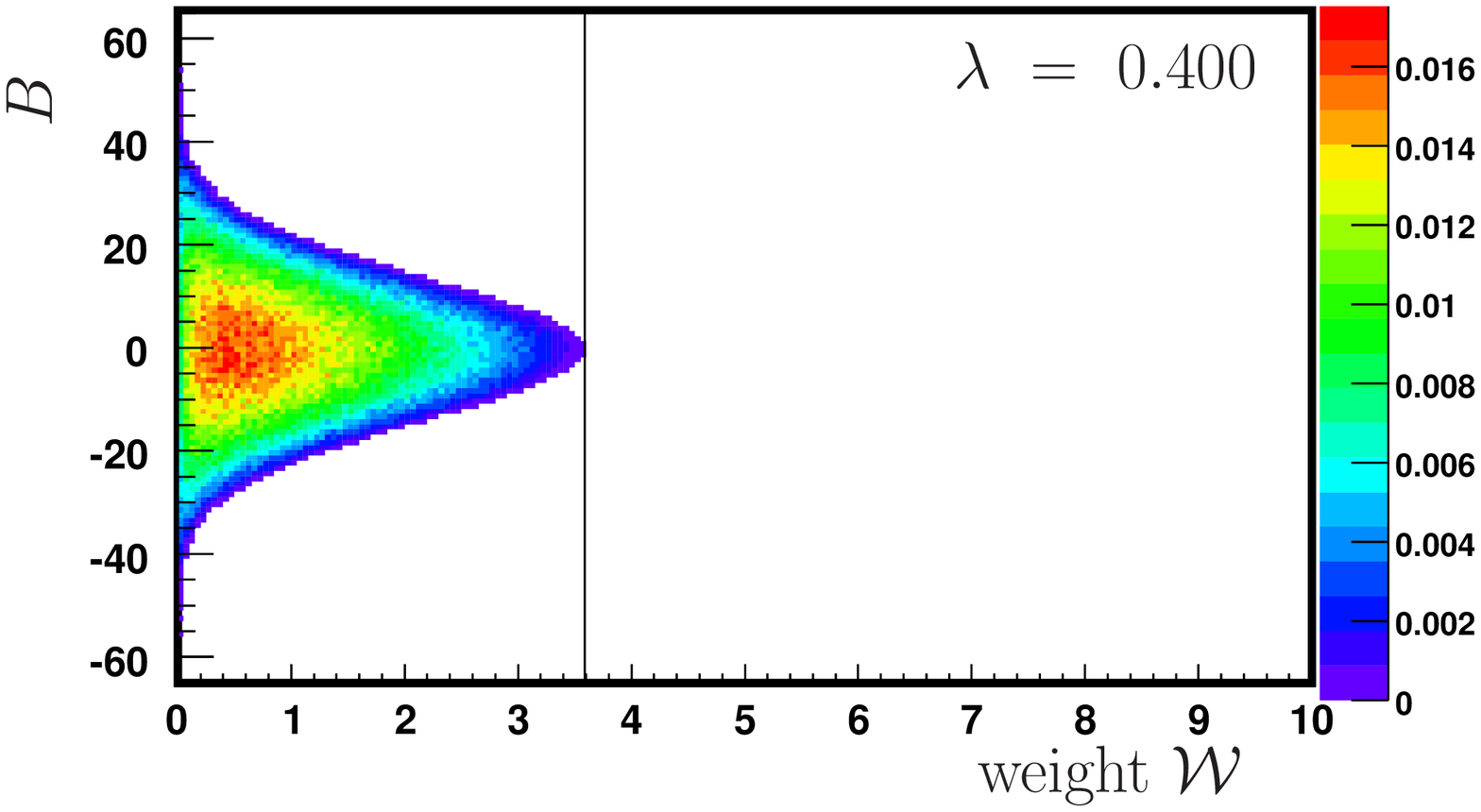,width=4.75cm,height=2.85cm}
  \epsfig{file=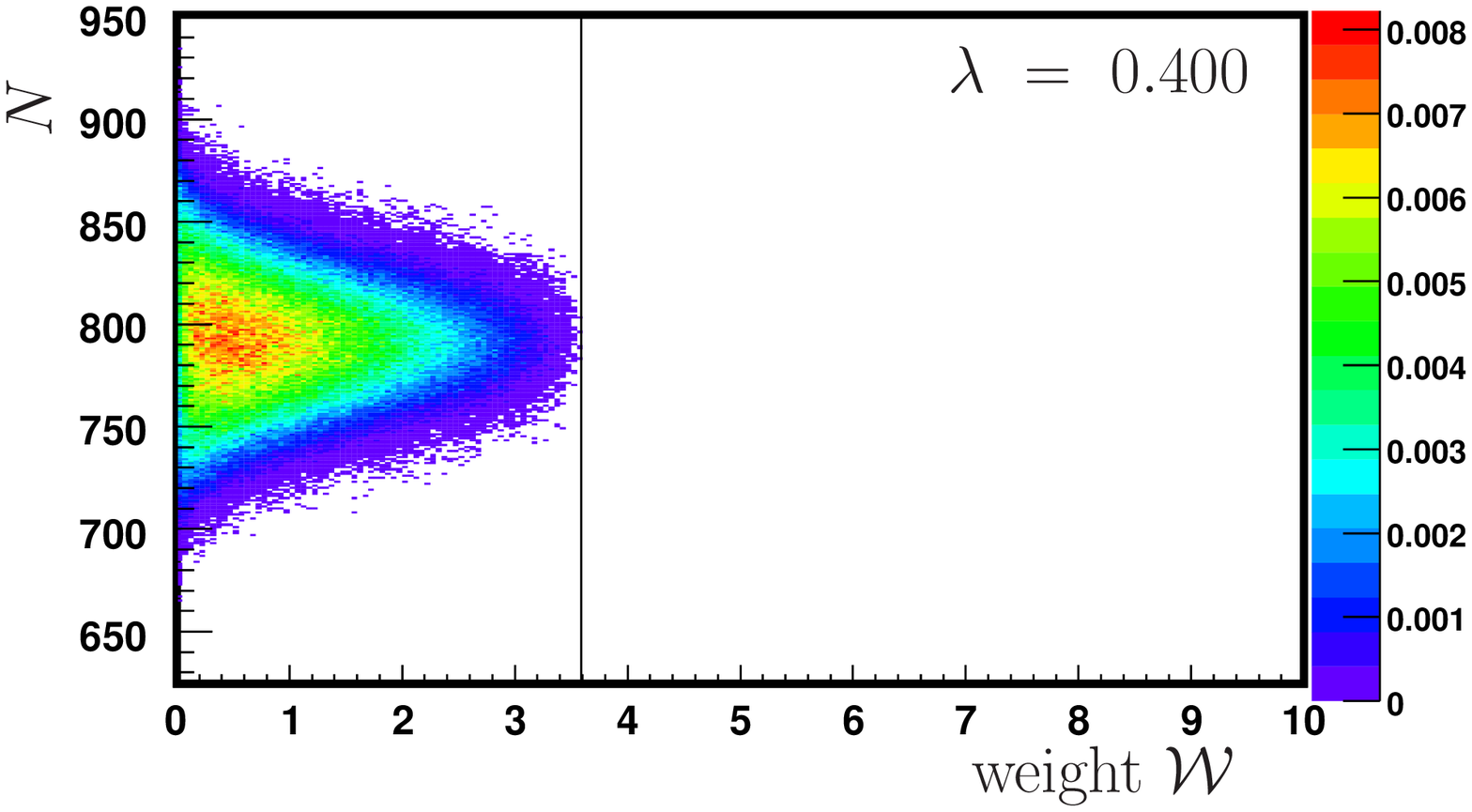,width=4.75cm,height=2.85cm}
  
  \epsfig{file=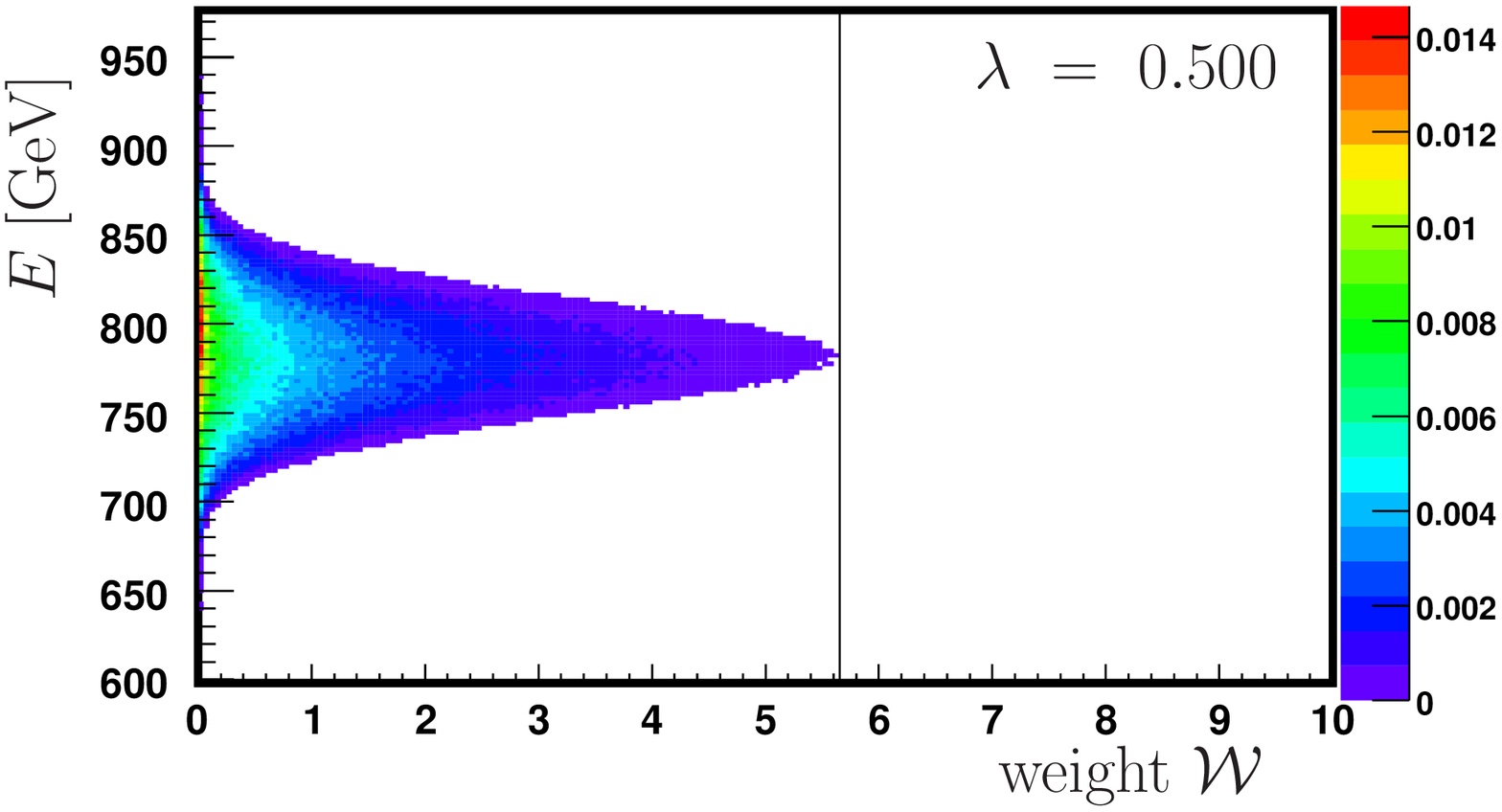,width=4.75cm,height=2.85cm}
  \epsfig{file=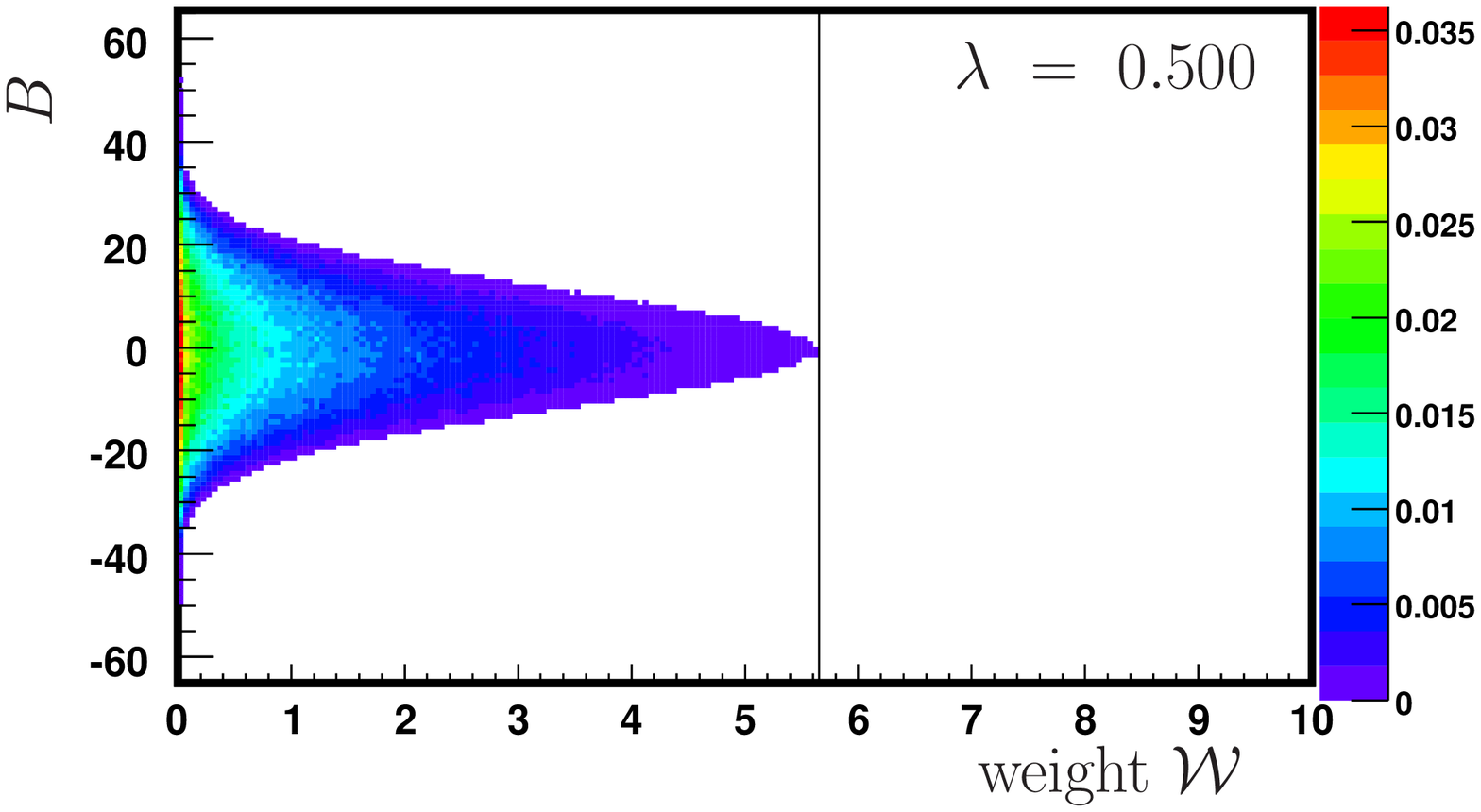,width=4.75cm,height=2.85cm}
  \epsfig{file=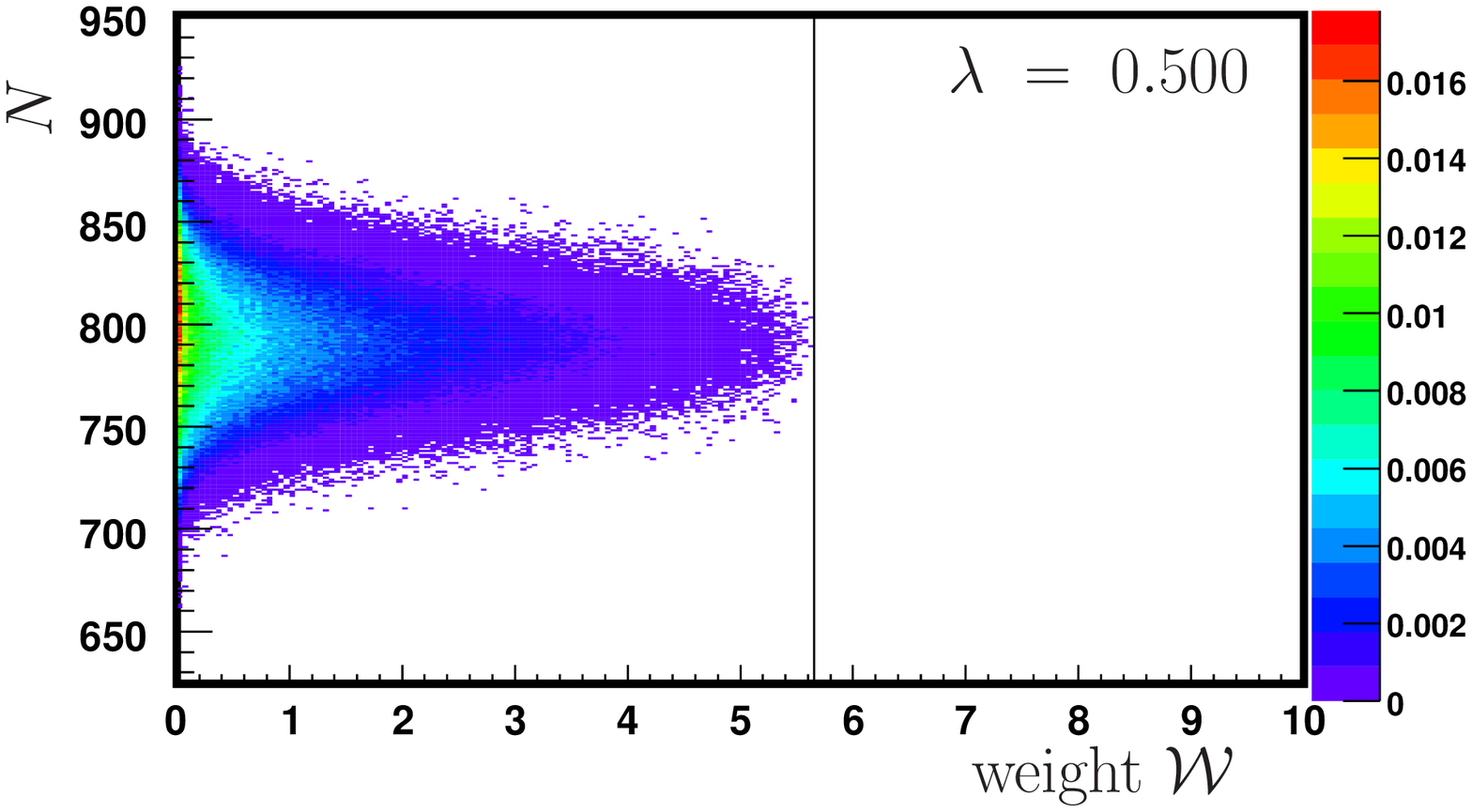,width=4.75cm,height=2.85cm}

  \epsfig{file=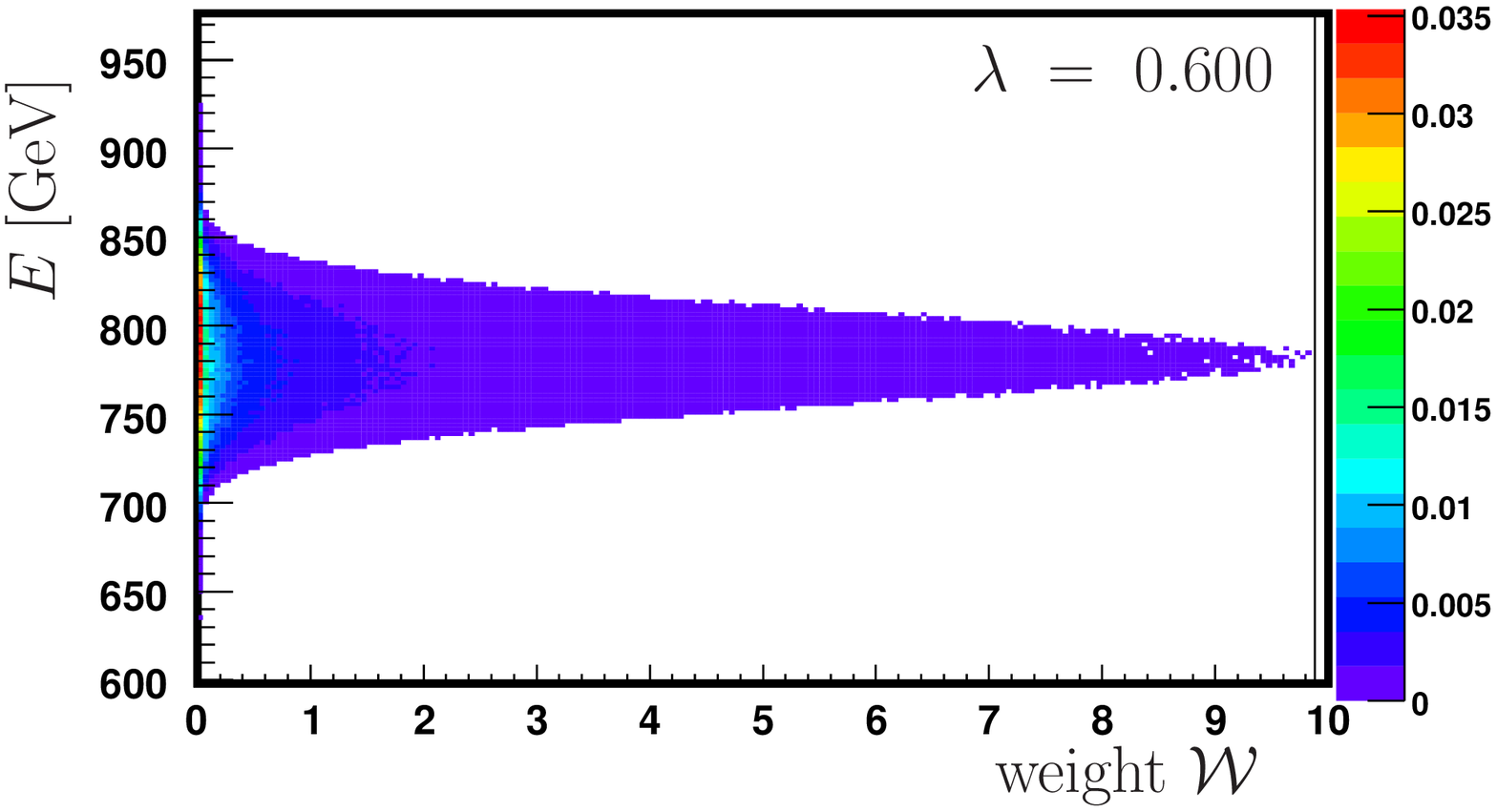,width=4.75cm,height=2.85cm}
  \epsfig{file=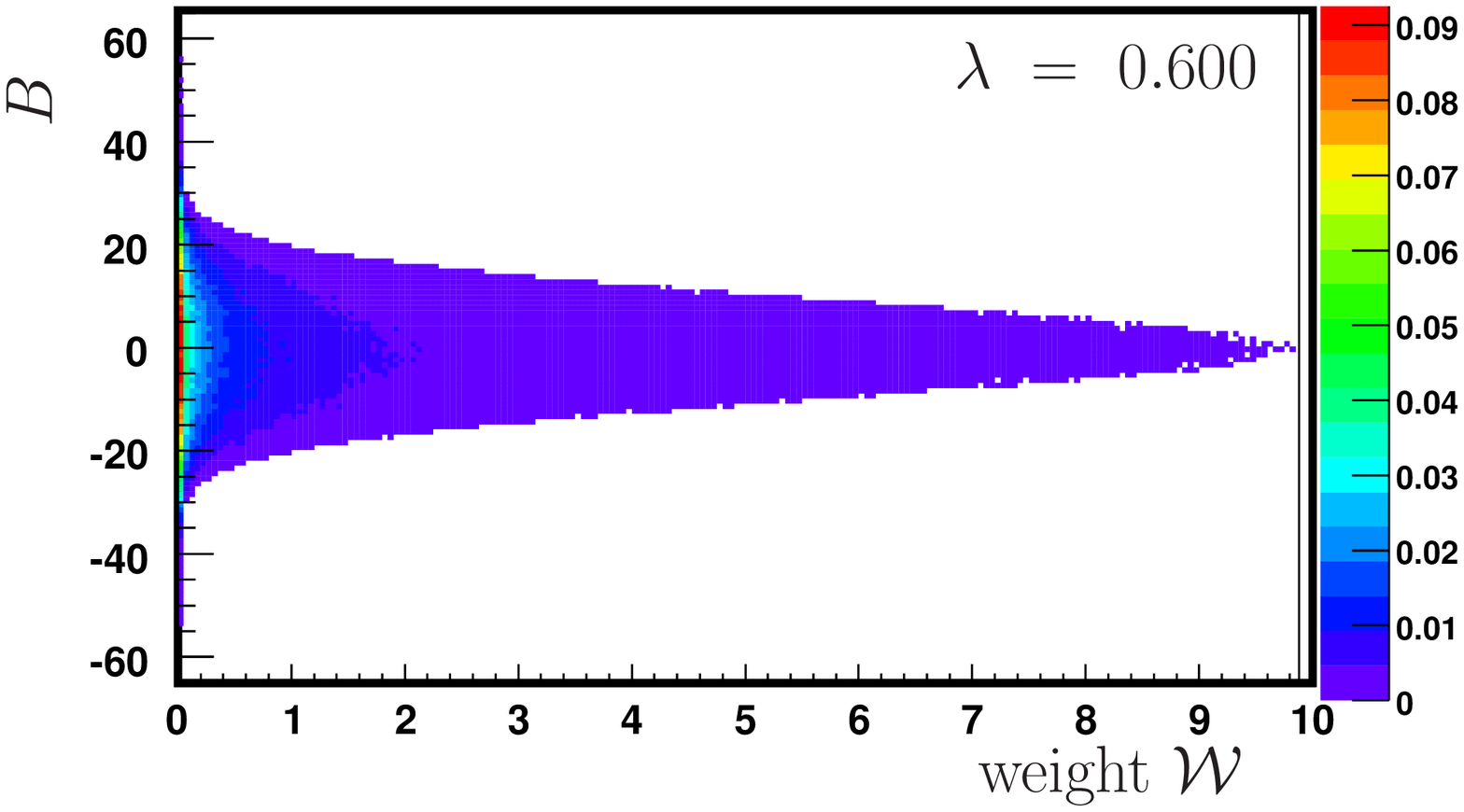,width=4.75cm,height=2.85cm}  
  \epsfig{file=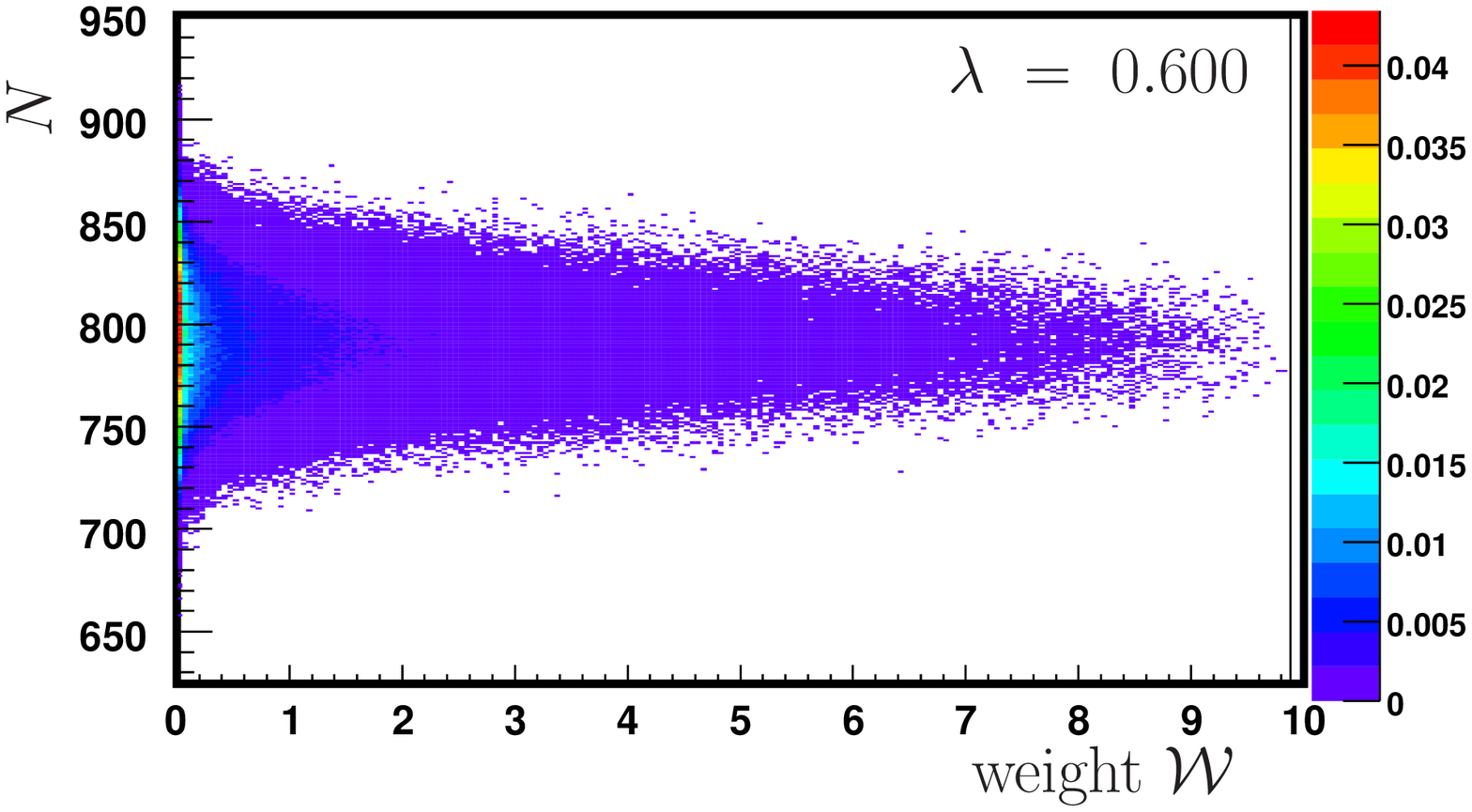,width=4.75cm,height=2.85cm}   
 
  \caption{Evolution of the weight factor distribution with the size of the bath~$\lambda = V_1/V_g$.
  The occurrence of certain weight factors is shown with the value of the extensive quantities
  energy~$E$~({\it left}), baryon number~$B$~({\it center}), and particle number~$N$~({\it right}).
  The extensive quantities $B$, $S$, $Q$, $E$, $P_z$ are considered for re-weighting. 
  Here~$5 \cdot 10^5$ events have been sampled for each value of~$\lambda$. 
  The solid vertical lines indicate the maximal weight~$w_n^{max}=\left( 1- \lambda \right)^{-L/2}$.}  
  \label{P_lambda_weight}
\end{figure}

\begin{figure}[ht!]
  \epsfig{file=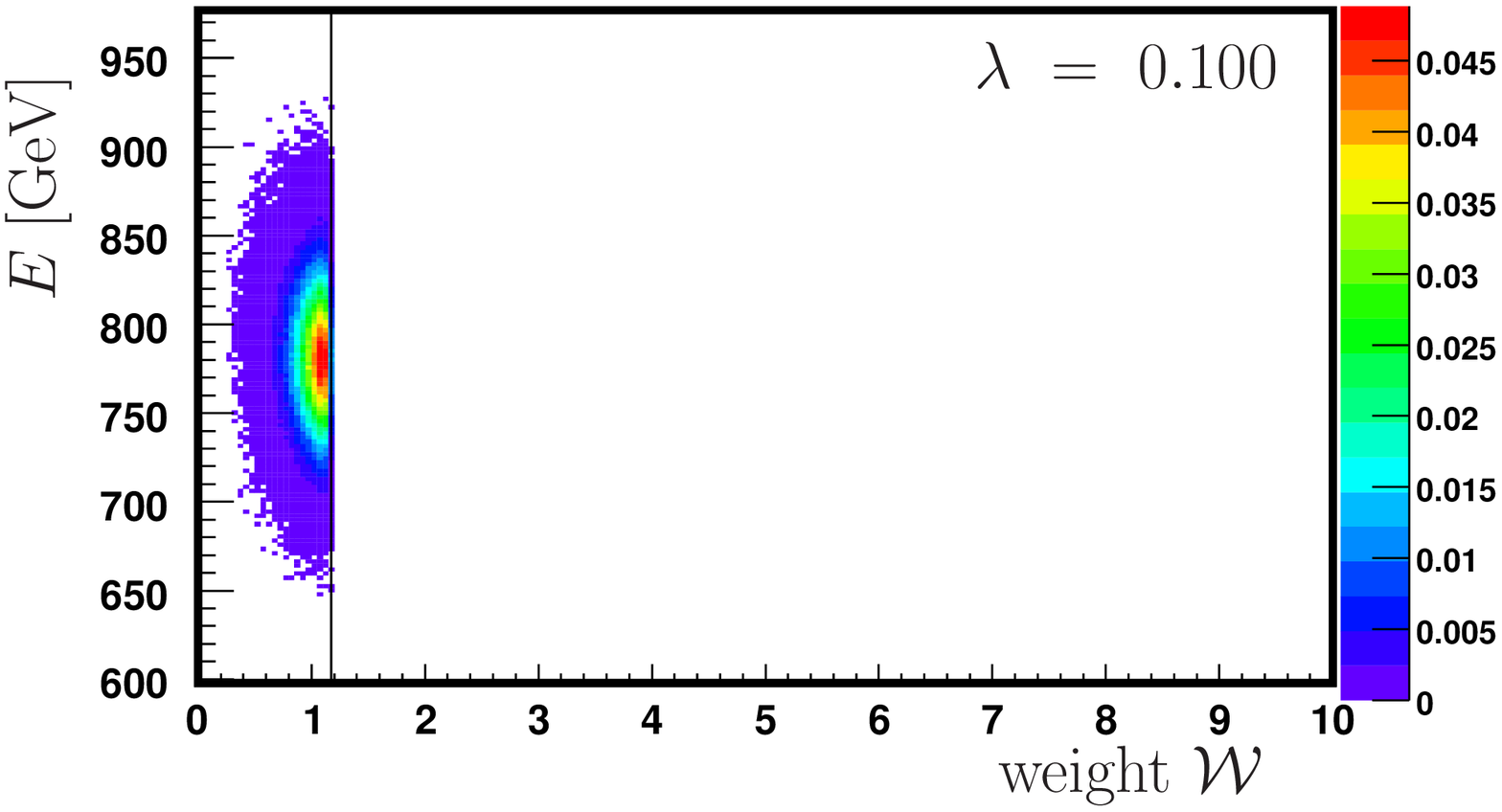,width=4.75cm,height=2.85cm}
  \epsfig{file=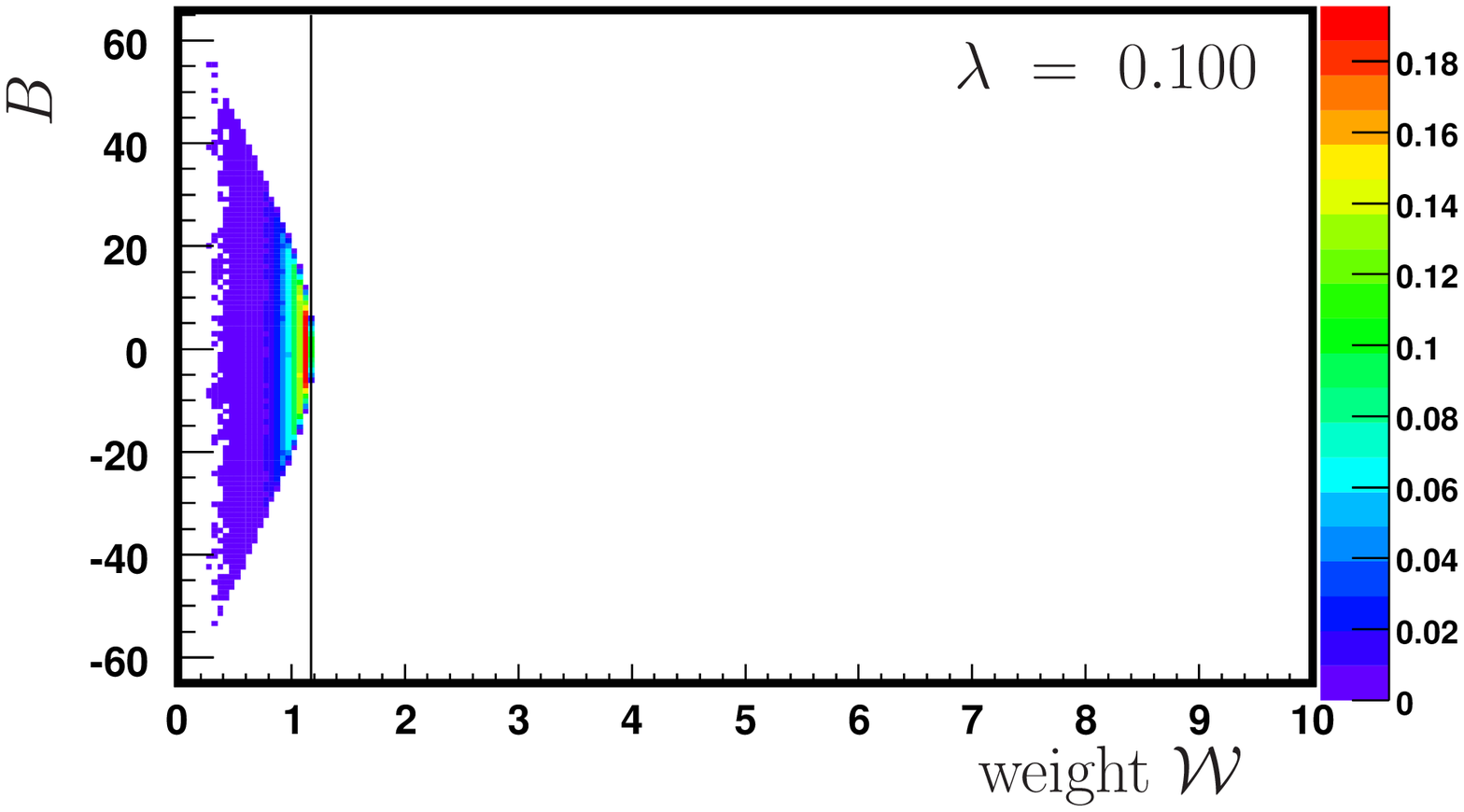,width=4.75cm,height=2.85cm}
  \epsfig{file=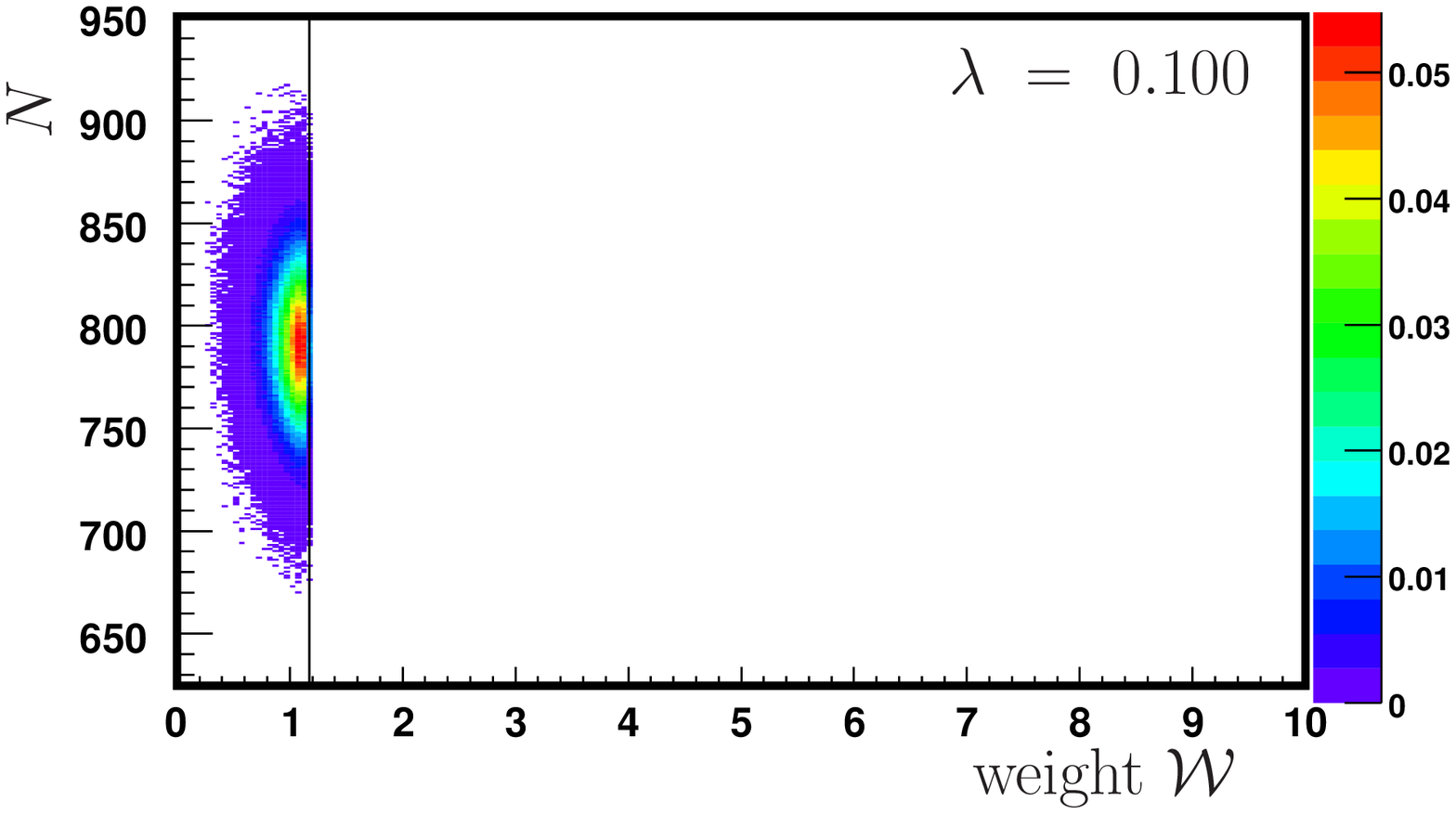,width=4.75cm,height=2.85cm}
  
  \epsfig{file=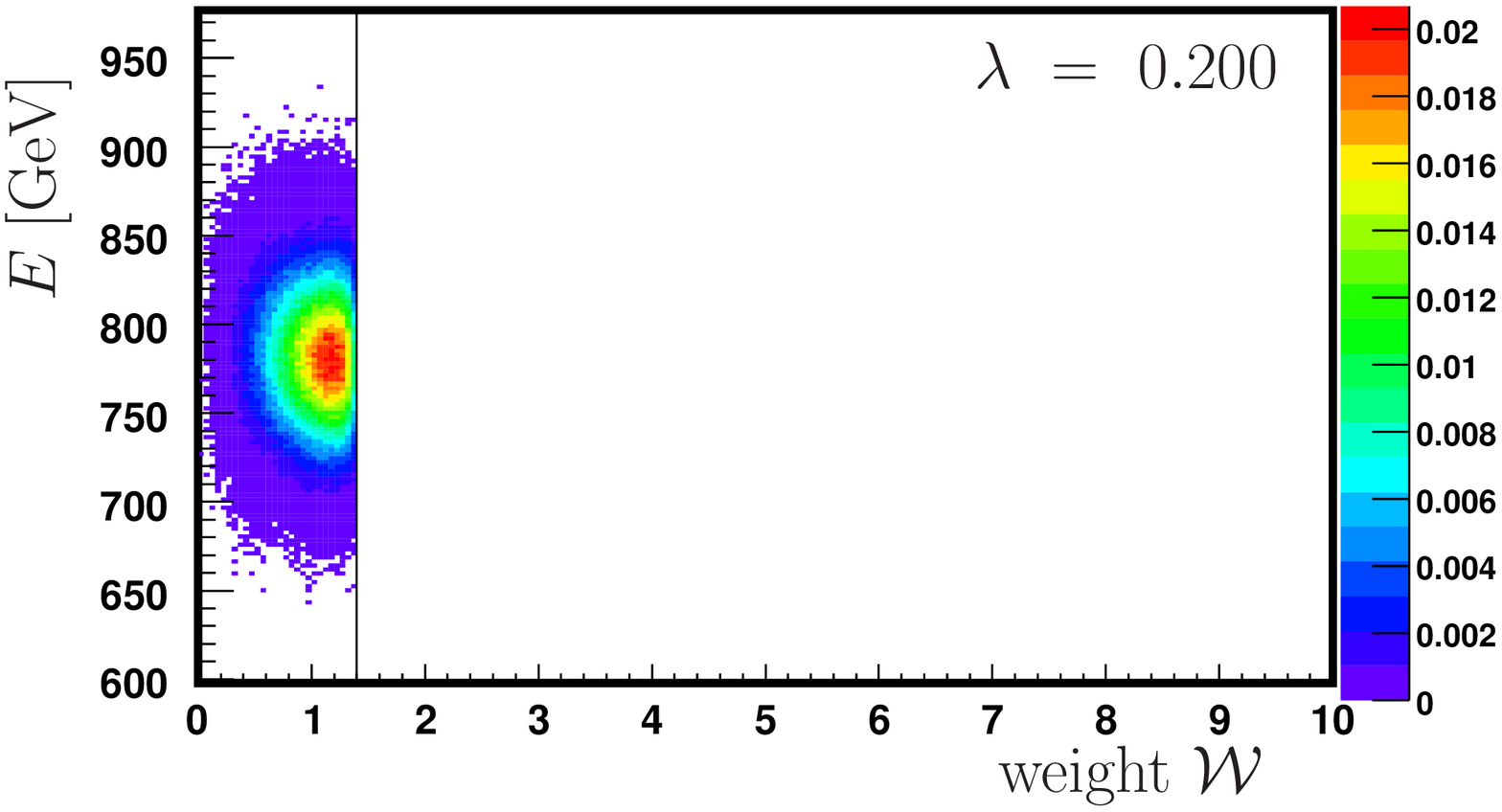,width=4.75cm,height=2.85cm}
  \epsfig{file=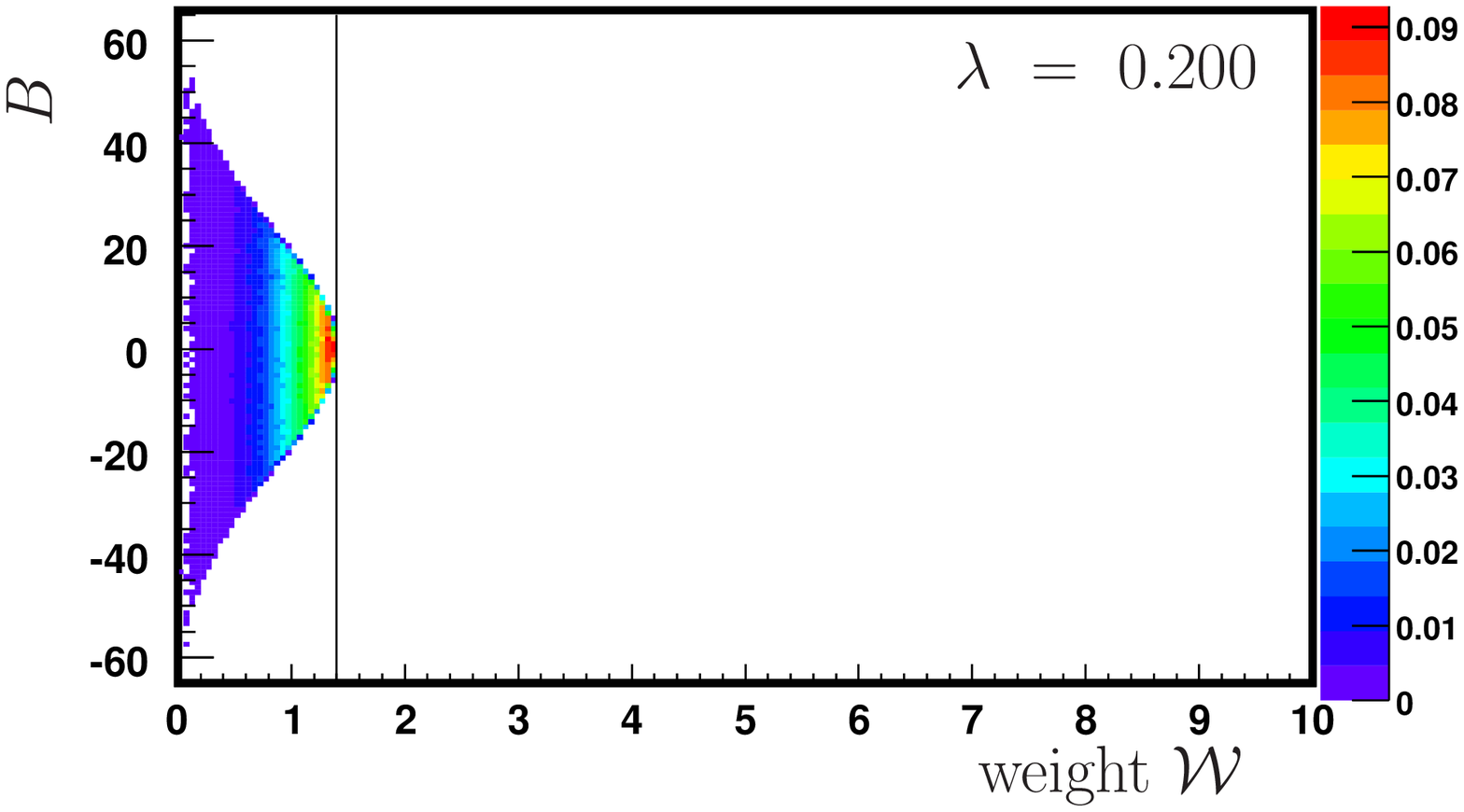,width=4.75cm,height=2.85cm}
  \epsfig{file=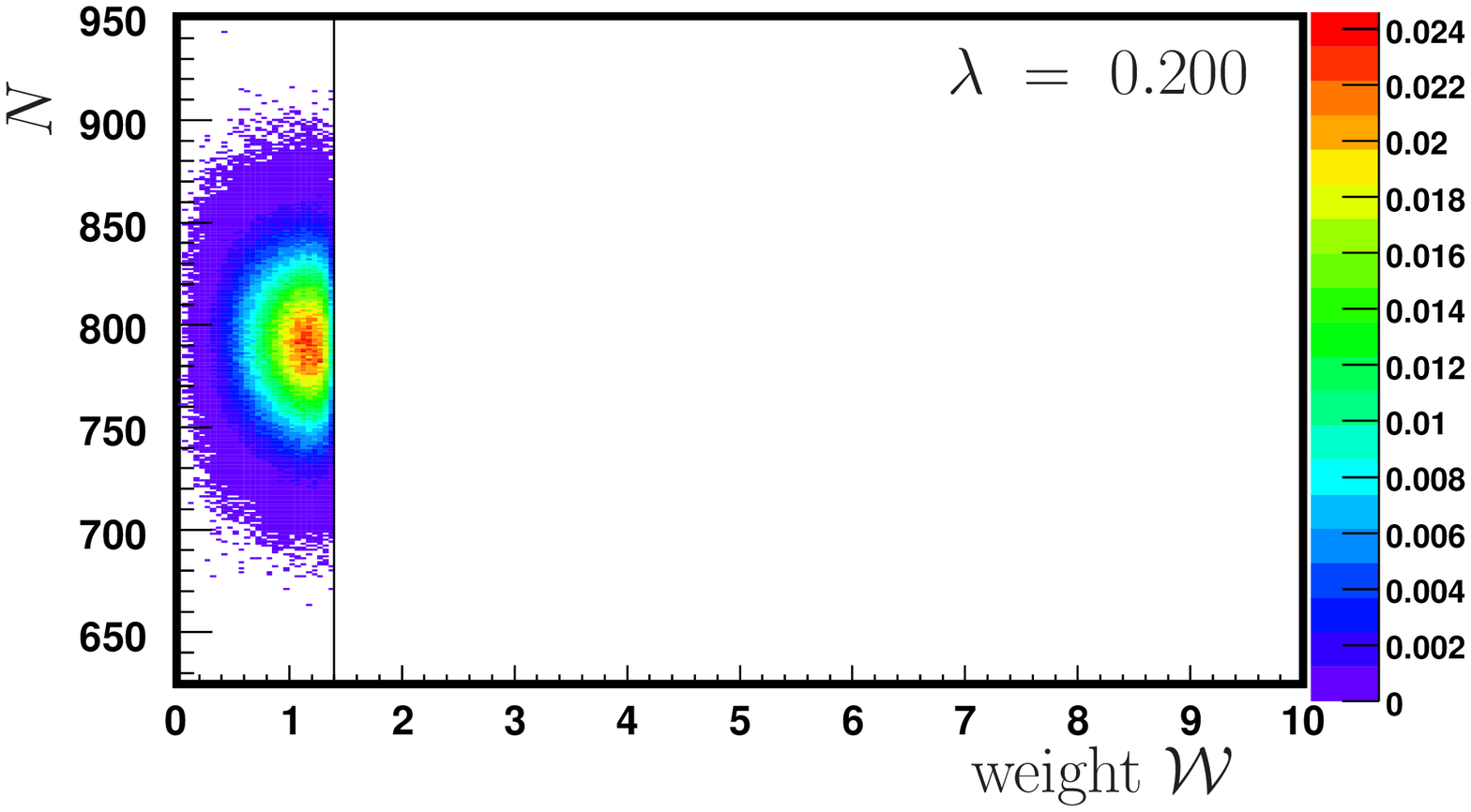,width=4.75cm,height=2.85cm}

  \epsfig{file=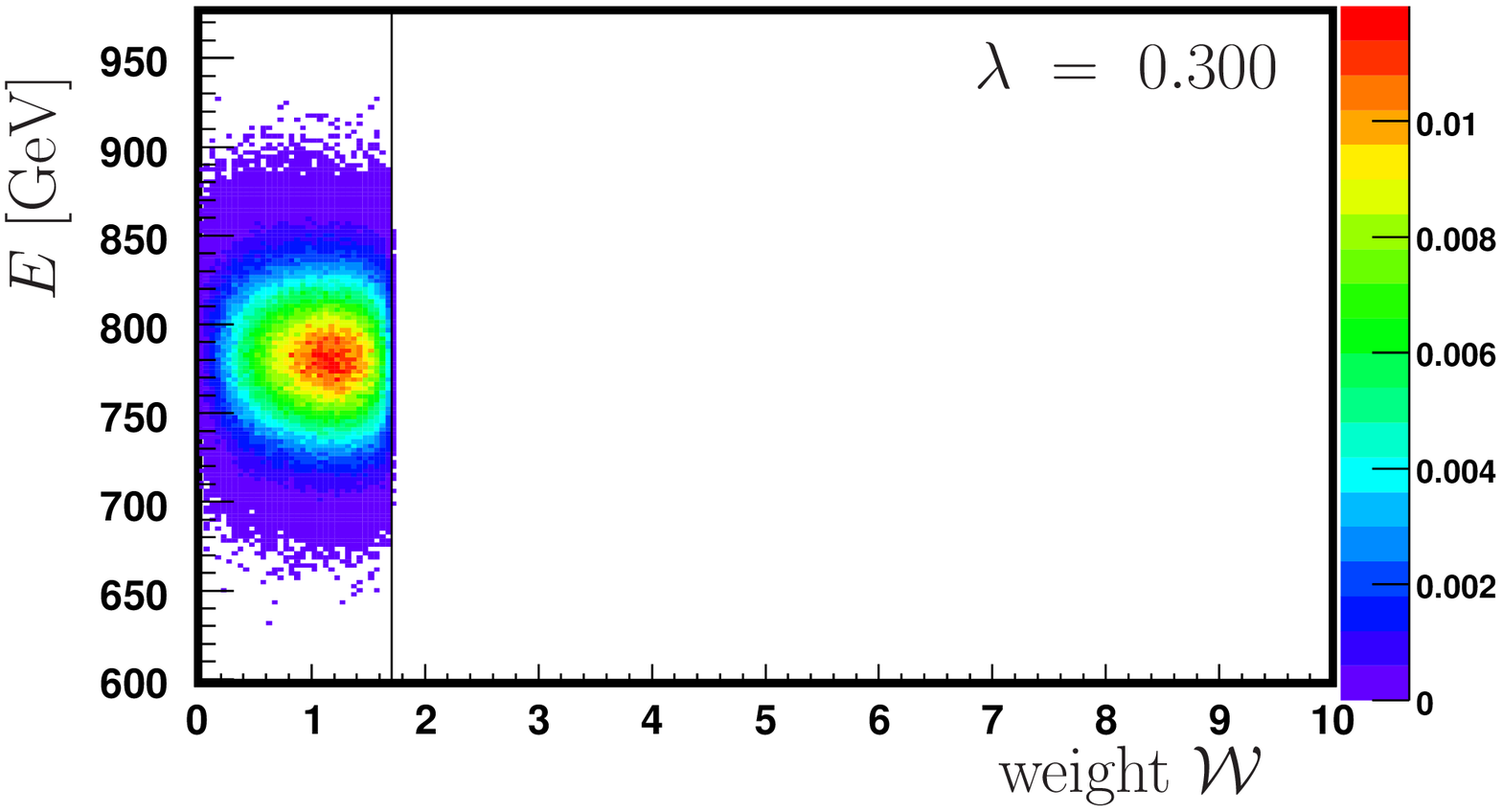,width=4.75cm,height=2.85cm}
  \epsfig{file=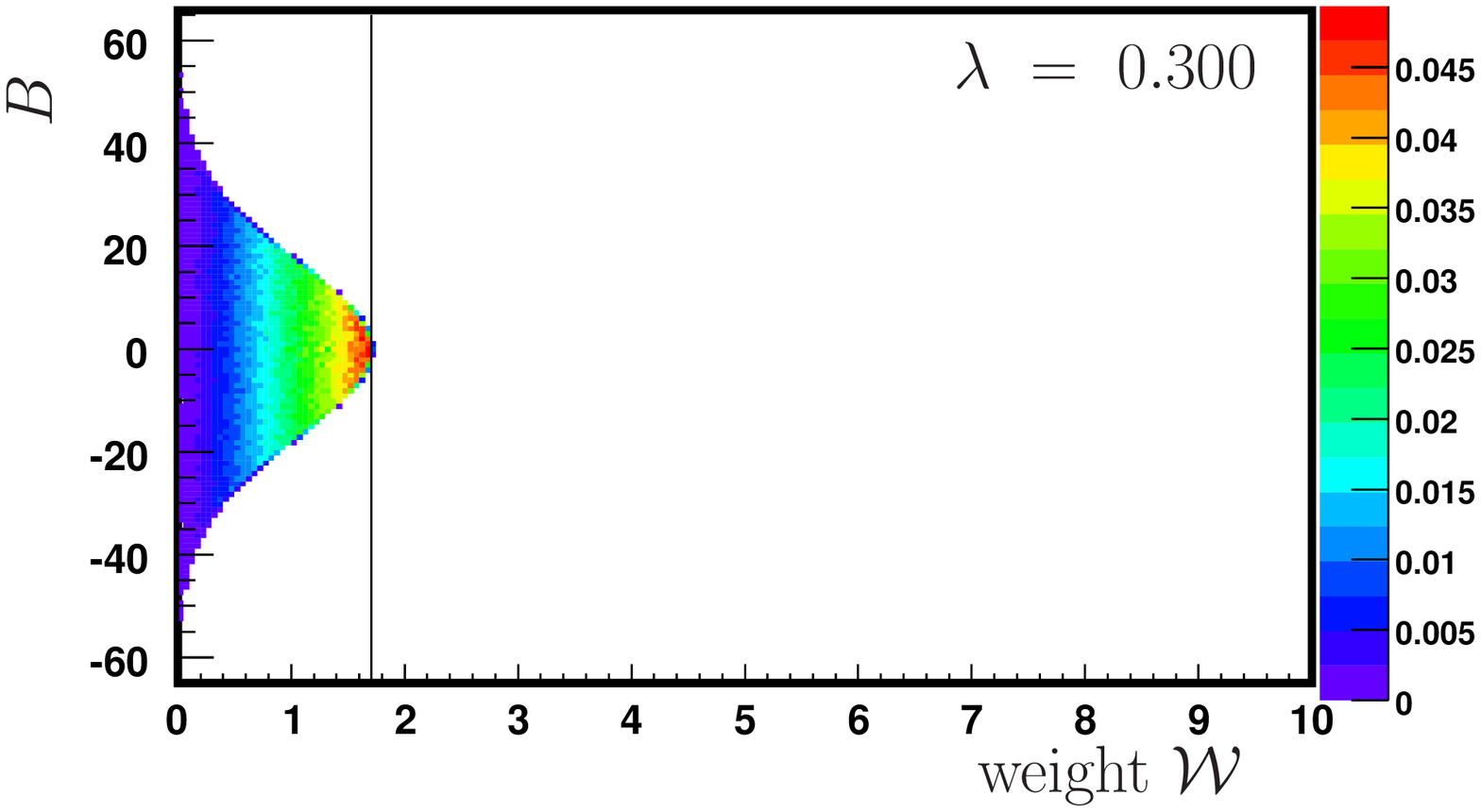,width=4.75cm,height=2.85cm}
  \epsfig{file=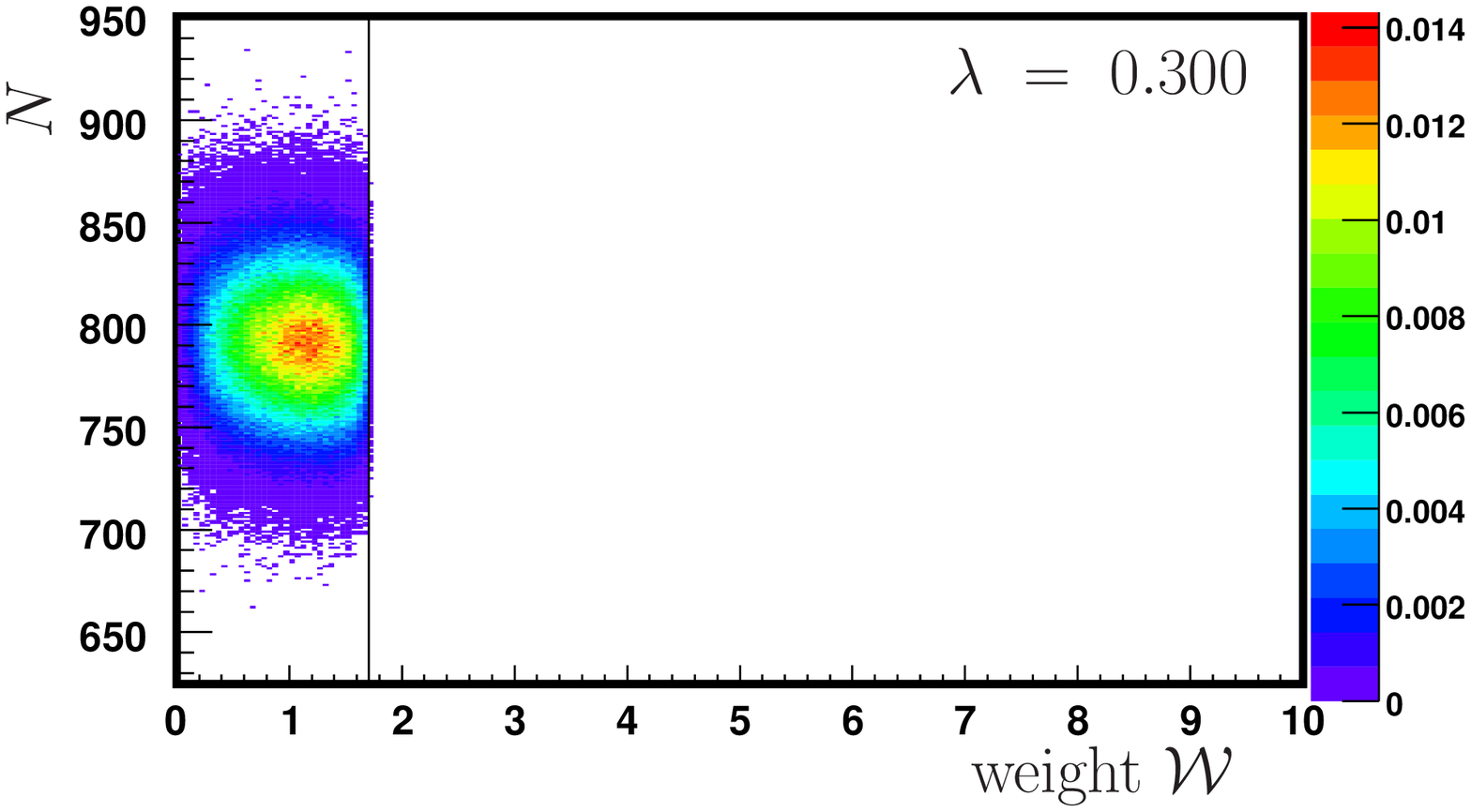,width=4.75cm,height=2.85cm}

  \epsfig{file=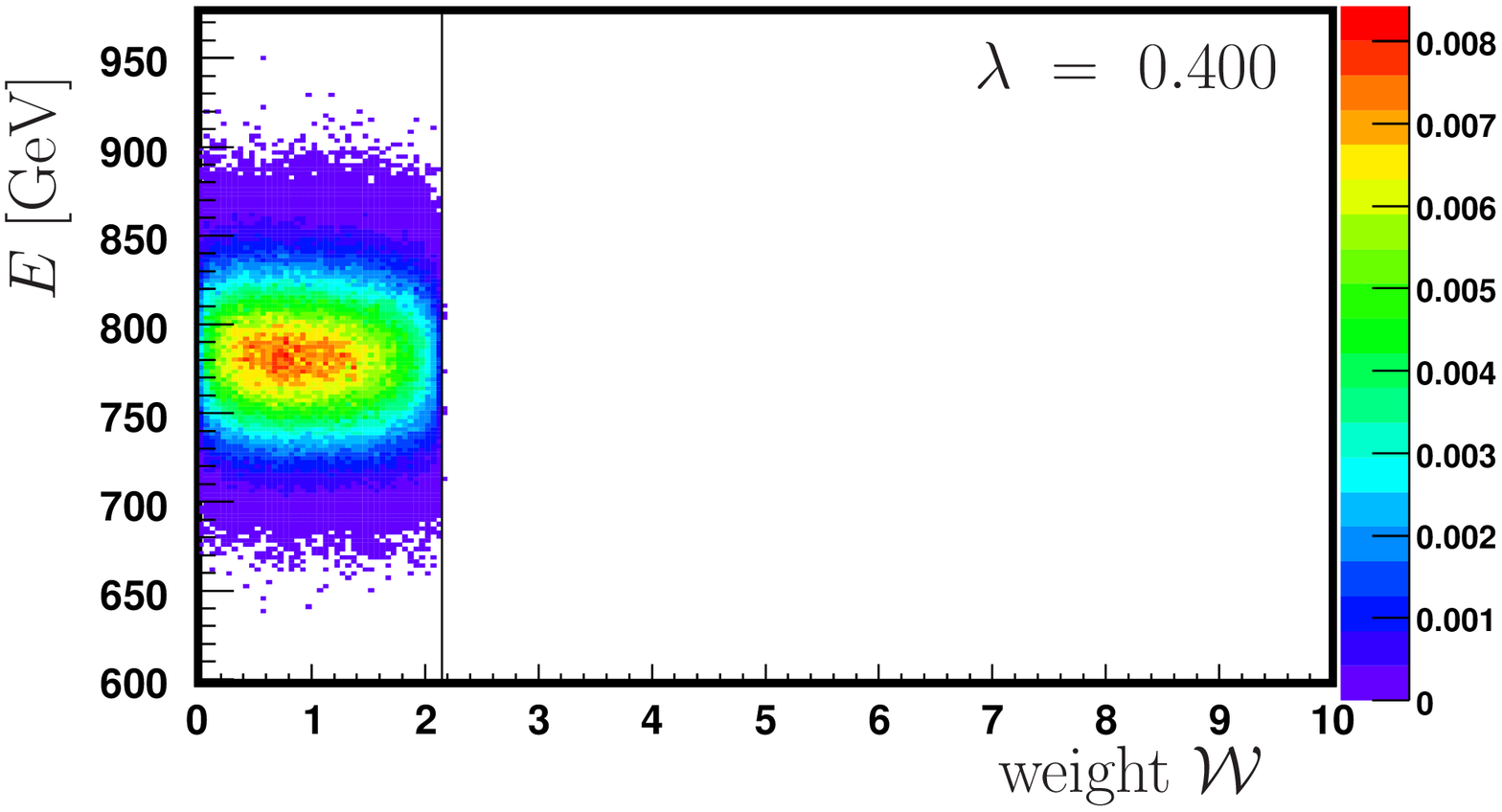,width=4.75cm,height=2.85cm}
  \epsfig{file=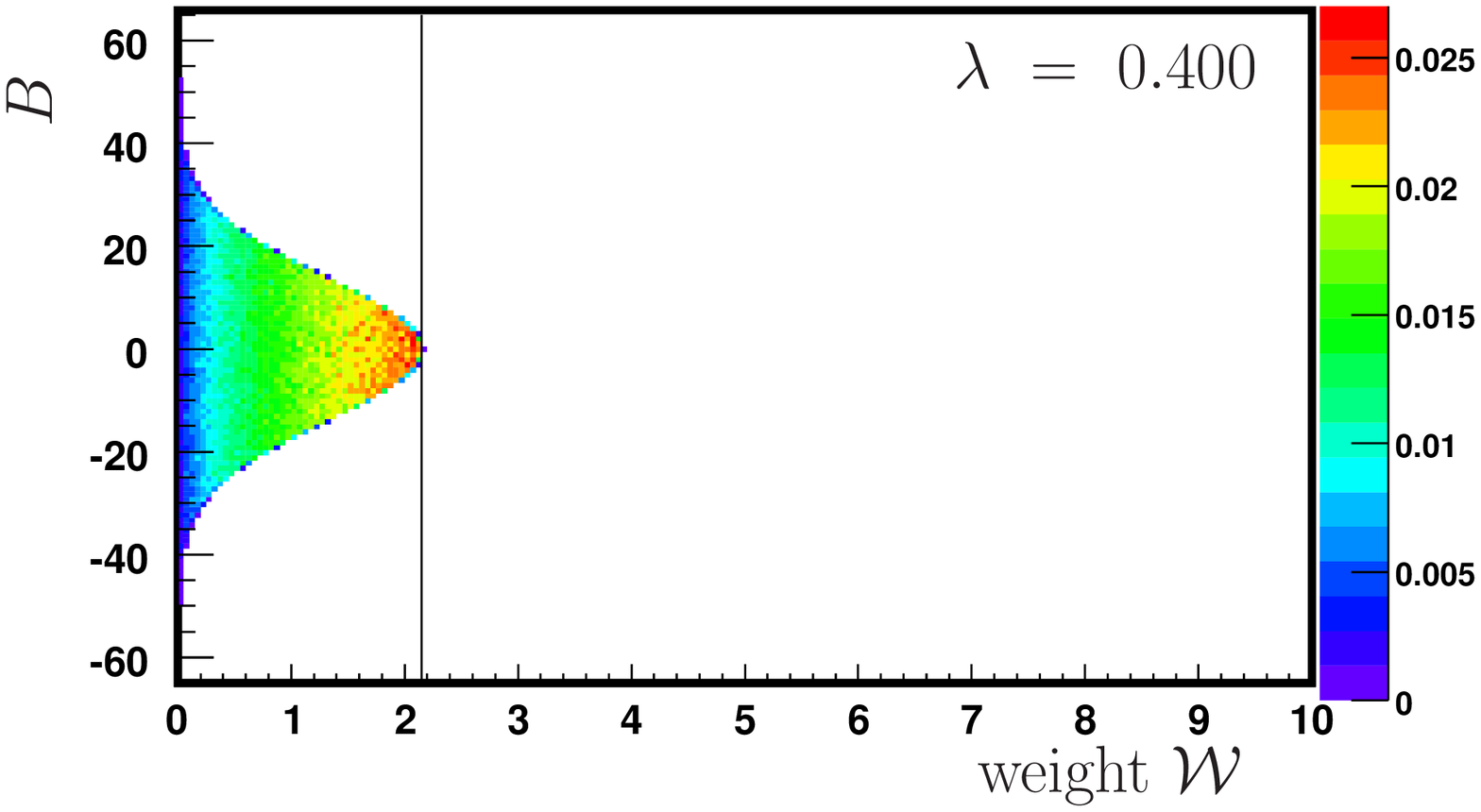,width=4.75cm,height=2.85cm}
  \epsfig{file=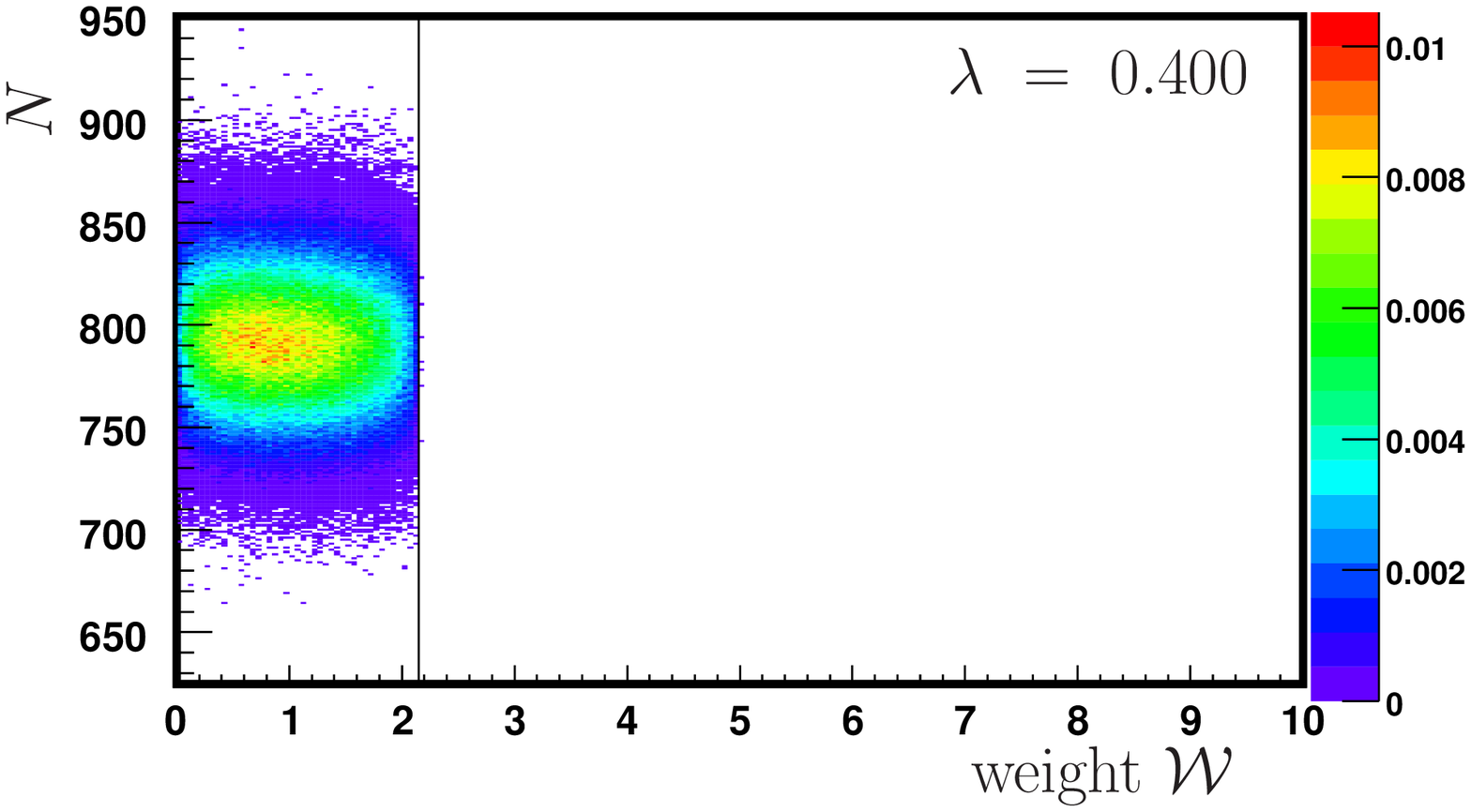,width=4.75cm,height=2.85cm}
  
  \epsfig{file=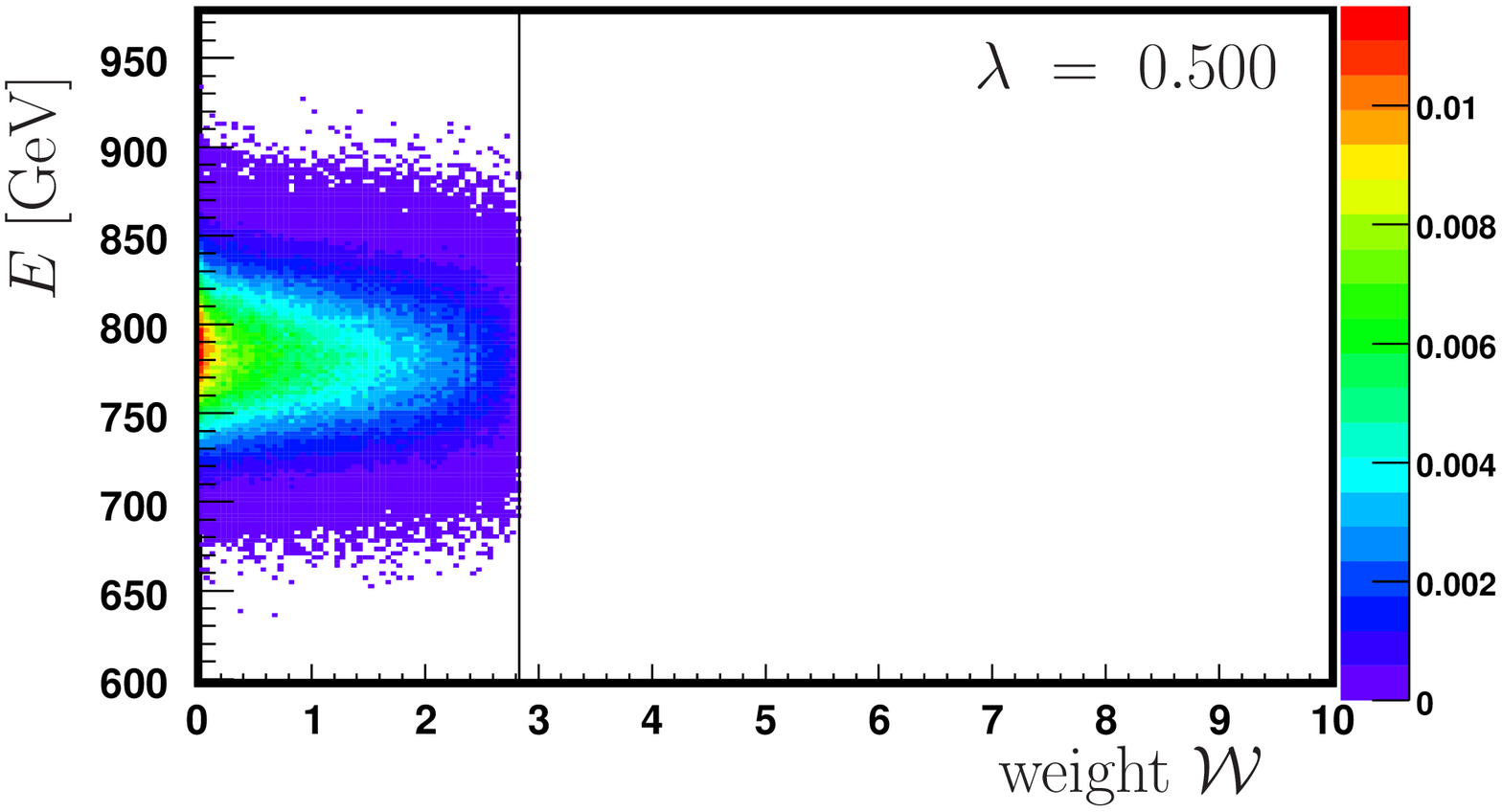,width=4.75cm,height=2.85cm}
  \epsfig{file=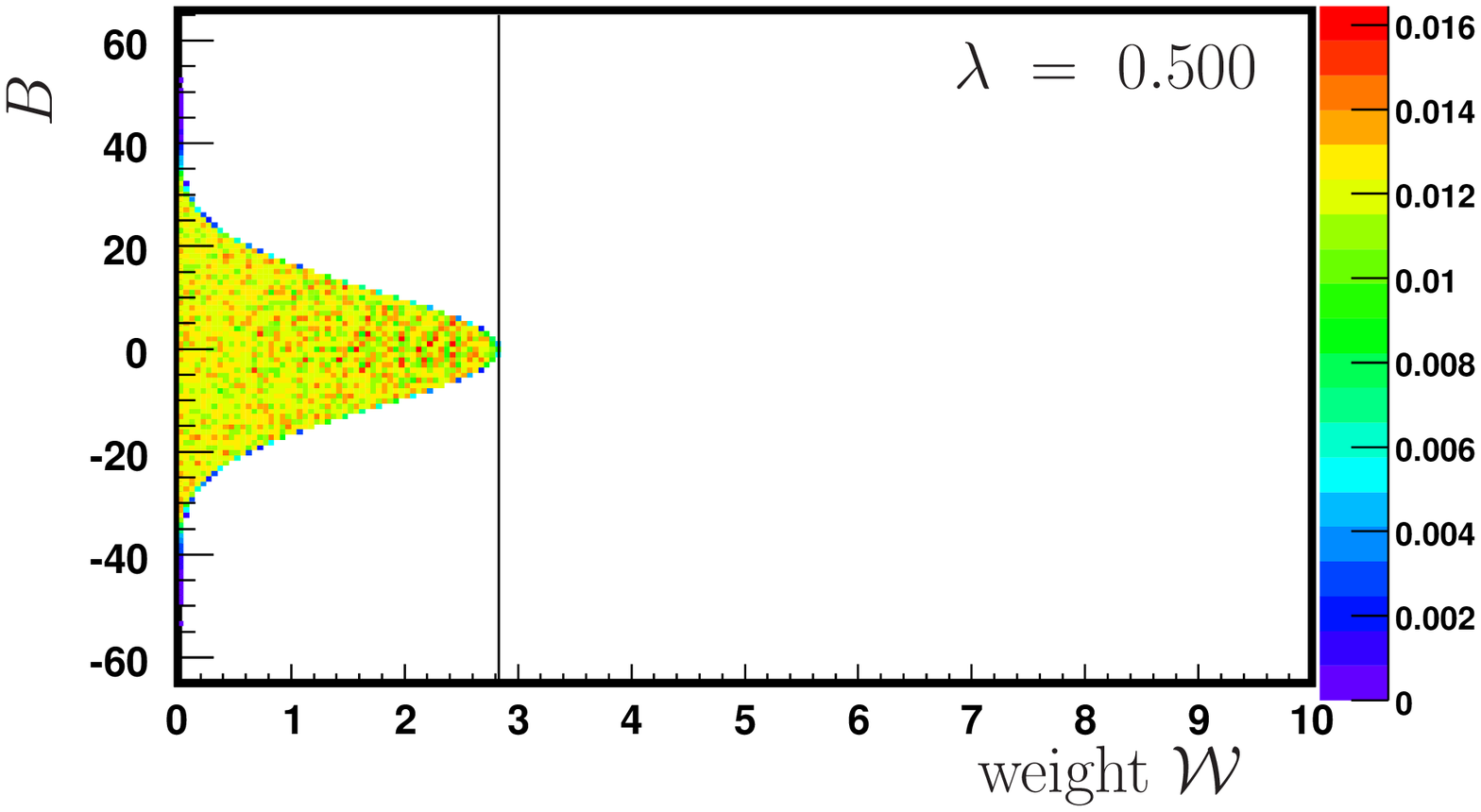,width=4.75cm,height=2.85cm}
  \epsfig{file=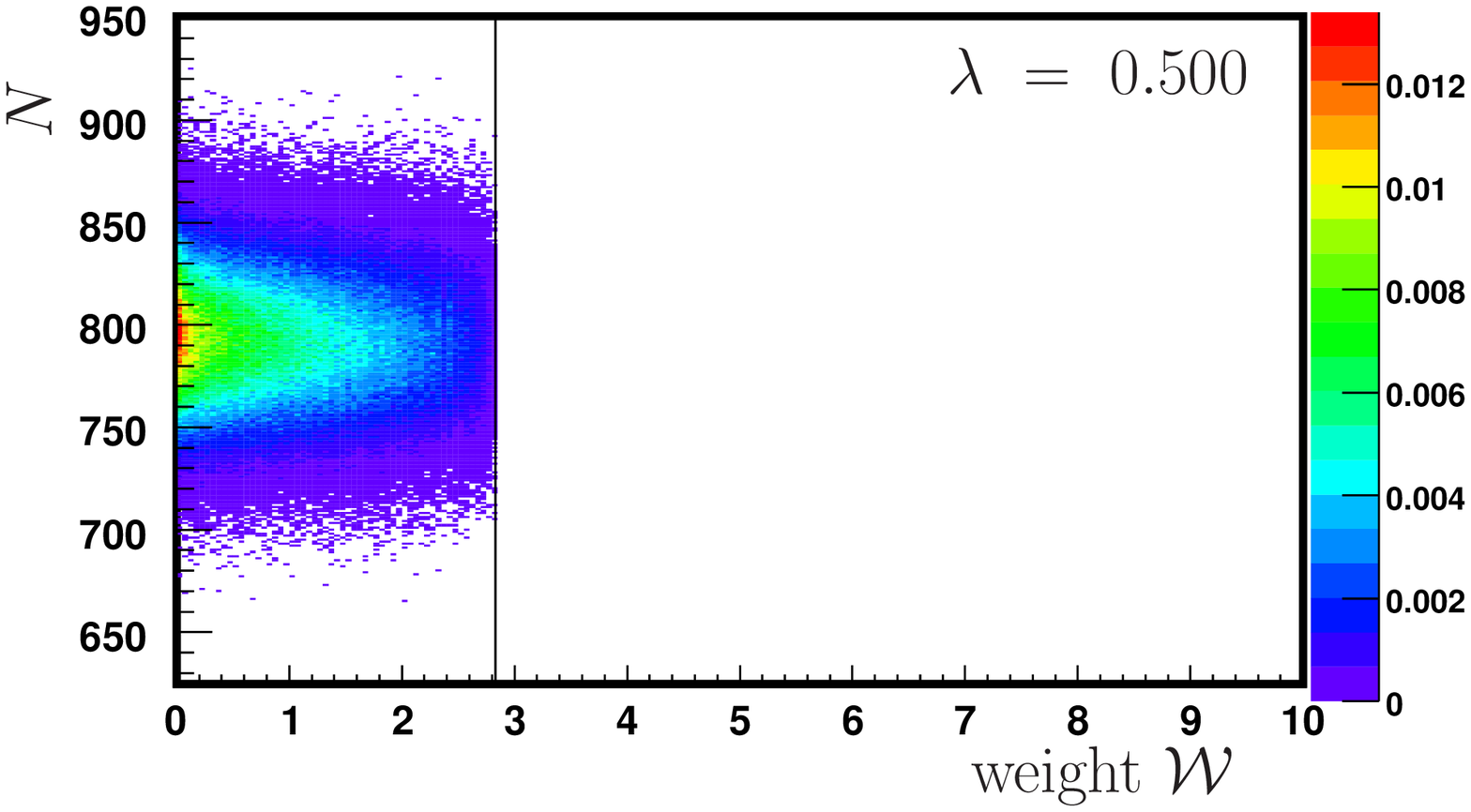,width=4.75cm,height=2.85cm}

  \epsfig{file=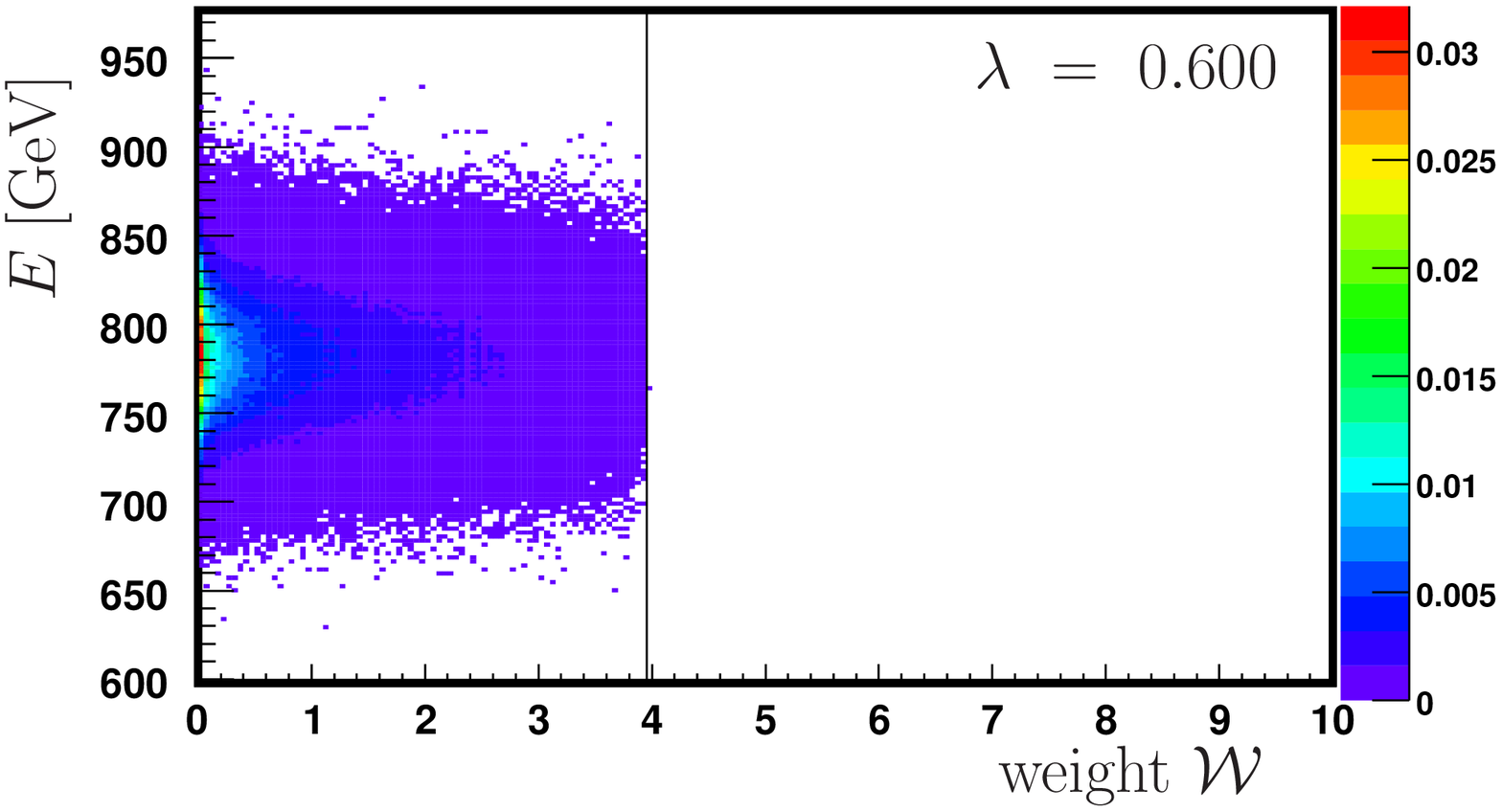,width=4.75cm,height=2.85cm}
  \epsfig{file=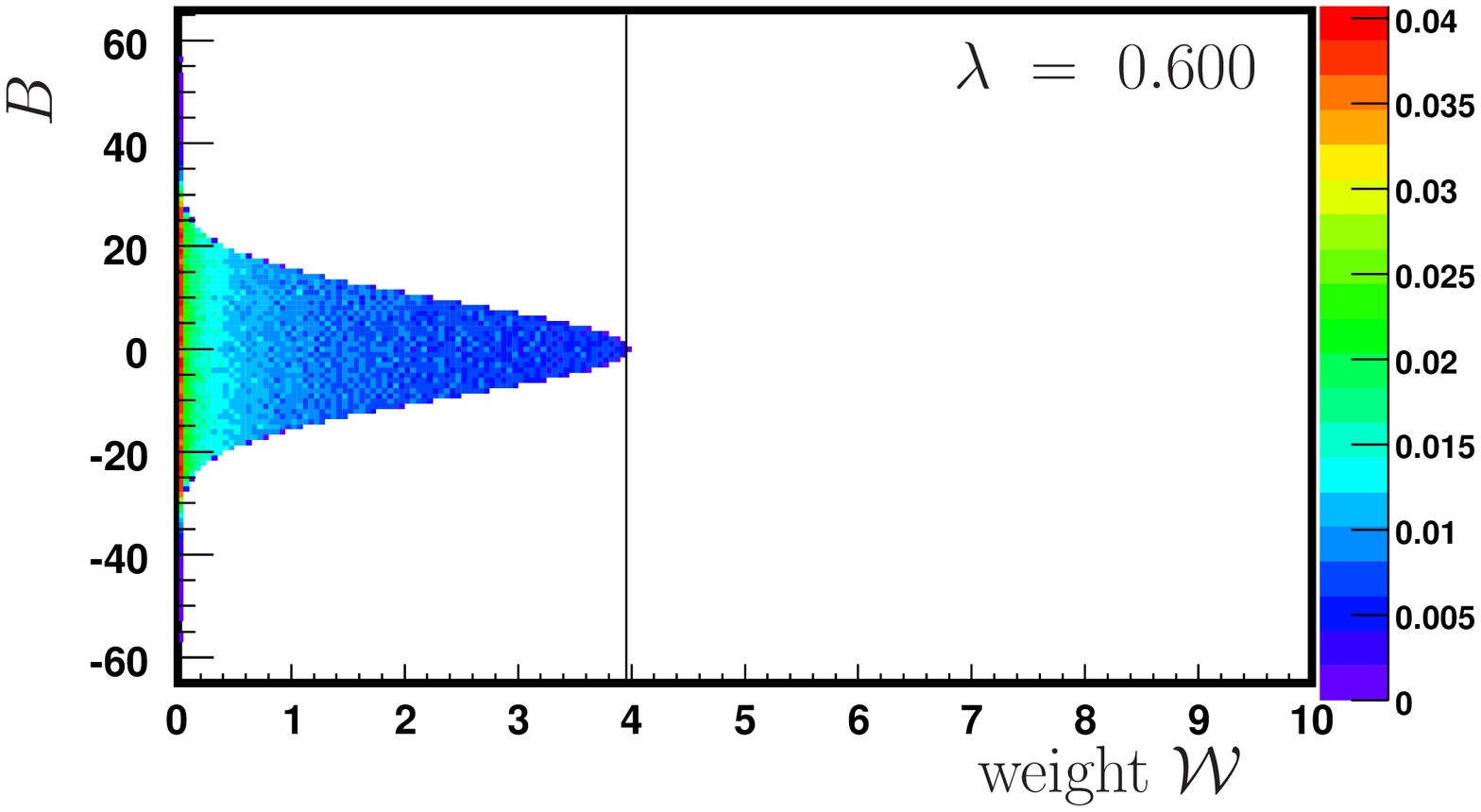,width=4.75cm,height=2.85cm}  
  \epsfig{file=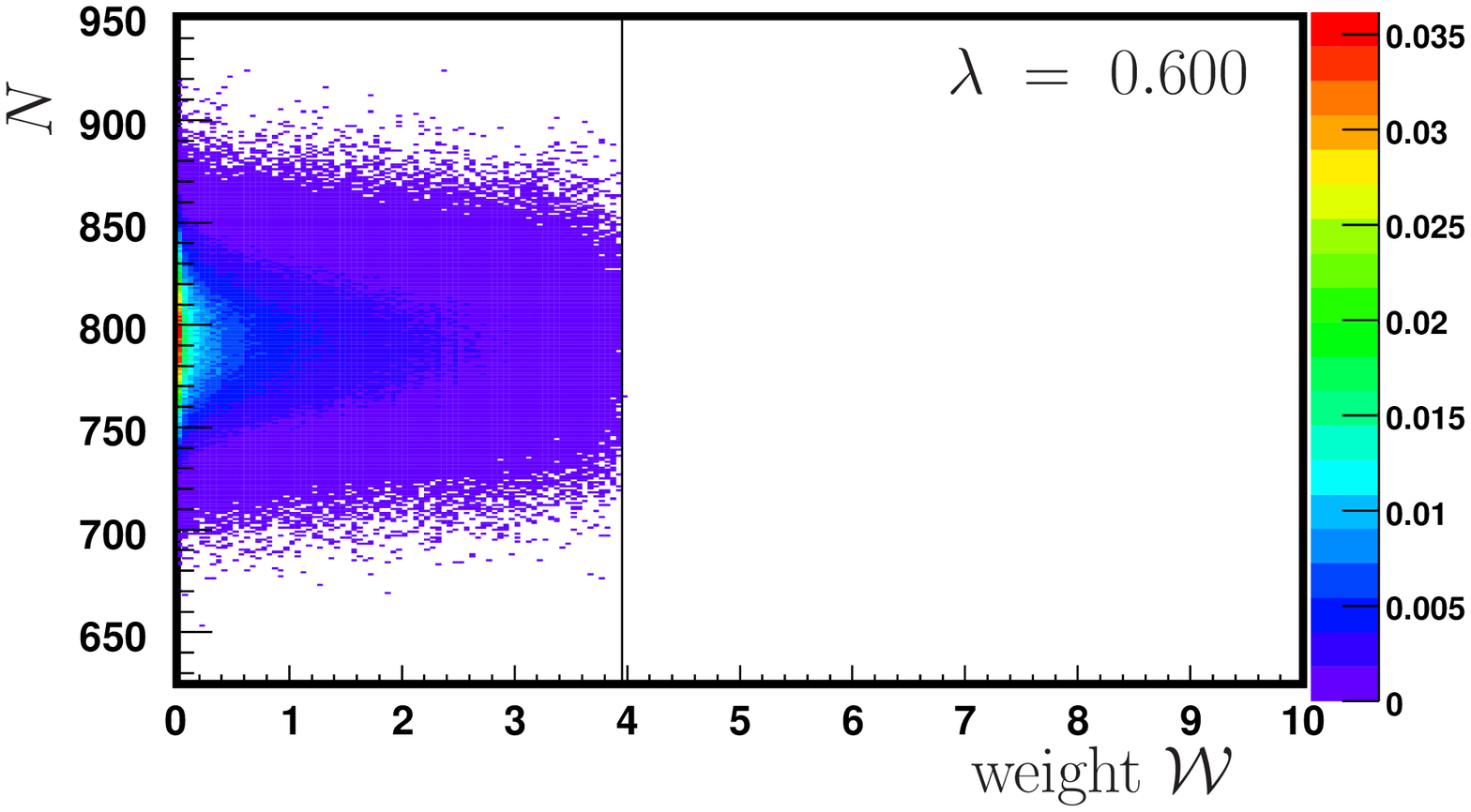,width=4.75cm,height=2.85cm}   
 
  \caption{Evolution of the weight factor distribution with the size of the bath~$\lambda = V_1/V_g$.
    The occurrence of certain weight factors is shown with the value of the extensive quantities
    energy~$E$~({\it left}), baryon number~$B$~({\it center}), and particle number~$N$~({\it right}).
    Only the extensive quantities $B$, $S$, $Q$ are considered for re-weighting. 
    Here~$5 \cdot 10^5$ events have been sampled for each value of~$\lambda$. 
    The solid vertical lines indicate the maximal weight~$w_n^{max}=\left( 1- \lambda \right)^{-L/2}$.}  
  \label{P_BSQ_lambda_weight}
\end{figure}

The Monte Carlo sample was therefore successively transformed. With decreasing size of the 
bath,~${\lambda \rightarrow 1}$, larger and larger weight is given to events in the immediate vicinity of the 
equilibrium expectation value, and smaller and smaller weight to events away from it. The distribution of extensive 
quantities considered for re-weighting (a multivariate normal distribution in the GCE in the large volume limit) 
hence gets contracted to a~$\delta$-function with vanishing variances and covariances. I.e., successively 
the properties of events which have very similar values of extensive quantities are highlighted. 
This will have a bearing on charge correlations and, in particular, multiplicity fluctuations and correlations which 
will be discussed in the following sections.

Considering the evolution of the weight factor distribution with the size of the bath~$\lambda = V_1/V_g$, 
Fig.(\ref{P_lambda_weight}) shows the occurrence of certain weight factors~$w_n$, Eq.(\ref{code_weight}), 
with the value of the extensive quantities energy~$E_{1,n}$~({\it left}),  baryon number~$B_{1,n}$~({\it center}), 
and particle number~$N_{1,n}$~({\it right}). 

\begin{wrapfigure}{r}{0.5\textwidth}
  \begin{center}
    \includegraphics[width=0.52\textwidth,height=6.5cm]{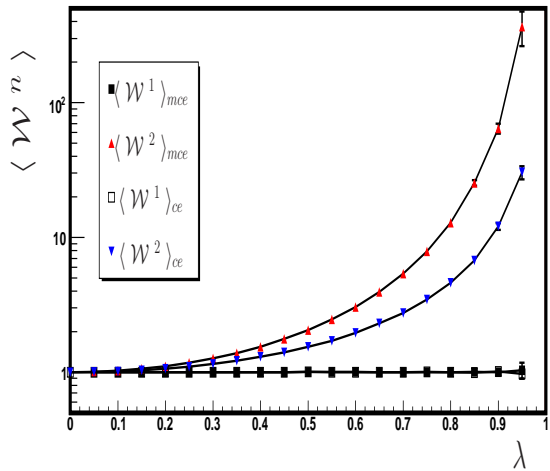}
  \caption{First and second moment of the weight factor Eq.(\ref{solved_curly_W}) as a 
  function of~$\lambda$. The rest as in Fig.(\ref{CwL_plot_fluc}).}
  \label{CwL_plot_weight_2nd_mom}
  \end{center}
\end{wrapfigure}

In the GCE,~$\lambda = 0$ all events have weight equal to unity as discussed in Chapter~\ref{chapter_montecarlo}. 
In the limit~${\lambda \rightarrow 1}$, i.e. approaching the MCE, one finds for quantities which are re-weighted,
that basically all events have weight equal to~$0$, except for those in the immediate vicinity of the equilibrium value. 
Not all events with the same value of energy~$E_{1,n}$ receive the same weight $\mathcal{W}^{Q_1^l;Q_g^l}_n$. The energy 
value might be close to the peak. If the baryon number (and strangeness, longitudinal momentum, etc.) value are now also 
close to their equilibrium values, the events obtain a higher weight, if the baryon value is far away, a lower weight. 
The largest weight an event can assume is, Eq.(\ref{solved_curly_W}), $w_n^{max}=\left( 1- \lambda \right)^{-L/2}$.
For intermediate~$\lambda$ a small hill, where most events are located, emerges. As~$\lambda$ is increased most events 
receive smaller and smaller weight and this small hill gets pushed to the left. The evolution of the baryon number 
distribution, Fig.(\ref{P_lambda_weight})~({\it center}) is similar.

The extensive quantity particle number is not re-weighted. Hence its distribution does not converge to a~$\delta$-function.
The connection between the extensive quantity~$N$ and other (re-weighted) quantities manifests itself indirectly. Events 
have certain energy, baryon number or strangeness. These quantities are correlated with particle number. Events with 
unusually large or small multiplicities now receive mostly small weight factors, because chance has it, that they also
have values of energy, net-baryon number or net-strangeness which are far from the equilibrium expectation value. 
But even in the limit $\lambda \rightarrow 1$ a certain width remains, and no sharp tip forms, as particle number 
continues to fluctuate in the MCE. In Fig.(\ref{CwL_plot_weight_2nd_mom}) the second moment of the weight factor, 
Eq.(\ref{solved_curly_W}) is shown as a function of~$\lambda$ for both the extrapolation to the MCE and to the CE. 
A large second moment~$\langle \mathcal{W}^2 \rangle$ 
implies a large statistical uncertainty and, hence, usually requires a larger sample. 

For comparison, the evolution of weight factors for an extrapolation to the CE limit is presented in 
Fig.(\ref{P_BSQ_lambda_weight}). Events are re-weighted only according to the values of their extensive quantities 
baryon number $B_{1,n}$, strangeness $S_{1,n}$, and electric charge $Q_{1,n}$. As in Fig.(\ref{P_lambda_weight}), 
the distributions for the extensive quantities energy $E$, baryon number $B$, and particle multiplicity $N$ are shown for 
different values of $\lambda$.  The energy distribution, Fig.(\ref{P_BSQ_lambda_weight})~({\it left}), is much wider than 
the one in Fig.(\ref{P_lambda_weight})~({\it left}). 
The systems energy content continues to fluctuate, $\langle \left( \Delta E \right) \rangle \not= 0$, in the CE. 
The baryon number distribution Fig.(\ref{P_BSQ_lambda_weight})~({\it center}) on the other hand converges, as in the MCE, 
to a $\delta$-function. The particle number distribution Fig.(\ref{P_BSQ_lambda_weight})~({\it right}) is then also 
wider than for an extrapolation to the MCE, Fig.(\ref{P_lambda_weight})~({\it right}). 
The last observation is that, indeed, the maximal weight $w_n^{max}$ an event can receive (at some fixed value 
of $\lambda$) is smaller for the extrapolation to the CE, Fig.(\ref{P_BSQ_lambda_weight}), than for an extrapolation 
to the MCE, Fig.(\ref{P_lambda_weight}). 

\section{Discussion}
\label{sec_extrapol_discussion}

In this chapter fully phase space integrated grand canonical results have been extrapolated to the micro canonical limit. 
For this purpose samples of events for various values of~$\lambda = V_1/V_g$ have been iteratively generated, re-weighted, 
and analyzed. By construction of the weight factor~$\mathcal{W}$ the extrapolation proceeds in a systematic fashion such 
that, for instance, particle momentum spectra, as well as mean values of extensive quantities, remain unchanged.

The GCE distribution of extensive quantities considered for re-weighting converges to a~$\delta$-function. Mean values 
as well as correlation coefficients stay constant, while variances and covariances converge to 0. Events in the vicinity 
of the equilibrium expectation value are projected out, the rest is suppressed. 

Although the extensive quantities particle multiplicity~$N$ and total transverse momentum~$P_T$  are not re-weighted, 
their fluctuations and correlations are nevertheless strongly affected. The effects of decreasing size of the heat bath 
emerge because~$N$ and~$P_T$ are correlated with~$E$,~$Q$, etc., which are re-weighted.

As the system under investigation gets larger, a smaller fraction of events will have values 
of~$E_{1,n}$,$B_{1,n}$,~$Q_{1,n}$, etc. in the vicinity of a desired equilibrium value, and the straight sample and 
reject method becomes ever more inefficient. This method can handle large system sizes, due to additional information 
obtained from the extrapolation. However, as~$\lambda$ grows, so too does the statistical uncertainty. 
In the limit~$\lambda \rightarrow 1$, one approaches a sample-reject type of formalism. 
One cannot, therefore, directly obtain the micro canonical limit for the large system size studied here, 
as this is prohibited by available computing power. However, extrapolation to this limit is possible. It is pointed out in 
this context that the intermediate ensembles, between the limits of GCE and MCE, may be of phenomenological interest, too.

\chapter{Multiplicity Fluctuations and Correlations}
\label{chapter_multfluc_hrg}
In this chapter joint multiplicity distributions of charged hadrons are analyzed in different ensembles with limited 
acceptance in momentum space assumed. Multiplicity fluctuations and correlations are qualitatively affected by the 
choice of ensemble and are directly sensitive to the fraction of the system observed. For vanishing size of ones 
acceptance window, one would lose all information on how the multiplicities of any two distinct groups~$N_i$ and~$N_j$ 
of particles are correlated, and  measure~$\rho_{ij} =0$. This information, on the other hand, is to some extent 
preserved in~$\rho_{BS}$,~$\rho_{BQ}$, and~$\rho_{SQ}$, i.e. the way in which quantum numbers are correlated, if at least 
occasionally a particle is detected  during an experiment.

The Monte Carlo scheme is employed further. In Section~\ref{sec_multfluc_hrg_gce}, the joint distributions of positively 
and negatively charged particles in momentum bins~$\Delta p_{T,i}$ and~$\Delta y_i$ are constructed. Then, in turn, 
primordial and final state GCE results on the scaled variance~$\omega$, Eq.(\ref{omega}), and the correlation 
coefficient~$\rho$, Eq.(\ref{rho}), are extrapolated to the MCE limit in Section~\ref{sec_multfluc_hrg_mce}.
To complement the above study of micro canonical effects, canonical (charge conservation) effects on final state 
multiplicity fluctuations are studied in Section~\ref{sec_multfluc_hrg_ce}. Further discussion of MCE effects is 
presented in Chapter~\ref{chapter_multfluc_pgas}.

\section{Grand Canonical Ensemble}
\label{sec_multfluc_hrg_gce}

\begin{figure}[ht!]
  \epsfig{file=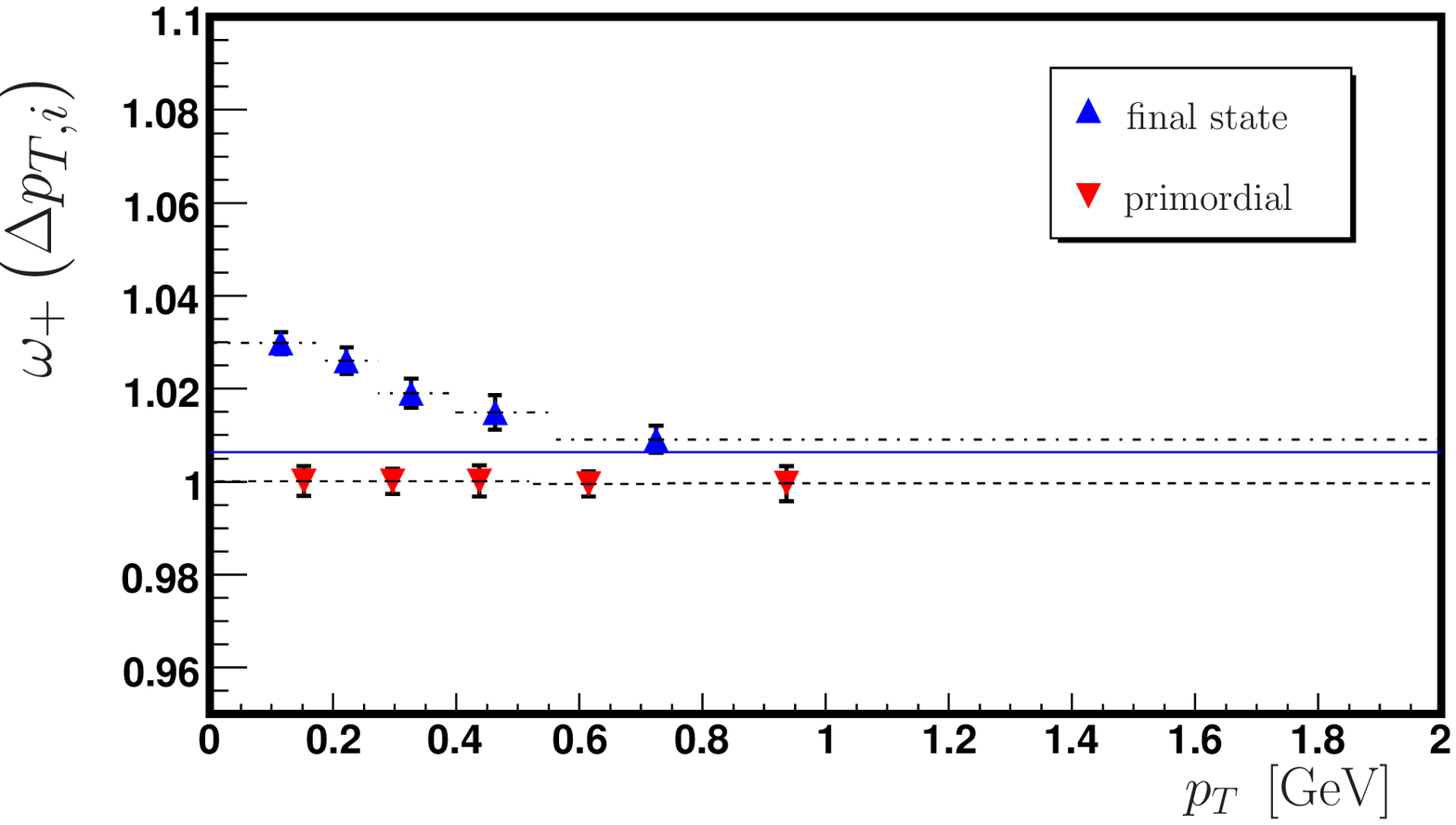,width=7.2cm,height=6.5cm}
  \epsfig{file=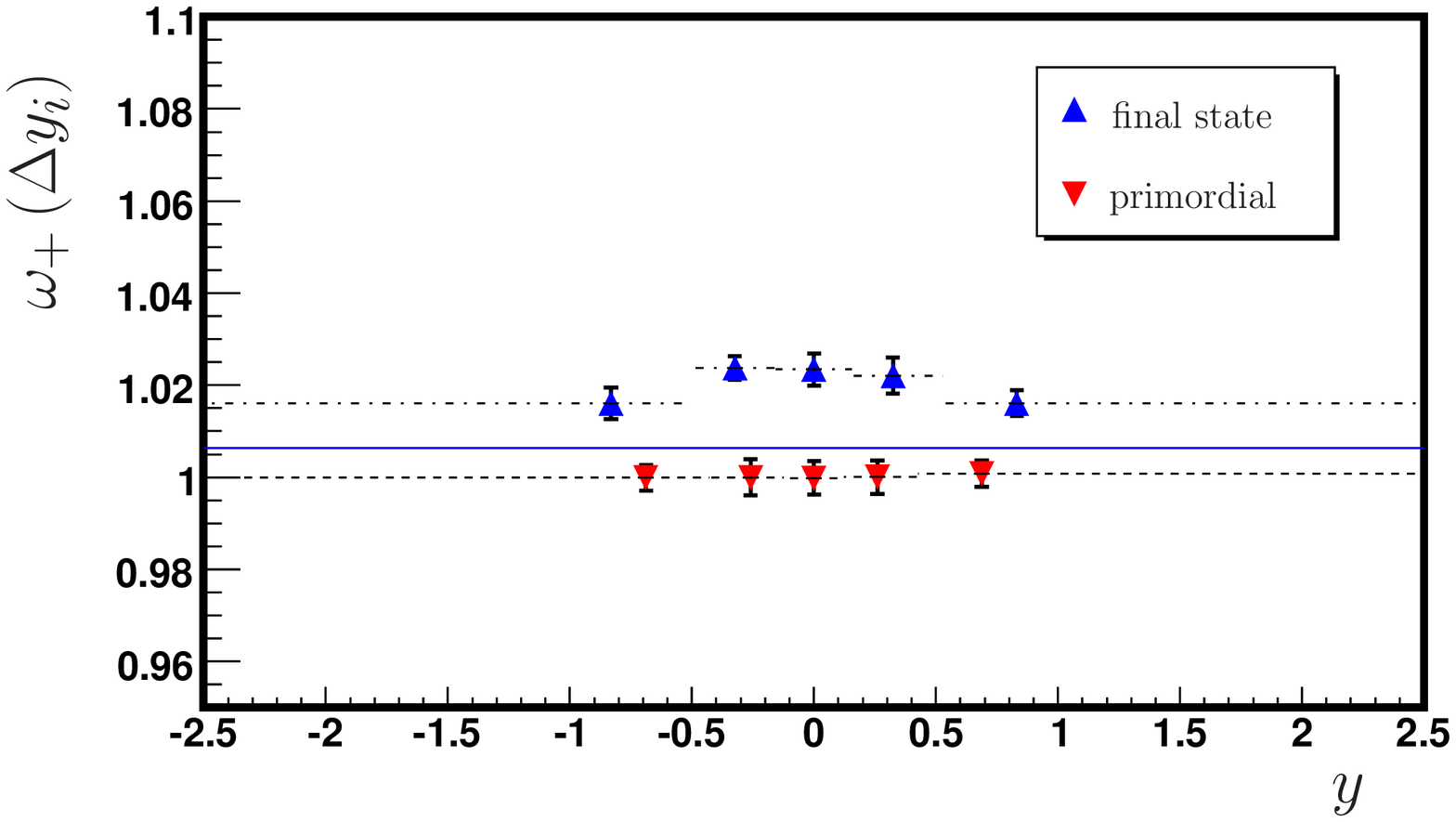,width=7.2cm,height=6.5cm}
  \caption{GCE scaled variance~$\omega_+$ of multiplicity fluctuations of positively charged hadrons, both primordial 
  and final state, in transverse momentum bins~$\Delta p_{T,i}$~({\it left}) and rapidity bins~$\Delta y_i$~({\it right}). 
  Horizontal error bars indicate the width and position of the momentum bins (And not an uncertainty!).
  Vertical error bars indicate the statistical uncertainty of~$20$ Monte Carlo runs of~$2\cdot 10^5$ events each.
  The markers indicate the center of gravity of the corresponding bin. 
  The solid line indicates the final state acceptance scaling estimate. }  
  \label{mult_pm_0000_omega}
\end{figure}

In Fig.(\ref{mult_pm_0000_omega}) the~$\Delta p_{T,i}$~({\it left}) and~$\Delta y_i$~({\it right}) dependence of the GCE 
scaled variance~$\omega_+$ of positively charged hadrons are shown, both primordial and final state. In the primordial 
Boltzmann case one finds no dependence of multiplicity fluctuations on the position and size of the acceptance window.
The observed multiplicity distribution is, within error bars, a Poissonian with scaled variance~$\omega_+=1$. In fact, 
in the primordial GCE Boltzmann case any selection of particles has~$\omega = 1$, as particle multiplicity,
as well as momenta, are sampled independently.

\begin{figure}[ht!]
  \epsfig{file=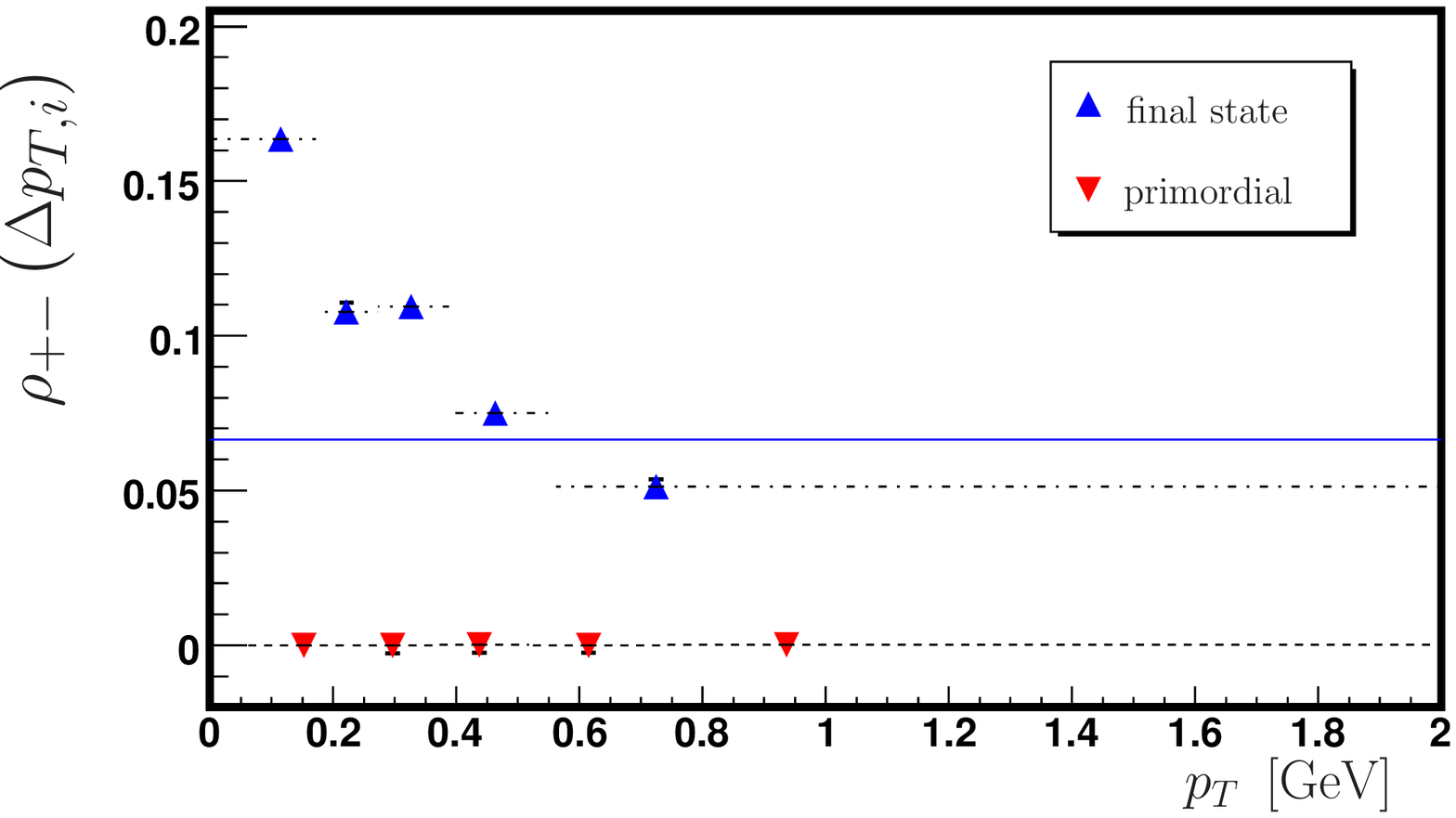,width=7.2cm,height=6.5cm}
  \epsfig{file=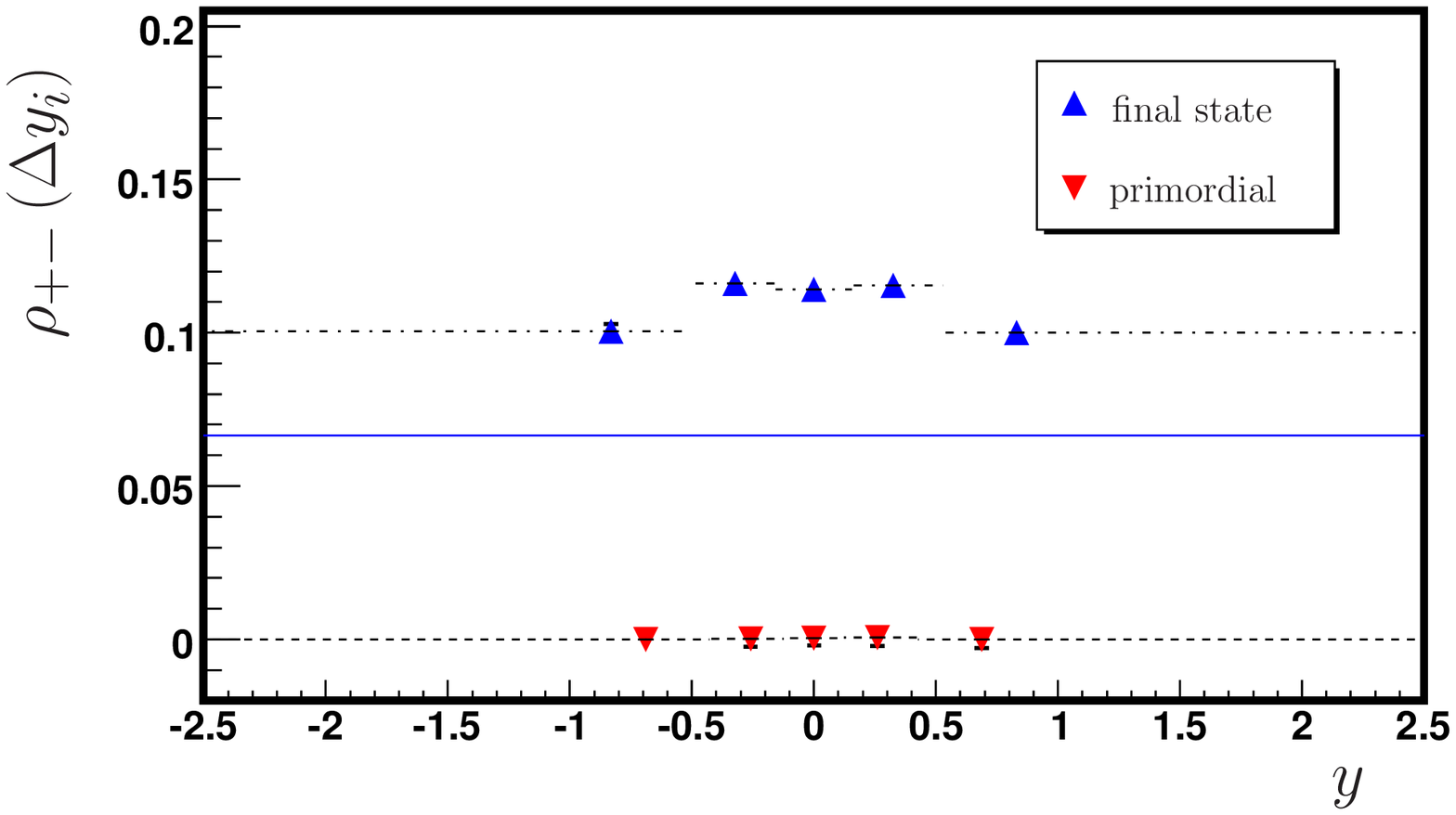,width=7.2cm,height=6.5cm}
  \caption{GCE multiplicity correlations~$\rho_{+-}$ between positively and negatively charged hadrons, both primordial 
  and final state, in transverse momentum bins~$\Delta p_{T,i}$~({\it left}) and rapidity bins~$\Delta y_i$~({\it right}). 
  The rest as in Fig.(\ref{mult_pm_0000_omega}).}  
  \label{mult_pm_0000_rho}
\end{figure}

\begin{figure}[ht!]
  \epsfig{file=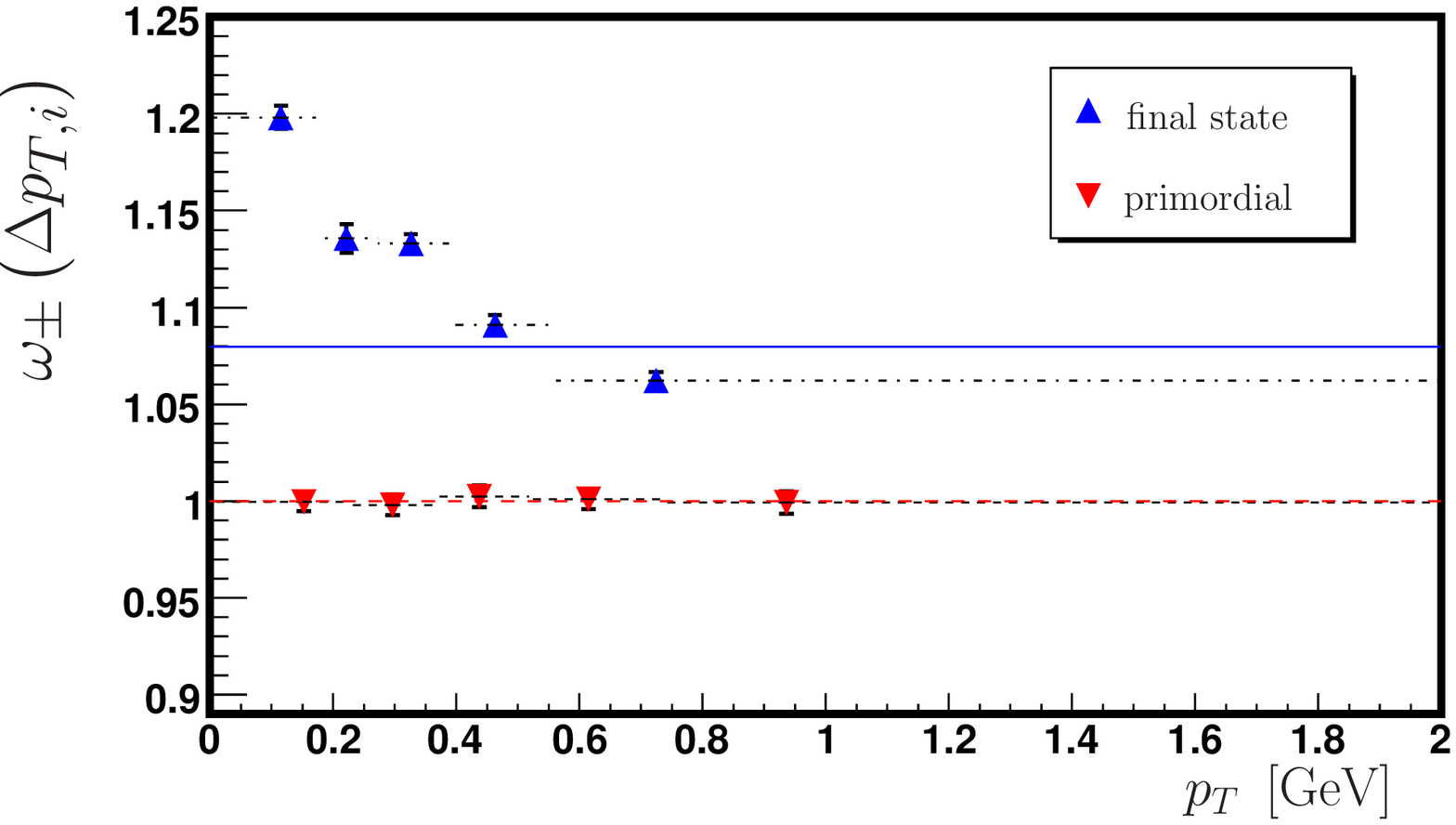,width=7.2cm,height=6.5cm}
  \epsfig{file=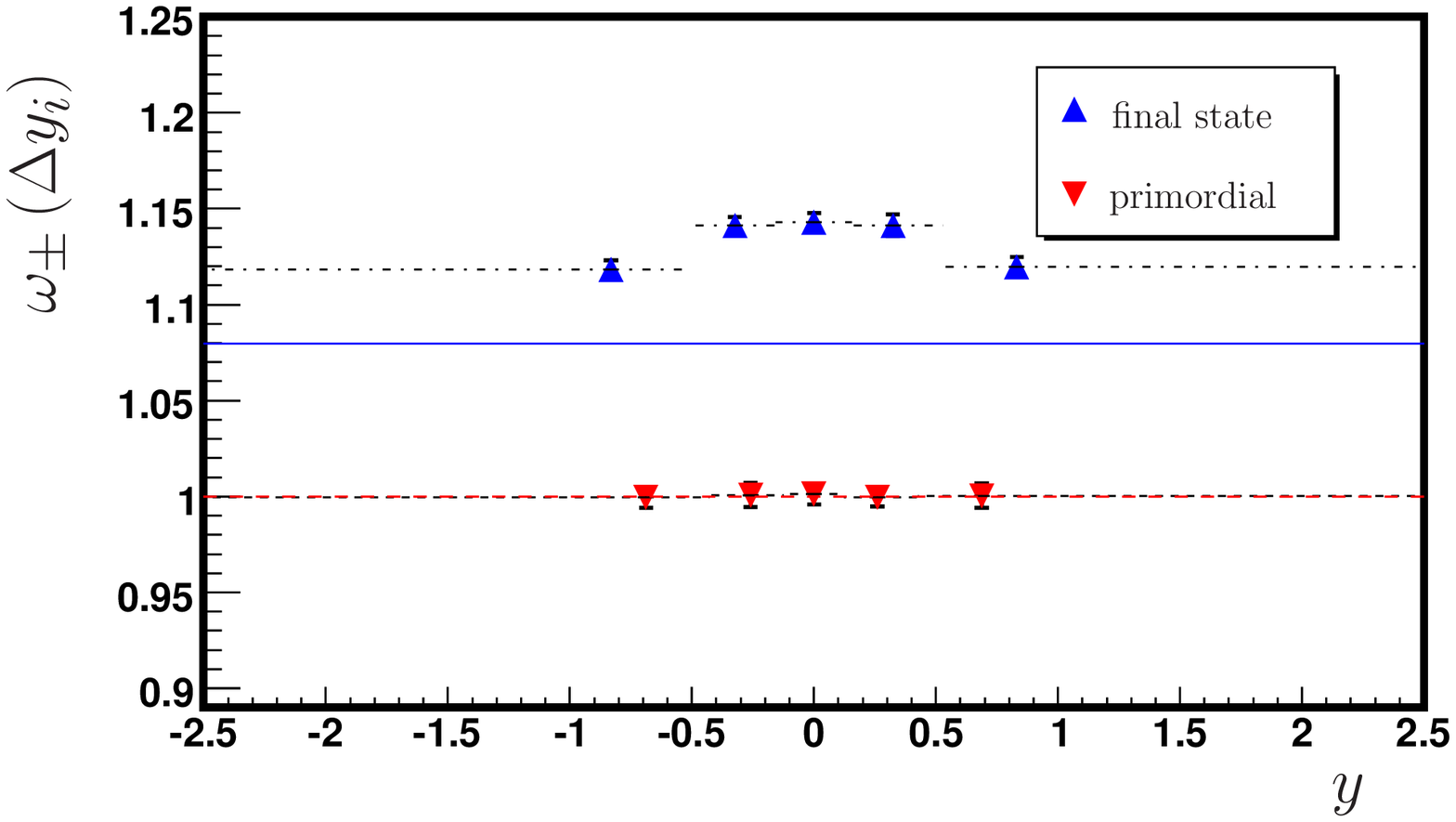,width=7.2cm,height=6.5cm}
  \caption{GCE scaled variance~$\omega_{\pm}$ of multiplicity fluctuations of all charged hadrons, both primordial and 
  final state, in transverse momentum bins~$\Delta p_{T,i}$~({\it left}) and rapidity bins~$\Delta y_i$~({\it right}). 
  The rest as in Fig.(\ref{mult_pm_0000_omega}). }  
  \label{mult_pm_0000_omega_CH}
\end{figure}

Fig.(\ref{mult_pm_0000_rho}) shows the~$\Delta p_{T,i}$~({\it left}) and~$\Delta y_i$~({\it right}) dependence of 
the GCE correlation coefficient~$\rho_{+-}$ between positively and negatively charged hadrons, both 
primordial and final state. In the primordial Boltzmann case one finds also no dependence of multiplicity correlations 
on the position and size of the acceptance window. The observed joint multiplicity distribution is a product of two 
Poissonians with correlation coefficient~$\rho_{+-}=0$. 

Fig.(\ref{mult_pm_0000_omega_CH}) presents the~$\Delta p_{T,i}$~({\it left}) and~$\Delta y_i$~({\it right}) dependence 
of the GCE scaled variance~$\omega_{\pm}$ of all charged hadrons, both primordial and final state. As the correlation
coefficient between positively and negatively charged hadrons is~${\rho_{+-}=0}$ in the primordial GCE one also 
finds~$\omega_{\pm} = \omega_+ = 1$.

Resonance decay is the only source of correlation in an ideal GCE Boltzmann gas. Neutral hadrons decaying into two 
hadrons of opposite electric charge are the strongest contributors to the correlation coefficient~$\rho_{+-}$. 
The chance that both (oppositely charged) decay products are dropped into the same momentum space bin is obviously 
highest at low transverse momentum (i.e. the correlation coefficient is strongest in~$\Delta p_{T,1}$).
The rapidity dependence is somewhat milder again, because heavier particles (parents) are dominantly produced around  
mid-rapidity and spread their daughter particles over a range in rapidity. One notes that the scaled variances and 
correlation coefficients in the respective acceptance bins in Figs.(\ref{mult_pm_0000_omega},\ref{mult_pm_0000_rho}) 
are generally larger than the acceptance scaling procedure\footnote{For the acceptance scaling approximation 
it is assumed that particles are randomly detected with a certain probability~$q=0.2$, independent of their momentum.} 
suggests, with the notable exception of~$\rho_{+-}(\Delta p_{T,5})$. 

There is a simple relation connecting the scaled variance of the fluctuations of all charged hadrons~$\omega_{\pm}$ to 
the fluctuations of only positively charged particles~$\omega_+$ via the correlation coefficient~$\rho_{+-}$ between 
positively and negatively charged hadrons in a neutral system:
\begin{equation}\label{omega_chr_eq}
\omega_{\pm} ~=~ \omega_+~\left(1~+~\rho_{+-}\right).
\end{equation}
The contribution of neutral parent particles to~$\omega_{\pm}$ is stronger than to~$\omega_+$, as both oppositely 
charged decay products go into the analysis rather than just one of them. Therefore, the effects of resonance decay 
on the~$\Delta p_{T,i}$ dependence of~$\omega_{\pm}$ are considerably enhanced compared to the one of~$\omega_+$, and 
generally~$\omega_{\pm} > \omega_+$, as the correlation coefficient~$\rho_{+-}$ remains positive in the final 
state GCE, in Fig.(\ref{mult_pm_0000_omega_CH}). Compared to this, the final state values of~$\omega_{\pm}$,~$\omega_+$ 
and~$\rho_{+-}$ remain rather flat with~$\Delta y_{i}$ in the GCE.

If one would construct now a larger and larger number of  momentum space bins of equal average particle multiplicities, 
one would successively lose more and more information about how multiplicities of distinct groups of particles are 
correlated, and approach the Poissonian limit also for final state particles.

\section{Micro Canonical Ensemble}
\label{sec_multfluc_hrg_mce}

In the very same way in which fully phase space integrated extensive quantities were extrapolated to the MCE limit 
in Chapter~\ref{chapter_extrapol}, now multiplicity fluctuations~$\omega_+$ and correlations~$\rho_{+-}$ in transverse 
momentum bins~$\Delta p_{T,i}$ and rapidity bins~$\Delta y_i$ for a hadron resonance gas are extrapolated from 
the GCE~$(\lambda =0)$ to the MCE ($\lambda \rightarrow 1$). Analytical primordial MCE results are obtained in the 
infinite volume approximation~\cite{Hauer:2007im,Hauer:2009nv}, providing some guidance as to asses the accuracy of the 
extrapolation scheme. For final state fluctuations and correlations in limited acceptance, on the other hand, 
no analytical results are available.

Mean values of particle numbers of positively charged hadrons~$\langle N_+ \rangle$ and negatively charged 
hadrons~$\langle N_- \rangle$ in the respective acceptance bins, defined in Table~\ref{accbins}, remain constant 
as~$\lambda$ goes from~$0$~to~$1$, while the variances~$\langle ( \Delta N_+)^2 \rangle$ 
and~$\langle ( \Delta N_-)^2 \rangle$, and covariance~$\langle \Delta N_+ \Delta N_- \rangle$ converge linearly to 
their respective MCE limits. The correlation coefficient~$\rho_{+-}$ between positively and negatively charged hadrons, 
on the other hand, will not approach its MCE value linearly, as discussed in Chapter~\ref{chapter_extrapol}.

\subsubsection{Primordial}
\label{none}

\begin{figure}[ht!]
  \epsfig{file=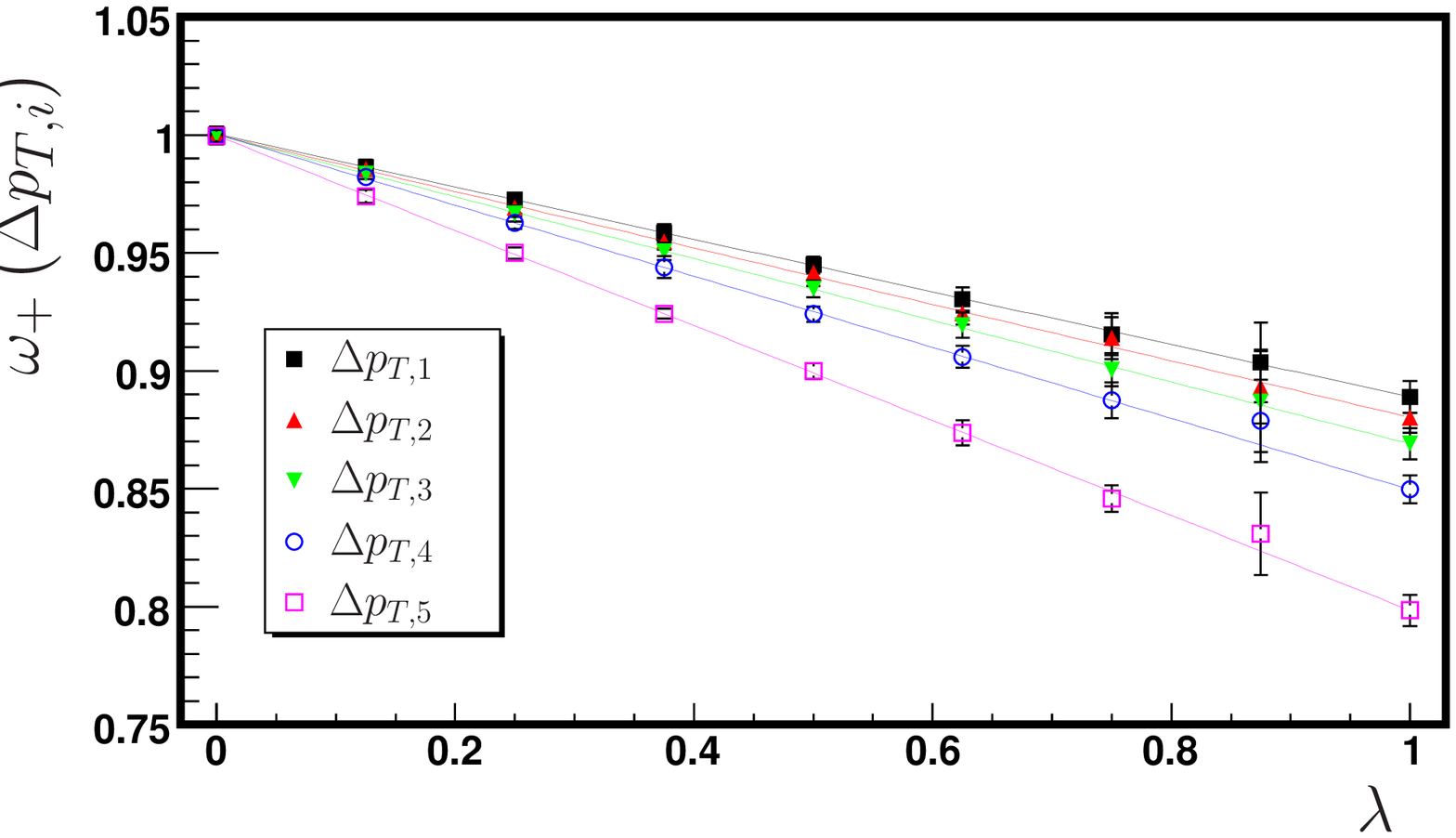,width=7.2cm,height=6.5cm}
  \epsfig{file=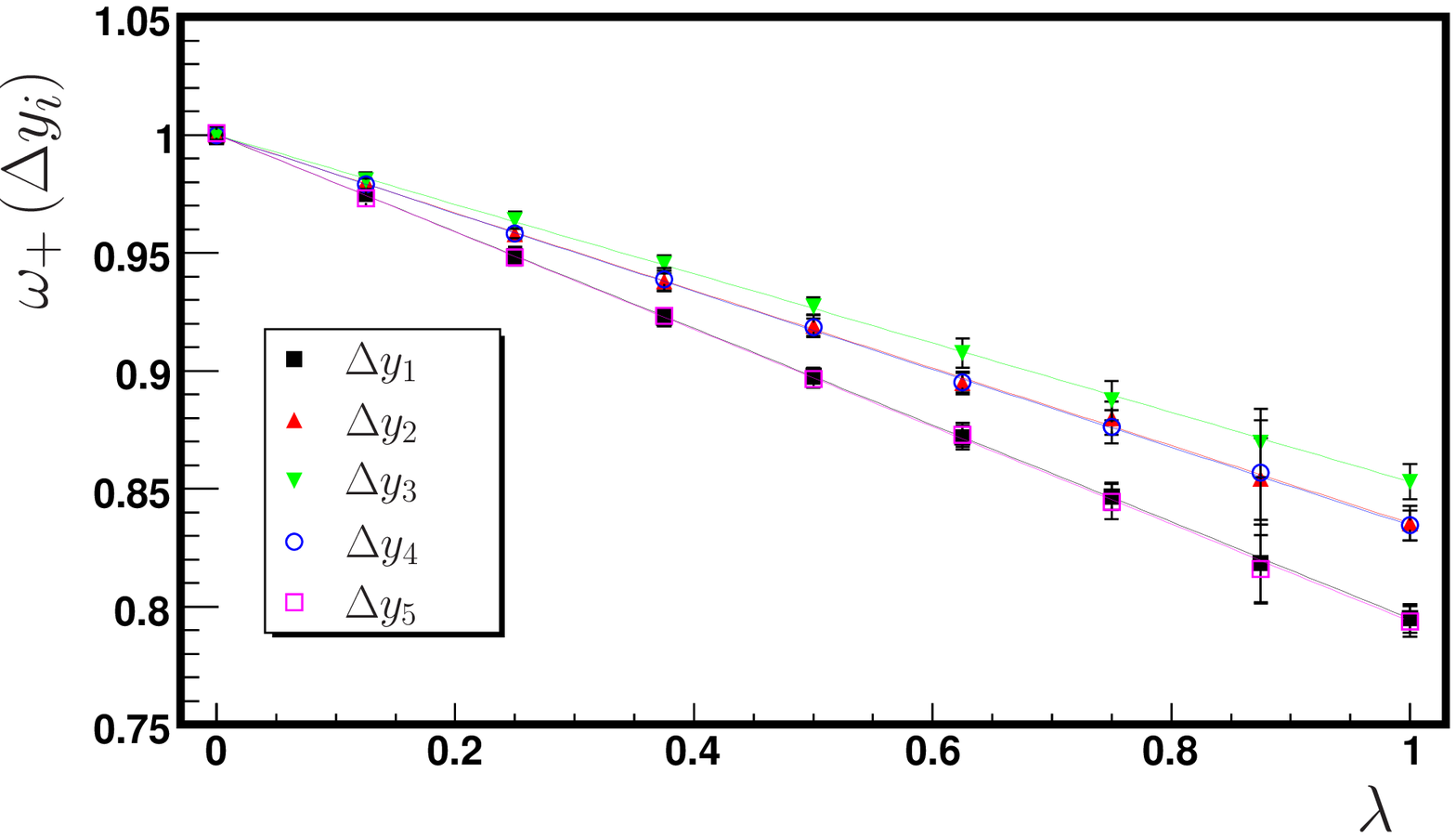,width=7.2cm,height=6.5cm}
  \caption{Evolution of the primordial scaled variance~$\omega_+$ of positively charged hadrons with the Monte Carlo 
  parameter~$\lambda = V_1/V_g$ for transverse momentum bins~$\Delta p_{T,i}$~({\it left}) and for rapidity 
  bins~$\Delta y_i$~({\it right}). The solid lines show an analytic extrapolation from GCE results ($\lambda =0$) to 
  the MCE limit ($\lambda \rightarrow 1$).
  Each marker and its error bar, except the last, represents the result of~$20$ Monte Carlo runs of~$2 \cdot 10^5$ events. 
  $8$ different equally spaced values of~$\lambda$ have been investigated.
  The last marker denotes the result of the extrapolation. }  
  \label{conv_lambda_omega_prim}
\end{figure}

\begin{figure}[ht!]
  \epsfig{file=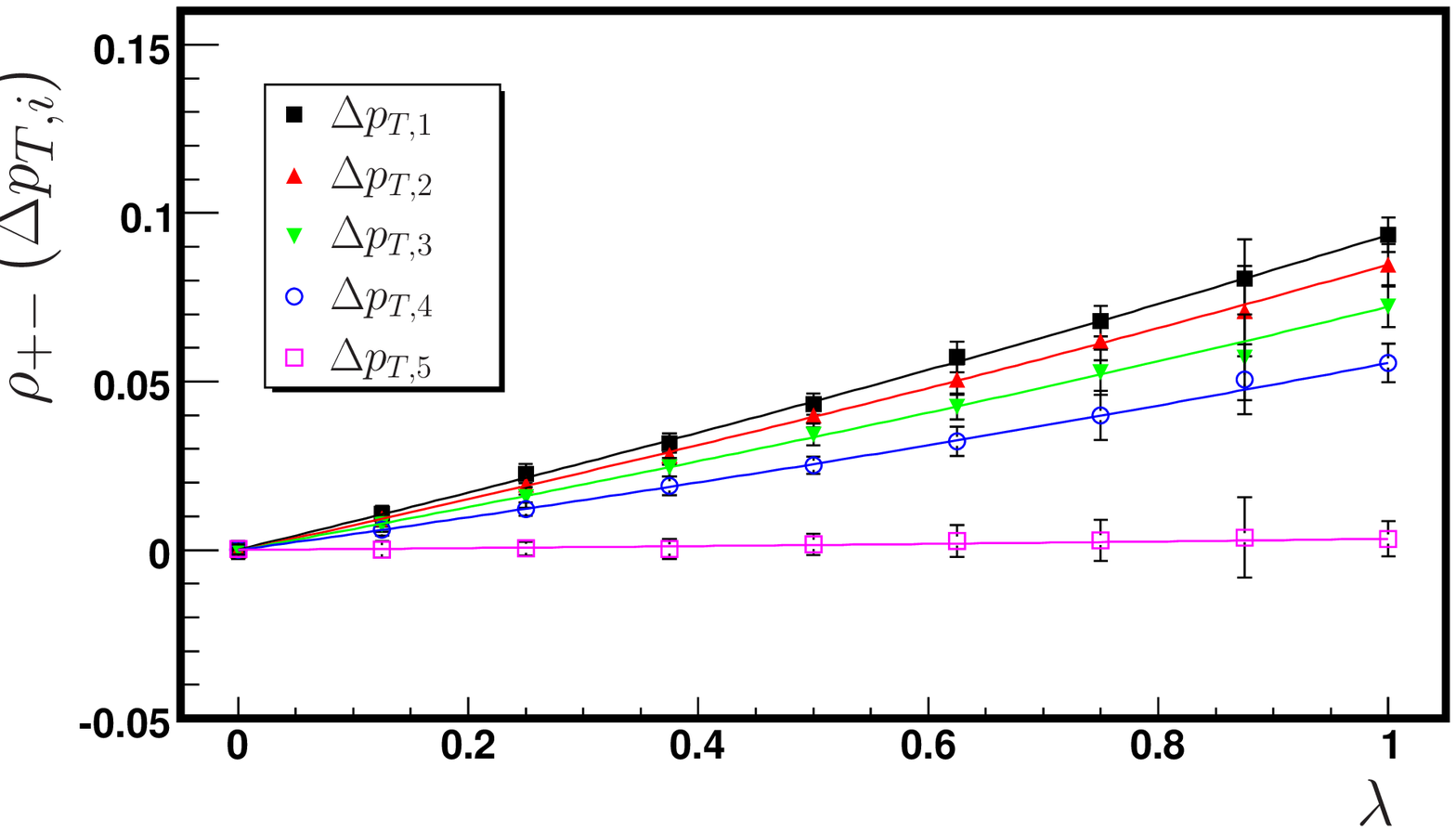,width=7.2cm,height=6.5cm}
  \epsfig{file=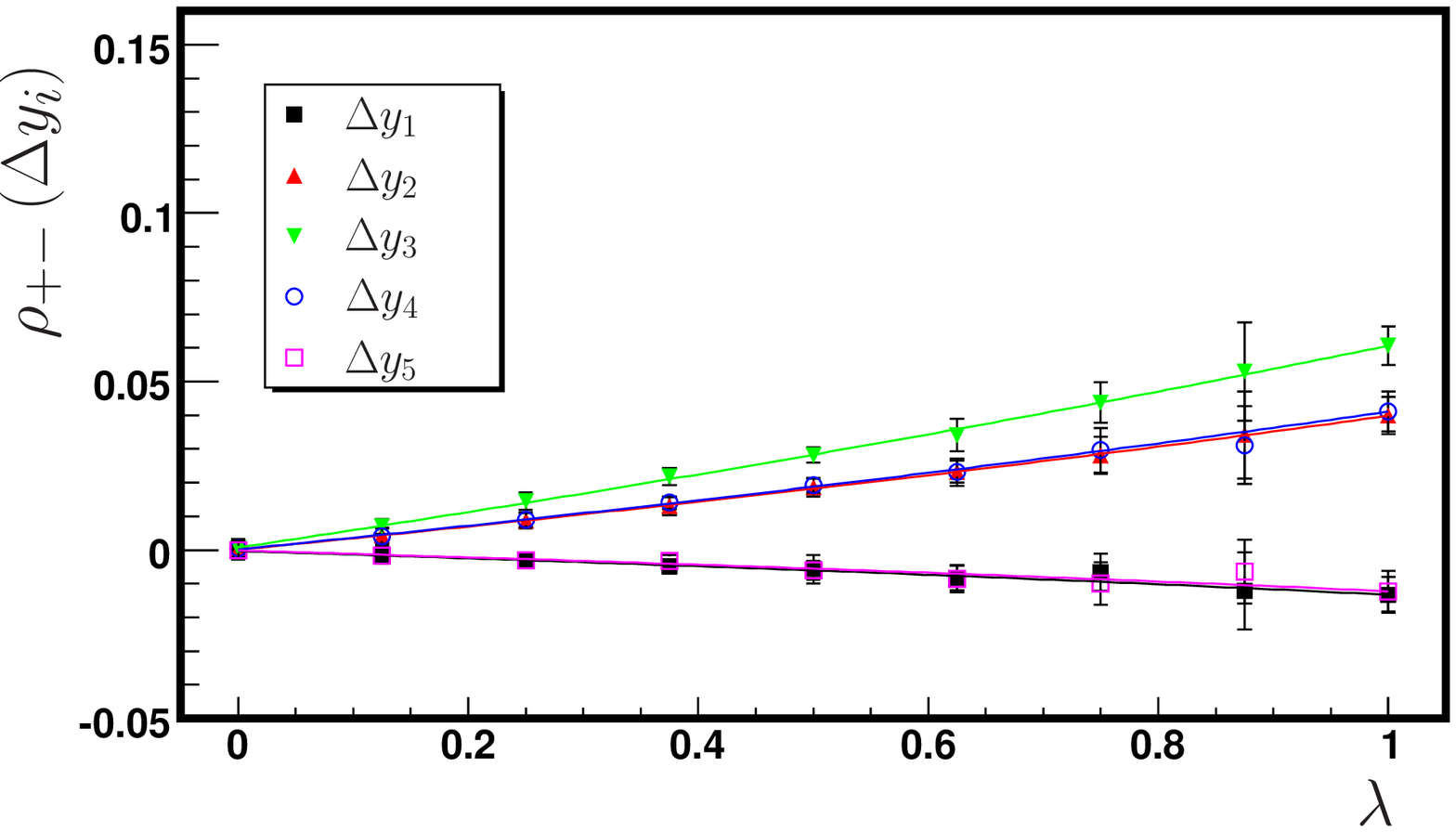,width=7.2cm,height=6.5cm}
  \caption{Evolution of the primordial correlation coefficient~$\rho_{+-}$ between positively and negatively charged 
  hadrons with the Monte Carlo parameter~$\lambda = V_1/V_g$ for transverse momentum bins~$\Delta p_{T,i}$~({\it left}) 
  and for rapidity bins~$\Delta y_i$~({\it right}).
  The rest as in Fig.(\ref{conv_lambda_omega_prim}).}  
  \label{conv_lambda_rho_prim}
\end{figure}

In Fig.(\ref{conv_lambda_omega_prim}) the primordial scaled variance~$\omega_+$ of positively charged hadrons in 
transverse momentum bins~$ \Delta p_{T,i}$~({\it left}) and rapidity bins~$\Delta y_i$~({\it right}) is shown as a 
function of the size of the bath~$\lambda= V_1/V_g$, while in Fig.(\ref{conv_lambda_rho_prim}) the dependence of 
the primordial correlation coefficient~$\rho_{+-}$ between positively and negatively charged hadrons in transverse 
momentum bins~$\Delta p_{T,i}$~({\it left}) and rapidity bins~$\Delta y_i$~({\it right}) is presented as a 
function of~$\lambda$.

The results of~$8 \cdot 20$ Monte Carlo runs of~$2 \cdot 10^5$ events each are summarized in 
Table~\ref{accbins_Mult_prim}. The system sampled was assumed to be neutral~$\mu_j = (0,0,0)$ and 
static~$u_{\mu}=(1,0,0,0)$ with local temperature~$\beta^{-1}=0.160$~GeV and a system volume of~$V_1=2000$~fm$^3$. 
8 different values of~$\lambda$ have been studied. The last marker~$(\lambda = 1)$ denotes the result of the 
extrapolation. Only primordial hadrons are analyzed. Values for both~$\Delta p_{T,i}$ and~$\Delta y_i$ bins are listed. 
Analytical numbers  are calculated according to the method developed in~\cite{Hauer:2007im,Hauer:2009nv}, using the 
acceptance bins defined in Table~\ref{accbins}, and are shown for comparison.
The effects of energy-momentum and charge conservation on primordial multiplicity fluctuations and correlations 
in finite acceptance will also be discussed in Chapter~\ref{chapter_multfluc_pgas}.

Firstly fully phase space integrated results, also later discussed in Chapters~\ref{chapter_phasediagram} 
and~\ref{chapter_freezeoutline}, are attended to. The scaled variance of multiplicity fluctuations is lowest in the 
MCE due to the requirement of exact energy and charge conservation, somewhat larger in the CE (see also 
Section~\ref{sec_multfluc_hrg_mce}), and largest in the GCE, as now all constraints on the micro states of the system 
have been dropped~\cite{Begun:2006jf,Begun:2006uu,Hauer:2007ju}. The fully phase space integrated MCE and CE correlation 
coefficients between oppositely charged particles are rather close to~$1$. Doubly charged particles allow for mild 
deviation, as also the~$\Delta^{++}$ resonance is counted as only one particle. 

\begin{table}[h!]\footnotesize
  \begin{center}
    \begin{tabular}{cccccc}
    \toprule
      \!\!Primordial\!\!\!\! & 
      $\Delta p_{T,1}$ & $\Delta p_{T,2}$ & $\Delta p_{T,3}$ & $\Delta p_{T,4}$ & $\Delta p_{T,5}$   \\
    \midrule
      $\omega^{gce}_+$ & 
      $1.000 \pm 0.002$\!\! & 
      $1.000 \pm 0.002$\!\! & $1.000 \pm 0.002$\!\! & 
      $1.000 \pm 0.002$\!\! & $1.000 \pm 0.002$\!\! \\
      $\omega^{mce}_+$ & 
      $0.889 \pm 0.007$\!\! & 
      $0.880 \pm 0.007$\!\! & $0.869 \pm 0.007$\!\! & 
      $0.850 \pm 0.006$\!\! & $0.798 \pm 0.007$\!\! \\
      $\omega^{mce,c}_+$ & $0.8886$ & $0.8802$ & $0.8682$ & $0.8489$ & $0.7980$ \\
    \midrule
      $\rho^{gce}_{+-}$ & 
      $0.000 \pm 0.002$\!\! & 
      $-0.000 \pm 0.002$\!\! & $-0.000 \pm 0.002$\!\! & 
      $0.000 \pm 0.002$\!\! & $0.000 \pm 0.001$\!\! \\
      $\rho^{mce}_{+-}$\!\! & 
      $0.094 \pm 0.005$\!\! & 
      $0.085 \pm 0.006$\!\! & $0.072 \pm 0.006$\!\! & 
      $0.056 \pm 0.006$\!\! & $0.003 \pm 0.005$\!\! \\
      $\rho^{mce,c}_{+-}$ & $0.0935$ & $0.0844$ & $0.0730$ & $0.0554$ & $0.0040$ \\
    \midrule[0.09em]
      \!\!Primordial\!\!\!\! & $\Delta y_1$  & $\Delta y_2$  & $\Delta y_3$  & $\Delta y_4$  
      & $\Delta y_5$ \\
    \midrule 
      $\omega^{gce}_+$ & 
      $1.000 \pm 0.002$\!\! & 
      $1.000 \pm 0.002$\!\! & $1.000 \pm 0.003$\!\! & 
      $1.000 \pm 0.002$\!\! & $1.000 \pm 0.002$\!\! \\
      $\omega^{mce}_+$ & 
      $0.795 \pm 0.006$\!\! & 
      $0.835 \pm 0.007$\!\! & $0.853 \pm 0.008$\!\! & 
      $0.834 \pm 0.006$\!\! & $0.794 \pm 0.007$\!\! \\
      $\omega^{mce,c}_+$ & $0.7950$ & $0.8350$ & $0.8521$ & $0.8351$ & $0.7949$\\
    \midrule 
      $\rho^{gce}_{+-}$ & 
      $-0.000 \pm 0.001$\!\! & 
      $0.000 \pm 0.002$\!\! & $0.001 \pm 0.002$\!\! & 
      $0.000 \pm 0.002$\!\! & $-0.000 \pm 0.002$\!\! \\
      $\rho^{mce}_{+-}$ & 
      $-0.013 \pm 0.005$\!\! & 
      $0.040 \pm 0.006$\!\! & $0.061 \pm 0.006$\!\! & 
      $0.041 \pm 0.006$\!\! & $-0.012 \pm 0.006$\!\! \\
      $\rho^{mce,c}_{+-}$ & $-0.0135$ & $0.0406$ & $0.0616$ & $0.0406$ & $-0.0135$ \\
    \bottomrule
    \end{tabular}
    \caption{Summary of the primordial scaled variance~$\omega_+$ of positively charged hadrons and the correlation 
    coefficient~$\rho_{+-}$ between positively and negatively charged hadrons in transverse momentum 
    bins~$\Delta p_{T,i}$ and rapidity bins~$\Delta y_i$. 
    Both the GCE result ($\lambda = 0$) and the extrapolation to MCE ($\lambda = 1$) are shown.
    The uncertainty quoted corresponds to~$20$ Monte Carlo runs of~$2 \cdot 10^5$ events (GCE) or is the result 
    of the extrapolation (MCE). Analytic MCE results~$\omega^{mce,c}_+$ and~$\rho^{mce,c}_{+-}$ are listed too.} 
    \label{accbins_Mult_prim}
  \end{center}
\end{table}

The transverse momentum dependence can be understood as follows: 
a change in particle number at high transverse momentum involves a large amount of energy, i.e., in order to balance 
the energy record, one needs to create (or annihilate) either a lighter particle with more kinetic energy, or two 
particles at lower~$p_T$. This leads to suppressed multiplicity fluctuations in high~$\Delta p_{T,i}$ bins compared to
low~$\Delta p_{T,i}$ bins. By the same argument, it seems favorable, due to the constraint of energy and charge 
conservation, to balance electric charge, by creating (or annihilating) pairs of oppositely charged particles,
predominantly in lower~$\Delta p_{T,i}$ bins, while allowing for a more un-correlated multiplicity distribution, i.e. 
also larger net-charge ($\delta Q = N_+-N_-$) fluctuations, in higher~$\Delta p_{T,i}$ bins. 

For the rapidity  dependence similar arguments hold. Here, however, the strongest role is played by longitudinal 
momentum conservation. A change in particle number at high~$y$ involves now, in addition to a large amount of energy, 
a large momentum~$p_z$ to be balanced. The constraints of global~$P_z$ conservation are, hence, felt least severely 
around~$|y| \sim 0$, and it becomes favorable to balance charge predominantly at mid-rapidity ($\rho_{+-}$ larger) and 
to allow for stronger multiplicity fluctuations ($\omega_+$ larger) compared to forward and backward rapidity bins.

In a somewhat casual way one could say: events of a neutral hadron resonance gas with values of extensive 
quantities~$B$,~$S$,~$Q$,~$E$ and~$P_z$ in the vicinity of~$\langle \mathcal{Q}_1^l \rangle$ have a tendency to have 
similar numbers of positively and negatively charged particles at low transverse momentum~$p_T$ and rapidity~$y$ 
and less strongly so at high~$p_T$ and~$|y|$.

\subsubsection{Final State}
\label{none}

Now the extrapolation of final state multiplicity fluctuations and correlations to the MCE limit is attended to. 
An independent Monte Carlo run for the same physical system was done, but now with only stable final state 
particles `detected'.

\begin{figure}[ht!]
  \epsfig{file=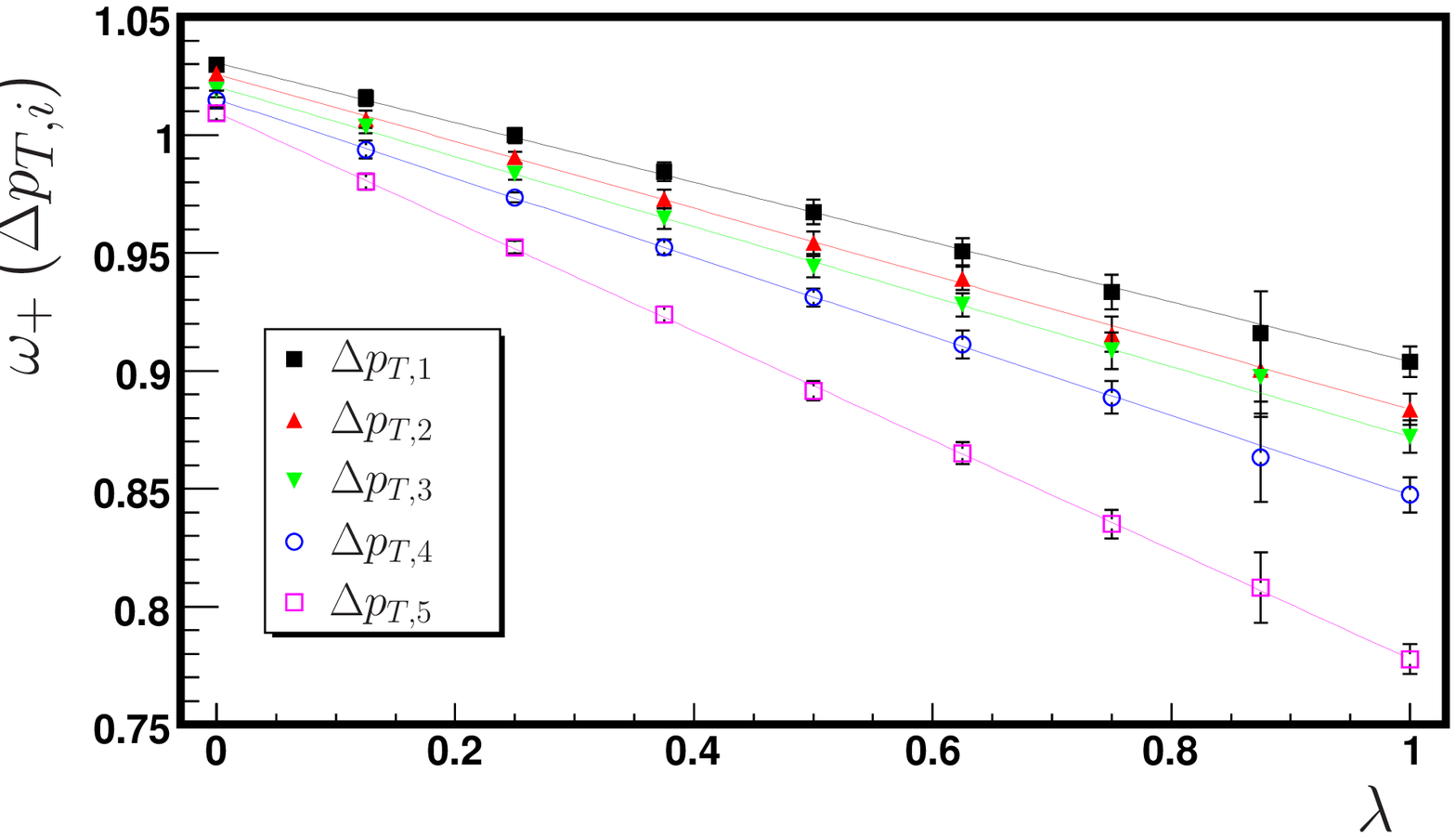,width=7.2cm,height=6.5cm}
  \epsfig{file=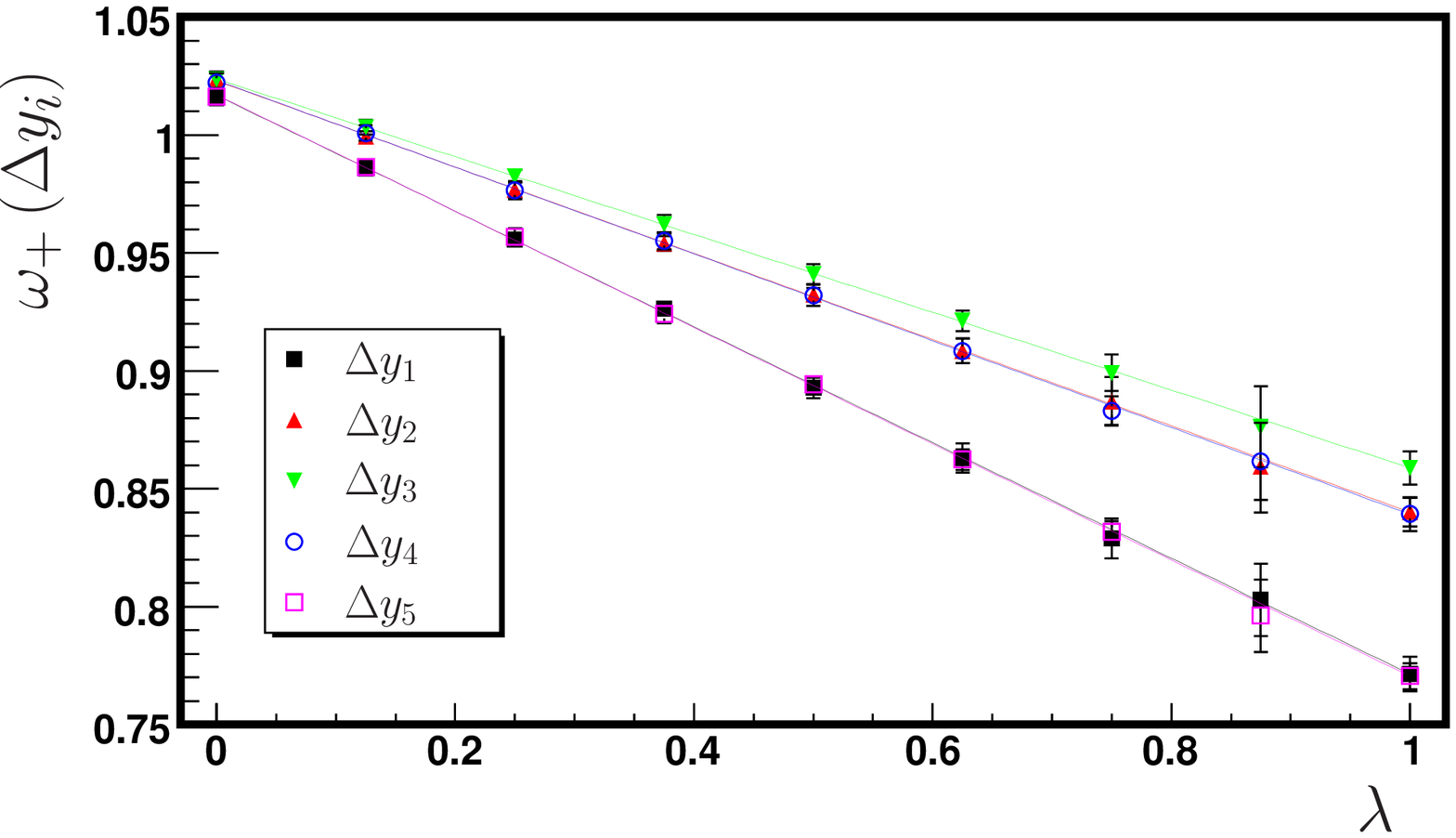,width=7.2cm,height=6.5cm}
  \caption{Evolution of the final state scaled variance~$\omega_+$ of positively charged hadrons with the Monte Carlo 
  parameter~$\lambda = V_1/V_g$ for transverse momentum bins~$\Delta p_{T,i}$~({\it left}) and for rapidity 
  bins~$\Delta y_i$~({\it right}). The solid lines show an analytic extrapolation from GCE results ($\lambda =0$) to 
  the MCE limit ($\lambda \rightarrow 1$).
  Each marker, except the last, represents the result of~$20$ Monte Carlo runs of~$2 \cdot 10^5$ events. 
  $8$~different equally spaced values of~$\lambda$ have been investigated.
  The last marker denotes the result of the extrapolation.}  
  \label{conv_lambda_omega_final}
\end{figure}

\begin{figure}[ht!]
  \epsfig{file=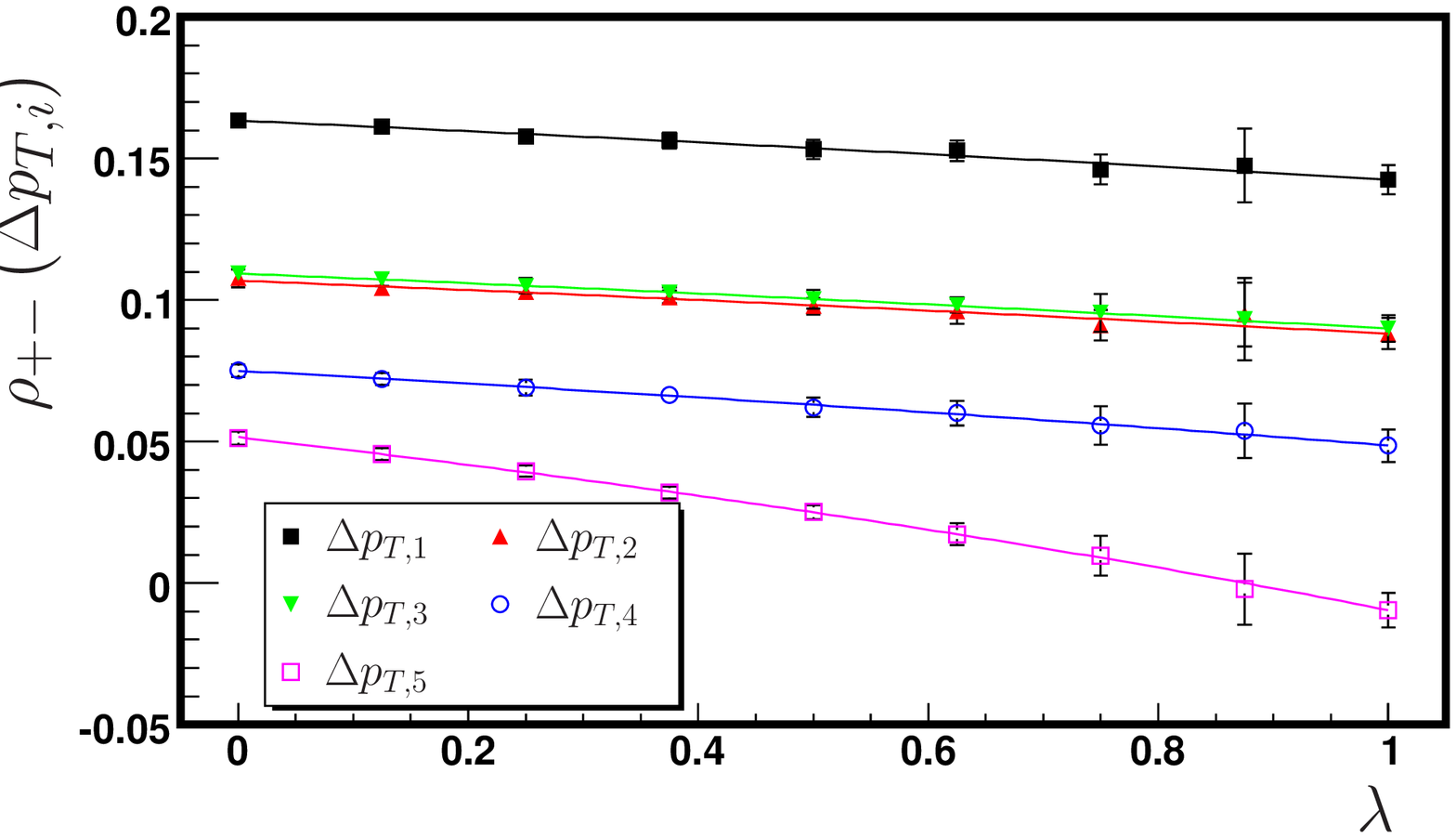,width=7.2cm,height=6.5cm}
  \epsfig{file=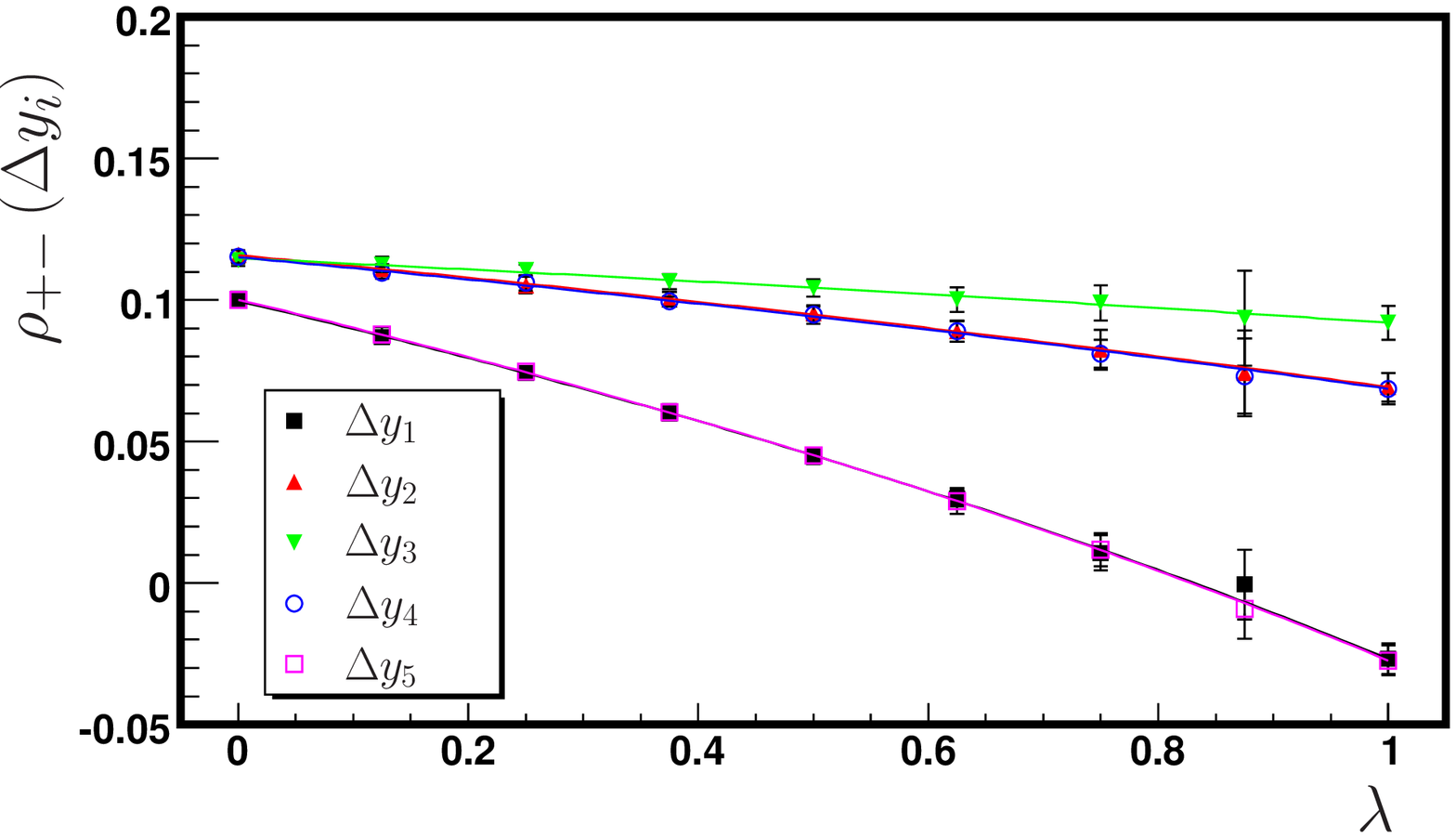,width=7.2cm,height=6.5cm}
  \caption{Evolution of the final state correlation coefficient~$\rho_{+-}$ between positively and negatively charged 
  hadrons with the Monte Carlo parameter~$\lambda = V_1/V_g$ for transverse momentum bins~$\Delta p_{T,i}$~({\it left}) 
  and for rapidity bins~$\Delta y_i$~({\it right}). The rest as in Fig.(\ref{conv_lambda_omega_final}).}  
  \label{conv_lambda_rho_final}
\end{figure}

In Fig.(\ref{conv_lambda_omega_final}) the final state scaled variance~$\omega_+$ of positively charged hadrons in 
transverse momentum bins~$\Delta p_{T,i}$~({\it left}) and rapidity bins~$\Delta y_i$~({\it right}) is shown as a 
function of~$\lambda$, while in Fig.(\ref{conv_lambda_rho_final}) the dependence of the final state correlation
coefficient~$\rho_{+-}$ between positively and negatively charged hadrons in transverse momentum 
bins~$\Delta p_{T,i}$~({\it left}) and rapidity bins~$\Delta y_i$~({\it right}) on the size of the 
bath~$\lambda = V_1 / V_g$ is shown.

The~$\Delta p_{T,i}$ and~$\Delta y_i$ dependence on~$\lambda$ of the final state MCE scaled variance~$\omega_+$ is 
qualitatively similar to that of the primordial versions, Fig.(\ref{conv_lambda_omega_prim}), and is essentially 
also explained by the arguments of the previous section. The effects of charge and energy-momentum conservation work very 
much the same way as before, and it still seems favorable to have events with wider multiplicity distributions at 
low~$p_T$ and low~$y$, and more narrow distributions at larger~$p_T$ and larger~$|y|$.  The dependence of the final state 
correlation coefficients~$\rho_{+-}$ on~$\lambda$, Fig.(\ref{conv_lambda_rho_final}), is a slightly different to the 
primordial case, Fig.(\ref{conv_lambda_rho_prim}). However, in the MCE limit, events still tend to have more similar 
numbers of oppositely charged particles at low~$p_T$ and low~$y$, than at large~$p_T$ and large~$|y|$.

The effects of resonance decay are qualitatively different in the MCE, CE, and GCE (see also 
Section~\ref{sec_multfluc_hrg_ce}). Again, firstly attending to fully phase space integrated multiplicity fluctuations 
discussed in~\cite{Begun:2006uu,Begun:2006jf} and Chapter~\ref{chapter_freezeoutline}. The final state scaled variance 
increases in the GCE and CE compared to the primordial scaled variance. Multiplicity fluctuations of neutral mesons 
remain unconstrained by (charge) conservation laws. 
However, they often decay into oppositely charged particles, which increases 
multiplicity fluctuations of pions, for instance. In the MCE, due to the constraint of energy conservation, the 
event-by-event fluctuations of primordial pions are correlated to the event-by-event fluctuations of, in general, 
primordial parent particles, and~$\omega^{final}< \omega^{prim}$ is possible in the MCE.

\begin{figure}[ht!]
  \epsfig{file=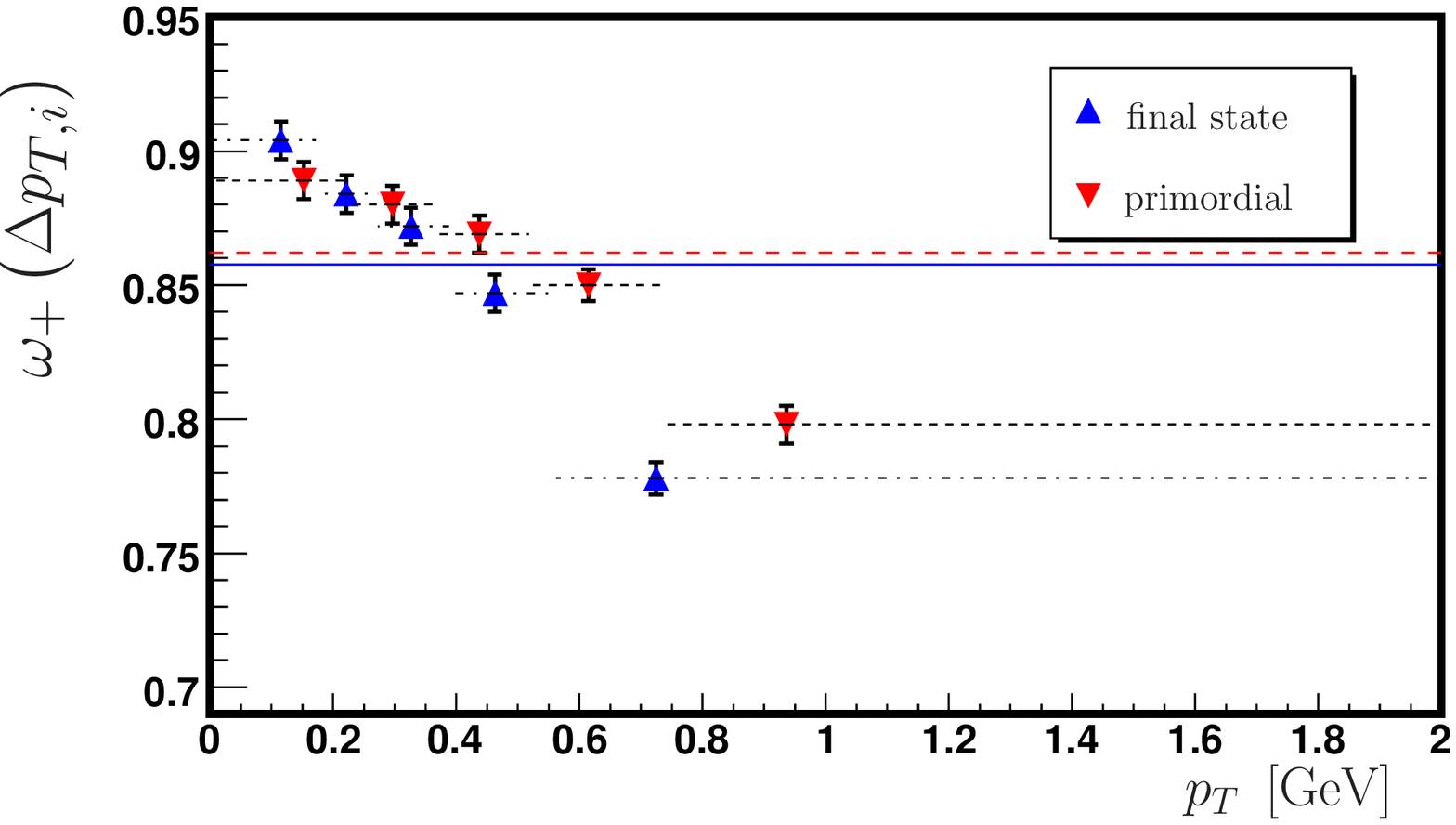,width=7.2cm,height=6.5cm}
  \epsfig{file=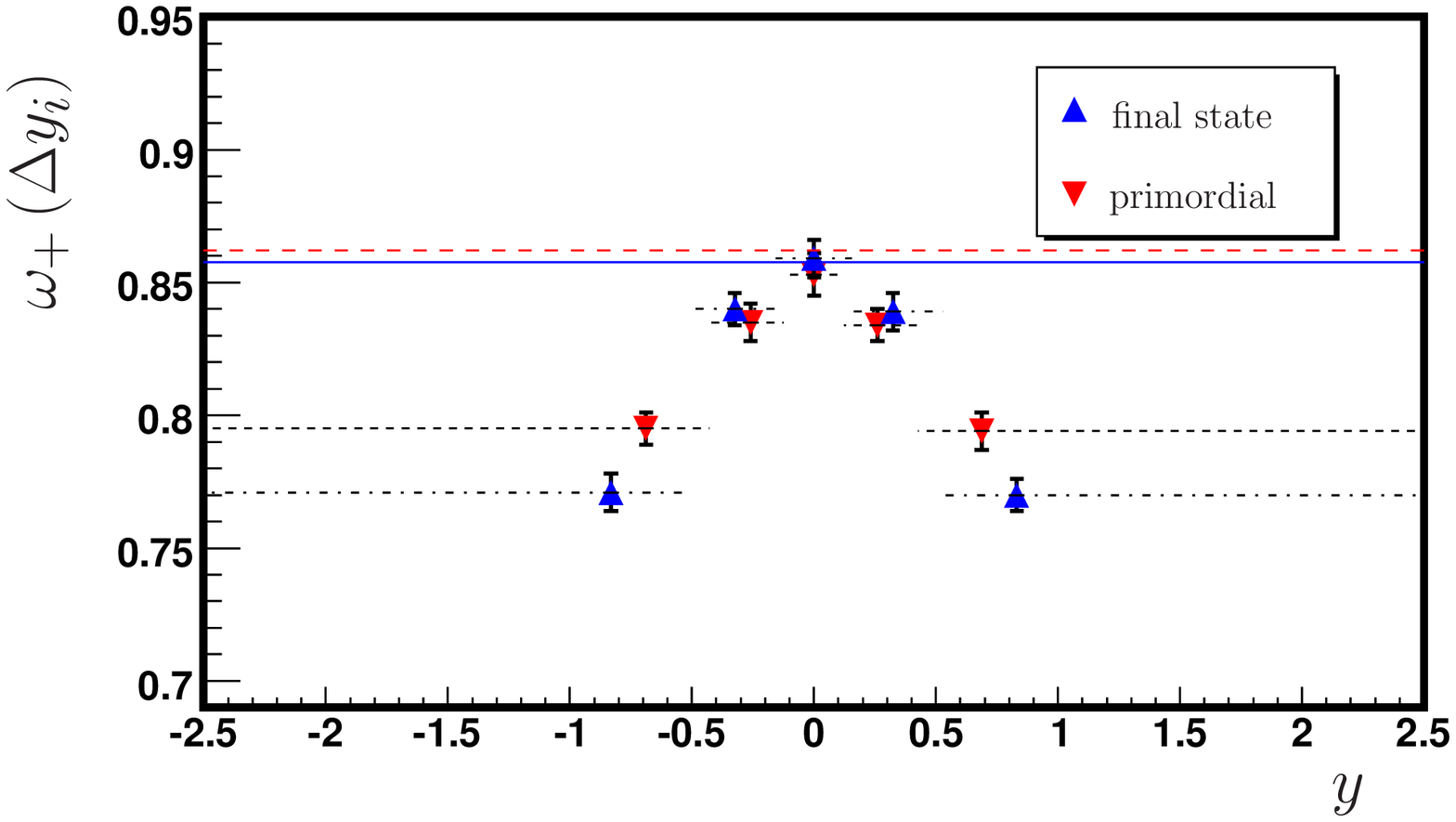,width=7.2cm,height=6.5cm}
  \caption{MCE scaled variance~$\omega_+$ of multiplicity fluctuations of positively charged hadrons, both primordial 
  and final state, in transverse momentum bins~$\Delta p_{T,i}$~({\it left}) and rapidity bins~$\Delta y_i$~({\it right}). 
  Horizontal error bars indicate the width and position of the momentum bins (And not an uncertainty!).
  Vertical error bars indicate the statistical uncertainty quoted in Table~\ref{accbins_Mult_final}.
  The markers indicate the center of gravity of the corresponding bin.
  The solid and the dashed lines show final state and primordial acceptance scaling estimates respectively.}
  \label{mult_pm_MCE_omega}
\end{figure}

\begin{figure}[ht!]
  \epsfig{file=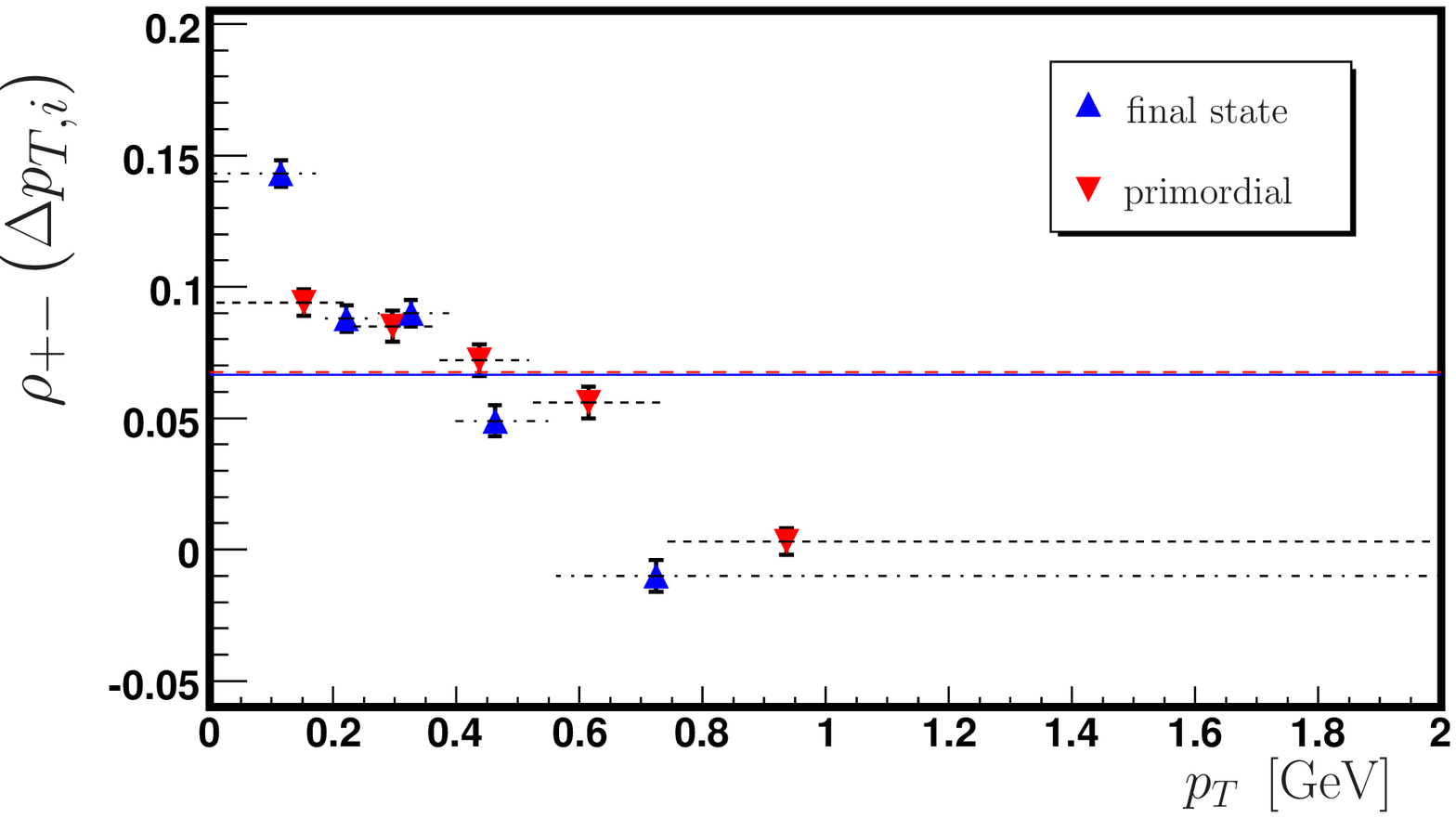,width=7.2cm,height=6.5cm}
  \epsfig{file=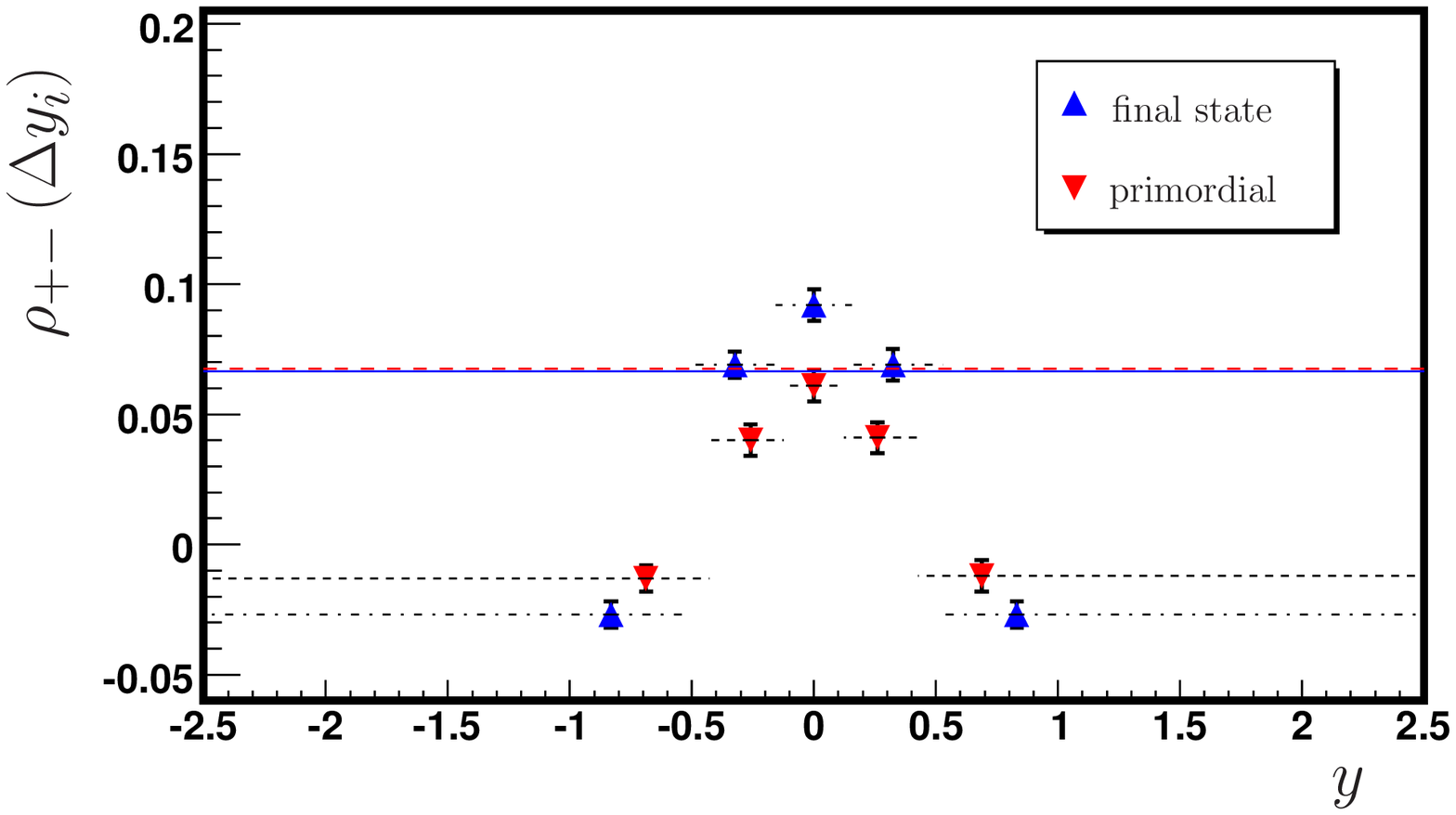,width=7.2cm,height=6.5cm}
  \caption{MCE multiplicity correlation coefficient~$\rho_{+-}$ between positively and negatively charged hadrons, 
  both primordial and final state, in transverse momentum bins~$\Delta p_{T,i}$~({\it left}) and rapidity 
  bins~$\Delta y_i$~({\it right}). The rest as in Fig.(\ref{mult_pm_MCE_omega}). }  
  \label{mult_pm_MCE_rho}
\end{figure}

\begin{figure}[ht!]
  \epsfig{file=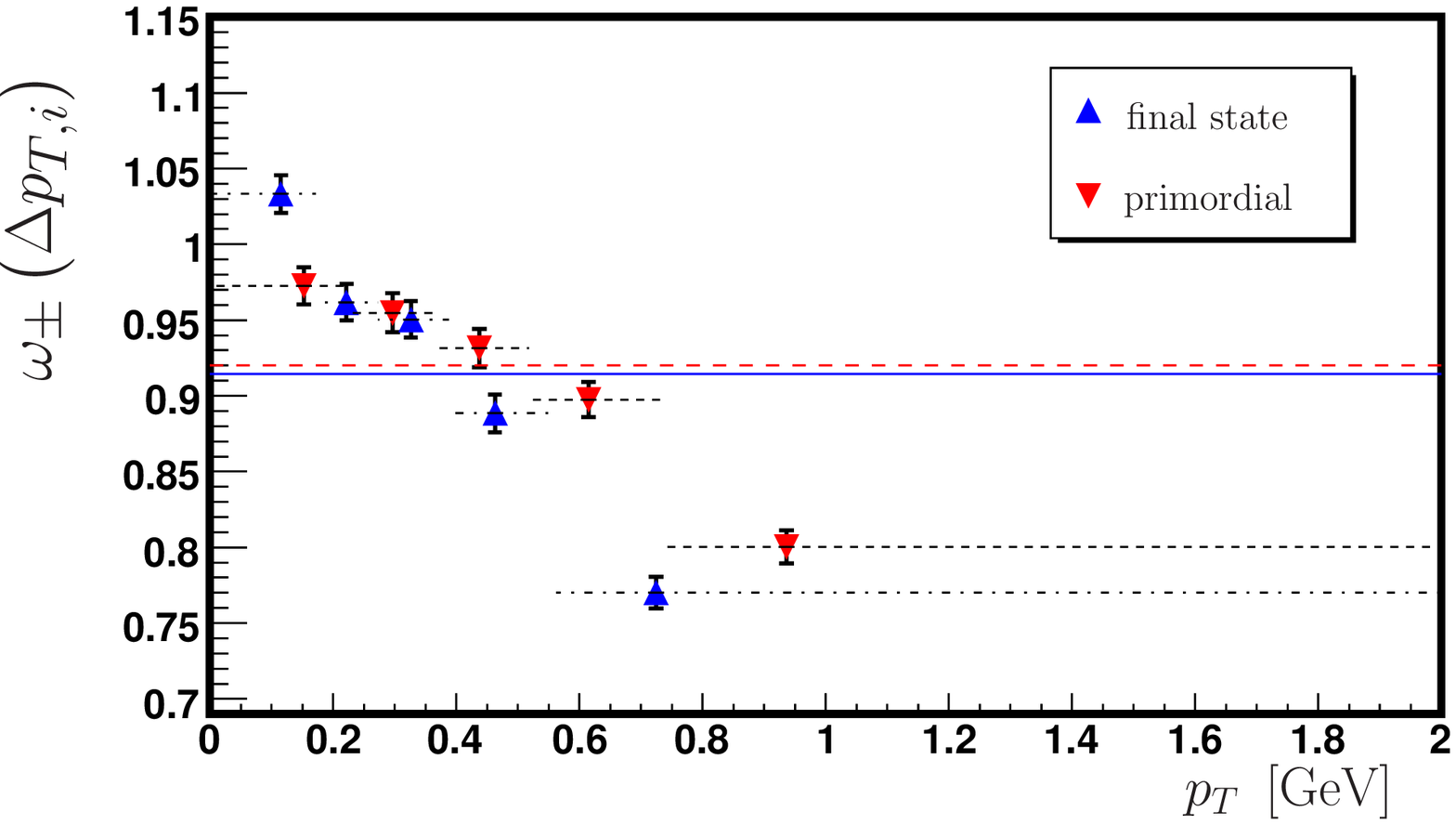,width=7.2cm,height=6.5cm}
  \epsfig{file=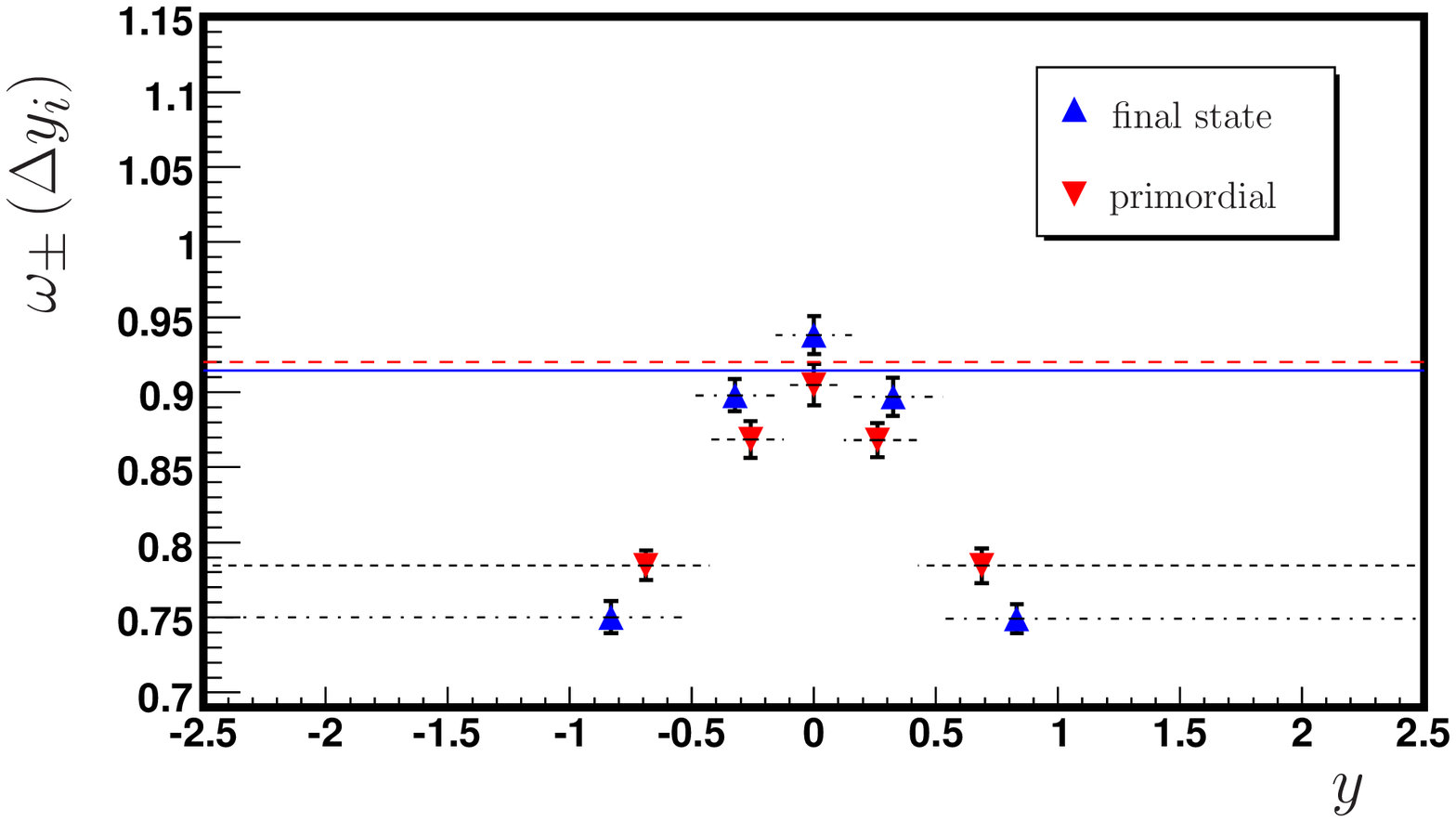,width=7.2cm,height=6.5cm}
  \caption{MCE scaled variance~$\omega_{\pm}$ of multiplicity fluctuations of all charged hadrons, both primordial and 
  final state, in transverse momentum bins~$\Delta p_{T,i}$~({\it left}) and rapidity bins~$\Delta y_i$~({\it right}). 
  The rest as in Fig.(\ref{mult_pm_MCE_omega}).}  
  \label{mult_pm_MCE_omega_CH}
\end{figure}

Figs.(\ref{mult_pm_MCE_omega}-\ref{mult_pm_MCE_omega_CH}) compare the final state~$\Delta p_{T,i}$~({\it left}) 
and~$\Delta y_i$~({\it right}) dependence of the MCE scaled variance of positively charged hadrons~$\omega_+$,
the MCE correlation coefficient between positively and negatively charged hadrons~$\rho_{+-}$, and the MCE scaled 
variance of all charged hadrons~$\omega_{\pm}$, respectively to their primordial counterparts. The results 
of~$8 \cdot 20$ Monte Carlo runs of~$2\cdot 10^5$ events each for a static and neutral hadron resonance gas 
 with~$T=0.160$~GeV are summarized in Table~\ref{accbins_Mult_final}. 

\begin{table}[h!]\footnotesize
  \begin{center}
    \begin{tabular}{cccccc}
    \toprule
      \!\!Final state\!\!\!\!& 
      $\Delta p_{T,1}$ & $\Delta p_{T,2}$ & $\Delta p_{T,3}$ & $\Delta p_{T,4}$ & $\Delta p_{T,5}$   \\
    \midrule
      $\omega^{gce}_+$\!\! & 
      $1.031 \pm 0.002$ & 
      $1.026 \pm 0.002$ & $1.020 \pm 0.002$ & 
      $1.015 \pm 0.002$ & $1.010 \pm 0.002$ \\
      $\omega^{mce}_+$\!\! & 
      $0.904 \pm 0.007$ & 
      $0.884 \pm 0.007$ & $0.872 \pm 0.007$ & 
      $0.847 \pm 0.007$ & $0.778 \pm 0.006$ \\
    \midrule
      $\rho^{gce}_{+-}$\!\! & 
      $0.163 \pm 0.001$ & 
      $0.107 \pm 0.001$ & $0.109 \pm 0.001$ & 
      $0.075 \pm 0.002$ & $0.052 \pm 0.002$ \\
      $\rho^{mce}_{+-}$\!\! & 
      $0.143 \pm 0.005$ & 
      $0.088 \pm 0.005$ & $0.090 \pm 0.005$ & 
      $0.049 \pm 0.006$ & $-0.010 \pm 0.006$ \\
    \midrule[0.09em]
      \!\!Final state\!\!\!\!& 
      $\Delta y_1$  & $\Delta y_2$  & $\Delta y_3$  & $\Delta y_4$ & $\Delta y_5$ \\
    \midrule
      $\omega^{gce}_+$\!\! & 
      $1.017 \pm 0.002$ & 
      $1.023 \pm 0.002$ & $1.024 \pm 0.002$ & 
      $1.023 \pm 0.003$ & $1.017 \pm 0.002$ \\
      $\omega^{mce}_+$\!\! & 
      $0.771 \pm 0.007$ & 
      $0.840 \pm 0.006$ & $0.859 \pm 0.007$ & 
      $0.839 \pm 0.007$ & $0.770 \pm 0.006$ \\
    \midrule 
      $\rho^{gce}_{+-}$\!\! & 
      $0.100 \pm 0.001$ & 
      $0.116 \pm 0.001$ & $0.115 \pm 0.002$ & 
      $0.115 \pm 0.002$ & $0.100 \pm 0.001$ \\
      $\rho^{mce}_{+-}$\!\! & 
      $-0.027 \pm 0.005$ & 
      $0.069 \pm 0.005$ & $0.092 \pm 0.006$ & 
      $0.069 \pm 0.006$ & $-0.027 \pm 0.005$ \\
    \bottomrule
    \end{tabular}
    \caption{Summary of the final state scaled variance~$\omega_+$ of positively charged hadrons and the correlation 
    coefficient~$\rho_{+-}$ between positively and negatively charged hadrons in transverse momentum 
    bins~$\Delta p_{T,i}$ and rapidity bins~$\Delta y_i$. Both the GCE result ($\lambda = 0$) and the extrapolation to 
    the MCE ($\lambda = 1$) are shown. The uncertainty quoted corresponds to~$20$ Monte Carlo runs of~$2 \cdot 10^5$ 
    events (GCE) or is the result of the extrapolation (MCE).} 
    \label{accbins_Mult_final}
  \end{center}
\end{table}

A few words to summarize Figs.(\ref{mult_pm_MCE_omega}-\ref{mult_pm_MCE_omega_CH}): 
resonance decay and (energy) conservation laws work in the same direction, as far as the transverse momentum dependence 
of the  scaled variances~$\omega_+$,~$\omega_{\pm}$  and the correlation coefficient~$\rho_{+-}$ is concerned. 
Both effects lead to increased multiplicity fluctuations and an increased correlation between the multiplicities of 
oppositely charged particles in the low~$p_T$ region, compared to the high~$p_T$ domain. Compared to this, 
the MCE~$\Delta y_i$ dependence of~$\omega_+$,~$\rho_{+-}$, and~$\omega_{\pm}$ is mainly dominated by global 
conservation of~$P_z$. Resonance decay effects, see also the GCE versions, 
Figs.(\ref{mult_pm_0000_omega}-\ref{mult_pm_0000_omega_CH}), act more equal across rapidity, than in transverse momentum.
 
Again, the scaled variance of all charged particles is found to be larger than the scaled variance of only positively 
charged hadrons~$\omega_{\pm} > \omega_+$, except for when~$\rho_{+-}<0$, i.e when the multiplicities of oppositely 
charged particles are anti-correlated, as for instance in~$\Delta p_{T,5}$,~$\Delta y_{1}$, and~$\Delta y_{5}$.  
In contrast to that, however narrowly,~$\omega_{\pm} > 1$ in the lowest transverse momentum bin~$\Delta p_{T,1}$. 
Acceptance scaling provides a good estimate of the average in transverse momentum. In rapidity, however the 
approximation somewhat overshoots the limited acceptance results, pointing to correlations 
amongst momentum bins, Chapter~\ref{chapter_multfluc_pgas}.

\section{Canonical Ensemble}
\label{sec_multfluc_hrg_ce}

To complement the discussion of multiplicity fluctuations and correlations for a hadron resonance gas in transverse 
momentum and rapidity segments, the canonical ensemble limit is studied next. Events will be re-weighted according 
to their extensive quantities baryon number~$B$, strangeness~$S$, and electric charge~$Q$. Conservation laws for 
energy~$E$ and longitudinal momentum~$P_z$ (as well as for transverse momenta~$P_x$ and~$P_y$) are hence not applied. 
Again some guidance for assessing the quality of the extrapolation is provided by  analytical methods. Two independent 
(final state and primordial)  Monte Carlo runs of~$8 \cdot 20 \cdot 2 \cdot 10^5$ events each have been generated. 
The same system as before is considered and the same acceptance cuts are applied. 

Resonance decay is a process respecting the conservation laws for charge, energy and momentum. So despite the fact 
that charge conservation laws do not correlate primordial particle momenta, these additional correlations are 
introduced into the final state momentum space dependence, as in the GCE.

\begin{figure}[ht!]
  \epsfig{file=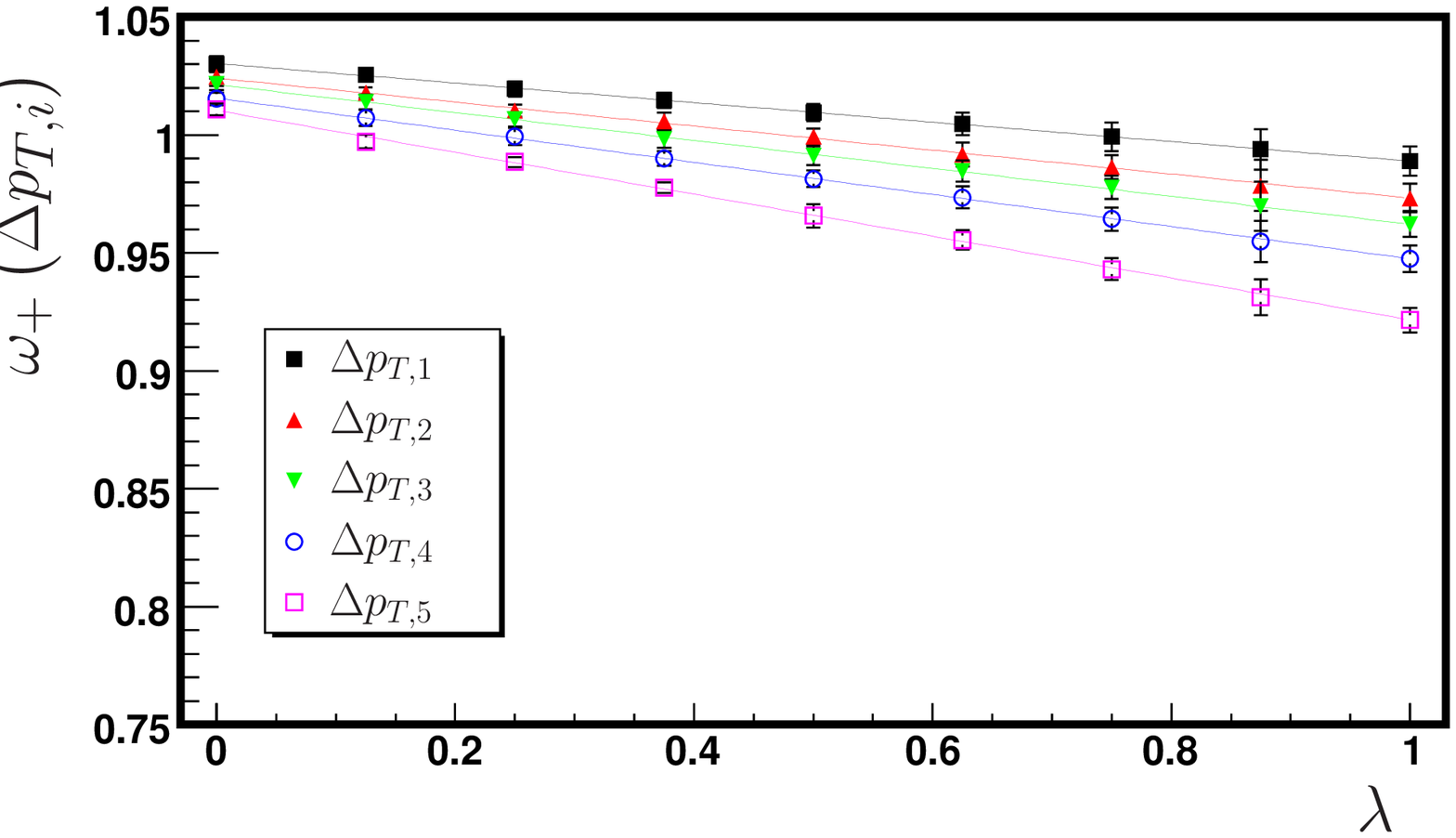,width=7.2cm,height=6.5cm}
  \epsfig{file=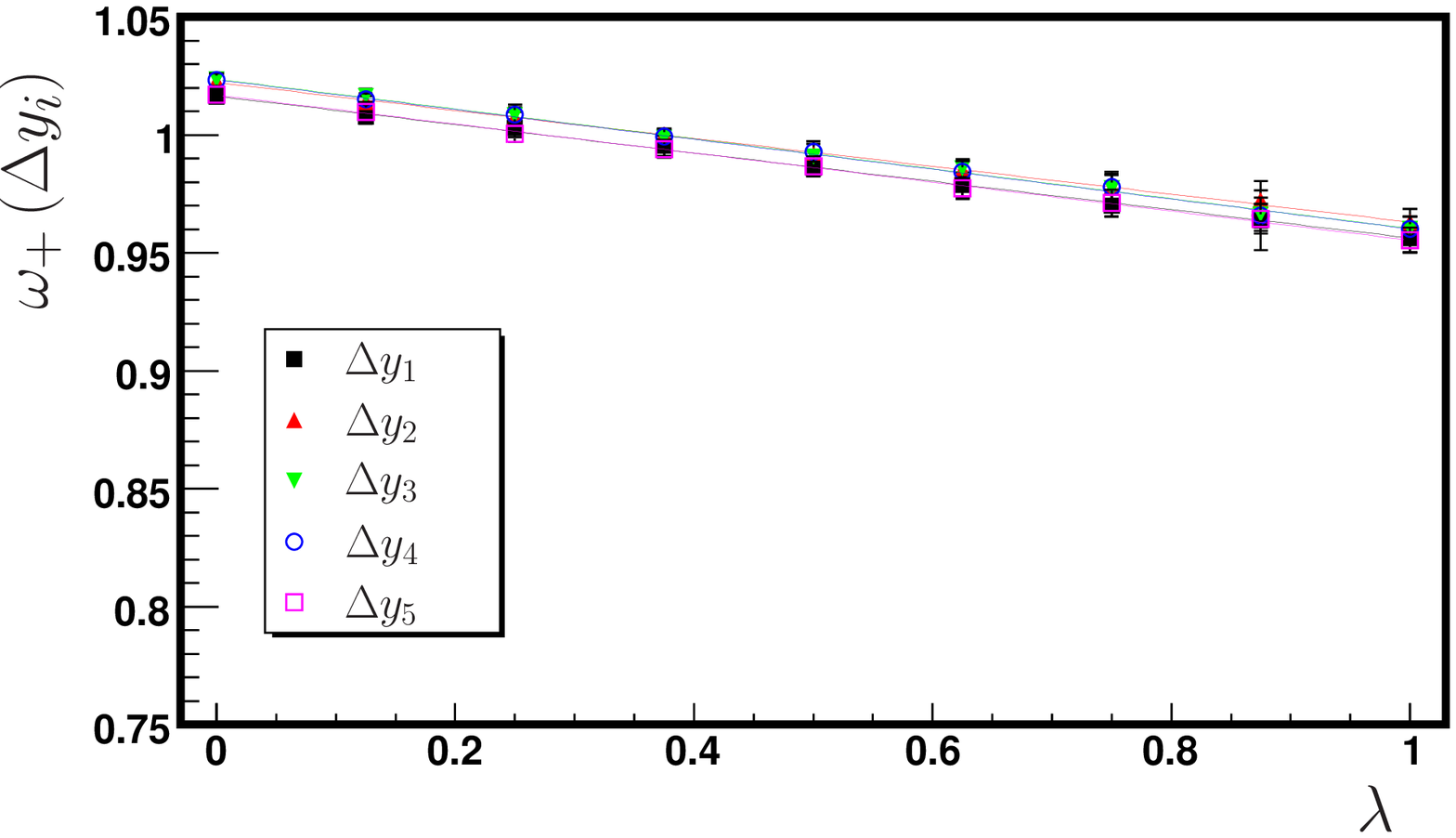,width=7.2cm,height=6.5cm}
  \caption{Evolution of the final state scaled variance~$\omega_+$ of positively charged hadrons with the Monte Carlo 
  parameter~$\lambda = V_1/V_g$ for transverse momentum bins~$\Delta p_{T,i}$~({\it left}) and for rapidity 
  bins~$\Delta y_i$~({\it right}). The solid lines show an analytic extrapolation from GCE results ($\lambda =0$) to 
  the CE limit ($\lambda \rightarrow 1$). Each marker, except the last, represents the result of~$20$ Monte Carlo runs 
  of~$2 \cdot 10^5$ events.~$8$~different equally spaced values of~$\lambda$ have been investigated.
  The last marker denotes the result of the extrapolation. }  
  \label{BSQ_conv_lambda_omega_final}
\end{figure}

\begin{figure}[ht!]
  \epsfig{file=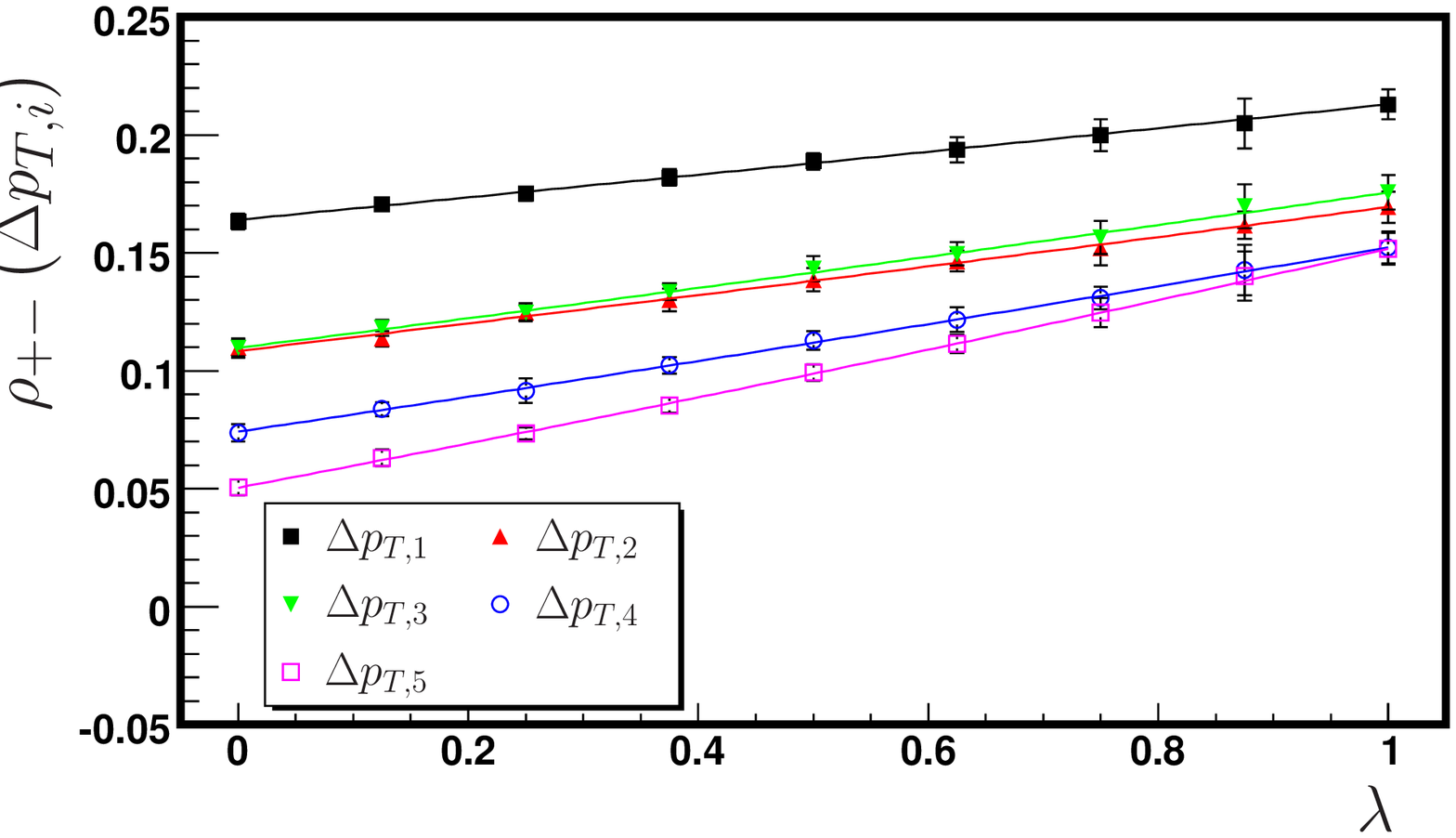,width=7.2cm,height=6.5cm}
  \epsfig{file=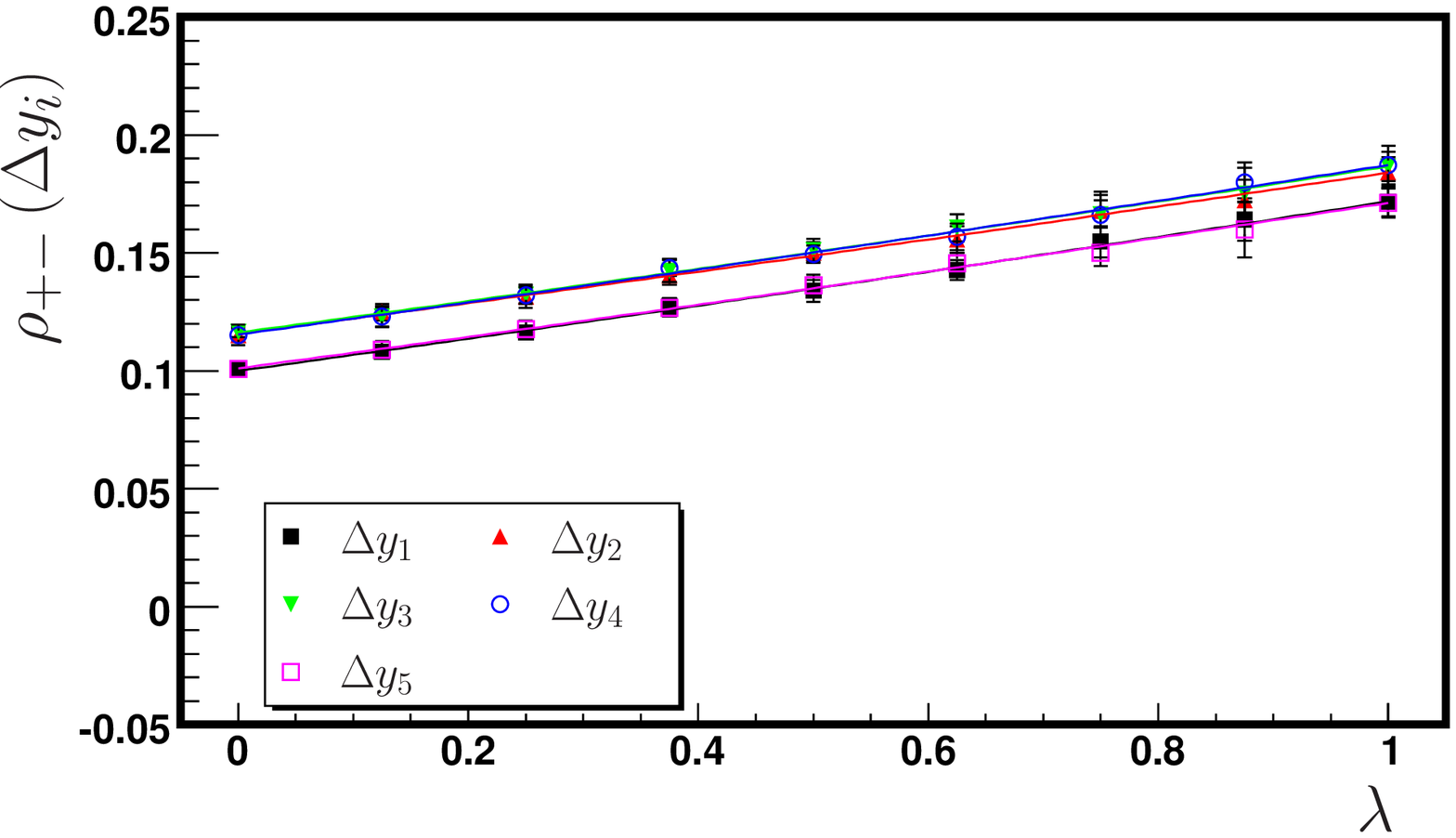,width=7.2cm,height=6.5cm}
  \caption{Evolution of the final state correlation coefficient~$\rho_{+-}$ between positively and negatively charged 
  hadrons with the Monte Carlo parameter~$\lambda = V_1/V_g$ for transverse momentum bins~$\Delta p_{T,i}$~({\it left}) 
  and for rapidity bins~$\Delta y_i$~({\it right}). The rest as in Fig.(\ref{BSQ_conv_lambda_omega_final}).}  
  \label{BSQ_conv_lambda_rho_final}
\end{figure}

\begin{figure}[ht!]
  \epsfig{file=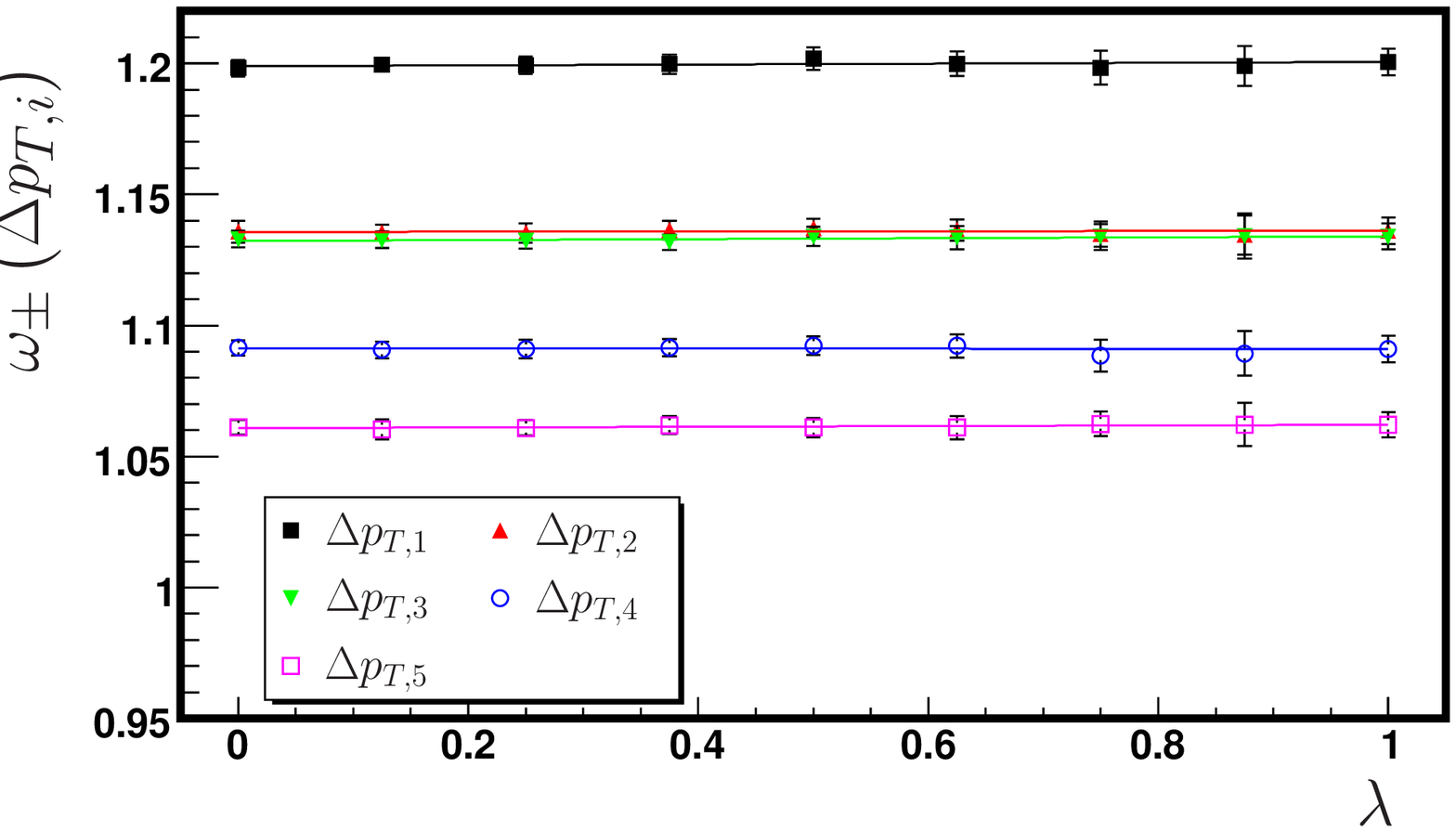,width=7.2cm,height=6.5cm}
  \epsfig{file=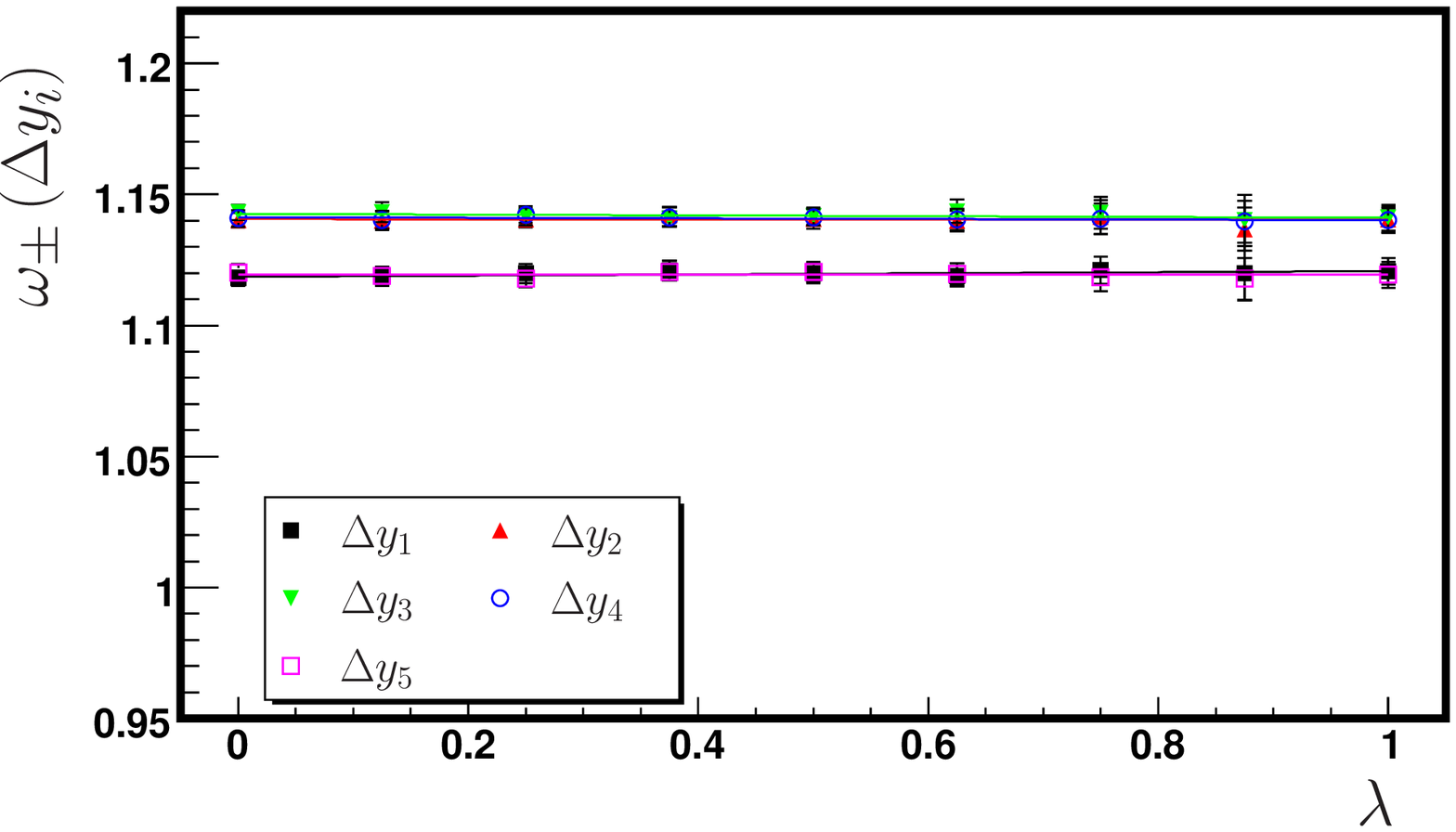,width=7.2cm,height=6.5cm}
  \caption{Evolution of the final state scaled variance ~$\omega_{\pm}$ between all charged hadrons with the Monte Carlo 
  parameter~$\lambda = V_1/V_g$ for transverse momentum bins~$\Delta p_{T,i}$~({\it left}) and for rapidity 
  bins~$\Delta y_i$~({\it right}). The rest as in Fig.(\ref{BSQ_conv_lambda_omega_final}).}  
  \label{BSQ_conv_lambda_omegaChr_final}
\end{figure}

In Figs.(\ref{BSQ_conv_lambda_omega_final}-\ref{BSQ_conv_lambda_omegaChr_final}) the final state scaled 
variance~$\omega_+$ of positively charged hadrons, the final state correlation coefficient~$\rho_{+-}$ between 
positively and negatively charged hadrons, and the final state scaled variance~$\omega_{\pm}$ of all charged hadrons, 
in transverse momentum bins~$\Delta p_{T,i}$~({\it left}) and rapidity bins~$\Delta y_i$~({\it right}) are shown as a 
function of~$\lambda$. Primordial results are not included.

\begin{figure}[ht!]
  \epsfig{file=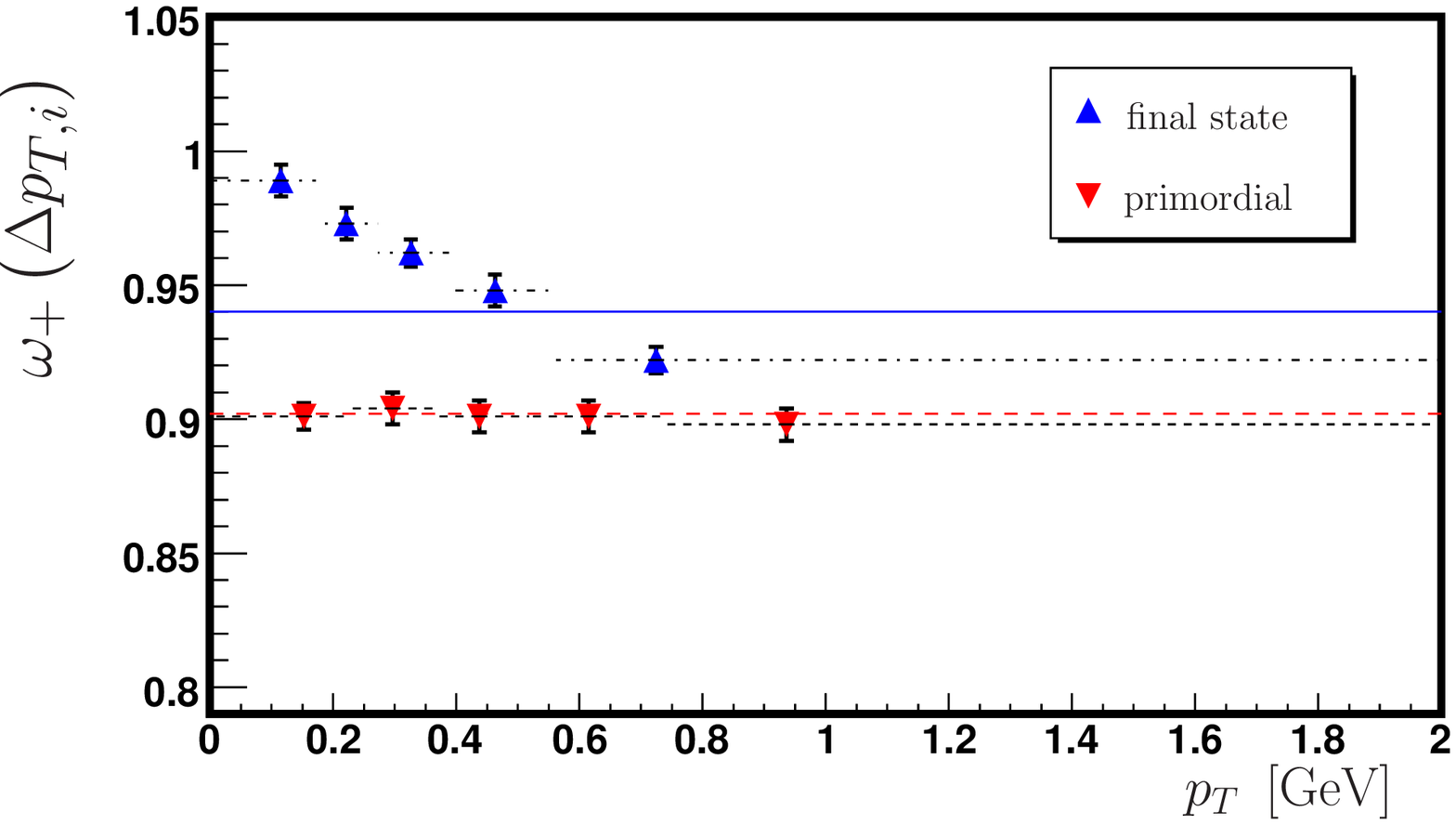,width=7.2cm,height=6.5cm}
  \epsfig{file=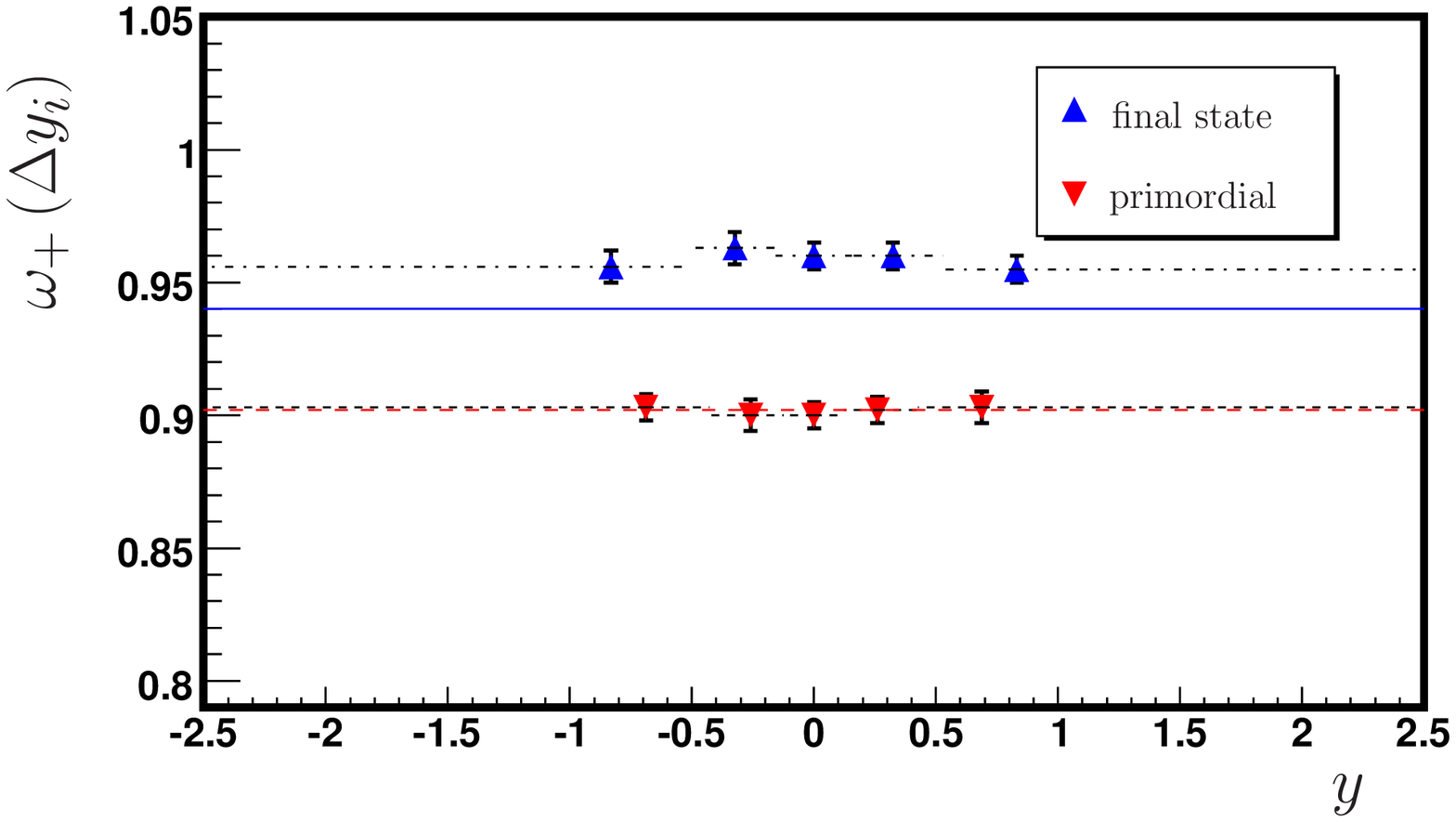,width=7.2cm,height=6.5cm}
  \caption{CE scaled variance~$\omega_+$ of multiplicity fluctuations of positively charged hadrons, both primordial 
  and final state, in transverse momentum bins~$\Delta p_{T,i}$~({\it left}) and rapidity bins~$\Delta y_i$~({\it right}). 
  Horizontal error bars indicate the width and position of the momentum bins (And not an uncertainty!).
  Vertical error bars indicate the statistical uncertainty quoted in Table~\ref{BSQ_accbins_Mult}.
  The markers indicate the center of gravity of the corresponding bin. 
  The solid and the dashed lines show final state and primordial acceptance scaling estimates respectively. }
  \label{mult_pm_CE_omega}
\end{figure}

\begin{figure}[ht!]
  \epsfig{file=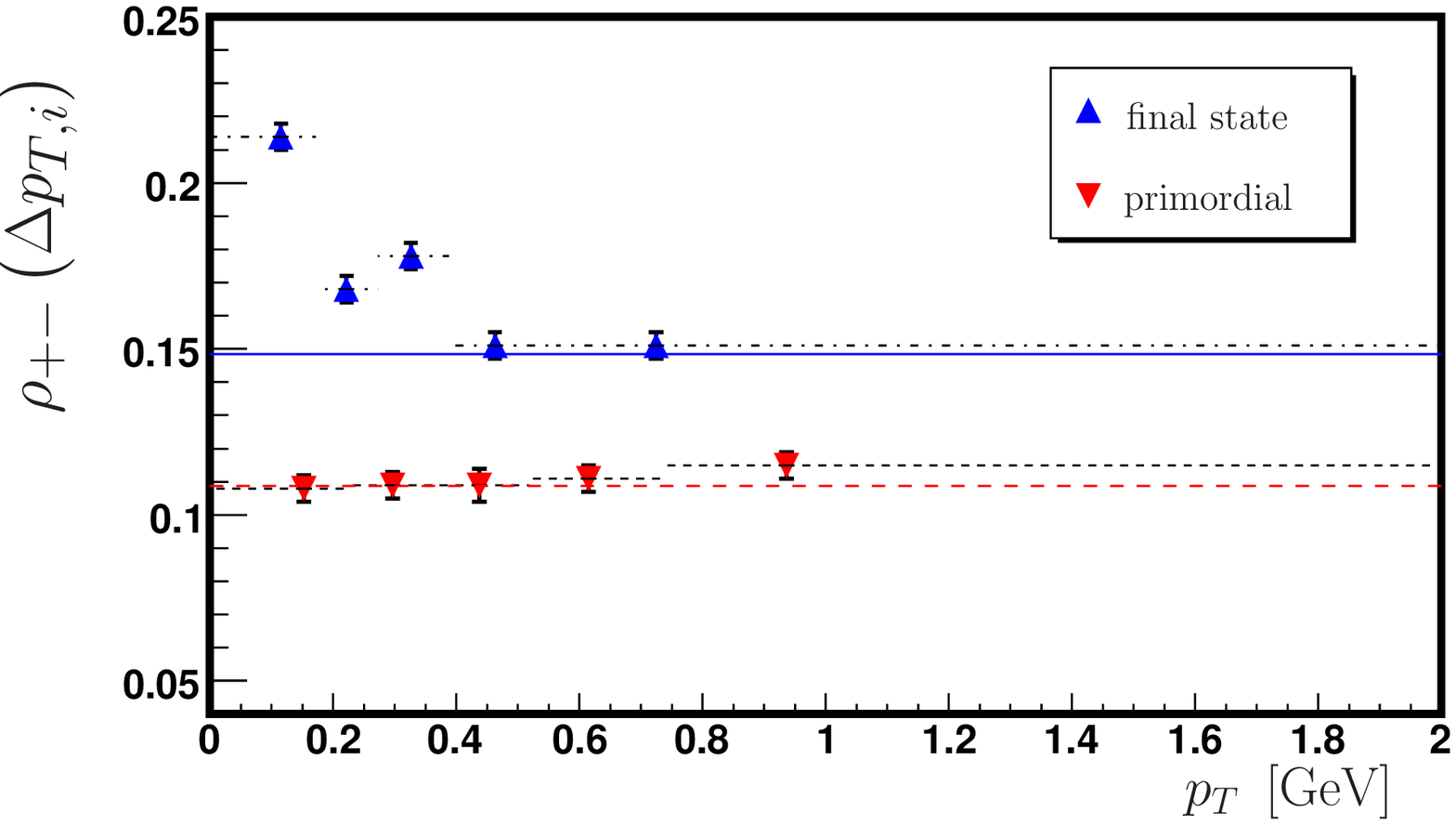,width=7.2cm,height=6.5cm}
  \epsfig{file=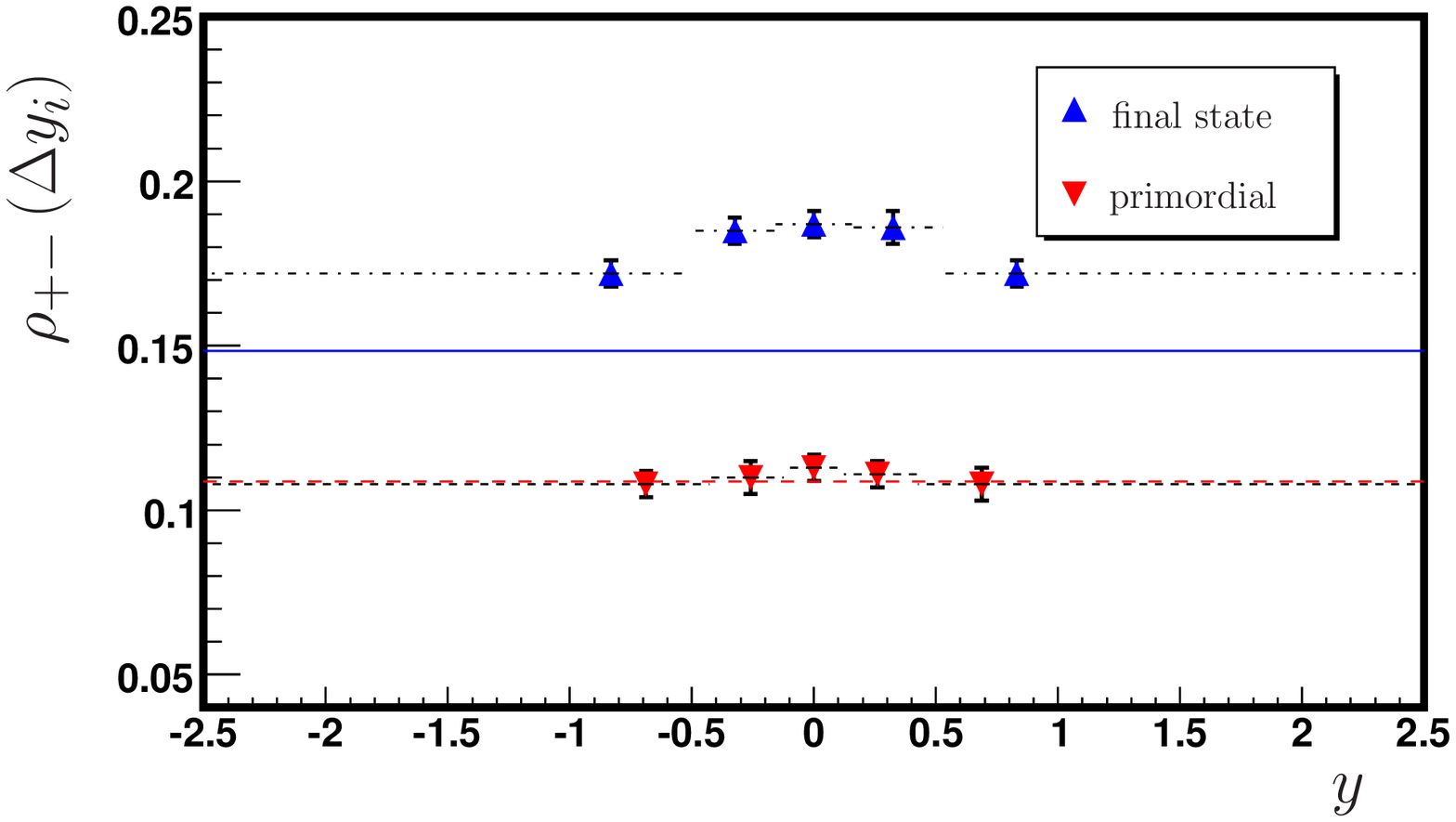,width=7.2cm,height=6.5cm}
  \caption{CE multiplicity correlation coefficient~$\rho_{+-}$ between positively and negatively charged hadrons, 
  both primordial and final state, in transverse momentum bins~$\Delta p_{T,i}$~({\it left}) and rapidity 
  bins~$\Delta y_i$~({\it right}). The rest as in Fig.(\ref{mult_pm_CE_omega}). }  
  \label{mult_pm_CE_rho}
\end{figure}

\begin{figure}[ht!]
  \epsfig{file=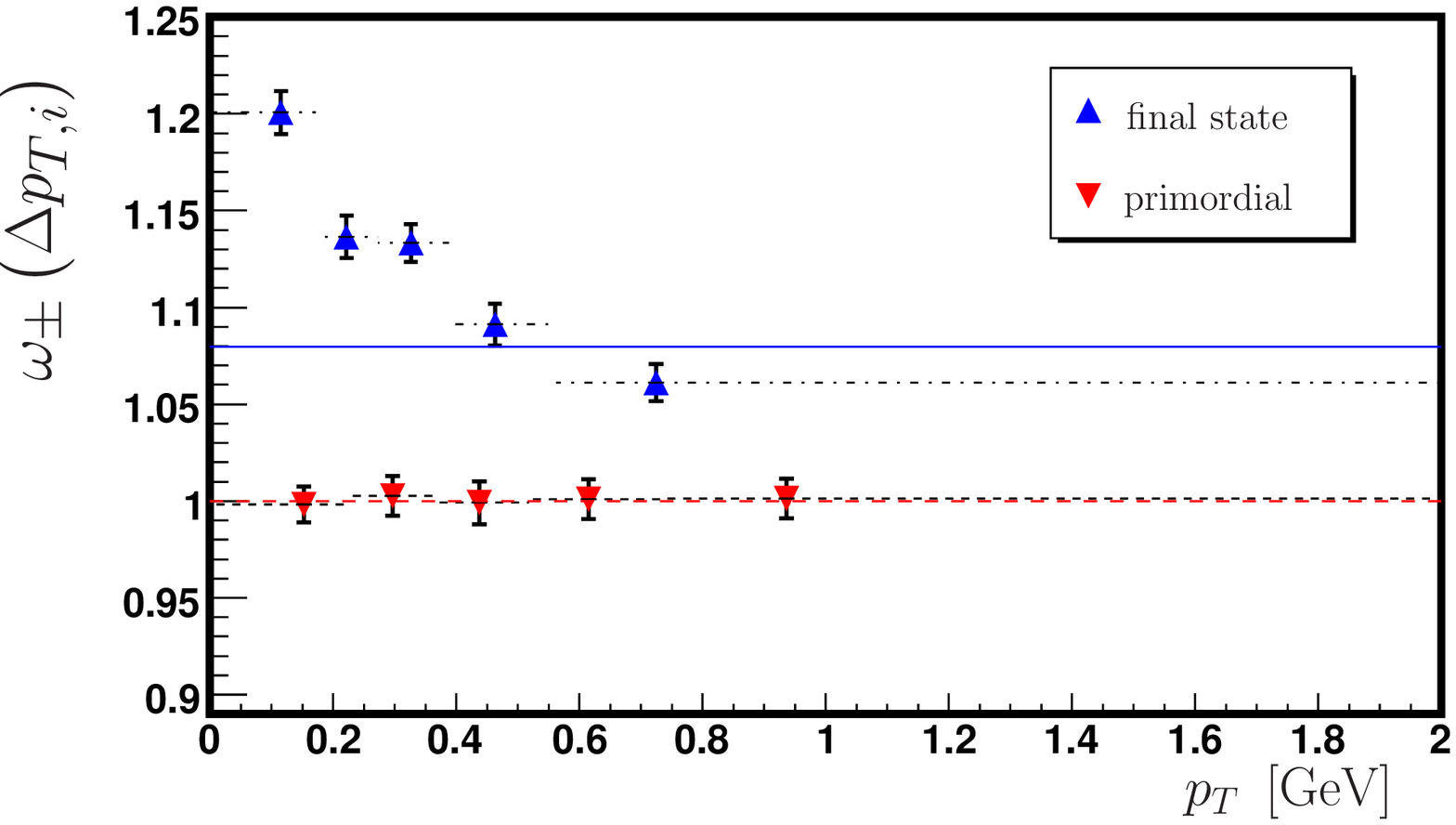,width=7.2cm,height=6.5cm}
  \epsfig{file=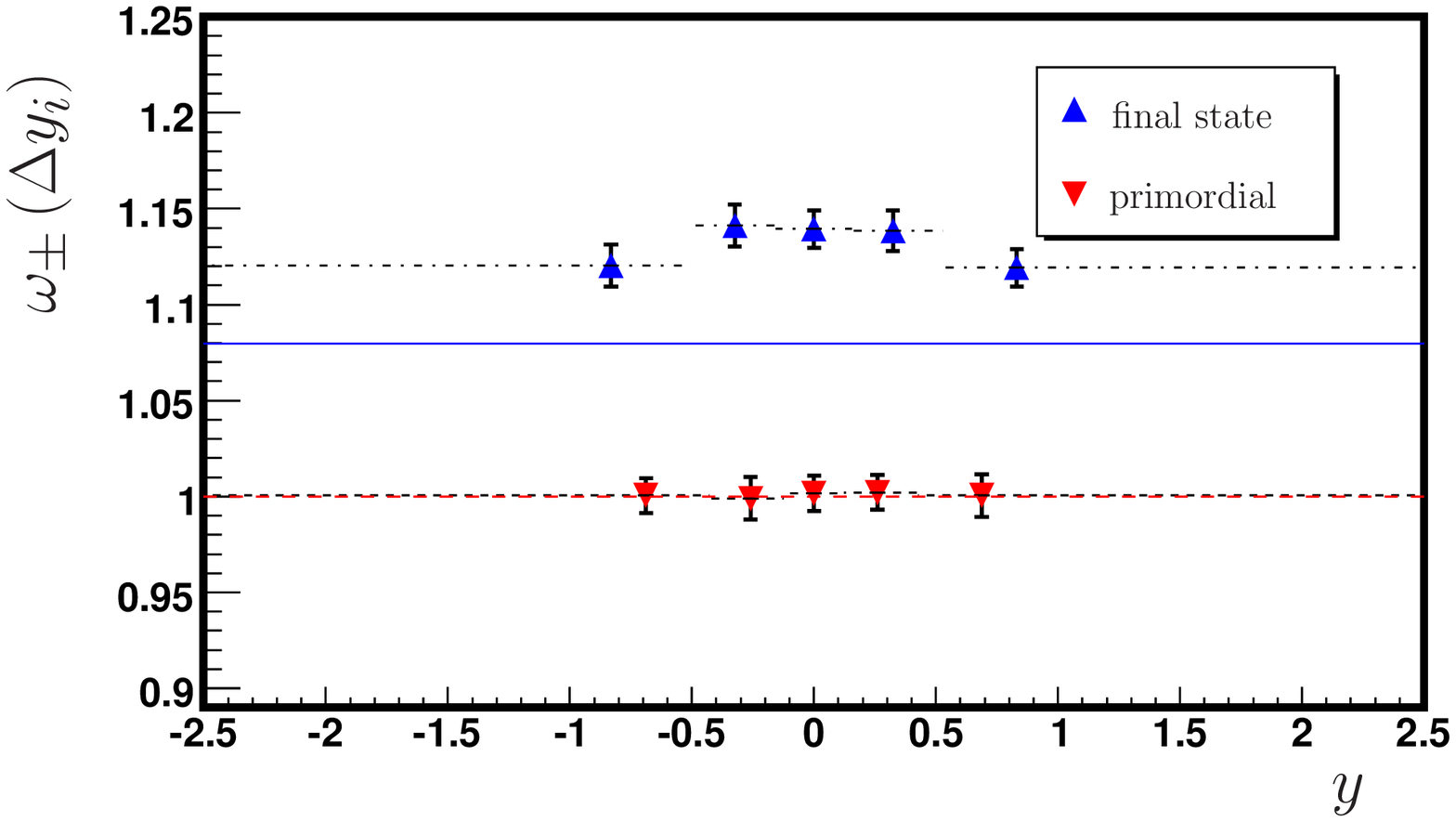,width=7.2cm,height=6.5cm}
  \caption{CE scaled variance~$\omega_{\pm}$ of multiplicity fluctuations of all charged hadrons, both primordial and 
  final state, in transverse momentum bins~$\Delta p_{T,i}$~({\it left}) and rapidity bins~$\Delta y_i$~({\it right}). 
  The rest as in Fig.(\ref{mult_pm_CE_omega}).
  }  
  \label{mult_pm_CE_omega_CH}
\end{figure}

Figs.(\ref{mult_pm_CE_omega}-\ref{mult_pm_CE_omega_CH}) compare the final state~$\Delta p_{T,i}$~({\it left}) 
and~$\Delta y_i$~({\it right}) dependence of the CE scaled variances~$\omega_+$, and~$\omega_{\pm}$, and the CE 
correlation coefficient~$\rho_{+-}$ respectively to their primordial counterparts. The results of~$8 \cdot 20$ Monte 
Carlo runs of~$2\cdot 10^5$ events each for a static and neutral hadron resonance gas with temperature~$T=0.160$~GeV are 
summarized in Table~\ref{BSQ_accbins_Mult}. 

Firstly the primordial scenario is considered. Charge conservation does not introduce any correlations in momentum space. 
Primordial results are thus independent of where in momentum space the acceptance window is located. Only the fraction 
of the system observed matters. One therefore finds the scaled variance and the correlation coefficient the same in 
all~$\Delta p_{T,i}$ and~$\Delta y_i$ bins. All values converge to the one suggested by the CE acceptance scaling 
approximation with acceptance probability~$q=0.2$ (i.e. $5$ bins of equal average particle number content). 
Fully phase space integrated values~$\omega$ and~$\rho$ can be found 
in~\cite{Begun:2006jf,Begun:2006uu,Gorenstein:2008et,Konchakovski:2009at} and later on in 
Chapters~\ref{chapter_phasediagram} and~\ref{chapter_freezeoutline}.

The primordial scaled variance~$\omega_+$ of positively charged hadrons drops linearly to its CE limit as the size of 
the charge bath is reduced. Here one also finds~$\omega_+^{mce} < \omega_+^{ce}$, however almost equal in~$\Delta p_{T,1}$, 
in limited acceptance. Combined energy and charge conservation leads to more constrained multiplicity fluctuations.
The primordial correlation coefficient~$\rho_{+-}$ between positively and negatively charged particles, on the other 
hand, rises as the size of the bath decreases,~$\lambda \rightarrow 1$. The correlation coefficient is indeed positive, 
as one would expect from charge conservation. In full phase space ~$\rho_{+-}^{ce}$ and~$\rho_{+-}^{mce}$ are rather 
close to unity, as discussed earlier. However the correlation coefficient in limited acceptance appears weaker in the 
MCE than in the CE,~$\rho_{+-}^{mce} < \rho_{+-}^{ce}$. Energy conservation de-correlates the joint multiplicity 
distribution.
In a primordial neutral CE system one finds the width of the distribution of all charged 
hadrons, Eq.(\ref{omega_chr_eq}),~$\omega_{\pm} \simeq 1~$, i.e. essentially a Poissonian, in any momentum 
bin~\cite{Begun:2004zb,Begun:2006jf}. Fluctuations of the number of all charged particles appear to be unconstrained 
by charge conservation and, thus, to yield the same result as in the GCE~\cite{Begun:2004gs}. But also, like 
for~$\omega$, the observed value is independent of the position and shape of the acceptance window. This is in contrast 
to primordial MCE calculations where the~$\omega$ and~$\rho$ show a distinctive dependence on the bins~$\Delta p_{T,i}$ 
and~$\Delta y_i$. See Figs.(\ref{conv_lambda_omega_prim}-\ref{conv_lambda_rho_prim}).

No additional correlations are introduced into the momentum space dependence of the final state multiplicity 
distribution, apart from the well-known correlations due to resonance decay. The CE final state scaled variance~$\omega_+$ 
in transverse momentum bins in Fig.(\ref{mult_pm_CE_omega})~({\it left}) shows qualitatively a similar behavior as in 
the GCE, Fig.(\ref{mult_pm_0000_omega})~({\it left}), and in the MCE, Fig.(\ref{mult_pm_MCE_omega})~({\it left}).
The scaled variance decreases as charge conservation is turned on, however not as strongly as in the MCE. 
Neutral hadrons lead to enhancement mostly at low~$p_T$, hence~$\omega_+^{ce}$ is larger in~$\Delta p_{T,1}$ than 
in~$\Delta p_{T,5}$. As resonance decay works more evenly across rapidity, one finds the~$\Delta y_{i}$ dependence 
mostly flat in Fig.(\ref{mult_pm_CE_omega})~({\it right}). 

\begin{table}[h!]\footnotesize
   \begin{center}
      \begin{tabular}{cccccc}
      \toprule
        Primordial & $\Delta p_{T,1}$ & $\Delta p_{T,2}$ & $\Delta p_{T,3}$ & $\Delta p_{T,4}$ 
        & $\Delta p_{T,5}$   \\
      \midrule
        $\omega^{ce}_+$ & $0.901 \pm 0.005$ & 
        $0.904 \pm 0.006$ & $0.901 \pm 0.006$ & 
        $0.901 \pm 0.006$ & $0.898 \pm 0.006$ \\
        $\rho^{ce}_{+-}$ & $0.108 \pm 0.004$ & 
        $0.109 \pm 0.004$ & $0.109 \pm 0.005$ & 
        $0.111 \pm 0.004$ & $0.115 \pm 0.004$ \\
      \midrule[0.09em]      
        Primordial & $\Delta y_1$  & $\Delta y_2$  & $\Delta y_3$  & $\Delta y_4$  
        & $\Delta y_5$ \\
      \midrule 
        $\omega^{ce}_+$ & $0.903 \pm 0.005$ & 
        $0.900 \pm 0.006$ & $0.900 \pm 0.005$ &   
        $0.902 \pm 0.005$ & $0.903 \pm 0.006$ \\
        $\rho^{ce}_{+-}$ &  $0.108 \pm 0.004$ & 
        $0.110 \pm 0.005$ & $0.113 \pm 0.004$ & 
        $0.111 \pm 0.004$ & $0.108 \pm 0.005$ \\
      \midrule[0.09em] 
        Final state& $\Delta p_{T,1}$ & $\Delta p_{T,2}$ & $\Delta p_{T,3}$ & $\Delta p_{T,4}$ 
        & $\Delta p_{T,5}$   \\
      \midrule
        $\omega^{ce}_+$ & $0.989 \pm 0.006$ & 
        $0.973 \pm 0.006$ & $0.962 \pm 0.005$ &   
        $0.948 \pm 0.006$ & $0.922 \pm 0.005$ \\
        $\rho^{ce}_{+-}$ & $0.214 \pm 0.004$ & 
        $0.168 \pm 0.004$ & $0.178 \pm 0.004$ & 
        $0.151 \pm 0.004$ & $0.151 \pm 0.004$ \\
      \midrule[0.09em]
        Final state& $\Delta y_1$  & $\Delta y_2$  & $\Delta y_3$  & $\Delta y_4$  
        & $\Delta y_5$ \\
      \midrule 
        $\omega^{ce}_+$ & $0.956 \pm 0.006$ & 
        $0.963 \pm 0.006$ & $0.960 \pm 0.005$ & 
        $0.960 \pm 0.005$ & $0.955 \pm 0.005$ \\
        $\rho^{ce}_{+-}$ & $0.172 \pm 0.004$ & 
        $0.185 \pm 0.004$ & $0.187 \pm 0.004$ & 
        $0.186 \pm 0.005$ & $0.172 \pm 0.004$ \\
      \bottomrule      
    \end{tabular}
    \caption{Summary of the final state and primordial scaled variance~$\omega_+$ of positively charged hadrons and 
    the correlation coefficient~$\rho_{+-}$ between positively and negatively charged hadrons in transverse momentum 
    bins~$\Delta p_{T,i}$ and rapidity bins~$\Delta y_i$. 
    Both the GCE result ($\lambda = 0$) and the extrapolation to CE ($\lambda = 1$) are shown. Analytical primordial CE 
    values are~$\omega_+^{ce} = 0.9019$ and~$\rho_{+-}^{ce} = 0.1088$, while final state acceptance scaling 
    values are~$\omega_+^{ce} = 0.9401$ and~$\rho_{+-}^{ce} = 0.1485$. The uncertainty quoted corresponds to~$20$ Monte 
    Carlo runs of~$2 \cdot 10^5$ events (GCE) or is the result of the extrapolation (CE).} 
    \label{BSQ_accbins_Mult}
  \end{center}
\end{table}

The CE final state correlation coefficient~$\rho_{+-}$ in transverse momentum bins 
in Fig.(\ref{mult_pm_CE_rho})~({\it left}) qualitatively shows an equally similar behavior as in the GCE, 
Fig.(\ref{mult_pm_0000_rho})~({\it left}), and in the MCE, Fig.(\ref{mult_pm_MCE_rho})~({\it left}). However with a 
notably stronger correlation at larger~$p_T$ bins in Fig.(\ref{mult_pm_CE_rho})~({\it left}) than in GCE and MCE.
The extrapolations in Figs.(\ref{conv_lambda_rho_final})~({\it left}) to the final state MCE and in 
Fig.(\ref{BSQ_conv_lambda_rho_final})~({\it left}) to the final state CE correlation coefficient, show hence different 
behavior. Starting from the final state GCE values the correlation coefficient grows as the charge bath is removed, but 
drops when both, charge and heat bath, are removed. Resonance decay introduces a positive correlation. 
Charge conservation makes this trend stronger. In the MCE energy conservation works in the opposite direction. 
The~$\Delta y_i$ dependence is again mild in the CE, Fig.(\ref{mult_pm_MCE_rho})~({\it right}). 

The final state CE values of~$\omega_{\pm}$ in Fig.(\ref{mult_pm_CE_omega_CH}) can be compared to their GCE counterparts 
in Fig.(\ref{mult_pm_0000_omega_CH}). As charge conservation does not introduce any correlations in momentum space, and 
as the effect of charge conservation is not seen in~$\omega_{\pm}$, also final state values are equal (within error bars) 
to the GCE results. GCE and CE values for~$\omega_{\pm}$ are the same in full phase space for a neutral system (both 
primordial and final state). As charge conservation ensures (in the absence of energy conservation) a positive 
correlation coefficient~$\rho_{+-}$, one finds~$\omega_+ < \omega_{\pm}$ in the CE in any momentum bin. The acceptance 
scaling estimate gives a good average in transverse momentum bins~$\Delta p_{T,i}$, while under-predicting correlations 
and fluctuations in rapidity bins~$\Delta y_i$, due to correlations introduced by resonance decay.

\section{Discussion}
\label{sec_multfluc_hrg_dis}

GCE joint distributions of particle multiplicity were extrapolated to the MCE and CE limits. 
A few remarks attempt to summarize. The final state transverse momentum dependence~$\Delta p_{T,i}$ of multiplicity 
fluctuations and correlations are qualitatively similar in MCE, CE, and GCE. On the other hand, their (final state)
rapidity dependence~$\Delta y_i$ exhibits only a very mild dependence in GCE and CE, which is in contrast to the MCE 
(with momentum conservation), where effects are strong. In the primordial case there is no momentum space dependence 
of multiplicity fluctuations and correlations in ensembles without energy conservation. 
Resonance decay trends are enhanced in the MCE. Charge conservation effects cease in a neutral system, and thus final 
state values of~$\omega_{\pm}$ are the same in GCE and CE, even after resonance decay in limited acceptance,
which provides a valuable cross check with analytical solutions. 

The extrapolation scheme works rather accurate and efficient. The statistical error on the `data` points grows 
as~$\lambda \rightarrow 1$. The extrapolation helps greatly to keep the statistical uncertainty on the MCE (or CE) limit 
low, which can be seen from a comparison of the right most markers on the extrapolation lines. The last point and its error 
bar denote the result of a linear extrapolation of variances and covariances, while the second to last data point and 
its error bar are the result of~$20$ Monte Carlo runs with~$\lambda = 0.875$. The analytical MCE (or CE) values are well 
within error bars of extrapolated Monte Carlo results, and agree surprisingly well, given the large number 
of `conserved' quantities (5 and 3, respectively) and a relatively small sample size 
of~$8 \cdot 20 \cdot 2 \cdot 10^5 = 3.2 \cdot 10^7$ events. In a sample-reject type 
of approach this sample size would yield a substantially larger statistical error, as only events with exact values of 
extensive quantities are kept for the analysis. As the system size is increased, a sample-reject formalism, hence, 
becomes increasingly inefficient, while the extrapolation method still yields good results. Error bars are mildly smaller 
for an CE extrapolation than for an extrapolation to the MCE, due to a smaller number of conserved quantities. 
For further discussion see Appendices~\ref{appendix_convstudy} and~\ref{appendix_cbg}.

The qualitative picture presented in Fig.(\ref{mult_pm_MCE_omega}) could be compared to similar analysis of UrQMD 
transport simulation data~\cite{Lungwitz:2007uc}, or recently published NA49 data on multiplicity fluctuations in 
limited momentum bins~\cite{Alt:2007jq}. 
Both (data and transport simulation) appear qualitatively similar to the MCE scenario, which could hint at a potentially 
strong role of conservation laws on fluctuation and correlation observables. The arguments used are general enough
to believe they might hold as `{\it rules of thumb}' in non-equilibrium situations too. If, in one event, one has a 
larger (smaller) number of particles in a momentum bin with larger and positive~$p_z$, than also a larger (smaller) 
number should go in the negative~$p_z$ direction. This effect should be considerable as long as the momentum
bins contain a not too small fraction of the particles of the whole event, say~$20\%$ as here. Or even 
smaller ($11\%$) as in Chapter~\ref{chapter_multfluc_pgas}. Or yet smaller ($<10\%$) as in the UrQMD transport 
simulation~\cite{Lungwitz:2007uc}.

It should be hard to assess the importance of conservations laws in simulations, let alone in data. 
This model is simple as one can follow things step by step and turn on and off certain effects. 
This is much harder in considerabely more complicated transport simulations, where individual sources of 
correlations can not that easily be pinned down. 

The systems to be studied are in general not neutral, so mean values of positively and negatively charged hadrons are 
not the same, hence the spectra will also vary. 
A simultanious measurement of the momentum space dependence of~$\omega_-$,~$\omega_+$, and~$\rho_{\pm}$, and 
their mean values~$\langle N_- \rangle$, and~$\langle N_+ \rangle$ should help to disentangle effects. 
For instance~$\omega_+ > \omega_{\pm}$ in a high momentum bin, while~$\omega_+ < \omega_{\pm}$ in a low momentum bins 
is hard to do, except with energy conservation.

\chapter{Micro Canonical Ensemble}
\label{chapter_multfluc_pgas}
In this chapter the MCE is studied more closely, as it appears to be the most interesting amongst the standard ensembles. 
A simplified physical system is chosen to allow for smoother discussion. Owing to available analytical 
solutions~\cite{Hauer:2009nv} Fermi-Dirac (FD) and Bose-Einstein (BE) effects are included into the analysis. 
The Monte Carlo scheme only facilitated  Maxwell-Boltzmann (MB) statistics. For the examples a gas with three 
degenerate massive particles (with positive, negative and zero charge) with mass~${m=0.140}$~GeV in three different 
statistics (MB, FD, BE) has been chosen. Thus examples, in particular the FD case, are a little academic in the sense 
that there is no fermion of this mass. In a hadron resonance gas, discussed in the previous chapters, the lightest 
fermion is the nucleon for which quantum effects are probably negligible. 
  
In Section~\ref{sec_multfluc_pgas_onebin} multiplicity fluctuations and correlations in transverse momentum 
bins~$\Delta p_{T,i}$ and rapidity bins~$\Delta y_i$ are discussed. Section~\ref{sec_multfluc_pgas_acrossbins} will 
turn to multiplicity correlations between bins disconnected in momentum space ($y_A$ and~$y_B$, or~$p_{T,A}$ 
and~$p_{T,B}$), and correlations between bins separated by some distance~$\phi_{gap}$ or~$y_{gap}$ in azimuth and rapidity. 

\section{Fluctuations and Correlations within one Momentum Bin}   
\label{sec_multfluc_pgas_onebin}   

The properties of a static thermal system will be discussed first. Joint distributions of multiplicities~$N_+$ and~$N_-$ 
are measured in limited bins of transverse momentum~$\Delta p_{T,i}$ and of rapidity~$\Delta y_i$. Results will, 
in particular, be compared to the acceptance scaling approximation employed 
in~\cite{Begun:2004gs,Begun:2006jf,Begun:2006uu}, which assumes random observation of particles with a certain 
probability~$q$, regardless of particle momentum (see also Appendix~\ref{appendix_accscaling}). Corresponding results   
for scaled variance in MB statistics can also be found in~\cite{Hauer:2007im}.   
\subsubsection{Static System}   

\begin{figure}[ht!]   
  \epsfig{file=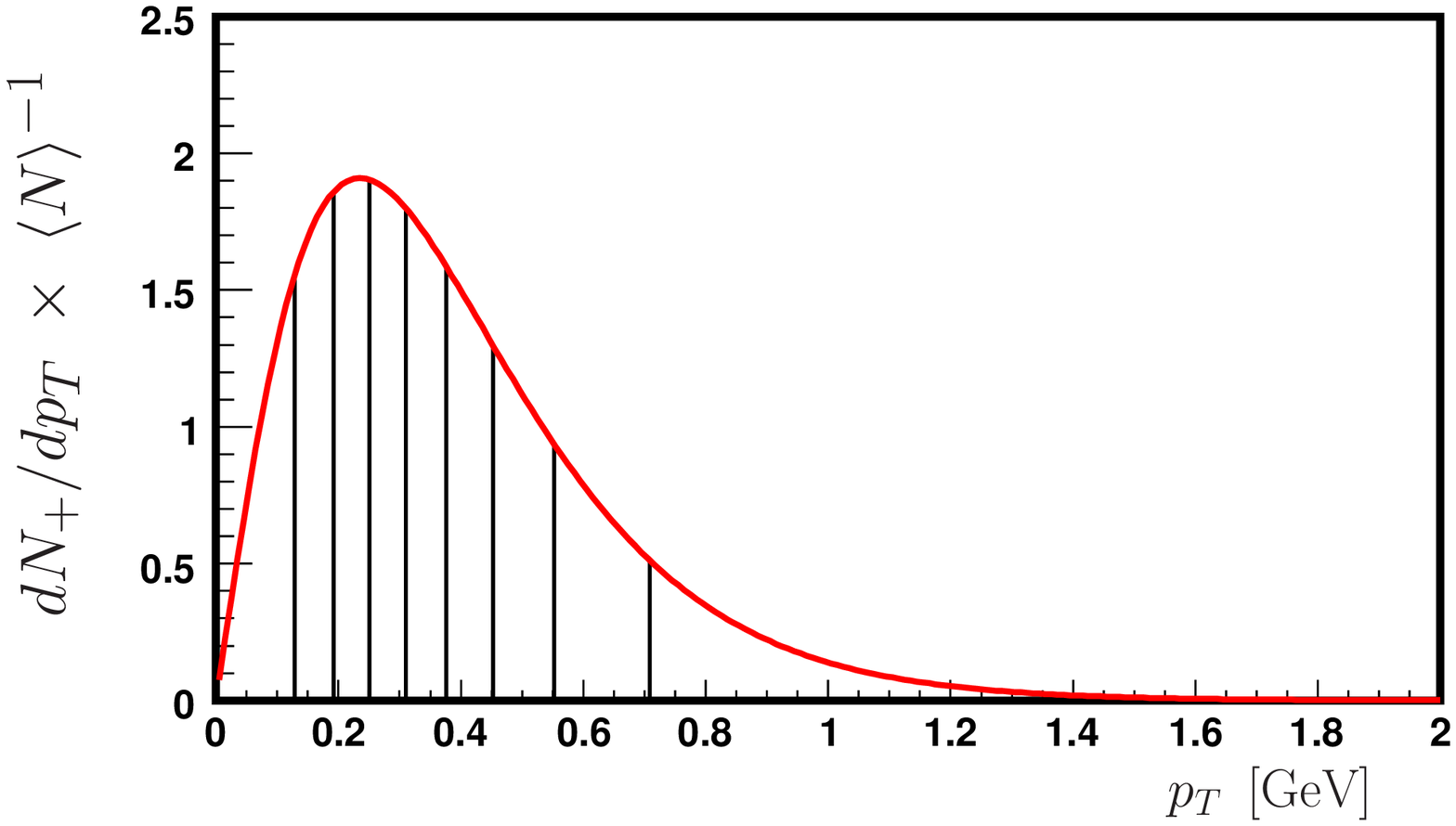,width=7.2cm,height=6.5cm}   
  \epsfig{file=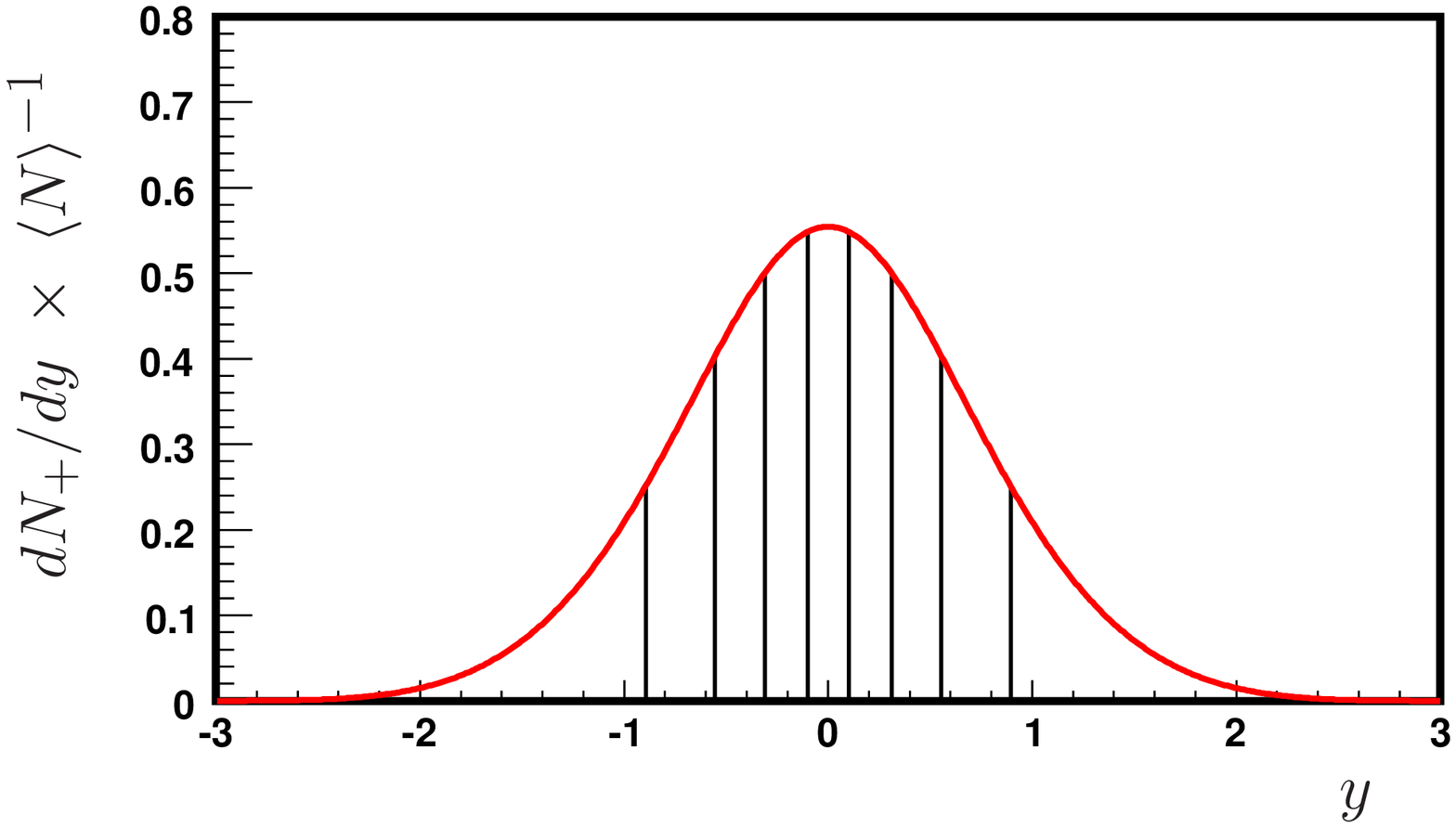,width=7.2cm,height=6.5cm}   
  \caption{Differential particle spectra for a `pion gas' at~$T=0.160$~GeV. Transverse momentum spectrum~({\it left}) 
  and rapidity spectrum~({\it right}). Both curves are normalized to unity. The bins are constructed such that each bin 
  contains~$1/9$ of the total yield.}    
  \label{Boltzmannspectra}   
\end{figure}

The momentum spectra are assumed to be ideal GCE spectra due to the large volume approximation. 
In Fig.(\ref{Boltzmannspectra}) transverse momentum and rapidity spectra are shown for MB statistics. BE and FD 
statistics yield similar spectra, unless chemical potentials are large. Momentum bins~$\Delta p_{T,i}$
and~$\Delta y_i$ are indicated by drop-lines.

\begin{figure}[ht!]   
  \epsfig{file=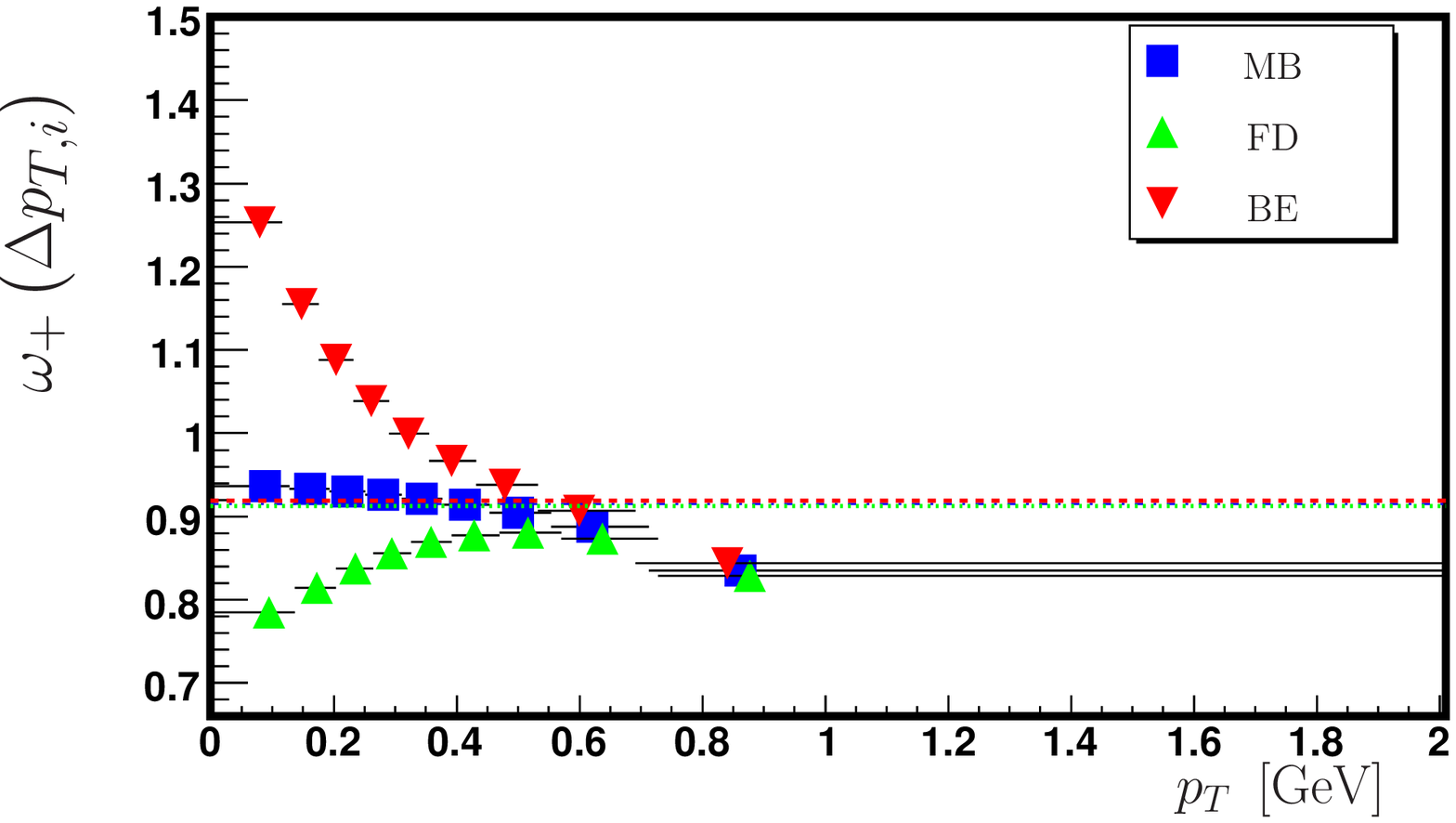,width=7.2cm,height=6.5cm}   
  \epsfig{file=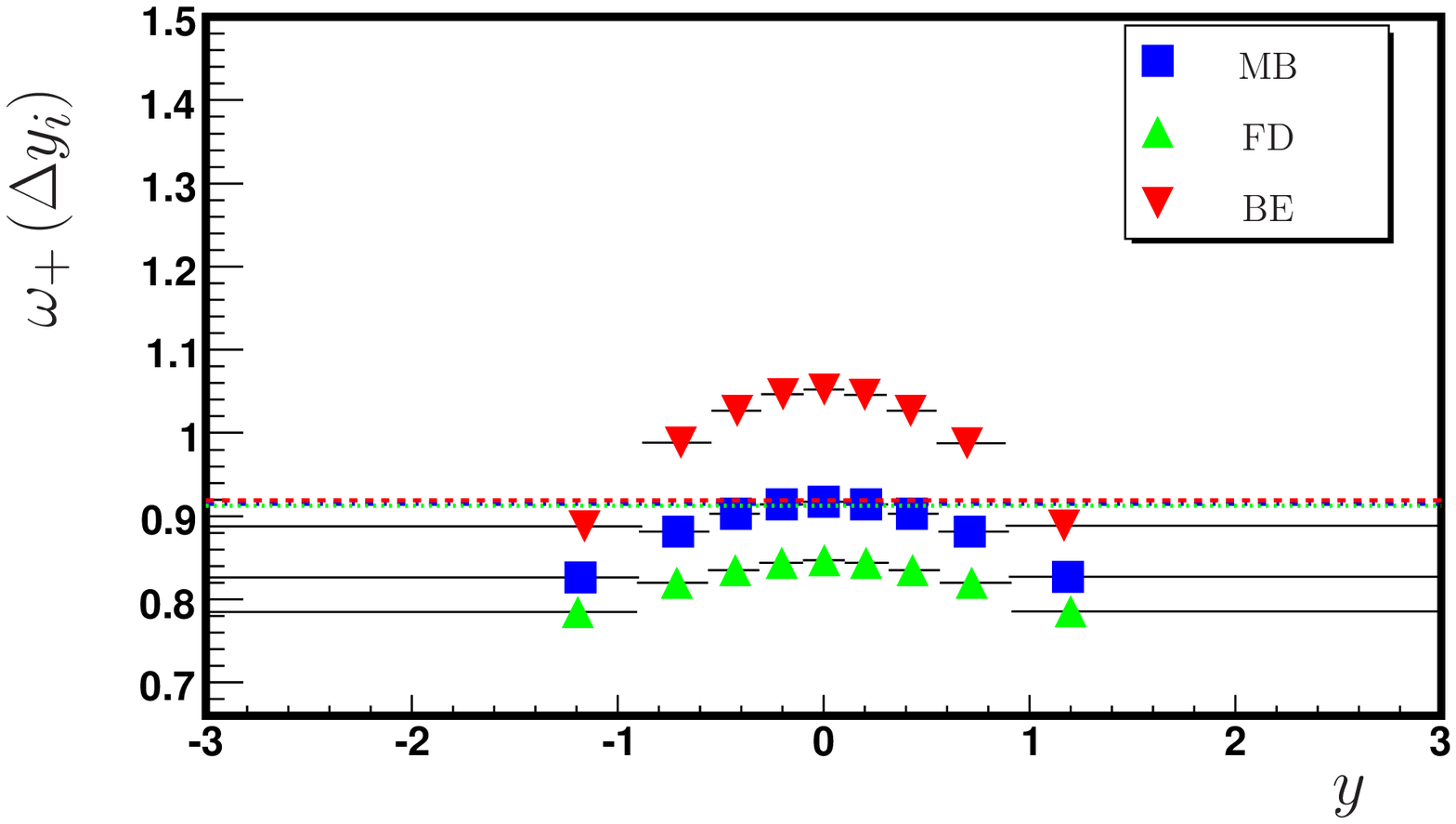,width=7.2cm,height=6.5cm}   
  \caption{Transverse momentum~({\it left}) and rapidity dependence~({\it right}) of the MCE scaled variance of 
  positively charged particles at {$T=0.160$~GeV}, for a MB (blue), FD (green), BE (red) `pion gas'  at zero charge 
  density. Momentum bins are constructed such that each bin contains the same fraction~$q$ of the average~$\pi^+$   
  yield. The horizontal bars indicate the width of the~$\Delta p_{T,i}$ or~$\Delta y_i$ bins, while the marker indicates 
  the position of the center of gravity of the corresponding bin. Dashed lines indicate acceptance scaling results.}     
  \label{StaticSource_dodp}   
\end{figure}

In Fig.(\ref{StaticSource_dodp}) the scaled variance of positively charged particles~$\omega_+$ is presented within 
different transverse momentum bins~$\Delta p_{T,i}$~({\it left}) and rapidity bins~$\Delta y_i$~({\it right}).  
The scaled variance in limited bins of momentum space is more sensitive  to the choice of particle statistics than the 
spectra would suggest. BE and FD effects are particularly strong in momentum space bins in which occupation numbers 
are large. Hence, at the low momentum tail one finds suppression of fluctuations for FD and enhancement for BE, while 
at the high momentum tail, one finds~$\omega_{BE} \simeq \omega_{MB} \simeq \omega_{FD}$, 
Fig.(\ref{StaticSource_dodp})~({\it left}). 

\begin{figure}[ht!]   
  \epsfig{file=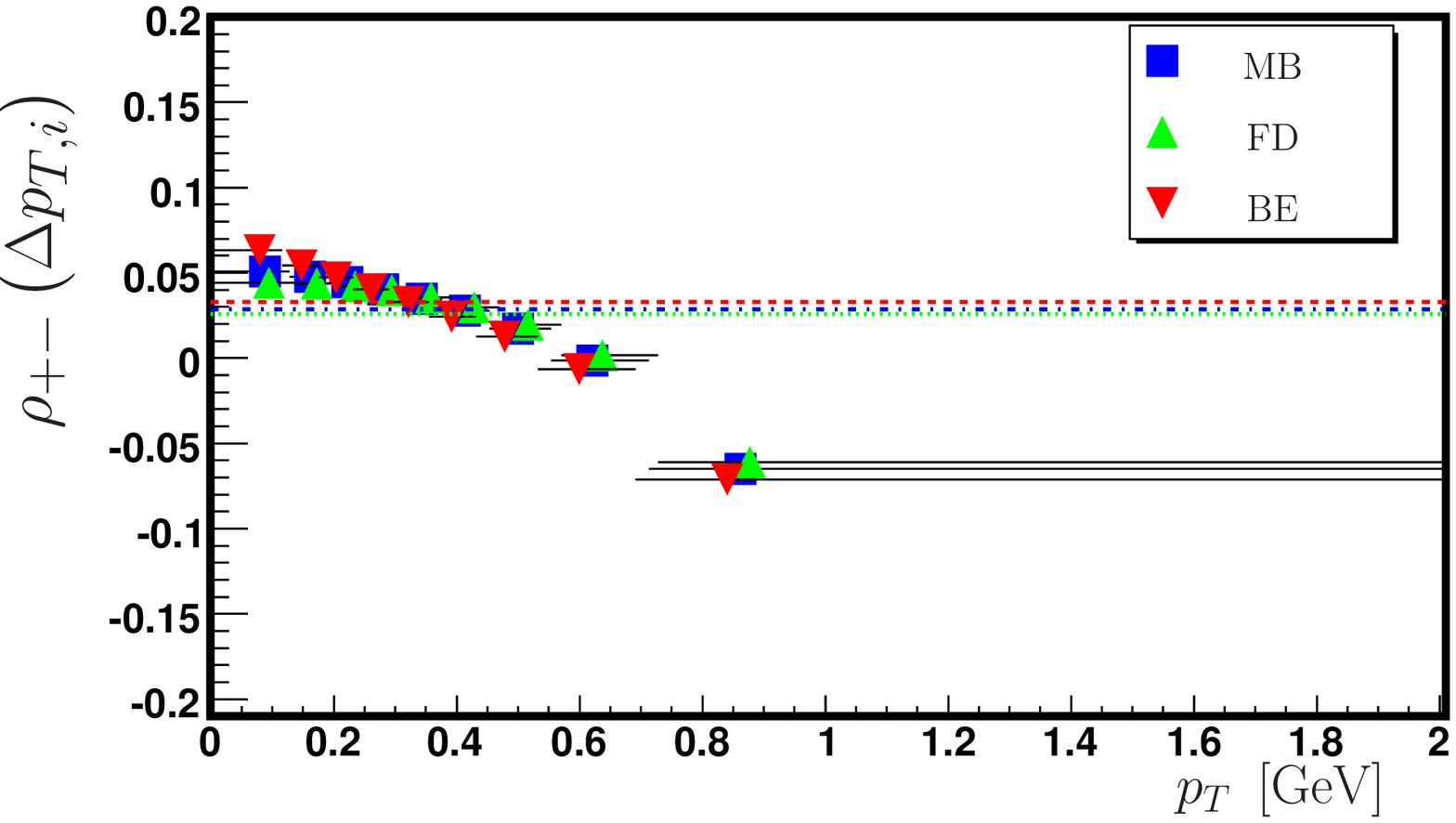,width=7.2cm,height=6.5cm}   
  \epsfig{file=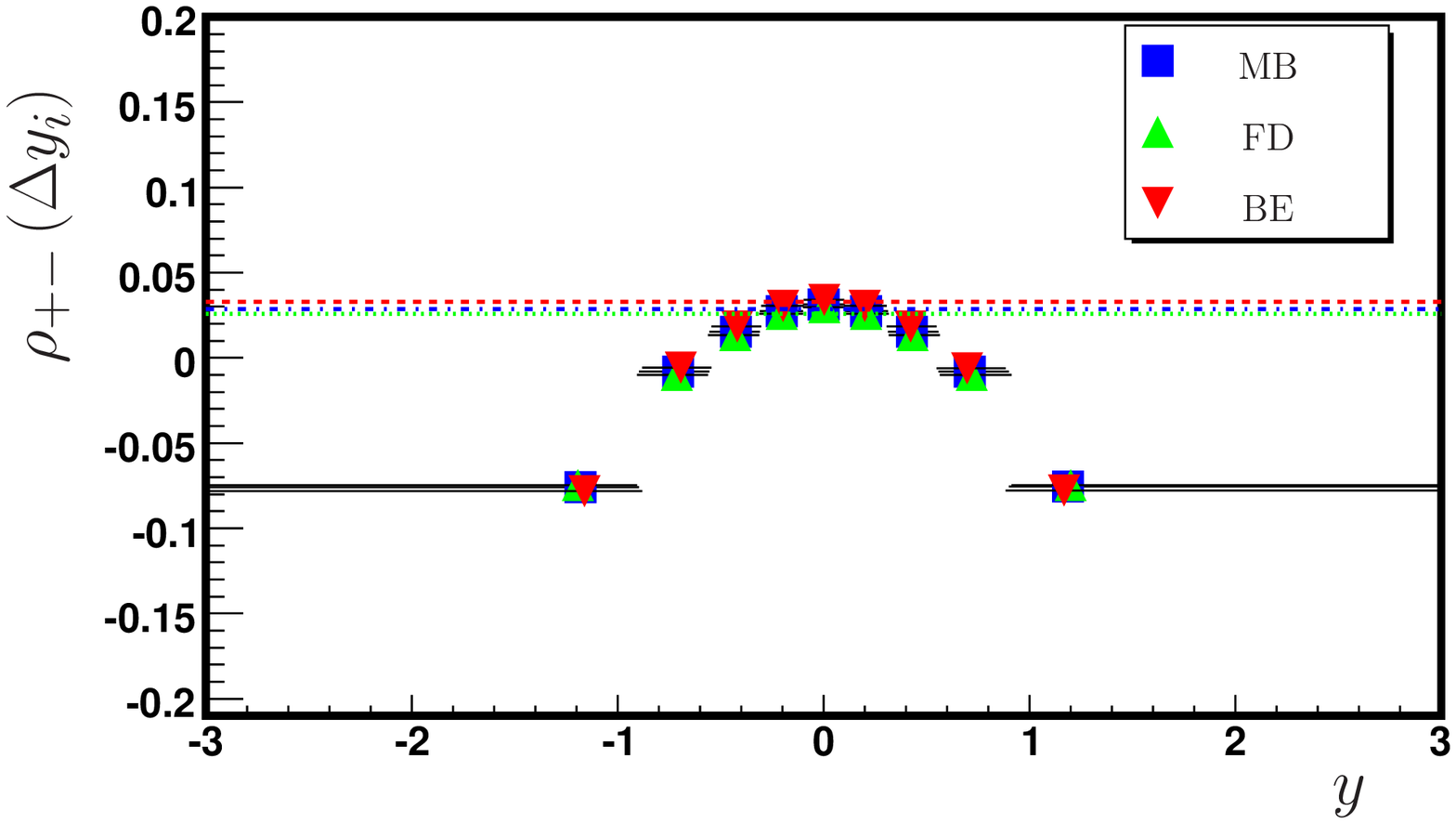,width=7.2cm,height=6.5cm}   
  \caption{Same as Fig.(\ref{StaticSource_dodp}), but for the MCE correlation coefficient between positively and 
  negatively charged particles. Dashed lines indicate acceptance scaling results.}     
  \label{StaticSource_drhodp}   
\end{figure}   

The rapidity dependence, Fig.(\ref{StaticSource_dodp})~({\it right}), has a different behavior. The reason is, 
that in any~$\Delta y_i$ bin there is some contribution from a low~$p_T$ tail of the differential momentum 
spectrum~$dN / dy / dp_{T}$ where quantum statistics effects are pronounced. This leads to a clear separation of the 
curves and one finds~$\omega_{BE}~>~\omega_{MB}~>~\omega_{FD}$. In contrast to this, the fully phase space 
integrated (all particles observed) scaled variance is rather insensitive to the choice of statistics~\cite{Begun:2005iv}  
(unless chemical potentials are large). There are in fact three different `acceptance scaling` lines in 
Fig.(\ref{StaticSource_dodp}), which extrapolate the fully phase space integrated scaled variance to limited acceptance. 
The differences are however very small and all three lines lie practically on top of each other. 
  
Fig.(\ref{StaticSource_drhodp}) presents the correlation coefficient~$\rho_{+-}$ between positively and negatively 
charged particles in transverse momentum bins~$\Delta p_{T,i}$~({\it left}) and rapidity bins~$\Delta y_i$~({\it right}).   
The fully phase space integrated correlation coefficient between positively and negatively charged particles would 
be~$\rho_{+-} =1$ in the CE and MCE. In the GCE it would be~$0$. In the MB CE it would not show any momentum space 
dependence and would always be~$\rho_{+-}>0$. In the MCE the situation is qualitatively different: in low momentum bins 
particles are positively correlated, while in high momentum bins they can even be anti-correlated. Horizontal lines 
again indicate acceptance scaling. Quantum effects for the correlation coefficient remain small as there is no explicit 
local (quantum) correlations between particles of different charge.  

\begin{figure}[ht!]   
  \epsfig{file=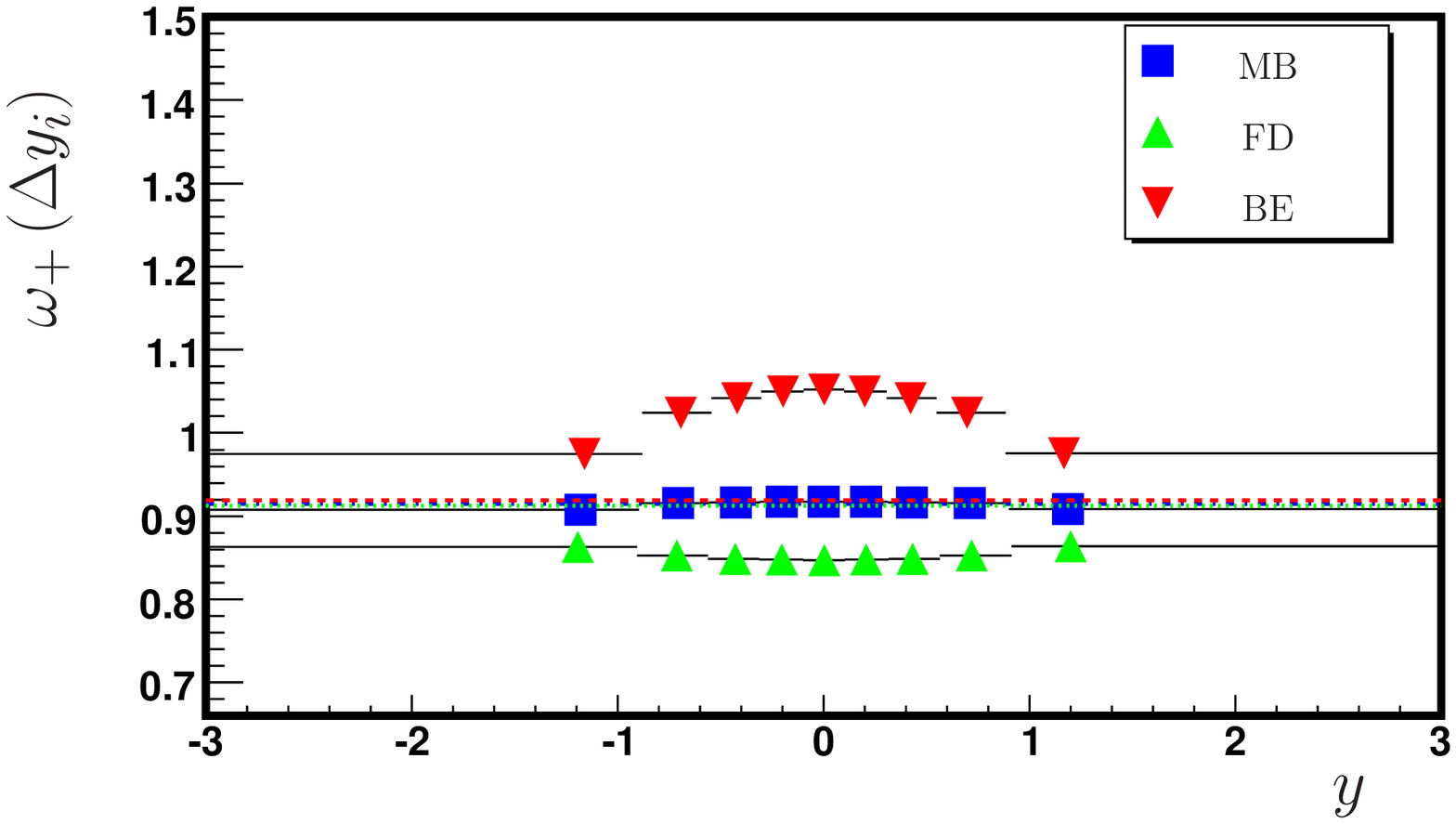,width=7.2cm,height=6.5cm}   
  \epsfig{file=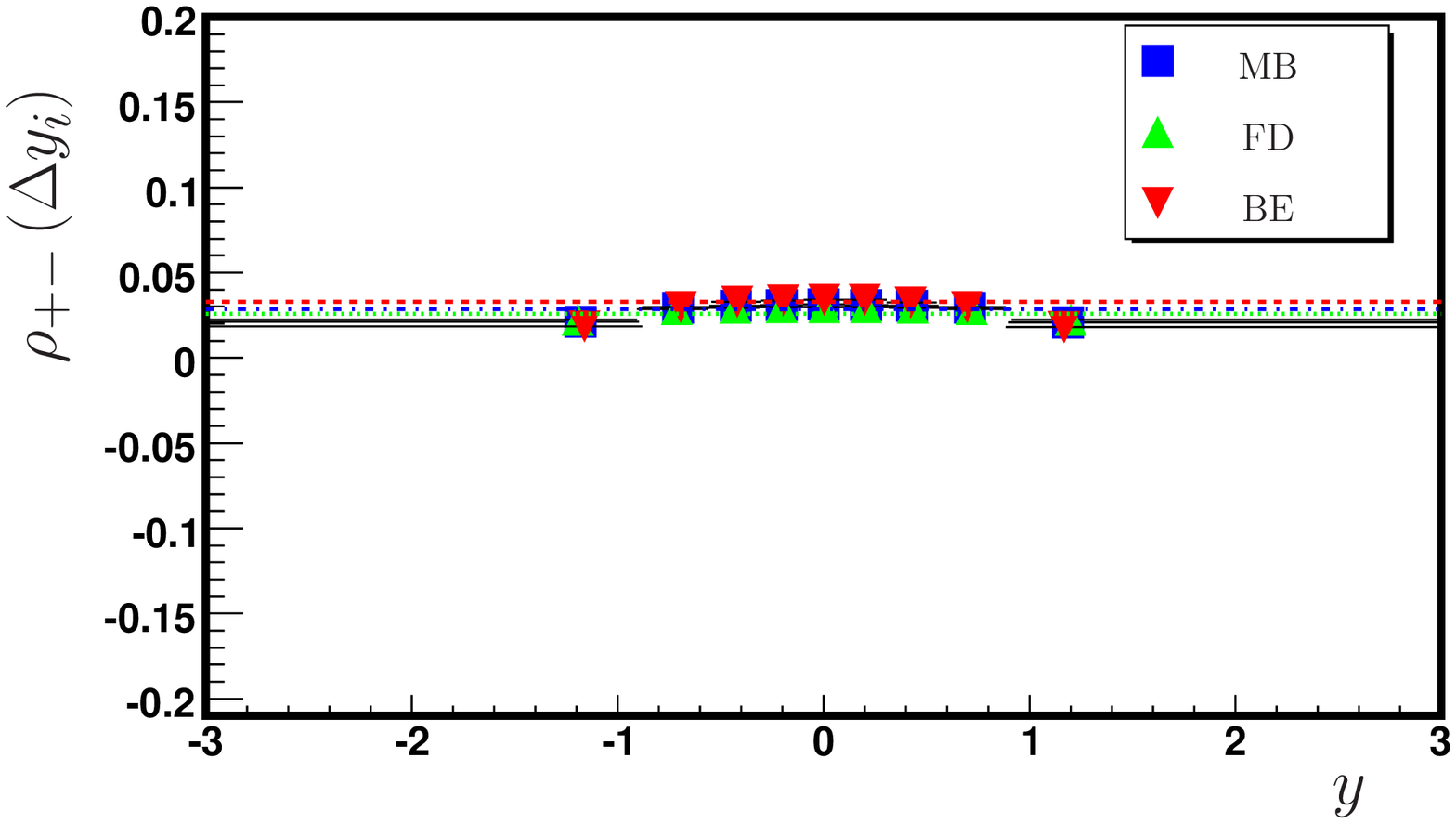,width=7.2cm,height=6.5cm}   
  \caption{Rapidity dependence of the scaled variance of positively charged particles~({\it left})
  and of the correlation coefficient between positively and negatively charged particles~({\it right}).
  Calculations are done for the same system as in Figs.(\ref{StaticSource_dodp},\ref{StaticSource_drhodp}), 
  however disregarding momentum conservation.}     
  \label{StaticSource_dxdy_nopz}   
\end{figure}

It should be stressed that the~$\Delta p_{T,i}$ dependence in Figs.(\ref{StaticSource_dodp},\ref{StaticSource_drhodp}) 
is a direct consequence of energy conservation. The~$\Delta y_i$ dependence of~$\omega_+$ and~$\rho_{+-}$, however, is 
due to joint energy and longitudinal momentum ($P_z$) conservation. Disregarding~$P_z$ conservation leads to a 
substantially milder~$\Delta y_i$ dependence, see Fig.(\ref{StaticSource_dxdy_nopz}).  

This behavior can be intuitively explained: in a low momentum bin it is comparatively easy to balance charge, as each 
individual particle carries little energy and momentum. In contrast to this, in a high momentum bin with, say an excess 
of positively charged particles, it is unfavorable to balance charge, as one would also have to have more than on average   
negatively charged particles, and each particle carries large energy and momentum. This leads to suppressed fluctuations 
and correlations in high momentum bins when compared to low momentum bins.  
  
\begin{figure}[ht!]
  \epsfig{file=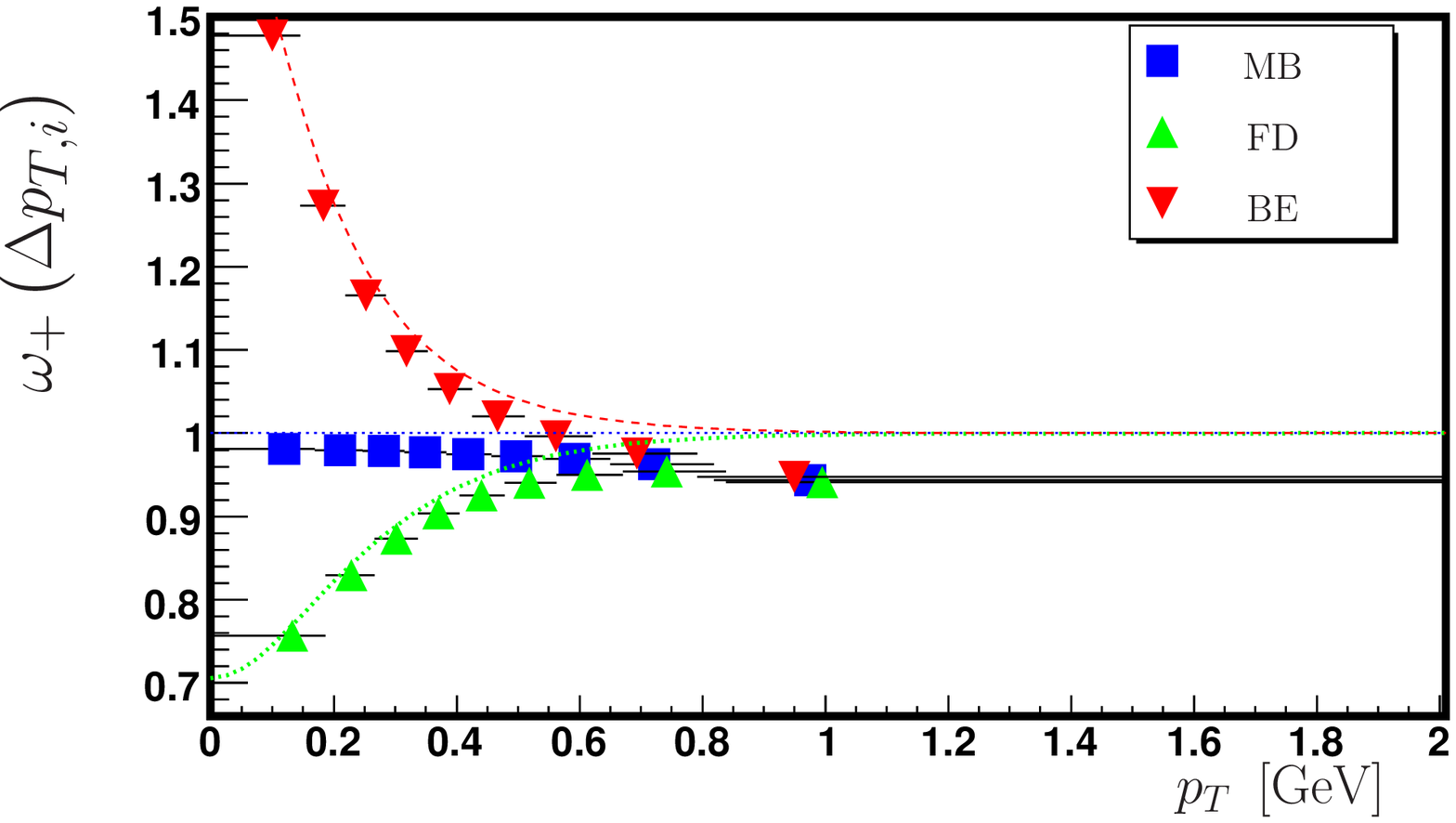,width=7.2cm,height=6.5cm}   
  \epsfig{file=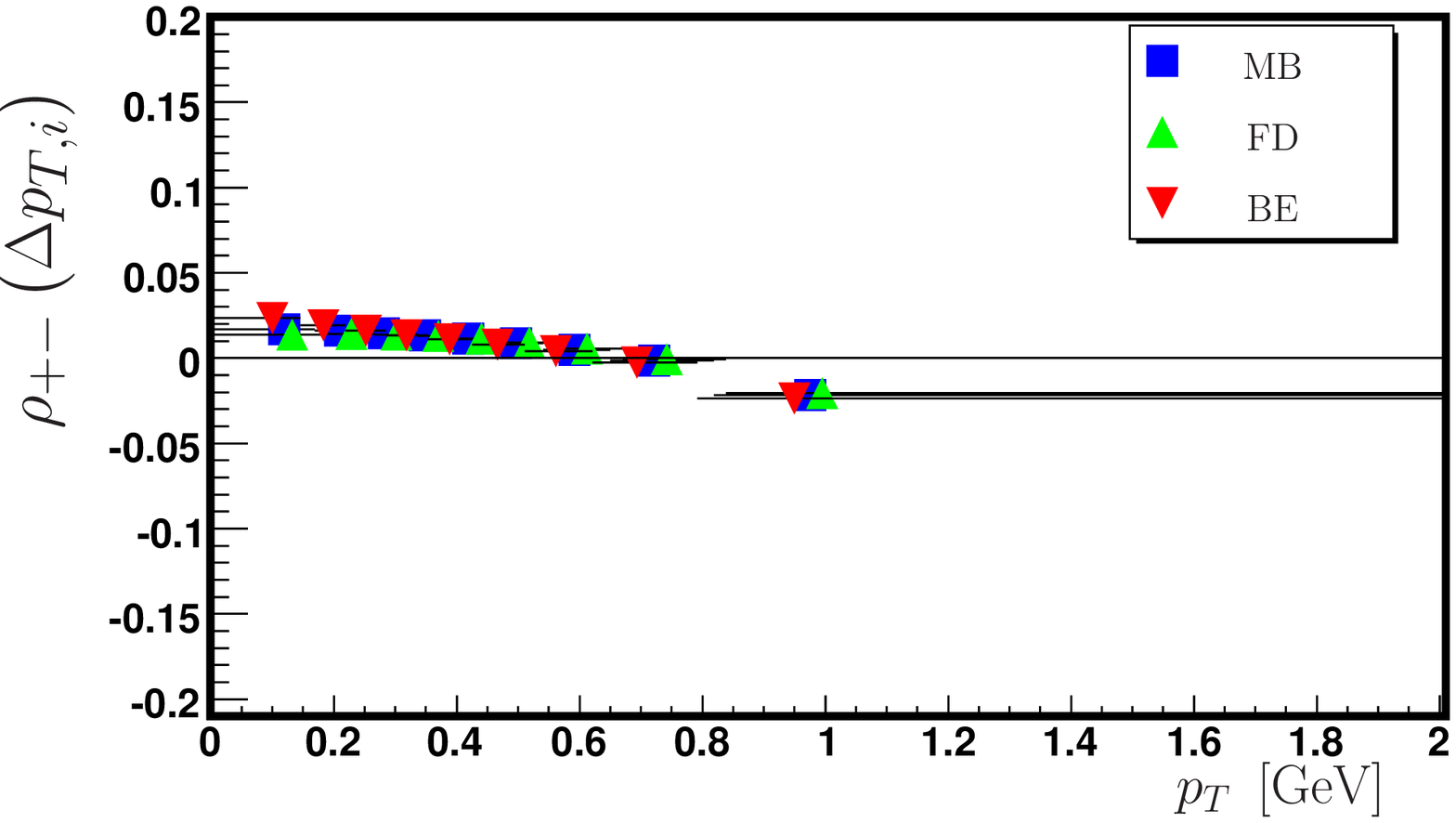,width=7.2cm,height=6.5cm}   
  \caption{Transverse momentum dependence of MCE scaled variance of negatively charged particles~({\it left}) and 
  the MCE correlation coefficient between positively and negatively charged particles~({\it right}). Only particles 
  in a mid-rapidity window~$|y|<0.3$ are measured.  Dashed lines denote the GCE result, Eq.(\ref{GCE_estimate}).}     
  \label{StaticSource_dodpt_VI}   
\end{figure}   

In a small mid-rapidity window, with~$|y|<0.3$, the effects of globally applied motional conservation laws cease to be 
important (see Fig.(\ref{StaticSource_dodpt_VI})). Local correlations due to BE and FD statistics begin to dominate, 
and MCE deviations from the GCE results, Eq.(\ref{GCE_estimate}), are relevant only for the highest momentum bins.    
In BE or FD statistics one finds for vanishing bin size~($\delta y,~\delta p_T$): 

\begin{equation}\label{GCE_estimate}   
\omega^{GCE}_{\delta} ~=~ \frac{\kappa_{2}^{N_{\delta},N_{\delta}}}{\kappa_{1}^{N_{\delta}}}  
~\simeq~ \frac{1}{1\pm e^{- \beta m_T \cosh y + \beta \mu}}~.  
\end{equation}   

BE and FD effects are strongest around mid-rapidity~$y=0$, especially at low~$p_T$, where occupation numbers are large. 
MCE calculations in Fig.(\ref{StaticSource_dodpt_VI}) are close to the GCE estimate Eq.(\ref{GCE_estimate}).   
In MB statistics  one finds only a weak~$\Delta p_T$ dependence in a small mid-rapidity window.  
Please note that the acceptance scaling procedure predicts a Poisson distribution with~$\omega_+ \rightarrow 1$ 
and~$\rho_{+-} \rightarrow 0$ for all three statistics in the limit of very small acceptance. 
  
\subsubsection{Collectively Moving System}   

As established a long time ago, in order to properly define the thermodynamics of a system with collective motion, 
the partition function needs to be Lorentz invariant~\cite{Pathria:1957fix_1,Touschek:1968fix_1}. 
The expectation values of observables need hence to transform according to the Lorentz transformation properties of 
these observables. In particular, the inverse temperature~$\beta =T^{-1}$ is promoted to a four-vector~$\beta^{\mu}$ 
(combining local temperature with collective velocity). The entropy, as well as particle multiplicities, 
remain Lorentz-scalars.

These requirements are in general not satisfied unless momentum conservation is put on an equal status with energy 
conservation. If the system is described by a MCE, then momentum should be conserved as well as 
energy~\cite{Pathria:1957fix_1,Touschek:1968fix_1}. If the system is exchanging energy with a bath, it needs to 
exchange momentum as well. 
 
For ensemble averages, neglecting these rules and treating momentum differently from energy is safe as long as the 
system is close to the thermodynamic limit, since there ensembles become equivalent. The same is not true for 
fluctuation and correlation observables, which remain ensemble-specific~\cite{Begun:2004gs}. For a system at rest, 
these requirements are not apparent since the net-momentum is zero. Statistical mechanics observables in a 
collectively moving system, however, lose their Lorentz invariance, if this is not maintained in the definition 
of the partition function.

To illustrate this point, the properties of a system moving along the~$z$-axis with a collective boost~$y_0$ are 
considered. The total energy of the fireball is then~${E=M \cosh(y_0)}$, while its total momentum is given 
by~${P_z=M \sinh(y_0)}$. The mass of the fireball in its rest frame is~$M=P^{\mu} u_{\mu}$. The system four-temperature 
is~$\beta^{\mu} = \beta u^{\mu}$. Local temperature and chemical potentials remain unchanged. The second rank tensor (or 
co-variance matrix)~$\kappa_2$, Eq.(\ref{kappa}), will be discussed in this section.

The second order cumulant~$\kappa_2$, Eq.(\ref{kappa}), is given by the second derivatives of the cumulant generating 
function with respect to the fugacities. Essentially this is the Hessian matrix of the function Eq.(\ref{Psi}), 
encoding the structure of its minima. The diagonal elements~$\kappa_2^{X,X}$ are the variances of the GCE distributions 
of extensive quantities~$X$. For example,~$\kappa_2^{N_A,N_A}$ measures the GCE variance of the distribution of particle   
multiplicity of species~$A$, while~$\kappa_2^{Q,Q}$ denotes the GCE electric charge fluctuations, etc. The 
off-diagonal~$ \kappa_2^{X,Y}$ elements give GCE co-variances of two extensive quantities~$X$ and~$Y$.    
  
For a boost along the~$z$-axis the general co-variance matrix for a relativistic gas with one conserved charge reads:     
\begin{align}   
\kappa_2 =    
\begin{pmatrix}   
\kappa_2^{N_A,N_A} & \kappa_2^{N_A,N_B} & \kappa_2^{N_A,Q} &   
\kappa_2^{N_A,E} & \kappa_2^{N_A,P_x} & \kappa_2^{N_A,P_y} & \kappa_2^{N_A,P_z} \\    
\kappa_2^{N_B,N_A} & \kappa_2^{N_B,N_B} & \kappa_2^{N_B,Q} &   
\kappa_2^{N_B,E} & \kappa_2^{N_B,P_x} & \kappa_2^{N_B,P_y} & \kappa_2^{N_B,P_z} \\    
\kappa_2^{Q,N_A} & \kappa_2^{Q,N_B} & \kappa_2^{Q,Q} & \kappa_2^{Q,E}    
& 0 & 0 &  0 \\    
\kappa_2^{E,N_A} & \kappa_2^{E,N_B} & \kappa_2^{E,Q} & \kappa_2^{E,E}    
& 0 & 0 & \kappa_2^{E,P_z}\\    
\kappa_2^{P_x,N_A} & \kappa_2^{P_x,N_B} & 0 & 0 & \kappa_2^{P_x,P_x} & 0 & 0 \\    
\kappa_2^{P_y,N_A} & \kappa_2^{P_y,N_B} & 0 & 0 & 0 & \kappa_2^{P_y,P_y} & 0 \\    
\kappa_2^{P_z,N_A} & \kappa_2^{P_z,N_B} & 0 & \kappa_2^{P_z,E} & 0 & 0 &  \kappa_2^{P_z,P_z}    
\end{pmatrix}~. \label{matrix}   
\end{align}   
Off-diagonal elements correlating a globally conserved charge with one of the momenta, i.e.~$\kappa_2^{Q,P_x}$, as well 
as elements denoting correlations between different momenta, i.e.~$\kappa_2^{P_x,P_y}$, are equal to zero due to 
antisymmetric momentum integrals.  The values of elements correlating particle multiplicity and momenta, 
i.e.~$\kappa_2^{N_A,P_x}$, depend strongly on the acceptance cuts applied. For fully phase space integrated multiplicity 
fluctuations and correlations these elements are equal to~$0$, again due to antisymmetric momentum integrals.  
  
It is instructive to review the transformation properties of~$\kappa$ under the Lorentz group:~$\kappa_2^{X,Y}$ contains 
the correlations between four-momenta~$P^{\mu}$ and, in general, (scalar) conserved quantities and particle 
multiplicities~$Q^j$. Hence, the elements~$\kappa_2^{P^{\mu},P^{\nu}}$, i.e.~$\langle \Delta P^{\mu} \Delta P^{\nu} \rangle$, 
will transform as a tensor of rank 2 under Lorentz transformations;~$\kappa_2^{P^{\mu},Q^j }$, 
i.e.~$ \langle \Delta P^{\mu} \Delta Q^j \rangle$, will transform as a vector (the rapidity distribution will 
simply shift); and the remaining~$\kappa_2^{Q^j,Q^k }$ will be scalars.  
    
\begin{figure}[ht!]   
  \epsfig{file=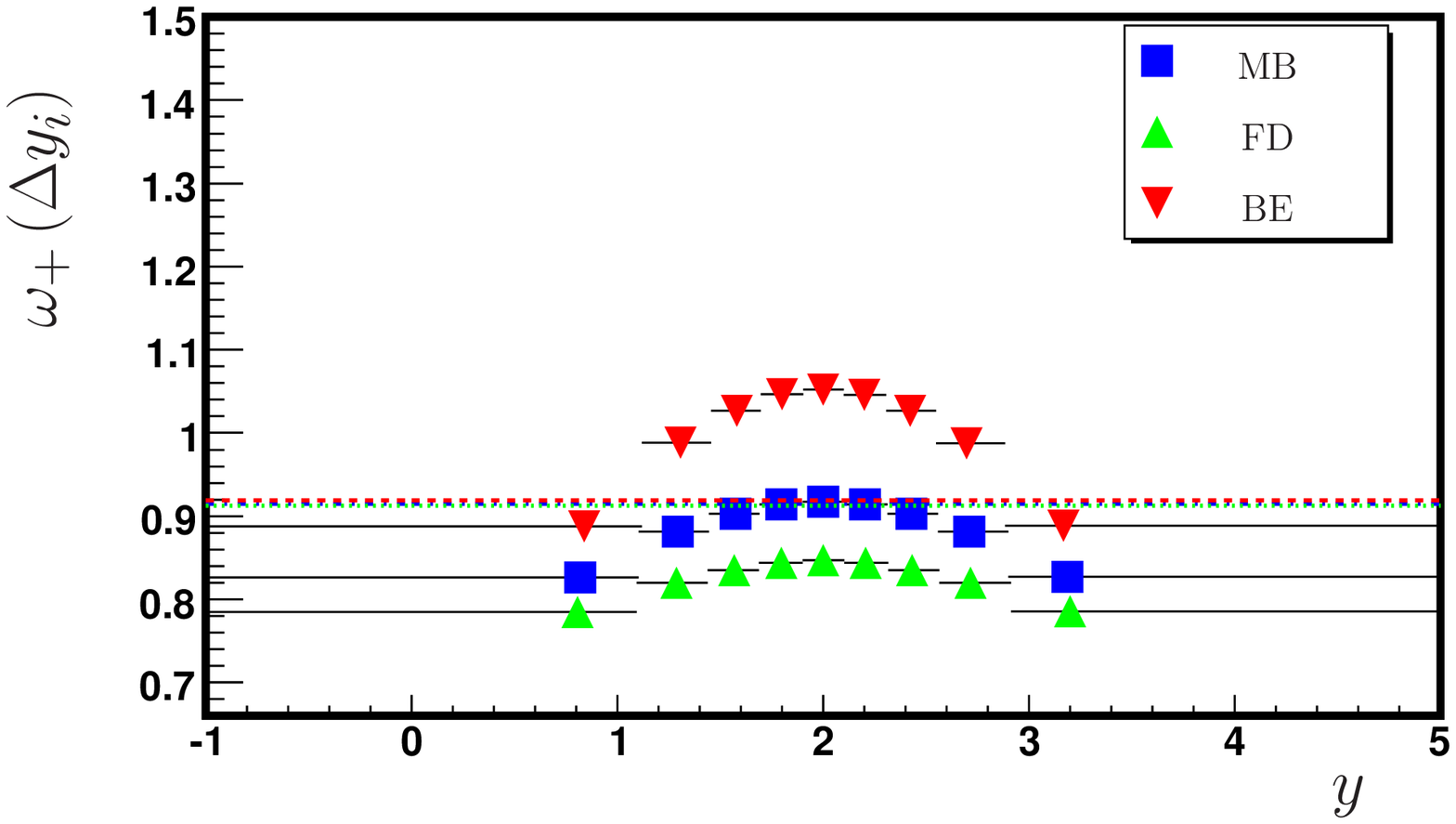,width=7.2cm,height=6.5cm}   
  \epsfig{file=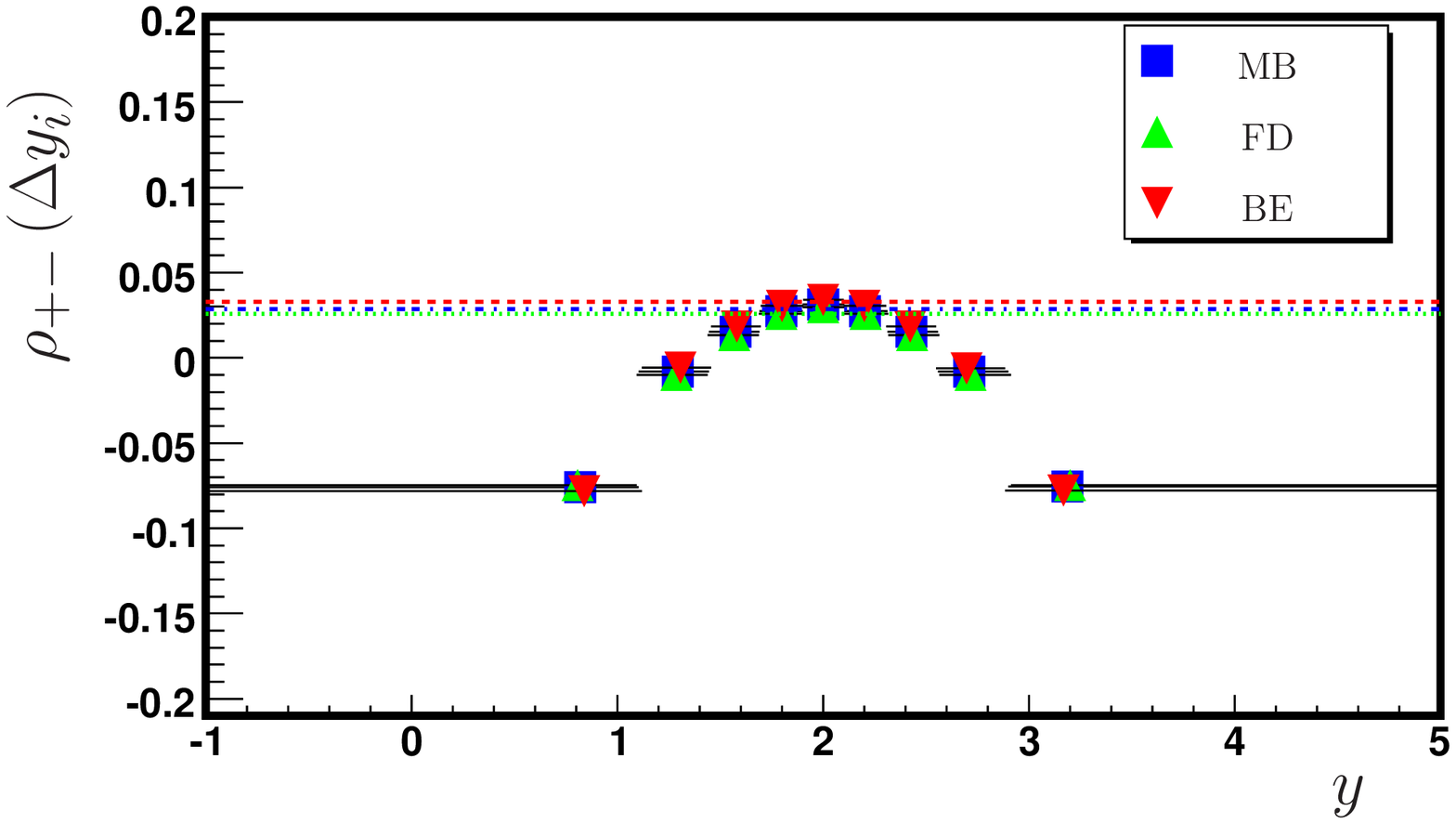,width=7.2cm,height=6.5cm}   
  \caption{Same as Fig.(\ref{StaticSource_dodp})~({\it right}) and Fig.(\ref{StaticSource_drhodp})~({\it right}), 
  but for a system moving with~$y_0 = 2$.}      
  \label{MovingSource_dodp}   
\end{figure}

\begin{figure}[ht!]   
  \epsfig{file=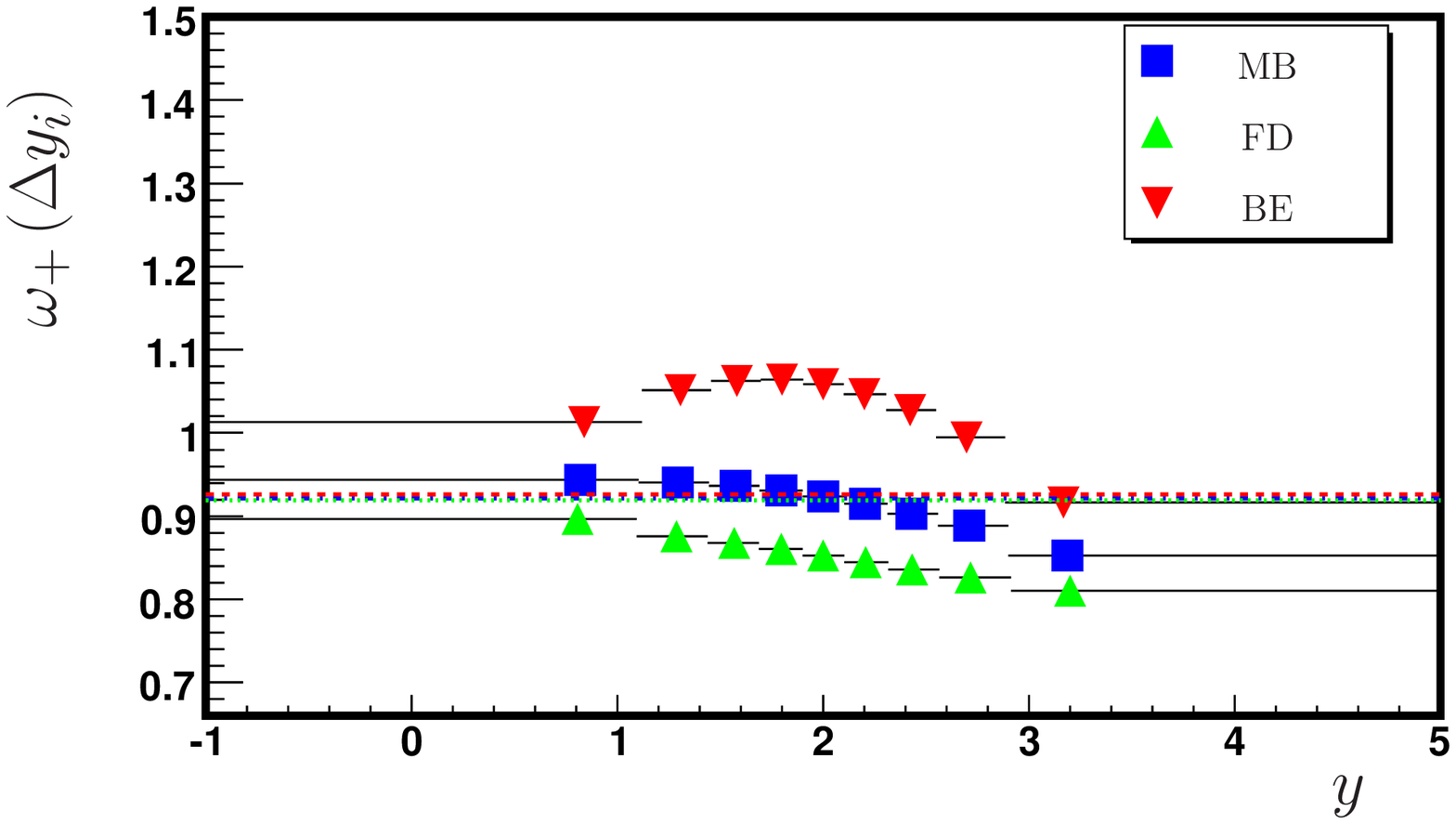,width=7.2cm,height=6.5cm}   
  \epsfig{file=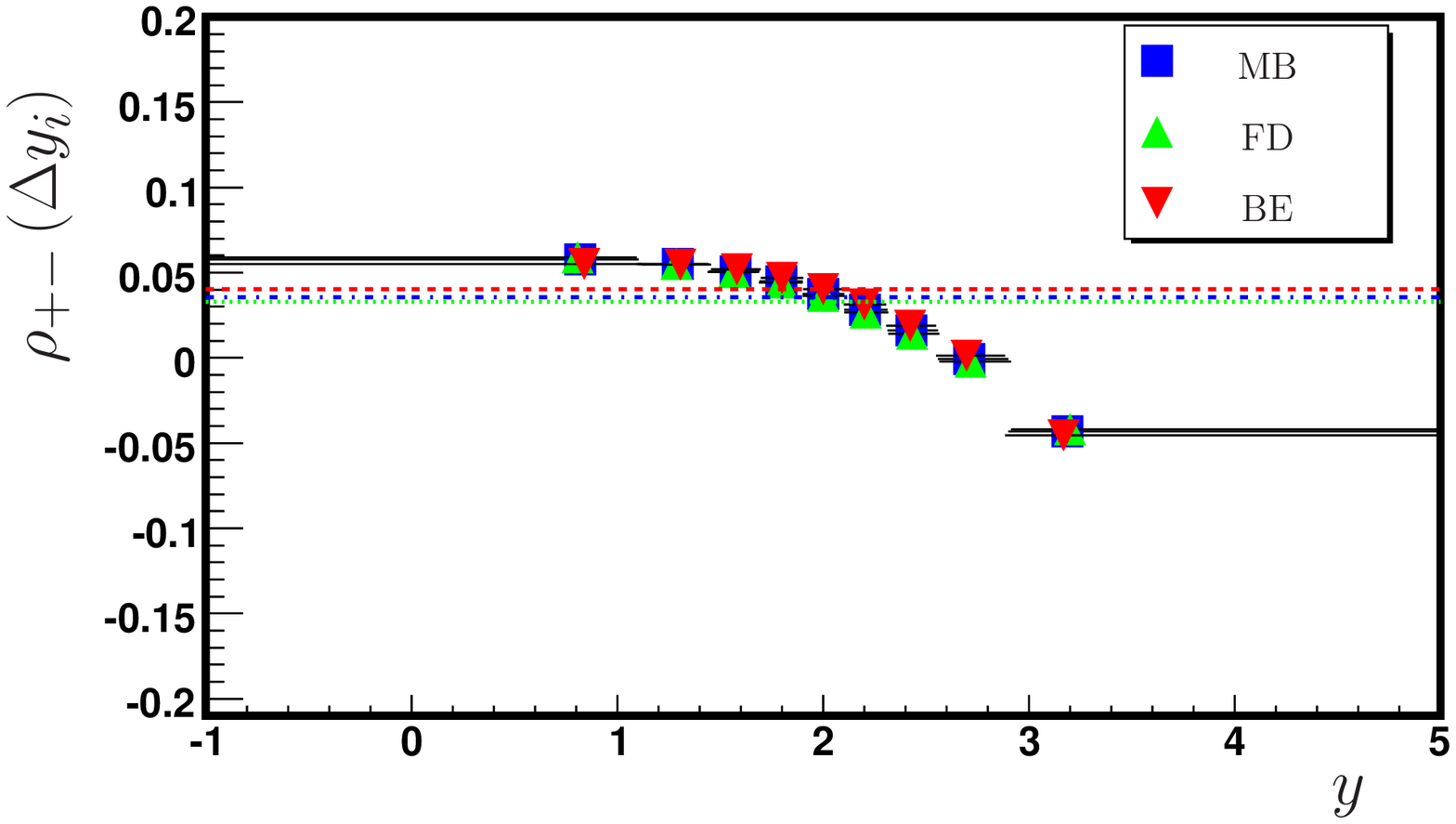,width=7.2cm,height=6.5cm}   
  \caption{Same as Fig.(\ref{MovingSource_dodp}), but without~$P_z$ exact conservation.}       
  \label{MovingSource_dodp_nopz}   
\end{figure}  

For a static system the one finds for the co-variances of energy and 
momenta~{$\kappa_2^{E,P_x}=\kappa_2^{E,P_y}=\kappa_2^{E,P_z}=0$}.  Under these two conditions, a static system and full 
particle acceptance, 
i.e.~$\kappa_2^{N_A,P_x}=\kappa_2^{N_B,P_x}=\kappa_2^{N_A,P_y}=\dots=0$, some eigenvalues of the matrix Eq.(\ref{matrix}) 
factorize, and momentum  conservation can be shown to drop out of the calculation~\cite{Hauer:2007im}.   
For a boost along the~$z$-axis (and arbitrary particle acceptance) non-vanishing 
elements~$\kappa_2^{P_z,E} = \kappa_2^{E,P_z} \not= 0$ appear, and ensure that the determinant of the 
matrix~$\kappa_2$, Eq.(\ref{matrix}), remains invariant against such a boost. Note that 
still~$\kappa_2^{P_x,E} = \kappa_2^{P_y,E} = 0$. Transverse momenta~$P_x$ and~$P_y$ remain un-correlated with energy~$E$.
   
In Fig.(\ref{MovingSource_dodp}) multiplicity fluctuations~({\it left}) and correlations~({\it right}) are shown for 
a system with boost~$y_0=2$. The rapidity spectrum of Fig.(\ref{Boltzmannspectra})~({\it right}) is simply 
shifted to the right by two units. The construction of  the acceptance bins is done as before. Multiplicity fluctuations 
and correlations within a rapidity bin~$\Delta y_i$ transform as a vector (i.e., its~$z$ component shifts in rapidity) 
as inferred from their Lorentz-transformation properties, provided both energy and momentum along the 
boost direction are conserved.      
  
This last point deserves attention because usually, starting from~\cite{Fermi:1950fix_1}, micro canonical calculations 
only conserve energy and not momentum. Imposing exact conservation for energy, and only average conservation of 
momentum will make the system non-Lorentz invariant, since in a different frame from the co-moving one energy and 
momentum will mix, resulting in micro state by micro state fluctuations in both momentum and energy. This result is 
not obvious, because energy-momentum is a vector of separately conserved currents. It is therefore natural to assume 
that these currents can be treated within different ensembles; they are, after all, conserved separately.
It must be kept in mind, however, that it is not energy or momentum, but particles that are exchanged between the 
system and any canonical or grand canonical bath. The amount of energy and momentum carried by each particle are 
correlated by the dispersion relation~\cite{Touschek:1968fix_1}.   

In the situation examined here (unlike in a Cooper-Frye formalism~\cite{Cooper:1974mv}, where the system is frozen out on 
a hyper surface, a space-time four-vector correlated with four-momentum) all time dependence within 
the system under consideration is absent due to the equilibrium assumption. Furthermore, the system is entirely thermal: 
the correlation between particle numbers when the system is sampled at different times is a~$\delta-$function, that 
stays a~$\delta-$function under all Lorentz-transformations.  Hence, unlike what happens in a Cooper-Frye freeze-out, 
energy-momentum and space-time do not mix in the partition function.  Together with the constraint from the particle 
dispersion relations, this means that different components of the four-momentum need to be treated by the same ensemble.  
   
This situation is explicitly shown in Fig.(\ref{MovingSource_dodp_nopz}). Here, multiplicity fluctuations and 
correlations are calculated for the same system as in Fig.(\ref{MovingSource_dodp}), but with exact conservation of only 
energy (and charge), but not momentum. In the co-moving frame of the system, the fluctuations and correlations are 
identical to Fig.(\ref{StaticSource_dxdy_nopz}). When the system is boosted the distribution changes (not only by a 
shift in rapidity, as required by Lorentz-invariance), and loses its symmetry around the system's average boost.
  
This last effect can be understood from the fact that momentum does not have to be conserved event-by-event, but 
energy does.  It is easier, therefore, to create a particle with less longitudinal momentum (energy) than the average, 
than with larger longitudinal momentum (energy), and still conserve energy overall. This leads to suppressed multiplicity 
fluctuations and a negative correlation coefficient for rapidity bins in the forward direction in comparison to rapidity 
bins in the backward direction. In Fig.(\ref{MovingSource_dodp}), where the system also needs to conserve momentum exactly, 
this enhancement is balanced by the fact that it will be more difficult to conserve momentum globally when particles 
having less momentum than the boost are created.  

\section{Correlations between disconnected Momentum Bins}   
\label{sec_multfluc_pgas_acrossbins}
Next, correlations between particles in different momentum bins are studied. `Long range correlations' between bins well 
disconnected in momentum have been suggested to arise from dynamical processes. Examples include color glass   
condensate~\cite{Armesto:2006bv,Lappi:2006fp}, droplet formation driven hadronization \cite{Torrieri:2007fb}, and phase 
transitions within a percolation-type mechanism~\cite{Armesto:2006bv,Bialas:1985jb}. The elliptic flow measurements, 
widely believed to signify the production of a liquid at RHIC~\cite{Adams:2005dq,Arsene:2004fa,Back:2004je,Adcox:2004mh}, 
are also, ultimately, correlations between particles disconnected in phase space (here, the azimuthal angle).  
  
As will be shown, conservation laws will also introduce such correlations between any two (connected or not) distinct 
regions of momentum space. No dynamical effects are taken into consideration (only an isotropic thermal system).  
  
\begin{figure}[ht!]   
  \epsfig{file=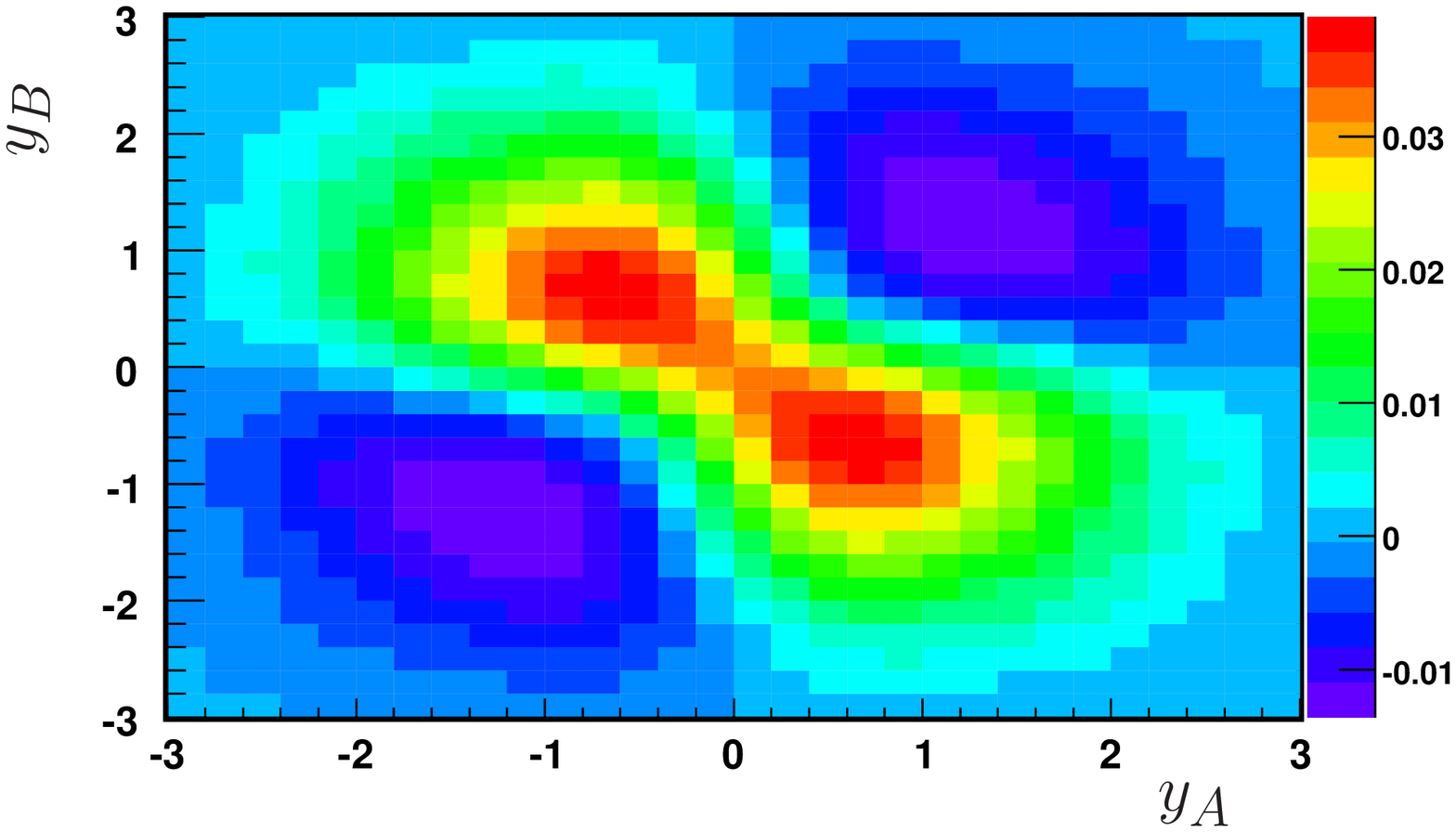,width=7.cm,height=6.5cm}   
  \epsfig{file=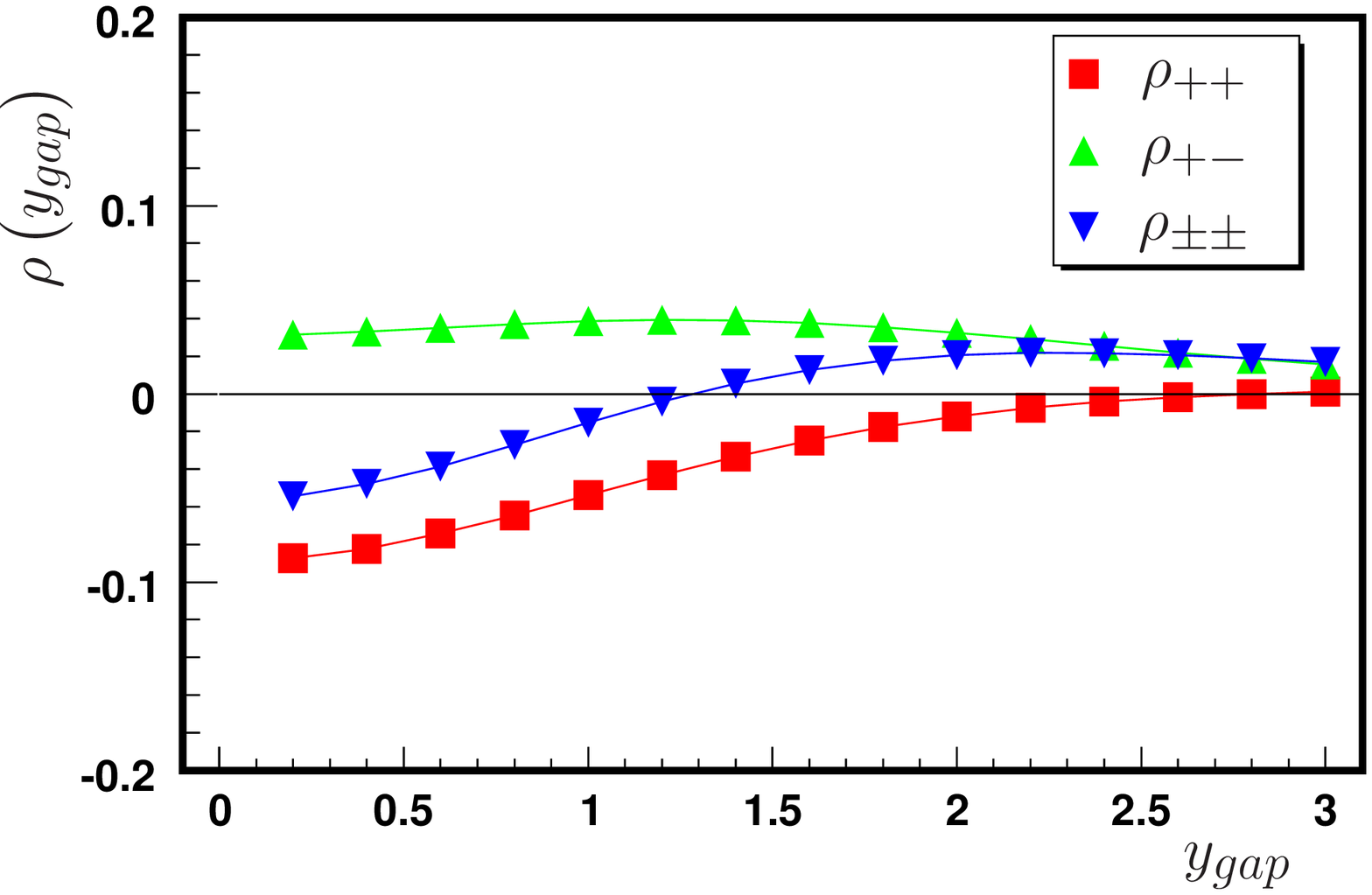,width=7.cm,height=6.5cm}   
  \caption{({\it Left}): The correlation coefficient~$\rho_{+-}$ between the multiplicity of positively charged particles 
  in a bin located at~$y_A$ with negatively charged particles in a bin located at~$y_B$, both with a 0.2 width in rapidity. 
  ({\it Right}): The correlation coefficients~$\rho_{+-}$,~$\rho_{++}$, and~$\rho_{\pm \pm}$ between multiplicities 
  in two bins separated by~$y_{gap}$ of like, unlike, and all charged particles. Both plots show MCE MB results.}  
  \label{FB_corr}  
\end{figure}

\begin{figure}[ht!]    
  \epsfig{file=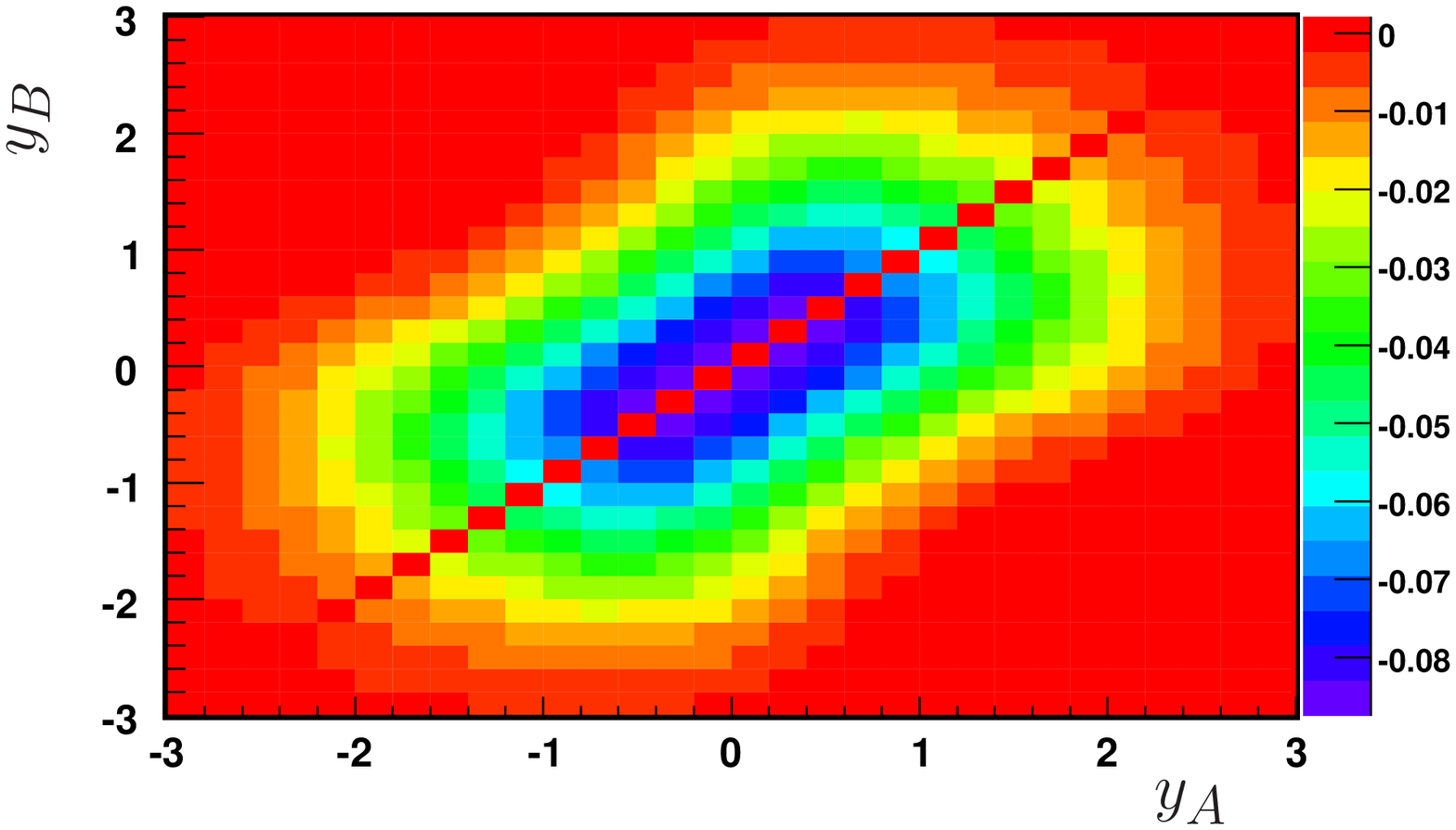,width=7.cm,height=6.5cm}   
  \epsfig{file=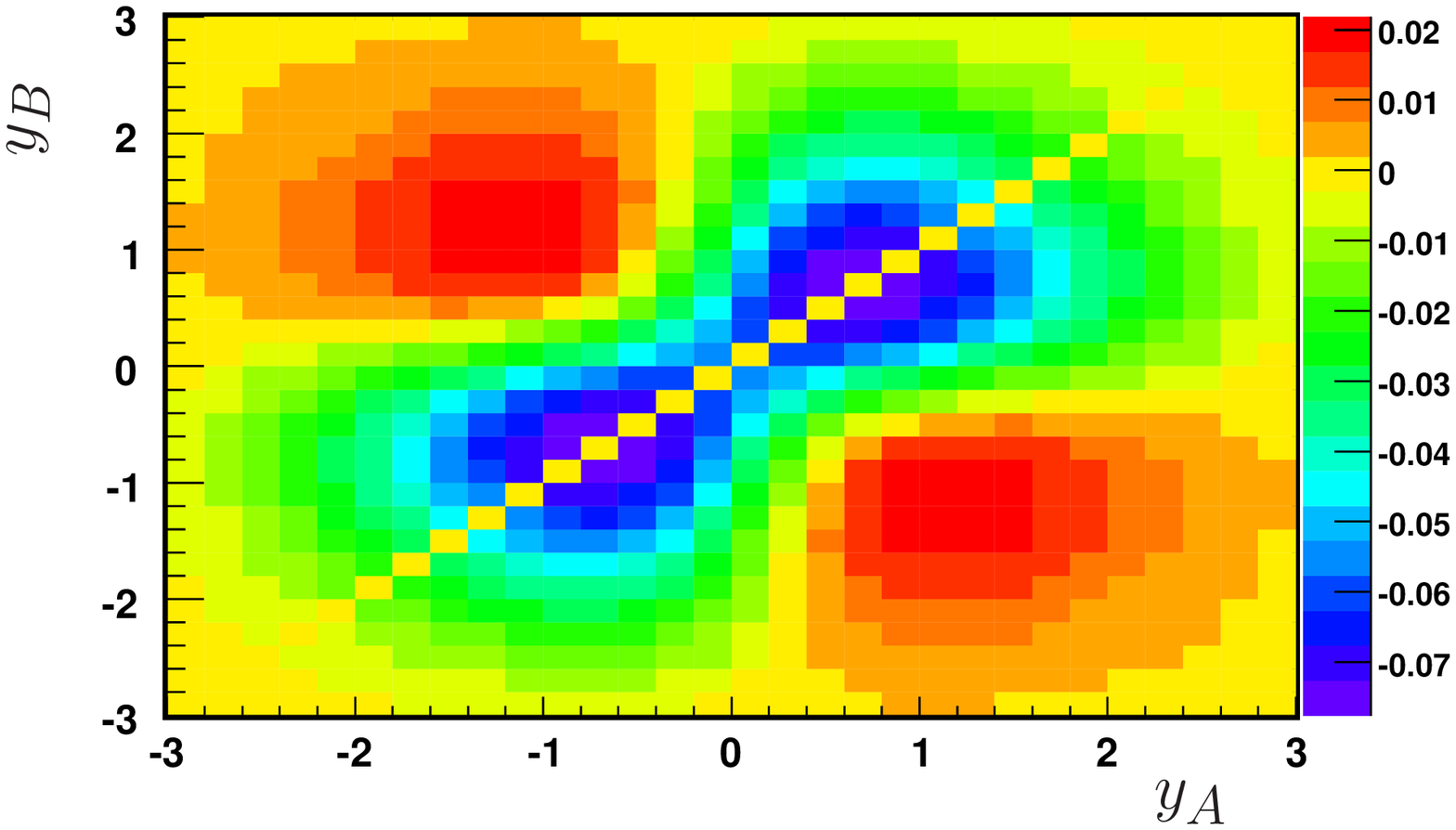,width=7.cm,height=6.5cm}   
  \caption{The correlation coefficient~$\rho_{++}$ between the multiplicity of positively charged particles in a bin 
  located at~$y_A$ with positively charged particles in a bin located at~$y_B$~({\it left}), and the correlation 
  coefficient~$\rho_{\pm \pm}$ between the multiplicity of all charged particles in a bin located at~$y_A$ with all 
  charged particles in a bin located at~$y_B$~({\it right}). Both plots show MCE MB results. }  
  \label{FB_corr_chr}  
\end{figure}

Considering correlations between the multiplicities of particles~$A$ and~$B$, within two bins, each centered around 
the rapidities~$y_A$ and~$y_B$, with (constant) widths~$\Delta y_A= \Delta y_B =0.2$, in Fig.(\ref{FB_corr})~({\it left}) 
the correlation coefficient~$\rho_{+-}$ between positively and negatively charged particles are shown as a function 
of~$y_A$ and~$y_B$. Fig.(\ref{FB_corr})~({\it right}) presents the correlation coefficients between 
like-charge~$\rho_{++}$, unlike-charge~$\rho_{+-}$, and all charged particles~$\rho_{\pm \pm}$, as a function   
of the separation of the two bin centers~${y_{gap} = y_A - y_B}$. In Fig.(\ref{FB_corr_chr}) the correlation 
coefficients~$\rho_{++}$~({\it left}) and~$\rho_{\pm \pm}$~({\it right}) are shown as a function of~$y_A$ and~$y_B$. 
The main diagonal contains multiplicities in the same momentum bin. In the case of~$\rho_{++}$ and~$\rho_{\pm \pm}$ this 
leads to a double counting of each particle and hence to~$\rho = 1$.

Energy conservation always leads to anti-correlation between different momentum space bins. Charge conservation leads 
to a positive correlation of unlike charged particles and anti-correlation of like-sign particles. Longitudinal momentum 
conservation, however, is responsible for the structure visible in Fig.(\ref{FB_corr})({\it left}). Having a small 
(large) number of particles of one charge in a bin with positive average longitudinal momentum, leads to a larger 
(smaller) number of particles of opposite charge in a bin with different but also positive~$P_z$, (blue dips). This makes 
also a state with smaller (larger) particle number with opposite average longitudinal momentum~$-P_z$ more likely (red 
hills). At large values of~$y_A$ the correlation coefficient~$\rho \simeq 0$ for any~$y_B$, because   
the yield~$\langle N_A \rangle$ in~$\Delta y_A$ is asymptotically vanishing. 
  
The correlation coefficients~$\rho_{++}$, Fig.(\ref{FB_corr_chr})({\it left}), and~$\rho_{\pm \pm}$, 
Fig.(\ref{FB_corr_chr})({\it right}), show strong anti-correlations around center-rapidity, as to be expected from 
charge, energy and momentum conservation. In the case of~$\rho_{++}$ the correlation stays negative. For~$\rho_{\pm \pm}$
one observes a positive correlation between the numbers of all charged particles going in opposite directions, while 
seeing an anti-correlation for particle multiplicities measured in the same direction. This behavior is similar to that 
in Fig.(\ref{FB_corr})({\it left}). For~$\rho_{+-}$ energy, momentum, and charge conservation favor pairs of oppositely 
charge particles in opposite directions over particles in the same direction. For like sign particles the correlation 
stays negative, however milder for bins located in opposite direction. For not charge separated multiplicities the effect 
of charge conservation cancels out and energy and momentum conservation determine the correlation function.

In Fig.(\ref{FB_corr})~({\it right}) the correlation coefficient is presented along the diagonal from top left to 
bottom right as a function of separation~${y_{gap} = y_A - y_B}$. Unlike-sign particles are positively correlated. 
Like-sign and all charged particles are negatively correlated  at small separation~$y_{gap}$. For large separation the 
correlation becomes asymptotically zero, because the yield is zero. However, please note that in 
particular~$\rho_{\pm\pm} > \rho_{+-}$ at large~$y_{gap}$. Here~$P_z$ conservation is indeed dominant.   

\begin{figure}[ht!]   
  \epsfig{file=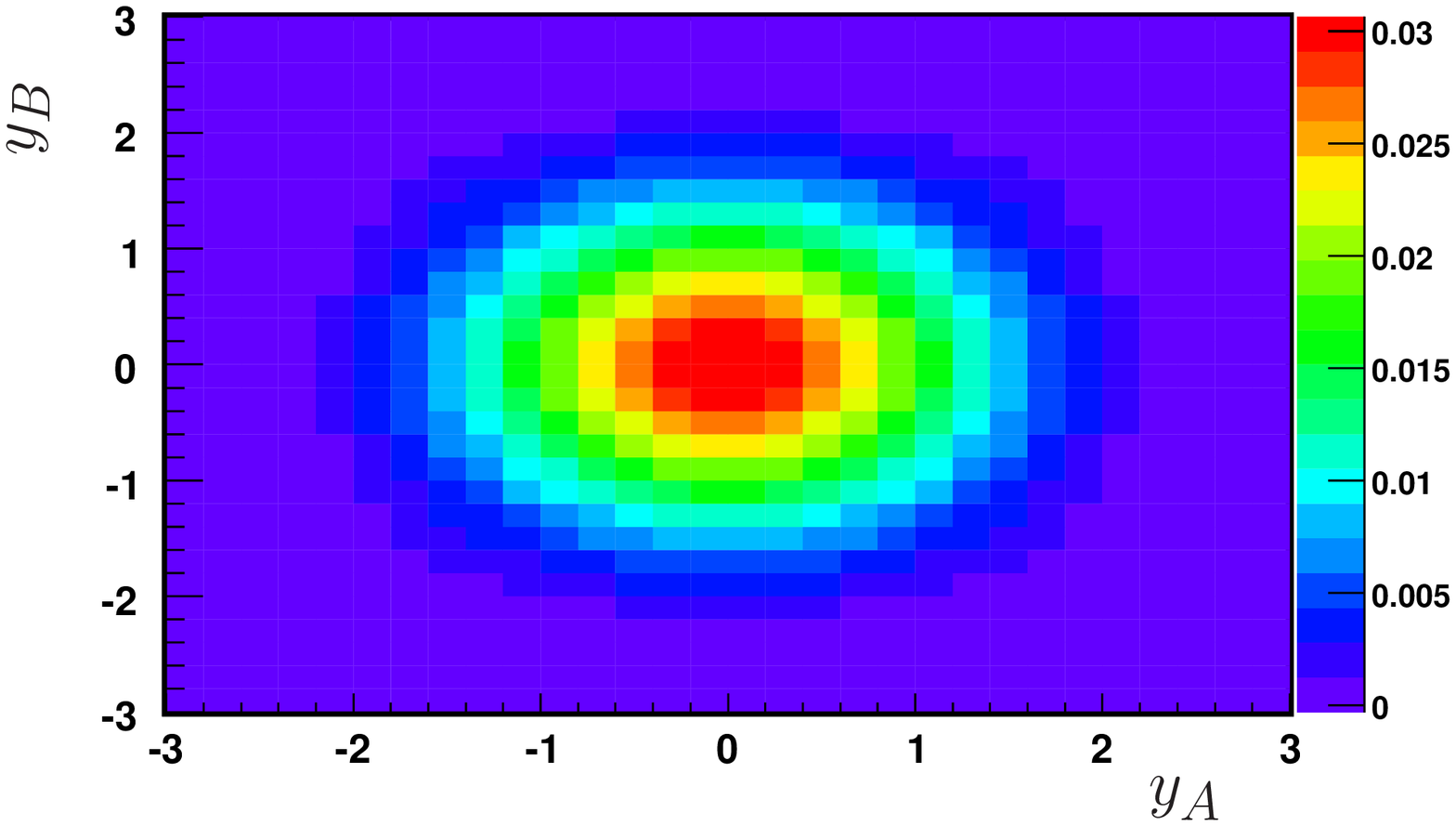,width=7.cm,height=6.5cm}   
  \epsfig{file=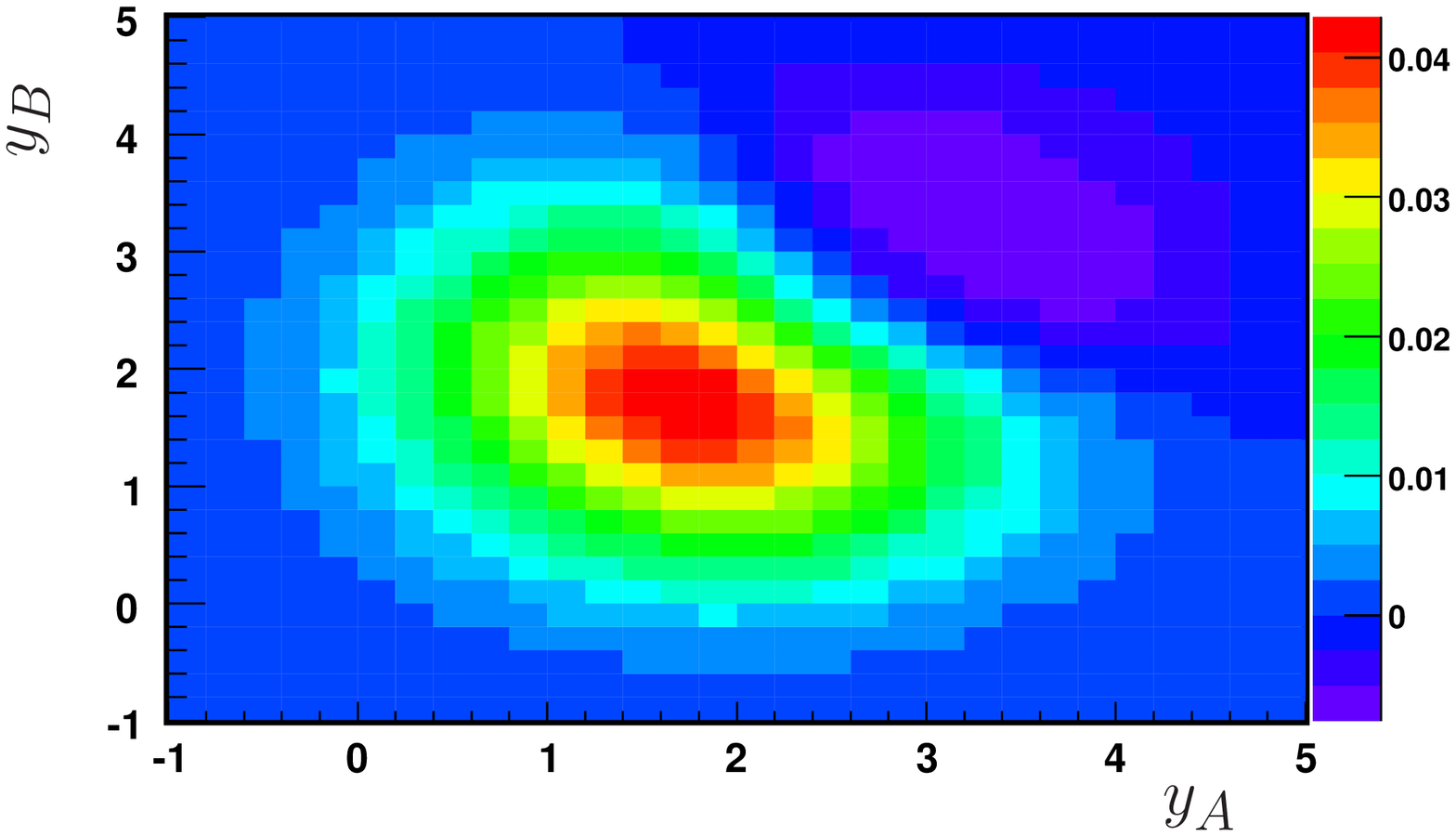,width=7.cm,height=6.5cm}   
  \caption{The correlation coefficient~$\rho_{+-}$ between the multiplicity of positively charged particles in a bin 
  located at~$y_A$ with negatively charged particles in a bin located at~$y_B$, without conservation of longitudinal 
  momentum~$P_z$, for a static source~({\it left}) and a source with a collective boost of~$y_0=2$~({\it right}). }
  \label{FB_corr_nopz}  
\end{figure}  

Fig.(\ref{FB_corr_nopz}) shows the correlation coefficient~$\rho_{+-}$ between positive and negative particles as   
a function of~$y_A$ and~$y_B$, this time disregarding momentum conservation, for a static source~({\it left}) and a 
source with a collective boost of~$y_0=2$~({\it right}). Disregarding~$P_z$ conservation destroys the particular 
structure in Fig.(\ref{FB_corr})~({\it left}) and leads to a single peak at the origin. The correlation is then 
insensitive to the momentum direction, and only sensitive to the energy content of a bin~$\Delta y$. The observables in 
Figs.(\ref{FB_corr},\ref{FB_corr_chr}) do not, Fig.(\ref{FB_corr_nopz})~({\it right}), transform under 
boosts ($y_{A,B} \rightarrow y_{A,B}-y_{0}$), unless momentum along the boost axis is exactly conserved.  

\begin{figure}[ht!]   
  \epsfig{file=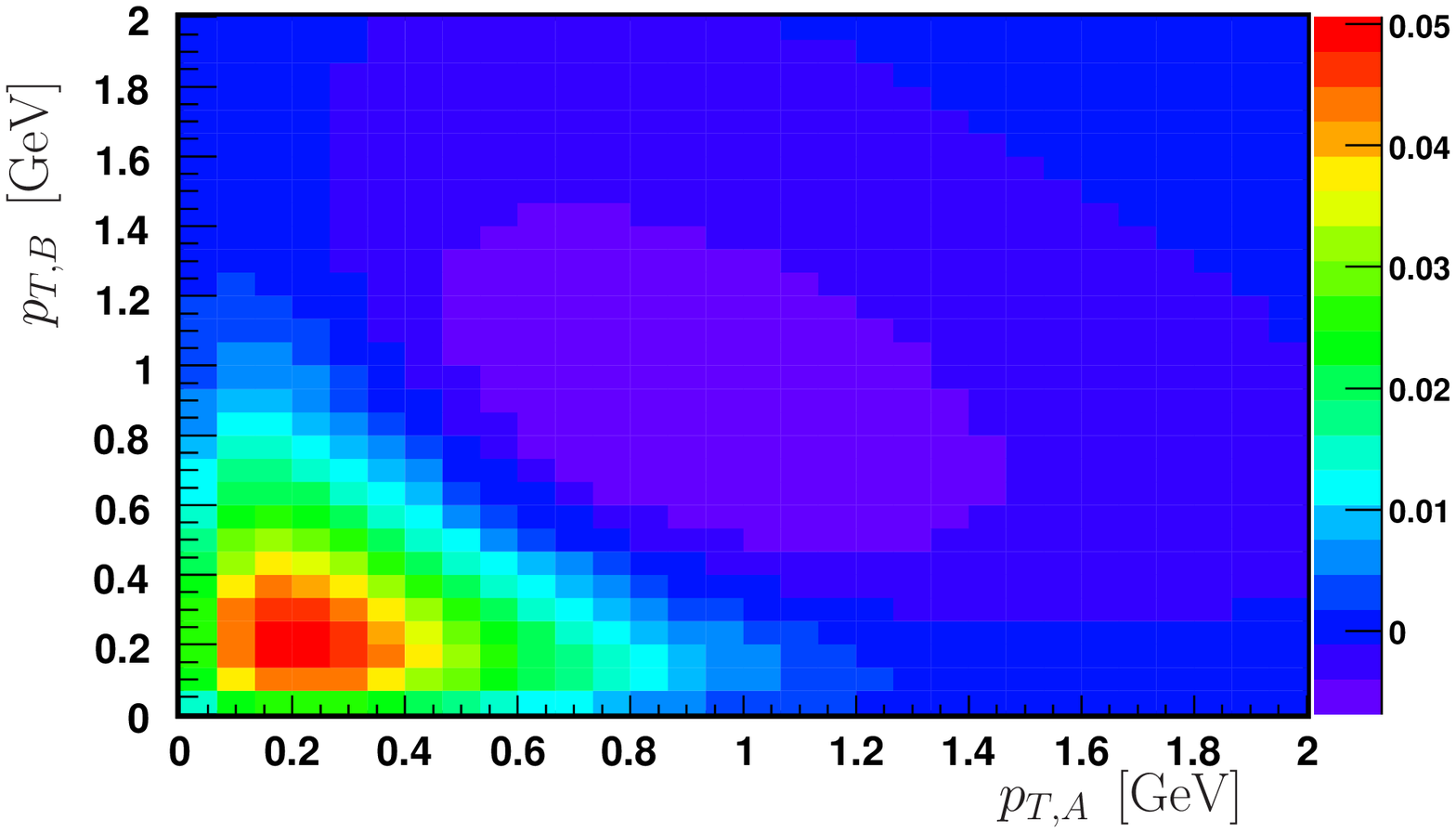,width=7.cm,height=6.5cm}   
  \epsfig{file=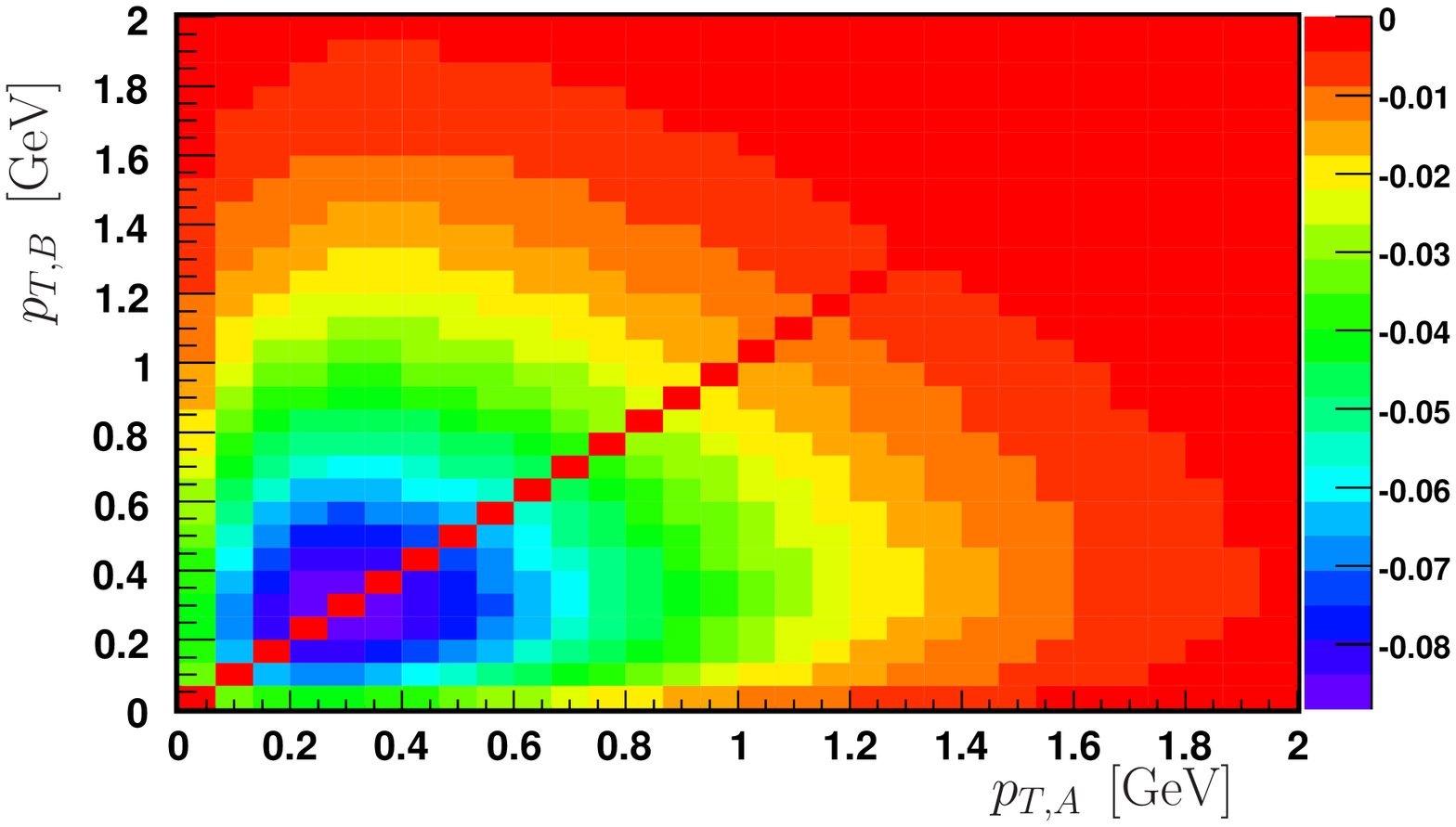,width=7.cm,height=6.5cm}   
  \caption{({\it Left:})~The correlation coefficient~$\rho_{+-}$ between the multiplicity of positively charged 
  particles in a bin located at~$p_{T,A}$ with negatively charged particles  in a bin located at~$p_{T,B}$.   
  ({\it Right:})~The correlation coefficient~$\rho_{++}$ between the multiplicity of positively charged particles in a 
  bin located at~$p_{T,A}$ with positively charged particles in a bin located at~$p_{T,B}$.}  
  \label{FB_corr_diff}  
\end{figure}

To complement the discussion of correlations across rapidity intervals, correlations between two distinct transverse 
momentum bins are studied. Here the discussion can be reduced to energy and charge conservation. 
Fig.(\ref{FB_corr_diff}) shows the correlation coefficient~$\rho_{+-}$ between the multiplicities of positively charged 
particles in transverse momentum bins of equal size ($\Delta p_T=0.2$~GeV) centered around~$p_{T,A}$ with the one of 
negatively charged particles in transverse momentum bins located at~$p_{T,B}$~({\it left}), and the correlation 
coefficient~$\rho_{++}$ between the multiplicities of positively charged particles in transverse momentum bins 
located at~$p_{T,A}$ with positively charged particles in bins located at~$p_{T,B}$~({\it right}). 

From charge conservation alone it follows that particles with electric charge of unlike sign are positively correlated,
while the multiplicities of particles of the same charge are negatively correlated. The main diagonal 
in Fig.(\ref{FB_corr_diff})~({\it right}) is excluded, as there~$\rho_{++}=1$ by construction. 
In Fig.(\ref{FB_corr_diff})~({\it left}) there is an island of weak anti-correlation. The correlation function has its 
maximum close to the peak of the transverse momentum spectrum, i.e. where the average particle yield is largest. 
In these bins, and charge, and in particular, energy conservation have strong effects. 

\begin{figure}[ht!]  
  \epsfig{file=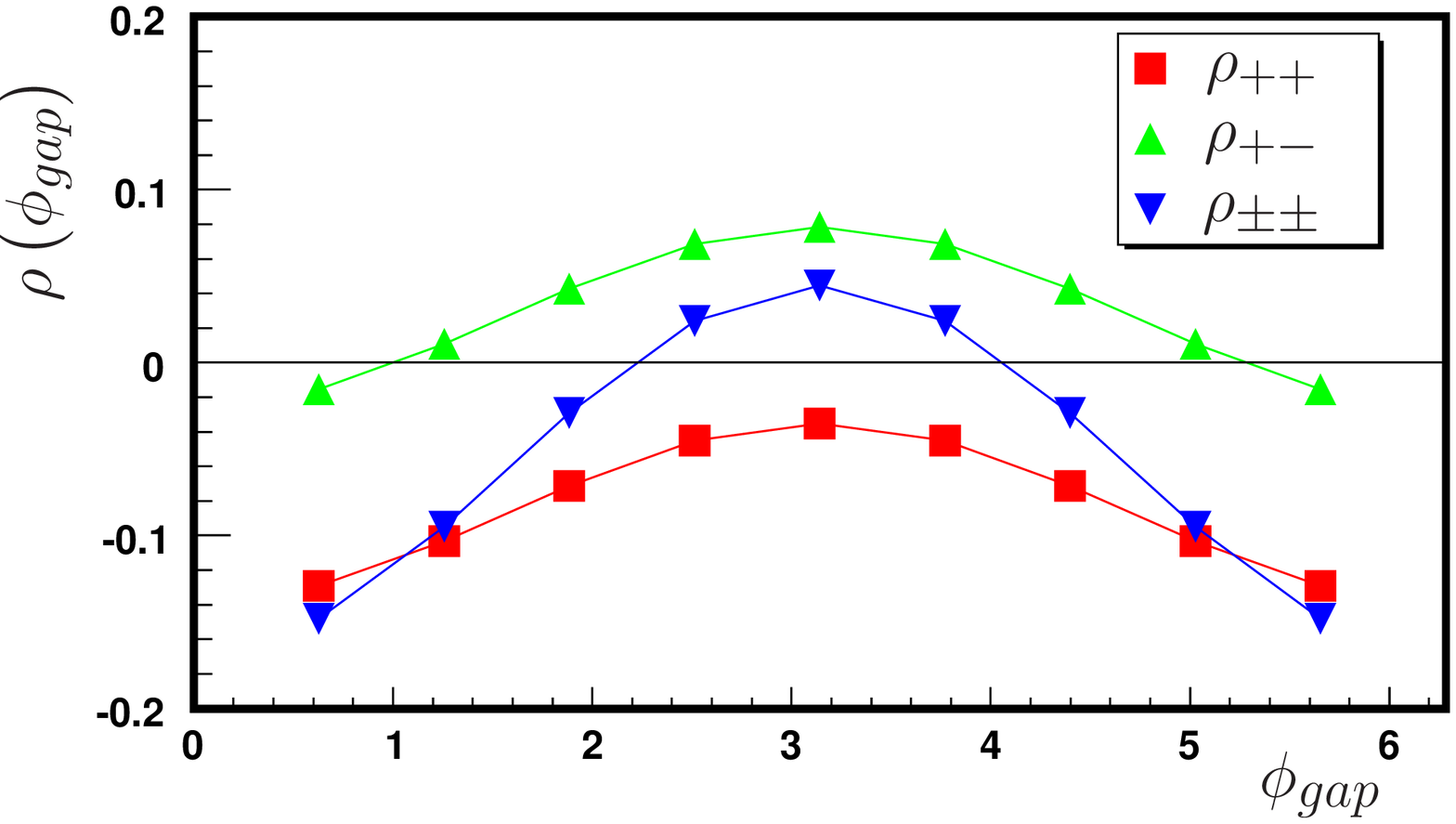,width=7.2cm,height=6.5cm}  
  \epsfig{file=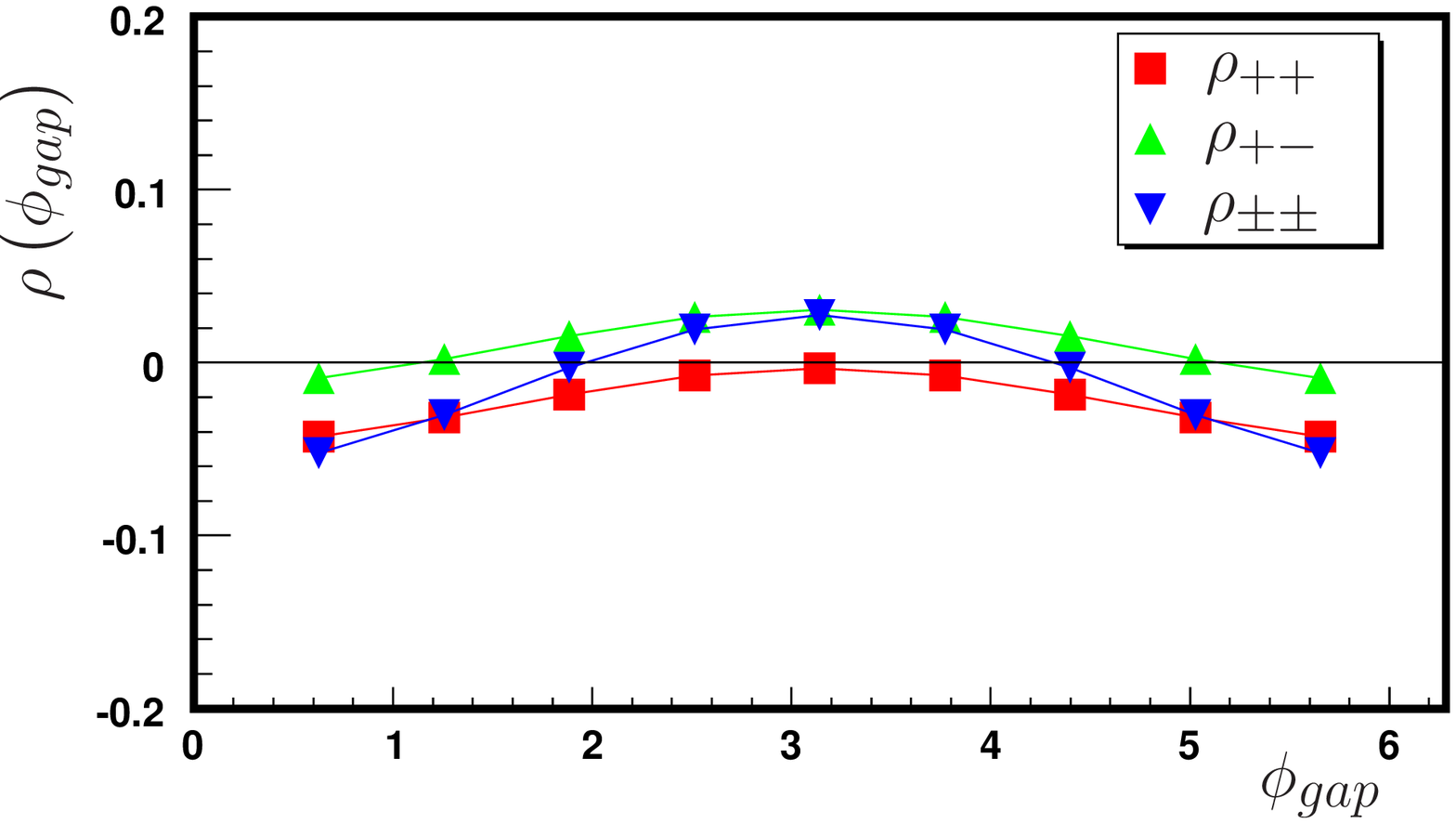,width=7.2cm,height=6.5cm}  
  \caption{The correlation coefficients of particles in distinct~$\Delta \phi$ bins as a function of 
  separation~$\phi_{gap}$ in azimuth. ({\it Left:})~integrated over all phase space. ({\it Right:})~only particles 
  with~$|y|<0.3$ are observed. Both plots show MCE MB results. No elliptic flow is considered.}    
  \label{Angular_corr}  
\end{figure}

Angular correlations, lastly, could arise due to conservation of transverse momenta~$P_x$ and~$P_y$. In 
Fig.(\ref{Angular_corr}) the correlation coefficient between particles  in different~$\Delta \phi$ bins are shown. 
The flat\footnote{Only globally equilibrated systems are considered, and elliptic flow is disregarded here.} angular 
spectrum~$dN / d\phi$ has been divided into~$10$ equal size bins and the correlation coefficient is presented   
as a function of separation of the centers of the corresponding bins.   

Considering Fig.(\ref{Angular_corr}) one firstly notes that when disregarding  exact conservation of~$P_x$ and~$P_y$ 
the correlation coefficients are insensitive to the distance~$\phi_{gap}$ of any two bins. Only the correlations due to 
energy and charge conservation affect the result. Charge conservation leads to correlation of unlike-sign particles and 
to anti-correlation of like-sign particles. Energy conservation always anti-correlates multiplicities in two bins. 
For~$\rho_{\pm \pm}$ the effect of charge conservation cancels for a neutral system, however, effects of energy-momentum 
conservation are stronger, as a larger number of particles (hence a larger part of the total system) is observed.  
  
Conservation of transverse momenta~$P_x$ and~$P_y$ is now responsible for the~$\phi_{gap}$ dependence of~$\rho$. The 
line of arguments is similar to the ones before: Observing a larger (smaller) number of particles in some bin at~$\phi_0$ 
implies that, in order to balance momenta~$P_x = P_y = 0$, one should also observe a larger (smaller) number of particles 
in the opposite direction~$\pi-\phi_0$. A larger (smaller) number of particles in a bin with~$\phi_{gap} = \pi/2$ would do 
little to help to balance momentum, but conflict with energy conservation. 

The observable effects of energy-momentum conservation are weaker, if the experimental acceptance only covers a 
mid-rapidity region of~$|y|<0.3$, as in Fig.(\ref{Angular_corr})~({\it right}), rather than the whole rapidity 
distribution, Fig.(\ref{Angular_corr})~({\it left}).

\section{Discussion}
\label{sec_multfluc_pgas_dis}

In this chapter multiplicity fluctuations and correlations in limited momentum bins for ideal relativistic gases have 
been discussed in the MCE in the thermodynamic limit. For the discussion a gas with three degenerate massive particles   
(positive, negative, neutral) in three different statistics (Maxwell-Boltzmann, Fermi-Dirac, Bose-Einstein) was chosen.  
  
For the width of multiplicity distributions in limited bins of momentum space a simple and intuitive picture emerges. 
In the Maxwell-Boltzmann approximation one finds a wider distribution for momentum bins with low average momentum 
when compared to bins with higher average momentum but same average particle number. This qualitative behavior is a 
direct consequence of energy and momentum conservation. The results in Fermi-Dirac and Bose-Einstein statistics, 
furthermore, show pronounced  effects at the low momentum tail of the spectrum.  
  
The correlation coefficient additionally shows a similar qualitative behavior. In bins with low average momentum the 
correlation coefficient between positively and negatively charged particles is indeed positive, as one would expect 
from charge conservation. However, in bins with large average momentum the effects of joint energy and momentum 
conservation can lead to anti-correlated distributions of unlike-charged particles.   
  
The role of exactly imposed motional conservation laws is particularly important for systems with collective velocity. 
Fluctuations and correlations transform  under boosts, provided momentum conservation along the boost direction is taken   
into account. This ensures, in particular, that they become boost-invariant if the underlying system is boost-invariant.  
  
Lastly, it was found that even in the thermodynamic limit long range correlations between disconnected regions in 
momentum space prevail. Multiplicities in different bins in rapidity, transverse momentum, or in azimuth, can have 
a non-zero correlation  coefficient.

\chapter{The Phase Diagram}
\label{chapter_phasediagram}
In this chapter the statistical properties of a hadron resonance gas are studied in their dependence on `freeze-out` 
parameters, and choice of ensemble. Fully phase space integrated extensive quantities are investigated, 
omitting limited acceptance effects, but still considering resonance decay.

Preceding chapters were mostly concerned with neutral and static gases. Attention was given, in particular, to the 
consequences of limited acceptance in momentum space. Here, fully phase space integrated fluctuations and correlations 
observables for a hadron resonance gas will be studied in their dependence on energy and net-baryon density.
For simplicity, remaining chemical potentials are chosen to be zero. Thus, mesons and anti-mesons contribute equally to 
the partition function. But as, for instance, the proton carries electric charge, and the~$\Lambda$ carries strangeness, 
the system is only for~$\mu_B=0$, i.e. equal baryon and anti-baryon contribution, also electric charge and strangeness 
neutral. For positive~$\mu_B$, while~$\mu_Q=\mu_S=0$, the system has positive electric charge density and negative 
strangeness density. For this discussion the phase space occupancy 
factors~\cite{Koch:1986ud,Letessier:1999rm,Rafelski:2000by} are taken to be~$\gamma_s=\gamma_q=1$. 
Nevertheless, the constant average energy per average particle 
multiplicity~$\langle E \rangle / \langle N \rangle \simeq 1$~GeV freeze-out line~\cite{Cleymans:1998fq} is shown for 
orientation. 

Above this freeze-out line the following discussion should be taken with a caveat mentioned:
First and foremost, a phase transition is expected to occur. Degrees of freedom should hence be deconfined quarks 
and gluons, not hadrons. Furthermore, even on the hadronic side, inclusion of Hagedorn states might become important 
close to the phase transition line~\cite{Andronic:2008gu,NoronhaHostler:2007jf}. The exponentially rising Hagedorn mass 
spectrum~\cite{Hagedorn:1965fix_1} then would lead to a diverging partition function, above a certain limiting 
temperature, and the phase is not even defined. This discontinuity could in principle be 
fixed~\cite{Gazdzicki:1998vd,Ferroni:2008ej}, by allowing for a transition to a bag model phase. Quantum number 
configurations or decay channels of these Hagedorn states or bags\footnote{The terms Hagedorn state, string, cluster, 
bag, or resonance are used equivalently in this chapter. The only property they are required to have is that they 
decay into hadrons.} are however unknown, and would have to be assumed. The chemical freeze-out line could be seen as 
an estimate of where in the phase diagram these effects would become important.

It is stressed, again, that no hadronic phase should exist above the freeze-out line. It should also be mentioned that, 
in principle, the intensive variables $T$ and $\mu_B$ are an unconventional choice to present MCE (or CE) correlation 
functions. Nevertheless, this choice of presentation was made, to allow for a comparison of statistical properties of 
different ensemble formulations at different energy and net-baryon densities. To highlight certain aspects, phase 
diagrams are extended to negative values of~$\mu_B$. 

In the~CE or~MCE charge correlation and fluctuations would be vanishing in full acceptance, since charges (or 
additionally energy and momentum) are exactly conserved in these ensembles. As discussed in 
Chapter~\ref{chapter_extrapol}, charge fluctuations are determined by the thermodynamic bath available, and remain, 
full acceptance assumed, unaffected by resonance decay. Charge, energy and momentum conservation is respected by each
decay process (weak decays omitted). In contrast to this, multiplicity correlation coefficients in the primordial 
GCE are equal to zero, while multiplicity fluctuations are Poissonian in the Boltzmann approximation. 
Correlations in the~GCE appear due to resonance decay, and are strongly modified, along with a suppression of 
multiplicity fluctuations, by conservation laws in the~CE or~MCE, see~Chapters~\ref{chapter_multfluc_hrg} 
and~\ref{chapter_multfluc_pgas}.

Acceptance effects are neglected here. The correlation coefficients and variances studied here, with full acceptance 
assumed, would also exhibit the qualitative behavior presented in Chapters~\ref{chapter_gce} for charge correlations, 
and Chapters~\ref{chapter_multfluc_hrg} and~\ref{chapter_multfluc_pgas} for multiplicity fluctuations and correlations. 

The~$T$-$\mu_B$ phase diagram will be explored for~GCE charge correlations and fluctuations in 
Section~\ref{sec_phasediagram_charcorr}, while in Sections~\ref{sec_phasediagram_multcorr} 
and~\ref{sec_phasediagram_multfluc} multiplicity correlations and fluctuations will be discussed in different ensembles.

\section{Charge Correlations and Fluctuations}
\label{sec_phasediagram_charcorr}

\begin{figure}[ht!]
  \begin{center}
    \epsfig{file=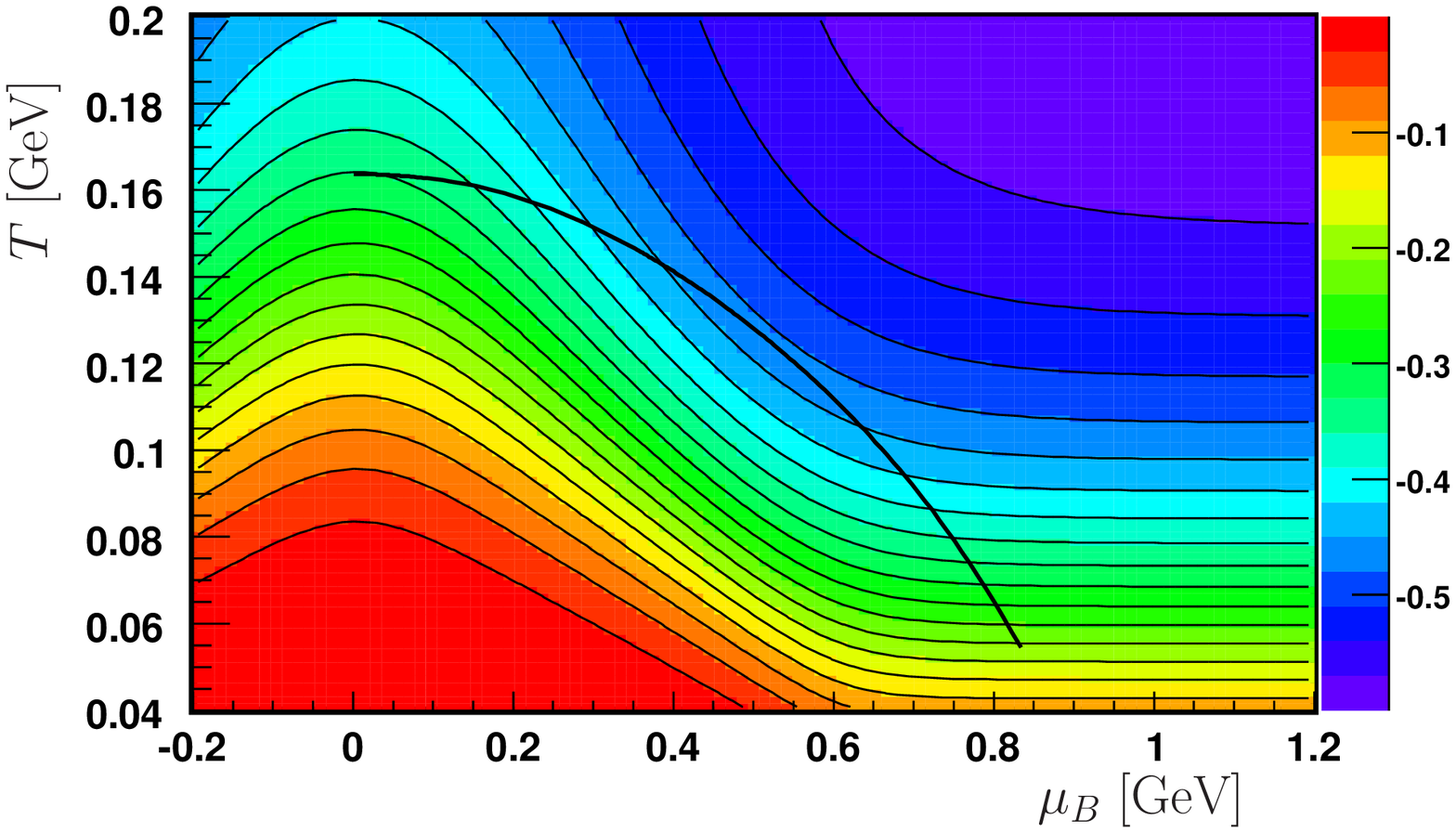,width=4.75cm,height=5.5cm}
    \epsfig{file=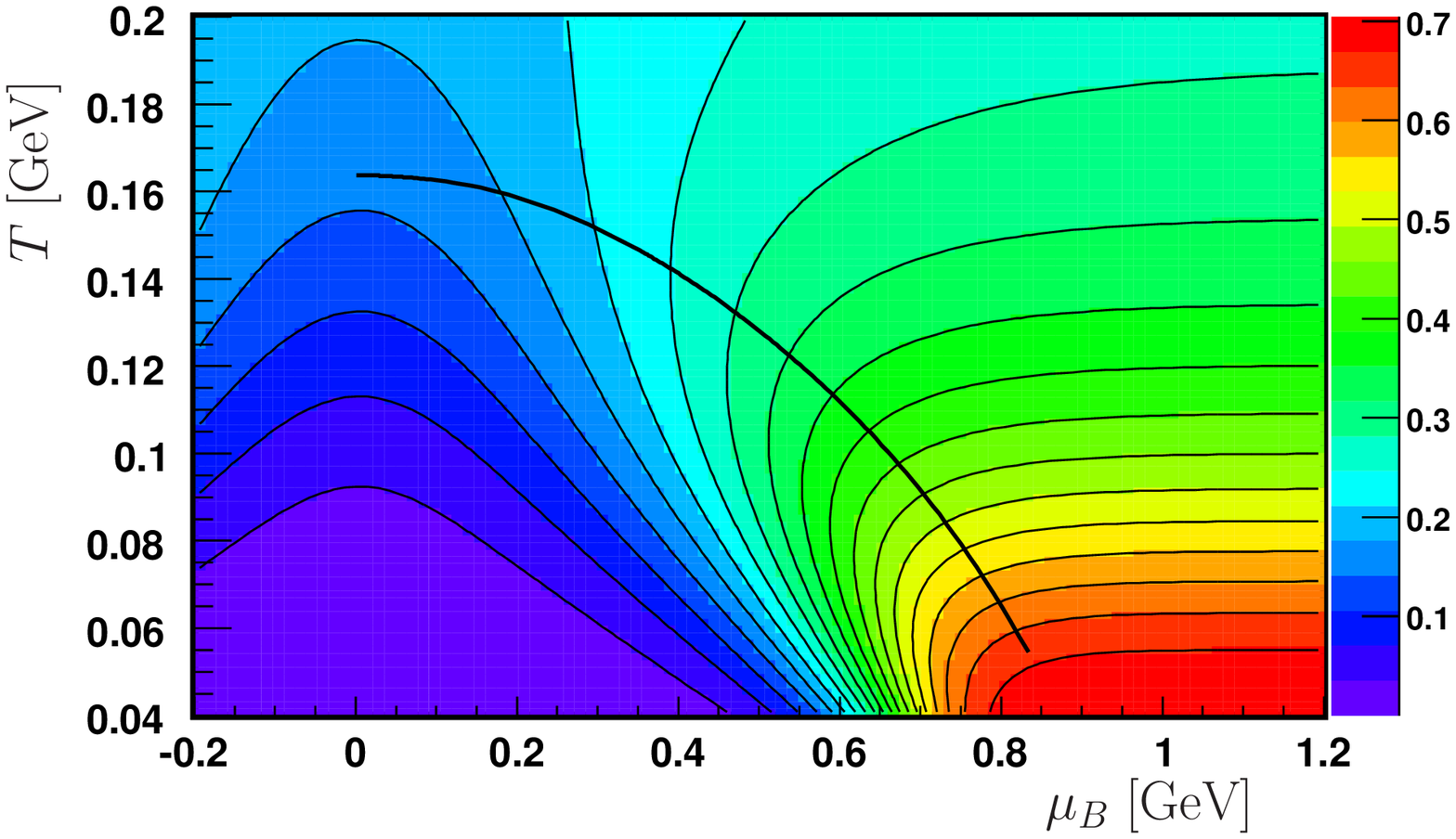,width=4.75cm,height=5.5cm}
    \epsfig{file=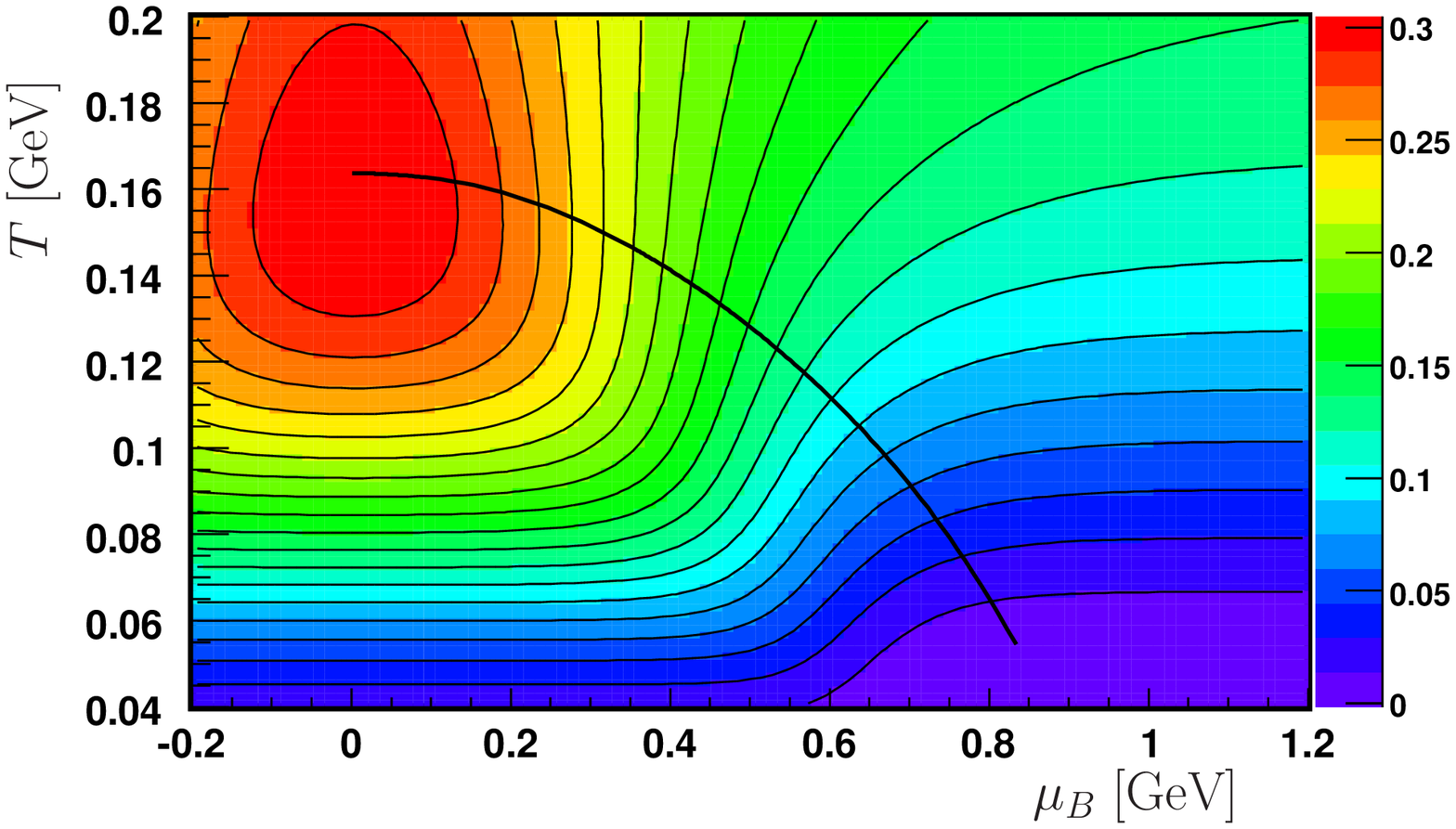,width=4.75cm,height=5.5cm}
    \caption{Phase diagram showing grand canonical charge correlation coefficients of baryon number and 
    strangeness~$\rho_{BS}$~({\it left}), baryon number and electric charge~$\rho_{BQ}$~({\it center}), and
    strangeness and electric charge~$\rho_{SQ}$~({\it right}). The solid line indicates 
    the~$\langle E \rangle/\langle N \rangle=1$~GeV chemical freeze-out line~\cite{Cleymans:1998fq}.
    } 
    \label{T_muB_GCE_corr}
  \end{center}
\end{figure}

In Fig.(\ref{T_muB_GCE_corr}) grand canonical correlation coefficients between charges are shown in their dependence 
on the freeze-out parameters~$T$ and~$\mu_B$. The correlation coefficients~$\rho_{BS}$, $\rho_{BQ}$, 
and $\rho_{SQ}$, as well as related variances and covariances are symmetric around~$\mu_{B}=0$ in the~$T$-$\mu_B$ 
phase diagram. The sum of baryon and anti-baryon yield is equal in systems with baryon chemical potentials of opposite 
sign, since here additionally remaining chemical potentials are chosen to be~$\mu_S=\mu_Q=0$. 
Further, one finds~$\langle \Delta B \Delta S \rangle$, 
$\langle \Delta B \Delta Q \rangle$, $\langle \left( \Delta Q \right)^2 \rangle$, etc. to be the same for~$\mu_B$ 
and~$-\mu_B$, due to equal contributions of particles and their anti-particles. The meson contribution then only 
depends on the temperature. The discussion will proceed by considering the four corners of the phase diagrams 
depicted in Fig.(\ref{T_muB_GCE_corr}), starting from the bottom left.

At low~$T$ and~$\mu_B$ the systems is mostly composed of pions. Accordingly one finds the correlation 
coefficients~$\rho_{BS} \simeq \rho_{BQ} \simeq \rho_{SQ} \simeq 0$ in the bottom left corners of phase diagrams shown 
in Fig.(\ref{T_muB_GCE_corr}). The system is too cold to populate states with hadrons which could mediate the correlation. 
Pions carry only electric charge, and thus cannot correlate to the strangeness and baryon number content of the system.

As~$T$ is increased, but still at small~$\mu_B$, the system starts to equally excite baryon and anti-baryon states.  
The correlation between baryon number and strangeness content of the system, $\rho_{BS}$, therefore grows negative due to 
emerging~$\Lambda$ and~$\overline{\Lambda}$ states. Compared to~$\rho_{BS}$, the baryon number electric charge 
correlation coefficient~$\rho_{BQ}$ grows more mildly. Protons and anti-protons are being equally produced, yet 
abundant pions de-correlate the quantum numbers~$B$ and~$Q$ more strongly than heavier kaons can de-correlate the quantum 
numbers~$B$ and~$S$. Lastly, in the top left corner of Fig.(\ref{T_muB_GCE_corr})~({\it right}), the strangeness 
electric charge correlation coefficient~$\rho_{SQ}$ has a peak. All  mesons and baryons (except the~$\Sigma^+$)
contribute positively to the covariance~$\langle \Delta S \Delta Q \rangle$. The combined contribution of particles and 
anti-particles at a given temperature is strongest in charge neutral matter. Only at rather large temperature 
heavy~$\Sigma^+$ resonances (and their anti-particles) start to again decrease~$\rho_{SQ}$.

\begin{figure}[ht!]
  \begin{center}
    \epsfig{file=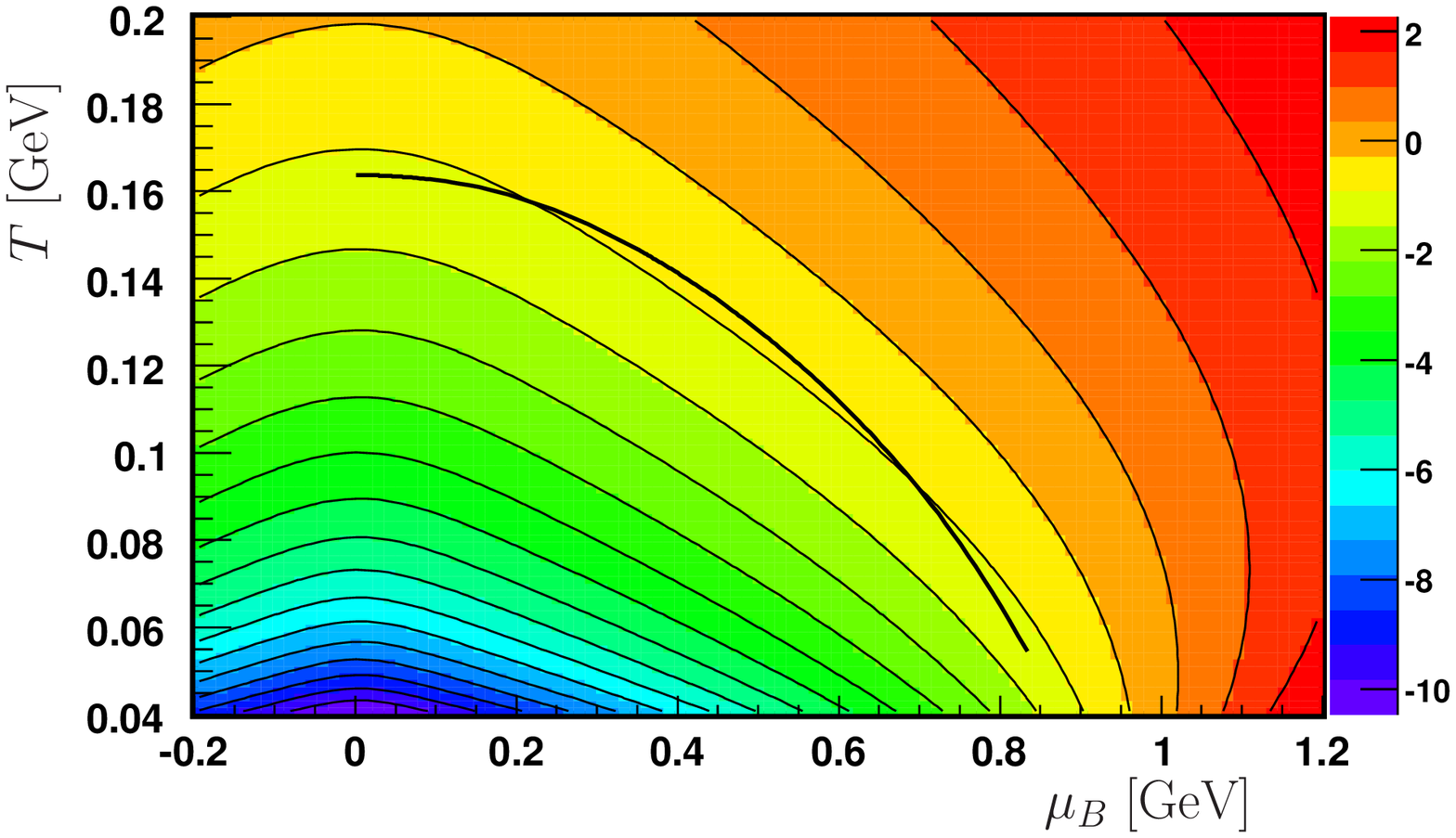,width=4.75cm,height=5.5cm}
    \epsfig{file=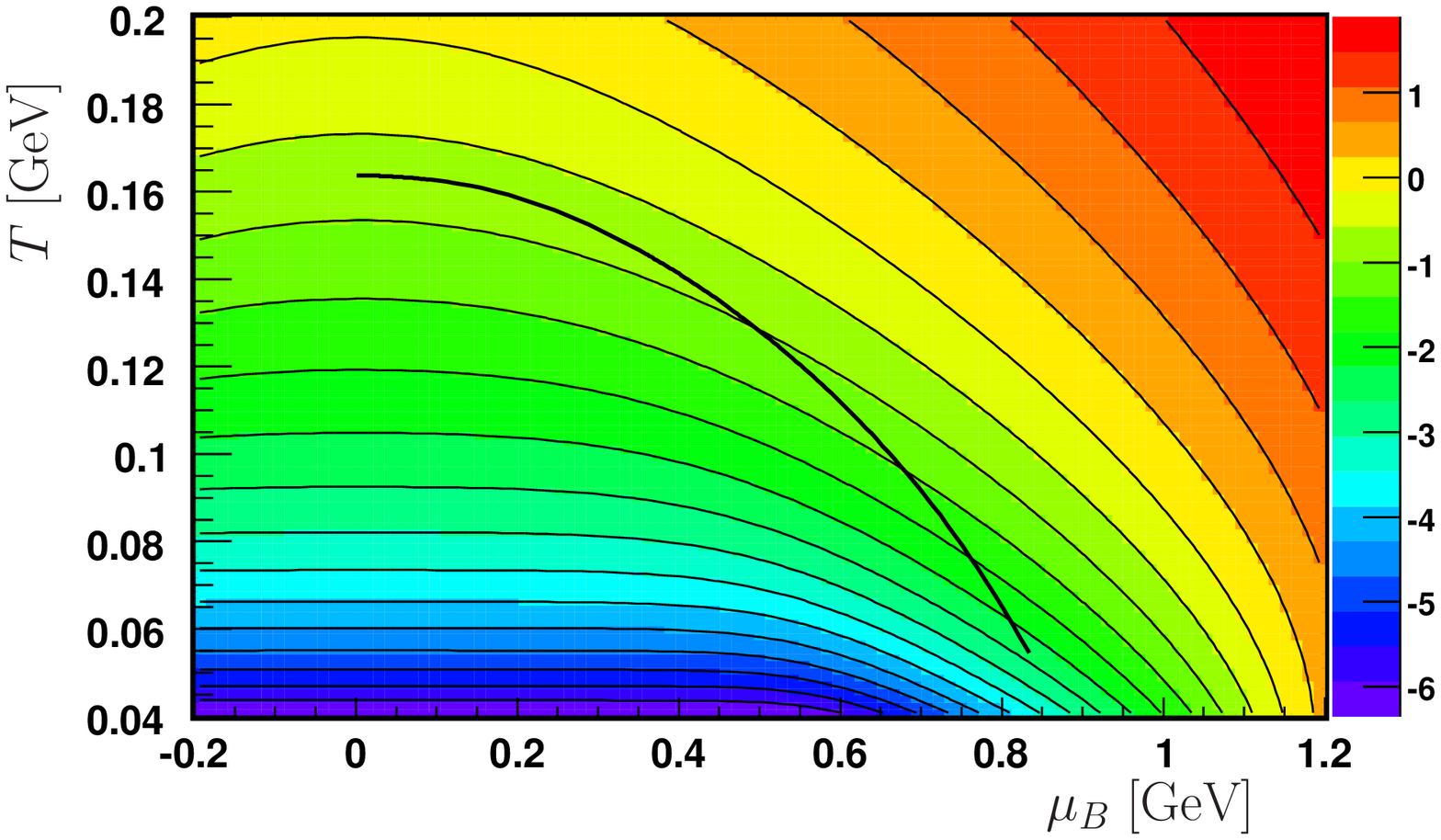,width=4.75cm,height=5.5cm}
    \epsfig{file=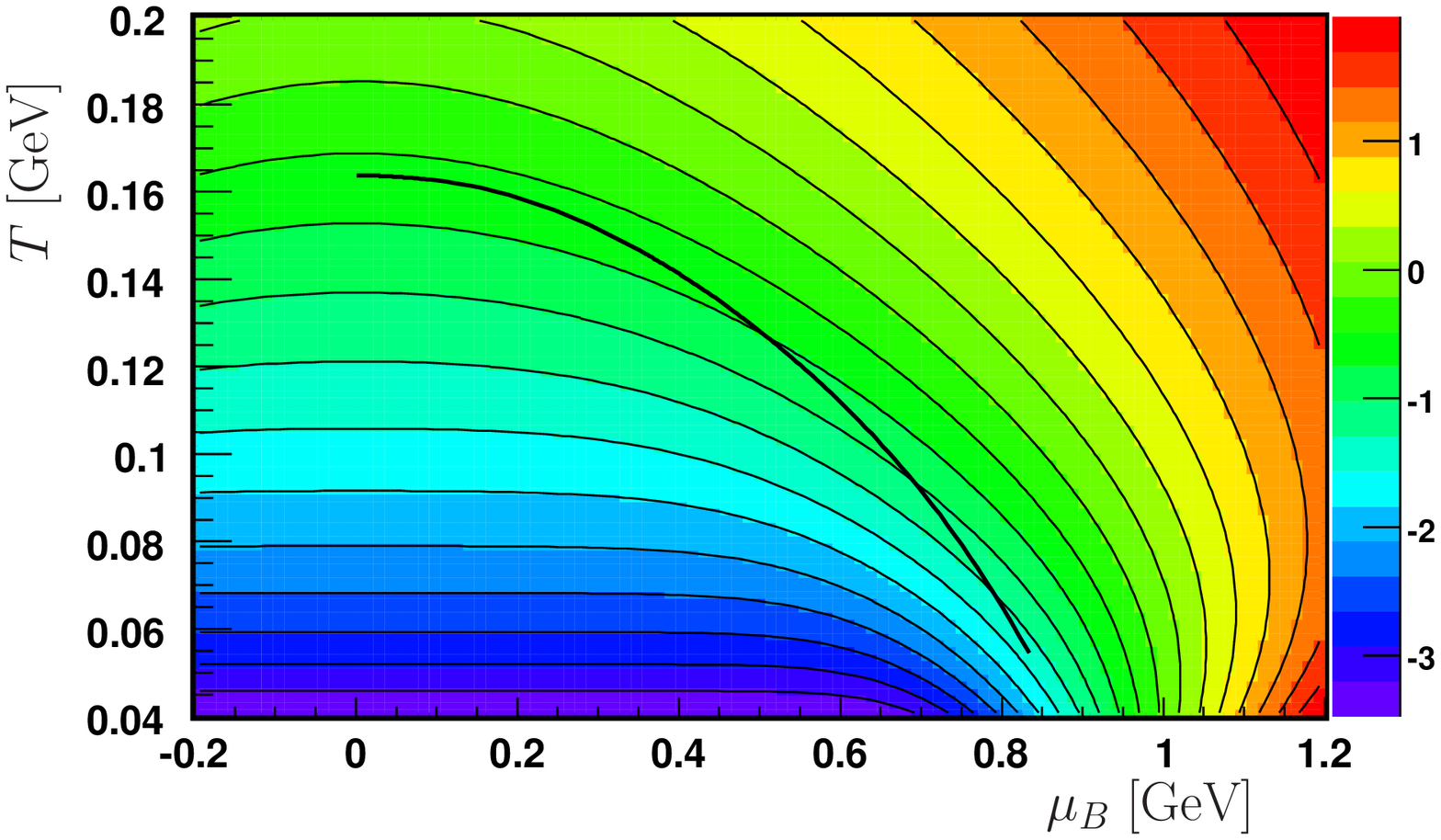,width=4.75cm,height=5.5cm}
    \caption{Phase diagram showing normalized grand canonical charge fluctuations of baryon 
    number~$\langle \left( \Delta B \right)^2 \rangle$~({\it left}), 
    strangeness~$\langle \left( \Delta S \right)^2 \rangle$~({\it center}), 
    and electric charge~$\langle \left( \Delta Q \right)^2 \rangle$~({\it right}). The color scale is logarithmic.
    } 
    \label{T_muB_GCE_fluc}
  \end{center}
\end{figure}

The bottom right corners in Fig.(\ref{T_muB_GCE_corr}) represent a hadronic phase of low temperature and large 
net-baryon density. The correlation coefficient of baryon number and strangeness~$\rho_{BS}$ grows in this direction 
due to~$\Lambda$ production. The system is on the other hand is too cold to produce kaons, which could de-correlate 
the quantum numbers~$B$ and~$S$, in sufficient numbers. As protons (and neutrons or light~$\Delta$s) are the dominant 
particles in this phase, the correlation between baryon number and electric charge~$\rho_{BQ}$ is particularly strong. 
And, as kaons are not sufficiently produced at low temperature (and the~$\Lambda$ is electrically neutral), $\rho_{SQ}$, 
the strangeness electric charge correlation coefficient, remains small at small~$T$. The increase of~$\rho_{SQ}$ for 
large $\mu_B$ is due to the onset of~$\Sigma$ production.

Finally, moving into the top right corners in Fig.(\ref{T_muB_GCE_corr}), with the above mentioned caveat in mind.
The correlation~$\rho_{BS}$ becomes strongly negative in hot net-baryon rich matter, as strangeness carrying baryons 
are produced abundantly. The correlation~$\rho_{BQ}$ stays modest as heavy negatively charged baryons emerge too, and 
like pions, de-correlate the systems electric net-charge from its net-baryon number content. Much for the same 
reasons~$\rho_{SQ}$ stays modest, too. Here heavy~$\Sigma^+$ resonances decrease the correlation.\\


In Fig.(\ref{T_muB_GCE_fluc}) grand canonical charge fluctuations are shown in their dependence on the freeze-out
parameters~$T$ and~$\mu_B$ on a logarithmic color scale. The variance is, unlike the correlation coefficient, an 
extensive quantity. The variances are therefore shown normalized to an unit volume. In a neutral system, with symmetric 
particle and anti-particle contribution to the partition function, charge fluctuations are largest. With increasing 
temperature, not only average occupation numbers, but also fluctuations in occupation numbers on individual momentum 
levels grow. Accordingly, net-charge fluctuations become stronger. For increasing~$\mu_B$, i.e. average baryon 
occupation numbers are enhanced, while occupation numbers of anti-baryons are suppressed by Boltzmann 
factors~$e^{\pm \mu_B \beta}$, charge fluctuations decrease. At low temperature, few particles are produced, 
and the net-charge content of the system cannot fluctuate much. As~$\mu_B$ is increased at low~$T$, baryon states are 
populated due to charge density. The net-charge content of the system can then fluctuate, by virtue of~$\mu_B$ induced 
particle density.

The baryon number variance~$\langle \left( \Delta B \right)^2 \rangle$ in Fig.(\ref{T_muB_GCE_fluc})~({\it left})
increases rapidly with baryon density and temperature. At low~$T$ and~$\mu_B$ essentially no baryon states are occupied, 
and hence their number cannot fluctuate strongly. The strangeness  variance~$\langle \left( \Delta S \right)^2 \rangle$
in Fig.(\ref{T_muB_GCE_fluc})~({\it center}) and electric charge variance~$\langle \left( \Delta Q \right)^2 \rangle$
in Fig.(\ref{T_muB_GCE_fluc})~({\it right}) are similar. Strangeness and electric charge density are introduced
indirectly via the baryons~p and~$\Lambda$. Although the strangeness and electric charge variances are initially 
larger, they grow more slowly than the baryon number variance. The mass difference between the lightest baryons and 
mesons becomes less important at very large temperature, and $\langle \left( \Delta B \right)^2 \rangle$ eventually 
overtakes $\langle \left( \Delta S \right)^2 \rangle$ and $\langle \left( \Delta Q \right)^2 \rangle$.

In direction from bottom left to top right in Figs.(\ref{T_muB_GCE_corr},\ref{T_muB_GCE_fluc}),
i.e. increasing both~$T$ and~$\mu_B$, a transition from a mesonic to baryonic matter~\cite{Cleymans:2005zx} occurs.
The variance~$\langle \left( \Delta B \right)^2 \rangle$ and the correlation coefficient~$\rho_{BS}$ grow as the entropy
of the system is increasingly carried by baryons. If one would instead follow the~$\langle E\rangle /\langle N \rangle$
freeze-out line one would observe a monotonous decrease in~$\rho_{BQ}$ as colliding beam energy is increased,
due to emerging pion production de-correlating the charges~$B$ and~$Q$. The correlation between~$S$ and~$Q$, $\rho_{SQ}$, 
on the other hand, increases, because kaons, and not~$\Lambda$, now carry the bulk of the system strangeness
at large center of mass energy. Along the freeze-out line~$\rho_{BS}$ has a strong minimum around~$T\sim 0.13$~GeV 
and~$\mu_B\sim0.5$~GeV. The strangeness and electric charge variances $\langle \left( \Delta S \right)^2 \rangle$ 
and $\langle \left( \Delta Q \right)^2 \rangle$ increase along this line, 
while $\langle \left( \Delta B \right)^2 \rangle$ stays more or less flat.

\section{Multiplicity Correlations}
\label{sec_phasediagram_multcorr}

In this section the correlation coefficients~$\rho_{\textrm{p} \pi^+}$ and~$\rho_{\textrm{p} \pi^-}$ between the 
event-by-event multiplicities of protons, $N_{\textrm{p}}$, and either of the charged pions, $N_{\pi^+}$
and~$N_{\pi^-}$ are studied for different ensembles. The discussion is then extended to multiplicity fluctuations of 
positively and negatively charged hadrons~$\omega_+$ and~$\omega_-$. The hadrons~p and~$\pi^{\pm}$ are the most abundant 
charged particles, and their relative yield determines structure of the~$T$-$\mu_B$ phase diagram to a large extent. 
In turn, GCE, CE, and MCE correlation functions will be compared. Resulting (primordial) correlations are not due to 
local interaction amongst constituents, but due to globally implemented conservation laws for energy and charge.  

\subsubsection{Grand Canonical Ensemble}

The multiplicities of any two distinct groups of primordial particles are uncorrelated in the grand canonical 
ensemble, $\rho=0$, due to the infinite thermodynamic bath assumption. Their joint distribution factorizes into a 
product of two Poissonians with scaled variance~$\omega=1$. In the final state grand canonical ensemble the proton and 
charged pion multiplicities are correlated by the decay of parent baryons~$N$. 
The following decay channels, containing protons and charged pions, are considered:
\begin{eqnarray}
\label{dec_chan_neg}
N^-~\; &\longleftrightarrow& ~\textrm{p} ~+~ \pi^- ~+~ \pi^-   \\
\label{dec_chan_neu}
N^0~\;\; &\longleftrightarrow& ~\textrm{p} ~+~ \pi^-  \\
\label{dec_chan_pos}
N^+~\; &\longleftrightarrow& ~\textrm{p} ~+~ \pi^+ ~+~ \pi^- \\
\label{dec_chan_dpos}
N^{++}\;\! &\longleftrightarrow& ~\textrm{p} ~+~ \pi^+ 
\end{eqnarray}
Baryonic resonances can decay into protons and charged pions taking channels of kind~(\ref{dec_chan_neg}) 
to~(\ref{dec_chan_dpos}). Negatively charged baryons~$N^-$, could only decay into negatively charged pions and a proton 
via~(\ref{dec_chan_neg}). The~$\Delta^-$ resonances, however, mostly decay via~$\Delta^-\rightarrow \textrm{n}+\pi^-$. 
Strangeness carrying~$N^-$ also cannot strongly populate the channel~(\ref{dec_chan_neg}), as either daughter particle, 
baryon or meson, has to pick up the strange quark. The channel~(\ref{dec_chan_neu}) for decay of neutral baryons~$N^0$ 
is also not too efficient. Two thirds of neutral baryons decay via~$N^0\rightarrow \textrm{n} + \pi^0$
due to ``Clebsch-Gordan'' coefficients. Although most low lying~$\Delta^+$ decay via the channel~(\ref{dec_chan_pos}), 
$\Delta^+\rightarrow \textrm{p} + \pi^0$, and do hence not contribute much to either correlation coefficient, 
heavier~$\Delta^+$ resonances, often produce several pions, and positively correlate p with both, $ \pi^+$ and~$\pi^-$.
Finally, channels (\ref{dec_chan_dpos}) are abundant due to a strong contribution of the~$\Delta^{++}(1232)$ resonance 
and its degenerate states.

\begin{figure}[ht!]
  \begin{center}
    \epsfig{file=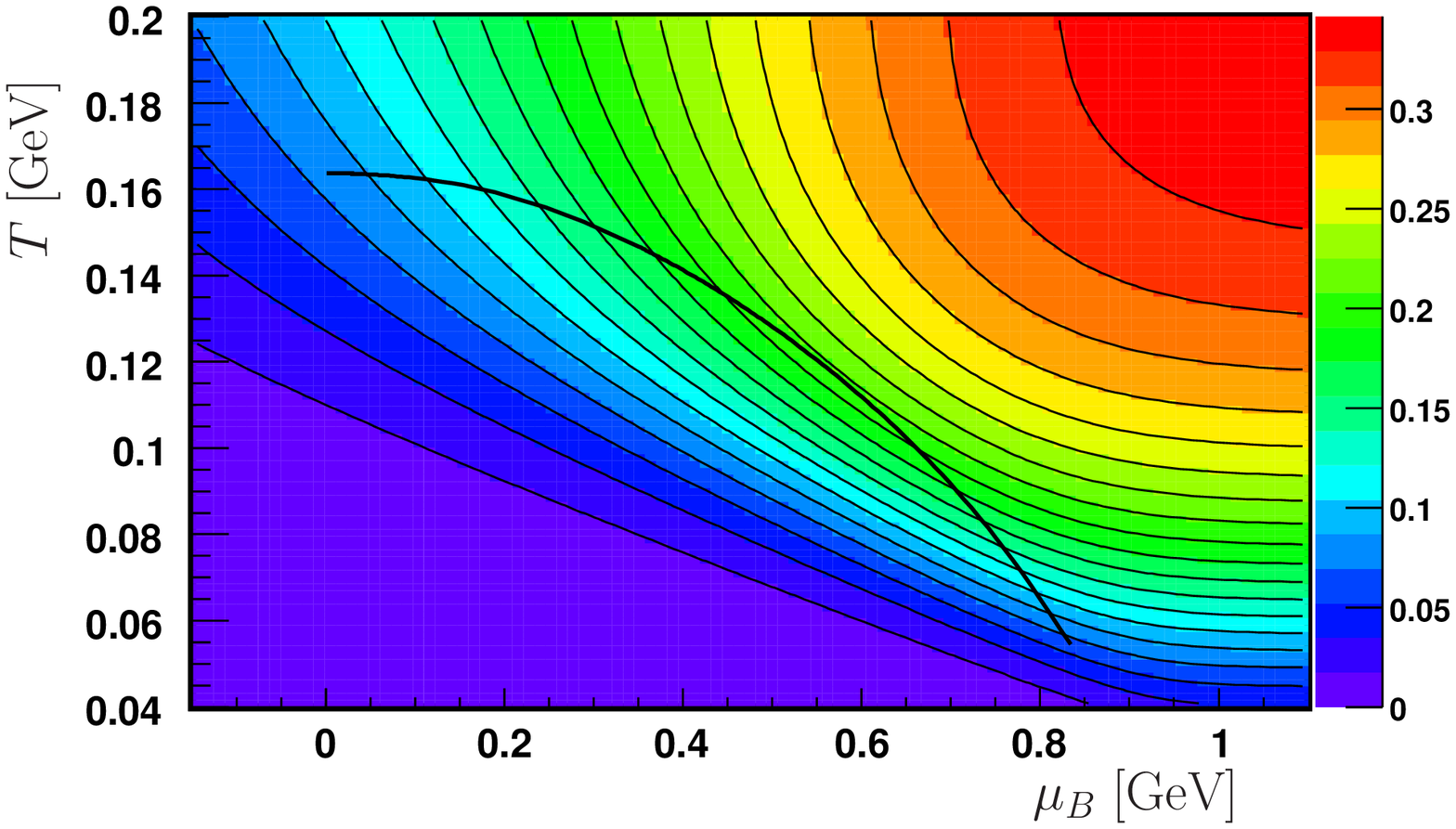,width=7.cm,height=6.5cm}
    \epsfig{file=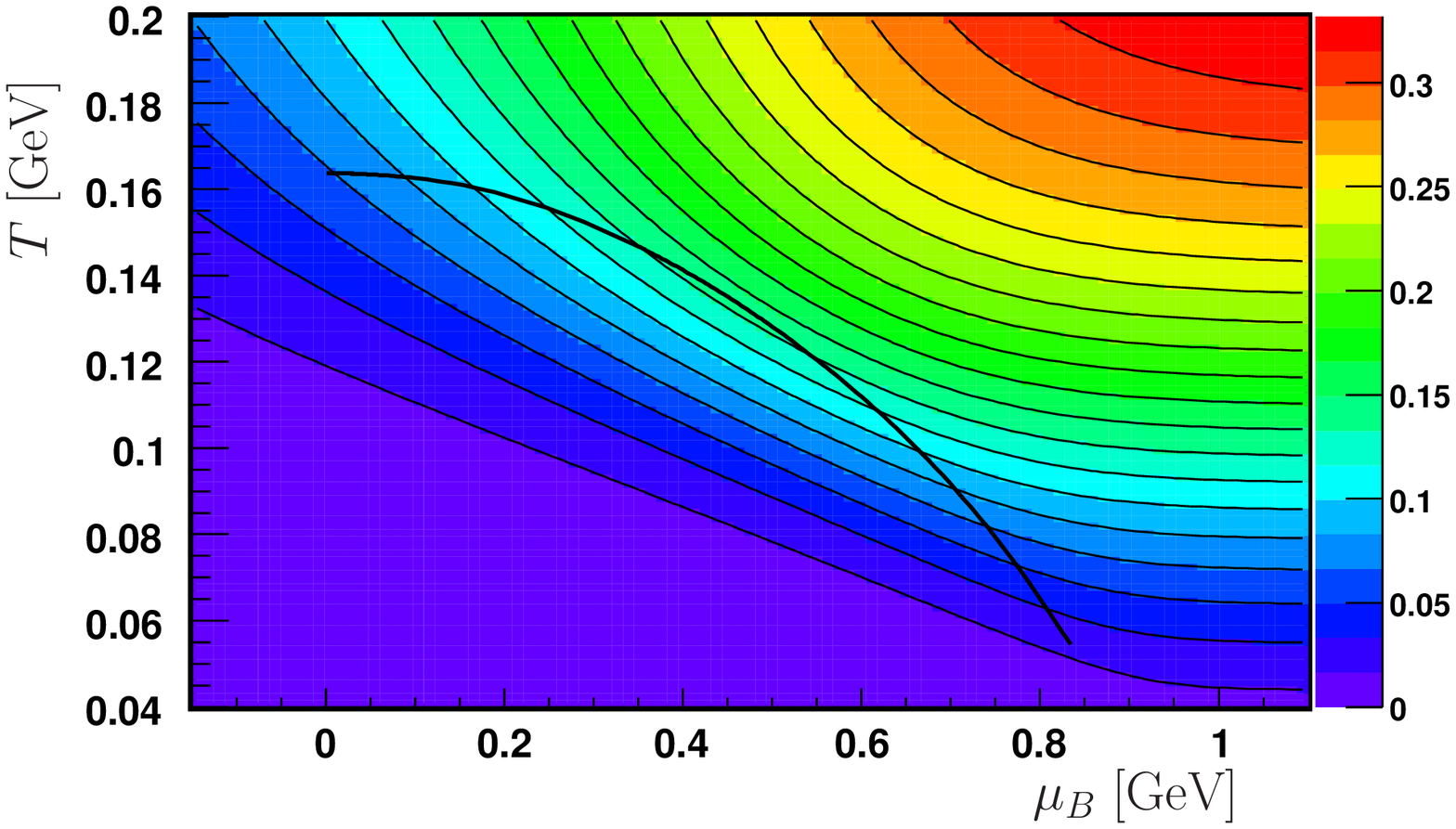,width=7.cm,height=6.5cm}
    \caption{
    Final state~GCE correlation coefficient between the multiplicities of protons and positively charged 
    pions~$\rho_{\textrm{p} \pi^+}$~({\it left}), and between the multiplicities of protons and negatively charged 
    pions~$\rho_{\textrm{p} \pi^-}$~({\it right}).} 
    \label{T_muB_rho_p_pi_GCE}
  \end{center}
\end{figure}

The reverse arrows indicate back-reactions. The strong interaction between the ground state hadrons proton, neutron, and 
pion is conceptually taken into account in the hadron resonance gas model by inclusion of resonance states which mediate 
the interaction~\cite{Dashen:1969ep,Dashen:1974xz}. In thermal and chemical equilibrium reaction rates are the same in 
both directions. The system is said to have attained ``detailed balance''. Only the existence of resonances states is 
relevant, and not how many particles are involved in each microscopic reaction. The ability of a non-equilibrium state 
to evolve into a chemical equilibrium state, on the other hand, is strongly increased by multi-hadronic (or Hagedorn) 
channels~\cite{NoronhaHostler:2007jf}.  

Hence, in the final state~GCE, Fig.(\ref{T_muB_rho_p_pi_GCE}), both, $ \pi^+$ and~$\pi^-$ are positively correlated 
with~p. Phase diagrams showing multiplicity correlations are generally not symmetric around~$\mu_B=0$. The correlation is 
particularly strong in hot baryonic matter. For negative~$\mu_B$ either correlation 
coefficient $\rho_{\textrm{p} \pi^+} \simeq \rho_{\textrm{p} \pi^-} \simeq 0$, as protons are then suppressed compared to 
anti-protons. Charged pions would now be strongly correlated with the more abundant anti-protons.

The fact that~$\pi^+$ are more strongly correlated via resonance decay with~p, than the~$\pi^-$, 
i.e.~$\rho_{\textrm{p} \pi^-} < \rho_{\textrm{p} \pi^+}$, can be understood considering reaction 
channels (\ref{dec_chan_neg})~to~(\ref{dec_chan_dpos}). Essentially one may add additional pions ($\pi^+$ +~$\pi^-$) to 
each channel. Channels (\ref{dec_chan_neg}) and (\ref{dec_chan_neu}) are not very effective at correlating~p and~$\pi^-$, 
since they are not very abundant. Channel (\ref{dec_chan_pos}) contributes equally to the~p-$\pi^-$ correlation 
and the~p-$\pi^+$ correlation, while channel (\ref{dec_chan_dpos}) strongly correlates~p and~$\pi^+$, yet hardly~p 
and~$\pi^-$. The influence of the~$\Delta^{++}(1232)$ can clearly be seen in a comparison of the bottom right corners of 
Fig.(\ref{T_muB_rho_p_pi_GCE}) ({\it left}) to ({\it right}).

\subsubsection{Canonical Ensemble}

\begin{figure}[ht!]
  \begin{center}
    \epsfig{file=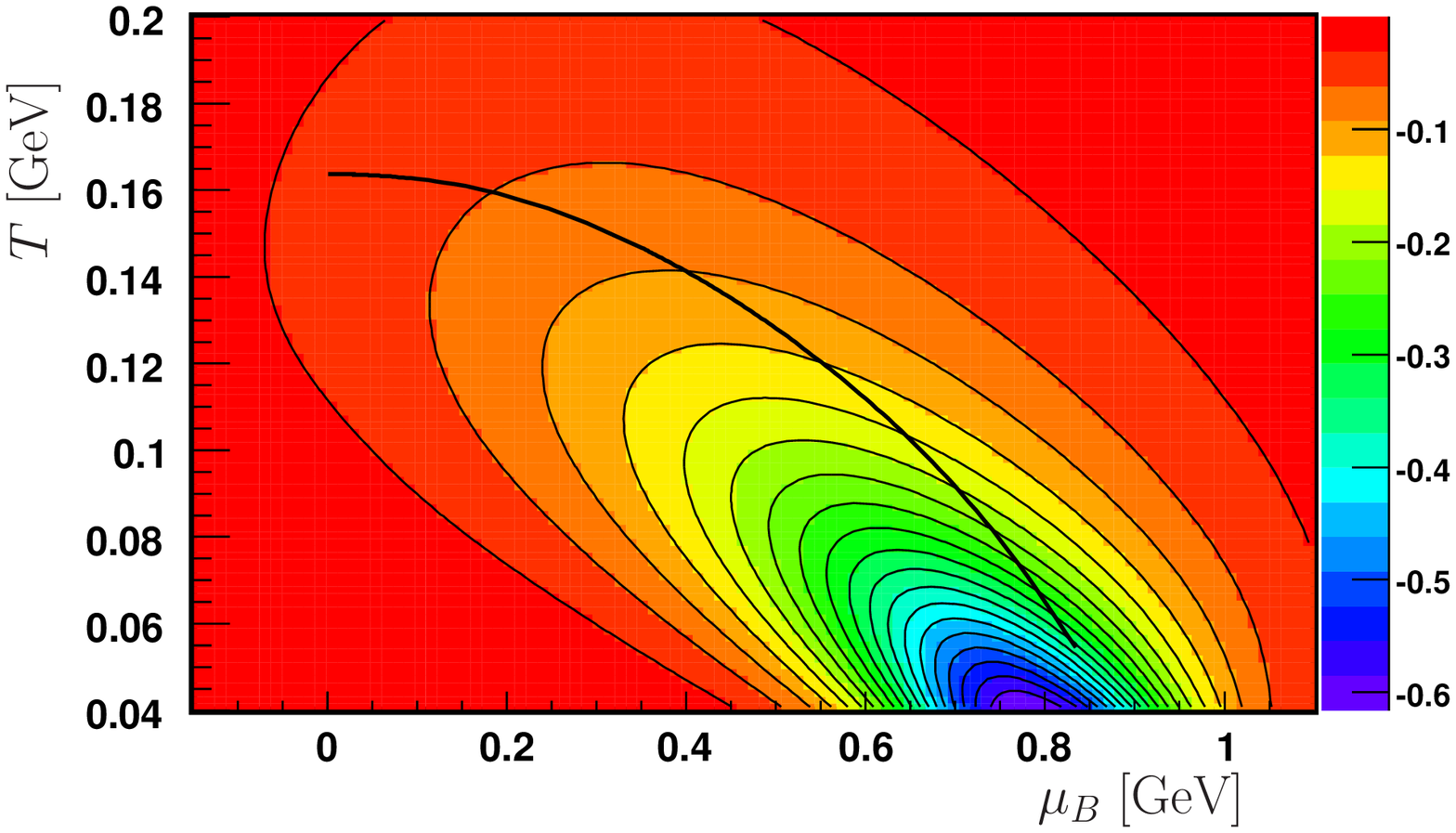,width=7.cm,height=6.5cm}
    \epsfig{file=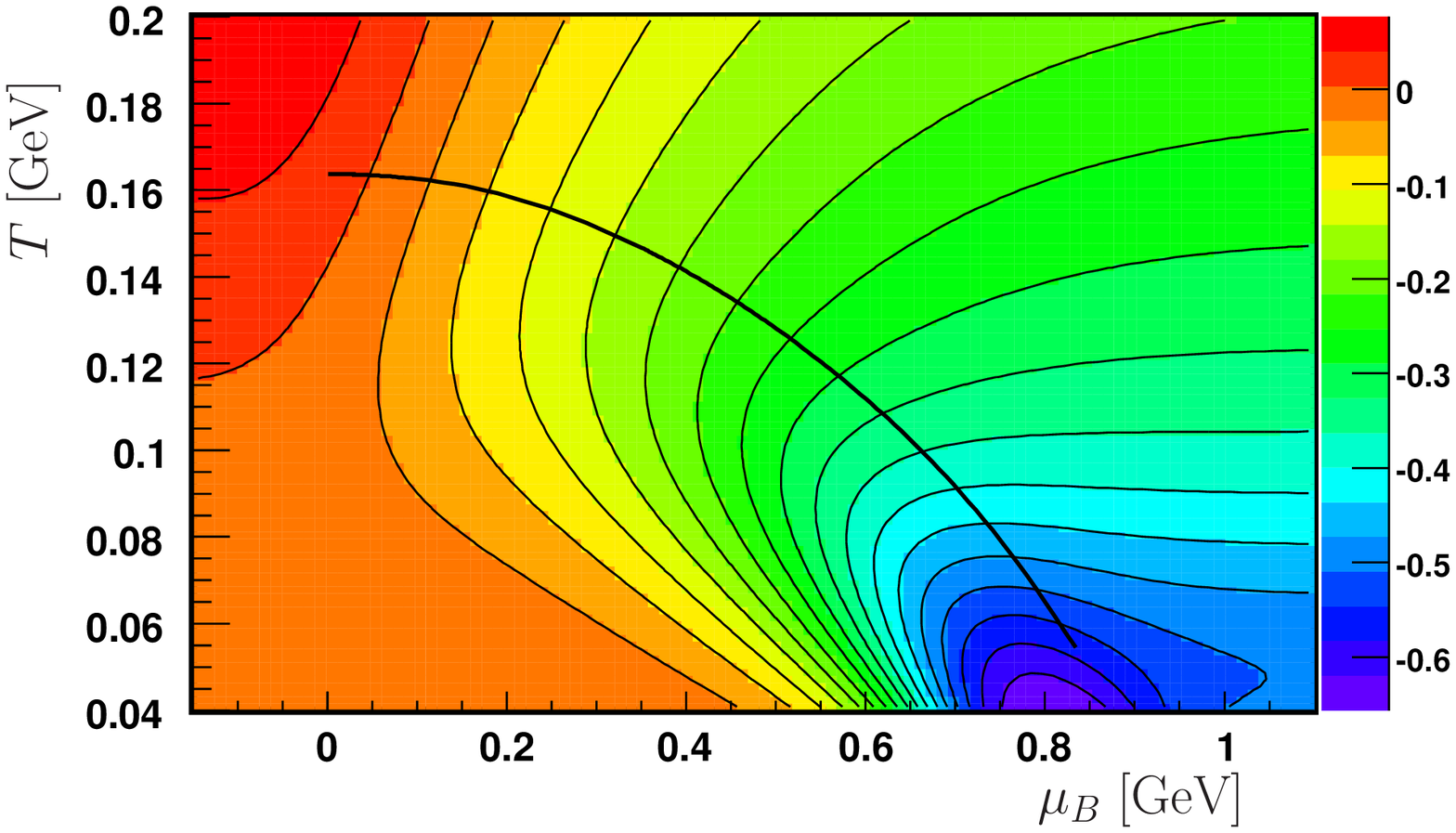,width=7.cm,height=6.5cm}
    \caption{
    CE correlation coefficient between the multiplicities of protons and positively charged 
    pions~$\rho_{\textrm{p} \pi^+}$. Both, primordial ({\it left}), and final state~({\it right}).} 
    \label{T_muB_rho_p_piplus_CE}
  \end{center}
\end{figure}

\begin{figure}[ht!]
  \begin{center}
    \epsfig{file=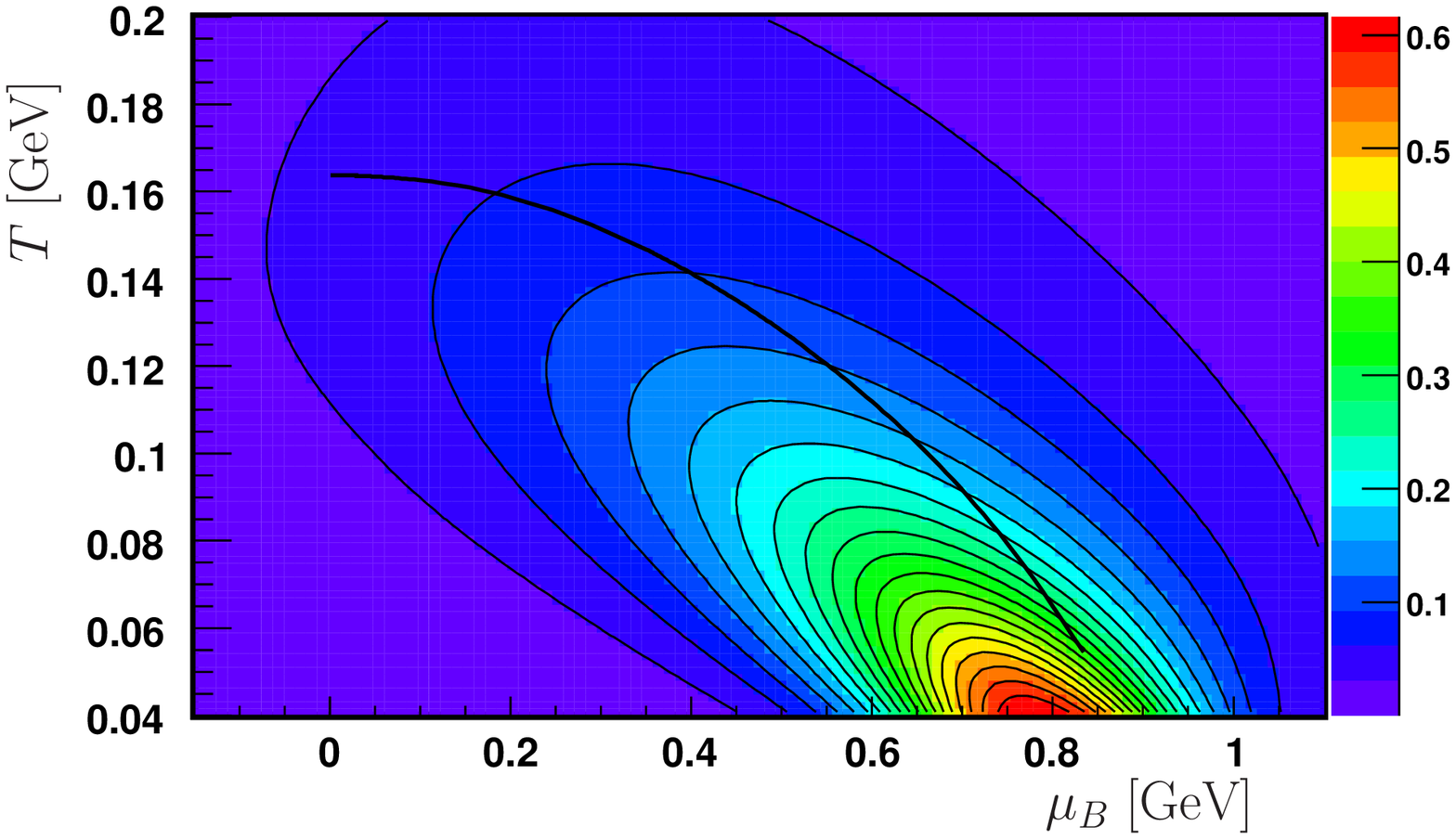,width=7.cm,height=6.5cm}
    \epsfig{file=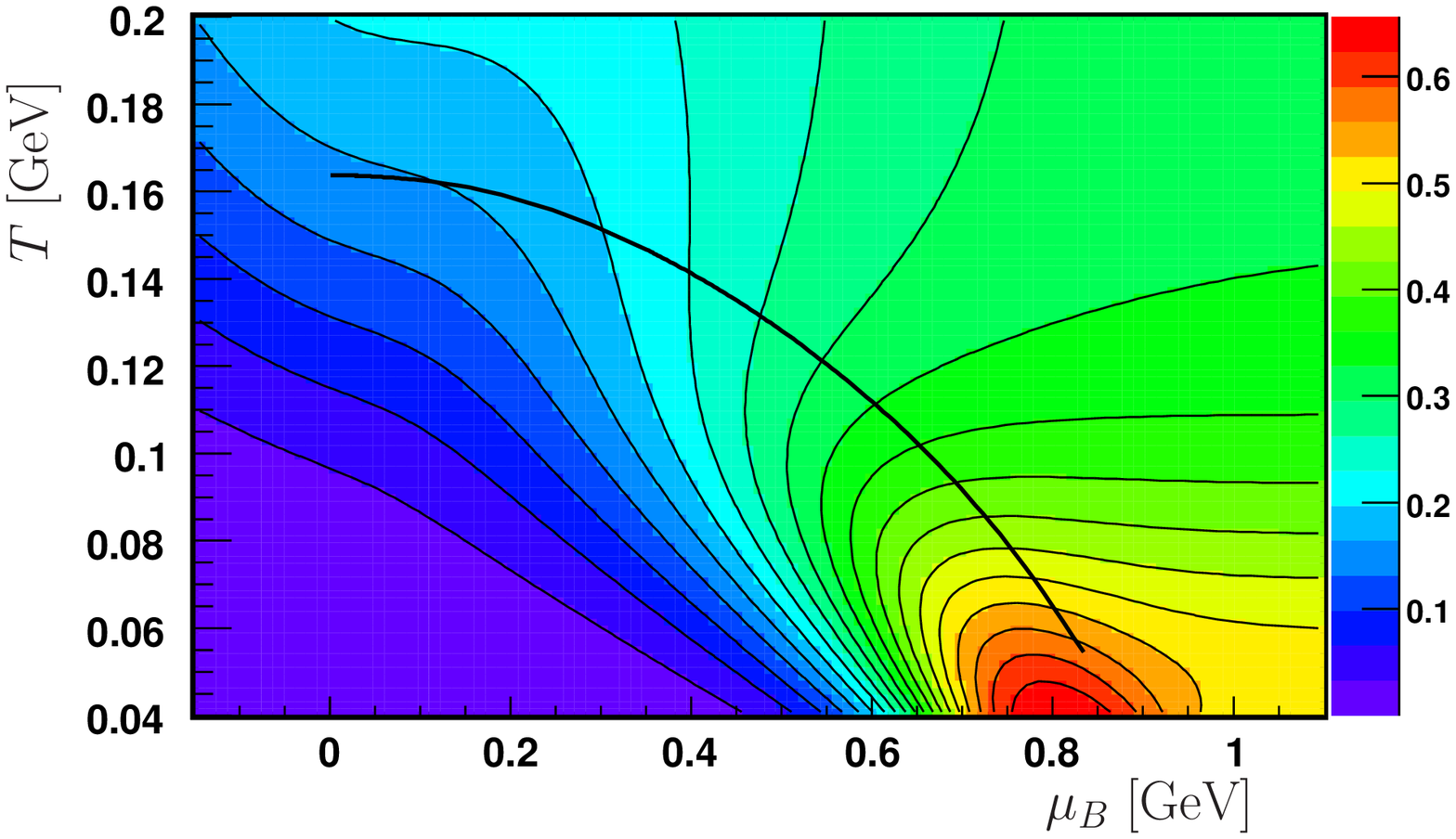,width=7.cm,height=6.5cm}
    \caption{
    CE correlation coefficient between the multiplicities of protons and negatively charged pions~$\rho_{\textrm{p} \pi^-}$. 
    Both, primordial~({\it left}), and final state~({\it right}). } 
    \label{T_muB_rho_p_piminus_CE}
  \end{center}
\end{figure}

The event-by-event multiplicities of~p and~$\pi^+$ are anti-correlated in the primordial~CE, 
Fig.(\ref{T_muB_rho_p_piplus_CE})~({\it left}), while the multiplicities of~p and~$\pi^-$ are positively correlated, 
Fig.(\ref{T_muB_rho_p_piminus_CE})~({\it left}), due to electric charge conservation. Additionally, multiplicity 
fluctuations of either species are suppressed in the~CE (see also Chapter~\ref{chapter_freezeoutline}).

The multiplicities of the hadrons~p and~$\pi^+$ are most strongly anti-correlated at low~$T$ and large~$\mu_B$. 
The multiplicity of baryons cannot fluctuate strongly in cold net-baryon rich matter, due to suppressed anti-baryon 
states, compare Fig.(\ref{T_muB_GCE_fluc})~({\it left}). Charge conservation now requires that once one removes (adds) 
a proton from (to) the system, to add (remove) a neutron and a positively charged 
pion, $\textrm{p} \leftrightarrow \textrm{n} + \pi^+$. The system is too cold to allow for strong baryon multiplicity 
fluctuations\footnote{Energy fluctuations in the~CE are also suppressed in comparison to the~GCE~\cite{Begun:2004zb}.}. 
In a neutral, but hot, hadron resonance gas the anti-correlation is weaker, since channels involving anti-baryons are 
open. In a hot and baryon rich phase, the appearance of heavy~$\Delta$ resonances makes the~p-$\pi$ channel less 
influential, and accordingly~$\rho_{\textrm{p} \pi^+} \simeq 0$ there.

For the final state~CE correlation coefficient~$\rho_{\textrm{p} \pi^+}$, in Fig.(\ref{T_muB_rho_p_piplus_CE})~({\it right}), 
the anti-correlation between~p and~$\pi^+$ extents to high~$\mu_B$ and~$T$. Primordial parent particles, in particular 
positively charged~$\Delta$ resonances, are anti-correlated, due to baryon number and electric charge conservation, 
with primordial~p and~$\pi^+$. These~$\Delta$ decay into both,~p and~$\pi^+$, and partially re-correlate their 
multiplicities, yet an residual anti-correlation remains. At low~$T$ and low~$\mu_B$  the correlation is then 
weaker, since the system is too cold to produce sufficient baryonic resonances which could correlate~p and~$\pi^+$ via 
their decay, i.e. the primordial and the final state scenario~$\rho_{\textrm{p} \pi^+}$ are similar in this region of the 
phase diagram. At high~$T$, but~$\mu_B \simeq 0$, the final state correlation coefficient is 
positive, $\rho_{\textrm{p} \pi^+} > 0$, unlike one would expect from electric charge conservation. In a neutral and hot 
hadron resonance gas, baryon and anti-baryon number fluctuations are sufficient, to de-correlate primordial~p 
and~$\pi^+$. The positive resonance decay contribution causes then an overall positive correlation.
In final state~GCE one always finds~$\rho_{\textrm{p} \pi^+} > 0$.

In the primordial~CE, one finds the correlation coefficient between the multiplicities of~p and~$\pi^-$ to be positive,
$\rho_{\textrm{p} \pi^-}>0$, Fig.(\ref{T_muB_rho_p_piminus_CE})~({\it left}), due to electric charge conservation. 
Their correlation is particularly strong for low~$T$ and large~$\mu_B$, since there baryon multiplicity fluctuations 
are strongly suppressed, making the channel~$\textrm{n} \leftrightarrow \textrm{p} + \pi^-$ dominant. 
The correlation is weaker elsewhere in the~$T$-$\mu_B$ phase diagram. The final state~$\rho_{\textrm{p} \pi^-}$, 
Fig.(\ref{T_muB_rho_p_piminus_CE})~({\it right}), is similar to the primordial scenario for low temperature.  
For larger $T$ and~$\mu_B$, resonance decay then again correlates the event-by-event multiplicities of~p and~$\pi^-$, 
when compared to the primordial scenario.

\subsubsection{Micro Canonical Ensemble}

\begin{figure}[ht!]
  \begin{center}
    \epsfig{file=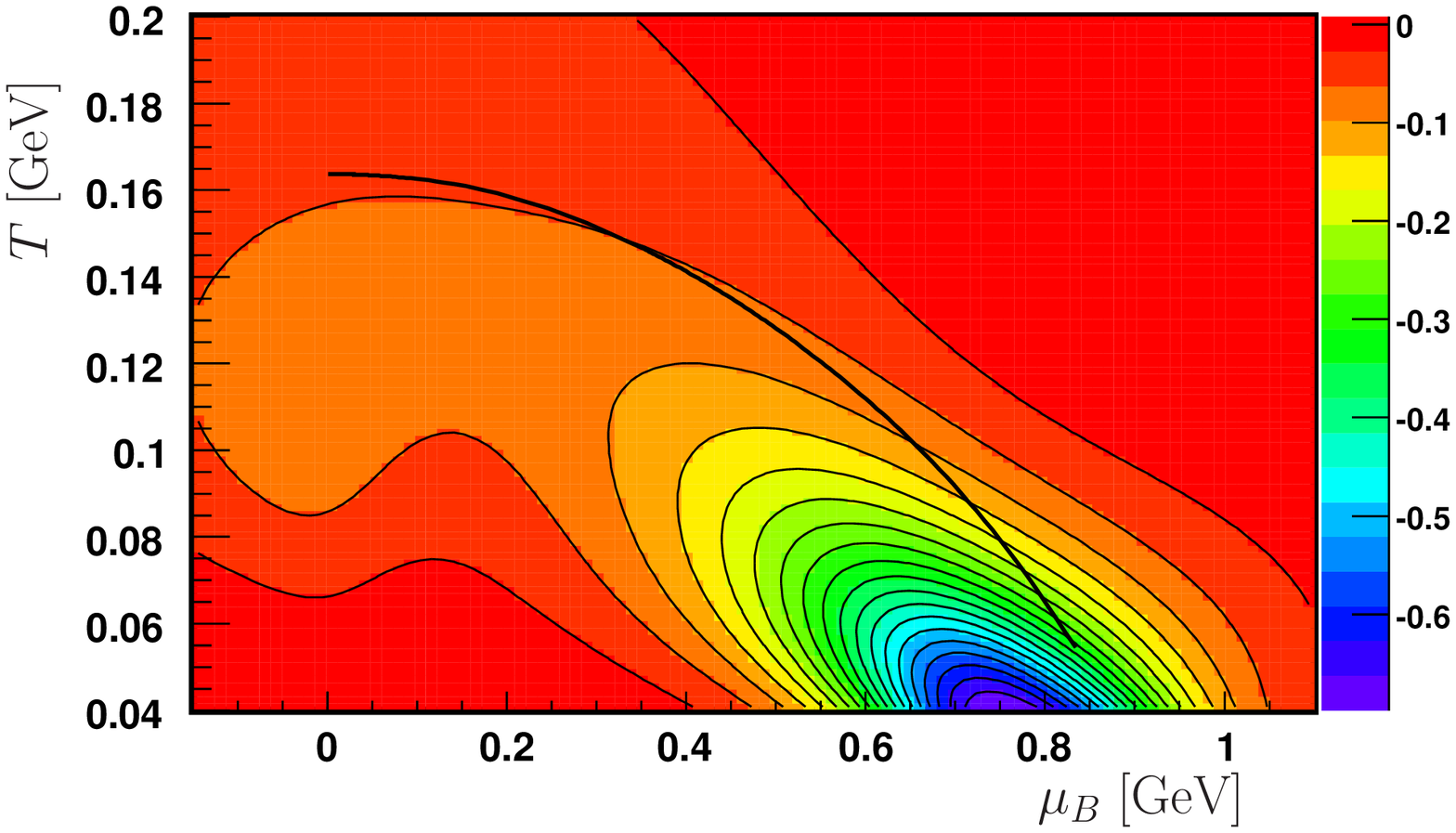,width=7.cm,height=6.5cm}
    \epsfig{file=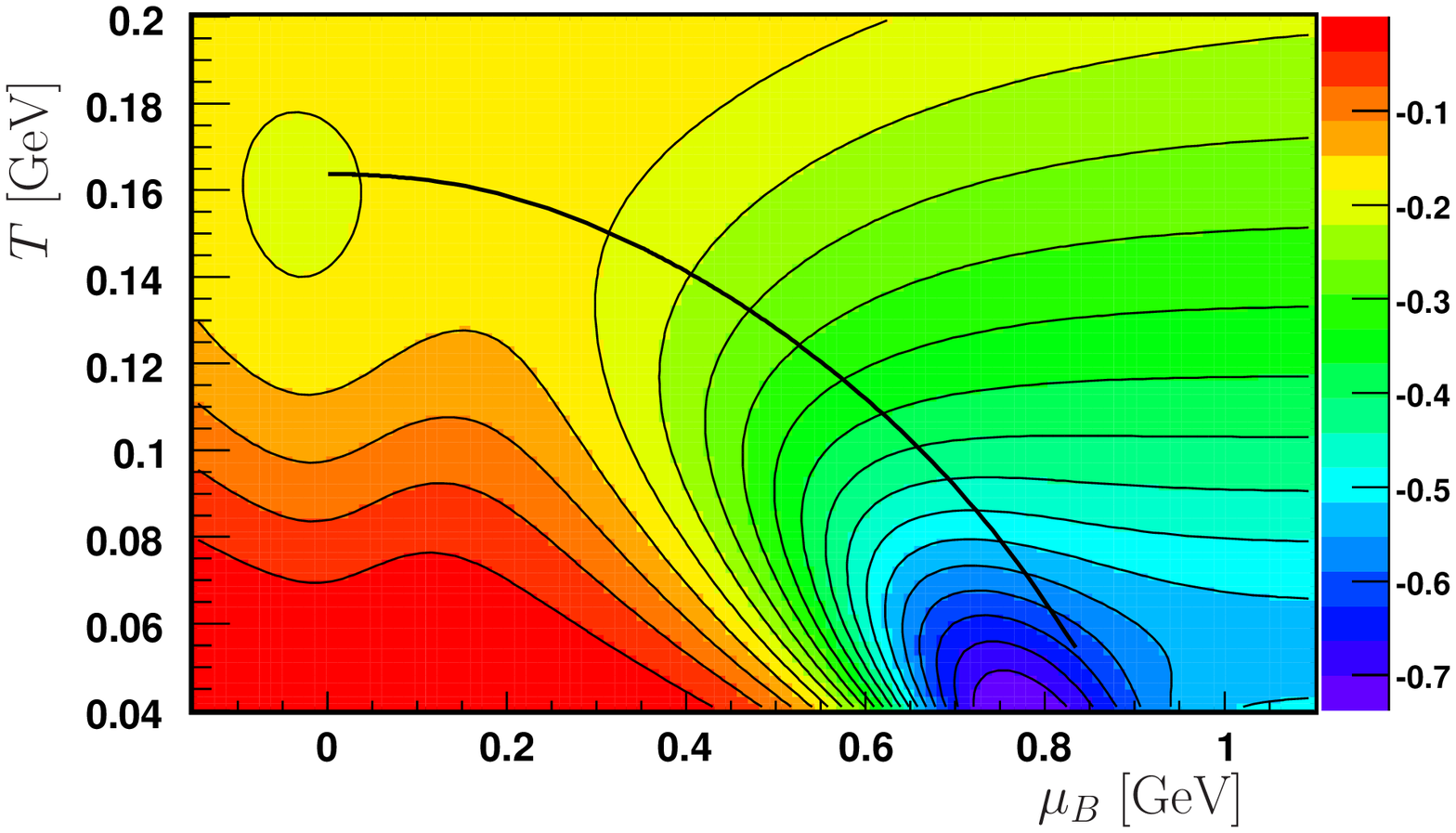,width=7.cm,height=6.5cm}
    \caption{
    MCE correlation coefficient between the multiplicities of protons and positively charged 
    pions~$\rho_{\textrm{p} \pi^+}$. Both, primordial~({\it left}), and final state~({\it right}).} 
    \label{T_muB_rho_p_piplus_MCE}
  \end{center}
\end{figure}

The primordial~MCE correlation coefficient~$\rho_{\textrm{p} \pi^+}$, Fig.(\ref{T_muB_rho_p_piplus_MCE})~({\it left}),
is generally weaker than its~CE counterpart~$|\rho_{\textrm{p} \pi^+}^{mce}| < |\rho_{\textrm{p} \pi^+}^{ce}|$, with
the exception of a deeper pocket in the bottom right corner of the phase diagram, denoting cold baryonic matter, 
where~$|\rho_{\textrm{p}\pi^+}^{mce}|>|\rho_{\textrm{p}\pi^+}^{ce}|$. 
The anti-correlation~$\textrm{p} \leftrightarrow \textrm{n} + \pi^+$ is made stronger at low temperature in the~MCE by 
combined energy and charge conservation. The sum of the multiplicities of~p and~n cannot vary much due to strongly 
suppressed~$\overline{\textrm{p}}$ and~$\overline{\textrm{n}}$ yields. The anti-correlation between the multiplicities 
of~p and~$\pi^+$ is strongly mediated by the~n multiplicity. At larger temperature other channels, 
like~$\Delta^+ \leftrightarrow \textrm{n} + \pi^+$, seem preferable for energy conservation. Charge conservation does 
not distinguish between these two channels, leading to~$|\rho_{\textrm{p} \pi^+}^{ce}| > |\rho_{\textrm{p} \pi^+}^{mce}|$
elsewhere in the phase diagram, except in the pocket.

\begin{figure}[ht!]
  \begin{center}
    \epsfig{file=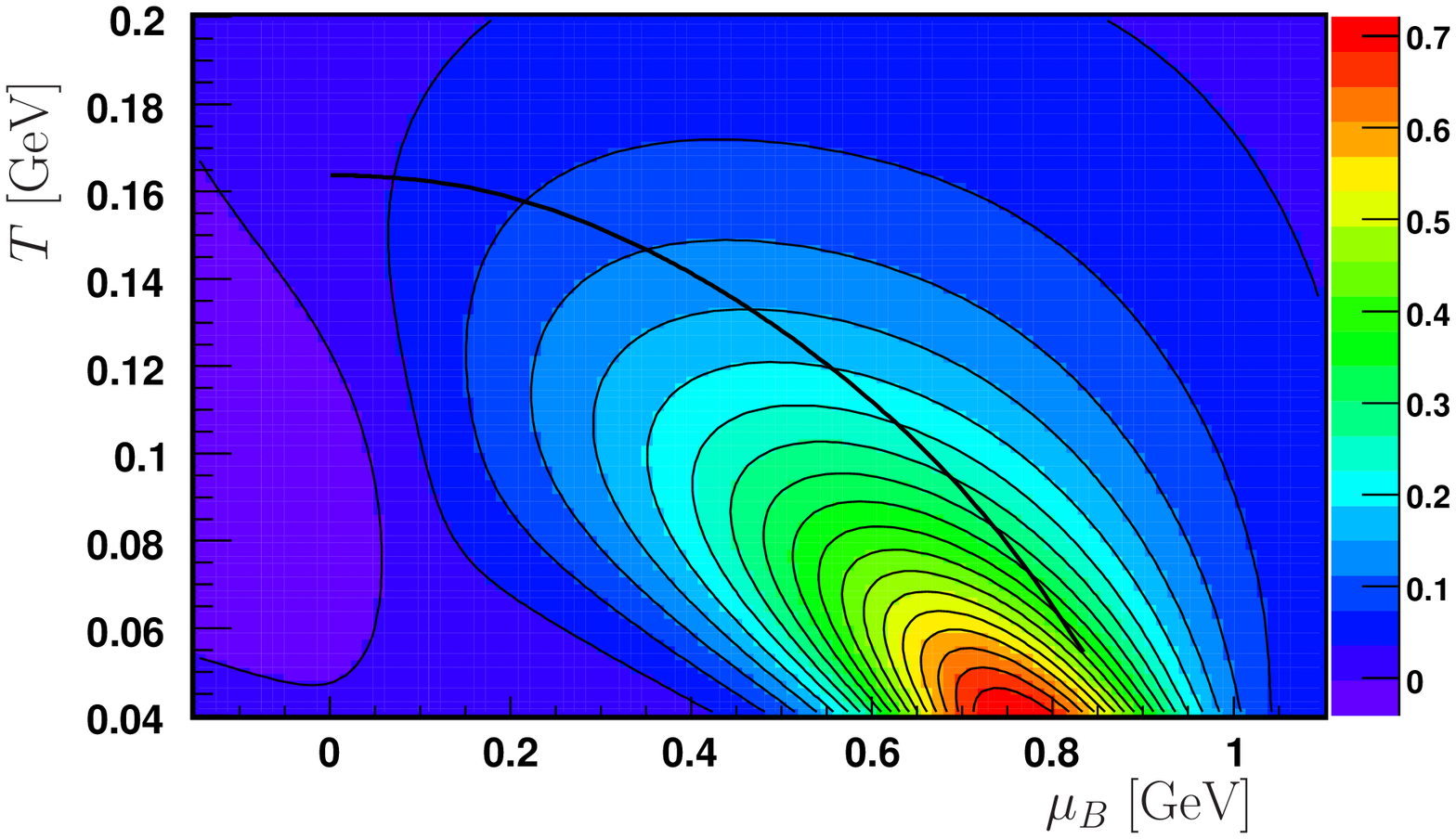,width=7.cm,height=6.5cm}
    \epsfig{file=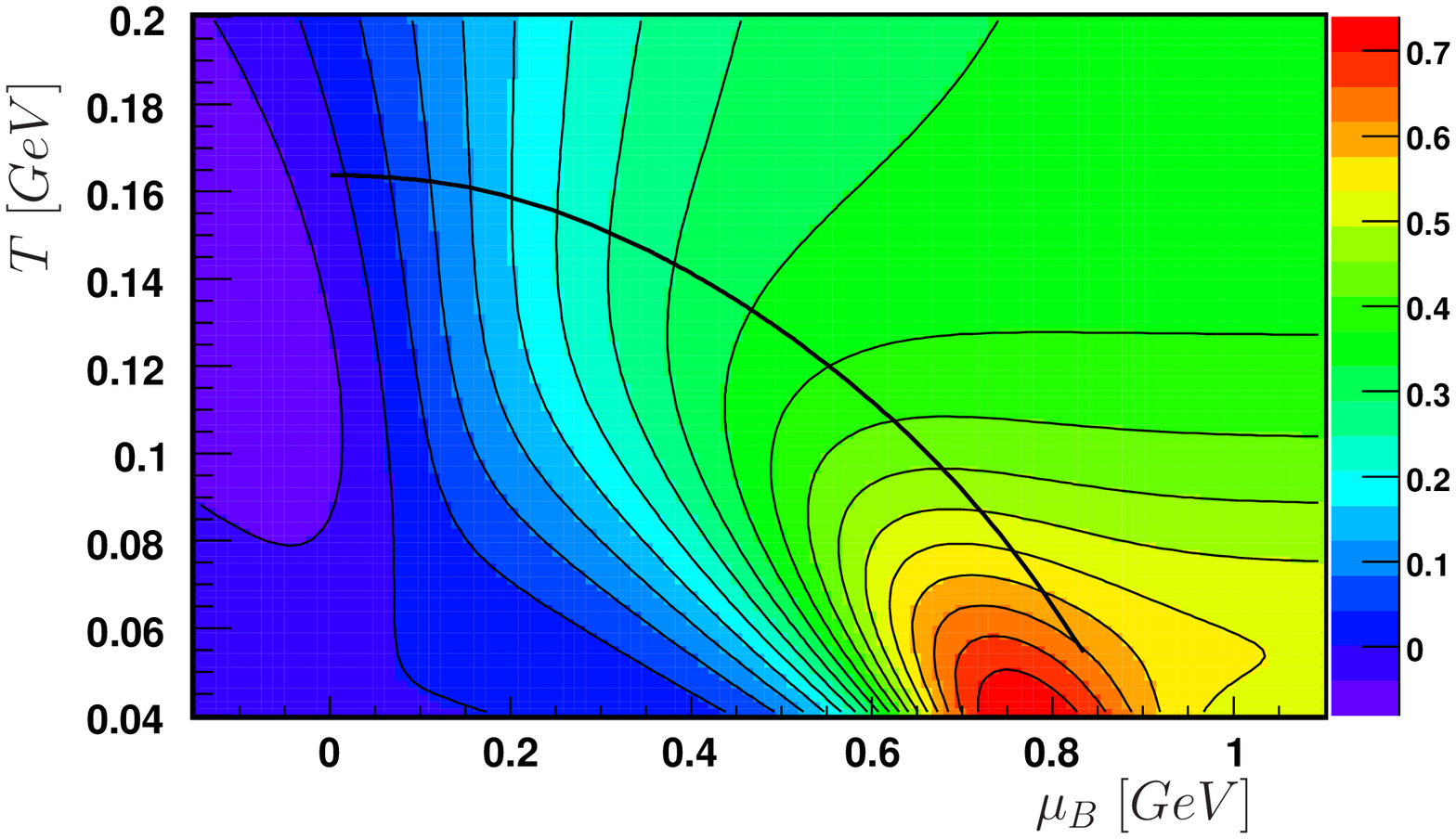,width=7.cm,height=6.5cm}
    \caption{
    MCE correlation coefficient between the multiplicities of protons and negatively charged 
    pions~$\rho_{\textrm{p} \pi^-}$. Both, primordial~({\it left}), and final state~({\it right}).} 
    \label{T_muB_rho_p_piminus_MCE}
  \end{center}
\end{figure}

The final state~MCE correlation coefficient~$\rho_{\textrm{p} \pi^+}$, Fig.(\ref{T_muB_rho_p_piplus_MCE})~({\it right}), 
is negative, $\rho_{\textrm{p} \pi^+}<0$, and notably stronger than in the final state~CE. Energy conservation, and hence 
anti-correlation between primordial parent particles and primordial~p and~$\pi^+$, is responsible. 

The primordial~MCE correlation coefficient~$\rho_{\textrm{p} \pi^-}$, Fig.(\ref{T_muB_rho_p_piminus_MCE})~({\it left}),
is again positive, $\rho_{\textrm{p} \pi^-}>0$, like in the~CE. The channel~$\textrm{n} \leftrightarrow \textrm{p} + \pi^-$ 
is not much affected by additional energy conservation, with the notable exception of a stronger positive 
correlation~$\rho^{mce}_{\textrm{p}\pi^-}>\rho^{ce}_{\textrm{p}\pi^-}$ in cold baryonic matter.

From a comparison to the final state~MCE correlation coefficient~$\rho_{\textrm{p} \pi^-}$, 
Fig.(\ref{T_muB_rho_p_piminus_MCE})~({\it right}), again a positive resonance decay contribution is visible. 
One notes that the~CE enhancement in net-baryon free and hot matter has vanished in the~MCE. Energy fluctuations in 
the~CE at a given temperature are strongest if the system is charge neutral. In the~MCE the energy content is fixed, 
and anti-correlation amongst primordial parent resonance and primordial~p and~$\pi^+$ balances the enhancement due 
to resonance decay.

\begin{figure}[ht!]
  \begin{center}
    \epsfig{file=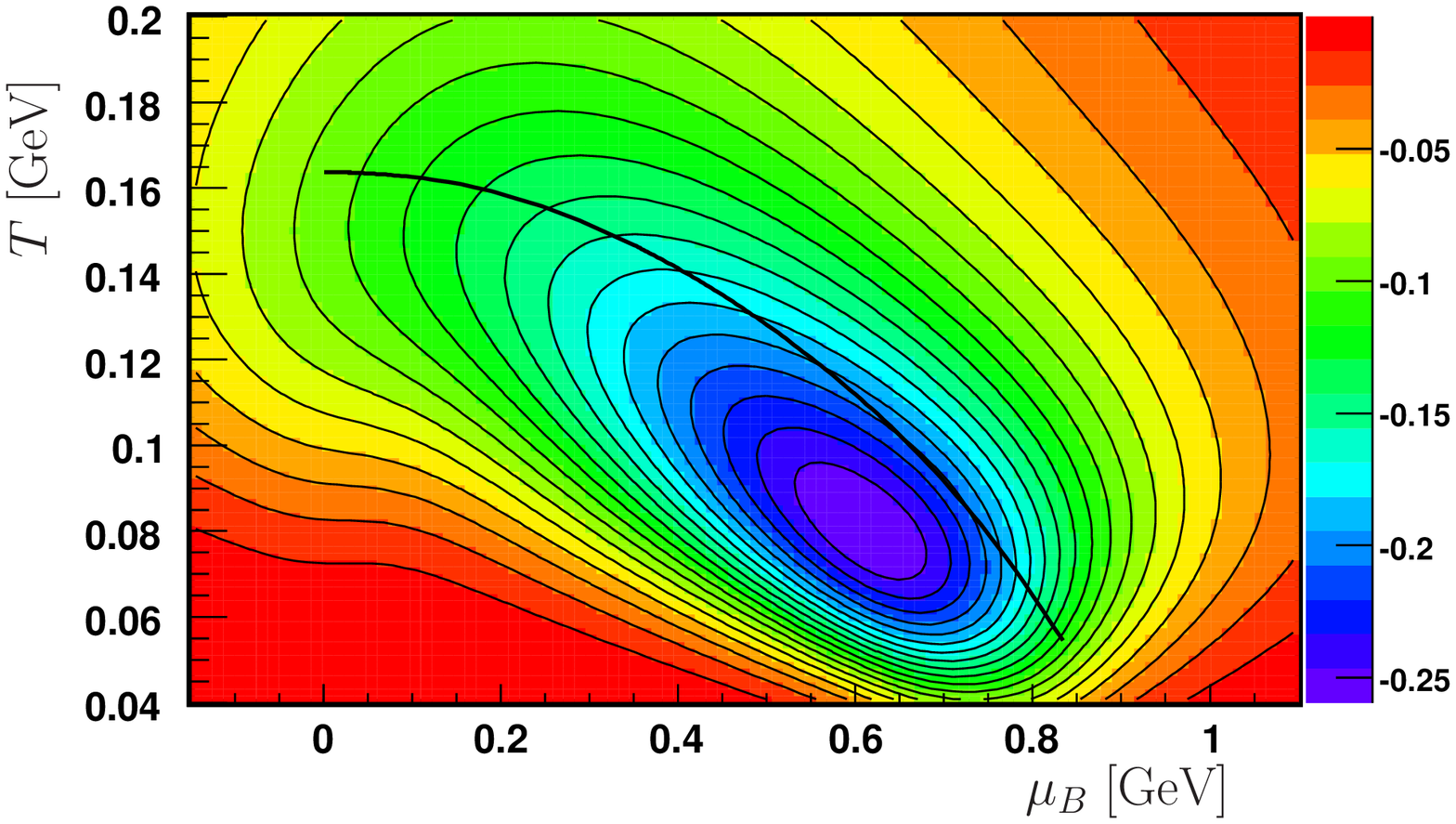,width=7.cm,height=6.5cm}
    \epsfig{file=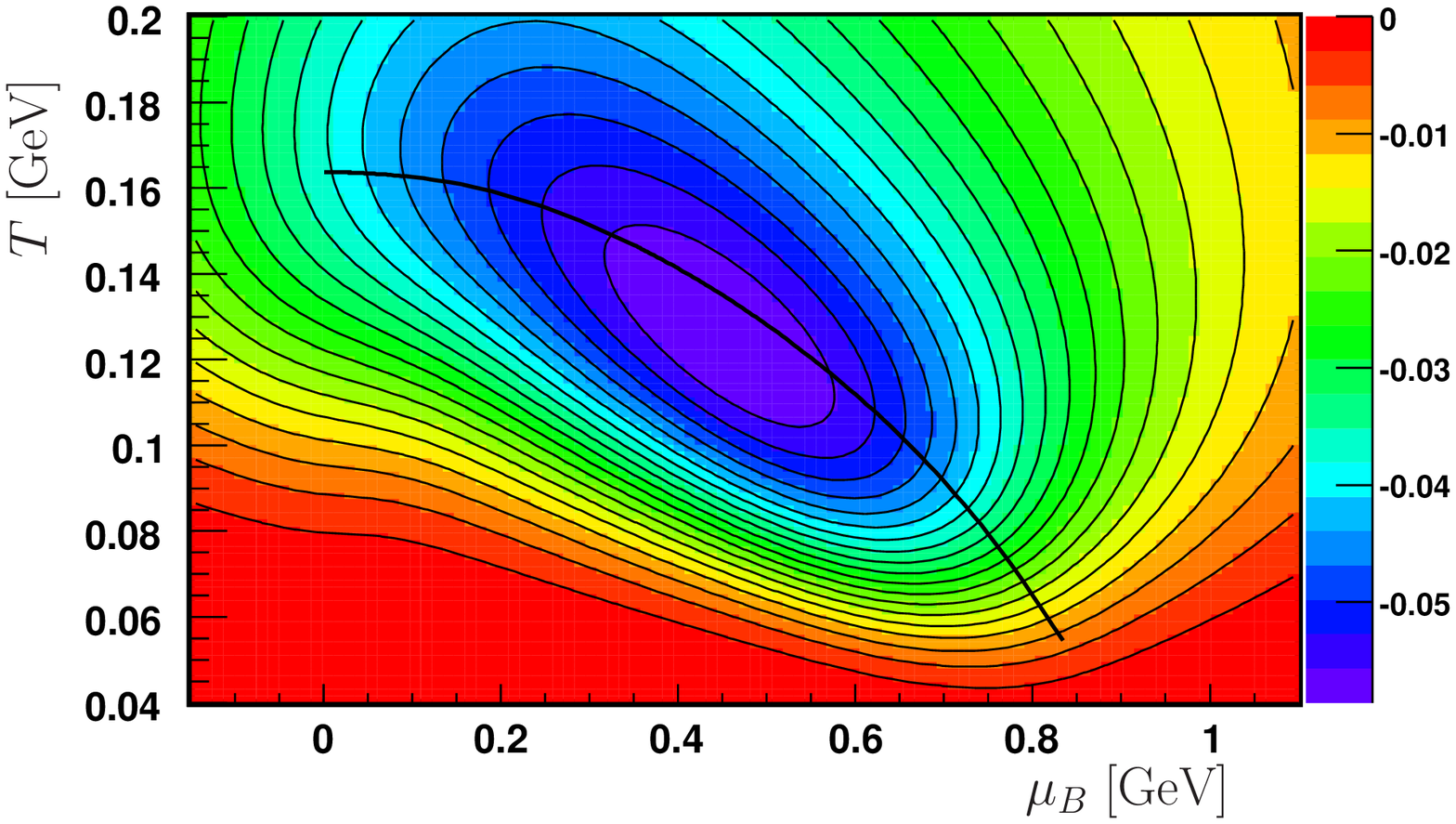,width=7.cm,height=6.5cm}
    \caption{
    Primordial MCE correlation coefficient between the multiplicities of the doublely charged $\Delta^{++}(1232)$
    and positively charged pions~$\rho_{\Delta^{++}(1232) \pi^+}$ ({\it left}), and between the multiplicities 
    of the doublely charged $\Delta^{++}(1600)$ and positively charged pions~$\rho_{\Delta^{++}(1600) \pi^+}$~({\it right}).} 
    \label{T_muB_rho_Delta_piplus_MCE}
  \end{center}
\end{figure}

To conclude the discussion of multiplicity correlations primordial MCE correlation coefficient between the multiplicities 
of the doublely charged $\Delta^{++}(1232)$ and positively charged pions~$\rho_{\Delta^{++}(1232) \pi^+}$, and between the 
multiplicities of the doublely charged $\Delta^{++}(1600)$ and positively charged pions~$\rho_{\Delta^{++}(1600) \pi^+}$, 
are shown in Fig.(\ref{T_muB_rho_Delta_piplus_MCE}) ({\it left}) and ({\it right}) respectively. 

The $\Delta$ resonances have larger rest-mass than the ground state nucleons~n and~p. Hence, they start to contribute 
to the system partition function only at larger $T$, and preferably at large $\mu_B$. The correlation between the 
event-by-event multiplicities of $\Delta$ resonances and $\pi$ mesons is the stronger the larger a fraction of the total 
energy density they make up. The maxima in Fig.(\ref{T_muB_rho_Delta_piplus_MCE}) ({\it left}) and ({\it right}) are hence 
in different positions. The heavier particle shows a weaker correlation. The correlation 
coefficients $\rho_{\Delta^{++}(1232) \pi^+}$ and $\rho_{\Delta^{++}(1600) \pi^+}$ should be compared to the primordial 
MCE $\rho_{\textrm{p}\pi^+}$ in Fig.(\ref{T_muB_rho_p_piplus_MCE}) ({\it left}). The CE yields a similar structure. 
In the GCE no primordial correlations appear in the ideal gas approximation.

\subsubsection{Meson Baryon Correlations}

\begin{figure}[ht!]
  \begin{center}
    \epsfig{file=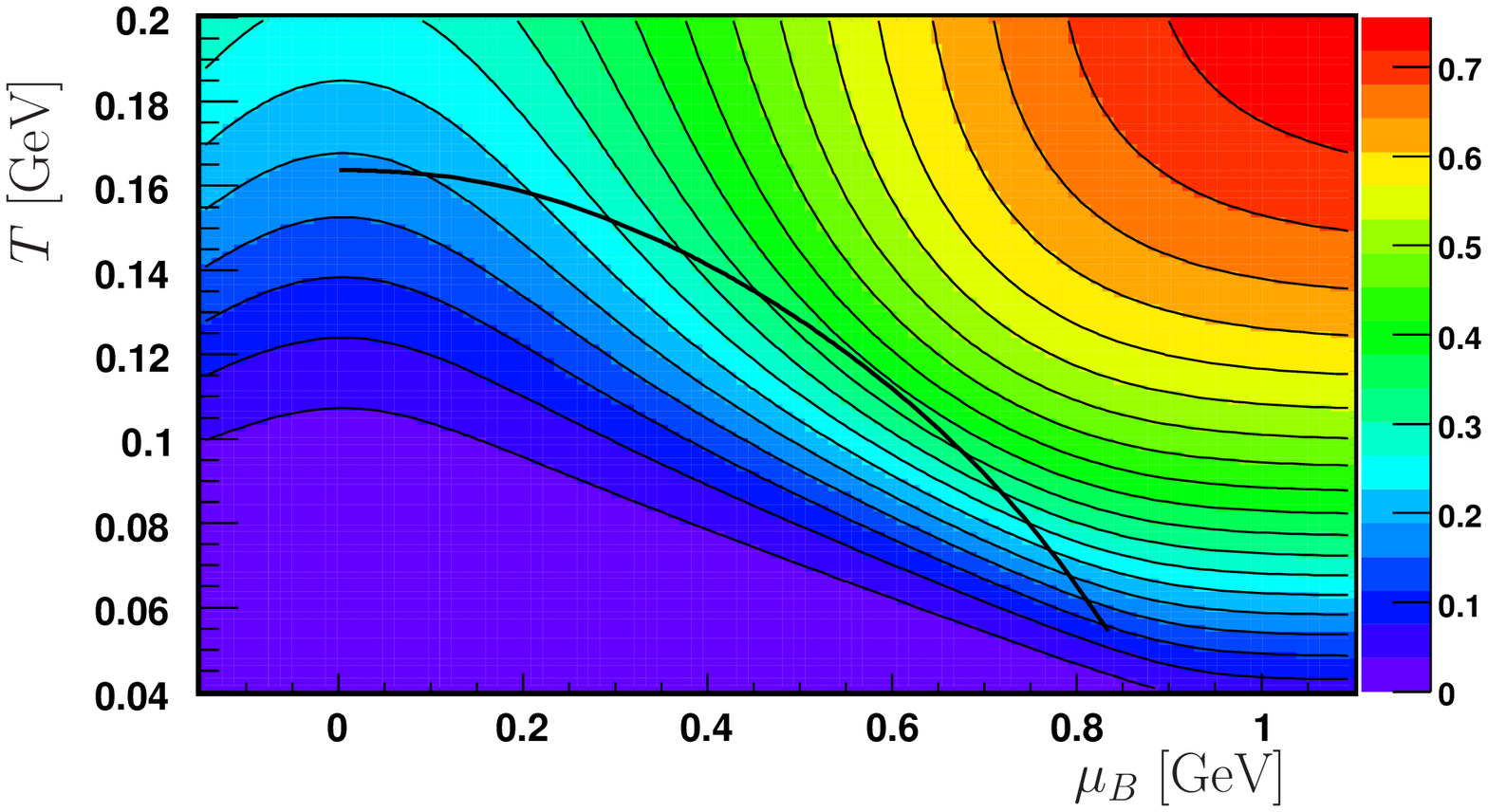,width=7.cm,height=6.5cm}
    \epsfig{file=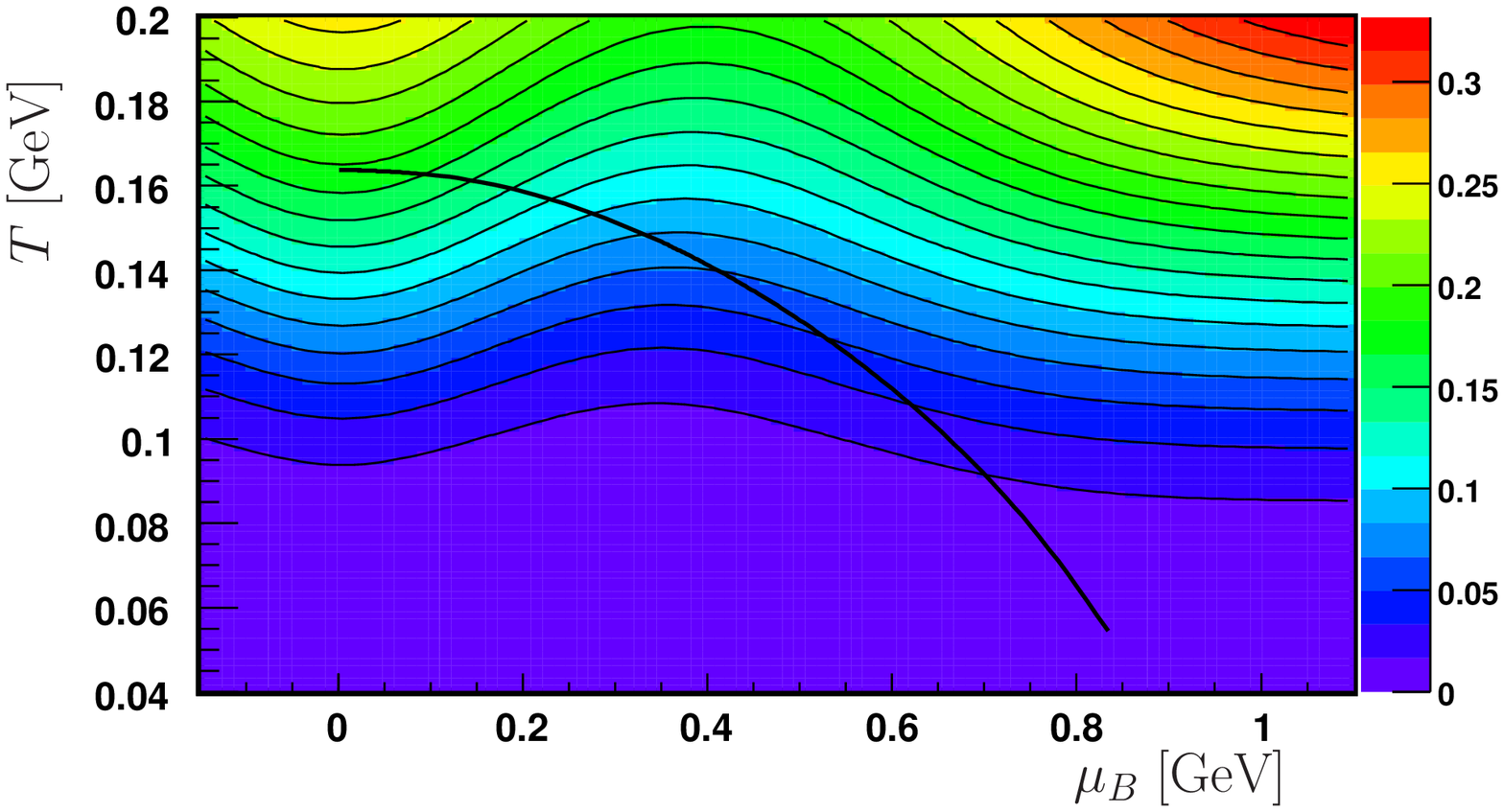,width=7.cm,height=6.5cm}
    \caption{   
    Final state  correlation coefficient~$\rho_{\mathcal{M}\mathcal{B}}$ between the event-by-event multiplicities of 
    mesons plus anti-mesons and baryon plus anti-baryons in the GCE~({\it left}) and CE~({\it right}).} 
    \label{T_muB_corr_MesBar_final}
  \end{center}
\end{figure}

\begin{figure}[ht!]
  \begin{center}
    \epsfig{file=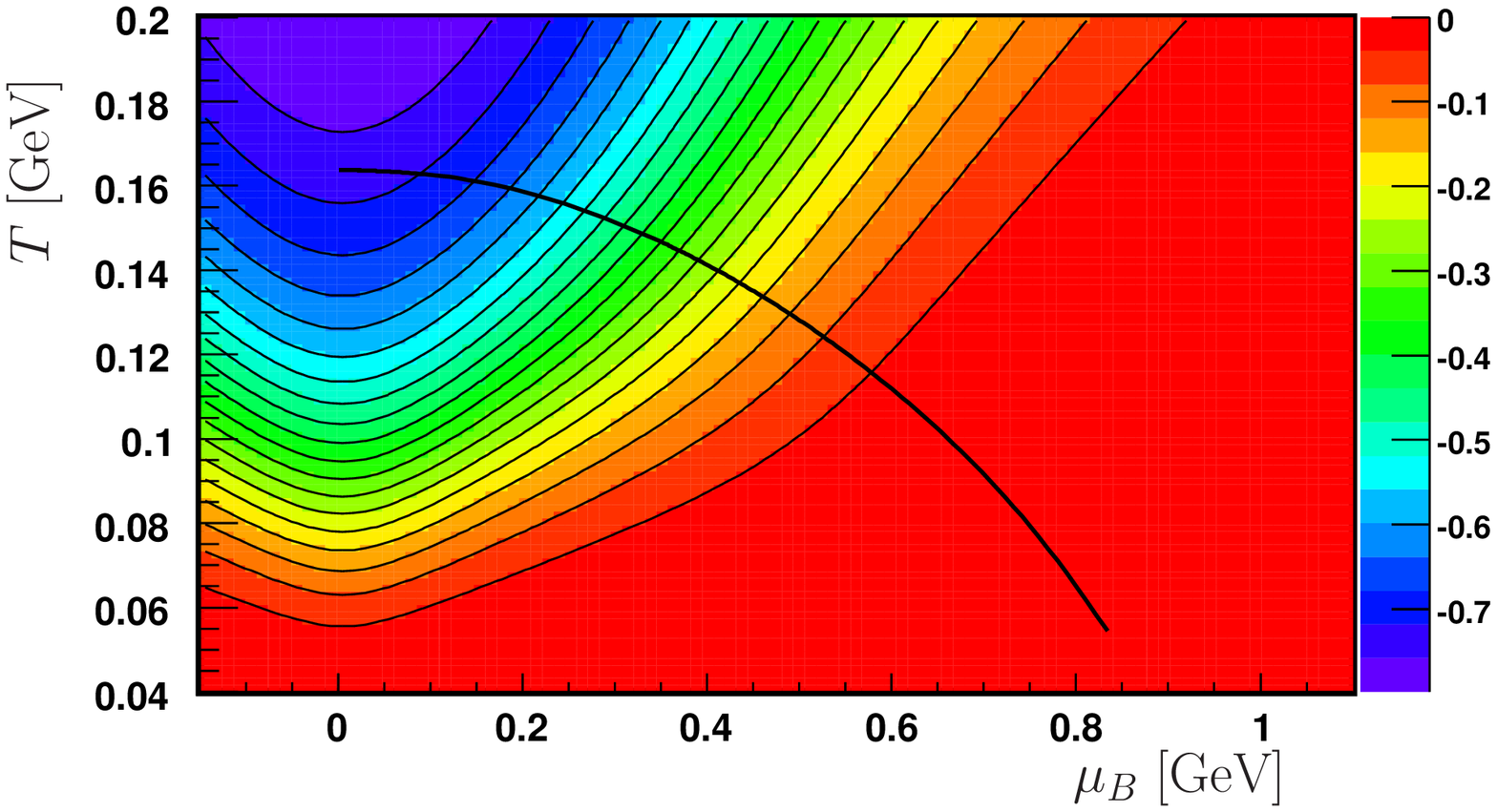,width=7.cm,height=6.5cm}
    \epsfig{file=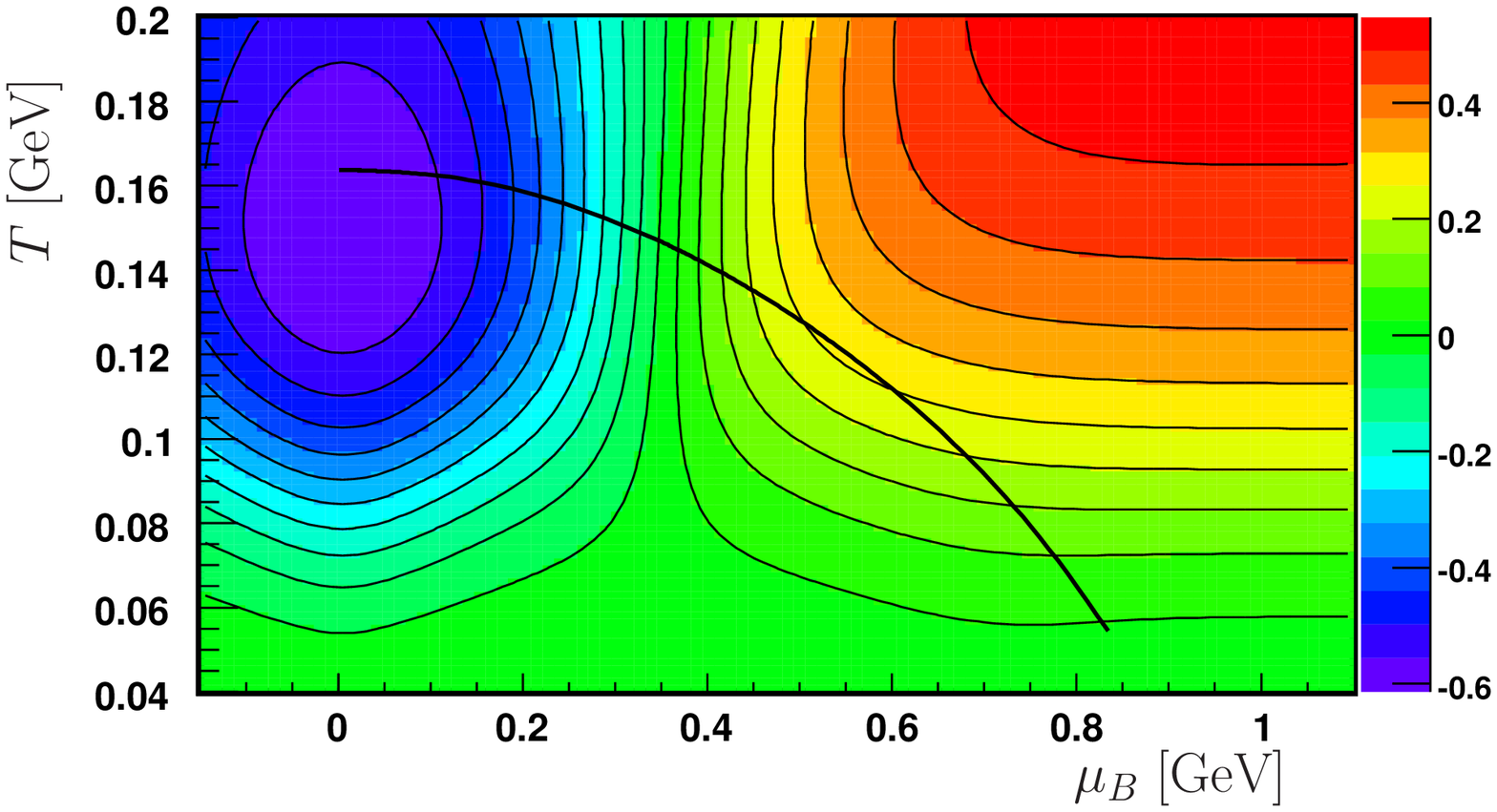,width=7.cm,height=6.5cm}
    \caption{
    MCE correlation coefficient~$\rho_{\mathcal{M}\mathcal{B}}$ between the event-by-event multiplicities of mesons plus 
    anti-mesons and baryon plus anti-baryons in the MCE, both, primordial~({\it left}) and final state~({\it right}).} 
    \label{T_muB_corr_MesBar_MCE}
  \end{center}
\end{figure}

Lastly, the correlation coefficients~$\rho_{\mathcal{M}\mathcal{B}}$ between the event-by-event multiplicities of mesons 
plus anti-mesons,~$N_{\mathcal{M}}=N_{\textrm{M}}+N_{\overline{\textrm{M}}}$, and the event-by-event multiplicities of baryon 
plus anti-baryons,~$N_{\mathcal{B}}=N_{\textrm{B}}+N_{\overline{\textrm{B}}}$, are considered in different ensembles. 
In Fig.(\ref{T_muB_corr_MesBar_final}) the temperature - baryon chemical potential phase diagram is shown for the 
final state correlation coefficient~$\rho_{\mathcal{M}\mathcal{B}}$ in the GCE~({\it left}) and CE~({\it right}).
In Fig.(\ref{T_muB_corr_MesBar_MCE}) the MCE correlation coefficient~$\rho_{\mathcal{M}\mathcal{B}}$ is shown,
both primordial~({\it left}) and final state~({\it right}).

No primordial correlation amongst distinct groups of hadrons exists in the GCE. Resonance decay is then again the only 
source of correlation, Fig.(\ref{T_muB_corr_MesBar_final})~({\it left}), between mesons and baryons. Yet, resonances 
essentially do not decay into baryon anti-baryon pairs,~$\textrm{B} + \overline{\textrm{B}}$. At low temperature,
where few resonances are formed, the correlation coefficient is then~$\rho_{\mathcal{M}\mathcal{B}} \sim 0 $. 
On the other hand, when the net-baryon density is large (positive or negative), either baryon or anti-baryons are 
strongly suppressed, and the multiplicities~$N_{\mathcal{M}}=N_{\textrm{M}}+N_{\overline{\textrm{M}}}$, are essentially 
correlated with $N_{\textrm{B}}$ ($\mu_B>0$) or $N_{\overline{\textrm{B}}}$ ($\mu_B<0$). 

Also in the CE primordial correlations are absent, due to the particular choice made for strangeness and electric 
charge chemical potentials~$\mu_Q=\mu_S=0$. Resonance decay enhancement, Fig.(\ref{T_muB_corr_MesBar_final})~({\it right}), 
is again visible at large~$T$ and~$\mu_B$. At large~$T$, but small~$\mu_B$, however the correlation coefficient is 
somewhat stronger (than in the GCE). The expectation value of the sum of baryons plus 
anti-baryons~$N_{\mathcal{B}}=N_{\textrm{B}}+N_{\overline{\textrm{B}}}$ is largest (at a given temperature) for 
charge neutral matter.  However, the strangeness and electric charge content of the baryonic sector is still fluctuating,
leading to a weak and positive correlation to the mesonic sector.

In the primordial MCE, Fig.(\ref{T_muB_corr_MesBar_MCE})~({\it left}) one observes a strong correlation 
between the event-by-event multiplicities of mesons plus anti-mesons and baryon plus anti-baryons. Unlike in the 
GCE and CE, the primordial MCE correlation coefficient is non-vanishing and negative,~$\rho_{\mathcal{M}\mathcal{B}} < 0$.
The difference between the  multiplicities of baryons and of anti-baryons, i.e. the system baryon number is 
conserved,~$B=N_{\textrm{B}}-N_{\overline{\textrm{B}}}$. Yet, fluctuations of their 
sum,~$N_{\mathcal{B}}=N_{\textrm{B}}+N_{\overline{\textrm{B}}}$, however release or bind a large amount of energy. 
But as energy is conserved in the MCE, the multiplicity~$N_{\mathcal{M}}=N_{\textrm{M}}+N_{\overline{\textrm{M}}}$ needs to 
adjust. Energy is spent either on the production of a baryon anti-baryon pair, or on the production of a few ($>2$) mesons 
and anti-mesons. This effect is particularly strong in hot and neutral matter. Resonance decay in the MCE, 
Fig.(\ref{T_muB_corr_MesBar_MCE})~({\it right}), leads again to a positive correlation  large~$T$ and~$\mu_B$ amongst 
final state hadrons, yet, a strong anti-correlation at large~$T$ and~$\mu_B \sim 0$ remains.

Following the~$\langle E \rangle / \langle N \rangle \simeq 1$~GeV freeze-out line of 
Fig.(\ref{T_muB_corr_MesBar_MCE})~({\it right}), one would observe a weak positive correlation at small~$\sqrt{s_{NN}}$, 
and a strong anti-correlation at large~$\sqrt{s_{NN}}$. In the SPS energy range a maximum might emerge. 
In particular, for the region of the phase diagram accessible to RHIC experiments, the MCE formulation implies that 
some events will be mostly composed of mesons and anti-mesons, while other events should be mostly composed of baryons and 
anti-baryons. The bulk of the events should be quite normal, but some events could be either predominantly mesonic 
(bosonic) or predominately baryonic (fermionic).

\section{Multiplicity Fluctuations}
\label{sec_phasediagram_multfluc}

\begin{figure}[ht!]
  \begin{center}
    \epsfig{file=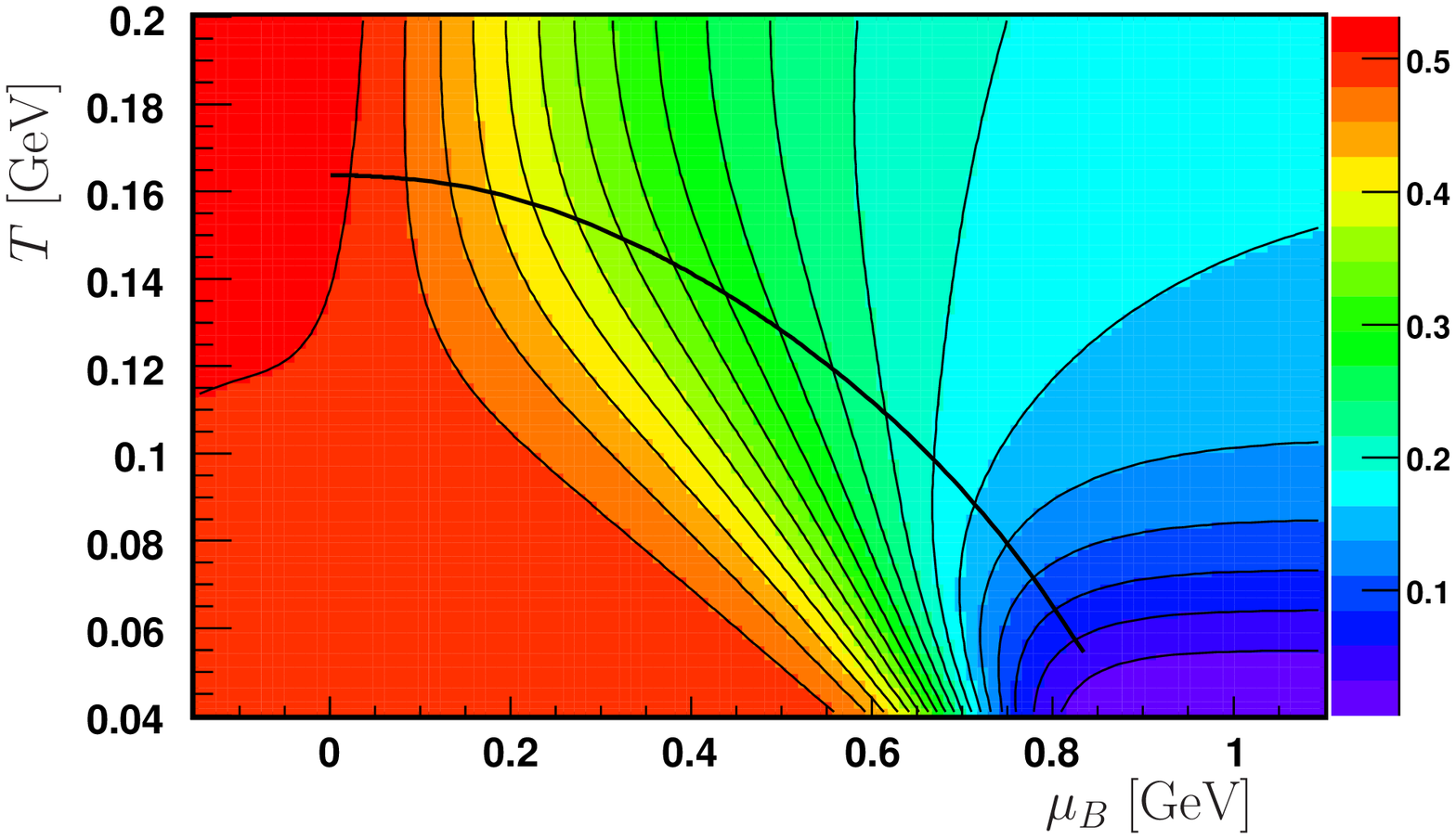,width=7.cm,height=6.5cm}
    \epsfig{file=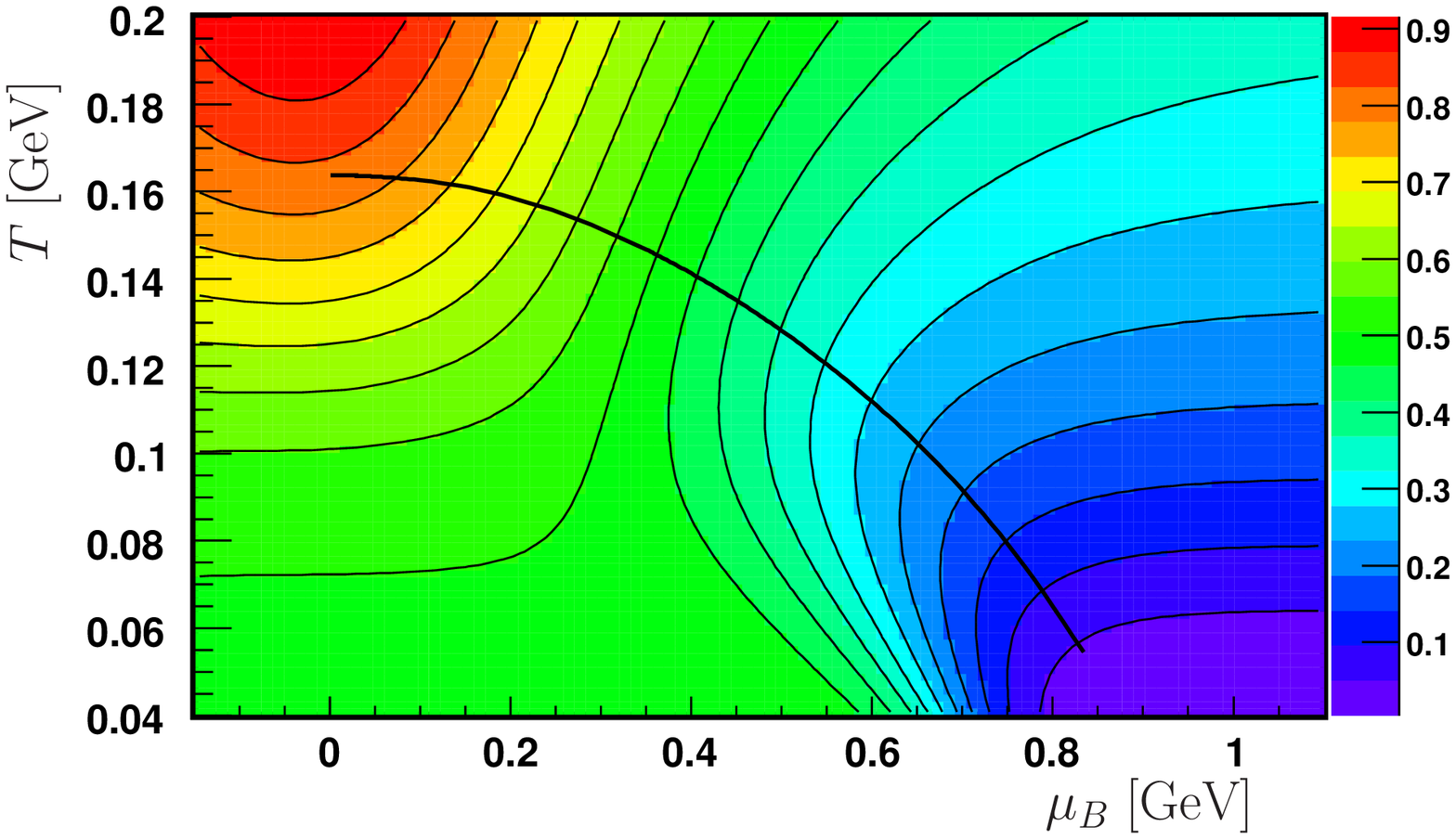,width=7.cm,height=6.5cm}
    \caption{
    CE fluctuations of the multiplicities of positively charged hadrons~$\omega_+$. 
    Both, primordial~({\it left}), and final state~({\it right}). } 
    \label{T_muB_fluc_pos_CE}
  \end{center}
\end{figure}

\begin{figure}[ht!]
  \begin{center}
    \epsfig{file=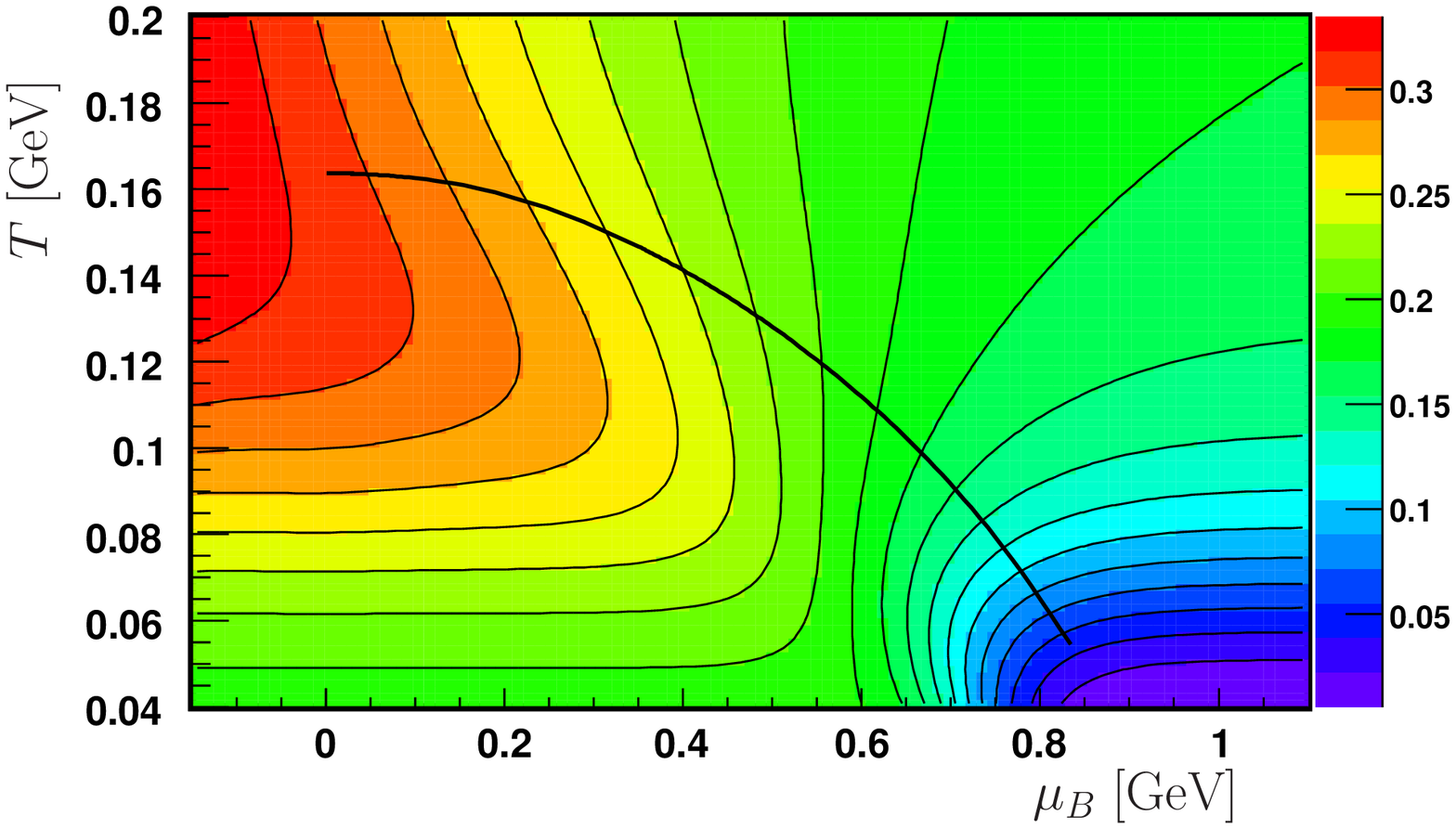,width=7.cm,height=6.5cm}
    \epsfig{file=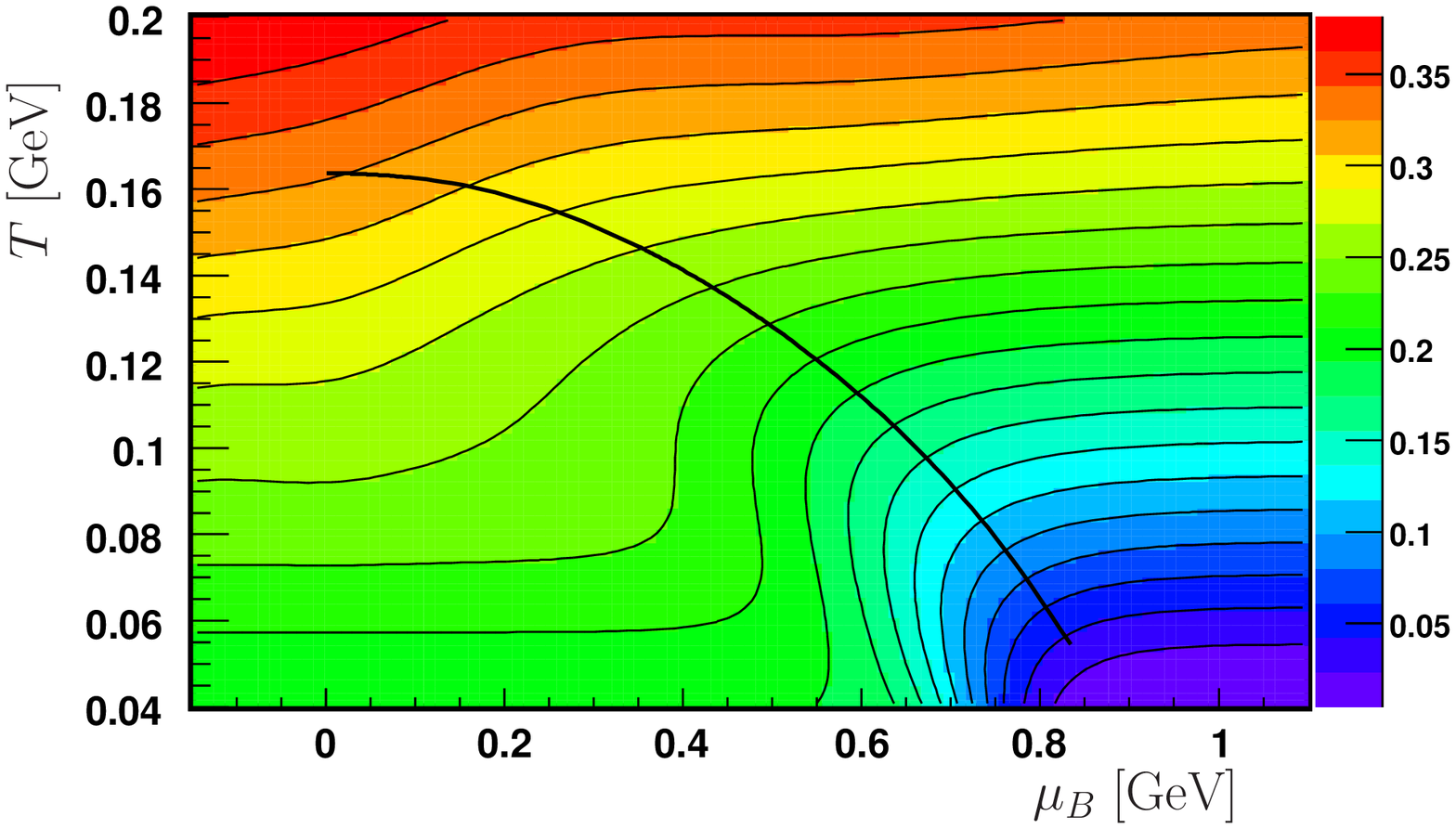,width=7.cm,height=6.5cm}
    \caption{
    MCE fluctuations of the multiplicities of positively charged hadrons~$\omega_+$. 
    Both, primordial~({\it left}), and final state~({\it right}).} 
    \label{T_muB_fluc_pos_MCE}
  \end{center}
\end{figure}

\begin{figure}[ht!]
  \begin{center}
    \epsfig{file=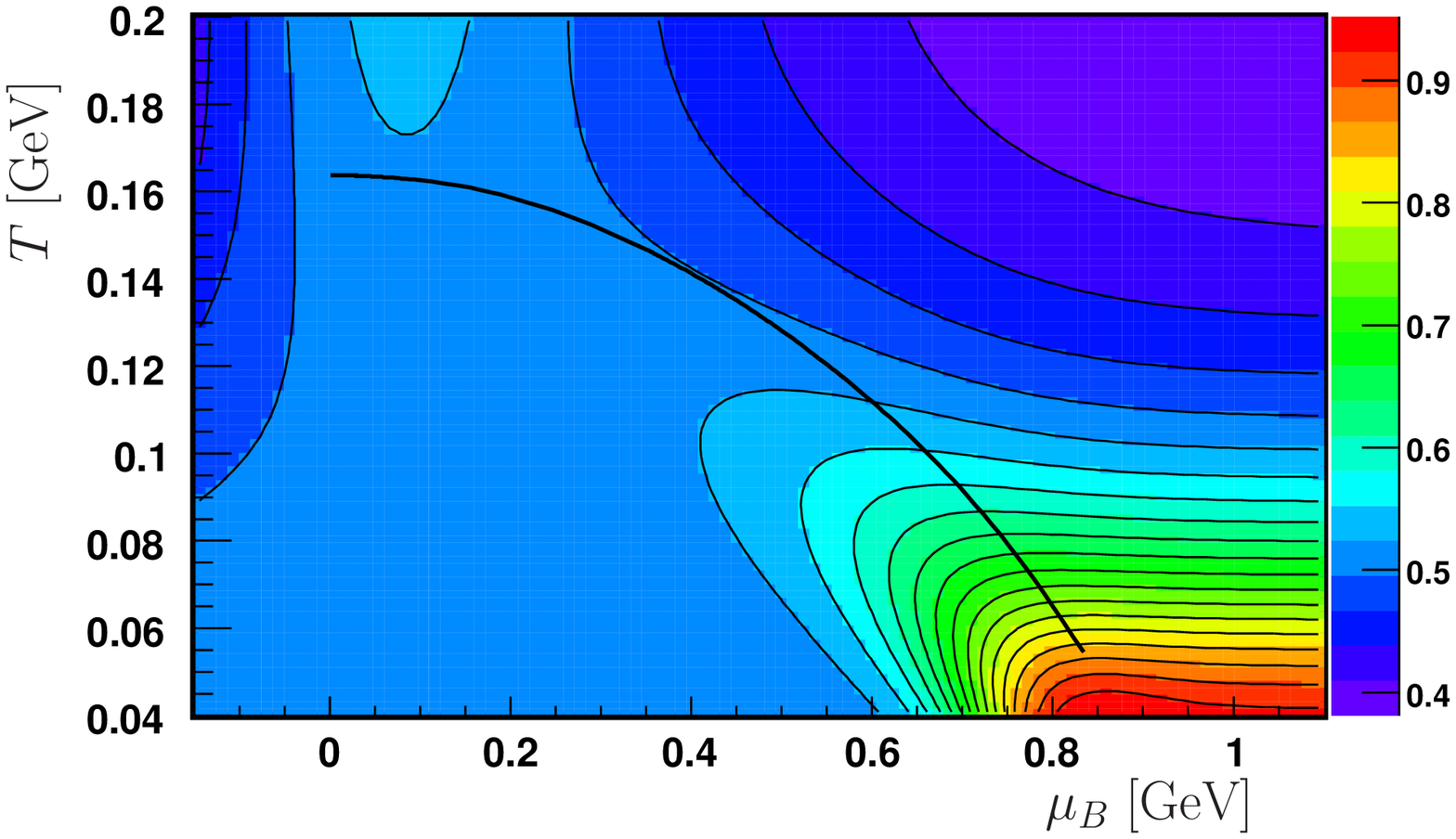,width=7.cm,height=6.5cm}
    \epsfig{file=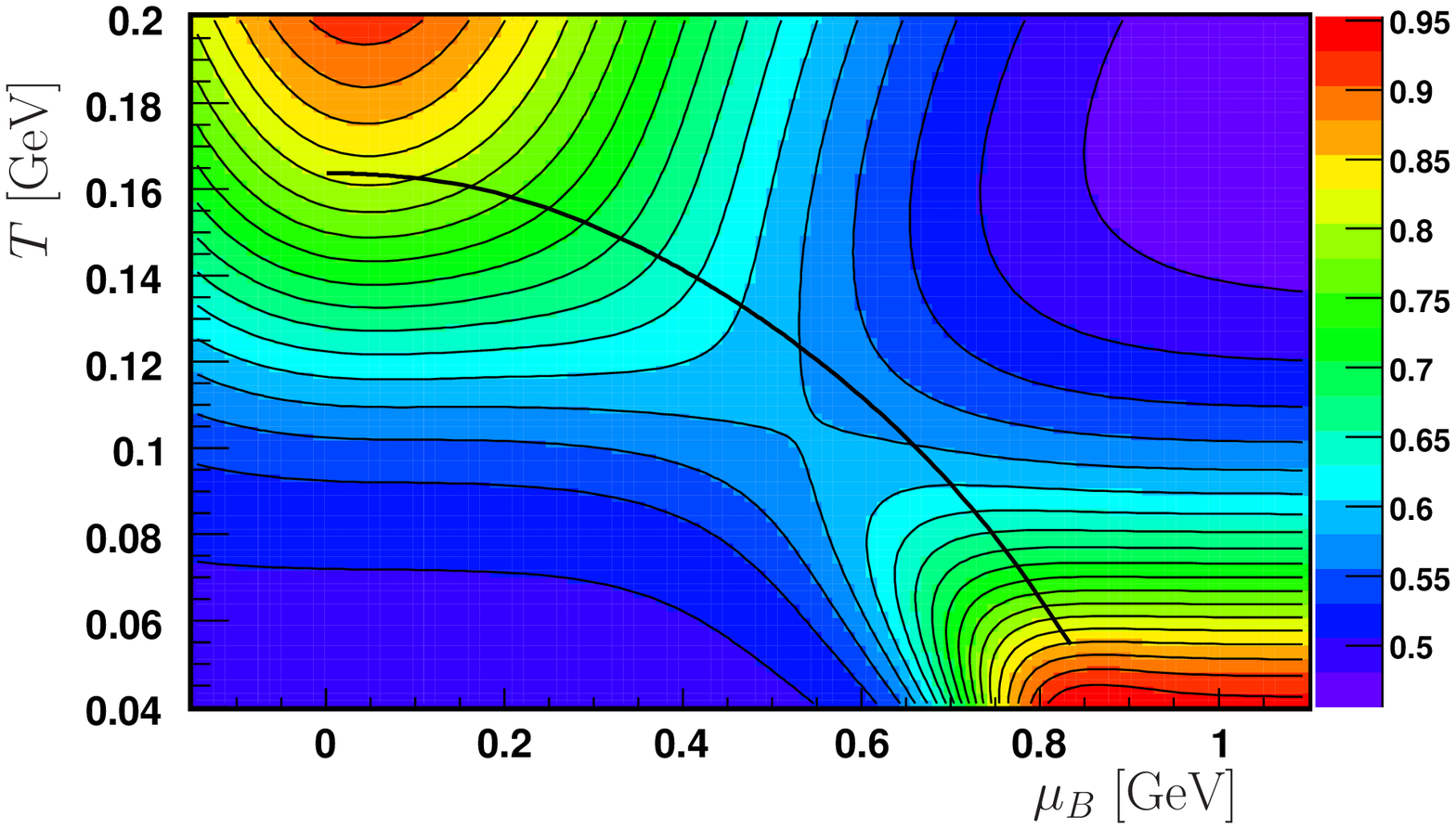,width=7.cm,height=6.5cm}
    \caption{   
    CE fluctuations of the multiplicities of negatively charged hadrons~$\omega^-$. 
    Both, primordial~({\it left}), and final state~({\it right}).} 
    \label{T_muB_fluc_neg_CE}
  \end{center}
\end{figure}

\begin{figure}[ht!]
  \begin{center}
    \epsfig{file=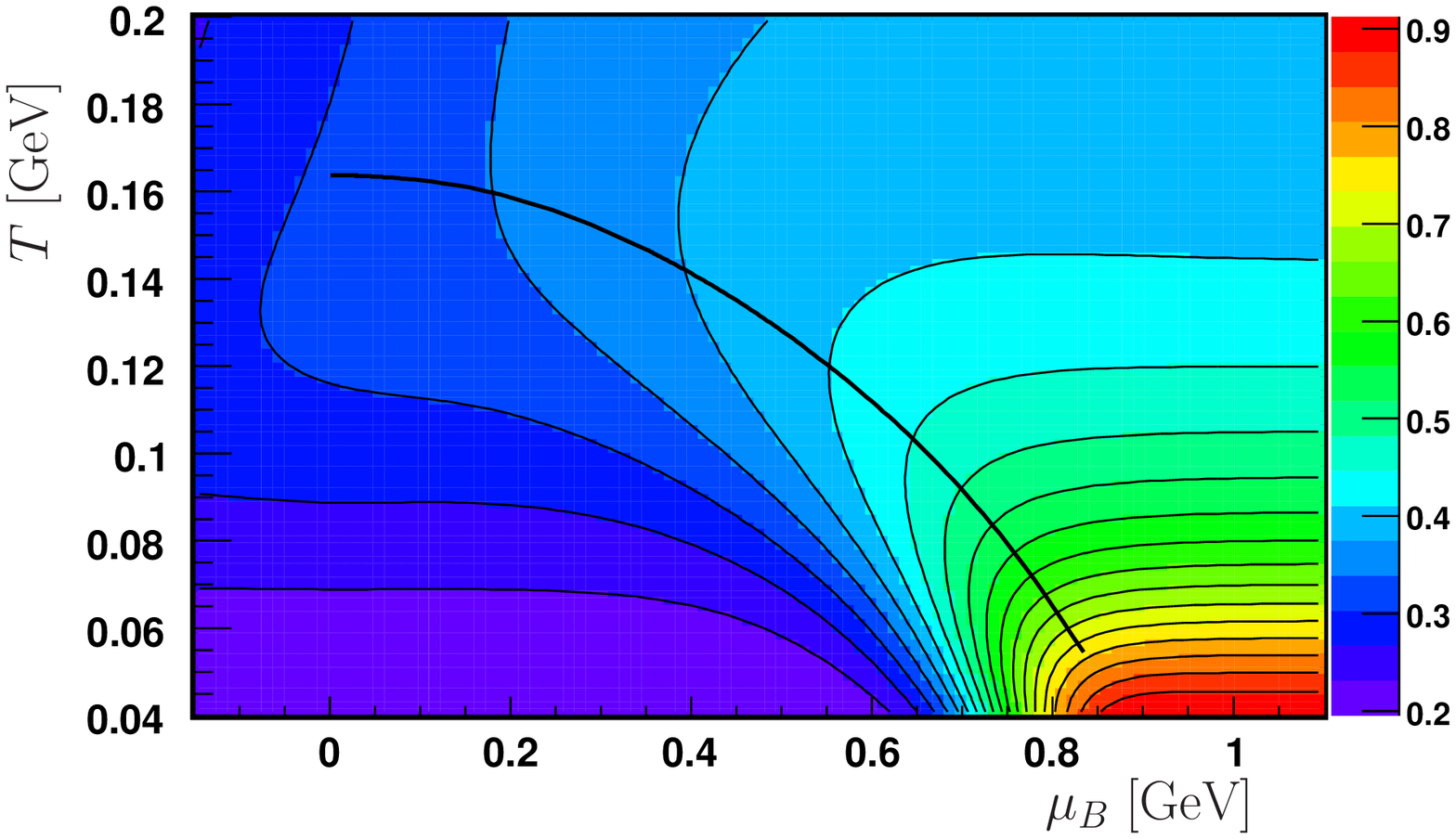,width=7.cm,height=6.5cm}
    \epsfig{file=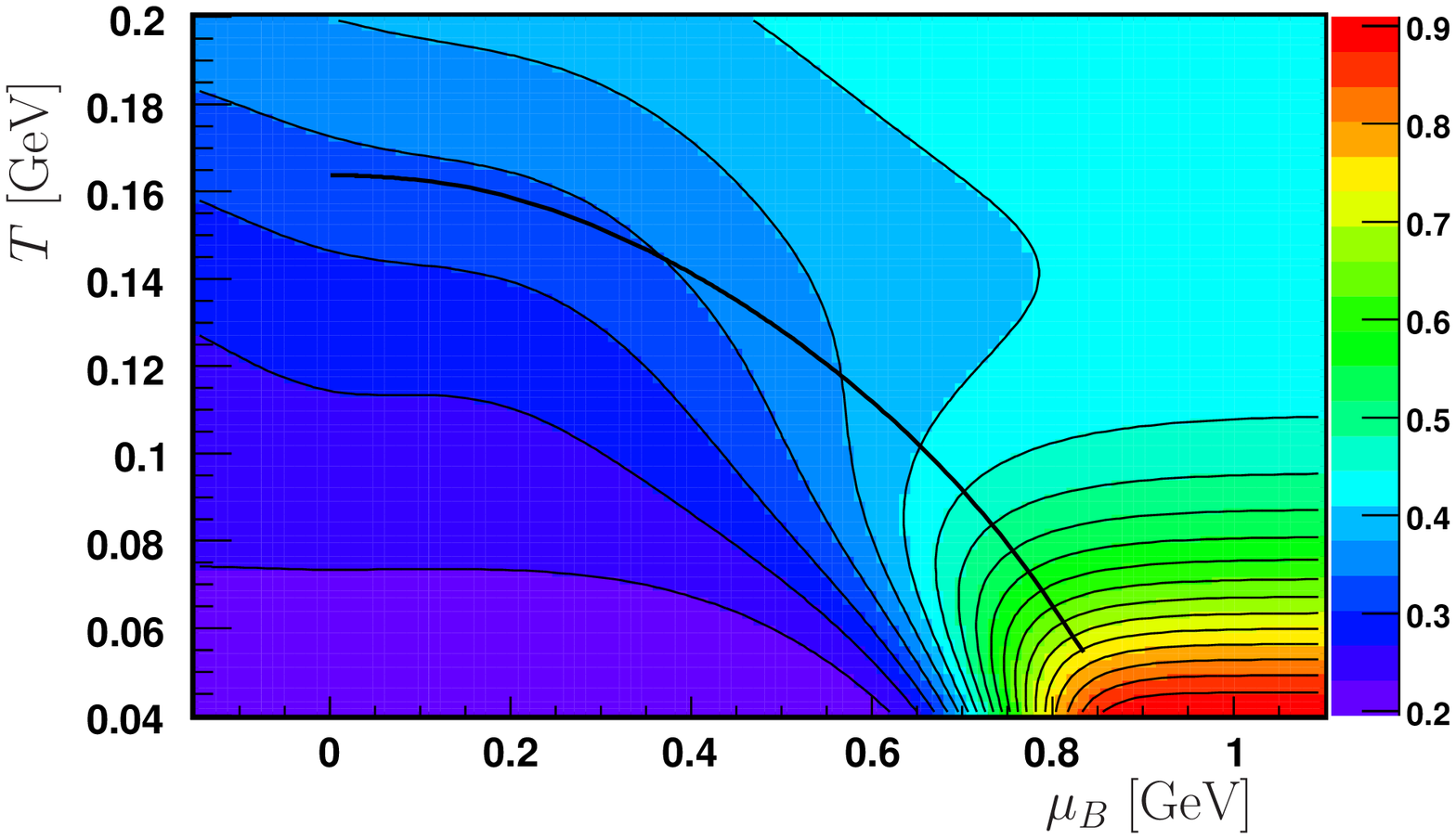,width=7.cm,height=6.5cm}
    \caption{
    MCE fluctuations of the multiplicities of negatively charged hadrons~$\omega^-$. 
    Both, primordial~({\it left}), and final state~({\it right}).} 
    \label{T_muB_fluc_neg_MCE}
  \end{center}
\end{figure}

In this section the phase diagrams for primordial and final state multiplicity fluctuations of positively and negatively 
charged hadrons~$\omega_+$ and~$\omega_-$ are compared in~CE and~MCE 
in {Figs.(\ref{T_muB_fluc_pos_CE}-\ref{T_muB_fluc_neg_MCE})}. 
GCE calculations will, in addition, be presented in Section~\ref{sec_freezeoutline_fol}. 

Some general comments attempt to summarize. At positive~$\mu_B$, also the electric charge density is positive.
The electric net-charge content of the system~${Q=N_+-N_-}$ is conserved. As now the average multiplicity of positively 
charged hadrons is larger than the one of negatively charged hadrons,~$\langle N_+ \rangle > \langle N_- \rangle$ 
(but as the variances ought to be 
equal,~$\langle \left( \Delta N_+ \right)^2 \rangle=\langle \left( \Delta N_- \right)^2 \rangle$) one finds the 
multiplicity distribution of positively charged hadrons to be more narrow than the one of negatively charged 
hadrons,~$\omega_+<\omega_-$. At negative baryon chemical potential~$\mu_B$, i.e. for negative electric charge density, 
the situation is reversed, and~$\omega^+>\omega^-$. Their fluctuations are equal for~$\mu_B \simeq 0$. 

With increasing temperature the system becomes more and more relativistic, and particle number increasingly fluctuates
in the~MCE. In the~CE this is already less apparent. A very mild temperature dependence of the scaled variance of 
multiplicity fluctuations arises only due to heavy hadrons carrying multiple charges. In a~CE with only one species of 
particles in Boltzmann approximation, the scaled variance in neutral matter 
is~$\omega^+ = \omega^- =0.5$~\cite{Begun:2004gs} independent of temperature. In the primordial Boltzmann~GCE fluctuations
are Poissonian independent of either thermal parameter. Note the different color scales in 
Figs.(\ref{T_muB_fluc_pos_CE}-\ref{T_muB_fluc_neg_MCE}).

Comparing primordial~CE Fig.(\ref{T_muB_fluc_pos_CE})~({\it left}) primordial~MCE 
Fig.(\ref{T_muB_fluc_pos_MCE})~({\it left}), one finds at low temperature~$\omega_+^{mce}<\omega_+^{ce}$.
In a cold gas particle multiplicity of positively charged hadrons cannot fluctuate much due to combined energy and 
charge conservation. At large~$T$ and~$\mu_B$, however very mildly,~$\omega_+^{mce}>\omega_+^{ce}$. Baryon multiplicity 
fluctuations induce meson fluctuations in hot baryonic matter, see Fig.(\ref{T_muB_corr_MesBar_MCE}). The net-baryon 
number is conserved,~$B=N_{B}-N_{\overline{B}}$. Yet, fluctuations in the sum of baryon number and anti-baryon 
number,~$N_{B}+N_{\overline{B}}$, release or bind a large fraction of the available energy of the system, with the meson 
multiplicity adjusting. In the CE, with only charge conservation, meson multiplicity fluctuations are less affected. 

A weak, yet generally useful, rule of thumb is:
Resonance decay in the~CE Fig.(\ref{T_muB_fluc_pos_CE})~({\it right}) increases~$\omega$, while in the~MCE 
Fig.(\ref{T_muB_fluc_pos_MCE})~({\it right}) decreases~$\omega$, when compared to their respective primordial 
scenarios~({\it left}). With the notable exception of very large~$T$. There it seems more economical to produce 
particles via decay of resonances, rather than thermally with large kinetic energy due to temperature. The correlation 
between primordial parents and stable daughters is then weaker. And~$\omega_+^{final}>\omega_+^{prim}$ in the~MCE
is possible. In final state one always finds~$\omega_+^{mce}<\omega_+^{ce}$. It is mentioned in this context, that even 
in the ultra-relativistic limit~$\omega^{mce}<\omega^{ce}$~\cite{Begun:2004pk}.

Having stated this rule of thumb, its caveat shall be discussed. Its validity depends strongly on possible cluster 
formation and their decay modes\footnote{In this context the terms cluster, string, Hagedorn state, or bag are used 
equivalently.}. Yet energy can only be spent once. Either on primordial particle production, or the formation of
a cluster, bag, string, or Hagedorn state. If particle production via cluster is more economical, i.e. more energy 
spent ``on mass'' and less ``on motion'', then final state fluctuations can be larger than primordial fluctuations in 
the~MCE. 

Now, turning to the fluctuations of negatively charged hadrons in the positive~$\mu_B$ half of the~$T$-$\mu_B$ phase 
diagram, the diagrams depicted in Figs.(\ref{T_muB_fluc_pos_CE},\ref{T_muB_fluc_pos_MCE}) could be connected to the 
reflected Figs.(\ref{T_muB_fluc_neg_CE},\ref{T_muB_fluc_neg_MCE}). The scaled variances of positively and negatively 
charged hadrons are equal at~$\mu_B=0$. For a change in sign of~$\mu_B$, the roles of~$\omega_+$ and~$\omega_-$ are 
interchanged, i.e. the phase diagrams are not symmetric around~$\mu_B=0$.

The scaled variance~$\omega_-$ is maximal where the ratio~$\langle N_- \rangle / \langle N_+ \rangle$ is minimal, 
i.e. when the electric charge density is strongly positive. For the primordial case, comparing~CE, 
Fig.(\ref{T_muB_fluc_neg_CE})~({\it left}) to~MCE, Fig.(\ref{T_muB_fluc_neg_MCE})~({\it left}), one finds 
generally~$\omega_-^{mce}<\omega_-^{ce}$. The final state enhancement at low~$\mu_B$ and large~$T$ in the~CE, 
Fig.(\ref{T_muB_fluc_neg_CE})~({\it right}) is due to resonance decay. For the final state~MCE, 
Fig.(\ref{T_muB_fluc_neg_MCE})~({\it right}) this enhancement is weaker, due to correlation of primordial parent 
resonances with stable particles via energy conservation. Similar to~$\omega_+$, at very large~$T$ and~$\mu_B$ one 
finds~$\omega_-^{mce}>\omega_-^{ce}$ in final state. Again, baryon multiplicity fluctuations induce meson multiplicity 
fluctuations.

\section{Discussion}
\label{sec_phasediagram_discussion}

The~$T$-$\mu_B$ phase diagram of the hadron resonance gas model has been explored in this chapter for fluctuation and 
correlation observables in their dependence on the chemical freeze-out parameters temperature and baryon chemical 
potential, omitting limited acceptance effects, but still considering resonance decay.

Grand canonical baryon number, strangeness and electric charge fluctuations and correlations are strongly sensitive to 
the degrees of freedom, i.e. different particle species, available to the system. The contributions of different 
hadron species is rather different in different corners of the~$T$-$\mu_B$ phase diagram. Mesons dominate at low 
temperatures. With increasing temperature and net-baryon density however baryonic degrees of freedom take a larger 
fraction of the system entropy. Two charges are strongly (anti-)correlated if particles which carry both quantum numbers 
are abundant. The same argument was essentially used to explain the momentum space dependence of charge fluctuations 
and correlations in Chapter~\ref{chapter_gce}.

Multiplicity correlations in the canonical and micro canonical ensembles follow suit. So is for instance the correlation 
between protons and charged pions strongly mediated by the neutron in cold baryonic matter. These correlations appear 
solely due to implementation of global conservation laws for charges and energy, and not due to a local interaction 
amongst constituent particles. With increasing temperature more channels, i.e. resonance formation and decay, become 
available to the statistical system, and subtle differences between canonical and micro canonical ensembles emerge.

Multiplicity fluctuations highlight then the connection between energy spent on resonance formation and energy spent 
on primordial particle production. The event-by-event multiplicities of stable particles and unstable parent resonances 
are anti-correlated in the micro canonical ensemble, while being almost unaffected in the canonical ensemble, and 
independent of each other in the grand canonical ensemble.

Above the freeze-out line, no hadronic phase should exist. Even in the presence of critical phenomena close to the 
phase transition, baseline contributions of the kind discussed here, and in previous chapters, should remain.

\chapter{The Chemical Freeze-out Line}
\label{chapter_freezeoutline}
In this chapter the results of the hadron resonance gas for the scaled variances and correlation coefficients 
are presented for freeze-out parameters following the constant average energy per average particle 
multiplicity~$\langle E \rangle / \langle N \rangle \simeq 1$~GeV chemical freeze-out line~\cite{Cleymans:1998fq}. 
Different ensemble predictions will be confronted with recent NA49 measurements of charged hadron multiplicity 
fluctuations. 

Mean hadron multiplicities in central heavy ion collisions at high energies are usually fitted within the GCE 
formulation of the hadron resonance gas model. The fit parameters then are: volume~$V$, temperature~$T$, baryon 
chemical potential~$\mu_B$, and saturation parameters~$\gamma_s$~\cite{Koch:1986ud} 
and~$\gamma_q$~\cite{Letessier:1999rm,Rafelski:2000by}. The former allows for non-equilibrium strange hadron yields,
while the latter describes non-equilibrium light quark numbers. For reviews of the model see 
Refs.~\cite{Cleymans:1997sw,Cleymans:1998yb,Averbeck:2000sn,BraunMunzinger:1994xr,BraunMunzinger:1995bp,BraunMunzinger:1999qy,Becattini:2003wp,Adams:2005dq}.

Recent discussion of system size and energy dependence of freeze-out parameters and comparison of different (chemical) 
freeze-out criteria can be found in 
Refs.~\cite{Cleymans:2005xv,Cleymans:1999st,Cleymans:1998fq,Becattini:2005xt,Andronic:2005yp}. 
The set of model parameters,~$V,T,\mu_B$, and~$\gamma_s$, chosen in the following should correspond to chemical 
freeze-out conditions in heavy ion collisions\footnote{The Bose-Einstein condensation model is not considered, as the 
analytic expansion used for the calculation here is inappropriate. However, one notes that energy conservation 
essentially prohibits diverging susceptibilities, if degrees of freedom have finite mass.}.
The evolution of model parameters with collision energy is taken from previous analysis of mean hadron multiplicity data. 
Hadron resonance gas model fits should then be able to provide a link between experimental control parameters, center 
of mass energy, and size of colliding ions, and the region probed in the phase diagram of strongly interacting matter.

\begin{wrapfigure}{r}{0.45\textwidth}
  \begin{center}
  \vspace{-0.4cm}
  \includegraphics[width=0.49\textwidth,height=6.1cm]{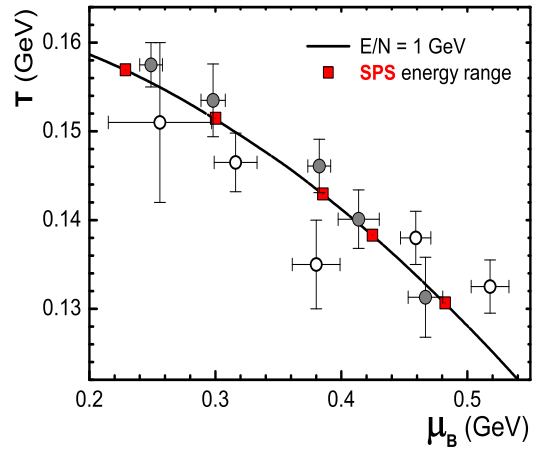}
  \end{center}
  \label{T_muB}
  \vspace{-0.6cm}
  \caption{Chemical freeze-out points in the~$T$-$\mu_B$ plane for central Pb+Pb collisions. The solid line indicates 
  the~$\langle E \rangle /\langle N \rangle = 1$~GeV freeze-out line. Square markers denote SPS beam energies 
  from~$20A$~GeV (right) to~$158A$~GeV (left). Full and open circles are the best fit parameters from 
  Refs.~\cite{Becattini:2005xt} and~\cite{Andronic:2005yp}, respectively. The Figure is taken from~\cite{Begun:2006uu}.}
\end{wrapfigure}

The dependence of the baryon chemical potential~$\mu_B$ on the collision energy is parametrized 
as~\cite{Cleymans:2005xv}:~$\mu_B \left( \sqrt{s_{NN}} \right) =1.308~\mbox{GeV}\cdot(1+0.273~ \sqrt{s_{NN}})^{-1}$,  
where the center of mass energy per nucleon pair,~$\sqrt{s_{NN}}$, is taken in units of GeV. 
The system is assumed to be net-strangeness free,~$S=0$, and to have the electric charge to baryon density ratio 
of the initial colliding nuclei,~$Q/B = 0.4$. 
These two conditions define the system's strangeness,~$\mu_S$, and electric charge,~$\mu _Q$, chemical potentials. 
The average energy per average particle yield~\cite{Cleymans:1998fq}~$\langle E \rangle/\langle N \rangle = 1$~GeV 
was chosen to fix the temperature~$T$. Finally, the strangeness saturation factor is 
parametrized~\cite{Becattini:2005xt},~$\gamma_s= 1 - 0.396~ \exp \left( - ~1.23~ T/\mu_B \right)$. 
The asymptotic (or large volume) solutions for the scaled variance and correlation coefficients are assumed. 
This determines all parameters of the model.

Here, some further details should be mentioned. Quantum statistics effects are considered, while the non-zero 
(Breit-Wigner) width of resonances is disregarded. The standard THERMUS particle table includes all known particles and 
resonances up to a mass of about~$2.5$~GeV and their respective decay channels. Heavy resonances do not always have 
well established decay channels. Branching ratios given in THERMUS are re-scaled to unity, where it was necessary to 
ensure global charge conservation. To make correspondence with NA49 data, both strong and electromagnetic decays should 
be taken into account, while weak decays should be omitted.

In Section~\ref{sec_freezeoutline_fol} multiplicity fluctuations and correlations are considered for different ensembles 
along the chemical freeze-out line. In Section~\ref{sec_freezeoutline_na49} a comparison to recent NA49 data on charged 
hadron multiplicity fluctuations is attempted. 

\section{Fluctuations and Correlations}
\label{sec_freezeoutline_fol}

\begin{figure}[ht!]
  \begin{center}
    \epsfig{file=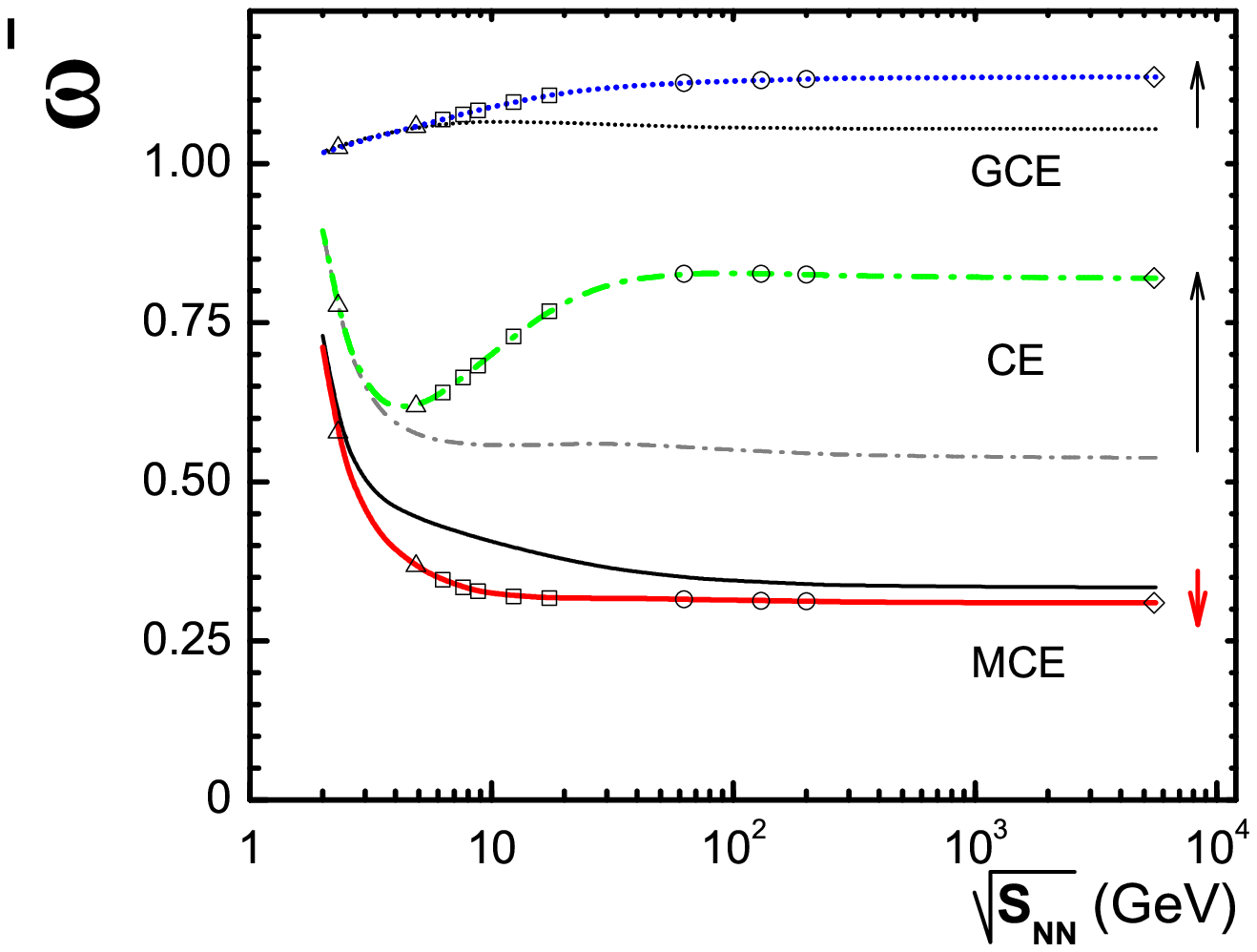,width=7.0cm,height=6.5cm}
    \epsfig{file=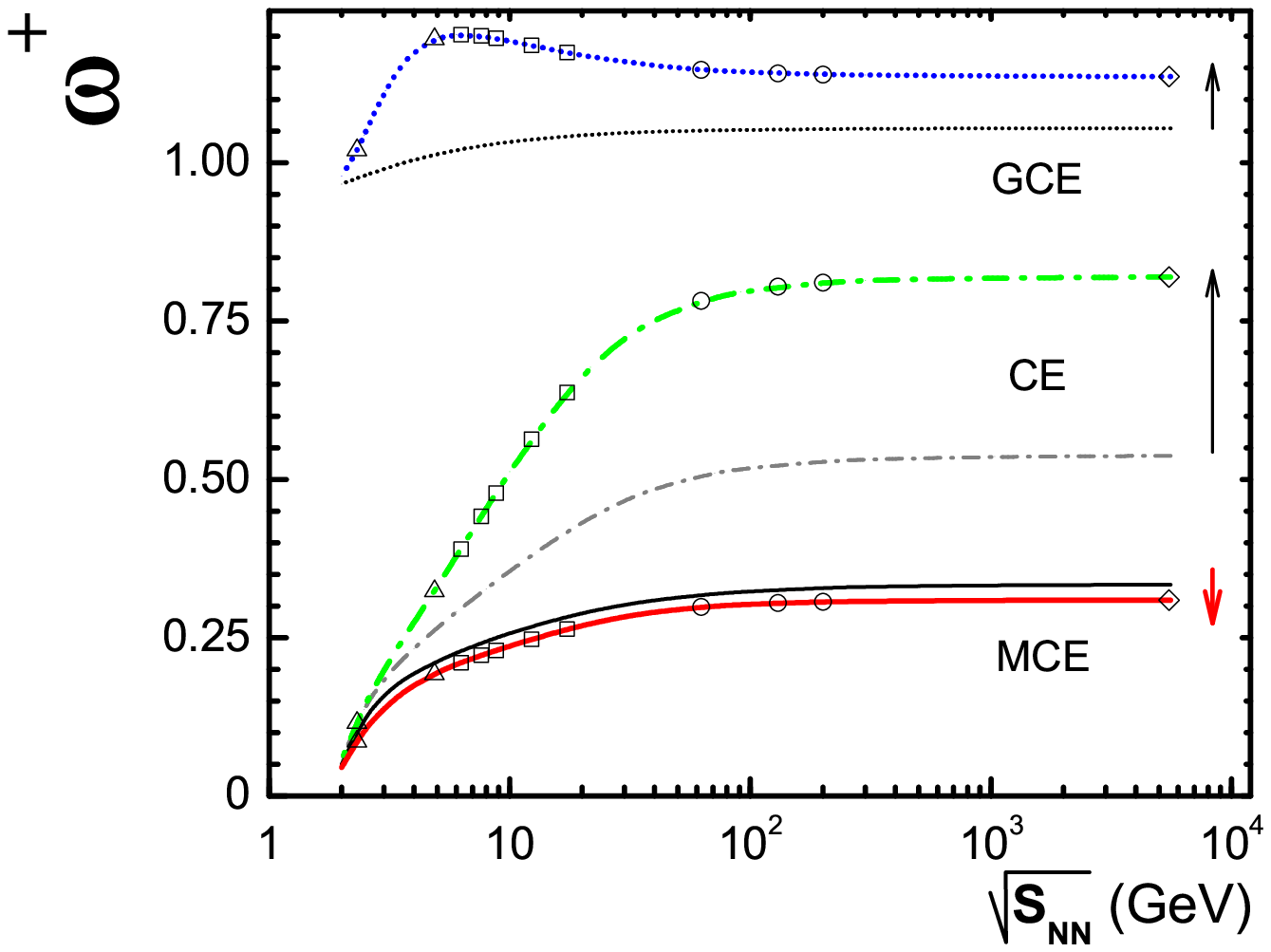,width=7.0cm,height=6.5cm}
    \caption{The scaled variances of negatively charged particles,~$\omega^-$~({\it left}), and the scaled variances 
    of positively charged particles,~$\omega^+$~({\it right}), both primordial and final state, along the
    chemical freeze-out line for central Pb+Pb (Au+Au) collisions. Different lines present GCE, CE, and MCE results. 
    Symbols on the lines for final state fluctuations correspond to specific collision energies, ranging from 
    SIS and AGS (triangles), SPS (squares), RHIC (circles), to LHC (diamonds). Arrows indicate the effect of resonance 
    decay. Figures are taken from~\cite{Begun:2006uu}.}
    \label{sqrtS_omega_p_m}
  \end{center}
\end{figure}

\begin{figure}[ht!]
  \begin{center} 
    \epsfig{file=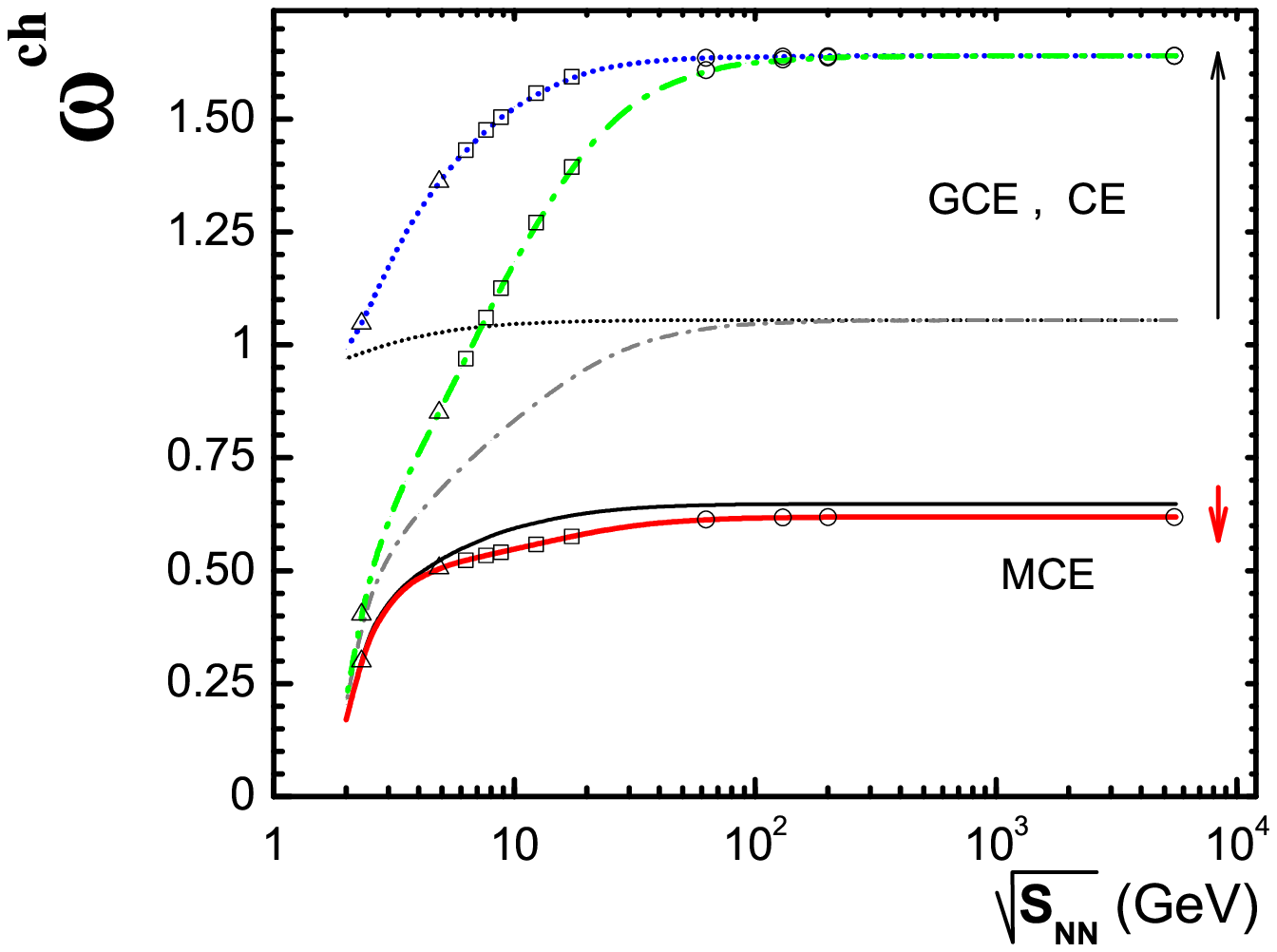,width=7.0cm,height=6.5cm}
    \epsfig{file=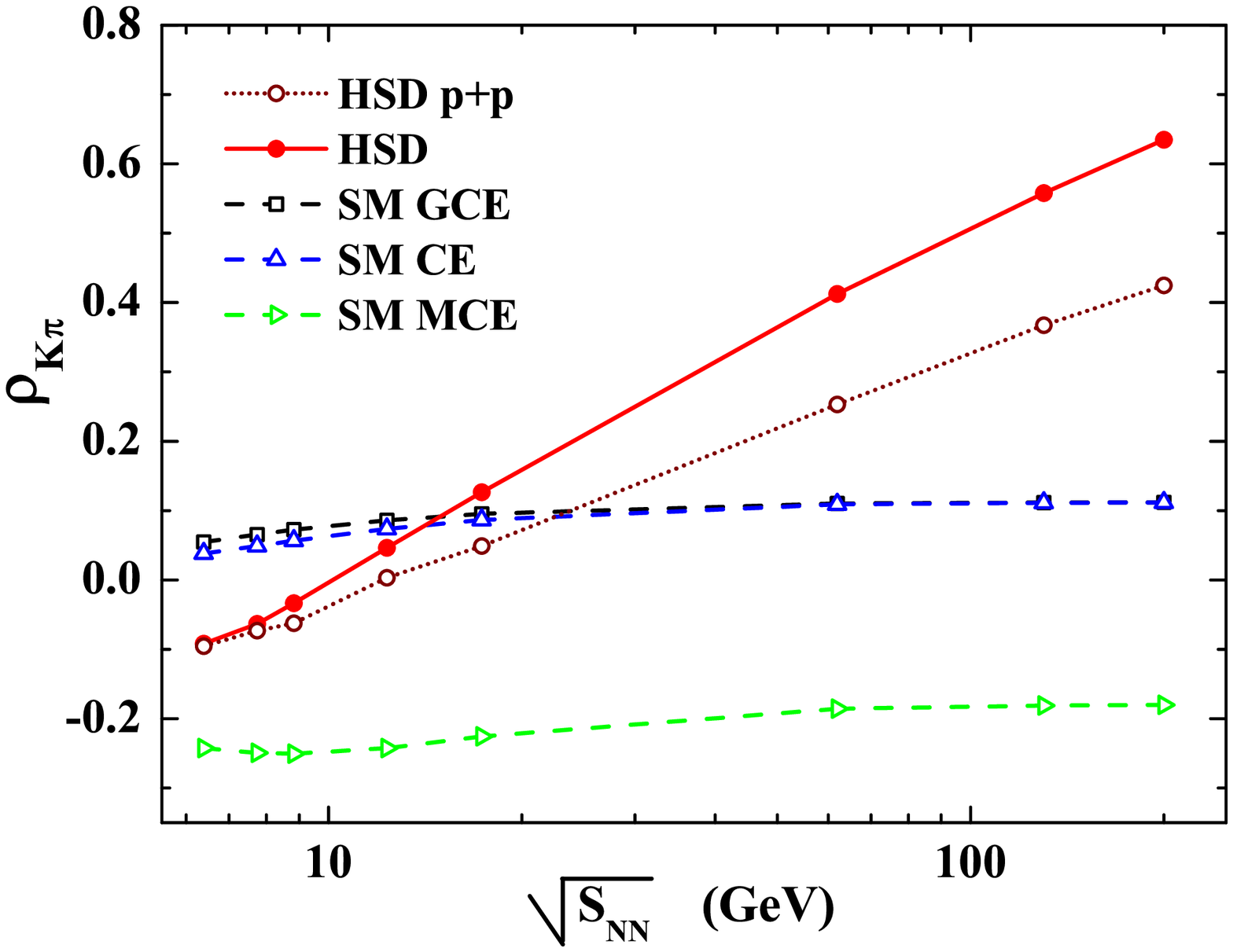,width=7.0cm,height=6.5cm}
    \caption{The scaled variances of all charged particles,~$\omega^{ch}$, both primordial and final state~({\it left}), 
    and the final state correlation coefficient~$\rho_{K \pi}$~({\it right}) along the chemical freeze-out line. 
    HSD transport model results for~$\rho_{K \pi}$ in~p+p and~Au+Au collisions are shown for comparison. 
    Figures are taken from~\cite{Begun:2006uu}~({\it left}) and~\cite{Gorenstein:2008et}~({\it right}).}
    \label{sqrtS_omega_ch_rho_Kpi}
  \end{center}
\end{figure}

In Fig.(\ref{sqrtS_omega_p_m}) the scaled variances of negatively charged hadrons,~$\omega^-$~({\it left}), and the 
scaled variances of positively charged hadrons,~$\omega^+$~({\it right}), are shown, both primordial and final state, 
along the chemical freeze-out line for most central Pb+Pb (Au+Au) collisions. 
In Fig.(\ref{sqrtS_omega_ch_rho_Kpi})~({\it left}) the scaled variances of all charged particles,~$\omega^{ch}$, 
is shown\footnote{Figures are taken form~\cite{Begun:2006uu}. In previous chapters the scaled variances of positively, 
negatively, and all charged hadrons were denoted as~$\omega_+$,~$\omega_-$, and~$\omega_{\pm}$. In this chapter these 
quantities are refered to as~$\omega^+$,~$\omega^-$, and~$\omega^{ch}$.}. 
The three standard ensembles are considered in turn, and some qualitative features are discussed. 

\begin{figure}[ht!]
  \begin{center}
    \epsfig{file=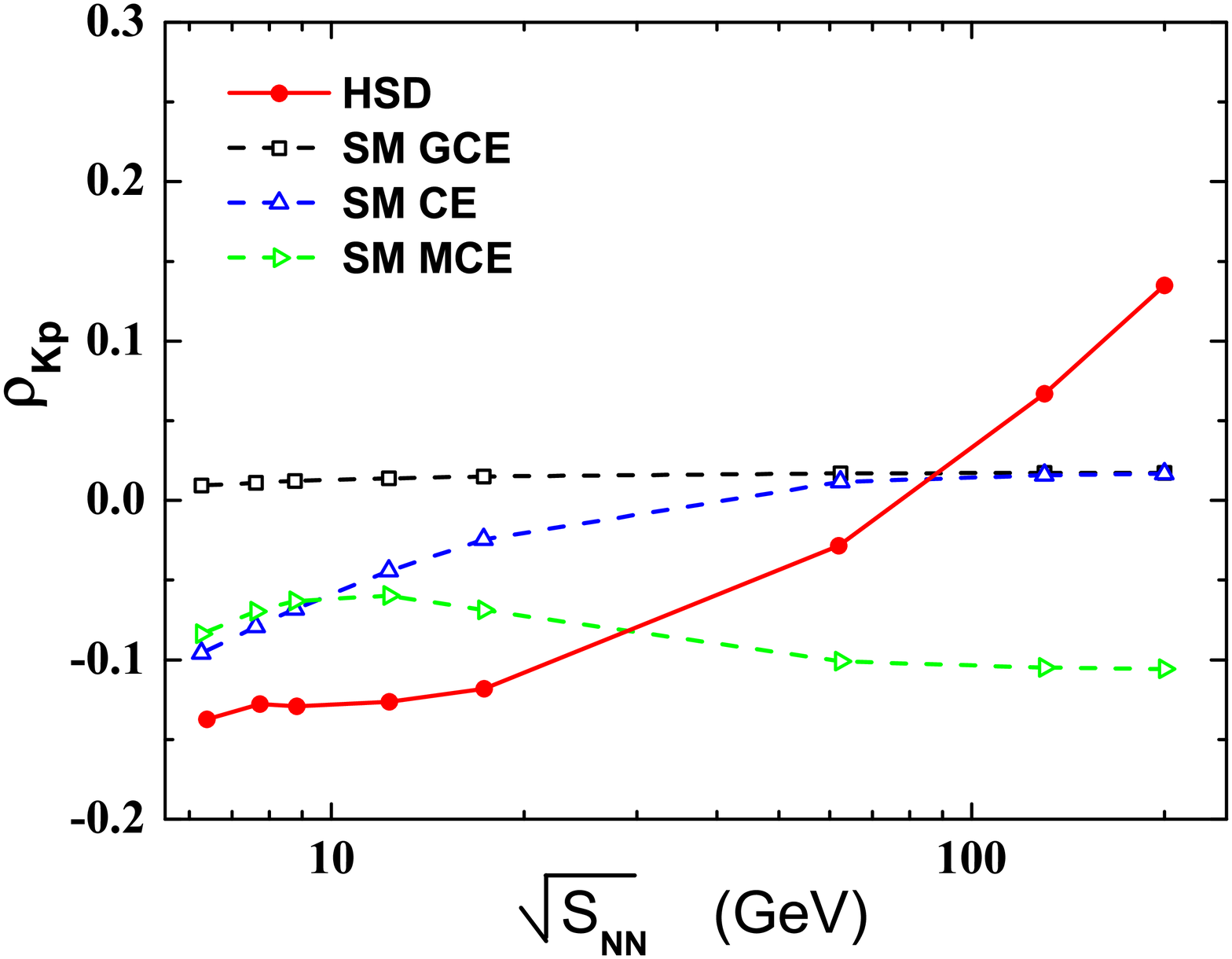,width=7.0cm,height=6.5cm}
    \epsfig{file=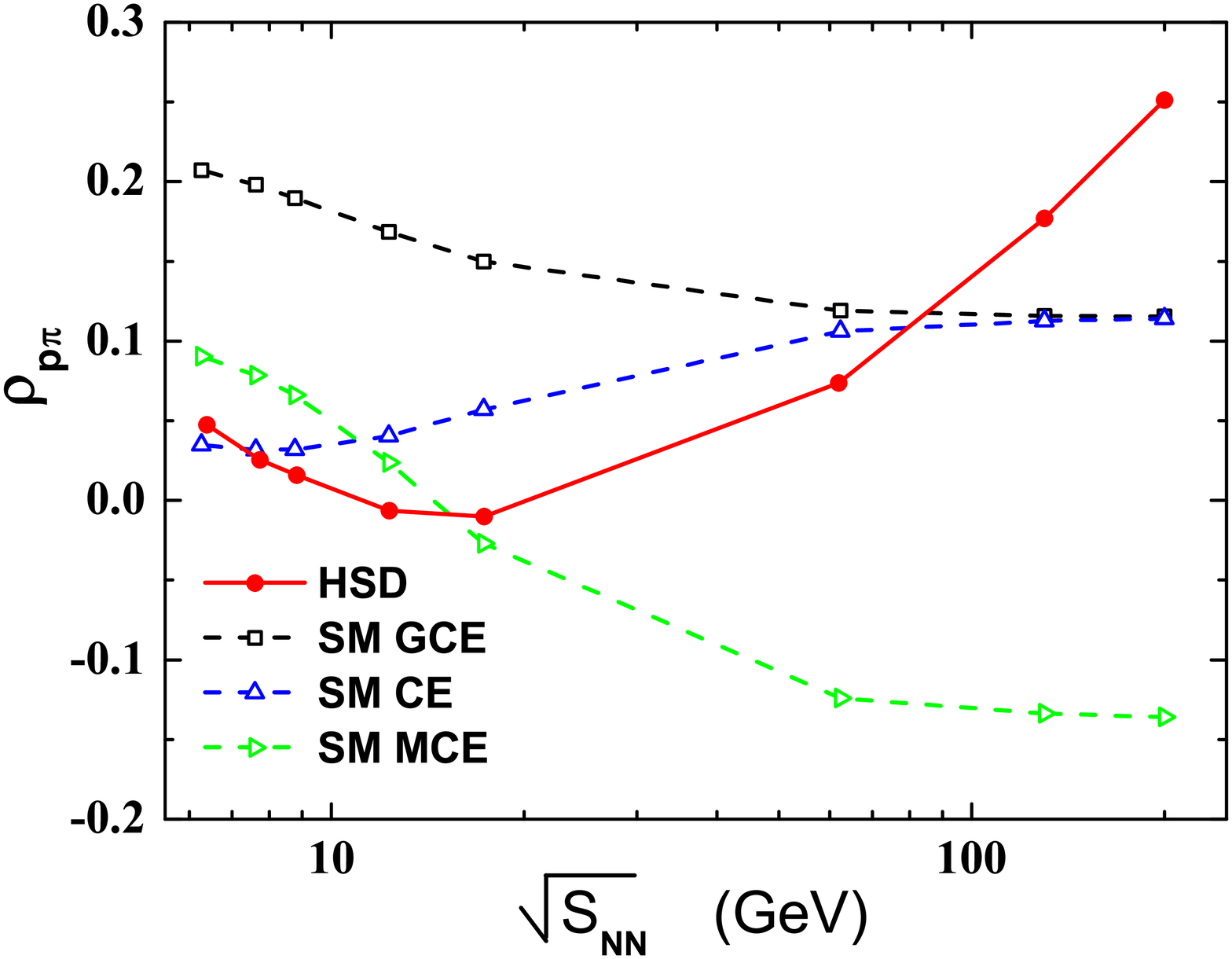,width=7.0cm,height=6.5cm}
    \caption{The final state correlation coefficient~$\rho_{K\textrm{p}}$~({\it left}), and the final state correlation 
    coefficient~$\rho_{\textrm{p} \pi}$~({\it right}) along the chemical freeze-out line. 
    HSD transport model results for~Au+Au collisions are shown for comparison. 
    Figures are taken from~\cite{Konchakovski:2009at}. } 
    \label{sqrtS_rho_Kp_ppi}
  \end{center}
\end{figure}

The effects of Bose-Einstein and Fermi-Dirac statistics are seen best in the primordial~GCE. At low temperatures most 
positively charged hadrons are protons, and Fermi-Dirac statistics dominate, leading to sub-Poissonian multiplicity 
fluctuations,~$\omega^{+}_{gce}, ~\omega^{ch}_{gce}<1$. On the other hand, in the limit of high temperature most charged 
hadrons are pions and the effects of Bose-Einstein statistics dominate. The multiplicity distribution is then wider than 
a Poissonian reference,~$\omega_{gce}^{\pm},~\omega_{gce}^{ch}>1$. Along the chemical freeze-out line,~$\omega_{gce}^-$ is 
always slightly larger than one, as~$\pi^-$ mesons are the most abundant negatively charged hadrons at, both, low and 
high temperatures, Fig.(\ref{sqrtS_omega_p_m})~({\it left}). 

The bump in~$\omega^+_{gce}$ for final state particles seen in Fig.(\ref{sqrtS_omega_p_m})~({\it right}) at small 
collision energies is due to correlated production of protons and~$\pi^+$ mesons from~$\Delta^{++}$ decay. This single 
resonance contribution dominates for~$\omega^+_{gce}$ at small collision energies (small temperatures), but becomes 
relatively unimportant at high collision energies. Along the freeze-out line the final state GCE correlation 
coefficient~$\rho_{\textrm{p} \pi^+}$, Fig.(\ref{T_muB_rho_p_pi_GCE}), has a maximum in the same region.
For the final state~$\omega^-_{gce}$ no bump is seen due to a weaker resonance decay contribution at low~$\sqrt{s_{NN}}$. 
For comparison, the GCE resonance decay enhancement of multiplicity fluctuations is strongest for all charged hadrons, 
Fig.(\ref{sqrtS_omega_ch_rho_Kpi})~({\it left}).
 
A minimum in~$\omega_{ce}^{-}$ for final particles is seen in Fig.~(\ref{sqrtS_omega_p_m})~({\it left}). 
This is due to two effects. As the number of negatively charged particles is relatively 
small,~$\langle N_-\rangle \ll \langle N_+\rangle$, at low collision energies, both CE suppression and the effect of 
resonance decay are small. With increasing~$\sqrt{s_{NN}}$, the CE effect alone leads to a decrease of~$\omega^-_{ce}$, 
but resonance decay leads to an increase of~$\omega^-_{ce}$. The combination of these two effects, CE suppression and 
resonance decay enhancement, leads to a minimum of~$\omega^-_{ce}$. The primordial~$\omega^-_{ce}$ decreases monotonously 
with~$\sqrt{s_{NN}}$, compare also Fig.(\ref{T_muB_fluc_neg_CE}). At large~$\sqrt{s_{NN}}$ the system is essentially 
charge neutral, CE suppression of~$\omega^+_{ce}$ and enhancement of~$\omega^-_{ce}$ therefore cease, 
and one finds~$\omega^-_{ce}=\omega^+_{ce}$ accordingly. Both, primordial and  final state,~$\omega_{ce}^{+}$ in 
Fig.~(\ref{sqrtS_omega_p_m})~({\it right}) hence increase with~$\sqrt{s_{NN}}$, compare Fig.(\ref{T_muB_fluc_pos_CE}).
For a charge neutral system the CE effects cancel exactly for the multiplicity fluctuations of all charged hadrons, 
both in final state and primordial,~$\omega_{ce}^{ch}= \omega_{gce}^{ch}$, Fig.~(\ref{sqrtS_omega_ch_rho_Kpi})~({\it left}). 
Compare also Fig.(\ref{BSQ_conv_lambda_omegaChr_final}).

As to be expected,~$\omega_{mce}<\omega_{ce}$, along the chemical freeze-out line. Energy conservation further suppresses 
particle multiplicity fluctuations. A particular feature of the MCE  is the additional suppression of multiplicity 
fluctuations after resonance decay, compare Figs.(\ref{T_muB_fluc_pos_MCE},\ref{T_muB_fluc_neg_MCE}). Energy conservation 
does not allow for strong multiplicity fluctuations at low temperature. With increasing~$\sqrt{s_{NN}}$ one 
finds~$\omega_{mce}^{ch}$ and~$\omega_{mce}^{+}$ to grow. On the other hand,~$\omega_{mce}^{-}$ drops. Negatively charged 
particles are outnumbered by positively charged particles at small collision energies. Therefore, as the variances of the 
distributions of positively and negatively charged hadrons are the same, one finds~$\omega_{mce}^{-}$ to grow towards 
smaller~$\sqrt{s_{NN}}$, essentially due to a dropping mean value~$\langle N^- \rangle$. One also notes, that the 
resonance decay enhancement of~$\omega_{ce}^{-}$ around~$\sqrt{s_{NN}}=10$~GeV is not seen for~$\omega_{mce}^{-}$, due to 
a strong correlation of primordial parent resonances with primordial hadrons resulting from global energy conservation. 
This effect is even stronger for~$\omega_{mce}^{ch}$.\\
 

The correlation coefficient~$\rho_{K \pi}$ between the event-by-event multiplicities of charged kaons,~$N_{K^+}+N_{K^-}$, 
and charged pions,~$N_{\pi^+}+N_{\pi^-}$, is shown in Fig.(\ref{sqrtS_omega_ch_rho_Kpi})~({\it right}) as a function of 
center of mass energy~$\sqrt{s_{NN}}$ for final state hadrons only. Fig.(\ref{sqrtS_rho_Kp_ppi}) additionally shows the 
correlation coefficient~$\rho_{K \textrm{p}}$~({\it left}) between the event-by-event multiplicities of charged kaons and 
protons plus anti-protons,~$N_{\textrm{p}}+N_{\overline{\textrm{p}}}$~({\it left}), and the correlation 
coefficient~$\rho_{\textrm{p} \pi}$~({\it right}) between the event-by-event multiplicities of protons plus anti-protons
and both charged pions\footnote{Figures are taken form~\cite{Gorenstein:2008et,Konchakovski:2009at}. In previous chapters 
the correlation coefficients denoted only the correlation between proton and either charged pion. In this chapter always 
both, proton and anti-proton, or both charged kaons, or both charged pions are correlated.}. The three correlation 
functions are discussed in turn.

In the final state GCE the correlation coefficient~$\rho_{K \pi}$ remains weak. A very mild increase at 
large~$\sqrt{s_{NN}}$ is due to rare resonances decay feeding this channel. In the final state CE~$\rho_{K \pi}$ is 
somewhat weaker than in the GCE. Charge conservation effects remain small for the particle selections of both charged 
kaons,~$N_{K^+}+N_{K^-}$, and both charged pions,~$N_{\pi^+}+N_{\pi^-}$. At large~$\sqrt{s_{NN}}$, i.e. for a neutral 
system,~$\rho^{gce}_{K \pi} \simeq \rho^{ce}_{K \pi}$, CE suppression ceases entirely. As to be expected, the event-by-event 
multiplicities of charged kaons and charged pions, are anti-correlated by energy conservation in the MCE. Due to a missing 
strong decay contribution, the correlation coefficient~$\rho^{mce}_{K \pi}$ stays negative at all~$\sqrt{s_{NN}}$.

The final state GCE correlation coefficient~$\rho_{K \textrm{p}}$ is also rather weak due to a weak resonance decay 
contribution. In the CE the event-by-event multiplicities of protons plus anti-protons and charged kaons are 
anti-correlated at low~$\sqrt{s_{NN}}$. Their anti-correlation,~$\rho_{K \textrm{p}}<0$, is mediated by correlation with 
primordial strangeness carrying baryons, in particular the~$\Lambda$, at large baryon chemical potential~$\mu_B$.  
With increasing~$\sqrt{s_{NN}}$ the temperature rises and~$\mu_B$ drops. Charge conservation effects cease again
for a neutral system,~$\rho_{K \textrm{p}}^{gce} \simeq \rho_{K \textrm{p}}^{ce}$, for the particle selections of both charged 
kaons,~$N_{K^+}+N_{K^-}$, and both charged pions,~$N_{\pi^+}+N_{\pi^-}$. At small~$\sqrt{s_{NN}}$ the results of CE and MCE 
are similar. At larger~$\sqrt{s_{NN}}$, however, the correlation~$\rho_{K \textrm{p}}^{mce}$ remains negative, again due to 
energy conservation.

Lastly, the final state correlation coefficient~$\rho_{\textrm{p}\pi}$ is considered along the chemical freeze-out line in 
different ensembles. In Section~\ref{sec_phasediagram_multcorr} the correlation coefficient~$\rho_{\textrm{p}\pi^+}$ 
and~$\rho_{\textrm{p}\pi^-}$ were discussed. Here, both proton and anti-proton and both charged pions are counted, while 
in Section~\ref{sec_phasediagram_multcorr} only the proton was correlated with either of the charged pions. In the final 
state GCE the correlation amongst not charge separated multiplicities is stronger than for charge separated 
multiplicities,~$\rho^{gce}_{\textrm{p}\pi}>\rho^{gce}_{\textrm{p}\pi^+},\rho^{gce}_{\textrm{p}\pi^-}$. 

The anti-proton multiplicity  is strongly suppressed at low~$\sqrt{s_{NN}}$, and hence their contribution 
to~$\rho_{\textrm{p}\pi}$ is weak. On the other hand, compare Fig.(\ref{T_muB_rho_p_pi_GCE}), one finds a strong resonance 
decay contribution at low~$\sqrt{s_{NN}}$. Approaching a neutral system, one finds all three final state GCE correlation 
coefficients to drop. Charge conservation, i.e. CE, effects again cease in a neutral system for the fluctuations of 
multiplicities~$N_{\textrm{p}}+N_{\overline{\textrm{p}}}$ and~$N_{\pi^+}+N_{\pi^-}$. At low~$\sqrt{s_{NN}}$ a weak, but 
positive, final state correlation remains, compare Figs.(\ref{T_muB_rho_p_piplus_CE},\ref{T_muB_rho_p_piminus_CE}). In 
the MCE, one finds a baryon density induced positive correlation. Anti-protons are essentially absent. The 
correlation coefficient~$\rho_{\textrm{p}\pi^+}$ is negative, while the correlation coefficient~$\rho_{\textrm{p}\pi^-}$ is 
positive. The correlation coefficient~$\rho_{\textrm{p}\pi}$ is then weakly positive due to abundant 
neutrons and $\Delta$ resonances, allwoing for~$\textrm{n}\leftrightarrow \textrm{p}+\pi^-$ 
and~$\Delta\leftrightarrow \textrm{p}+\pi$. At large~$\sqrt{s_{NN}}$, 
hence for a neutral system, the anti-correlation between primordial parent particles and primordial protons, anti-protons 
and charge pions takes over. For the charge separated counterpart, one finds on the other 
hand~${\rho_{\textrm{p}\pi^-}\sim \rho_{\textrm{p}\pi^+}\sim 0}$,  
compare Figs.(\ref{T_muB_rho_p_piplus_MCE},\ref{T_muB_rho_p_piminus_MCE}), for large~$T$ and~$\mu_B\simeq 0$.

The results of the relativistic transport model HSD on correlations coefficients between particle multiplicities 
of charged kaons, charged pions, and protons plus anti-protons are summarized 
in Refs.~\cite{Gorenstein:2008et,Konchakovski:2009at}. 


\section{Comparison with NA49 Data}
\label{sec_freezeoutline_na49}

The fluctuations in nucleus-nucleus collisions are studied on an event-by-event basis: a given quantity is measured for 
each collision and a distribution of this quantity is constructed for a selected sample of these collisions. In the 
following a few aspects are discussed. Details of the experimental setup and analysis can be found 
in~\cite{Lungwitz:2006cx,Alt:2007jq}.

\subsubsection{Centrality Selection}

The fluctuations in the number of nucleon participants provide a dominant contribution to hadron multiplicity fluctuations.
In the language of statistical models, fluctuations of the number of nucleon participants correspond to volume 
fluctuations caused by the variations in the collision geometry. Mean hadron multiplicities are proportional (in the 
large volume limit) to the volume, hence, volume fluctuations translate directly to the multiplicity fluctuations.
Thus, a comparison between  data and predictions of statistical models should be performed for results which correspond to
collisions with a fixed number of nucleon participants.

Due to experimental limitations it is only possible to measure the number of participants of the projectile 
nucleus,~$N_P^{proj}$, in fixed target experiments (e.g. NA49 at the CERN SPS). This is done by measuring the 
energy deposited in a downstream Veto calorimeter. A large fraction of this energy is due to projectile 
spectators~$N_S^{proj}$. Using baryon number conservation for the projectile nucleus ($A = N_P^{proj}+N_S^{proj}$) the 
number of projectile participants can be estimated. However, also a fraction of non-spectator particles, mostly protons 
and neutrons, contribute to the Veto energy~\cite{Lungwitz:2006cx,Alt:2007jq}. Furthermore, the total number of nucleons 
participating in the collision can fluctuate considerably even for collisions with a fixed number of projectile 
participants, especially for peripheral collisions (see Ref.~\cite{Konchakovski:2005hq,Konchakovski:2006aq}). This is due 
to fluctuations of the number of target participants. The consequences of the asymmetry in the event selection depend on 
the dynamics of nucleus-nucleus collisions (see Ref.~\cite{Gazdzicki:2005rr} for details). Still, for the most central 
Pb+Pb collisions selected by the number of projectile participants an increase of the scaled variance can be estimated to 
be smaller than a few \%~\cite{Gazdzicki:2005rr} due to the target participant fluctuations. In the following hadron 
resonance gas results will be compared with the data of the NA49 collaboration on the 1\% most central Pb+Pb collisions at 
20$A$-158$A$~GeV~\cite{Lungwitz:2006cx,Alt:2007jq}.

\subsubsection{Modeling of Acceptance}

In the experimental study of nuclear collisions at high energies only a fraction of all produced particles is registered.
Thus, the multiplicity distribution of the measured particles is expected to be different from the distribution of all 
produced particles. In general, in statistical models, the  correlations in momentum space are caused by resonance 
decays, quantum statistics and the energy-momentum conservation law. In this section these correlations are neglected 
and the procedure of Appendix~\ref{appendix_accscaling} is applied for different ensembles. This may be reasonably valid 
for~$\omega^+$ and~$\omega^-$, as most decay channels only contain one positively (or negatively) charged particle, but 
is certainly much worse for~$\omega^{ch}$, for instance due to  decays of neutral resonances into two charged particles. 
In order to limit correlations caused by resonance decays, the analysis is restricted to negatively and positively
charged hadrons.

The NA49 acceptance used for the fluctuation measurements is located in the forward hemisphere ($1<y(\pi)<y_{beam}$, 
where~$y(\pi)$ is the hadron rapidity calculated assuming pion mass, and shifted to the collision center of mass 
system~\cite{Lungwitz:2006cx,Alt:2007jq}). The acceptance probabilities for positively and negatively charged hadrons are 
approximately equal,~$q^+\approx q^-$, and the numerical values at different SPS energies 
are:~$q^{\pm}=0.038,~0.063,~0.085,~ 0.131,~0.163$ 
at~$\sqrt{s_{NN}}=6.27,~7.62,~8.77,~12.3,~17.3$~GeV, respectively.

 \subsubsection{Comparison with NA49 Data}
\begin{figure}[h!]
  \epsfig{file=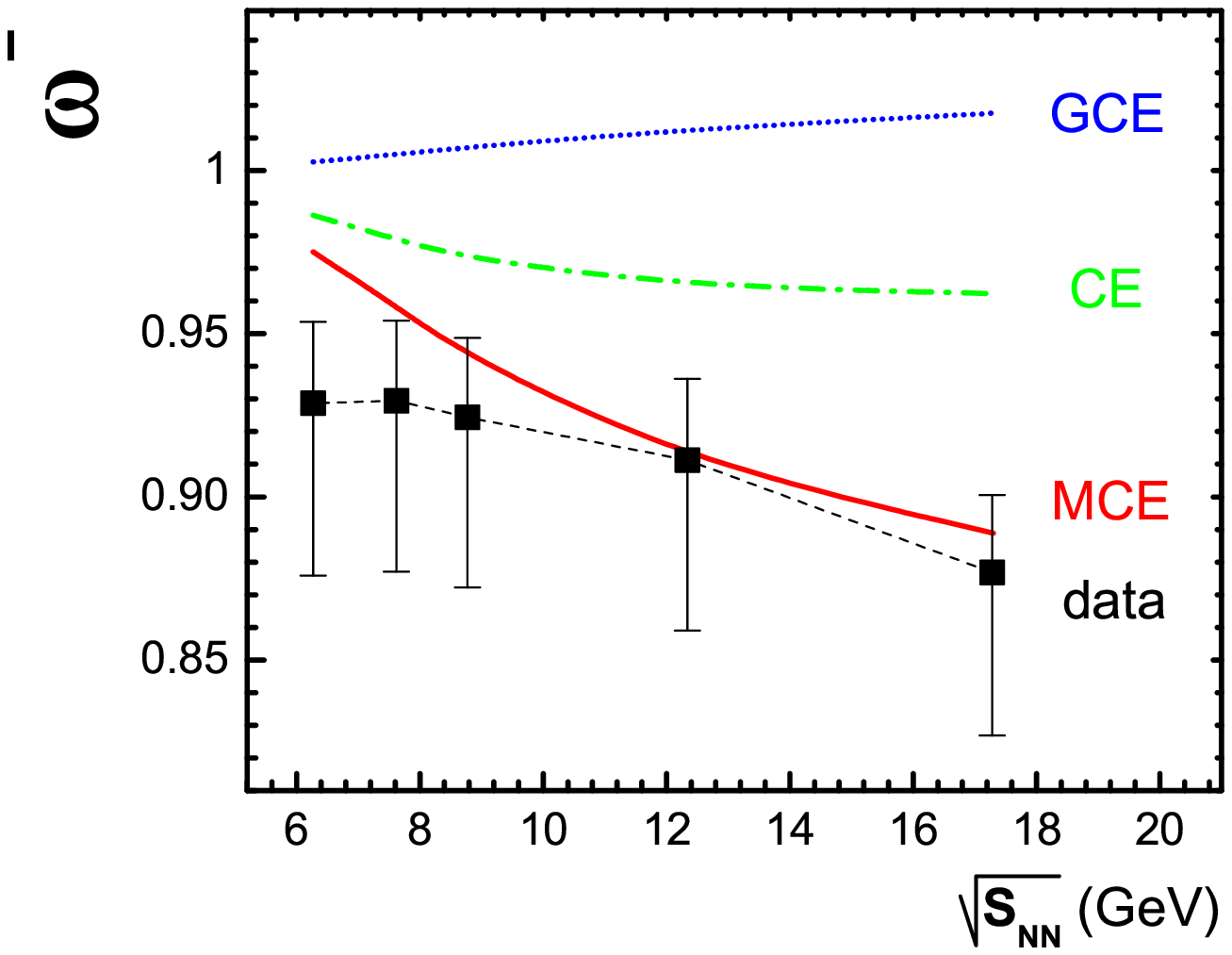,width=7.0cm}
  \epsfig{file=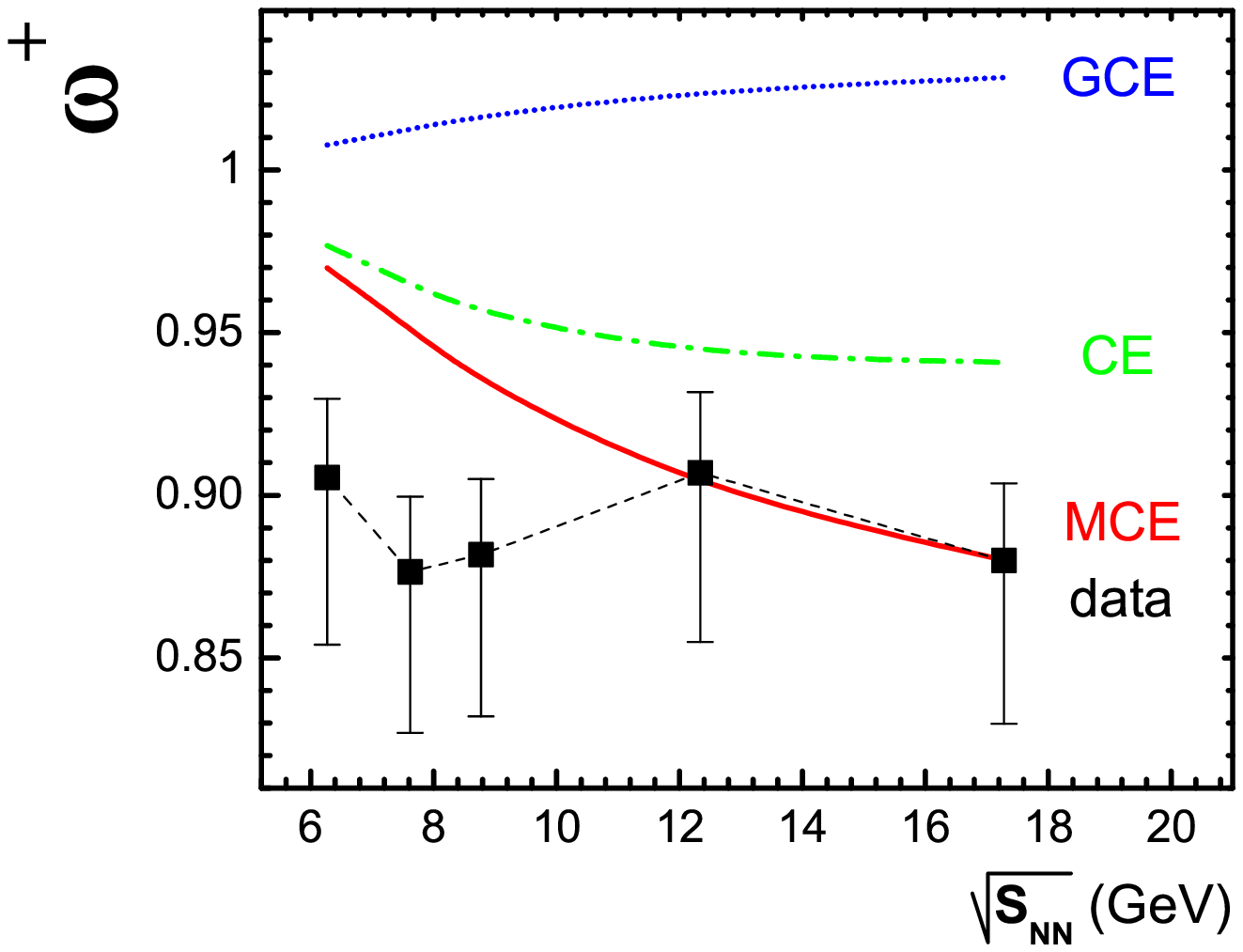,width=7.0cm}
  \caption{The scaled variances for negatively charged~({\it left}) and positively charged hadrons~({\it right}) along 
  the chemical freeze-out line for central Pb+Pb collisions at SPS energies. Full squares show the data of 
  NA49~\cite{Lungwitz:2006cx,Alt:2007jq}. Total (statistical+systematic) errors are indicated. Lines show the GCE, CE, 
  and MCE acceptance scaling estimates. Figures are taken from~\cite{Begun:2006uu}.}
  \label{stat-acc}
\end{figure}

Fig.(\ref{stat-acc}) presents the scaled variances~$\omega^-$ and~$\omega^+$, obtained from applying the acceptance 
procedure of Appendix~\ref{appendix_accscaling} to hadron resonance gas calculations in the GCE, CE, and MCE
shown in Fig.~(\ref{sqrtS_omega_p_m}). 

From Fig.(\ref{stat-acc}) it follows that the NA49 data for~$\omega^{-}$ and~$\omega^{+}$ extracted from 1\% of the most 
central  Pb+Pb collisions at all SPS energies are best described by the results of the hadron resonance gas model 
calculated within the MCE. The data reveal even stronger suppression of the particle number fluctuations. Given that 
the NA49 acceptance is located in the forward rapidity hemisphere, one would expect further suppression of multiplicity 
fluctuations compared to the acceptance scaling estimate employed here. See also Chapters~\ref{chapter_multfluc_hrg} 
and~\ref{chapter_multfluc_pgas}. \\

As discussed, the multiplicity distribution in statistical models in the full phase space and in the large volume
limit approaches a normal distribution. Particle detection is modeled by the simple procedure, 
Appendix~\ref{appendix_accscaling}, which is applicable to any form of full acceptance distribution~$P_{4\pi} (N)$. 
In the following multiplicity distributions in limited acceptance are discussed, and the statistical model results in 
different ensembles are compared with data on negatively and positively charged hadrons. The multiplicity distribution 
is, after the acceptance scaling procedure was applied, not Gaussian anymore. It is enough to mention that a Gaussian is
symmetric around its mean value, while the new distribution, in the limit of very small acceptance, converges to 
a Poissonian.

\begin{figure}[h!]
  \epsfig{file=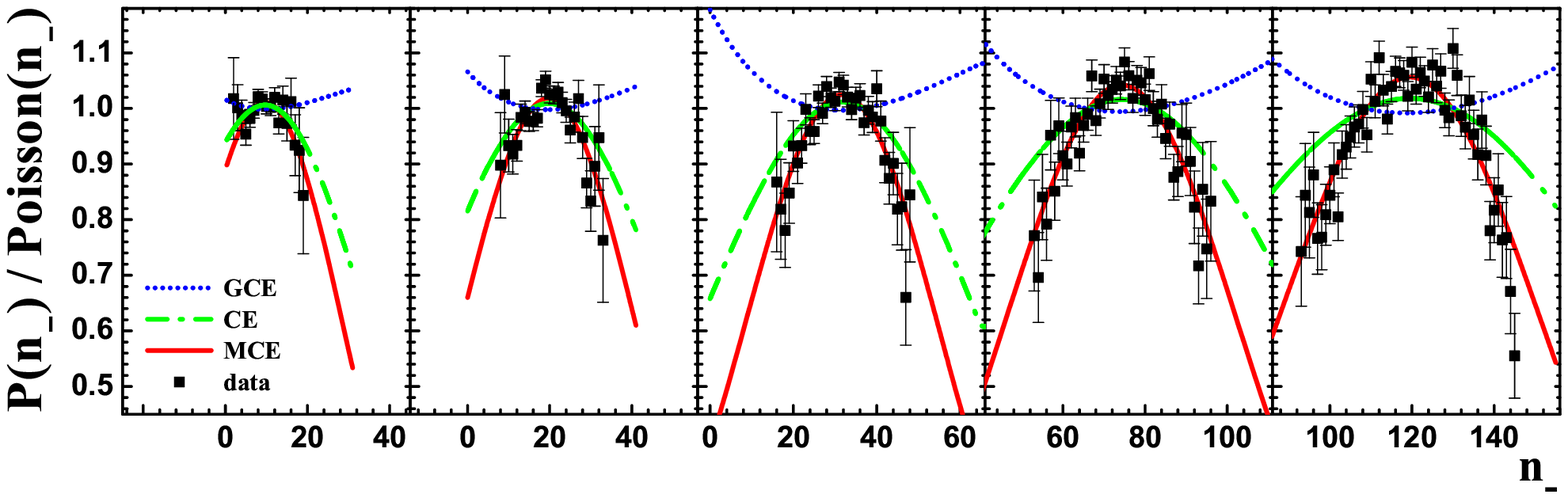,width=14.4cm}
  \epsfig{file=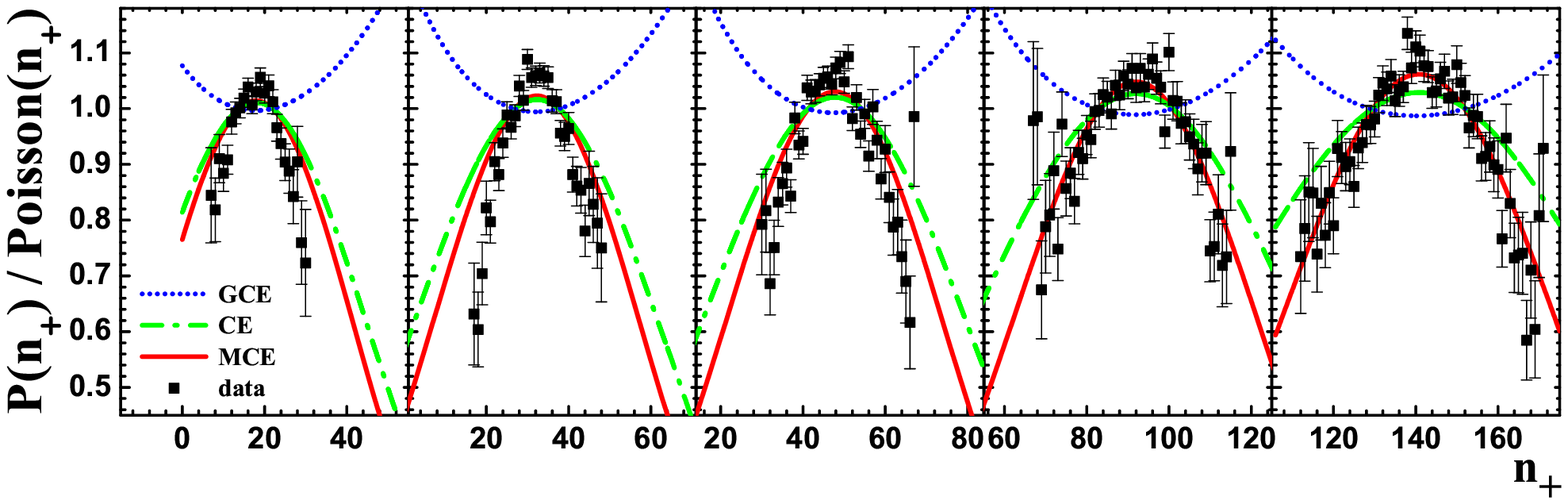,width=14.4cm}
  \caption{
  The ratio of the multiplicity distributions to Poisson ones for negatively charged~({\it top}) and positively charged 
  hadrons~({\it bottom}) produced in central (1\%) Pb+Pb collisions at 20$A$~GeV, 30$A$~GeV, 40$A$~GeV, 80$A$~GeV, 
  and 158$A$~GeV ({\it left}) to ({\it right}) in the NA49 acceptance. The experimental data (solid points) of 
  NA49~\cite{Lungwitz:2006cx,Alt:2007jq} are compared with the acceptance scaling estimates of the hadron resonance gas 
  model obtained within different statistical ensembles, the GCE (dotted lines), the CE (dashed-dotted lines), and the MCE 
  (solid lines). Figures are taken from~\cite{Begun:2006uu}. }
  \label{data2}
\end{figure}

In the following for each beam energy the volume is adjusted to fit the mean values for negatively ($V^-$) and 
positively ($V^+$) charged yields, separately. Note that values for the volume are about~$10-20$\% larger than the ones
in~\cite{Becattini:2005xt,Andronic:2005yp}, which were obtained using a much less stringent centrality selection (here 
only the~$1$\% most central data is analyzed). One finds that the volume parameters~$V^-$ and~$V^+$ deviate from each 
other by less than 10\%. Deviations of a similar magnitude are observed between the data on hadron yield systematics 
and hadron resonance gas model fits.

In order to allow for a detailed comparison of the distributions the ratios of data and model distributions to a 
Poissonian reference are presented in Fig.(\ref{data2}).  The results for negatively and positively charged hadrons
at 20$A$~GeV, 30$A$~GeV, 40$A$~GeV, 80$A$~GeV, and 158$A$~GeV are shown separately. The convex shape of the data reflects 
the fact that the measured distribution is significantly narrower than a Poissonian. This suppression of fluctuations
is observed for both charges, at all five SPS energies, and it is consistent with the results for the scaled variance 
shown and discussed previously. The GCE hadron resonance gas results are broader than the corresponding Poisson 
distribution. The ratio has a concave shape. Introduction of quantum number conservation laws leads to the convex 
shape and significantly improves agreement with the data. Further improvement is obtained by additional introduction 
of energy conservation.

\section{Discussion}
\label{sec_freezeoutline_discussion}

In this chapter multiplicity fluctuations and correlations have been analyzed for thermal parameter sets following the 
chemical freeze-out line for most central heavy ion collisions. The evolution of model parameters was taken from 
previous hadron resonance gas model comparison to experimental on average hadron production yields. These fits provide 
a connection between experimental control parameters center of mass energy and ion size, and the region of the phase 
diagram probed by the experiment.

Taking the model in its present form, a naive comparison the NA49 data on charged hadron multiplicity fluctuations has been
attempted. High resolution of the NA49 experimental data could in principle allow to distinguish between multiplicity 
fluctuations expected in hadron resonance gas model for different statistical ensembles. The measured spectra clearly
favor predictions of the micro canonical ensemble. Worse description is obtained for the canonical ensemble and
disagreement is seen considering the grand canonical one.

All calculations were performed in the thermodynamical limit which is a proper approximation for the considered
reactions. Thus, these results could be treated as an observation of the recently 
predicted~\cite{Begun:2004zb,Begun:2004gs,Begun:2004pk,Begun:2005ah,Begun:2005iv,Becattini:2005cc,Cleymans:2004iu,Cleymans:2005vn,Keranen:2004eu} 
suppression of multiplicity fluctuations due to conservation laws in relativistic gases in the large volume limit.
Similar observations have been made in detector physics~\cite{Fano:1947zz}, and in low temperature Bose-Einstein 
condensation~\cite{PhysRevE.54.3495}.

The experimental resolution of the NA49/61 detector for the measurement of enhanced fluctuations due to the onset of 
deconfinement can be increased by increasing acceptance. This will give a chance to observe, for example, the dynamical 
fluctuations discussed in Ref.~\cite{Gazdzicki:2003bb,Gorenstein:2003hk}. The observation of the MCE suppression effects 
of the multiplicity fluctuations by NA49 was possible only because a selection of a sample of collisions without 
projectile spectators. This selection seems to be possible only in the fixed target experiments. 

In collider experiments nuclear fragments which follow the beam direction cannot be measured, and another measure has 
to be used to determine which centrality class a given event (or collision) would fall into. Often, a charged hadron 
multiplicity reference distribution is employed. This measure (like any other) provides, however, only indirect access to 
the initial collision geometry.  

Likewise, the energy deposited in the interaction region, and hence the amount of energy available for particle 
production cannot be considered as fixed across a whole sample of heavy ion collisions. Yet, for each individual 
collision the energy content available for particle production and (collective) motion is fixed. Thus, even if a 
centrality class of collision events cannot be seen as a representative a micro canonical sample with the same values 
of global observables, the final state of each collision will in general fulfill energy-momentum and charge conservation.
One could then hope that the transverse momentum and rapidity dependence will not be washed out by a less stringent 
centrality selection, but merely shifted up or down.

More differential data on multiplicity fluctuations and correlations are required for further tests of the validity of 
the statistical models and observation of possible signals of the phase transitions. Simultaneous differential 
measurement of mean values~$\langle N_+ \rangle$, and~$\langle N_- \rangle$, 
variances~$\langle \left( \Delta N_+ \right)^2 \rangle$, and~$\langle \left( \Delta N_- \right)^2 \rangle$, and of the 
covariance~$\langle \Delta N_+ \Delta N_-  \rangle$ should lead to an improved understanding. 

The momentum space dependence for scaled variances demanded by the micro canonical ensemble is qualitatively also seen 
in (non-equilibrium) transport simulations~\cite{Lungwitz:2007uc} and recent data of the NA49 
collaboration~\cite{Alt:2007jq}. The experimental and theoretical 
study of multiplicity fluctuations should be complemented with a study of multiplicity correlation coefficients. 
This study could in principle be extended by a transverse momentum measurement. This would the be either a three by 
three covariance matrix, or if~$P_T$ is charge separated, a four by four covariance matrix. The~$\sqrt{s_{NN}}$ and ion 
size program \cite{Gazdzicki:2006fy} should be complemented by an acceptance effect study. Critical and quantum phenomena 
might more strongly feature at low transverse momentum.

\chapter{Summary}
\label{chapter_summary}
In this thesis the statistical properties of ideal relativistic hadronic equilibrium ensembles have been studied. Apart 
from the three standard canonical ensembles being discussed, a class of ensembles with finite thermodynamic bath has been 
introduced. The influence of global conservation laws, resonance decay, quantum statistics, finite acceptance in 
momentum space, and thermal freeze-out parameters on fluctuation and correlation observables of a sample of hadron 
resonance gas events have been considered. It was argued that the emerging picture would at least qualitatively apply to 
heavy ion physics.

In Chapter~\ref{chapter_clt} grand canonical joint distributions of extensive quantities were obtained by Fourier 
integration of the grand canonical partition function. An analytical expansion method for calculation of distributions 
at finite volume for the canonical as well as the micro canonical ensembles of the ideal relativistic hadron resonance gas
was presented. The introduction of temperature into the micro canonical partition function and chemical potentials into 
the canonical partition function have lead to the identification of the grand canonical partition function with the
characteristic function of associated joint probability distributions. Micro canonical and canonical multiplicity 
distributions could then be defined through conditional distributions, i.e. the  probability of finding a certain 
multiplicity, while other parameters (global charge or energy) were taken to be fixed. 

In Chapter~\ref{chapter_montecarlo} joint distributions of extensive quantities were then considered for statistical 
systems with finite rather than infinite thermodynamic bath. To introduce the scheme, a micro canonical system was 
conceptually divided into two subsystems. These subsystems were assumed to be in equilibrium with each other, and subject 
to the constraints of joint energy-momentum and charge conservation. Particles are only measured in one subsystem, while 
the second subsystem provides a thermodynamic bath. By keeping the size of the first subsystem fixed, while varying the 
size of the second, one can thus study the dependence of statistical properties of an ensemble on the fraction of the 
system observed, i.e. assess their sensitivity to globally applied conservation laws. The ensembles generated are 
thermodynamically equivalent in the sense that mean values in the observed subsystem remain unchanged when the size of 
the bath is varied, provided the combined system is sufficiently large. As the three standard canonical ensembles remain 
particular idealizations of physical systems, these intermediate ensembles might be of phenomenological interest, too.
These two chapters form the mathematical basis for analytical calculations and Monte Carlo simulations performed in 
this thesis.

The analysis of hadron resonance gas events started with the study of the grand canonical ensemble in 
Chapter~\ref{chapter_gce}. The grand canonical ensemble is considered to be the most accessible amongst the standard 
canonical ensembles. Due to the assumption of an infinite thermodynamic bath, occupation numbers of individual momentum 
levels are un-correlated with each other. Likewise, particle multiplicities of any two distinct groups of particles appear 
un-correlated. Together with occupation number fluctuations unconstrained, by global constraints, all extensive 
quantities, except for the volume, fluctuate on an event-by-event (or micro state-by-micro state) basis. Therefore, one 
finds the energy content and particle multiplicity to be strongly correlated, while the average energy per particle 
is un-correlated with multiplicity. The net-charge content of baryon number and strangeness in the system are correlated 
because some hadron species carry both charges. Different hadron species have different quantum numbers and follow 
different momentum spectra. The correlation between baryon number and strangeness then depends, like the one of energy 
and momentum, or energy and particle multiplicity, on which part of the momentum spectrum is accessible to measurement.

In Chapter~\ref{chapter_extrapol} joint distributions of extensive quantities were extrapolated to their micro canonical 
limit. For this iteratively samples of events were generated and analyzed for decreasing size of the thermodynamic bath. 
By construction, the distribution of extensive quantities considered for re-weighting converges to a $\delta$-function, 
while the positions of the mean values stayed constant. Whereas mean transverse momentum per particle and particle 
multiplicity were un-correlated in the grand canonical ensemble, this is not true anymore for systems with finite heat 
bath. Through successive focusing on events in the vicinity of a chosen equilibrium value the effects of global 
conservation laws became apparent. This proceeded in a systematic manner, allowing for extrapolation of observables 
from their grand canonical to their micro canonical limit. A caveat is that the statistical uncertainty, associated with 
finite samples, grows as the sample-reject limit is approached.  

In Chapter~\ref{chapter_multfluc_hrg} multiplicity fluctuations and correlations were studied for a neutral hadron 
resonance gas in limited acceptance in momentum space. The effects of resonance decay were considered, and the 
extrapolation scheme was applied to obtain the canonical and micro canonical ensemble limits. Comparison to analytical 
asymptotic solutions, available for primordial distributions in limited acceptance, suggests good agreement. The larger 
the number of conserved quantities, the larger the statistical uncertainty associated with finite samples of events. Yet, 
micro canonical effects are accurately reproduced by the Monte Carlo approach. As conservation laws are turned on, 
multiplicity fluctuations and correlations are modified. Momentum space effects on primordial multiplicity fluctuations 
and correlations arise due to conservation laws. For an ideal primordial grand canonical ensemble in the Boltzmann 
approximation (the original sample), multiplicity distributions are just uncorrelated Poissonians, regardless of the 
acceptance cuts applied, as particles are assumed to be produced independently. The requirement of energy-momentum and 
charge conservation leads to suppressed fluctuations and enhanced correlations between the multiplicities of two distinct 
groups of particles at the `high momentum' end of the momentum spectrum compared to the `low momentum' end of the 
momentum spectrum, provided some fraction of an isolated system is observed. Resonance decay does not change these trends.

Chapter~\ref{chapter_multfluc_pgas} was dedicated to the micro canonical ensemble. A simplified physical system was 
chosen to allow for smoother discussion. Owing to available analytical solutions, Fermi-Dirac and Bose-Einstein effects 
have been included into the analysis. Bose-Einstein enhancement and Fermi-Dirac suppression of multiplicity 
fluctuations are strong in momentum space segments where occupation numbers are large. This effect has been found to be 
much stronger than previous calculations of fully phase space integrated multiplicity fluctuations of Fermi-Dirac and 
Bose-Einstein systems suggested. For systems in collective motion is was found that the role of exactly imposed kinematic 
conservation laws is particularly important. Fluctuation and correlation observables transform under boosts, provided 
momentum conservation along the boost direction is taken into account. Lastly, it has been found that even in the 
thermodynamic limit long range correlations between disconnected regions in momentum space prevail. Multiplicities in 
different bins of rapidity, transverse momentum, or of azimuth, can have a non-vanishing correlation coefficient.   

In Chapter~\ref{chapter_phasediagram} the temperature - baryon chemical potential phase diagram of the hadron resonance 
gas model has been explored. Grand canonical charge correlations and fluctuations are different in different corners of the 
phase diagram. Like in the acceptance analysis, the correlation between two charges is strong when particles carrying 
both charges are abundant. At low temperature and baryon chemical potential mesons dominate, at high temperature and 
baryon chemical potential baryons take over. Fluctuations and correlations of charges systematically evolve accordingly. 
Multiplicity fluctuations and correlations in canonical and micro canonical ensembles follow suit. The influence of a 
particular conservation law on multiplicity fluctuations is strong if the analyzed species is abundant. A comparison 
between ensembles with and without energy, charge, or momentum conservation reveals subtle differences, such as for 
instance resonance decay in canonical and micro canonical ensembles. Resulting (primordial) multiplicity correlations 
are not due to local interactions amongst constituents, but due to globally implemented conservation laws for energy and 
charge.  

In Chapter~\ref{chapter_freezeoutline} multiplicity fluctuations and correlations were analyzed for thermal parameter 
sets following the chemical freeze-out line. Model parameters were taken from previous hadron resonance gas model 
comparisons to experimental on average hadron production yields. These fits provide a connection between experimental 
control parameters center of mass energy and ion size and the region of the phase diagram probed by an experiment. 
A first comparison to available experimental data suggests good agreement with hadrons resonance gas calculations. 
In particular the micro canonical formulation of the model seems to accurately reproduce qualitative features.


The calculations presented here provide qualitative effects affecting  multiplicity fluctuations and correlations. These 
effects arise solely from statistical mechanics and conservation laws. It will be interesting to see whether these 
qualitative effects are visible in further experimental measurements of the momentum dependence of fluctuations and 
correlations. If so, these effects might well be of similar magnitude to the signals for new physics. Disentangling 
them from dynamical correlations will then be an important, and likely non-trivial task. Equilibrium correlations are 
residual correlations which remain after the system has `forgotten` its history. If the system breaks up before 
equilibrium is attained, then correlations due to the initial configuration will remain. Yet, even if the system
stays far from any equilibrium point throughout its lifetime, correlations due to global conservation laws remain. 


\renewcommand{\chaptermark}[1]{\markboth{#1}{}} 

\appendix 

\chapter{Partition Function} 
\label{appendix_partfunc}
This section serves to provide a connection between~Eqs.(\ref{P_one}-\ref{P_three})   
and~Eq.(\ref{Curly}); namely to prove the following relation:  
\begin{equation}\label{Connection}  
\mathcal{Z}^{P^{\mu},Q^j}(V,\beta,u_{\mu},\mu_j) =   
e^{- P^{\mu} u_{\mu} \beta} ~ e^{Q^j \mu_j \beta} ~ Z(V,P^{\mu},Q^j)~,  
\end{equation}  
where~$Z(V,P^{\mu},Q^j)$ is the standard MCE partition function for a system of volume~$V$, 
collective four-momentum~$P^{\mu}$ and a set of conserved Abelian charges~$Q^j$, as worked out 
in~\cite{Becattini:2003ft,Becattini:2004rq}. 
The MCE partition function~$Z(V,P^{\mu},Q^j)$ counts the number of micro states consistent with this   
set of fixed extensive quantities. 
Likewise, one could interpret the partition function~$\mathcal{Z}^{P^{\mu},Q^j}(V,\beta,u_{\mu},\mu_j)$ as the number of 
micro states with the same set of extensive quantities for a GCE with local inverse temperature~$\beta$, 
four-velocity~$u_{\mu}$, and chemical potentials~$\mu_j$.  
  
The starting point for this calculation is~Eq.(\ref{Curly}):  
\begin{eqnarray}\label{Curly_app}  
\mathcal{Z}^{P^{\mu},Q^j}(V,\beta,u_{\mu},\mu_j) &=&   
\int \limits_{-\pi}^{\pi} \frac{d^J\phi}{\left( 2\pi \right)^J}   
~e^{-iQ^j \phi_{j}}    
~\int \limits_{-\infty}^{\infty} \frac{d^4\alpha }{\left( 2\pi \right)^4}    
~ e^{-iP^{\mu} \alpha_{\mu}}  \nonumber\\
&&\times~\exp \Bigg[V \Psi \left(\beta,u_{\mu},\mu_j; \alpha_{\mu},\phi_j \right) \Bigg]~.  
\end{eqnarray}  
Taking a closer look at the exponential of~Eq.(\ref{Curly_app}), one can spell out the single partition 
functions,~Eq.(\ref{psi}), of the cumulant generating function,~Eq.(\ref{Psi}), after having used the 
substitutions~Eqs.(\ref{subst_1}) and (\ref{subst_2}):  
\begin{equation}  
\exp \left[ \sum_i \frac{Vg_i}{\left(2 \pi \right)^3} \int d^3p \ln \left( 1 \pm   
e^{-p_i^{\mu}\left(\beta u_{\mu} - i\alpha_{\mu} \right)}   
e^{q_i^j \left(\beta \mu_j + i\phi_j \right)}   
\right)^{\pm 1} \right]~.  
\end{equation}  
Expanding the logarithm yields:  
\begin{equation}\label{Psi_app}  
\exp \left[ \sum_i \frac{Vg_i}{\left(2 \pi \right)^3} \int d^3p   
\sum \limits_{n_i=1}^{\infty}  \frac{\left(\mp 1 \right)^{n_i}}{n_i}  
e^{-n_i~\! p_i^{\mu}\left(\beta u_{\mu} - i\alpha_{\mu} \right)}   
e^{n_i~\! q_i^j \left(\beta \mu_j + i\phi_j \right)}   
 \right]~.  
\end{equation}  
Replacing now the momentum integration in~Eq.(\ref{Psi_app}) by the usual summation over individual 
momentum levels~$\frac{V}{\left(2 \pi \right)^3} \int d^3p \rightarrow \sum_{k_{n_i}}$ gives:  
\begin{equation}  
\exp \left[ \sum_i \sum \limits_{n_i=1}^{\infty} \sum \limits_{k_{n_i}}   
 \frac{g_i \left(\mp 1 \right)^{n_i}}{n_i}  
e^{-n_i~\! p_{k_{n_i}}^{\mu}\left(\beta u_{\mu} - i\alpha_{\mu} \right)}   
e^{n_i~\! q_i^j \left(\beta \mu_j + i\phi_j \right)}   
 \right]~.  
\end{equation}  
Finally, expanding the exponential results in:  
\begin{eqnarray}  
&&\mathcal{Z}^{P^{\mu},Q^j}(V,\beta,u_{\mu},\mu_j) =   
\int \limits_{-\pi}^{\pi} \frac{d^J\phi}{\left( 2\pi \right)^J}   
~e^{-iQ^j \phi_{j}}    
~\int \limits_{-\infty}^{\infty} \frac{d^4\alpha }{\left( 2\pi \right)^4}    
~ e^{-iP^{\mu} \alpha_{\mu}}   
~\prod_i \prod \limits_{n_i=1}^{\infty} \prod \limits_{k_{n_i}} 
\sum_{c_{k_{n_i}} =0}^{\infty}  \nonumber \\  
 && \qquad \times \frac{1}{c_{k_{n_i}}!}  
\left( \frac{g_i \left(\mp 1 \right)^{n_i}}{n_i} \right)^{c_{k_{n_i}}}  
e^{-c_{k_{n_i}} n_i~\! p_{k_{n_i}}^{\mu}\left(\beta u_{\mu} - i\alpha_{\mu} \right)}   
e^{c_{k_{n_i}} n_i ~\! q_i^j \left(\beta \mu_j + i\phi_j \right)}~.  
\end{eqnarray}  
Only sets of numbers~$\{c_{k_{n_i}}\}$ which meet the requirements:  
\begin{equation}  
\sum_i \sum \limits_{n_i=1}^{\infty} \sum \limits_{k_{n_i}}   
c_{k_{n_i}} n_i ~ p_{k_{n_i}}^{\mu} = P^{\mu}~,
\qquad  \textrm{and} \qquad  
\sum_i \sum \limits_{n_i=1}^{\infty} \sum \limits_{k_{n_i}}  
c_{k_{n_i}} n_i ~ q_i^j = Q^j~,  
\end{equation}  
have a non-vanishing contribution to the integrals. 
Therefore one can pull these factors in front of the integral:  
\begin{eqnarray}  
&&\mathcal{Z}^{P^{\mu},Q^j}(V,\beta,u_{\mu},\mu_j) =   
e^{- P^{\mu} u_{\mu} \beta} ~ e^{Q^j \mu_j \beta}   
\int \limits_{-\pi}^{\pi} \frac{d^J\phi}{\left( 2\pi \right)^J}   
~e^{-iQ^j \phi_{j}}    
~\int \limits_{-\infty}^{\infty} \frac{d^4\alpha }{\left( 2\pi \right)^4}    
~ e^{-iP^{\mu} \alpha_{\mu}}  \nonumber \\  
 &&\quad \times 
\prod_i \prod \limits_{n_i=1}^{\infty} \prod \limits_{k_{n_i}} \sum_{c_{k_{n_i}} =0}^{\infty}  
\frac{1}{c_{k_{n_i}}!} \left( \frac{g_i \left(\mp 1 \right)^{n_i}}{n_i} \right)^{c_{k_{n_i}}}  
e^{ i c_{k_{n_i}} n_i ~\! p_{k_{n_i}}^{\mu}  \alpha_{\mu} }~
e^{ i c_{k_{n_i}} n_i ~\! q_i^j \phi_j } ~.  
\end{eqnarray}  
Reverting the above expansions one returns to the definition of~$Z(V,P^{\mu},Q^j)$   
from Refs.~\cite{Becattini:2003ft,Becattini:2004rq} times the Boltzmann factors:  
\begin{eqnarray}  
&&\mathcal{Z}^{P^{\mu},Q^j}(V,\beta,u_{\mu},\mu_j) ~=~   
e^{- P^{\mu} u_{\mu} \beta} ~ e^{Q^j \mu_j \beta}   
\int \limits_{-\pi}^{\pi} \frac{d^J\phi}{\left( 2\pi \right)^J}   
~e^{-iQ^j \phi_{j}}    
~\int \limits_{-\infty}^{\infty} \frac{d^4\alpha }{\left( 2\pi \right)^4}    
~ e^{-iP^{\mu} \alpha_{\mu}}  \nonumber \\  
&& \qquad \qquad \times~
\exp \left[ \sum_i \frac{Vg_i}{\left(2 \pi \right)^3} \int d^3p \ln \left( 1 \pm   
e^{ i p_i^{\mu}\ \alpha_{\mu} }~ e^{ i q_i^j \phi_j }   
\right)^{\pm 1} \right]~,  
\end{eqnarray}  
which proves~Eq.(\ref{Connection}). Therefore, writing for the GCE distribution of extensive quantities:  
\begin{equation}   
P_{gce}(P^{\mu},Q^j) ~=~ \frac{e^{- P^{\mu} u_{\mu} \beta} ~ e^{Q^j \mu_j \beta}~    
Z(V,P^{\mu},Q^j)}{Z(V,\beta,u_{\mu},\mu_j)} ~=~   
\frac{\mathcal{Z}^{P^{\mu},Q^j}(V,\beta,u_{\mu},\mu_j)}{Z(V,\beta,u_{\mu},\mu_j)}~,  
\end{equation}  
provides the promised connection between~Eqs.(\ref{P_one}-\ref{P_three}) and~Eq.(\ref{Curly}).

\chapter{Second Derivative Test} 
\label{appendix_sdt}
Considering the integrand of~Eq.(\ref{Curly}), after change of notation~Eq.(\ref{AllQVector}), one 
may use short hand notation:
\begin{equation}\label{int}
I(\theta) ~\equiv~ e^{- i \mathcal{Q}^l \theta_l} 
\exp \Big[V~ \Psi \left( \theta \right) \Big]~.
\end{equation}
Taking the first derivative, i.e. the gradient, yields:
\begin{eqnarray}
\frac{\partial I(\theta)}{\partial \theta_l} ~=~
I(\theta) ~\times~ \Big[ ~V ~\frac{\partial \Psi
  (\theta) }{\partial \theta_l}  ~-~ i Q^l~\Big]~.
\end{eqnarray}
At the origin, using the definition of the cumulant tensor,~Eq.(\ref{kappa}), one finds: 
\begin{eqnarray}\label{grad}
\frac{\partial I(\theta)}{\partial \theta_l} 
\Bigg|_{\theta_l = 0_l} 
~=~ Z_{GCE} ~\times~ i ~\Big[ ~V ~\kappa_1^l - Q^l~\Big]~.
\end{eqnarray}
The GCE partition function~$Z_{GCE}=\exp \left[V \Psi \left( 0 \right) \right]$ is the~$0^{th}$ cumulant.
Therefore, the real part of~Eq.(\ref{grad}) is identical to zero. In the large volume limit the 
imaginary part is also zero, for~$Q^l = V \kappa_1^l$. Hence,~Eq.(\ref{int}) has a stationary point 
at~$\theta_l=0_l$. A second derivative test~\cite{Abramowitz:1965fix_1,Gradshteyn:2000fix_1} should be 
done in order to distinguish relative minimum, relative maximum or saddle point. 
Now, taking the second derivative, i.e. calculating the Hessian at the origin yields:
\begin{eqnarray}\label{hesse}
\frac{\partial I(\theta)}{\partial \theta_l ~\partial \theta_m} 
\Bigg|_{{}_{\theta_m =\;\; 0_m}^{\theta_l~ =~ 0_l}} \!\! \!\!= -Z_{GCE}  ~\times
\Bigg[ \Big[ V\kappa_1^l ~-~  Q^l~\Big]~ \Big[ V\kappa_1^m ~-~  Q^m~\Big] 
~+~ V \kappa_2^{m,l} \Bigg] ~. 
\end{eqnarray}
Here the imaginary part of~Eq.(\ref{hesse}) is equal to zero, while in the large  
volume limit (choosing intensive variables such that~$Q^l = V \kappa_1^l$)
 the real part converges to:
\begin{equation}\label{hesse2}
\frac{\partial I(\theta)}{\partial \theta_l ~\partial \theta_m}
\Bigg|_{{}_{\theta_m =\;\; 0_m}^{\theta_l~ =~ 0_l}} ~= -~Z_{GCE}  ~\times~ V \kappa_2^{m,l} 
\end{equation}
In order  to have a maximum for the integrand~$I(\theta)$ at the origin, the Hessian,~Eq.(\ref{hesse2}), 
has to be negative definite. Therefore,~$\kappa_2^{m,l}$ has to be positive definite. For a symmetric matrix 
it is enough to demand that all eigenvalues are positive. 
This is the case for all situations considered in this thesis. 
The derivatives are proportional to the GCE partition function~$Z_{GCE}$. 
In conclusion, the integrand of~Eq.(\ref{Curly}) is strongly peaked at~$\theta_l = 0_l$ in the thermodynamic limit.

\chapter{The Cumulant Tensor} 
\label{appendix_cumulant}
In this section the~$\kappa$-tensor Eq.(\ref{kappa}) is examined.
Only a static source in  full acceptance is considered here. 
Resonance decay effects are included.
For calculation of asymptotic multiplicity fluctuations and correlations, integrated over full momentum space, 
it is not necessary to take  momentum conservation into account. 
The choice of ensemble naturally defines the cumulants needed for calculations. 
Limited acceptance and moving sources, requiring momentum conservation,
are considered in Refs.\cite{Hauer:2007im,Hauer:2009nv}.

The analogs of the matrix~$\kappa_2$ and the vector~$\kappa_1$, however in different notation, 
were used in both previously published methods for calculation of scaled variance under the thermodynamic limit; 
the micro-correlator approach, see~\cite{Begun:2004zb,Begun:2004pk,Begun:2005ah,Begun:2005iv,Begun:2006jf,Begun:2006gj}, 
and saddle point expansion method~\cite{Becattini:2005cc}. 

\subsubsection{Cumulant structure}
Cumulants of order one give GCE expectation values, hence average baryon, strangeness, electric charge, 
and particle density. The second cumulant contains information about GCE fluctuations of some quantity 
(diagonal elements), as well as correlations between different quantities (off-diagonal elements). 
The first two cumulants are:
\begin{equation}
\kappa_1^{l_1} =  
\left( -i \frac{\partial}{\partial \theta_{l_1}} \right) 
\Psi ~\Bigg|_{\theta_l ~=~ 0_l} ~,
\end{equation}    
and    
\begin{equation} 
\kappa_2^{l_1,l_2} =  
\left( -i \frac{\partial}{\partial \theta_{l_1}} \right) 
\left( -i \frac{\partial}{\partial \theta_{l_2}} \right) 
\Psi ~\Bigg|_{\theta_l ~=~  0_l} ~.
\end{equation}
The index~$l$ denotes one of~$L$ conserved quantities which form the vector~$\mathcal{Q}^l$.  
For the asymptotic solution to the distributions~$P(\mathcal{Q}^l)$ the first and second cumulants are sufficient. 
The set of conserved quantities could be~$B$,~$S$,~$Q$,~$E$, with additionally particle multiplicities~$N_A$ 
and~$N_B$ included, hence:
\eq{
\kappa_1^{l_1} =
\begin{pmatrix}
\kappa_1^{N_A}, & \kappa_1^{N_B}, & \kappa_1^B, & \kappa_1^S, & \kappa_1^Q , & \kappa_1^E 
\end{pmatrix}~, \label{CEvector}} 
and 
\eq{ \kappa_2^{l_1,l_2} = 
\begin{pmatrix} 
\kappa_2^{N_A,N_A} & \kappa_2^{N_A,N_B} & \kappa_2^{N_A,B} & \kappa_2^{N_A,S} & \kappa_2^{N_A,Q} & \kappa_2^{N_A,E} \\
\kappa_2^{N_B,N_A} & \kappa_2^{N_B,N_B} & \kappa_2^{N_B,B} & \kappa_2^{N_B,S} & \kappa_2^{N_B,Q} & \kappa_2^{N_B,E} \\ 
\kappa_2^{B,N_A} & \kappa_2^{B,N_B} & \kappa_2^{B,B} & \kappa_2^{B,S} & \kappa_2^{B,Q} & \kappa_2^{B,E} \\ 
\kappa_2^{S,N_A} & \kappa_2^{S,N_S} & \kappa_2^{S,S} & \kappa_2^{S,S} & \kappa_2^{S,Q} & \kappa_2^{S,E} \\ 
\kappa_2^{Q,N_A} & \kappa_2^{Q,N_Q} & \kappa_2^{Q,Q} & \kappa_2^{Q,S} & \kappa_2^{Q,Q} & \kappa_2^{Q,E} \\ 
\kappa_2^{E,N_A} & \kappa_2^{E,N_E} & \kappa_2^{E,E} & \kappa_2^{E,S} & \kappa_2^{E,Q} & \kappa_2^{E,E} 
\end{pmatrix} ~.\label{CEmatrix} }

For clarity some elements are explicitly given. A general shorthand notation for the derivatives is used:
\begin{equation}\label{psi_deriv}
\psi^{\left(a,b;c \right)}_i =  \left(\pm 1 \right)^{a+1} ~
\frac{g_i}{\left(2\pi \right)^3} \int d^3p~ ~ \varepsilon_i^c ~ \frac{\left(
e^{-\left( \varepsilon_i -\mu_i\right) \beta  } \right)^a} {\left( 1 \pm
  e^{-\left( \varepsilon_i -\mu_i\right) \beta } \right)^b}\;.
\end{equation} 
The mean values of the electric charge density~$\kappa_1^{Q}$ and energy density~$\kappa_1^{E}$ are:
\begin{eqnarray} 
\kappa_1^{Q} 
&=& 
\left( -i \frac{\partial}{\partial \theta_{Q}}\right) 
\Psi ~\Bigg|_{\theta_l = 0_l} 
~=~ \sum_i q_i~
\psi^{\left(1,1;0 \right)}_i~,  \label{k_1_Q} \\
\kappa_1^{E} 
&=&  
\left( -i \frac{\partial}{\partial \theta_E}\right) 
\Psi ~\Bigg|_{\theta_l = 0_l} 
~=~ \sum_i~
\psi^{\left(1,1;1 \right)}_i ~.\label{k_1_E}
\end{eqnarray}
For instance~$\kappa_1^{Q}=0$ in an electric charge neutral system, due to equal contributions of
particles and anti-particles with~$q_i \rightarrow -q_i$. 
The final state mean value~$\kappa_1^{N_A}$ of particle density of species~$A$ is given by:
\begin{eqnarray}
\kappa_1^{N_A} 
&=& 
\left( -i \frac{\partial}{\partial \theta_{N_A}}\right)
\Psi ~\Bigg|_{\theta_l = 0_l} 
~=~ \sum_i \left[ \sum_{n_A}~\Gamma_{i}^{n_A} ~ n_A \right] ~ 
\psi^{\left(1,1;0 \right)}_i ~,\label{k_1_NA}
\end{eqnarray}
More about resonance decay and the definition of decay channels~$\Gamma_{i}^{n_A}$ later in this section.
In case two quantities are un-correlated, as for example primordial~$\pi^+$ multiplicity and globally 
conserved strangeness ($\pi^+$ does not carry strangeness), then the corresponding 
elements~$\kappa_2^{N_A,S}=\kappa_2^{S,N_A}=0$.     
Energy fluctuations~$\kappa_2^{E,E}$, correlations between baryonic charge and energy~$\kappa_2^{B,E}$, and
correlations between strangeness and baryonic charge~$\kappa_2^{S,B}$, are given by: 
\begin{eqnarray}
\kappa_2^{E,E} 
&=&  
\left( -i \frac{\partial}{\partial \theta_E}\right)^2 
\Psi ~\Bigg|_{\theta_l = 0_l} 
~=~  \sum_i~
\psi^{\left(1,2;2 \right)}_i~, \label{k_2_EE}\\
\kappa_2^{B,E} 
&=&  
\left( -i \frac{\partial}{\partial \theta_E}\right) 
\left( -i \frac{\partial}{\partial \theta_B}\right)  
\Psi ~\Bigg|_{\theta_l = 0_l} 
~=~ \sum_i ~b_i~ 
\psi^{\left(1,2;1\right)}_i~,\label{k_2_BE} \\
\kappa_2^{S,B} 
&=&
\left( -i \frac{\partial}{\partial \theta_S}\right)
\left( -i \frac{\partial}{\partial \theta_B}\right)
\Psi ~\Bigg|_{\theta_l = 0_l}  
~=~ \sum_i s_i ~ b_i ~ 
\psi^{\left(1,2;0 \right)}_i~.  \label{k_2_SB}
\end{eqnarray}
Here for instance~$\kappa_2^{B,E}=0$ for a neutral system, yet in general~$\kappa_2^{S,B} \not= 0$.
The correlations between particle number~$N_A$ and baryonic charge~$\kappa_2^{B,N_A}$, 
or energy~$\kappa_2^{E,N_A}$ are:
\begin{eqnarray}
\kappa_2^{B,N_A} 
&=&
\left( -i \frac{\partial}{\partial \theta_B}\right) 
\left( -i \frac{\partial}{\partial \theta_{N_A}}\right)
\Psi ~\Bigg|_{\theta_l = 0_l} \nonumber \\ 
&=&  \sum_i ~ b_i 
\left[ \sum_{n_A}~ \Gamma_{i}^{n_A} ~ n_A \right] 
\psi^{\left(1,2;0 \right)}_i~,\label{k_2_BNA} \\
\kappa_2^{E,N_A} 
&=&  
\left( -i \frac{\partial}{\partial \theta_E}\right) 
\left( -i \frac{\partial}{\partial \theta_{N_A}}\right)
\Psi ~\Bigg|_{\theta_l = 0_l} \nonumber \\ 
&=& \sum_i 
\left[ \sum_{n_A}~ \Gamma_{i}^{n_A} ~ n_A \right] 
\psi^{\left(1,2;1\right)}_i~. \label{k_2_ENA}
\end{eqnarray}
If, for instance~$N_A=N_{\textrm{p}}+N_{\overline{\textrm p}}$, 
then~$\kappa_2^{B,N_A}=0$ in a neutral system, as contribution of particles and 
anti-particles cancel each other out, and baryon number conservation does not 
affect multiplicity fluctuations of~$N_A=N_{\textrm{p}}+N_{\overline{\textrm p}}$.
Fluctuations in the density of particle species~$A$,~$\kappa_2^{N_A,N_A}$, are:
\begin{eqnarray}
\kappa_2^{N_A,N_A} 
&=& 
\left( -i \frac{\partial}{\partial \theta_{N_A}}\right)^2  
\Psi ~\Bigg|_{\theta_l = 0_l}  \nonumber \\ 
&=& \sum_i 
\left[ \sum_{n_A}~ \Gamma_{i}^{n_A} ~ n_A^2 \right] ~
\psi^{\left(1,1;0 \right)}_i  ~-~ 
\left[ \sum_{n_A}~ \Gamma_{i}^{n_A} ~ n_A \right]^2 ~ 
\psi^{\left(2,2;0 \right)}_i~. \label{k_2_NANA}
\end{eqnarray}
Lastly, correlation between the multiplicities~$N_A$ and~$N_B$ are only due to resonance decay: 
\begin{eqnarray}
\kappa_2^{N_A,N_B} 
&=& 
\left( -i \frac{\partial}{\partial \theta_{N_A}}\right)
\left( -i \frac{\partial}{\partial \theta_{N_B}}\right)  
\Psi ~\Bigg|_{\theta_l = 0_l} \nonumber \\ 
&=& \sum_i
\left[ \sum_{n_A,n_B}~ \Gamma_{i}^{n_A,n_B} ~ n_A n_B \right] ~
\psi^{\left(1,1;0 \right)}_i  \nonumber \\ 
&& - 
\left[ \sum_{n_A} \Gamma_{i}^{n_A} ~ n_A \right] ~ 
\left[ \sum_{n_B} \Gamma_{i}^{n_B} ~ n_B \right] ~
\psi^{\left(2,2;0 \right)}_i~. \label{k_2_NANB}
\end{eqnarray}

\subsubsection{Resonance Decay}
In this section  resonance decay is  included analytically directly into the system partition function. 
This has proved far more efficient than the definition of a generating function~\cite{Begun:2006jf,Begun:2006uu}, 
which requires a rather cumbersome calculation of all possible primordial correlators. 
Particle decay is itself a random process. Nevertheless, one can assign a particular volume in
phase space, given by the value of its single particle partition function~$\psi_i$, Eq.(\ref{psi}), 
to one type of resonance~$i$. 
Resonance decay will now populate this volume in phase space according to some weight factor, the
branching ratio, for each of the possible decay modes. This weight can be assigned  to the particle type(s) 
one is set to investigate. Based on the assumption that detected particles are drawn in the form of a random sample
from all final state particles, e.g. disregarding correlation in momentum space, this procedure leads to the 
acceptance scaling approximation employed in Refs.\cite{Begun:2004gs,Begun:2006jf,Begun:2006uu}.
(see also Appendix \ref{appendix_accscaling}) 
Conservation laws can be imposed on the primordial state (rather than the final state),
since decay channels, which are experimentally measured, do not only obey
charge conservation, but all relevant conservation laws (omitting weak decays).

For calculation of final state distributions, one needs to determine {\it final} branching ratios of a resonance 
type into only stable particles. As an example the decay channel~$A \rightarrow B + X$ is considered with branching
ratio\footnote{The  the letter~$\Gamma$ is used instead of the more conventional~$Br$ or~$br$ to denote 
branching ratios.}~$\Gamma_{A\rightarrow B + X} = a$. Resonance~$B$ could  itself be unstable and subsequently decay 
via the channel~$B \rightarrow Y +Z$ with branching ratio~$\Gamma_{B  \rightarrow Y + Z} = b$. 
So the {\it  final} branching ratio is 
defined~$\Gamma_{A \rightarrow X+ Y+ Z} = \Gamma_{A \rightarrow B +X}~\Gamma_{B\rightarrow Y + Z} = ab$. 
Decay tables in~\cite{Begun:2006jf,Begun:2006uu,Gorenstein:2008et,Konchakovski:2009at}
have also been generated according to this prescription.

For resonances it seems economical to define further {\it absolute} branching ratios~$\Gamma_i^{n_A,n_B}$ as 
the sum over all {\it final} decay channels of resonance~$i$ with given numbers~$n_A$ and~$n_B$ of selected daughters. 
Hence~$\Gamma_i^{2,0}$ is the sum over all {\it final} decay  channels with two daughter particles~$A$ and none 
of species~$B$ which are of interest, i.e. two positively charged particles in case one wants to calculate~$\omega^+$. 
As a consequence of this definition, branching ratios~$\Gamma_i^{n_a,n_B}$ will depend on which~$\omega$ or~$\rho$ one 
is set to calculate.
 
For the final state one has to take all {\it absolute} decay channels of resonance type~$i$ into a 
number~$n_A$ and~$n_B$ of selected stable particles into account. 
For the sake of a common treatment for all particles and resonances are assigned a `decay` channel to stable particles 
as well, either~$\Gamma_i^{1,0} = 1$  if selected, or~$\sum_{n_B} \Gamma_l^{0,n_B} = 1$ if not selected. 
The single particle partition function then reads: 
\begin{eqnarray} 
\psi_i \left( \theta_E,\theta_j,\theta_{N_A},\theta_{N_B};\Gamma_i^{n_A,n_B} \right) 
&=&
\frac{g_i}{\left( 2 \pi \right)^3} 
\int d^3p \ln \Bigg[ 1 \pm 
e^{ - \left( \varepsilon_i-\mu_i\right)\beta} e^{i \varepsilon_i \theta_E}
e^{i q_i^j  \theta_j} \nonumber \\
&& 
\Big[ \sum_{n_A,n_B} \; \Gamma_i^{n_A,n_B} \; 
e^{i  n_A  \theta_{N_A}}  e^{i  n_B  \theta_{N_B}} 
\Big] \Bigg]^{\pm 1}~,
 \end{eqnarray}
where the form of the vector~$q_i^j$ depends on ones choice of ensemble. 
For instance, one could have~$q_i^j = (q_i,b_i,s_i)$ for a hadron resonance gas in the CE with three conserved charges, 
or~$ q_i^j = (q_i,b_i,s_i,\varepsilon_i)$ in the MCE without momentum conservation. 
The sum over all decay channels of the selected types needs to be one: 
\begin{equation}\label{Decay1}
\sum_{n_A,n_B} \; \Gamma_i^{n_A,n_B} = 1~.
\end{equation}
This is somewhat of a practical challenge, since decay chains of heavier resonances are not always well 
established~\cite{PDG:2002fix_1} and respective thermal models 
codes~\cite{Torrieri:2004zz,Torrieri:2006xi,Wheaton:2004qb,Kisiel:2005hn} struggle to implement this. 
There are several ways to deal with this, the two extreme ones are 1) rescale all known channels according 
to Eq.(\ref{Decay1}), to unity, or 2) assign the missing fraction to the 'channel'~$\Gamma_i^{0,0}$, e.g. 
to the channel without stable particles of interest.

\chapter{Distributions}
\label{appendix_multmist}
From the assumption that the distribution~$P_{gce}(\mathcal{Q}^l,N_A,N_B)$ of a GCE with~$L$ 
charges~$\mathcal{Q}^l$ and particle numbers of species~$A$ and~$B$ converges to a multivariate normal 
distribution, it also follows that the marginal distribution~$P_{gce}(\mathcal{Q}^l)$, as well as the 
conditional distribution~$P_{gce}(N_A,N_B|\mathcal{Q}^l)$, are normal distributions. 
Hence,~$P_{mce} (N_A,N_B)$ should have a good approximation in a bivariate normal 
distribution~$P_{BND}(N_A,N_B)$ in the large volume limit (where particle multiplicity can 
be appropriately treated as continuous):  
\begin{eqnarray}\label{BND}   
P_{BND}(N_A,N_B) &=& \frac{1}{2\pi V \sqrt{\sigma_A^2 \sigma_B^2 \left(1-\rho^2   
    \right)}} \\   
&\times& \exp \Bigg[ -\frac{1}{2V} \Bigg[    
    \frac{\left(\Delta N_A\right)^2}{\sigma_A^2 \left(1-\rho^2 \right) }   
      - 2\rho \frac{\left(\Delta N_A \right)\left(\Delta N_B \right)}{\sigma_A   
        \sigma_B \left(1-\rho^2 \right) }    
        +  \frac{\left(\Delta N_B \right)^2}{\sigma_B^2 \left(1-\rho^2 \right)   
        } \Bigg] \Bigg]~, \nonumber   
\end{eqnarray}   
where~$\Delta N_X = N_X - \langle N_X \rangle$, with~$X=A,B$ and:   
\begin{eqnarray}   
V \sigma_A^2 &\equiv& \langle N_A^2 \rangle ~-~ \langle N_A \rangle^2~, \\   
V \sigma_B^2 &\equiv& \langle N_B^2 \rangle ~-~ \langle N_B \rangle^2~, \\    
V \sigma_{AB} &\equiv& \langle N_A N_B \rangle ~-~ \langle N_A \rangle \langle N_B   
\rangle~.    
\end{eqnarray}    
Here~$V \sigma_A^2$ and~$V \sigma_B^2$ are the variances of the marginal distributions of 
particle multiplicities~$N_A$ and~$N_B$. The term~$V \sigma_{AB}$ is called the co-variance.  

The MCE joint multiplicity distribution~$P_{mce}(N_A,N_B)$ is conveniently expressed by the ratio of 
two GCE joint distributions, i.e. is given by the conditional probability distribution:   
\begin{eqnarray}   
P_{mce}(N_A,N_B) &=& P_{gce}(N_A,N_B|\mathcal{Q}^l)~, \\   
&=&\frac{P_{gce}(N_A,N_B,\mathcal{Q}^l)}{P_{gce}(\mathcal{Q}^l)}~.   
\end{eqnarray}   
In the thermodynamic limit the distributions~$P_{gce}(N_A,N_B,\mathcal{Q}^l)$ and~$P_{gce}(\mathcal{Q}^l)$   
can be approximated by multivariate normal distributions, Eq.(\ref{MND}). 
The extended vector~$\mathcal{Q}^j$ summarizes particle numbers~$N_A$ and~$N_B$ and extensive 
quantities~$\mathcal{Q}^l$ to a vector of length~$J=2+L$, hence~$\mathcal{Q}^j=(N_A,N_B,\mathcal{Q}^l)$.
The vector denoting the deviation of the mean values~$\Delta \mathcal{Q}^j=\mathcal{Q}^j-V\kappa_1^j$  would 
then read:  
\begin{equation}   
\Delta \mathcal{Q}^j
~=~ \left(\Delta N_A,~ \Delta N_B,~ \Delta \mathcal{Q}^l \right)~.  
\end{equation}   
Evaluating the multivariate normal distribution, Eq.(\ref{MND}), around its peak for~($\mathcal{Q}^l$) yields:  
\begin{equation}   
\Delta \mathcal{Q}^j 
~=~ \left(\Delta N_A,~ \Delta N_B, ~0^l~\right)~.  
\end{equation}   
The vector Eq.(\ref{xi-var}) then becomes:   
\begin{align}   
\xi  ^j
~=~ V^{-1/2} ~   
\begin{pmatrix}   
\lambda_{1,1} ~\Delta N_A + \lambda_{1,2} ~ \Delta N_B \\    
\lambda_{2,1} ~\Delta N_A + \lambda_{2,2} ~ \Delta N_B \\    
\lambda_{3,1} ~\Delta N_A + \lambda_{3,2} ~ \Delta N_B \\    
\lambda_{4,1} ~\Delta N_A + \lambda_{4,2} ~ \Delta N_B \\    
\lambda_{5,1} ~\Delta N_A + \lambda_{5,2} ~ \Delta N_B \\    
\dots \\    
\end{pmatrix}~,   
\end{align}   
where~$\lambda_{i,j}$ are the elements of the matrix Eq.(\ref{sigmatensor}). Therefore:  
\begin{equation}\label{xijxij}  
\xi^j ~ \xi_j  
~=~ V^{-1}~\Big[ \left( \Delta   
  N_A \right)^2 \sum \limits_{j=1}^J ~    
\lambda_{j,1}^2 ~+~2\left( \Delta N_A \right) \left( \Delta N_B \right) \sum   
\limits_{j=1}^J ~     
\lambda_{j,1} \lambda_{j,2}   ~+~ \left( \Delta N_B \right)^2 \sum   
\limits_{j=1}^J ~ \lambda_{j,2}^2 \Big]~,  
\end{equation}   
with~$J=2+L=9$, where~$L=3+4$ for a MCE hadron resonance gas with momentum conservation. The micro canonical 
joint multiplicity distribution of particle species~$A$ and~$B$ can thus be written as:  
\begin{equation}\label{PCE}   
 P_{mce}(N_A,N_B) ~=~  \frac{1}{\left(2\pi V\right)} 
 ~\frac{\det |\sigma_N|}{\det |\sigma|} 
 ~\exp \left[-\frac{1}{2}~ \xi^j~ \xi_j \right]~,  
\end{equation}   
where~$ \sigma_N$ is the 7-dimensional inverse sigma tensor of the distribution~$P_{gce}(\mathcal{Q}^l)$. 
Comparing this to a bivariate normal distribution, Eq.(\ref{BND}), one finds:    
\begin{eqnarray} \label{term_A}  
\sum \limits_{j=1}^J ~ \lambda_{j,1}^2  &=& \frac{1}{\sigma_A^2 \left(1-\rho^2   
  \right)} ~=~ A  ~,\\ \label{term_B}  
\sum \limits_{j=1}^J ~ \lambda_{j,2}^2  &=& \frac{1}{\sigma_B^2 \left(1-\rho^2   
  \right)} ~=~ B  ~,\\ \label{term_C}  
\sum \limits_{j=1}^J ~ \lambda_{j,1} \lambda_{j,2} &=& -   
~\frac{\rho}{\left(1-\rho^2\right) \sigma_A \sigma_B}  ~=~ - C~.  
\end{eqnarray}   
After short calculation one finds for the co-variances:  
\begin{eqnarray}   
\sigma_A^2 &=& \frac{B}{AB-C^2}~, \\   
\sigma_B^2 &=& \frac{A}{AB-C^2}~, \\   
\sigma_{A,B} &=& \frac{C}{AB-C^2}~,    
\end{eqnarray}   
and additionally for the correlation coefficient, Eq.(\ref{rho}):   
\begin{eqnarray}   
\rho ~=~¸\frac{\sigma_{A,B}}{\sigma_A~\sigma_B} ~=~ \frac{C}{\sqrt{AB}}~,   
\end{eqnarray}   
where the terms~$A,B,C$ are given by Eqs.(\ref{term_A} - \ref{term_C}). For the normalization 
in Eq.(\ref{PCE}) (from a comparison with Eq.(\ref{BND})) one finds:    
\begin{eqnarray}   
\frac{\det|\sigma_N|}{\det| \sigma |} &=& 
\frac{1}{\sigma_A \sigma_B \sqrt{(1-\rho^2)}} ~=~ \sqrt{AB}~.  
\end{eqnarray}   
An analytic formula for the scaled variance, without the need to invert 
matrices, can be found in~\cite{Hauer:2007ju}.

\chapter{Acceptance Scaling}   
\label{appendix_accscaling}
To illustrate the `acceptance scaling` procedure employed in~\cite{Begun:2004gs,Begun:2006jf,Begun:2006uu}  
uncorrelated acceptance of particles of species~$A$ and~$B$ is assumed. Particles are measured, or observed, with 
probability~$q$ regardless of their momentum. The distribution of measured particles~$n_A$, when a total 
number~$N_A$ is produced, is then given by a binomial distribution:  
\begin{equation}  
P_{acc} \left(n_A|N_A \right)~=~ q^{n_A} ~\left(1-q \right)^{N_A-n_A}   
\binom{N_A}{n_A}~.  
\end{equation}  
The same acceptance distribution is used for particles of species~$B$. Independent of the original multiplicity 
distribution~$P(N_A,N_B)$, the moments of the measured particle multiplicity distribution are defined by:  
\begin{equation}  
\langle n_A^a \cdot n_B^b \rangle ~\equiv~ \sum_{n_A,n_B} ~\sum_{N_A,N_B}~ n_A^a ~ n_B^b ~   
P_{acc} \left(n_A|N_A \right) ~ P_{acc} \left(n_B|N_B \right)~P(N_A,N_B)~.  
\end{equation}  
For the first moment~$\langle n_A \rangle$ one finds:  
\begin{equation}  
\langle n_A \rangle ~=~ q~\langle N_A \rangle~.  
\end{equation}  
The second moment~$\langle n_A^2 \rangle$ and the correlator~$\langle n_A \cdot n_B \rangle$  are given by:  
\begin{eqnarray}  
\langle n_A^2 \rangle &=& q^2 \langle N_A^2 \rangle ~+~ q\left(1-q \right)  
\langle N_A \rangle~, \\  
\langle n_A \cdot n_B \rangle &=& q^2 \langle N_A \cdot N_B \rangle~.  
\end{eqnarray}  
For the scaled variance~$\omega^A_q$ of observed particles one now finds:  
\begin{eqnarray}\label{acc_omega}  
\omega^A_q &=& \frac{\langle n_A^2 \rangle - \langle n_A \rangle^2}  
{\langle n_A \rangle} ~=~ 1~-~q~+q~\omega^A_{4\pi}~,  
\end{eqnarray}  
where~$\omega^A_{4\pi}$ is the scaled variance of the distribution if all particles of   
species~$A$ are observed. Lastly, the correlation coefficient~$\rho_q$ is:  
\begin{equation}  
\rho_q ~=~ \frac{\langle \Delta n_A \Delta n_B \rangle}  
{\sqrt{\langle \left( \Delta n_A \right)^2 \rangle ~   
\langle \left( \Delta n_B \right)^2 \rangle  }}~,  
\end{equation}  
with~$\langle \Delta n_A \Delta n_B \rangle = \langle n_A \cdot n_B \rangle -   
\langle n_A \rangle \langle n_B \rangle$, and~$\langle \left( \Delta n_A \right)^2 \rangle  
= \langle n_A^2 \rangle - \langle n_A \rangle^2$.  
Substituting the above relations, one finds after a short calculation:  
\begin{equation}\label{acc_rho_comp}  
\rho_q ~=~ \rho_{4\pi}~q~\sqrt{\omega^A_{4\pi} \omega^B_{4\pi}}   
\Bigg[  
q^2\omega^A_{4\pi} \omega^B_{4\pi}   
+ q(1-q)\omega^A_{4\pi}  
+ q(1-q)\omega^B_{4\pi}   
+ (1-q)^2  
\Bigg]^{-1/2}.  
\end{equation}  
In case~$\omega^A_{4\pi}= \omega^B_{4\pi}=\omega_{4\pi}$, Eq.(\ref{acc_rho_comp}) simplifies to:  
\begin{equation}\label{acc_rho}  
\rho_q ~=~ \rho_{4\pi}~\frac{q~\omega_{4\pi}}{1~-~q~+~q~\omega_{4\pi}}~.  
\end{equation}  
Both lines are independent of the mean values~$\langle N_A \rangle$ and~$\langle N_B \rangle$.

\chapter{Convergence Study}   
\label{appendix_convstudy}
Not only for the sake of completeness, the convergence of various quantities with the sample size, 
i.e. the number of events,~$N_{events}$, in our Monte Carlo scheme is discussed in this section. 
Here only final state (stable against electromagnetic and weak decays) particles are analyzed. 
A closer look is taken at the data sub-set of~$20 \cdot 2 \cdot 10^5$ events, with~${\lambda=V_1/V_g=0.875}$ for 
the size of the bath, which already has been discussed in Chapter~\ref{chapter_multfluc_hrg}. 

There is a degree of freedom at so how to estimate the statistical uncertainty on the 
moments of a distribution of observables of a finite sample. The approach taken
here is straight forward, but could, however, certainly be improved.

\begin{figure}[ht!]
  \epsfig{file=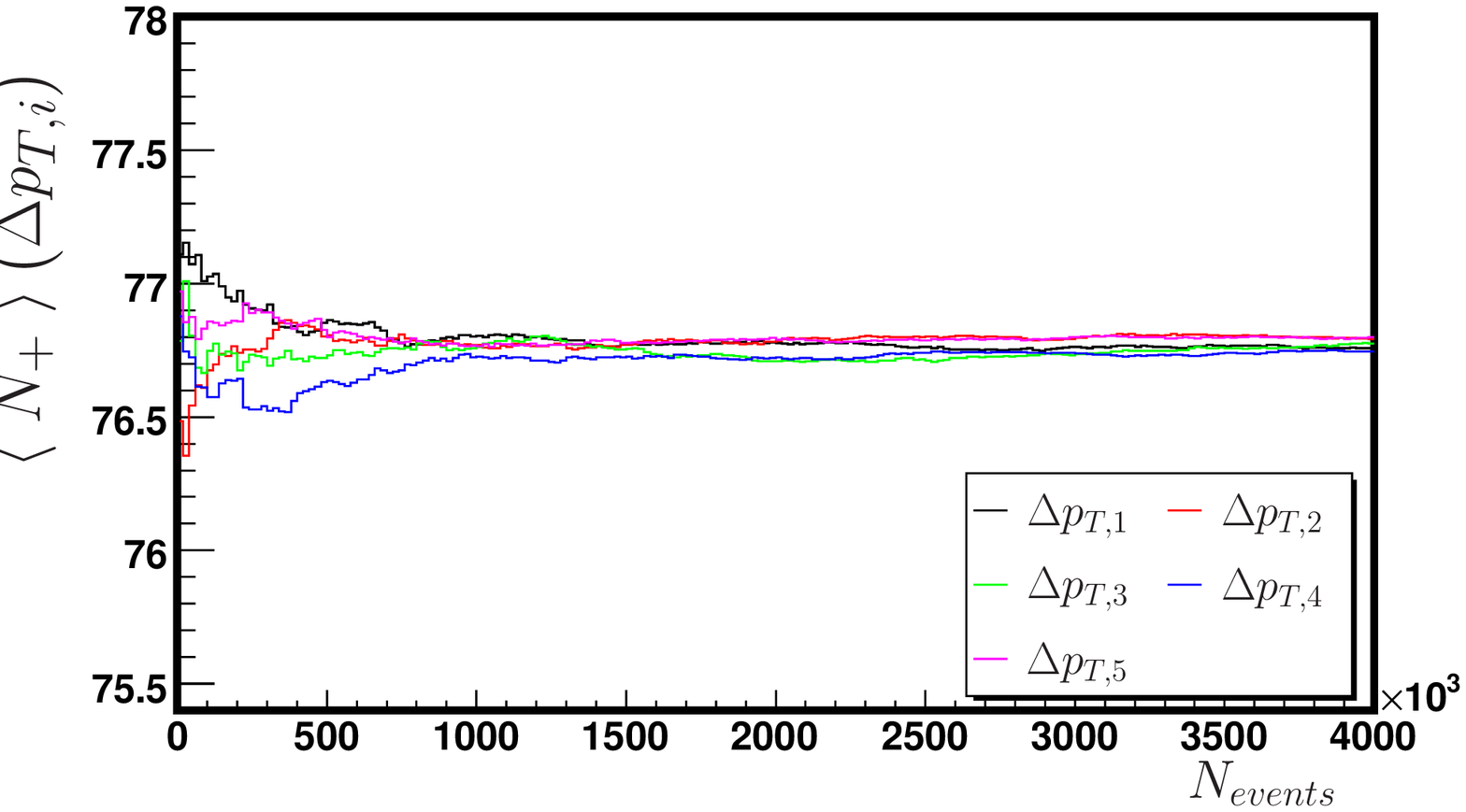,width=7.2cm,height=6.5cm}
  \epsfig{file=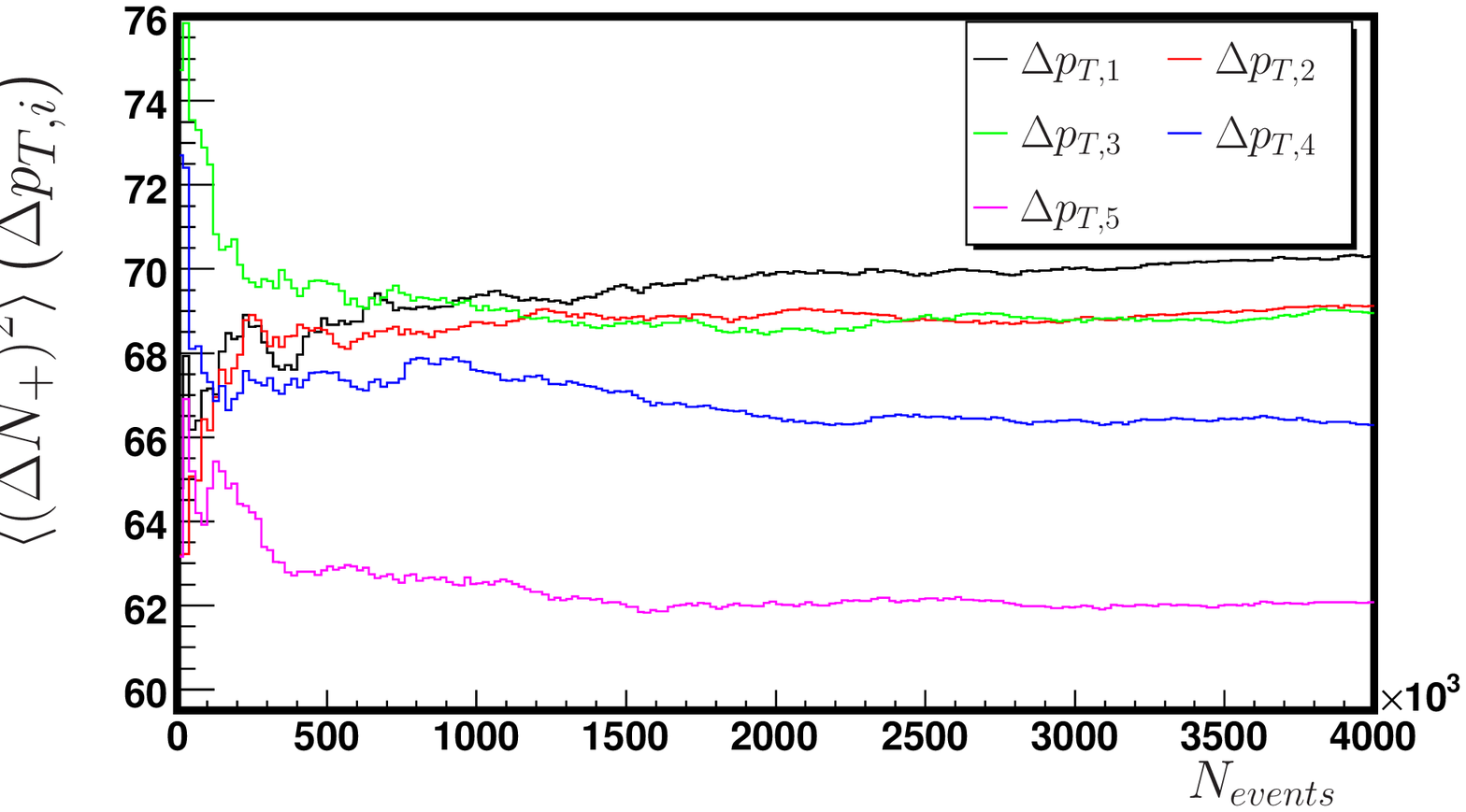,width=7.2cm,height=6.5cm}
  \caption{Step histogram showing the convergence of the mean values~$\langle N_+ \rangle$ ({\it left})
    and variances~$\langle \left( \Delta N_+ \right)^2 \rangle$ ({\it right}) for positively 
    charged final state hadrons in transverse momentum bins~$\Delta p_{T,i}$ for 
    a hadron resonance gas with~$\lambda = V_1/ V_g =0.875$.}  
  \label{stephist_N_DeltaNSq}
\end{figure}

\begin{figure}[ht!]
  \epsfig{file=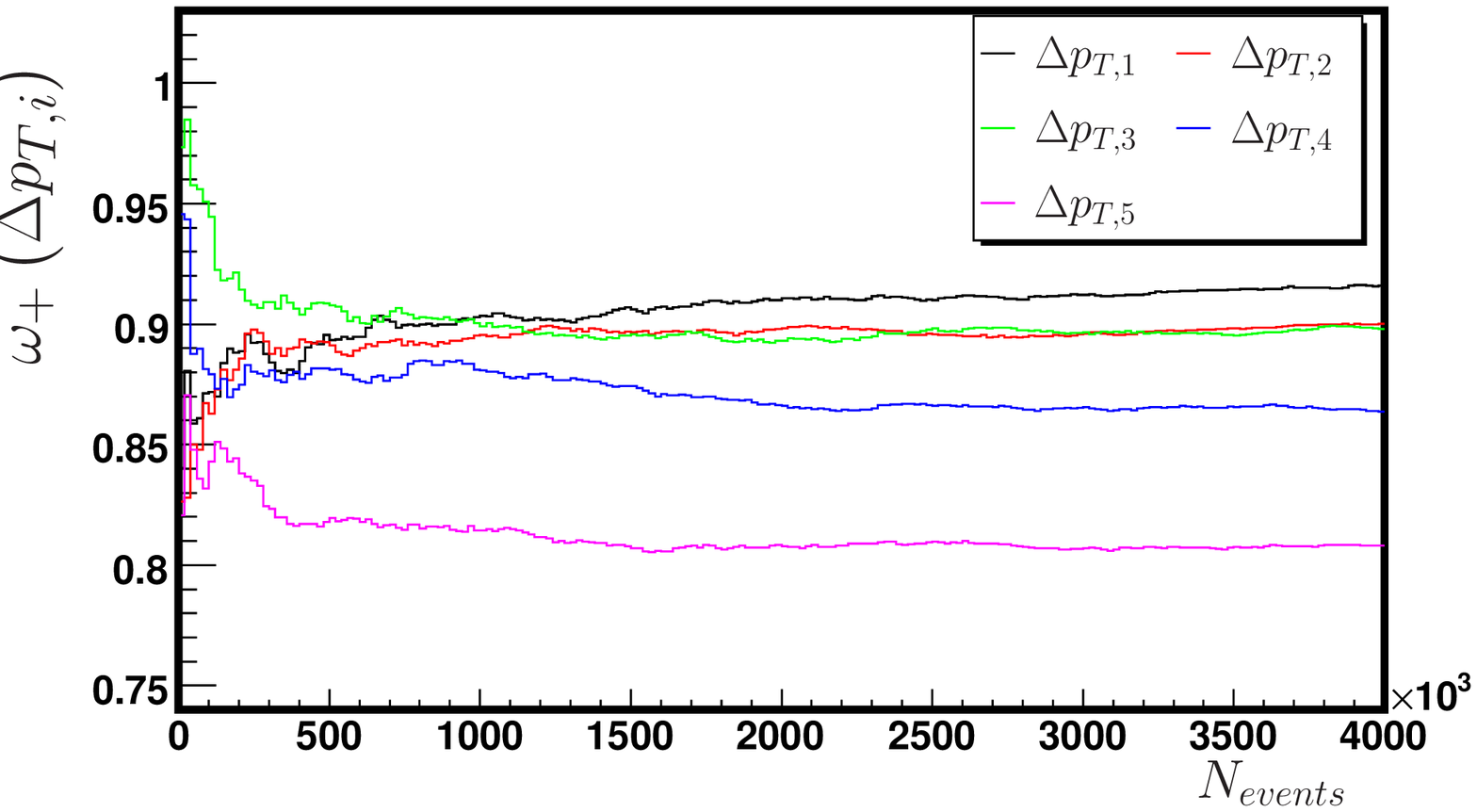,width=7.2cm,height=6.5cm}
  \epsfig{file=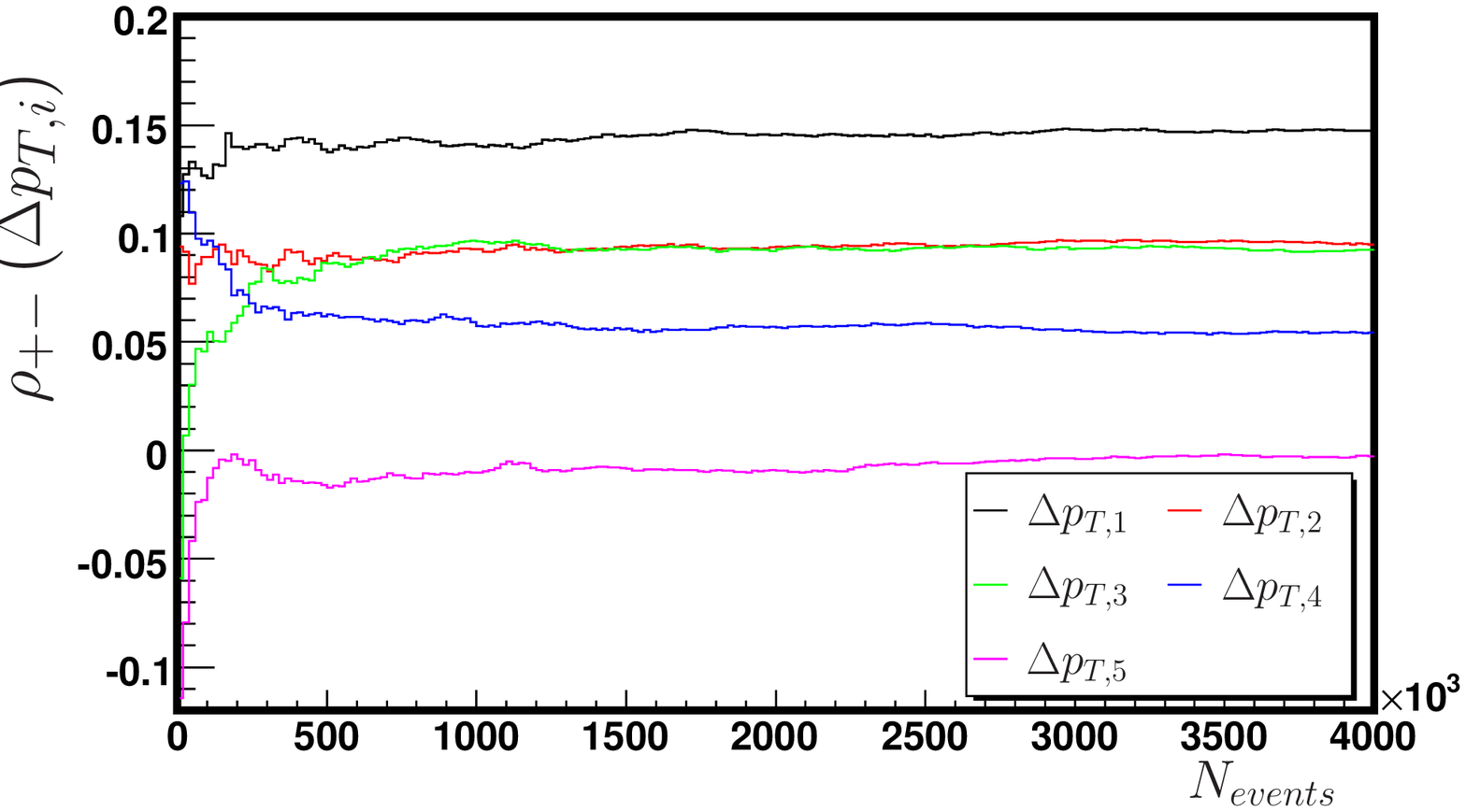,width=7.2cm,height=6.5cm}
  \caption{Step histogram showing the convergence of the scaled variance~$\omega_+$ ({\it left})
    of positively charged hadrons and the correlation coefficient~$\rho_{+-}$ between positively 
    and negatively charged hadrons ({\it right}) in transverse momentum bins~$\Delta p_{T,i}$ for 
    a final state hadron resonance gas with~$\lambda = V_1/ V_g =0.875$.}  
  \label{stephist_omega_rho}
\end{figure}

\begin{figure}[ht!]
  \epsfig{file=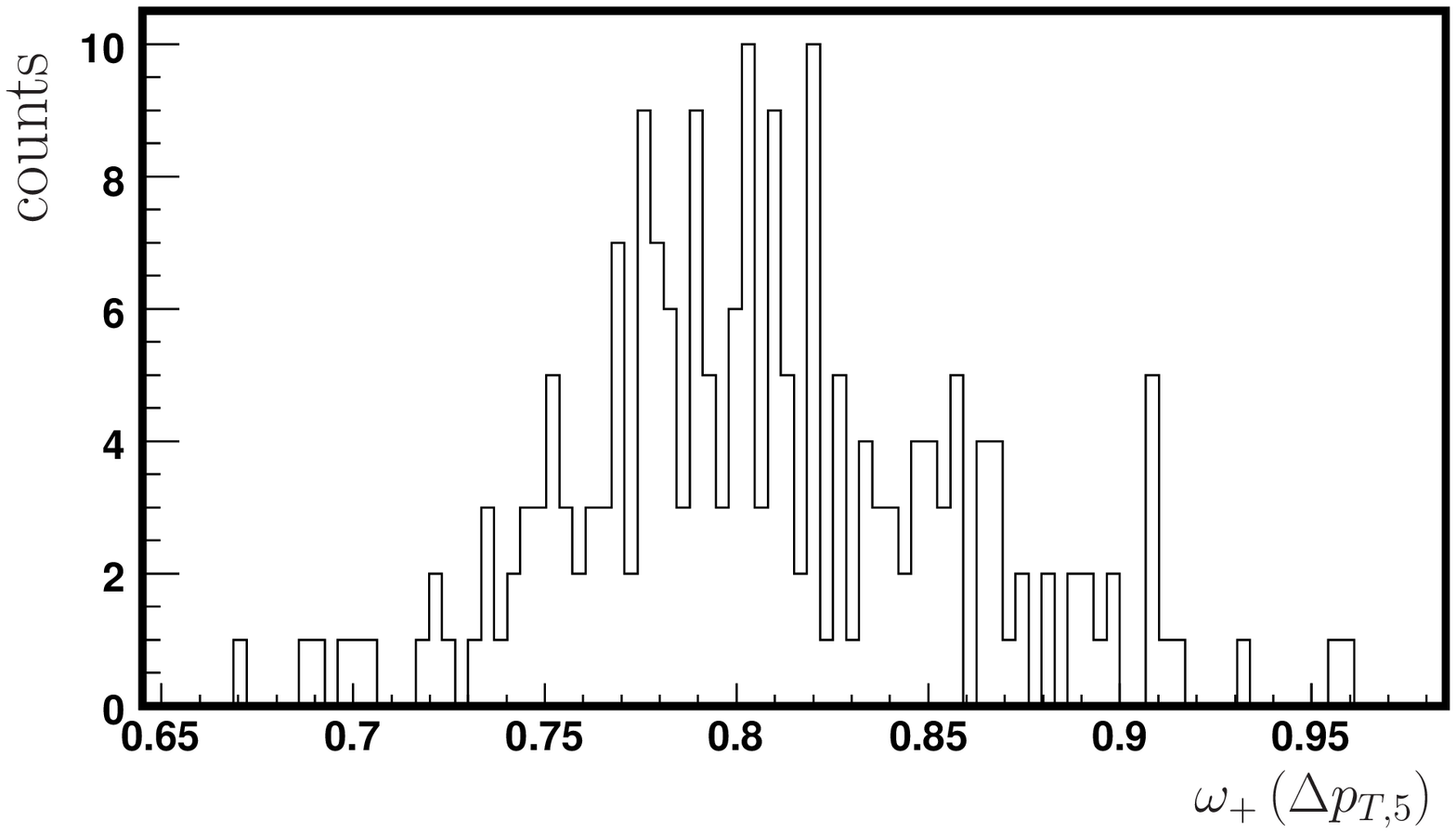,width=7.2cm,height=6.5cm}
  \epsfig{file=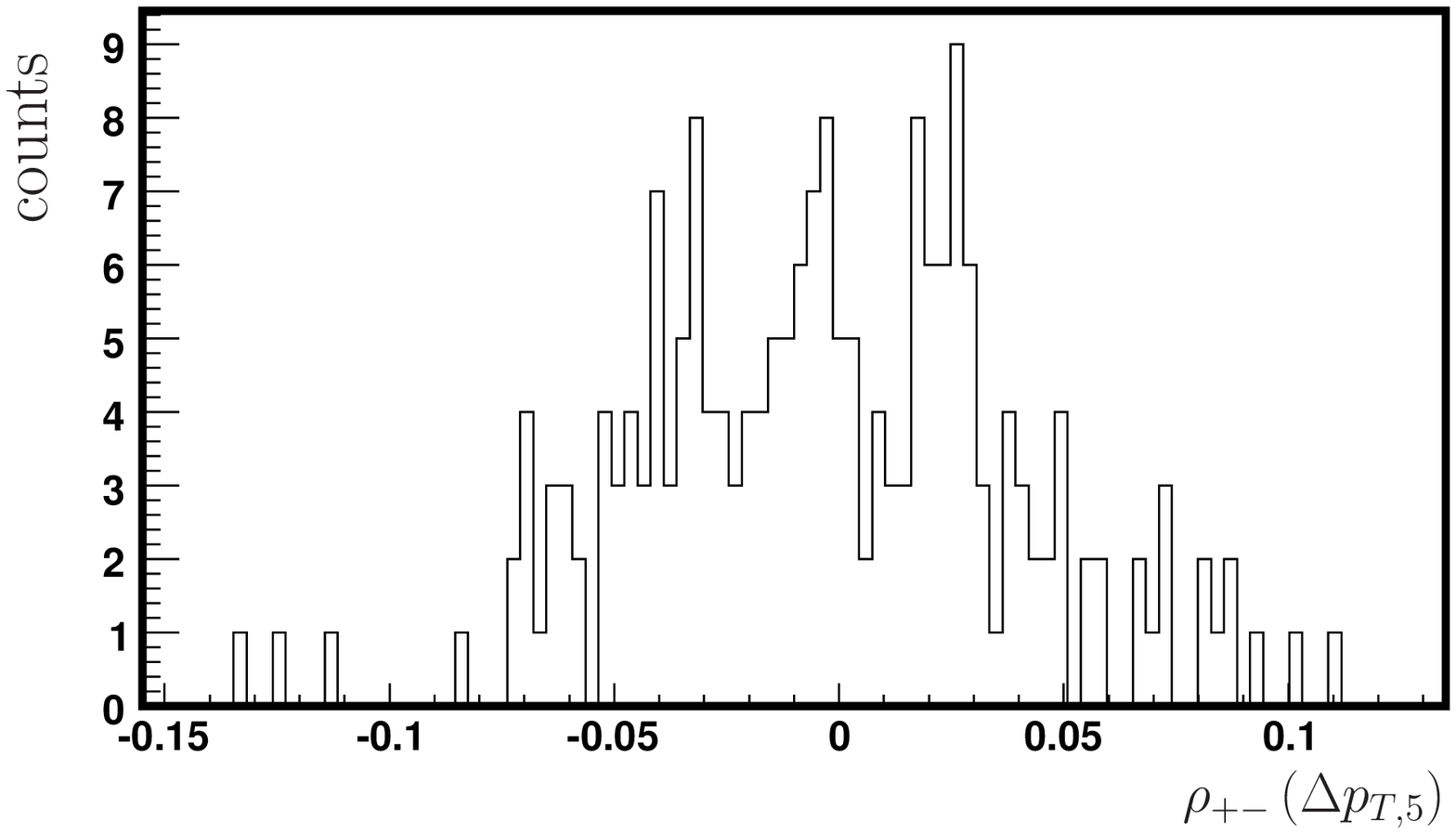,width=7.2cm,height=6.5cm}
  \caption{Histogram showing the results for the scaled variance~$\omega_+$ ({\it left})
    of positively charged hadrons and the correlation coefficient~$\rho_{+-}$ between positively 
    and negatively charged hadrons ({\it right}) in the transverse momentum bin~$\Delta p_{T,5}$ for 
    a final state hadron resonance gas with~$\lambda = V_1/ V_g =0.875$. 
    200 Monte Carlo runs of~$2\cdot 10^4$ events each are analyzed.}  
  \label{stat_error}
\end{figure}

In Fig.(\ref{stephist_N_DeltaNSq}) the evolution of the mean values~$\langle N_+ \rangle$ ({\it left}) 
and the variances~$\langle (\Delta N_+)^2\rangle$ ({\it right}) of the distributions of positively charged 
hadrons for the five transverse momentum bins~$\Delta p_{T,i}$, defined in Table~\ref{accbins}, with the sample 
size is shown. Mean values of particle multiplicities in respective bins are in rather good approximation
equal to each other, but are, however, not identical due to finite resolution on the underlying 
momentum spectrum, even for~$\lambda=0.875$ (bins were constructed using GCE events from an independent run). 
Variances converge steadily and are different in different bins.
The event output was iteratively stored in histograms, which were then evaluated after steps of~$2\cdot10^4$~events. 

In Fig.(\ref{stephist_omega_rho}) the evolution of the scaled variance~$\omega_+$ of positively charged final 
state particles ({\it left}) and the correlation coefficient~$\rho_{+-}$ between positively and negatively 
charged particles ({\it right}) with the sample size is shown.
The results for the respective transverse momentum bins can be compared to the second to last markers 
Figs.(\ref{conv_lambda_omega_final},\ref{conv_lambda_rho_final}) ({\it left}), which denote the corresponding 
results of grouping the same data into~$20$ Monte Carlo sets of~$2 \cdot 10^5$ events each.

\begin{figure}[ht!]
  \epsfig{file=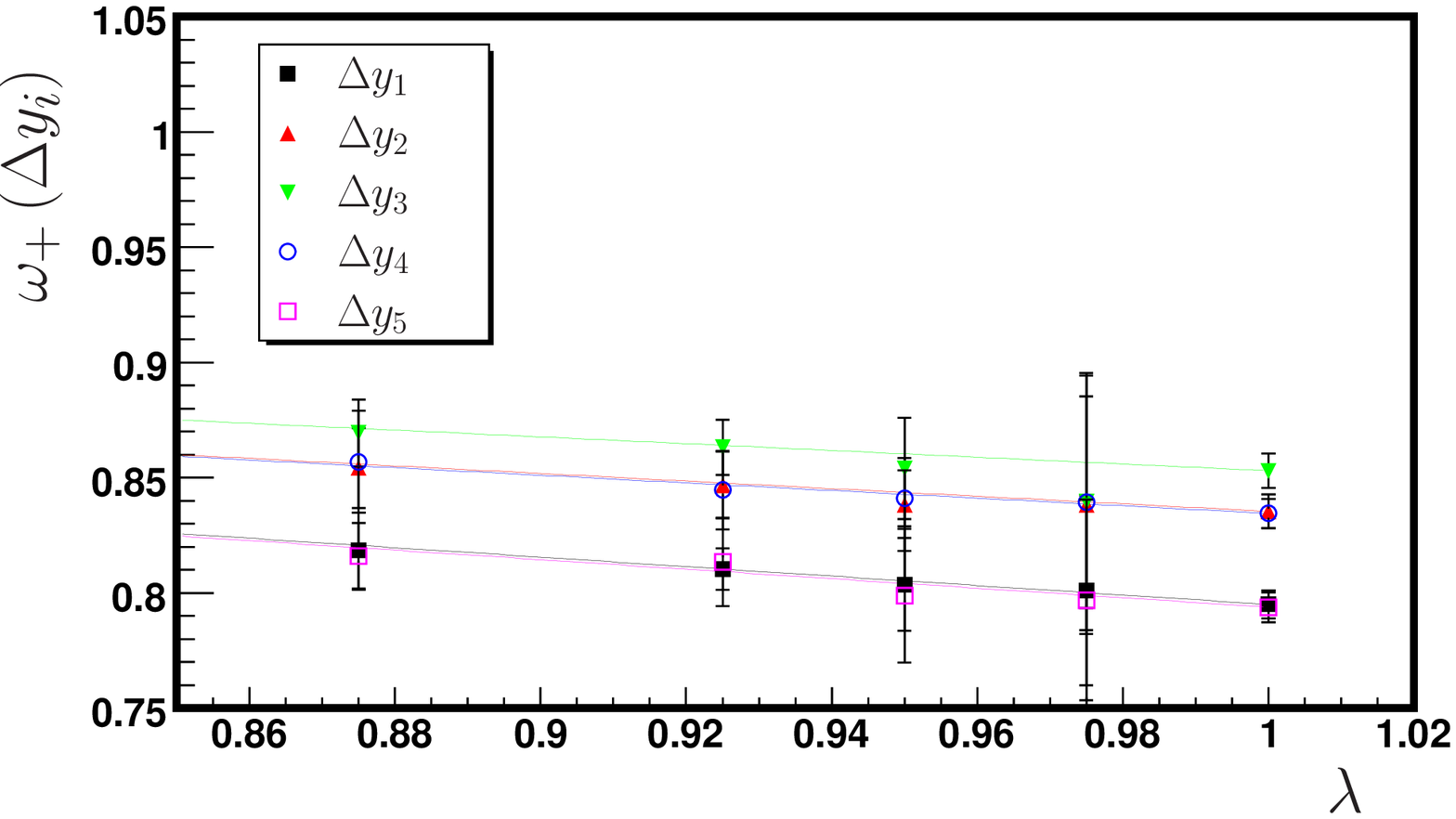,width=7.2cm,height=6.4cm}
  \epsfig{file=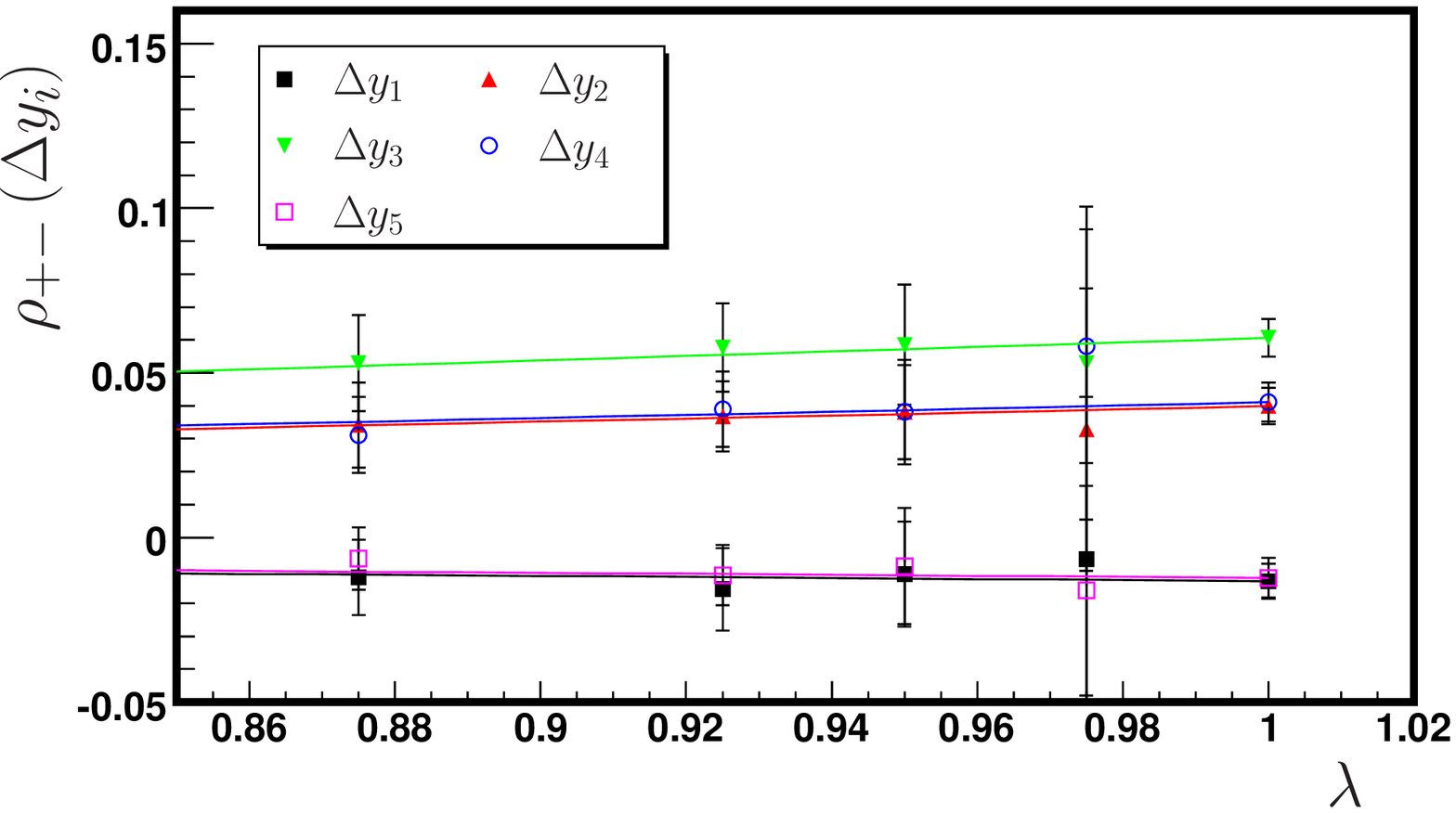,width=7.2cm,height=6.4cm}
  \caption{Evolution of the primordial scaled variance~$\omega_+$ of positively 
    charged hadrons ({\it left}) and the primordial correlation coefficient~$\rho_{+-}$
    between positively and negatively charged hadrons ({\it right}) with the Monte Carlo 
    parameter~$\lambda = V_1/V_g$ in different rapidity bins~$\Delta y_i$. 
    The solid lines show an analytic extrapolation from GCE results ($\lambda =0$)
    to the MCE limit ($\lambda \rightarrow 1$).
    The~$5$ leftmost markers and their error bars represent the results of~$20$ Monte Carlo runs 
    of~$2 \cdot 10^5$ events. 
    Three additional values of~$\lambda$ have been investigated with~$20$ Monte Carlo runs 
    of~$1 \cdot 10^7$ events. 
    The~$5$ rightmost markers denote the results of the extrapolation.
   }  
  \label{largelambda}
\end{figure}

The distribution of scaled variances of positively charged particles~$\omega_+$ ({\it left}) and correlation 
coefficients between positively and negatively charged particles~$\rho_{+-}$ ({\it right}), resulting from 
grouping again the same data set into~$200$ samples of~$2\cdot 10^4$ events each are shown in Fig.(\ref{stat_error}). 
The transverse momentum bin~$\Delta p_{T,5}$ for a final state hadron resonance gas 
with~$\lambda = V_1/ V_g =0.875$ was chosen.

Monte Carlo results for~$\lambda = 0.875$ of the analysis shown in Fig.(\ref{stat_error}), are for the scaled 
variance~$\omega_+ (\Delta p_{T,5}) = 0.8069  \pm 0.0514$, and the correlation 
coefficient~$\rho_{+-} (\Delta p_{T,5}) = -0.0026 \pm 0.0421$. 
They are nicely scattered around the mean values, denoted by the bottom lines in 
Fig.(\ref{stephist_omega_rho}),~$\omega_+ (\Delta p_{T,5}) = 0.8082$, and~$\rho_{+-} (\Delta p_{T,5}) = -0.0028$, 
respectively.

They are also compatible with the analysis shown in Figs.(\ref{conv_lambda_omega_final},\ref{conv_lambda_rho_final}),
of Chapter~\ref{chapter_multfluc_hrg},~$\omega_+ (\Delta p_{T,5}) = 0.8081 \pm 0.0149$, 
and~$\rho_{+-} (\Delta p_{T,5}) = -0.0022 \pm 0.0125$, at the same value of~$\lambda$. 
The comparatively large statistical error on the analysis in Fig.(\ref{stat_error}) is due to the splitting 
up into many small sub-samples. The mean values of different analysis agree rather well.

Lastly, in Fig.(\ref{largelambda}) the results of additional Monte Carlo runs are shown for values 
of~$\lambda$ closer to unity. Here~$20$ additional runs of~$1 \cdot 10^7$ primordial events 
for~$\lambda =0.925$,~$0.950$, and~$0.975$ were performed.
As discussed above, error bars diverge, but convergence seems to be rather good. The additional data has 
not been used for the extrapolation, so it can serve as an un-biased cross-check.

\chapter{The Canonical Boltzmann Gas}   
\label{appendix_cbg}
An analytical and instructive example is the canonical classical relativistic particle anti-particle gas discussed 
in~\cite{Begun:2004zb,Begun:2004gs,Cleymans:2004iu,Cleymans:2005vn}. 
This example is used to show that, although the procedure is formally independent of ones choice of Lagrange 
multipliers, it is most efficient for those defined by Maxwell's relations. Starting off with Eqs.(\ref{eq_one}), 
this section will discuss, in turn, the first and second moments of the multiplicity distribution of particles, and 
the first four moments of the Monte Carlo weight factor. 

The canonical partition function~$Z_{N_1}(V_1,\beta,Q_1)$ of a system with volume~$V_1$, temperature~${T=\beta^{-1}}$, 
charge~$Q_1$, particle number~$N_1$, and anti-particle number~$M_1=N_1-Q_1$,  is given by:
\begin{equation}
Z_{N_1}(V_1,\beta,Q_1) ~=~ \frac{\left( V_1 \psi \right)^{N_1}}{N_1!} 
~ \frac{\left( V_1 \psi \right)^{N_1-Q_1}}{\left( N_1 -Q_1 \right)!}~.
\end{equation}
The single particle partition function in Boltzmann approximation is given by 
Eq.(\ref{psi_meanN}),~${\psi=\frac{g}{2\pi^2}~m^2~\beta^{-1}~K_2\left(m \beta\right)}$. 
The canonical partition function with arbitrary particle number, but still fixed charge~$Q_1$, is obtained by:
\begin{equation}
Z(V_1,\beta,Q_1) ~=~ \sum_{N_1=Q_1}^{\infty} ~ Z_{N_1}(V_1,\beta,Q_1)  
~=~ I_{Q_1} \left(2~V_1~ \psi \right)~.
\end{equation}
Here~$I_{Q_1}$ is a modified Bessel function. Temperature is the same in both subsystems; the bath and 
the observable part. The partition function of the bath is therefore:
\begin{equation}
Z(V_2,\beta,Q_2) ~=~  I_{Q_2} \left( 2~ V_2~\psi \right)~.
\end{equation}
Imposing the constraints~$V_2=V_g-V_1$, and~$Q_2=Q_g-Q_1$, similar to Eq.(\ref{constraint_Q}), one 
finds~\cite{Abramowitz:1965fix_1} for the canonical partition function, Eq.(\ref{PF_combined}), of the combined system:
\begin{equation}
Z(V_g,\beta,Q_g) ~=~ 
\sum_{Q_1= - \infty}^{\infty}  I_{Q_1} \left(2~V_1~ \psi \right) 
~I_{Q_g-Q_1} \Big(2 \left( V_g -V_1\right)\psi \Big) 
~=~ I_{Q_g} \left(2~V_g~ \psi \right)~,
\end{equation}
as required. The weight factor is then:
\begin{equation}\label{appendix_W}
  W(V_1,Q_1;V_g,Q_g|\beta) ~=~ 
  \frac{ I_{Q_g-Q_1} \big( 2 \left(V_g-V_1\right) \psi \big) }{ 
    I_{Q_g}\left(2V_g \psi \right)}~.
\end{equation}
Analogous to Eq.(\ref{basic}) one obtains for the joint particle multiplicity and charge distribution:
\begin{eqnarray}\label{BP_Q1N1_Qg}
P(Q_1,N_1) &=~& W(V_1,Q_1;V_g,Q_g|\beta) ~ Z_{N_1}(V_1,\beta,Q_1)~.
\end{eqnarray}

\subsubsection{Monte Carlo Weight}
Next, introducing Eq.(\ref{Pgce}), the joint GCE distribution of charges and particle multiplicity:
\begin{equation}
P_{gce}(Q_1,N_1)~=~ \frac{e^{Q_1 \mu \beta }}{Z(V_1,\beta,\mu)}
~Z_{N_1}(V_1,\beta,Q_1)~.
\end{equation}
The Monte Carlo weight, Eq.(\ref{simpleW}), is then given by:
\begin{equation}\label{appendix_W_G}
\mathcal{W}^{Q_1;Q_g}(V_1;V_g|\beta,\mu)~\equiv~ 
W(V_1,Q_1;V_g,Q_g|\beta)~Z(V_1,\beta,\mu) ~ e^{-Q_1 \mu \beta}~.
\end{equation}
In accordance with Eq.(\ref{thetrick}), the distribution Eq.(\ref{BP_Q1N1_Qg}) is then equivalently written as:
\begin{eqnarray}\label{BP_Q1N1_Qg_G}
P(Q_1,N_1) &=~& \mathcal{W}^{Q_1;Q_g}(V_1;V_g|\mu,\beta)
~P_{gce}(Q_1,N_1)~.
\end{eqnarray}
The GCE partition function is:
\begin{equation}
Z(V_1,\beta,\mu) ~=~ \sum \limits_{Q_1=-\infty}^{\infty} e^{Q_1 \mu \beta  }
~ Z(V_1,\beta,Q_1) ~=~ \exp \big[ V_1 2 \cosh(\beta \mu) \big]~.
\end{equation}

\subsubsection{Moments of Distributions}
The moments of the multiplicity distributions Eq.(\ref{BP_Q1N1_Qg}) or Eq.(\ref{BP_Q1N1_Qg_G}) are given by:
\begin{equation}\label{mom_N}
\langle N_1^n \rangle ~\equiv~ \sum \limits_{N_1=0}^{\infty} 
\sum \limits_{Q_1=-\infty}^{\infty} ~N_1^n ~P(N_1,Q_1)~.
\end{equation}
Additionally, the moments of the weight, Eq.(\ref{appendix_W}), are defined through:
\begin{equation}\label{mom_W}
\langle W^n \rangle ~\equiv~ \sum \limits_{N_1=0}^{\infty} 
\sum \limits_{Q_1=-\infty}^{\infty} 
~\Big[  W(V_1,Q_1;V_g,Q_g|\beta) \Big]^n ~  Z_{N_1}(V_1,\beta,Q_1)~,
\end{equation}
and of the Monte Carlo weight, Eq.(\ref{appendix_W_G}): 
\begin{equation}\label{mom_gen_W}
\langle\mathcal{W}^n \rangle ~\equiv~ \sum \limits_{N_1=0}^{\infty} 
\sum \limits_{Q_1=-\infty}^{\infty} 
~\Big[ \mathcal{W}^{Q_1;Q_g}(V_1;V_g|\beta,\mu) \Big]^n ~   P_{gce}(Q_1,N_1)~.
\end{equation}
Next, attending to the first two moments of the multiplicity distribution, substituting Eq.(\ref{BP_Q1N1_Qg}), 
or Eq.(\ref{BP_Q1N1_Qg_G}) into Eq.(\ref{mom_N}) yields:
\begin{equation}\label{mean_N}
\langle N_1 \rangle ~=~ \left( V_1 \psi \right) ~ \frac{
  I_{Q_g-1}\left(2V_g \psi \right)}{ I_{Q_g}\left(2V_g \psi \right)}~,
\qquad \textrm{and}
\end{equation}
\begin{equation}\label{N_Sq}
\langle N^2_1 \rangle ~=~ \left( V_1 \psi \right) ~ \frac{
  I_{Q_g-1}\left(2V_g \psi \right)}{ I_{Q_g}\left(2V_g \psi \right)} ~+~
\left( V_1 \psi \right)^2 ~ \frac{ I_{Q_g-2}\left(2V_g \psi \right)}{
  I_{Q_g}\left(2V_g \psi \right)}~.
\end{equation}
Canonical suppression of yields and fluctuations acts on the global volume~$V_g$. In the GCE the first two moments 
are~$\langle N_1 \rangle =  V_1 \psi e^{\mu \beta }$,
and~$\langle N^2_1 \rangle = \langle N_1 \rangle^2 + \langle N_1 \rangle$, respectively.  The CE limit is obtained 
by~$V_g \rightarrow V_1$, and~$Q_g = \langle Q_1 \rangle$. Substituting Eq.(\ref{mean_N}) and Eq.(\ref{N_Sq}) 
into Eq.(\ref{omega}), and using Eq.(\ref{lambda_def}),~$\lambda = V_1/V_g$, yields:
\begin{eqnarray}\label{svar}
\omega &=& \lambda ~ \omega_{ce} +
\left(1-\lambda \right) ~ \omega_{gce}~,
\end{eqnarray} 
where the CE scaled variance~$\omega_{ce}$ of the combined system is given by~\cite{Begun:2004zb,Begun:2004gs}: 
\begin{equation}\label{omega_ce}
\omega_{ce} ~=~ 1 ~-~ \left(V_g \psi \right) ~ \left[ ~
\frac{ I_{Q_g-1}\left(2V_g \psi \right)}{ I_{Q_g}\left(2V_g \psi \right)}
~-~ \frac{ I_{Q_g-2}\left(2V_g \psi \right)}{ I_{Q_g-1}\left(2V_g \psi \right)}
~\right] ~,
\end{equation}
and~$\omega_{gce}=1$ is the GCE scaled variance, as the particle number distribution is a Poissonian. 

The Monte Carlo scheme is now applied to an observable subsystem of volume~$V_1=50$~fm$^3$ embedded into a 
system of volume~$V_g=75$~fm$^3$, charge~$Q_g=10$, and temperature~$T=\beta^{-1}=0.160$~GeV. 
Particles and anti-particles have mass~$m=0.140$~GeV and degeneracy factor~$g=1$. The average charge content in 
the observable subsystem is then~$\langle Q_1 \rangle \simeq 6.667$. The mean particle multiplicity, 
Eq.(\ref{mean_N}), is~$\langle N_1 \rangle \simeq 7.335$, and the scaled variance of particle number fluctuations, 
Eq.(\ref{svar}), is~$\omega \simeq 0.3896$. The GCE in~$V_1$ is sampled for various values of~$\mu_Q$.
The Monte Carlo weight, Eq.(\ref{appendix_W_G}), is then employed to transform these samples to have the statistical 
properties required by Eq.(\ref{BP_Q1N1_Qg}) or Eq.(\ref{BP_Q1N1_Qg_G}). For each value of~$\mu_Q$ fifty samples of
$2000$ events each have been generated to allow for calculation of a statistical uncertainty estimate. 

\begin{figure}[ht!]
  \epsfig{file=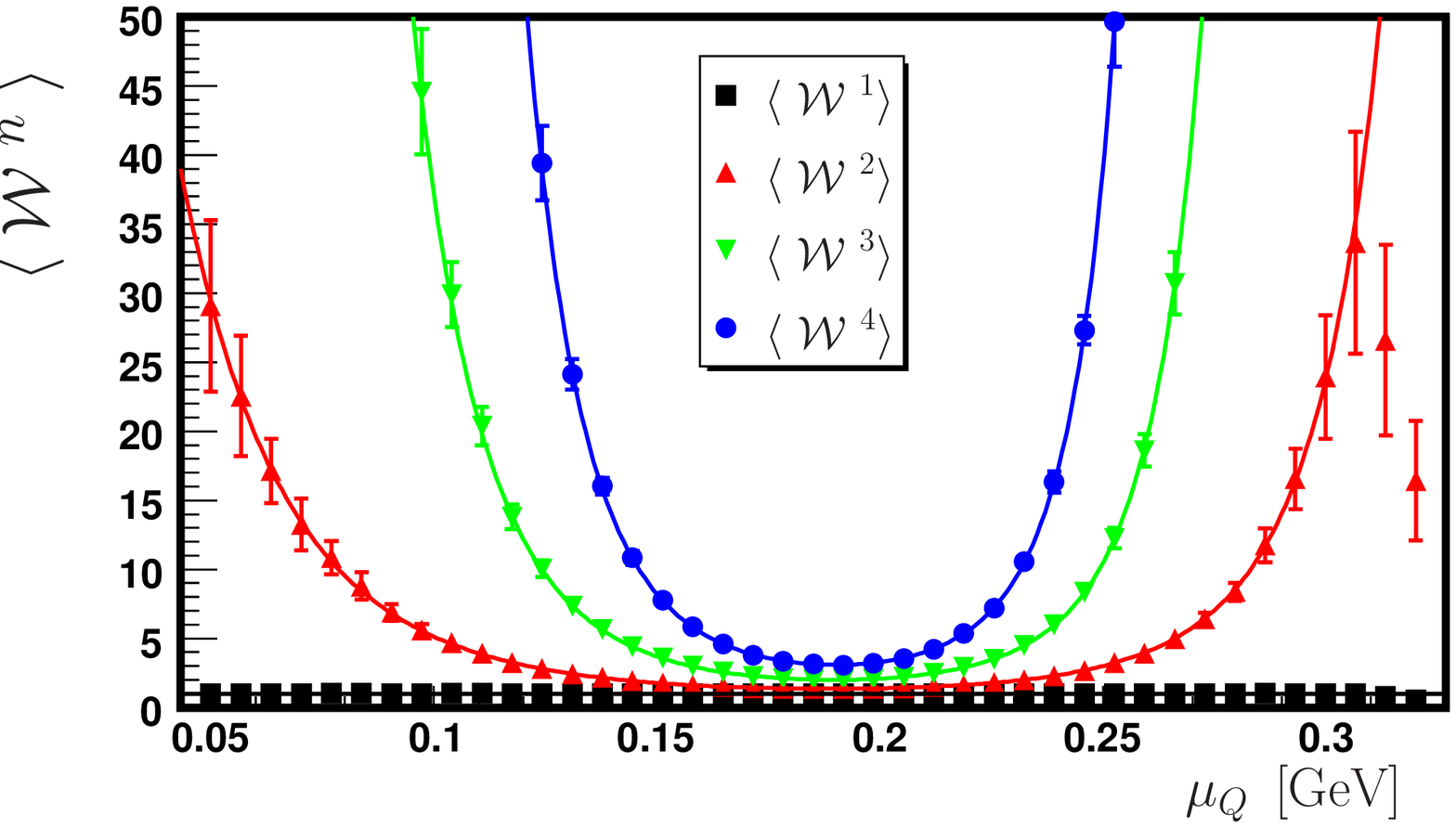,width=7.2cm,height=6.5cm}
  \epsfig{file=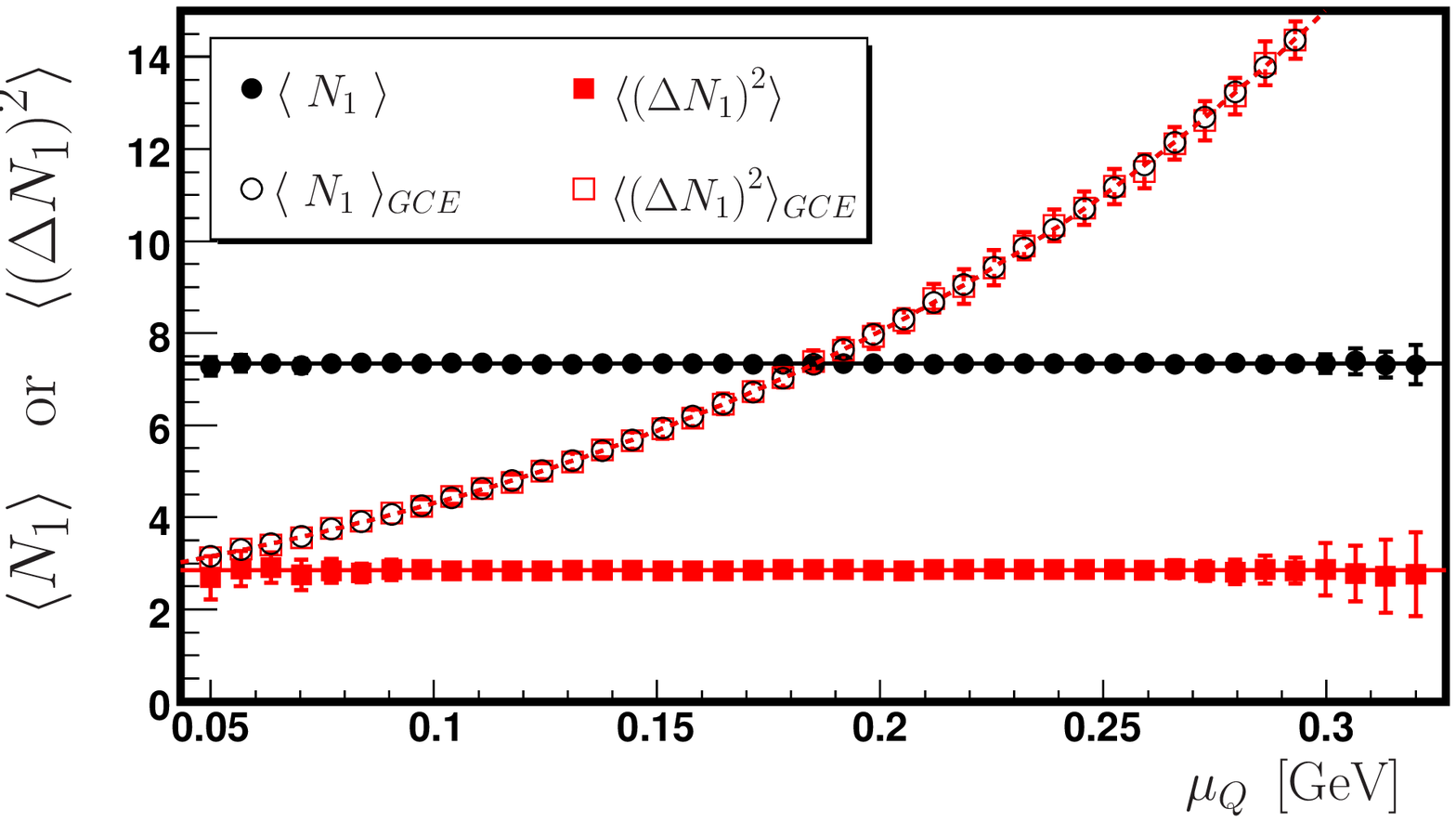,width=7.2cm,height=6.5cm}
  \caption{ The first four moments of the Monte Carlo weight,
    Eq.(\ref{appendix_W_G}) ({\it left}) and the first two moments 
    of multiplicity distributions ({\it right})
    , as described in the text.
   }  
  \label{weight_moments_mu}
\end{figure}

In Fig.(\ref{weight_moments_mu}) ({\it right}) mean value~$\langle N_1 \rangle$ and 
variance~$\langle \left( \Delta N_1 \right)^2 \rangle$ of the particle multiplicity distribution of the 
original GCE samples for different values of chemical potential~$\mu_Q$ are shown in open symbols. 
The closed symbols denote mean value and variance of these samples after the transformation Eq.(\ref{appendix_W_G}) 
was applied. Independent of the original sample the result stays (within error bars) the same. However,
the statistical error is lowest for a chemical potential close to:
\begin{equation}\label{GCE_mu}
\mu_Q ~=~ T ~ \sinh^{-1}\left(\frac{Q_g}{2V_g\psi} \right)~,
\end{equation}
i.e. when the initial sample is already similar (at least in terms of mean values) to the desired sample. 
This is reflected in the moments of the Monte Carlo Weight factor, Fig.(\ref{weight_moments_mu}) ({\it left}). 
Higher moments have a strong minimum around~$\mu_Q=0.1896$~GeV, i.e. the weights are most homogeneously
distributed amongst events, and most efficient used is made of them.

 
 
\newpage 
\phantomsection 
\addcontentsline{toc}{chapter}{Bibliography} 
\bibliographystyle{unsrt-jaakko.bst}
\bibliography{Bibliography.bib} 
               
 
\end{document}